\documentclass[11pt, twoside]{andres}%	UW Thesis

\usepackage{amssymb,latexsym,revsymb,mathbbol}
\usepackage{amsmath,amsbsy}
\usepackage{epsfig,bm}
\usepackage{graphicx,comment}
\usepackage[figuresright]{rotating}

\usepackage{alltt}  % 

\newcommand{\veps}{{\vec\epsilon}}
\newcommand{\vepsprime}{{\vec\epsilon\, '}}
\newcommand{\vsigma}{{\vec\sigma}}

\newcommand{\epp}{\epsilon^\prime}
\newcommand{\xpt}[0]{{$\chi$PT}}
\newcommand{\qxpt}[0]{{Q$\chi$PT}}
\newcommand{\pqxpt}[0]{{PQ$\chi$PT}}
\newcommand{\hbxpt}[0]{{HB$\chi$PT}}
\newcommand{\str}[0]{\rm str}
\DeclareMathOperator{\tr}{tr}

\newcommand{\beq}{\begin{equation}}
\newcommand{\eeq}{\end{equation}}
\newcommand{\bea}{\begin{eqnarray}}
\newcommand{\eea}{\end{eqnarray}}

\newcommand{\nn}{\nonumber}
\newcommand{\benn}{\begin{displaymath}}
\newcommand{\eenn}{\end{displaymath}}
\newcommand{\D}{{\mathcal D}}

\renewcommand{\l}{\left}

\newcommand{\x}{{\mathbf{x}}}

\begin{document}
% Greek Letters
\def\a{{\alpha}}
\def\b{{\beta}}
\def\d{{\delta}}
\def\D{{\Delta}}
\def\e{{\epsilon}}
\def\g{{\gamma}}
\def\G{{\Gamma}}
\def\k{{\kappa}}
\def\l{{\lambda}}
\def\L{{\Lambda}}
\def\m{{\mu}}
\def\n{{\nu}}
\def\w{{\omega}}
\def\O{{\Omega}}
\def\S{{\Sigma}}
\def\s{{\sigma}}
\def\t{{\tau}}
\def\th{{\theta}}
\def\x{{\xi}}

\def\ol#1{{\overline{#1}}}

%slash's
\def\Dslash{D\hskip-0.65em /}
\def\dslash{{\partial\hskip-0.5em /}}
\def\vslash{{\rlap \slash v}}
\def\qbar{{\overline q}}

% Jargon
\def\CPT{{$\chi$PT}}
\def\QCPT{{Q$\chi$PT}}
\def\PQCPT{{PQ$\chi$PT}}
\def\tr{\text{tr}}
\def\str{\text{str}}
\def\diag{\text{diag}}
\def\order{{\mathcal O}}
\def\vit{{\it v}}
\def\vD{\vit\cdot D}
\def\am{\alpha_M}
\def\bm{\beta_M}
\def\gm{\gamma_M}
\def\smb{\sigma_M}
\def\smt{\overline{\sigma}_M}
\def\tb{{\tilde b}}

\def\mc#1{{\mathcal #1}}

% Fields
\def\Bbar{\overline{B}}
\def\Tbar{\overline{T}}
\def\cBbar{\overline{\cal B}}
\def\cTbar{\overline{\cal T}}
\def\pq{(PQ)}

\def\eqref#1{{(\ref{#1})}}

% Useful Math
\def\loarrow#1{\overleftarrow{#1}}
\def\roarrow#1{\overrightarrow{#1}}

% ==========   Preliminary pages
%

\prelimpages
 
%
%
%	title page
%
%
\Title{{\bf{\Huge Topics in Effective Field Theory \\
	for \\
	Lattice QCD}}}
\Author{ {\Large Andr\'{e} Walker-Loud}}
\Year{May 2006}
\Program{Physics}

{\Degreetext{A dissertation submitted in partial fulfillment of\\
  the requirements for the degree of}}
\titlepage%	comment me out for arXiv

%\setcounter{footnote}{0}
 
%
%
%	signature page
%
%

\Chair{Martin J. Savage}{Professor}{Physics}

%\Signature{David B. Kaplan}
%\Signature{Martin J. Savage}
%\Signature{Stephen R. Sharpe}
%\signaturepage%	comment me out for arXiv

%
%
%	quoteslip
%
%
%\setcounter{page}{-1}%	comment me out for arXiv
%  \thesisquoteslip
%\doctoralquoteslip%	comment me out for arXiv
%  \doctoralabstractquoteslip

%
% ----- abstract
%

\setcounter{page}{-1}
\abstract{%

Quantitatively understanding hadronic physics from first principles requires numerical solutions to QCD, known as lattice QCD, which are necessarily performed with a finite spatial extent at finite lattice spacing.  Current simulations employ quark masses that are heavier than those in nature, as lighter quarks are computationally too costly, the physical systems simulated are not significantly larger than the lattice spacing nor are they significantly smaller than the lattice size.  These approximations, or lattice artifacts, modify the observable quantities of interest.  To make a rigorous connection between the physical world and current lattice QCD simulations, these lattice artifacts must be understood.  The tool to systematically understand these effects is effective field theory.

In this work, we extend and apply effective field theory techniques to systematically understand a subset of these lattice artifacts in addition to the underlying physics of interest.  Where possible, we compare to existing lattice QCD.  In particular, we extend the heavy baryon Lagrangian to the next order in partially quenched chiral perturbation theory and use it to compute the masses of the lightest spin-1/2 and spin-3/2 baryons to next-to-next-to leading order.  We then construct the twisted mass chiral Lagrangian for baryons and apply it to compute the lattice spacing corrections to the baryon masses simulated with twisted mass lattice QCD.  

We extend computations of the nucleon electromagnetic structure to account for finite volume effects, as these observables are particularly sensitive to the finite extent of the lattice.  We resolve subtle peculiarities for lattice QCD simulations of polarizabilities and we show that using background field techniques, one can make predictions for the 4 spin-dependent nucleon polarizabilities, quantities which are difficult to access experimentally.

We then discuss the two-pion system in finite volume, determining the exponentially small volume corrections necessary for lattice determinations of the scattering parameters.  We also determine the lattice spacing artifacts that arise for a mixed-action lattice simulation of the two-pion system with Ginsparg-Wilson valence quarks and staggered sea quarks.  We show that the isospin 2 scattering length has a near continuum like behavior, differing from the chiral perturbation theory calculation by a computable difference.
}

%
% ----- contents & etc.
%
\tableofcontents
\listoffigures
\listoftables
 
%
% ----- glossary 
%
\chapter*{Glossary}\label{glossary}
\addcontentsline{toc}{chapter}{Glossary}
\thispagestyle{plain}
%
%\chapter*{glossary}\label{glossary}

\begin{glossary}

%
%	A
%
%\item[LQCD] lattice QCD
%
%	B
%
%\item[LQCD] lattice QCD
%
%	C
%
\item[\textbf{$\chi$PT}] chiral perturbation theory
%
%	D
%
%\item[LQCD] lattice QCD
%
%	E
%
%\item[LQCD] lattice QCD
%
%	F
%
\item[\textbf{FV}] finite volume
%
%	G
%
%\item[LQCD] lattice QCD
%
%	H
%
\item[\textbf{HADRON}] composite particle made of quarks and gluons
\item[\textbf{HB$\chi$PT}] heavy baryon chiral perturbation theory
%
%	I
%
%\item[\textbf{IV}] infinite volume
%
%	J
%
%\item[LQCD] lattice QCD
%
%	K
%
%\item[LQCD] lattice QCD
%
%	L
%
\item[\textbf{LEC}] low energy constant; the coefficients of operators in the effective field theory
\item[\textbf{LQCD}] lattice QCD
\item[\textbf{LO}] leading order
%
%	M	
%
\item[\textbf{MA}] mixed action
%
%	N
%
\item[\textbf{NDA}] naive dimensional analysis
\item[\textbf{NLO}] next-to leading order
\item[\textbf{NNLO}] next-to-next-to leading order
%
%	O
%
%\item[LQCD] lattice QCD
%
%	P
%
\item[\textbf{PQ$\chi$PT}] partially quenched chiral perturbation theory
%
%	Q	
%
\item[\textbf{QCD}] quantum chromodynamics
\item[\textbf{Q\CPT}] quenched chiral perturbation theory
%
%	R
%
\item[\textbf{RPI}] reparameterization invariance
%
%	S	
%
%\item[QCD] quantum chromodynamics
%
%	T	
%
\item[\textbf{tmLQCD}] twisted mass lattice QCD
\item[\textbf{tm$\chi$PT}] twisted mass chiral perturbation theory
%
%	U	
%
%\item[QCD] quantum chromodynamics
%
%	V	
%
%\item[QCD] quantum chromodynamics
%
%	W	
%
%\item[QCD] quantum chromodynamics
%
%	X	
%
%\item[QCD] quantum chromodynamics
%
%	Y	
%
%\item[QCD] quantum chromodynamics
%
%	Z	
%
%\item[QCD] quantum chromodynamics

\end{glossary}

%
%	Dedication
%
\dedication{I would like to dedicate this thesis to my advisor, Martin Savage, who has continuously  pushed and encouraged me.  Under his guidance these last few years, my curiosity about physics has turned into a real passion (obsession).  From him I have learned something I have always known; there is no substitute for hard work, dedication and self punishment ... I mean discipline.}

%
% ----- acknowledgments
%
\acknowledgments{I would like to begin by acknowledging the members of the Nuclear Theory group at the University of Washington for providing a very stimulating, encouraging and supportive environment to do research.  I have had the best experience here and would not trade it for anything.  I would also like to thank the members of my reading committee, for agreeing to read this thesis, and mostly because there doors were always open for my questions, an opportunity I took many advantages of.  I also thank the two post-docs across the hall who have really had no end of my questions, and provided me with the needed support and beer to get through longer days.  At least I can say I have given them some form of entertainment and cheap coffee.  And I also want to thank all the other members of the physics department at the University of Washington for making this a great place to do physics and hang out.

I would also like to thank my family, who have always helped me stay a little more grounded than otherwise, and who have also been very understanding as I have been only 5 miles away, but often absent.  I also want to thank all the members of the Settle JuJutsu Club who have provided a great means to relieve stress and help me stay sane.

I also would like to thank all of my collaborators, who have been great friends as well as research partners.  The research has definitely been more exciting being able to share it with other people who are so enthusiastic to do physics.

And last but not least, I would like to thank Erin B. Warnock for all of her motivational impertinent counsel which has kept me going these last few years.}

%
% end of the preliminary pages

%\input{arXiv_prelim}

%
% ==========      Text pages
%

\textpages

% ========== Chapter 1: LQCD and EFT
 \chapter{Lattice QCD and Effective Field Theory}\label{chap:NPLQCD}

\pagenumbering{arabic}
\setcounter{page}{1}

Quantum chromodynamics (QCD) is the fundamental theory of the nuclear strong force, describing the interactions between quarks and gluons, which bind into the hadrons we observe in nature; the proton, the neutron, the pions \textit{etc.}  QCD is a non-Abelian gauge theory~\cite{Yang:1954ek} based upon the $SU(3)$ \textit{color} group~\cite{Fritzsch:1972jv,Fritzsch:1973pi}.  The non-Abelian nature of the theory gives it the property of \textit{asymptotic freedom}~\cite{Politzer:1973fx,Gross:1973id}, which is  necessary to describe the observed \textit{Bjorken} scaling~\cite{Bjorken:1969ja,Panofsky:1968pb} in deep-inelastic scattering processes, in which at high momentum transfer the individual \textit{partons}, or hadron constituents, act like nearly free, point-like objects~\cite{Feynman:1969ej}.  Conversely, at lower energy scales relevant to nuclear physics, $\mu \sim 1$~GeV, the coupling between the quarks and gluons becomes strong, $\a_s(\mu) \sim \mc{O}(1)$, and our well established perturbative treatment of a quantum field theory (QFT) breaks down.  Currently, we still lack a closed-form solution to QCD, making a rigorous description of the nuclear physics governing the world around us directly from QCD difficult, to say the least.

A conceptual breakthrough in our understanding of field theories~\cite{Wilson:1974sk} has lead to a lattice formulation of quantum field theory, in which spacetime is treated discretely instead of continuously.  This technique naturally lends itself to numerical solutions which can be applied to the non-perturbative regime of QCD, and in fact, lattice QCD (LQCD) is currently our only known solution to this non-perturbative physics.  With this powerful technique, we can ask a number of previously unanswerable questions about the world around us.  It is now well accepted that QCD gives rise to quark confinement, as evidenced by the lack of any direct observation of isolated quarks and also by the quark potential plots determined with lattice QCD.   Lattice techniques have also allowed us to gain at least a qualitative understanding of the hadron spectrum we observe experimentally.  However, we would also like to know if we can use QCD to explain the structure of hadrons?  Does QCD contribute to the observed CP violations in the universe (in addition to the known CP violations of the \textit{electroweak} theory~\cite{Fermi:1934sk,Fermi:1934hr,Lee:1956qn,Wu:1957my,Feynman:1958ty,Sudarshan:1958vf,Sakurai:1958,Glashow:1961tr,Weinberg:1967tq,Salam:1964ry,Higgs:1964ia,Higgs:1964pj,Higgs:1966ev,Arnison:1983rp,Arnison:1983mk,'tHooft:1971rn,'tHooft:1972fi}), or do we need a CP violating extension to the Standard Model?  Can we understand the formation and structure of the nuclei, \textit{deuterium} to \textit{uranium}, from the fundamental parameters of the QCD Lagrangian?  These are just a few examples of many interesting questions which are actively being pursued with LQCD.  It has only been in these last few years that we have seen serious quantitative progress being made to answer these questions, which has largely been possible through the advances in available computing power.

The QCD Lagrangian, along with the electroweak interactions, is responsible for all of hadronic physics, and thus in principle expressible as a function of only 4 fundamental QCD parameters, $f(\L_{QCD},m_u,m_d,m_s)$, along with the fundamental electroweak parameters.  Here, $\L_{QCD}$ is the dynamically generated QCD scale where roughly speaking the QCD coupling becomes strong.  The light quark masses, which dominate the infrared hadronic physics, are given by $m_u , m_d$ and $m_s$ (the other three quarks are heavy enough that they can be integrated out of the theory).  Lattice QCD, in principle, is a tool we can use to numerically determine these functions, at least for systems with a small number of hadrons.   LQCD simulations turn out to be very computationally demanding, requiring some of the world's fastest super computers.  Even still, approximations must be made to perform simulations in a finite amount of time.  

All LQCD simulations are necessarily performed with a finite spatial extent (finite box) and at finite lattice spacing.  Typically, the box sizes are not significantly larger than the physical system of interest, the lattice spacing is not significantly smaller than the physical system, and currently, the quark masses are larger than those of nature.  These approximations, or lattice artifacts, modify the observable quantities of interest.  Therefore, to make a rigorous connection between the physical world and current lattice QCD simulations, these lattice artifacts must be understood.  The tool to systematically understand these effects is effective field theory (EFT).  In this thesis, we contribute formal steps in the direction of understanding hadronic physics from lattice QCD.  We accomplish this by developing, extending and applying various effective field theory techniques to hadronic observables.  Much of this work has been published previously and we collect it here in three categories.

%%%%%%%%%%%%%%%%%%%%%%%%%%%%%%%%%%%%%%%%%%%%%%%%%%%
%
%  Baryon Masses
%
%%%%%%%%%%%%%%%%%%%%%%%%%%%%%%%%%%%%%%%%%%%%%%%%%%%
\subsection*{Baryon Masses}

Using what is known as heavy baryon chiral perturbation theory (HB\CPT)~\cite{Jenkins:1990jv,Jenkins:1991ne}, one can compute the masses of the lowest lying spin-$\frac{1}{2}$ and spin-$\frac{3}{2}$ baryons, as functions of the light quark masses.  In this way, one can fit the derived mass formulae to lattice computations of the baryon spectrum, and determine the mass of the nucleon as an explicit function of the quark masses.  Why do we want to know the mass of the nucleon from LQCD?  In addition to providing information about HB\CPT, it turns out that knowing the mass of the nucleon as a function of the quark masses, in particular the strange quark mass, allows one to constrain the possibility of kaon or hyperon condensation in neutron stars~\cite{Kaplan:1986yq,Nelson:1987dg}.  Additionally, the expansion parameter of HB\CPT\ is not as small as that of purely mesonic chiral perturbation theory (\CPT), and it is therefore important to push the determination of the baryon masses to higher orders so one can test the convergence of HB\CPT.  To this end, we extend the heavy baryon Lagrangian to next-to-next-to leading order (NNLO) in HB\CPT\ as well as extensions of this theory for LQCD, and compute the quark mass dependence of the spin-$\frac{1}{2}$ and spin-$\frac{3}{2}$ baryons to 
$\mc{O}(m_q^2)$.  This work has been published;

\begin{itemize}
\item
Andr\'{e} Walker-Loud,
{\it Octet baryon masses in partially quenched chiral perturbation theory},
Nucl.\ Phys.\ A {\bf 747} (2005) 476.

\item
Brian C.\ Tiburzi and Andr\'{e} Walker-Loud,
{\it Decuplet baryon masses in partially quenched chiral perturbation theory},
Nucl.\ Phys.\ A {\bf 748} (2005) 513.

\item
Brian C.\ Tiburzi and Andr\'{e} Walker-Loud,
{\it Strong isospin breaking of the nucleon and delta masses on the lattice},
Nucl.\ Phys.\ A {\bf 764} (2006) 274.

\end{itemize}

One exciting new lattice discretization technique is known as \textit{twisted mass LQCD}~\cite{Frezzotti:2000nk}.  We construct the \textit{twisted mass} heavy baryon Lagrangian which incorporates the lattice artifacts arising from the twisted mass lattice action.  We then apply this Lagrangian to compute the effects on the nucleon and delta spectrum arising from the twisting.  We also discuss the subtle but very interesting consequences of performing twisted mass LQCD simulations which incorporate both isospin violation in the up and down quark masses as well as the twisted mass effects.  This work has been published;

\begin{itemize}

\item
Andr\'{e} Walker-Loud and Jackson M.\ S.\ Wu,
{\it Nucleon and delta masses in twisted mass chiral perturbation theory},
Phys.\ Rev.\ D {\bf 72}, 014506 (2005).

\end{itemize}

%%%%%%%%%%%%%%%%%%%%%%%%%%%%%%%%%%%%%%%%%%%%%%%%%%%
%
%  Nucleon EM structure
%
%%%%%%%%%%%%%%%%%%%%%%%%%%%%%%%%%%%%%%%%%%%%%%%%%%%
\subsection*{Electromagnetic Structure of the Nucleon}

The structure of the nucleon is interesting in its own right, from the nucleon parton~\cite{Feynman:1969ej} distributions, to the spin content of the nucleon~\cite{Ji:1996ek,Ji:1996nm} to the electromagnetic moments and polarizabilities of the nucleon.  Each of these is a measure of the content and structure of the nucleon, as well as a necessary building block to understanding the structure of nuclei.  In this work we show how to determine the polarizabilties of the nucleon from lattice QCD calculations using background field techniques.  We show that the polarizabilities are particularly sensitive to the finite size of the box used in the simulation.  We address other subtleties which arise for LQCD determinations of the nucleon polarizabilities and additionally, we show how LQCD in conjunction with effective field theory can be used to predict the spin polarizabilities of the 
nucleon~\cite{Ragusa:1993rm,Guichon:1995pu,Holstein:1999uu}, which are difficult to access experimentally (in fact, there are 4 spin polarizabilities, and currently only two linear combinations of them are measured).
This work has been published;

\begin{itemize}

\item
William Detmold, Brian C.\ Tiburzi and Andr\'{e} Walker-Loud,
{\it  Electromagnetic and spin polarisabilities in lattice QCD},
to be published in Phys.\ Rev.\ D (2006).

\end{itemize}

%%%%%%%%%%%%%%%%%%%%%%%%%%%%%%%%%%%%%%%%%%%%%%%%%%%
%
%  Two-Hadron Interactions
%
%%%%%%%%%%%%%%%%%%%%%%%%%%%%%%%%%%%%%%%%%%%%%%%%%%%
\subsection*{Two-hadron interactions}

To understand hadronic physics from LQCD, in particular nuclear physics, one must understand multi-hadron systems with LQCD.  A first step in this direction is the two-hadron system and in this work we focus on the two-pion system, which is both formally and numerically easier than the two-nucleon system.  We study both the volume dependence as well as the lattice spacing dependence of the two-pion system.  An elegant method has been developed for extracting two-particle infinite volume scattering parameters from finite volume energy levels of the two-particle system~\cite{Hamber:1983vu,Luscher:1986pf,Luscher:1990ux,Maiani:1990ca,Rummukainen:1995vs,Beane:2003yx,Beane:2003da,Kim:2005gf}.  We compute the exponentially suppressed volume corrections to this relation for the two-pion system and show that these effects will become important in the next generation of lighter pion masses simulated with lattice QCD.  We also compute the lattice spacing and partial quenching corrections to the isospin 2 ($I=2$), two-pion scattering length, in particular for a mixed-action simulation employing Ginsparg-Wilson valence quarks~\cite{Ginsparg:1981bj} and staggered sea quarks~\cite{Susskind:1976jm}.  We demonstrate that a very nice cancellation of lattice spacing and partial quenching effects occurs if one expresses the scattering length in the effective field theory in terms of correlation functions one measures in the lattice QCD simulations instead of in terms of the bare parameters of the theory.  These results explain the remarkable success of the recent determination of the $I=2$ $\pi\pi$ scattering length from a dynamical mixed action lattice QCD simulation~\cite{Beane:2005rj}.

Finally, based upon the insights gained from determining the lattice spacing and partial quenching effects mentioned above, we demonstrate the utility of mixed-action simulations employing Ginsparg-Wilson valence quarks.  In particular, we show that through the one-loop order in the effective theory, the lattice spacing effects which arise can all be absorbed by multiplicative redefinitions of the coefficients of the continuum theory.  We also demonstrate how at the next order in the chiral expansion, this renormalization scheme breaks down, and in particular, for staggered sea quarks, we show how the \textit{taste}-breaking enters these mixed action simulations.  Most of this work has been previously published, however the final chapter of this thesis is work in progress;
\begin{itemize}
\item
Paulo F. Bedaque, Ikuro Sato and Andr\'{e} Walker-Loud,
{\it  Finite volume corrections to $\pi\pi$ scattering},
Phys.\ Rev.\ D {\bf 73}, 074501 (2006).
\item
Jiunn-Wei Chen, Donal O'Connell, Ruth S. Van de Water and Andr\'{e} Walker-Loud,
{\it  Ginsparg-Wilson pions scattering in a sea of staggered quarks},
Phys.\ Rev.\ D {\bf 73}, 074510 (2006).
\item
Jiunn-Wei Chen, Donal O'Connell and Andr\'{e} Walker-Loud,
{\it  Mixed-action simulations with Ginsparg-Wilson valence quarks},
to be published.
\end{itemize}

%%%%%%%%%%%%%%%%%%%%%%%%%%%%%%%%%%%%%%%%%%%%%%%%%%%
%
%	LQCD
%
%%%%%%%%%%%%%%%%%%%%%%%%%%%%%%%%%%%%%%%%%%%%%%%%%%%
\section{Lattice QCD \label{sec:LQCD}}

Here we do not pretend to give a thorough introduction or review of lattice QCD.  We aim only to review some key concepts and introduce common jargon useful to understanding this work.  There are numerous reviews and books describing lattice QCD (and lattice QFT) in the literature.  A few references the author finds particularly useful are~\cite{Sharpe:1993wt,Rothe:1997kp}.

As we mentioned in the previous section, lattice QCD (or more generally lattice field theory), is a powerful numerical technique which allows us to study the non-perturbative regime of field theories, in particular, low-energy QCD.  Lattice theories are formulated with the path integral formulation~\cite{Feynman:1948ur} of quantum field theories.  The generating function for QCD is
\begin{equation}
	\mc{Z} = \int\, D A_\mu\, D q\, D \ol{q}\ \textrm{exp}
		\left( i S_G [A_\mu] + i \int d^4 x\, \ol{q} \left( i\Dslash - m_q \right) q \right)\, ,
\end{equation}
where $q$ is a quark field, $m_q$ is the quark mass matrix, $A_\mu$ are the gluons, $S[A_\mu]$ is the standard Yang-Mills actions, $\ol{q}\, \Dslash\, q$ is the gauge-interaction between the quarks and gluons, and the functional integral is performed over the infinite set of possible field configurations of the quarks and gluons.  The idea is then to discretize spacetime, which regularizes the theory by providing a natural \textit{ultraviolet} (UV) cut-off set by the inverse lattice spacing, $a^{-1}$.  Additionally, one provides a finite spatial extent to the universe, with some user specified boundary conditions (most commonly periodic), and this then provides a natural \textit{infrared} (IR) cut-off of the theory, set by the inverse box size, $L^{-1}$.  In this way, the generating function is now a finite dimensional integral and one can think about using computers to evaluate it.  Unfortunately, this is still not possible as there are an infinite set of field configurations, and in Minkowski space, the integral is performed over the phase of an exponential, and thus every configuration is important, with large cancellations happening amongst configurations which are far from the classical action.

To get around this problem, we can Wick rotate to Euclidean space, $t \rightarrow -i\t$, such that the functional integral becomes an integral over a real valued function, which is now weighted by exponentially damped field configurations away from the field configurations which give rise to the minimum of the action,
\begin{equation}\label{eq:EuclideanPathInt}
	\mc{Z}_E = \int\, D A_\mu\, D q\, D \ol{q}\ \textrm{exp}
		\left( -S_G [A_\mu] -\int d^4 x\, \ol{q} \left( \Dslash + m_q \right) q \right)\, .
\end{equation}
One can then apply \textit{Monte Carlo} techniques to determine a set of field configurations which give the dominant contributions to Eq.~\eqref{eq:EuclideanPathInt}, and in this way limit the number of field configurations needed to estimate physical quantities.  To do this, one first does the integral over the quark degrees of freedom and uses the resulting fermionic determinant as part of the weighting in the Monte Carlo updating routines.  In this way, the expectation value of some physical observable is given by
\begin{equation}
	\langle\ \mc{O}\ \rangle = \frac{1}{N_{configs}} \sum_i^{N_{configs}}\ \mc{O}[U_i]\, ,
\end{equation}
where $\mc{O}[U_i]$ is a functional of the latticized gluon gauge fields, $U_i$, and the selection of these gauge fields is weighted by
\begin{equation*}
	\textrm{Det} \left[ \Dslash +m_q \right]\ \textrm{exp} \left( - S_G[U] \right)\, .
\end{equation*}
In the limit of an infinite set of gauge configurations, $N_{configs} \rightarrow \infty$, this Monte Carlo averaging converges to the exact answer.  To make these ideas more concrete, and to introduce some lattice jargon, we shall discuss these ideas in more detail with a specific example, the pion two-point correlation function.
%
%	fig: two-point pion correlator
%
\begin{figure}[t]
\center
% include rotate option (angle = -90) only for compressed eps figures
\includegraphics[width=0.6\textwidth]{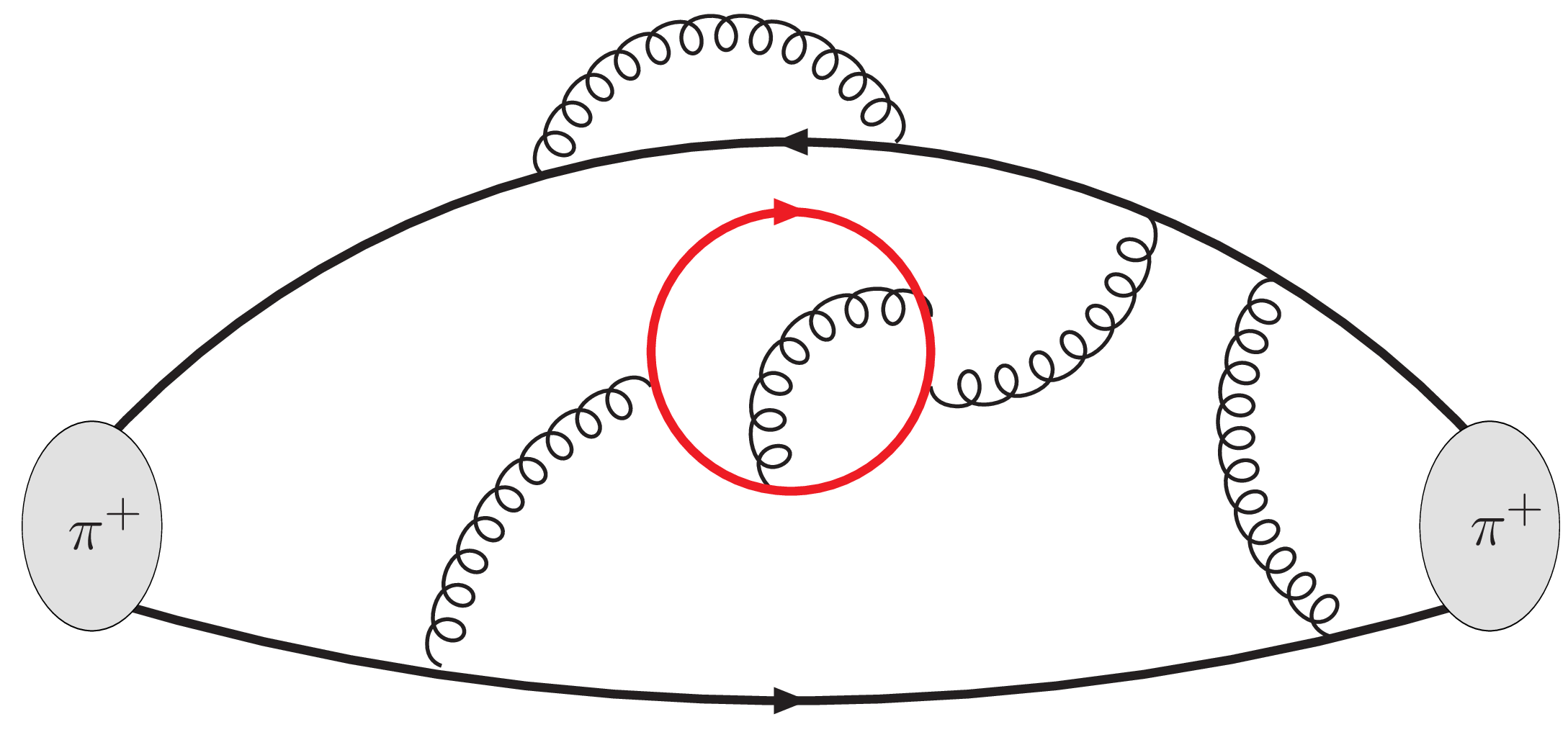}
\caption[Cartoon of a pion two-point correlation function.]{\label{fig:pionCorrelator} The pion two-point correlation function.  In this figure the curly lines represent gluons, and the grey blobs are sources/sinks for the $\pi^+$.  The outer solid lines (black online), which are connected to the source and sink are \textit{valence} quarks.  The closed inner grey lines (red online) represent dynamical \textit{sea} quark - antiquark pairs.  The fermionic determinant in Eq.~\eqref{eq:pionCorrelator} involves the dynamical sea quarks, while the operators which make up the sources and sinks involve valence quarks.}
\end{figure}

The first thing to do is construct and interpolating field for the pion which has an overlap with the physical pion in the lattice simulation.  This interpolating field will excite all the states of the theory which have the same quantum numbers as the pion, but as the pion is the lightest mode of the theory, it will dominate the correlator.  Let us consider the $\pi^+$, for which an interpolating field is given by
\begin{equation}
	\pi^+ (\vec x, \t) = \ol{d}(\vec{x},\t)\, \g_5\, u(\vec{x},\t)\, ,
\end{equation}
where $d$ and $u$ are the down and up quarks respectively, and $\g_5$ is used as the pions are pseudo-scalars.  In Figure~\ref{fig:pionCorrelator}, we provide a cartoon of the pion two-point function, for which the correlator is given by
\begin{align}\label{eq:pionCorrelator}
	\langle\ \pi^-(x_E)\ \pi^+(0)\ \rangle
	&= \frac{1}{\mc{Z_E}} \int D A\ \textrm{Det}\left[ \Dslash_s +m_s \right]
		\tr \Big( \mc{G}_u(0,x_E)\, \g_5\, \mc{G}_d(x_E,0)\, \g_5\, \Big)
		\textrm{exp}\left( -S_G[A] \right) \nonumber\\
	&\rightarrow 	\frac{1}{N_{configs}} \sum_i^{N_{configs}}
		\tr \Big( \mc{G}_u(0,x_E)\, \g_5\, \mc{G}_d(x_E,0)\, \g_5\, \Big) [U_i]\, ,
\end{align}
where the up and down quark propagators are given by
\begin{equation}\label{eq:quarkProp}
	\mc{G}(x,y) = \left[ \Dslash_v +m_v \right]_{xy}^{-1}\, .
\end{equation}
The labels, `$v$' and `$s$' stand for \textit{valence} and \textit{sea} respectively.  The valence quarks are those which are tied to the source and sink, and thus their masses appear in the propagators, Eq.~\eqref{eq:quarkProp}, arising from the sink and source operators.  The sea quarks are the dynamical quarks which appear in quantum loop fluctuations and thus their masses appear in the fermion determinant of Eq.~\eqref{eq:pionCorrelator}.  The most expensive part of lattice QCD simulations is computing the fermion determinant, as it is a non-local operator, for which the computation time scales roughly as $t \sim m_s^{-2.5}$.  This is in contrast to the computation time of the quark propagators, which scale roughly as $t \sim m_v^{-1}$.  

The process of generating of gauge configurations, in which the sea quarks are simulated, is independent of the process of constructing interpolating fields and computing correlation functions, in which the valence quarks are simulated.  There is therefore no technical reason that masses of the valence and sea quarks must be the same.  Moreover, the discretization methods (or Dirac operators) used in the valence and sea sectors can also be different.  However, there are \textit{unphysical} consequences for using different masses or Dirac operators, the most notable being a loss of unitarity.  This is simple to understand as with either different valence and sea masses or Dirac operators, the internal quarks of some correlation function (the sea quarks) are not the same as the external quarks (the valence quarks) as is required by unitarity and the \textit{optical theorem}.

Before discussing the motivations of using these unitarity violating theories, we will first define some common lattice jargon, which for convenience we list in Table~\ref{t:LQCDjargon}.
%%%%%%%%%%%%%%%%%%%%%%%%%%%%%%%%%%%%%%%%%%%%%%%%%%%
%
%	table: Lattice Jargon
%
%%%%%%%%%%%%%%%%%%%%%%%%%%%%%%%%%%%%%%%%%%%%%%%%%%%
\begin{table}
\caption[Lattice jargon]{\label{t:LQCDjargon} Common lattice QCD jargon.}
\center
\begin{tabular}{| c | c | c |}
\hline
$\Dslash_s \neq \Dslash_v$ & $\begin{array}{c} m_s \neq m_v \\ m_s = m_v \end{array}$ 
	& Mixed Action (MA) or hybrid  lattice QCD \\
\hline
$\Dslash_s = \Dslash_v$ & $m_s \neq m_v$ & partially quenched QCD (PQQCD) \\
\hline
$\Dslash_s = \Dslash_v$ & $m_s = m_v$ & QCD \\
\hline
NA & $m_v \neq 0\ ,\ m_s = \infty$ & quenched QCD (QQCD) \\
\hline
\end{tabular}
\end{table}
%%%%%%%%%%%%%%%%%%%%%%%%%%%%%%%%%%%%%%%%%%%%%%%%%%%
%
%%%%%%%%%%%%%%%%%%%%%%%%%%%%%%%%%%%%%%%%%%%%%%%%%%%
When the sea and valence Dirac operators are not equal, this is known as \textit{mixed action} (MA) or \textit{hybrid} lattice QCD.  Mixed action lattice simulations can never be tuned exactly to the QCD point as in any simulation, there will be different lattice spacing corrections to the valence and sea quarks which will result in different quark masses and dispersion relations even if the mass parameters are tuned equal.  However, for sufficiently small lattice spacings, one can incorporate these lattice artifacts into the effective field theory (EFT) description of the mixed action LQCD simulation,%
\footnote{Of course, for arbitrarily small lattice spacings, these differences arising from the different lattice actions should become negligible, which practically means smaller than the statistical uncertainty of a given quantity.  This would also imply that the need for a mixed action effective field theory description would not be necessary, but today and for the foreseeable future, these differences must be incorporated into the MA effective theory relevant for a given MA lattice simulation.} 
and use these expressions to remove the lattice spacing effects, allowing an extrapolation to the QCD point.  We shall discuss a particular example in some detail in Chapter~\ref{chap:pipiMA}.

A simulation with the same Dirac operator but different quark masses in the sea and valence sectors is known as partially quenched QCD (PQQCD).  This theory is also unitarity violating, but has the advantage over MA simulations that at the level of the simulations, one can always set the valence and sea quark masses equal, and thus perform numerical simulations without any sicknesses.  The important point here is that at the level of simulations, one knows that there is a limit, $m_s = m_v$ in which one is simulating QCD, the unitarity conserving theory.  This also implies that the low energy effective theory of PQQCD must also contain this limit, for the same tuning of valence and sea quark masses.  One can then perform simulations away from this limit, match the EFT description of observable quantities to the lattice correlation functions, and later by hand take the QCD limit, thus recovering the physics of interest.%
\footnote{The belief of this author is that this same methodology works for the MA theories, with the combined continuum limit (in which the Dirac operators become equivalent), despite not being able to tune to the QCD point in the simulations.  This can only be verified or disproven by a matching of the MA effective field theory to MA LQCD simulations, and then performing the continuum, QCD limits.}

Lastly, one can set the sea quark masses to infinity, such that they can be integrated out of the theory.  This is known as quenched QCD (QQCD).  The motivation of this theory is that up until a few years ago, it was numerically too expensive to simulate dynamical quarks in a reasonable amount of time.  It has also been found that QQCD typically introduces an uncontrolled error of $\lesssim 20\%$.  Thus, QQCD provides an inexpensive check of methodology and can provide approximate predictions for physical observables.  We stress that there is no rigorous connection of QQCD to QCD, meaning there is no systematic way to correct the \textit{quenched approximation} and include the effects of the neglected sea quarks.

The motivation to consider these partially quenched simulations is related to the extreme cost of full scale lattice QCD computations (which are presently measured in tera-flop years).  The first reason is that as stated above, it is numerically much cheaper to simulate valence quarks as compared to sea quarks.  If one can make more use of the expensive gauge configurations containing the sea quarks, by using a number of different valence quark masses for each configuration, then one obviously gains much in terms of computing resources.  Additionally, it is important to reduce the masses of the quarks used in the simulations, as it has only been in these last few years that the quarks masses have been in the chiral regime~\cite{Bernard:2002yk,Beane:2004ks} such that one can make use of effective field theory techniques to compare the lattice QCD simulations to the physical world.  By reducing the valence quark masses more than the sea quark masses, one can further probe the chiral regime, increasing the region of overlap between the parameter space where lattice simulations are performed and where one has confidence in the effective field theories~\cite{Sharpe:2000bc}.

The motivation to consider mixed action simulations, which are generalizations of partially quenched simulations, is also related to the numerical cost of full scale lattice QCD simulations.  Additionally, the MIMD Lattice Computation (MILC) collaboration has made there lattice source code and gauge configurations publicly available~\cite{Bernard:2001av}.  For reasons which we will discuss in some detail in Chapters~\ref{chap:pipiMA} and \ref{chap:MAGW}, it is beneficial to use a different discretization technique for the valence quarks, than the technique used to generate the MILC configurations.  Two groups in particular have determined some very impressive physical results using these mixed action techniques~\cite{Negele:2004iu,Bonnet:2004fr,Beane:2005rj,Edwards:2005ym,Beane:2006mx,Beane:2006pt,Beane:2006fk}.

There are good reasons to believe that the low energy effective theories of these unitarity violating lattice theories fully capture the unitarity violations introduced by either different masses for the sea and valence quarks, and also for the use of different discretization techniques in the sea and valence sectors.  Provided this is true,%
\footnote{We stress again, that this can only be tested by performing MA simulations, and comparing them to the EFT predicted form of correlation functions.}%
then the EFTs for these lattice simulations provide a rigorous means to describe the unitarity violations in the region where the quark mass and lattice spacing corrections can all be treated perturbatively, and thus remove their contribution to the lattice correlation functions and extract the physics of interest.  We will come back to this issue below in Section~\ref{sec:PQChPT} as well as Chapters~\ref{chap:pipiMA} and \ref{chap:MAGW}.  We now move on to describe some of the effective field theories necessary to understand lattice QCD simulations today, and extrapolate these LQCD computations to the physical world.

%%%%%%%%%%%%%%%%%%%%%%%%%%%%%%%%%%%%%%%%%%%%%%%%%%%
%
%	Effective Field Theory
%
%%%%%%%%%%%%%%%%%%%%%%%%%%%%%%%%%%%%%%%%%%%%%%%%%%%
\section{Effective Field Theory}

What is an effective field theory (EFT)?%
\footnote{As with our introduction to LQCD, we do not pretend to give a comprehensive review of effective field theories.  We merely summarize the main ideas and introduce the EFTs necessary to understand the work in this thesis.  The following is a list of EFT references the author has found invaluable in the course of pursuing his degree in physics,~\cite{Manohar:1995xr,Kaplan:1995uv,Manohar:1996cq,Phillips:2002da,Scherer:2002tk,Kaplan:2005es}.}
Effective field theories are based upon simple ideas with powerful consequences.  The underlying idea is that physical phenomena at low-energies (long wave-lengths) are not sensitive to the details of the high-energy (short wave-length) structure of particles or fields.  We use these ideas everyday in physics.  In freshman undergraduate physics courses we teach that the change in gravitational potential energy of a particle of mass $m$ raised a height $h$ above the surface of the Earth is given by
\begin{equation}
	\D U =  m\, g\, h\, ,
\end{equation}
where $g$ is the acceleration of a massive particle induced by the Earth's gravitational field.  This is actually the leading term in an \textit{effective} expansion of the height of the object over the radius of the Earth, $R$.  The actual change in potential energy has the form
\begin{align}\label{eq:Gravity}
	\D U &= m\, g\, h \frac{R}{R+h} \nonumber\\
		& \approx m\, g\, h\, \left( 1 - \frac{h}{R} + \mc{O} \left( \frac{h^2}{R^2} \right) \right)\, .
\end{align}
This is a simple example of a classical effective theory (no quantum loop effects), of which there are many.  Classical Newtonian mechanics is an effective theory of quantum mechanics for distances large compared to quantum length scales, set by $\hbar$.  The pattern we see from these two examples is that there is a separation of scales, in these cases length scales, which sets a range of validity of the theory.  For heights comparable to the radius of the earth, Eq.~\eqref{eq:Gravity} is no longer a good approximation of the true change in potential, as the neglected terms, $(h/R)^2 \sim 1$ are no longer small, and must be included.  However, for $h<<R$, Eq.~\eqref{eq:Gravity} is a very good approximation to the change in potential, and moreover, the neglected terms give us a rigorous estimate upon the theoretical uncertainty we have introduced by not including them.%
\footnote{Actually, in this example, it is an exact uncertainty as the coefficient of the neglected term is exactly known.}
Let us define a small scale that serves as an expansion parameter for the effective theory,
\begin{equation}\label{eq:epsilon}
	\varepsilon_G \equiv \frac{h}{R}\, .
\end{equation}
We see that $\varepsilon$ both acts as a small parameter, allowing computations to arbitrary precision (powers of $\varepsilon$), and also is an indicator of the range of validity of the theory.  These are characteristic features of effective theories, which also apply to quantum field theories.

When we apply these ideas to quantum field theories, there are a few more steps to the process.  The first task is to identify the appropriate degrees of freedom relevant to the physics questions we are interested in asking, \textit{i.e.} what are the particles one would like to describe?  Next, we must identify all the symmetries (and approximate symmetries) the underlying, or fundamental theory exhibits.  We then construct a Lagrangian which includes all linearly independent operators consistent with the underlying symmetries (this is often an infinite number of operators).  The power of the EFT treatment is the separation of scales, or the expansion parameters, such as Eq.~\eqref{eq:epsilon} in the gravitation potential example.  Often, there will be more than one ratio of relevant scales, let us call them $\varepsilon_a$ and $\varepsilon_b$.  All the operators in the effective Lagrangian will have associated powers of these expansion parameters.  One then decides to which precision in $\varepsilon_a$ and $\varepsilon_b$ we wish to know the physics, and computes all processes which contribute to this order of precision.  For instance, in our gravitational potential energy example, the expansion parameter for heights on the order of 10 meters is $\varepsilon_G \sim 1.6 \times 10^{-6}$, and thus the first term in the series is a \textit{very} good approximation.  The (often infinite) set of operators in the effective Lagrangian we need are thus truncated by this desired precision, and we have a theoretical estimate on the neglected physics, the first order in $\varepsilon_a$ and $\varepsilon_b$ not included.

If we know the underlying theory, as in the case of the gravitational potential example, then the coefficients of the operators in the effective theory can be exactly determined by matching the effective theory to the underlying theory at an energy scale both theories converge in, and requiring the physics to be the same.  If, however the underlying theory is unknown, as is the case with extensions to the Standard Model, or if the underlying theory is not computable at energy scales of interest, as with QCD, then there is no way \textit{a priori} to know the coefficients of the operators in the effective theory, and they must be determined by fitting expressions of physical observables computed with the EFT either to experimental data or in the case of QCD, to lattice QCD simulations.  The predictive power of an EFT is that these \textit{a priori} undetermined coefficients of the effective operators are universal.  Once they have been determined by comparing to one set of experiments, they can be used to make predictions about other experimental quantities not used to constrain them.

This is a crucial component of EFTs.  As one has included all operators consistent with the symmetries in the effective Lagrangian, the EFT provides a \textit{model independent} description of the physics in the region of applicability of the theory.  The quantum loop contributions arising from the effective operators encode the long distance, or infrared (IR) physics, while the coefficients of the higher dimensional operators in the theory, known as low-energy-constants (LECs), encode the short distance, or ultra-violet (UV) physics.  The IR physics is dominated by the lightest degrees of freedom in the theory, and at a given order are generally known, while the UV physics at a given order in $\varepsilon_a$ and $\varepsilon_b$ is dominated by the highest dimensional operators included in the computation.  The LECs of these operators are generally unknown and must be determined to have predictive power.
We now move on to introduce the low energy effective theory of QCD, chiral perturbation theory.

%%%%%%%%%%%%%%%%%%%%%%%%%%%%%%%%%%%%%%%%%%%%%%%%%%%
%
%	Chiral Perturbation Theory
%
%%%%%%%%%%%%%%%%%%%%%%%%%%%%%%%%%%%%%%%%%%%%%%%%%%%
\subsection{\textbf{Chiral Perturbation Theory}}

The most well known effective field theory is the Fermi theory of weak interactions~\cite{Fermi:1934sk,Fermi:1934hr}.  Perhaps the next most well known EFT is chiral perturbation theory (\CPT) which was introduced by Weinberg~\cite{Weinberg:1967kj,Weinberg:1968de,Weinberg:1978kz} and systematically developed by Gasser and Leutwyler~\cite{Gasser:1983yg,Gasser:1984gg}.  One can not derive \CPT\ from QCD as that would require an exact closed-form solution of the QCD equations of motion.  However, with one phenomenologically based assumption, one can construct the most general Lagrangian consistent with the symmetries of the QCD Lagrangian.
The quark content of the QCD Lagrangian is%
\footnote{We should also include the theta term in the Lagrangian, $\theta \frac{g^2}{32 \pi^2}G_{\mu\nu} \tilde{G}^{\mu\nu}$, as this term can be rotated into a phase of the quark masses by a chiral transformation.  This theta term gives rise to CP violations in QCD and also to the neutron electric dipole moment (EDM), which has yet to be found experimentally~\cite{Harris:1999jx}.  One can also theoretically understand the neutron EDM using effective field theory methods~\cite{Crewther:1979pi,Pich:1991fq,Borasoy:2000pq,Hockings:2005cn,O'Connell:2005un} and lattice QCD~\cite{Aoki:1989rx,Aoki:1990ix,Guadagnoli:2002nm,Shintani:2005xg,Berruto:2005hg}.  However, in this thesis, we do not explore any CP violating processes, and thus shall not include this term.}
\begin{align}\label{eq:QCD}
	\mc{L}_{QCD} &= \sum_{i,j=u,d,s} \, \bar{q}_i\, \Big[ i \Dslash -m_Q \Big]_{ij}\, q_j \nonumber\\
		&= \sum_{i=u,d,s} \, \Big[ \bar{q}_{i,L}\, i\Dslash_{ij}\, q_{j,L} 
			+ \bar{q}_{i,R}\, i\Dslash_{ij}\, q_{i,R}
			+\bar{q}_{i,L}\, m_{Q_{ij}}\, q_{j,R} + \bar{q}_{i,R}\, m_{Q_{ij}}\, q_{j,L} \Big]\, ,
\end{align}
where we use the quark mass matrix
\begin{equation}
	m_Q = \begin{pmatrix}
			m_u \\
			& m_d \\
			& & m_s \\
		\end{pmatrix}\, ,
\end{equation}
and the left and right handed quark fields are defined as
\begin{equation}
	q_{L} \equiv \frac{1- \g_5}{2}q\quad,\quad q_{R} \equiv \frac{1+\g_5}{2}q\, .
\end{equation}
We can see from Eq.~\eqref{eq:QCD} that the kinetic term of the QCD Lagrangian is invariant under independent $L$ and $R$ chiral transformations, where $L,R \in SU(3)_{L,R}$.  The quark mass term explicitly breaks this chiral symmetry of $SU(3)_L \otimes SU(3)_R$ down to the vector subgroup.  However, for sufficiently small quark masses, as compared to $\L_{QCD}$, this explicit breaking can be treated perturbatively.  In the zero quark mass limit, the QCD Lagrangian exhibits an exact chiral symmetry.  However, in nature we observe that the parity partner of the nucleon is significantly heavier than the nucleon.  We also observe that all the observed hadrons composed of the three light quarks, $H$, have masses $m_H \sim \mc{O}(1 \textrm{ GeV})$, except for a handful of pseudo scalars, the pions ($m_\pi \simeq 135$~MeV), the kaons ($m_K \simeq 498$~MeV) and the eta ($M_\eta \simeq 548$~MeV).  This phenomenological evidence leads us to postulate that the vacuum of QCD spontaneously breaks the global chiral symmetry of the QCD Lagrangian, Eq.~\eqref{eq:QCD} down to the vector subgroup~\cite{Vafa:1983tf}, \begin{equation}
	SU(N_f)_L \otimes SU(N_f)_R \longrightarrow SU(N_f)_V\, ,
\end{equation}
creating the 8 pseudo-Nambu-Goldstone modes~\cite{Nambu:1960xd,Goldstone:1961eq,Goldstone:1962es} which encode the long distance fluctuations about the many possible minima of the QCD vacuum.  In the zero quark mass limit, these 8 mesons would be exact Nambu-Goldstone modes, and thus massless.  The inclusion of the quark mass parameters, however, gives rise to the non-vanishing mass of these mesons.

In the following, we will consider the theory with two light quarks, the up and down, as well as the theory with three quarks, the up, down and strange.  The mass of the strange quark is significantly heavier than the up and down quarks but still lighter than $\L_{QCD}$.  The inclusion of the strange quark in the theory is thus necessary, but the convergence of the chiral expansion is not as good.  This is especially true for the theory including heavy baryons, as the expansion parameter of heavy baryon \CPT\ (HB\CPT) scales with the meson mass, and not the meson mass squared as in the purely mesonic theory.  Therefore, sometimes it will be prudent to explicitly include the strange quark, and other times it will be more beneficial to include its effects indirectly through modifications of the LECs.  This can only be determined phenomenologically through a comparison of the two theories with their respective LECs fit either to experiment or lattice QCD simulations.  We will show this explicitly in the next two sections in which we first focus on the two-flavor theory, and then the three-flavor theory.

%%%%%%%%%%%%%%%%%%%%%%%%%%%%%%%%%%%%%%%%%%%%%%%%%%%
%
%	SU(2)
%
%%%%%%%%%%%%%%%%%%%%%%%%%%%%%%%%%%%%%%%%%%%%%%%%%%%
\subsubsection{\textbf{Two Flavors: The Pions} \label{sec:SU2}}

The hypothesis is that quarks condense in the QCD vacuum,
\begin{equation}\label{eq:condensate}
	\langle 0 |\, \ol{q}_R^j\, q_L^i\, | 0 \rangle = \l\, \d^{ij}\, ,
\end{equation}
where $i,j$ are flavor indices and $\l$ is a dimension-3 quantity whose scale is set by $\L_{QCD}$,
$\l \sim (\L_{QCD})^3$.  Under a chiral transformation, 
\begin{align}\label{eq:SigmaDef}
	\langle 0 |\, \ol{q}_R^j\, q_L^i\, | 0 \rangle \longrightarrow 
		\langle 0 |\, \ol{q}_R^k\, R_k^{\dagger\, j}\, L^{i}_{\ l}\, q_L^{l}\, | 0 \rangle 
		&= \l\, (LR^\dagger)^{ij} \nonumber\\
		&\equiv \l\, \S^{ij}\, ,
\end{align}
and we see that this quark condensate is only invariant under the vector subgroup, $L=R=V$, for which $\S^{ij} = \d^{ij}$.  Otherwise, for $L \neq R$, $\S^{ij}$ corresponds to a different vacuum of QCD than that of Eq.~\eqref{eq:condensate}, and in the massless quark limit, these different vacua would be degenerate.  The spontaneous breaking of the chiral group then would give rise to exact Nambu-Goldstone bosons representing the long wavelength fluctuations of the QCD vacuum, $\S$.  In the case of $SU(2)$, for example, three generators are broken by the vacuum, giving rise to 3 pions.  We then wish to parameterize these bosons with a linear combination of the broken generators~\cite{Coleman:1969sm,Callan:1969sn}.  The choice is not unique, (although all forms must give the same physics) and here we choose the non-linear sigma model form.  We replace $\S$ with a non-linear realization of the pions, 
\begin{equation}\label{eq:Sigma}
	\S = \textrm{exp} \left( \frac{2i\phi}{f} \right) \quad \textrm{with}\quad
		\phi = \begin{pmatrix}
			\frac{\pi^0}{\sqrt{2}} & \pi^+ \\
			\pi^- & -\frac{\pi^0}{\sqrt{2}}
		\end{pmatrix}\, ,
\end{equation}
where we use the normalization $f \simeq 132$~MeV.  From Eq.~\eqref{eq:SigmaDef}, one can see that the $\S$-field transforms under the chiral group as
\begin{equation}\label{eq:SigTransform}
	\S \longrightarrow L\, \S\, R^\dagger\, .
\end{equation}
To construct the most general chiral Lagrangian, we must first include the quark mass parameters by utilizing spurion analysis, in which one first treats the quark mass matrix as a spurion field which transforms under the chiral group as
\begin{equation}
	m_Q \rightarrow L\, m_Q\, R^\dagger\, .
\end{equation}
Setting $m_Q = \textrm{diag}(m_u,m_d)$,%
\footnote{The choice of the real, diagonal quark mass matrix is not necessary but is the most natural choice for obvious reasons.  It is this choice, which then leads dynamically to the vacuum energy being minimized for $\S \propto \mathbf{1}$.  However, any complex choice is allowed, and explicitly breaks the chiral symmetry group.  The different choices are all related by chiral transformations.}
explicitly breaks the chiral symmetry down to the vector subgroup, $SU(2)_V$.  The most general Lagrangian consistent with chiral symmetry at leading order is then
\begin{equation}\label{eq:ChPTLO}
	\mc{L} = \frac{f^2}{8} \tr \left( \partial_\mu \S \partial^\mu \S^\dagger \right)
		+\frac{f^2\, B_0}{4} \tr \left( \S m_Q^\dagger + m_Q \S^\dagger \right)\, ,
\end{equation}
where the LEC $B_0$ is defined by
\begin{equation}
	B_0 = \textrm{lim}_{m_Q \rightarrow 0}\ 
		\frac{| \langle\, \bar{q} q\, \rangle |}{f^2}\, .
\end{equation}
The normalization of the first term in Eq.~\eqref{eq:ChPTLO} is fixed by the canonical normalization for a scalar kinetic operator, which is seen by expanding the $\S$-field in powers of $\phi / f$.  To quadratic order in $\phi$, Eq.~\eqref{eq:ChPTLO} gives,%
\footnote{We see that the square of the meson mass is related linearly to the quark masses, and the chiral condensate, in agreement with the Gell-Mann--Oakes--Renner prediction~\cite{Gell-Mann:1968rz}.  However, it is only recently that this relation has been phenomenologically confirmed by precisely determining the LECs of the corrections to the pion mass which arise at the next order in the chiral expansion~\cite{Ananthanarayan:2000ht,Colangelo:2000jc,Colangelo:2001sp,Colangelo:2001df,Pislak:2001bf,Pislak:2003sv}.  This rules out the alternative description of nature using a generalized chiral perturbation theory in which the chiral condensate is not the leading order parameter of chiral symmetry breaking~\cite{Stern:1993rg,Knecht:1995tr,Knecht:1995ai}.}
\begin{equation}
	\mc{L} = | \partial \pi^+ |^2 + \frac{1}{2} (\partial \pi^0)^2
		+B_0 (m_u +m_d)\, \left(\, \pi^+ \pi^- +\frac{1}{2} \pi^0 \pi^0\, \right) +\dots
\end{equation}
and thus the pion masses are given at leading order by
\begin{equation}\label{eq:piMasses}
	m_\pi^2 = B_0 (m_u +m_d)\, .
\end{equation}
The theory gets much more interesting when one includes the quantum loops arising from this Lagrangian.  To be consistent with EFT techniques, we then need to include tree-level terms arising at the next order, treating them as the same order as the loop corrections from the LO Lagrangian.  The most general Lagrangian at the next order is given by
\begin{align}\label{eq:ChPTNLO}
	\mc{L}^{(4)} =& 
		\frac{\ell_1}{4} \left[ \tr\left( \partial_\mu \S^\dagger \partial^\mu \S \right) \right]^2 
		+\frac{\ell_2}{4} \tr \left( \partial_\mu \S^\dagger \partial_\nu \S \right)
			\tr \left( \partial^\mu \S^\dagger \partial^\nu \S \right) \nonumber\\
		& +\frac{\ell_3\, B_0^2}{4} \left[ \tr \left( m_Q \S^\dagger + \S m_Q \right) \right]^2 
		 +\frac{\ell_4\, B_0}{4} \tr \left( \partial_\mu \S \partial^\mu \S^\dagger \right)
		 	\tr \left( m_Q \S^\dagger + \S m_Q^\dagger \right)\, ,
\end{align}
where the coefficients~\cite{Scherer:2002tk} are related to the original LECs of Gasser and Leutwyler~\cite{Gasser:1983yg} by
\begin{equation}
	\ell_1 = \ell_1^{GL}\quad ,\quad
	\ell_2 = \ell_2^{GL}\quad, \quad
	\ell_3 = \ell_3^{GL} +\ell_4^{GL} \quad,\quad
	\ell_4 = \ell_4^{GL}\, .
\end{equation}
The loop effects typically give contributions proportional to $\left( \frac{1}{(4\pi f)^2} \right)^n$, where $n$ is the number of loops.  Naive dimensional analysis~\cite{Manohar:1983md} and phenomenological evidence suggest that the chiral symmetry breaking scale is $\L_\chi \sim 4\pi f$, and thus the chiral corrections are expected to scale as powers of the expansion parameter,
\begin{equation}
	\varepsilon^2 \sim \frac{m_\pi^2}{(4\pi f)^2} \sim \frac{p^2}{(4\pi f)^2}\, .
\end{equation}
This is the promised EFT expansion parameter, and we can see that in the real world with $m_\pi \simeq 135$~MeV, this is expected to be a good expansion, with corrections being on the order of $1 \%$ for $SU(2)$ \CPT.  This will not be as good with the explicit inclusion of the strange mesons, as the kaon mass is significantly heavier than the pion masses.  We will come back to this shortly.

The Lagrangian~\eqref{eq:ChPTLO} can also be used to compute the leading $\pi\pi$ scattering amplitudes, which give the prediction of the scattering lengths by Weinberg using current algebra~\cite{Weinberg:1966kf}, with the $I=2\, \pi\pi$ scattering amplitude, for example, given by
\begin{equation}
	\mc{T}_2 = -\frac{2}{f_\pi^2} \frac{3s - 6m_\pi^2}{3}\, .
\end{equation}
We shall discuss the $\pi\pi$ scattering in much more detail in Chapters~\ref{chap:pipiFV} and \ref{chap:pipiMA}.

%%%%%%%%%%%%%%%%%%%%%%%%%%%%%%%%%%%%%%%%%%%%%%%%%%%
%
%	SU(3)
%
%%%%%%%%%%%%%%%%%%%%%%%%%%%%%%%%%%%%%%%%%%%%%%%%%%%
\subsubsection{\textbf{Three Flavors}}

As will be a recurring theme in this Chapter, the formalism for the three flavor theory is identical to the two-flavor theory, the difference being the group algebra.  There are some simplifications which occur in the two-flavor theory, which do not hold for the three flavor theory.  These will become clear in this section.

The LO Lagrangian, Eq.~\eqref{eq:ChPTLO}, takes the same form as in the two flavor theory, with the $\phi$ field now defined as
\begin{equation}
	\phi = \begin{pmatrix}
		\frac{\pi^0}{\sqrt{2}}+\frac{1}{\sqrt{6}}\eta & \pi^+ & K^+\\
		\pi^- & -\frac{\pi^0}{\sqrt{2}} +\frac{1}{\sqrt{6}}\eta & K^0 \\
		K^- & \ol{K^0} & -\frac{2}{\sqrt{6}} \eta
		\end{pmatrix}\, .
\end{equation}
In the isospin limit ($m_u=m_d$), the $\pi^0$ and $\eta$ fields do not mix, and the masses of the 8 pseudo-Nambu--Goldstone mesons are given by
\begin{align}
	m_\pi^2 &= B_0 (m_u+m_d) \nonumber\\
	m_K^2 & = B_0 (m_{u,d} +m_s) \nonumber\\
	m_\eta^2 &= \frac{B_0}{3}(m_u +m_d +4m_s)\, .
\end{align}

The next difference from the two-flavor theory occurs when constructing the NLO chiral Lagrangian.  It has the same general form as Eq.~\eqref{eq:ChPTNLO}, but there are more simplifications one can make in the two-flavor theory, as $SU(2)$ has a real representation.  The most general Lagrangian consistent with the symmetries is given by~\cite{Gasser:1984gg}
\begin{align}\label{eq:ChPTNLOSU3}
	\mc{L}^{(4)} =&\ L_1\, \left[ \tr \left( \partial_\mu \S \partial^\mu \S^\dagger \right) \right]^2
		+L_2\, \tr \left( \partial_\mu \S \partial_\nu \S^\dagger \right)
			\tr \left( \partial^\mu \S \partial^\nu \S^\dagger \right) \nonumber\\
		&+L_3\, \tr \left( \partial_\mu \S \partial^\mu \S^\dagger \partial_\nu \S \partial^\nu \S^\dagger 
			\right)
		+2 B_0\, L_4\, \tr \left( \partial_\mu \S \partial^\mu \S^\dagger \right)
			\tr \left( m_Q \S^\dagger +\S m_Q^\dagger \right) \nonumber\\
		&+ 2B_0\, L_5\, \tr \left( \partial_\mu \S \partial^\mu \S^\dagger
			\left( m_Q \S^\dagger +\S m_Q^\dagger \right) \right)
		+4 B_0^2\, L_6\, \left[ \tr \left( m_Q \S^\dagger +\S m_Q^\dagger \right) \right]^2 \nonumber\\
		&+4 B_0^2\, L_7\, \left[ \tr \left( m_Q \S^\dagger -\S m_Q^\dagger \right) \right]^2
		+ 4 B_0^2\, L_8\, \tr \left( m_Q \S^\dagger m_Q \S^\dagger
			+\S m_Q^\dagger \S m_Q^\dagger \right)\, .
\end{align}
One can match this $SU(3)$ Lagrangian onto the $SU(2)$ Lagrangian of Eqs.~\eqref{eq:ChPTLO} and \eqref{eq:ChPTNLO}, and learn that the $SU(2)$ LECs depend logarithmically upon the strange quark mass.  This is important for comparing $SU(2)$ \CPT\ calculations to lattice QCD simulations which include the strange quark in the simulation.  We now move on to show how to include heavy baryons into the chiral Lagrangian.

%%%%%%%%%%%%%%%%%%%%%%%%%%%%%%%%%%%%%%%%%%%%%%%%%%%
%
%	HBChPT
%
%%%%%%%%%%%%%%%%%%%%%%%%%%%%%%%%%%%%%%%%%%%%%%%%%%%
\subsection{\textbf{Heavy Baryon Chiral Perturbation Theory}}

To systematically include the baryons into the chiral Lagrangian, we use heavy baryon chiral perturbation theory (HB\CPT) first formulated in Refs.~\cite{Jenkins:1990jv,Jenkins:1991ne}.  The idea is based upon heavy quark effective theory (HQET)~\cite{Georgi:1990um,Manohar:2000dt} in which ones performs a velocity dependent phase transformation of the Lagrangian to separate the hard and soft modes of the theory.  This amounts to studying the fluctuations about the heavy mass parameter.  The velocity dependent fields are
\begin{equation}\label{eq:Bv}
	B_v (x) = \frac{1 +\vslash}{2} e^{i M \vit \cdot x}\, B(x)\, ,
\end{equation}
where $M$ is the mass of the heavy field and $\vit_\mu$ is the four-velocity of the baryon, $B$.  The momentum of the baryon can be parameterized as
\begin{equation}\label{eq:RPImomentum}
	p_\mu = M v_\mu +k_\mu\, ,
\end{equation}
where $k_\mu$ is the off-shell momentum of the baryon.  The effect of this parameterization is to transform the standard Dirac Lagrangian for the heavy baryon of spin-$\frac{1}{2}$ in the following way,
\begin{equation}
	\ol{B}\, \left( i \dslash -M \right)\, B \rightarrow
		\ol{B}_v\, i\dslash\, B_v + \mc{O} \left( \frac{1}{M} \right)\, .
\end{equation}
From Eq.~\eqref{eq:Bv}, it is easy to verify that derivatives acting on $B_v$ bring down powers of the off-shell momentum, $k$, rather than the full baryon momentum, $p$.  Thus, higher derivative operators in the EFT are suppressed by powers of the heavy mass $M$, and a consistent power counting emerges.%
\footnote{With the standard Dirac Lagrangian for the baryons, the power counting problem arises when one considers loop graphs and higher dimensional operators of the effective theory.  The baryon mass is not small compared to the chiral symmetry breaking scale and thus one can not power count loop graphs with internal baryons, as they will be proportional to powers of $M / \L_\chi \sim \mc{O}(1)$.  The same problem occurs for higher dimensional operators, as one must include all operators with arbitrary powers of $M / \L_\chi$, which are not suppressed compared to the leading mass operator, $\ol{B}\, M\, B$.  For an alternative approach to resolving this problem, see Refs.~\cite{Becher:1999he,Fuchs:2003qc}.}
Heavy baryon \CPT\ is applicable in the limit that the pion masses, momentum and the off-shellness of the baryon are small compared to the heavy baryon mass (which is approximately the chiral symmetry breaking scale),
\begin{equation}
	\frac{m_\pi}{M} \sim \frac{q}{M} \sim \frac{\vit \cdot k}{M} \ll 1\, .
\end{equation}	

We also must include the lowest lying spin-$\frac{3}{2}$ baryons in the theory as they play a very important phenomenological role in nucleon (or octet baryon) properties~\cite{Jenkins:1990jv,Jenkins:1991ne,Jenkins:1992ts,Butler:1992ci,Butler:1992pn,Bernard:1993nj,Butler:1994ej,Lebed:1994yu,Lebed:1994gt,Banerjee:1995bk,Hemmert:1996rw,Hemmert:1997tj,Beane:2002wn,Pascalutsa:2002pi,Pascalutsa:2003zk,Beane:2004ra,Wies:2006rv}.  However, there are questions which arise because these states are unstable, undergoing strong decay with a lifetime of $\t \sim 10^{-23}$~seconds (except for the $\Omega^-$ which decays weakly), and thus one can not define unique S-matrix elements for them.  The effects of these states are not appropriately included by only absorbing their contributions to observable quantities into the LECs of the nucleon (or octet baryon) chiral Lagrangian for energies on the order of the pion mass.  This is because the mass splitting between these resonant states and the spin-$\frac{1}{2}$ baryons is small compared to $\L_\chi$, and thus they are easily excited.  This leads to large contributions to the spin-$\frac{1}{2}$ baryon observables from these spin-$\frac{3}{2}$ baryons, and more importantly, in scattering processes, a rapid oscillation of the scattering phase shift for energies near the mass of the resonance.  Moreover, for the vast majority of lattice QCD simulations to date, the pion masses are heavy enough that the spin-$\frac{3}{2}$ states are actually stable.  We will come back to this issue in Section~\ref{sec:DeltaMasses}.

The inclusion of the spin-$\frac{3}{2}$ fields into HB\CPT\ requires the addition of an additional mass parameter, the mass splitting between the $T$ and $B$ fields in the chiral limit, $\D = M_T - M_B$.  This mass parameter is dynamically generated from the interaction of the quarks with the gluons, and is independent of the quark mass.  The need for $\D$ arises from the velocity transformation of Eq.~\eqref{eq:Bv}, as one can not simultaneously remove both the nucleon and the delta masses (or octet and decuplet baryon masses).%
\footnote{This can be accomplished for the full octet or decuplet masses, as in each case there is a symmetry protecting their common masses in the chiral limit.  However, with the inclusion of both sets of fields simultaneously, there is no symmetry relating $M_B$ to $M_T$, and thus one must choose a mass to phase away, the simplest choice being the lightest mass.  We note that in the large $N_C$ limit, $\D \rightarrow 0$, and thus the $B$ and $T$ fields become degenerate~\cite{Witten:1979kh}.  This is not relevant to our current discussion however.}
To see the explicit effect upon the HB\CPT\ Lagrangian, we will now focus on the 2-flavor theory, then the 3-flavor theory and lastly the partially quenched versions of these theories.

%%%%%%%%%%%%%%%%%%%%%%%%%%%%%%%%%%%%%%%%%%%%%%%%%%%
%
%	Nucleons and deltas
%
%%%%%%%%%%%%%%%%%%%%%%%%%%%%%%%%%%%%%%%%%%%%%%%%%%%
\subsubsection{\textbf{Nucleons and Deltas} \label{sec:NDLagrangian}}

Heavy matter fields (nucleons, deltas, heavy mesons etc.) transform under the vector subgroup (parity even subgroup) of the full chiral group.  For example the nucleon fields, which are a doublet under $SU(2)_V$,%
\footnote{Here, and in the rest of this work, we always drop the subscript, ``$\vit$" on the heavy baryon fields, as we never use the full field defined in Eq.~\eqref{eq:Bv}.}
\begin{equation}\label{eq:N}
	N = \begin{pmatrix}
		p \\ n
		\end{pmatrix}\, ,
\end{equation}
transform as 
\begin{equation}\label{eq:NucTransform}
	N_i \rightarrow U_{ij}\, N_j\, .
\end{equation}
When including heavy matter fields in \CPT\ it is useful to introduce the field
\begin{equation}
	\xi^2 = \S\, ,
\end{equation}
where $\S$ is defined in Eq.~\eqref{eq:Sigma}.  Using the chiral transformation properties of $\S$, Eq.~\eqref{eq:SigTransform}, one can deduce that under an arbitrary $SU(2)_L \otimes SU(2)_R$ transformation,
\begin{equation}\label{eq:xiTransform}
	\xi \rightarrow L\, \xi\, U^\dagger = U\, \xi\, R^\dagger\, ,
\end{equation}
where $L,R$ are elements of $SU(2)_{L,R}$ respectively and $U$ is an element of the chiral group.  In the case of a vector transformation, $L=R=V$, then $U=V$ as well.  Otherwise, $U$ is a complicated function of the pion fields in $\S$ as well as space-time.  It is also now clear that the choice of nucleon (or other matter field) is not unique, as the field
\begin{equation}
	N^\prime = \xi\, N \rightarrow L N^\prime\, ,
\end{equation}
transforms identically to the field in Eq.~\eqref{eq:NucTransform} under the vector subgroup~\cite{Coleman:1969sm,Callan:1969sn,Manohar:1995xr,Manohar:1996cq}.  Thus, any choice of nucleon field which transforms as Eq.~\eqref{eq:NucTransform} under the vector subgroup is a good choice.  The difference between these choices is simply a field redefinition of the nucleon field, and must give the same S-matrix elements.  When constructing the most general Lagrangian, it is useful to introduce the chiral \textit{vector} and \textit{axial-vector} fields,
\begin{align}\label{eq:VandA}
	\mc{V}_\mu =& \frac{1}{2} \left( \xi\, \partial_\mu\, \xi^\dagger + \xi^\dagger\, \partial_\mu\, \xi \right)
		\rightarrow U\, \mc{V}_\mu\, U^\dagger -\partial_\mu\, U\, U^\dagger\, ,\nonumber\\
	\mc{A}_\mu =& \frac{i}{2} \left( \xi\, \partial_\mu\, \xi^\dagger - \xi^\dagger\, \partial_\mu\, \xi \right)
		\rightarrow U\, \mc{A}_\mu\, U^\dagger\, .
\end{align}
The most general nucleon pion Lagrangian is then constructed from the $N$ and $\xi$ fields.  At LO this results in
\begin{equation}\label{eq:NucleonLO}
	\mathcal{L} = \ol{N}\, i \vit \cdot D\, N
			+2\, \a_M \ol{N} \mc{M}_+ N + 2\, \s_M \ol{N}N\, \tr (\mc{M}_+),
\end{equation}
where the chiral covariant derivative is given by
\begin{equation}
	(D_\mu\, N)_i = \partial_\mu\, N_i +\mc{V}_i^{\ j}\, N_j\, .
\end{equation}
In the isospin limit, $\mc{M}_+ \propto \mathbb{1}_{SU(2)}$ and thus the Lagrangian reduces to
\begin{equation}
	\mathcal{L} = \ol{N}\, i \vit \cdot D\, N + 2\, \tilde{\s}_M \ol{N}N\, \tr (\mc{M}_+)
\end{equation}
where $\tilde{\s}_M = \frac{1}{2}\a_M +\s_M$.  For generality, we will always use Eq.~\eqref{eq:NucleonLO} when computing the masses but will often specialize to the isospin limit.  

We also want to include the delta fields, which are spin-$\frac{3}{2}$ and isospin-$\frac{3}{2}$.  To incorporate the spin degrees of freedom we use a Rarita-Schwinger field~\cite{Rarita:1941mf}, $T^\mu$, satisfying the constraint $\g_\mu T^\mu = 0$.  To include the flavor degrees of freedom, we embed the delta fields in a rank-3 totally symmetric flavor tensor, which transforms under the chiral group as
\begin{equation}\label{eq:Ttransform}
	T_{ijk} \rightarrow U_i^{\ l}\, U_j^{\ m}\, U_k^{\ n}\, T_{lmn}\, ,
\end{equation}
with the normalization,
\begin{equation}\label{eq:Tnorm}
	T_{111} = \D^{++} \quad,\quad
	T_{112} = \frac{1}{\sqrt{3}} \D^{+} \quad,\quad
	T_{122} = \frac{1}{\sqrt{3}} \D^{0} \quad,\quad
	T_{222} = \D^{-}\, .
\end{equation}
The full LO Lagrangian including the delta fields is
\begin{align}\label{eq:HBChPTLO}
	\mathcal{L} =& \ol{N}\, i \vit \cdot D\, N
		+2\, \a_M \ol{N} \mc{M}_+ N + 2\, \s_M \ol{N}N\, \tr (\mc{M}_+) \nonumber\\
		&-i \ol{T}^\mu\, \vit \cdot D T_\mu +\D_0\, \ol{T}^\mu T_\mu
		+2\g_M\, \ol{T}^\mu\, \mc{M}_+\, T_\mu
		-2\ol{\s}_M\, \ol{T}^\mu T_\mu\, \tr ( \mc{M}_+ )
\end{align}
Here, $\D_0$ is leading order mass splitting between the nucleons and deltas in the chiral limit.  There is some ambiguity in the definition of $\D_0$ and what we denote as the renormalized mass splitting in the chiral limit, $\D$.  We will clarify this ambiguity in Section~\ref{sec:Nmass}.  This Lagrangian gives rise to the velocity dependent nucleon and delta propagators
\begin{align}
	\mc{G}_N(k) =&\ \frac{i}{k\cdot v +i\e}\, , \nonumber\\ 
	\mc{G}_T^{\mu\nu}(k) =&\ \frac{i P^{\mu\nu}}{k\cdot v -\D +i\e}\, ,
\end{align}
where $P^{\mu\nu}$ is a spin projector which projects out the spin-$\frac{3}{2}$ components of the Rarita-Schwinger field, and is given in d-dimensions (in the heavy baryon formalism) by
\begin{equation}
	P^{\mu\nu} = v^\mu v^\nu - g^{\mu\nu} -\frac{4}{d-1}S^\mu\, S^\nu\, ,
\end{equation}
where $S_\mu$ is the Pauli-Lubanski spin vector~\cite{Jenkins:1990jv,Jenkins:1991ne}.  

The chiral covariant derivative acting on the delta fields is
\begin{equation}\label{eq:CovDerivativeT}
	(D^\mu T)_{ijk} = \partial^\mu T_{ijk} + (\mc{V}^\mu)_{i}{}^{i'} T_{i'jk} 
		+ (\mc{V}^\mu)_{j}{}^{j'} T_{ij'k} + (\mc{V}^\mu)_{k}{}^{k'} T_{ijk'}\, .
\end{equation}

We also must include higher order terms which include the coupling of the nucleon and deltas to the axial field (these are the operators which give rise to the ``pion cloud" of the nucleon),
\begin{align}\label{eq:NNpi}
	\mathcal{L}=&\ 2 g_A \, \ol{N} \, S \cdot \mc{A} \, N 
		+ g_{\D N} \left( \ol{T}^\mu\, \mc{A}_\mu N + \ol{N} \mc{A}^\mu T_\mu \right)
		+ 2 g_{\D \D} \ol{T}^\mu \, S \cdot \mc{A} \, T_\mu\, .
\end{align}

This is the necessary formalism to compute many baryonic observables to LO and NLO in the chiral expansion.  Before applying this formalism to the baryon masses in Chapter~\ref{chap:BMasses}, we shall first discuss the 3 flavor theory and the partially quenched extensions of these theories.  Most of the formalism caries over from the 2 flavor theory to the others, and so we shall only highlight the differences as compared to the above discussion.

%%%%%%%%%%%%%%%%%%%%%%%%%%%%%%%%%%%%%%%%%%%%%%%%%%%
%
%	Octet and Decuplet
%
%%%%%%%%%%%%%%%%%%%%%%%%%%%%%%%%%%%%%%%%%%%%%%%%%%%
\subsubsection{\textbf{Octet and Decuplet Baryons}}

When considering the octet and decuplet baryons~\cite{Gell-Mann:1962xb,Gell-Mann:1964nj}, the only technical difference from the two-flavor theory is the group structure.  There are some simplifications which occur for $SU(2)$ as it has a real representation, which do not hold for $SU(3)$.  These will become clear in this section.  Another difference is that the inclusion of the strange quark, and the resulting mesons and baryons, makes the convergence of the HB\CPT\ more questionable.  For two-flavors, the expansion parameter is $m_\pi / \L_\chi \sim 15\%$ for the physical pion mass, whereas $m_K / \L_\chi \sim 40\%$.  In this sense it is more important to get a precise determination of the $SU(3)$ HB\CPT\ LECs to really test the convergence of the theory.

\bigskip

The octet baryons transform as an \textbf{8} under the $SU(3)_V$ flavor group,
\begin{equation}\label{eq:Btransform}
	B \rightarrow U\, B\, U^\dagger\, 
\end{equation}
and can be expressed in a traceless 2-dimensional matrix,
\begin{equation}\label{eq:B}
	B = \begin{pmatrix}
		\frac{1}{\sqrt{6}}\Lambda + \frac{1}{\sqrt{2}}\Sigma^0 & \Sigma^+ & p\\
		\Sigma^- & \frac{1}{\sqrt{6}}\Lambda -\frac{1}{\sqrt{2}}\Sigma^0 & n\\
		\Xi^- & \Xi^0 & -\frac{2}{\sqrt{6}}\Lambda\\
	\end{pmatrix}\, .
\end{equation}
The spin-$\frac{3}{2}$ decuplet baryons are still constructed with a Rarita-Schwinger field and a totally symmetric flavor tensor, $T^\mu_{ijk}$, and the same normalization of Eq.~\eqref{eq:Tnorm}, with $T_{113} = \frac{1}{\sqrt{3}} \S^{*+}$ etc.  The decuplet transforms as a \textbf{10} under the $SU(3)_V$ flavor group, as in Eq.~\eqref{eq:Ttransform}, except the flavor indices run from 1--3 instead of 1--2 as in the two-flavor theory.  The most general LO Lagrangian for the octet and decuplet fields is%
\footnote{We stress here that sometimes we will use the same notation for LECs in the three-flavor as in the two-flavor theory, however, they are different, and one should not confuse them.  This problem could be avoided with larger alphabets.}
\begin{align}\label{eq:HBChPTSU3}
	\mc{L} =&\ \tr \left( \ol{B}\, i v \cdot D\, B \right)
		+2b_D\, \tr \left( \ol{B} \left\{ \mc{M}_+, B \right\} \right)
		+2 b_F\, \tr \left( \ol{B} \left[ \mc{M}_+, B \right] \right)
		+2\s_M\, \tr \left( \ol{B} B \right)\, \tr \left( \mc{M}_+ \right) \nonumber\\
		&- \left( \ol{T}^\mu \left[ i v \cdot D -\D_0 \right] T_\mu \right)
		+2\g_M \left( \ol{T}^\mu\, \mc{M}_+\, T_\mu \right)
		-2\ol{\s}_M \left( \ol{T}^\mu T_\mu \right)\, \tr \left( \mc{M}_+ \right)\, .
\end{align}
The chiral covariant derivative on the decuplet field has the same form as in Eq.~\eqref{eq:CovDerivativeT}, while on the octet fields, is defined as
\begin{equation}
	D_\mu\, B = \partial_\mu B + \left[ \mc{V}_\mu, B \right]\, .
\end{equation}

The leading axial coupling of the mesons to the baryons is given by
\begin{align}\label{eq:BBpi}
	\mc{L} =&\ 2D\, \tr \left( \ol{B} S^\mu \left\{ \mc{A}_\mu, B \right\} \right)
		+2F\, \tr \left( \ol{B} S^\mu \left[ \mc{A}_\mu, B \right] \right)
		+2\mc{H}\, \left( \ol{T}^\mu\, S \cdot \mc{A}\, T_\mu \right) \nonumber\\
		&+ \mc{C} \left( \ol{T}^\mu\, \mc{A}_\mu B + \ol{B} \mc{A}_\mu \, T^\mu \right)\, .
\end{align}
Matching onto the two-flavor theory of Eq.~\eqref{eq:NNpi}, one learns
\begin{equation}
	g_A = D+F\quad,\quad g_{\D\D} = \mc{H}\quad,\quad g_{\D N} = \mc{C}\,  .
\end{equation}
We see that the $SU(3)$ HB\CPT\ Lagrangian, Eqs.~\eqref{eq:HBChPTSU3} and \eqref{eq:BBpi}, has more operators involving the spin-$\frac{1}{2}$ baryons than the corresponding Lagrangian in $SU(2)$.  These are the differences arising from the group structure we have mentioned.  Using the Lagrangian, Eq.~\ref{eq:HBChPTSU3}, one can confirm the Gell-Mann--Okubo (GMO) mass relation amongst the octet baroyns~\cite{Gell-Mann:1962xb,Okubo:1961jc}, which is violated at the one-loop order in HB\CPT\ from the interactions in Eq.~\eqref{eq:BBpi}.  We now move on to describe the partially quenched theories.

%%%%%%%%%%%%%%%%%%%%%%%%%%%%%%%%%%%%%%%%%%%%%%%%%%%
%
%	Partially Quenched ChPT
%
%%%%%%%%%%%%%%%%%%%%%%%%%%%%%%%%%%%%%%%%%%%%%%%%%%%
\subsection{\textbf{Partially Quenched Chiral Perturbation Theory} \label{sec:PQChPT}}

Partially quenched chiral perturbation theory (PQ\CPT) is the low energy EFT of partially quenched lattice QCD simulations, as described in Section~\ref{sec:LQCD}.  For LQCD simulations in which all the quark masses of the sea and valence sectors only perturbatively break chiral symmetry, the sicknesses which arise from the lack of unitarity can be systematically studied and understood with PQ\CPT%
\footnote{In the author's opinion, this is a great example of the power of effective field theory techniques.  Using EFT, one is not limited to studying rigorously defined quantum field theories, but can also study theories in which some of the cornerstone foundations of a quantum field theory are broken, eg. unitarity.  One can most likely never rigorously prove that partially quenched simulations away from the QCD limit do not produce uncontrolled effects, but partially quenched EFT is the only tool we have to study these systems, and to date, they have very successfully described partially quenched LQCD simulations.}
~\cite{Bernard:1993sv,Sharpe:1997by,Golterman:1997st,Sharpe:2000bc,Sharpe:2001fh,Savage:2001jw,Chen:2001yi,Beane:2002vq,Sharpe:2003vy,Beane:2002nu,Beane:2002np}.
To construct the chiral Lagrangians in the previous sections, we made one assumption about the symmetry breaking pattern induced by the QCD vacuum, that the chiral symmetry group is broken down to the vector subgroup.  To construct the partially quenched effective field theories, we will have to make a similar assumption and additionally one other assumption.  The second assumption is that when all the quark masses can be treated perturbatively, the unitarity violation introduced by having different masses in the sea and valence sectors of the PQLQCD action is completely encoded in the low energy effective field theory, partially quenched chiral perturbation theory (PQ\CPT).  This assumption is motivated by the fact that in the limit the sea and valence quark masses are degenerate, one recovers QCD.  As with the assumption about the chiral symmetry breaking induced by the QCD vacuum (and the PQQCD vacuum), this assumption can not be proven, and so must be checked by comparing PQ\CPT\ expressions to PQQCD lattice correlation functions, and verifying wether the predicted behavior is observed.  To date, there have been no signatures in PQQCD lattice simulations that PQ\CPT\ is somehow failing to capture the unitarity problems introduced by the sick theory.  The real test however, will come in the next few years as the pions masses are pushed further into the chiral regime~\cite{Bernard:2002yk,Beane:2004ks}.

\bigskip

We will now construct the PQ chiral Lagrangian.  Where possible, we will be general and use $N_v$ for the number of valence quarks and $N_s$ for the number of sea quarks, and will provide specific examples to help clarify certain points.  To cancel the contributions from closed valence quark loops, we introduce fictitious \textit{ghost} quarks, which are spin-$\frac{1}{2}$, but given Bose-statistics.%
\footnote{There is an alternative method to partially quench which is known as the \textit{replica method}~\cite{Damgaard:2000gh}.}
The mass of the ghost quarks is set exactly equal to the corresponding valence quark mass, resulting in a complete cancellation of the valence loops (which is manifested as a ghost-quark determinant identical in value but inverse to the fermionic determinant of the valence quarks in the path integral formulation of the theory).  The quark content of the PQQCD Lagrangian is
\begin{equation}\label{eq:PQQCD}
	\mc{L}^{PQ} = \sum_{j,k}^{2N_v+N_s} \ol{Q}^j\, \left( i \Dslash -m_Q \right)_{j}^{\ k}\, Q_k\, .
\end{equation}
As in Eq.~\eqref{eq:QCD}, for $m_Q << \L_{QCD}$, the PQQCD Lagrangian exhibits an approximate
$SU(N_v+N_s|N_v) \otimes SU(N_v+N_s|N_v)_R$ \textit{graded}~\cite{BahaBalantekin:1981qy} chiral symmetry.  One then assumes the PQQCD vacuum spontaneously breaks this symmetry down to the vector subgroup, giving rise to $(2N_v+N_s)^2 -1$ pseudo-Nambu-Goldstone mesons.  Then, following the construction of the pions in Section~\ref{sec:SU2}, we construct the partially quenched mesons representing the lowest energy excitations about the PQQCD vacuum.

%%%%%%%%%%%%%%%%%%%%%%%%%%%%%%%%%%%%%%%%%%%%%%%%%%%
%
%	PQ Mesons
%
%%%%%%%%%%%%%%%%%%%%%%%%%%%%%%%%%%%%%%%%%%%%%%%%%%%
\subsubsection{\textbf{Mesons}}

At leading order, the form of the partially quenched chiral Lagrangian for the mesons has almost the exact same form as in the $SU(2)$ or $SU(3)$ theory, with trace operators over flavor indices replaced with super-trace operators.  In addition, there are operators involving the singlet field, 
$\Phi_0 \propto \str(\Phi)$ which contribute to the leading two-point functions for the flavor diagonal meson fields.  The Lagrangian is
\begin{equation}\label{eq:PQChPT}
	\mc{L} = \frac{f^2}{8} \str \left( \partial_\mu \S \partial^\mu \S^\dagger \right)
		+\frac{ f^2\, B_0}{4} \str \left( m_Q \S^\dagger + \S m_Q^\dagger \right)
		+\a_\Phi\, \partial_\mu \Phi_0\, \partial^\mu\, \Phi_0
		-m_0^2 \Phi_0^2\, .
\end{equation}
The $\S$ field here contains many more mesons than in the two or three flavor theory, 
\begin{equation}
	\S = {\rm exp} \left( \frac{2 i \Phi}{f} \right) ,  \;\;\; 
	\Phi = \begin{pmatrix}
			M & \chi^\dagger\\
			\chi & \tilde{M}\\
		\end{pmatrix}\, .
\label{eq:sigma}
\end{equation}
The matrices $M$ and $\tilde{M}$ contain bosonic mesons while $\chi$
and $\chi^\dagger$ contain fermionic mesons. Specifically,
\begin{align}\label{eq:mesons}
	M &=\begin{pmatrix}
			\eta_u & \pi^+ & K^+ & \phi_{uj} & \phi_{ul} & \phi_{ur} \\
			\pi^- & \eta_d & K^0 & \phi_{dj} & \phi_{dl} & \phi_{dr} \\
			K^- & \ol{K}^0 & \eta_s & \phi_{sj} & \phi_{sl} & \phi_{sr} \\
			\phi_{ju} & \phi_{jd} & \phi_{js} & \eta_j & \phi_{jl} & \phi_{jr} \\
			\phi_{lu} & \phi_{ld} & \phi_{ls} & \phi_{lj} & \eta_l & \phi_{lr} \\
			\phi_{ru} & \phi_{rd} & \phi_{rs} & \phi_{rj} & \phi_{rl} & \eta_{r} \\
		\end{pmatrix}\quad ,\quad 
	\tilde M = \begin{pmatrix}
			{\tilde \eta}_u & {\tilde \pi}^+ & \tilde{K}^+ \\
			{\tilde \pi}^- & {\tilde \eta}_d & \tilde{K}^0 \\
			\tilde{K}^- & \tilde{\ol{K}}^0 & \tilde{\eta}_{s} \\
		\end{pmatrix}
	\nonumber\\
	\chi &= \begin{pmatrix}
		\phi_{\tilde{u} u} & \phi_{\tilde{u} d} & \phi_{\tilde{u} s} & \phi_{\tilde{u} j} & \phi_{\tilde{u} l} 
			& \phi_{\tilde{u} r} \\
		\phi_{\tilde{d} u} & \phi_{\tilde{d} d} & \phi_{\tilde{d} s} & \phi_{\tilde{d} j} & \phi_{\tilde{d} l} 
			& \phi_{\tilde{d} r} \\
		\phi_{\tilde{s} u} & \phi_{\tilde{s} d} & \phi_{\tilde{s} s} & \phi_{\tilde{s} j} & \phi_{\tilde{s} l} 
			& \phi_{\tilde{s} r} \\
		\end{pmatrix}.
\end{align}
The upper $N_v \times N_v$ block of $M$
contains the usual mesons composed of a valence quark and anti-quark.  The lower $N_s \times N_s$ block of $M$ contains the sea quark-antiquark mesons and the off-diagonal block elements of $M$ contain bosonic mesons of mixed valence-sea type.  Expanding the Lagrangian~\eqref{eq:PQChPT} to quadratic order in the meson fields, one finds that for a meson composed of quark and anti-quark of flavor $Q$ and $Q^\prime$, the LO mass of the mesons is given by
\begin{equation}
	m_{QQ^\prime}^2 = B_0 (m_Q +m_{Q^\prime})\, .
\end{equation}

As in QCD, PQQCD has a strong axial anomaly, and thus the mass of the singlet field, $m_0$ can be taken to be on the order of the chiral symmetry breaking scale, $m_0 \rightarrow \L_\chi$.  In this limit, the two point function between the flavor diagonal mesons, the $\eta$ fields, deviate from a simple, single-pole form.  The momentum space two-point correlator between the $\eta_a$ and $\eta_b$ fields is found at LO to be~\cite{Sharpe:2001fh}
\begin{equation}\label{eq:etaProp}
	\mc{G}_{\eta_a \eta_b}(p^2) =
		\frac{i \e_a \d_{ab}}{p^2 -m_{\eta_a}^2 +i\e}
		- \frac{i}{N_f} \frac{\prod_{k=1}^{N_f}(p^2 -\tilde{m}_{k}^2 +i\e)}
			{(p^2 -m_{\eta_a}^2 +i\e)(p^2 -m_{\eta_b}^2 +i\e) \prod_{k^\prime=1}^{N_f-1}
				(p^2 -\tilde{m}_{k^\prime}^2 +i\e)},
\end{equation}
where
\begin{equation}
	\e_a = \left\{ \begin{array}{ll}
			+1& \text{for a = valence or sea quarks}\\
			-1 & \text{for a = ghost quarks}\,.
			\end{array} \right.
\end{equation}
In Eq.~\eqref{eq:etaProp}, $k$ runs over the flavor neutral states
($\phi_{jj}, \phi_{ll}, \phi_{rr}$) and $k^\prime$ runs over the
mass eigenstates of the sea sector.  Here, we wish to show how the unitarity violations and partial quenching can be parameterized.  To the author's knowledge, the following discussion was first brought up in Ref.~\cite{Beane:2002nu,Beane:2002np} as a means to quantify these unitarity violations in the two-nucleon system.  In Ref.~\cite{Chen:2005ab}, this discussion was then expanded to the meson sector for both the two quark flavor as well as three quark flavor theory.  There is additionally a very similar discussion in Ref.~\cite{Sharpe:2000bc} regarding the same issues.  We refer the reader to Chapter~\ref{chap:pipiMA} for the full details, but here highlight the main idea.  When computing correlation functions with the partially quenched (or mixed action) theory, one can use a basis of fields other than the $\eta_{uu}$, $\eta_{dd}$ \textit{etc.} which naturally arise from Eq.~\eqref{eq:PQChPT}.  For example, in a theory with two valence quarks, and two sea quarks, in the isospin limit, one can instead work with the fields,
\begin{equation}
	\pi^0 = \frac{1}{\sqrt{2}} \left( \eta_u - \eta_d \right)\quad , \quad
	\bar{\eta} = \frac{1}{\sqrt{2}} \left( \eta_u +\eta_d \right)\, .
\end{equation}
The usefulness of this basis choice becomes very clear when one constructs the leading propagators for these fields
\begin{align}
	\mc{G}_{\pi^0}(p^2) &= \frac{i}{p^2 -m_\pi^2 +i\e},\label{eq:PionProp} \\
	\mc{G}_{\bar{\eta}}(p^2) &= \frac{i}{p^2 -m_\pi^2 +i\e} 
			-i \frac{p^2 -m_{jj}^2}{(p^2 -m_\pi^2 +i\e)^2} \nonumber\\
		&= \frac{i \D_{PQ}^2}{(p^2 -m_\pi^2 +i\e)^2} \label{eq:EtaBarProp}\, .
\end{align}
We see that the propagator for the $\pi^0$ has the exact form as in \CPT, which is not surprising as in the isospin limit, the $\pi^0$ is an isospin 1 particle, and thus has no overlap with the vacuum (times $\g_5$).  However, the $\ol{\eta}$-propagator is proportional to the partially quenched breaking parameter 
\begin{equation}\label{eq:DeltaPQ}
	\D_{PQ}^2 =  m_{sea}^2 - m_{valence}^2\, ,
\end{equation}
where $m_{sea}$ is the mass of a pion composed of two sea quarks and $m_{valence}$ is the mass of a pion composed of two valence quarks.  The double-pole structure of the $\ol{\eta}$-propagator is a consequence of the unitarity violations introduced in the partially quenched theory.  In the theory with two valence and two sea quarks, all unitarity violating terms can be traced back to loop graphs involving this $\ol{\eta}$-field.  We explicitly see here that these effects vanish in the QCD limit, $m_{sea} = m_{valence}$.  A similar change of basis can be performed for the two flavor theory away from the isospin limit, and also for the theory with three valence and sea quarks, although in these cases, the algebra is more involved.  However, this parameterization, Eq.~\eqref{eq:EtaBarProp} and the equivalent parameterizations for the more involved theories provides a means to quantify the partially quenched breaking, or unitarity violations which arise in these sick theories.  We will demonstrate a clear example of the usefulness of this in Chapter~\ref{chap:pipiMA}.

%%%%%%%%%%%%%%%%%%%%%%%%%%%%%%%%%%%%%%%%%%%%%%%%%%%
%
%	PQ Baryons
%
%%%%%%%%%%%%%%%%%%%%%%%%%%%%%%%%%%%%%%%%%%%%%%%%%%%
\subsubsection{\textbf{Baryons}}

In PQ$\chi$PT the baryons are composed of three quarks, $Q_iQ_jQ_k$,
where $i-k$ can be valence, sea or ghost quarks. 
One decomposes the irreducible representations of $SU(N_s+N_v | N_v)_V$ into
irreducible representations of $SU(N_v)_{val} \otimes SU(N_s)_{sea}
\otimes SU(N_v)_{ghost} \otimes U(1)$.  The method for including the
octet and decuplet baryons into PQ$\chi$PT is to use the interpolating
field~\cite{Labrenz:1996jy,Chen:2001yi}:
\begin{equation}
	\mc{B}_{ijk}^\g \sim
		\left(Q_i^{\a,a} Q_j^{\beta,b} Q_k^{\g,c}-Q_i^{\a,a} Q_j^{\g,c} Q_k^{\beta,b}\right)
		\epsilon_{abc}(C\g_5)_{\a \beta}.
\end{equation}
We then require that when the indices $i,j,k,$ are restricted to the valence sector, 
$\mc{B}_{ijk} = B_{ijk}$, where
\begin{align}
	B_{ijk} &= \frac{1}{\sqrt{6}}\left( \e_{ij}\, N_k +\e_{ik}\, N_j \right)
		&\text{for $SU(4|2)$,} \nonumber\\
	B_{ijk} &= \frac{1}{\sqrt{6}}  \left( \e_{ijl}\, B_k^{\ l} + \e_{ikl}\, B_j^{\ l} \right)
		&\text{for $SU(6|3)$,}
\end{align}	
where $N_i$ is defined in Eq.~\eqref{eq:N}, $B_i^{\ j}$ is defined in Eq.~\eqref{eq:B} and $\e_{ij}$ and $\e_{ijk}$ are the completely anti-symmetric 2 and 3 dimensional tensors respectively.  Thus the octet baryons are contained as an $(\mathbf{8,1,1})$ in the $\mathbf{240}$ representation of 
$SU(3)_{val} \otimes SU(3)_{sea} \otimes SU(3)_{ghost} \otimes U(1)$~\cite{Savage:2001dy,Chen:2001yi}, while the nucleons are contained as an $(\mathbf{2,1,1})$ in the $\mathbf{70}$ representation of 
$SU(2)_{val} \otimes SU(2)_{sea} \otimes SU(2)_{ghost} \otimes U(1)$~\cite{Beane:2002vq}.
In addition to the conventional baryons composed of valence quarks, $\mc{B}_{ijk}$
also contains baryon fields composed of sea and ghost quarks.  In this
paper we only need the baryons which contain at most one
sea or ghost quark, and these states have been explicitly constructed
in Ref.~\cite{Savage:2001dy,Chen:2001yi} for $SU(6|3)$ and in Ref.~\cite{Beane:2002vq} for 
$SU(4|2)$.  

Under the interchange of flavor indices, one finds~\cite{Labrenz:1996jy}:
\begin{equation}
\mc{B}_{ijk} = (-)^{1+\eta_j \eta_k}\mc{B}_{ikj}
           \quad {\rm and} \quad
  \mc{B}_{ijk} + (-)^{1+\eta_i \eta_j}\mc{B}_{jik}
                 + (-)^{1+\eta_i \eta_j + \eta_j \eta_k + \eta_i
                   \eta_k} \mc{B}_{kji}
                 =0\, ,
\end{equation}
where the commuting or anti-commuting nature of the quark fields are handled with the grading factors which are defined as
\begin{equation}
	\eta_k = \left\{ 
		\begin{array}{cl}
		1 & \text{for } k=\textrm{ valence or sea quark} \\
		0 & \text{for } k=\textrm{ ghost quark}
		\end{array} 
	\right. .
\end{equation}

Similarly, one can construct the spin-$\frac{3}{2}$ baryons.  The decuplet baryons are embedded as an 
$(\mathbf{10,1,1})$ in the $\mathbf{138}$ representation of 
$SU(3)_{val} \otimes SU(3)_{sea} \otimes SU(3)_{ghost} \otimes U(1)$~\cite{Savage:2001dy,Chen:2001yi} and the delta baryons are embedded as an $(\mathbf{4,1,1})$ of the $\mathbf{40}$ representation of 
$SU(2)_{val} \otimes SU(2)_{sea} \otimes SU(2)_{ghost} \otimes U(1)$~\cite{Beane:2002vq}.
An interpolating field for the spin-$\frac{3}{2}$ baryons is
\begin{equation}
  \mc{T}_{ijk}^{\a,\mu} \sim
      \left( Q_i^{\a,a} Q_j^{\beta,b} Q_k^{\g,c}
            +Q_i^{\beta,b} Q_j^{\g,c} Q_k^{\a,a}
            +Q_i^{\g,c} Q_j^{\a ,a} Q_k^{\beta,b}
      \right)
         \epsilon_{abc} \left(C\g^\mu \right)_{\beta \g}.
\end{equation}
We require that $\mc{T}_{ijk} = T_{ijk}$ defined in Eq.~\eqref{eq:Tnorm}, when the indices $i,j,k$ are
restricted to the valence sup-space.  In addition to the conventional decuplet resonances composed of
valence quarks, $\mc{T}_{ijk}$ contains fields with sea and ghost
quarks, which have also been constructed in Ref.~\cite{Savage:2001dy,Chen:2001yi} for $SU(6|3)$ and Ref.~\cite{Beane:2002vq} for $SU(4|2)$.  Under the interchange of flavor indices, one finds that
\begin{equation}
\mc{T}_{ijk} = (-)^{1 + \eta_i \eta_j}\mc{T}_{jik} = 
                 (-)^{1 + \eta_j \eta_k}\mc{T}_{ikj}\, .
\end{equation}
Under $SU(6|3)_V$, both $\mc{B}_{ijk}$ and ${\cal
  T}_{ijk}$ transform as \cite{Labrenz:1996jy}
\begin{equation}
\mc{B}_{ijk} \rightarrow
       (-)^{\eta_l (\eta_j +\eta_m)+(\eta_l + \eta_m)(\eta_k +
         \eta_n)} U_i^{\hskip 0.3em l} U_j^{\hskip 0.3em m} 
                  U_k^{\hskip 0.3em n}
               \mc{B}_{lmn}\, .
\end{equation}

\subsubsection{\textbf{Leading order partially quenched heavy baryon Lagrangian} \label{sec:LOPQHBChPT}}

The form of the heavy baryon Lagrangian in PQ\CPT\ is independent of whether one is working with the $SU(4|2)$ or the $SU(6|3)$ theory, at least through $\mc{O}(m_q^2)$.  Therefore, we shall construct both Lagrangians simultaneously, but it is important to remember that the values of the LECs will differ between the two theories.  This is because in the $SU(4|2)$ theory, the effects of the valence, ghost and sea quarks are implicitly contained in modifications of the $SU(4|2)$ LECs.  One difference between PQ theory and the unquenched theories is the addition of the grading factors which keep track of the commuting or anti-commuting nature of a given baryon.  In the following Lagrangian, we shall use the flavor contractions defined in Ref.~\cite{Labrenz:1996jy}.  For a matrix $Y^i_{\ j}$ acting in flavor-space, the contractions are
\begin{align}\label{eq:PQcontractions}
	\left( \ol{\mc{B}}\, \mc{B} \right) &= \ol{\mc{B}}^{kji}\, \mc{B}_{ijk}
	&,& \quad
	\left( \ol{\mc{T}}^\mu\, \mc{T}_\mu \right) = 
		\ol{\mc{T}}^{\mu,kji}\, \mc{T}_{\mu,ijk}\, , \nonumber\\
	\left( \ol{\mc{B}}\, Y\, \mc{B} \right) &= 
		\ol{\mc{B}}^{kji}\, Y_i^{\ l}\, \mc{B}_{ljk}
	&,& \quad
	\left( \ol{\mc{T}}^\mu\, Y\, \mc{T}_\mu \right) = 
		\ol{\mc{T}}^{\mu,kji}\, Y_i^{\ l}\, \mc{T}_{\mu,ljk}\, , \nonumber\\
	\left( \ol{\mc{B}}\, \mc{B}\, Y \right) &=
		(-)^{(\eta_i+\eta_j)(\eta_k+\eta_l)}\, 
		\ol{\mc{B}}^{kji}\, Y_k^{\ l}\, \mc{B}_{ijl}
	&,&\quad
	\left( \ol{\mc{B}}\, Y^\mu\, \mc{T}_\mu \right) =
		\ol{\mc{B}}^{kji}\, (Y^\mu)_i^{\ l}\, \mc{T}_{\mu,ljk}\, .
\end{align}
Using these flavor contractions, the LO partially quenched heavy baryon Lagrangian is given by~\cite{Savage:2001jw,Chen:2001yi,Beane:2002vq,Labrenz:1996jy}
\begin{align}
\mc{L}^{PQ} =&\ 
	\left( \ol{\mc{B}}\, i\vD\ \mc{B} \right)\ 
	+2\a_M^{(PQ)}\, \left( \ol{\mc{B}}\, \mc{B} \mc{M}_+ \right)
	+2\b_M^{(PQ)}\, \left( \ol{\mc{B}}\, \mc{M}_+ \mc{B} \right)
	+2\s_M^{(PQ)}\, \left( \ol{\mc{B}}\, \mc{B} \right) {\rm str}(\mc{M}_+) \nonumber\\
& -\left(\ol{\mc{T}}^{\mu}\left[\ i\vD-\Delta\right]\mc{T}_\mu \right)
	+2\g_M^{(PQ)}\, \left(\ol{\mc{T}}^\mu \mc{M}_+ \mc{T}_\mu \right) 
	-2\ol{\s}_M^{(PQ)}\, \left(\ol{\mc{T}}^\mu \mc{T}_\mu \right) {\rm str}(\mc{M}_+) \nonumber\\
& +2\a^{(PQ)}\, \left(\ol{\mc{B}}\, S^\mu \mc{B} \mc{A}_\mu \right) 
	+ 2\b^{(PQ)} \left(\ol{\mc{B}}\, S^\mu \mc{A}_\mu \mc{B} \right) 
	+ 2{\cal H}^{(PQ)} \left( \ol{\mc{T}}^\nu S^\mu \mc{A}_\mu \mc{T}_\nu \right) \nonumber\\
& +\sqrt{\frac{3}{2}}{\mc{C}}^{(PQ)} \Big[ \left( \ol{\mc{T}}^\nu \mc{A}_\nu \mc{B}\right)
	+ \left( \ol{\mc{B}}\, \mc{A}_\nu \mc{T}^\nu \right) \Big]
\label{eq:leadlagPQ}\, .
\end{align}
One can then determine the relation between these PQ LECs and the LECs in the two and three flavor theories by restricting the flavor indices in Eq.~\eqref{eq:leadlagPQ} to the valence sector.  This leads to the relations between the $SU(2)$ Lagrangians, Eqs.~\eqref{eq:HBChPTLO}, \eqref{eq:NNpi} and the $SU(4|2)$ PQ Lagrangian,
\begin{align}
	\a_M =&\ \frac{2}{3}\a_M^{(4|2)} -\frac{1}{3}\b_M^{(4|2)}
	&,& \qquad \g_M^{(2)} = \g_M^{(4|2)}\, , \nonumber\\
	\s_M^{(2)} =&\ \s_M^{(4|2)} +\frac{1}{6}\a_M^{(4|2)} +\frac{2}{3}\b_M^{(4|2)}
	&,& \qquad \ol{\s}_M^{(2)} = \ol{\s}_M^{(4|2)}\, , \nonumber\\
	g_A =&\ \frac{1}{3} \left( 2\a^{(4|2)} -\b^{(4|2)} \right) 
	&,&\qquad g_{\D \D} = \mc{H}^{(4|2)}\, , \nonumber\\
	g_1 =& \frac{1}{3} \left( \a^{(4|2)} +4\b^{(4|2)} \right)
	&,&\qquad g_{\D N} = -\mc{C}^{(4|2)}\, .
\end{align}
Similarly, when matching the $SU(6|3)$ PQ Lagrangian to the $SU(3)$ Lagrangians of Eqs.~\eqref{eq:HBChPTSU3} and \eqref{eq:BBpi}, one finds the relations,
\begin{align}
	b_D =&\ \frac{1}{4} \left( \a_M^{(6|3)} -2\b_M^{(6|3)} \right)
	&,&\qquad \g_M^{(3)} = \g_M^{(6|3)}\, , \nonumber\\
	b_F =&\ \frac{1}{12} \left( 5\a_M^{(6|3)} +2\b_M^{(6|3)} \right)
	&,& \qquad \ol{\s}_M^{(3)} = \ol{\s}_M^{(6|3)}\, , \nonumber\\
	\s_M^{(3)} =&\ \s_M^{(6|3)} +\frac{1}{6}\a_M^{(6|3)} +\frac{2}{3}\b_M^{(6|3)}
	&,&\qquad \mc{H} = \mc{H}^{(6|3)}\, , \nonumber\\
	D =&\ \frac{1}{4} \left( \a^{(6|3)} -2\b^{(6|3)} \right) 
	&,&\qquad \mc{C} = -\mc{C}^{(6|3)}\, ,\nonumber\\
	F =&\ \frac{1}{12} \left( 5\a^{(6|3)} +2\b^{(6|3)} \right)\, .
\end{align}

With this formalism, we shall now proceed to further develop the Lagrangians as needed, and apply these effective field theory techniques to understand various baryonic observables.  We begin with a detailed look at the masses of the lowest lying spin-$\frac{1}{2}$ and spin-$\frac{3}{2}$ baryons in HB\CPT\ and the partially quenched analogues.

% ========== Chapter 2: Baryon Masses
\chapter{Baryon Masses}\label{chap:BMasses}

Perhaps the simplest question we can ask is what is the spectrum of hadron states which emerge from QCD?  A quick glance at the Particle Data Book~\cite{Eidelman:2004wy} shows a very rich spectrum.  A significant understanding of these hadrons can be made by using the approximate 
$SU(3)_{\textrm{Flavor}}$ symmetry of QCD, the most famous example being the Gell-Mann--Okubo mass relations amongst the octet baryons~\cite{Okubo:1961jc,Ne'eman:1961cd,Gell-Mann:1962xb} and the Gell-Mann--Oakes--Renner relation for the octet mesons~\cite{Gell-Mann:1968rz}.  However, we would like to be able to make a rigorous connection between the QCD Lagrangian and the observed particle spectrum, to arbitrary precision.  This can be accomplished for the lightest hadrons (with a given set of quantum numbers) by utilizing lattice QCD in conjunction with chiral perturbation theory, as described in Chapter~\ref{chap:NPLQCD}.

We are interested in extending our understanding of the masses of the lowest lying baryons of 
spin-$\frac{1}{2}$ and spin-$\frac{3}{2}$ to NNLO, or $\mc{O}(m_q^2)$, in the chiral expansion.  This is important for several reasons.  Firstly, as HB\CPT\ is an asymptotic series, there is no guarantee that it is a convergent series, and so it is always good to push the order to which we know the theory to higher orders and test the convergence.  Moreover, the expansion parameter of HB\CPT\ is not as good as \CPT\ with purely mesons, and so the expansion is expected to have larger corrections order by order.  Thus, to do precision mass calculations (or precise determinations of any baryon observable) it is necessary to compute to higher orders in the effective expansion as compared to mesonic quantities.  Therefore, the focus of this chapter is the computation of the $\mc{O}(m_q^2)$ mass corrections to the above mentioned baryons, using heavy baryon \CPT\ and partially quenched HB\CPT.

To accomplish this task, we must first construct the heavy baryon chiral Lagrangian to the next relevant order, beyond the known orders of Eqs.~\eqref{eq:HBChPTLO}~and~\eqref{eq:NNpi}.  We first do this for the two-flavor chiral Lagrangian (up and down), and then for the theory with three flavors (up, down and strange), and finally extend both of these Lagrangians to their partially quenched versions.  This chapter is based upon the work in Refs.~\cite{Walker-Loud:2004hf,Tiburzi:2004rh,Tiburzi:2005na}.

%%%%%%%%%%%%%%%%%%%%%%%%%%%%%%%%%%%%%%%%%%%%%%%%%%%
%
%	SU(2) Masses
%
%%%%%%%%%%%%%%%%%%%%%%%%%%%%%%%%%%%%%%%%%%%%%%%%%%%
\section{Nucleon and Delta Masses in $SU(2)$ \CPT \label{sec:Nmass}}

\subsection*{\textbf{Review of nucleon mass in \CPT }}

We begin by reviewing the nucleon mass to NLO in the chiral expansion, to set conventions and to motivate the need for knowing the NNLO correction to the baryon masses.  It is useful to parameterize the nucleon mass as follows,
\begin{equation}\label{eq:NucMassExp}
     M_{N_i} = M_0 \left(\D \right) - M_{N_i}^{(1)}\left(\D,\mu \right)
                - M_{N_i}^{(3/2)}\left(\D,\mu \right)
                - M_{N_i}^{(2)}\left(\D,\mu \right) + \ldots
\end{equation}
The leading contribution to the nucleon mass, $M_0$, is generated by the non-perturbative dynamical interactions between the quarks and gluons and is independent of the quark masses; this parameter is often referred to as the nucleon mass in the chiral limit.  The parameter, $\D$, is the quark-mass independent splitting between the delta mass and the nucleon mass, and is phenomenologically determined to be $\D \sim 293$~MeV.%
\footnote{We note that the experimental mass splitting between the deltas and nucleons is determined to be $\D_{phys} = 293$~MeV~\cite{Eidelman:2004wy}, but that this value includes quark mass dependence as well.  The difference, however, between using this value for $\D$ in \CPT\ and the true chiral limit value of $\D$ is a higher order effect.}
This mass splitting arises dynamically from QCD and vanishes in the large $N_C$ limit~\cite{Witten:1979kh}.  The terms, $M_{N_i}^{(n)}$, are contributions to the mass of the $i^{th}$ nucleon of the order $(m_q)^n$.  In Eq.~\eqref{eq:NucMassExp}, $\mu$ is the renormalization scale which arises when regulating divergent loop diagrams.  Throughout this chapter, we use dimensional regularization with a modified minimal subtraction ($\ol{\rm   MS}$) scheme, in which we consistently subtract terms proportional to 
\begin{equation}\label{eq:MSbar}
	\frac{2}{4-d} - \gamma_E + 1 + \log 4 \pi\, ,
\end{equation}
where $d$ is the number of space-time dimensions.  

There are a few points worth noting regarding the chosen form of the mass expansion.  First, the minus signs in Eq.~\eqref{eq:NucMassExp} are purely for notational convenience.  Second, the renormalization scale dependence appears at each order because we are treating the LECs as polynomial functions of $\Delta$.  This is motivated by our interest in comparing \CPT\ expressions to lattice QCD correlation functions in which the quark masses (but not $\D$) can be varied.  Moreover, as we will show, because one does not have the ability to vary $\D$, either in nature or with lattice QCD, there is some ambiguity in how to include the finite analytic contributions proportional to this mass splitting.%
\footnote{We note as an aside, that if we instead were to
treat this dependence explicitly and expand the nucleon mass in powers of 
$q$, where $\Delta \sim q$ and $ m_q \sim q^2$, the mass (and all baryonic observables) would then take the form
\begin{equation}
	M_{N_i} = m_0  -  m_{N_i}^{(2)} - m_{N_i}^{(3)} - m_{N_i}^{(4)} + \ldots
\nonumber,\end{equation}
with $m_{N_i}^{(n)}$ as the renormalization-scale independent mass contribution 
to the $i^{th}$ nucleon strictly of the order $q^n$. The parameter $m_0$ is
nucleon mass in the chiral limit, but only to leading order in $\Delta$, i.e. $m_0 = M_0(\D=0)$.}
We will now make this explicit by reviewing the chiral contributions to the nucleon mass.

\begin{figure}
\center
\begin{tabular}{ccc}
\includegraphics[width=0.15\textwidth]{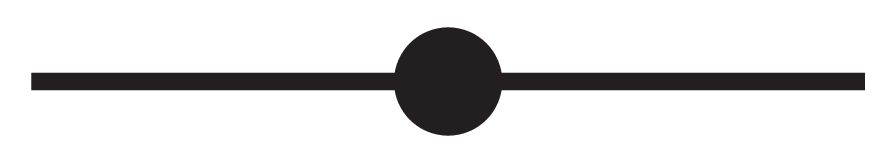} &
\includegraphics[width=0.3\textwidth]{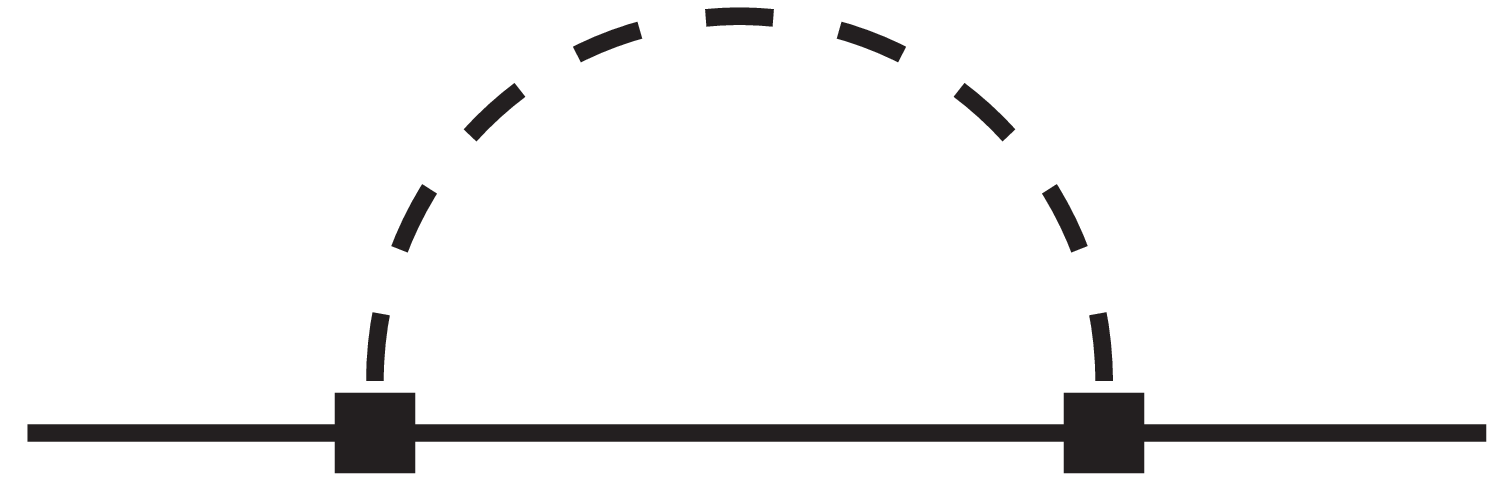} &
\includegraphics[width=0.3\textwidth]{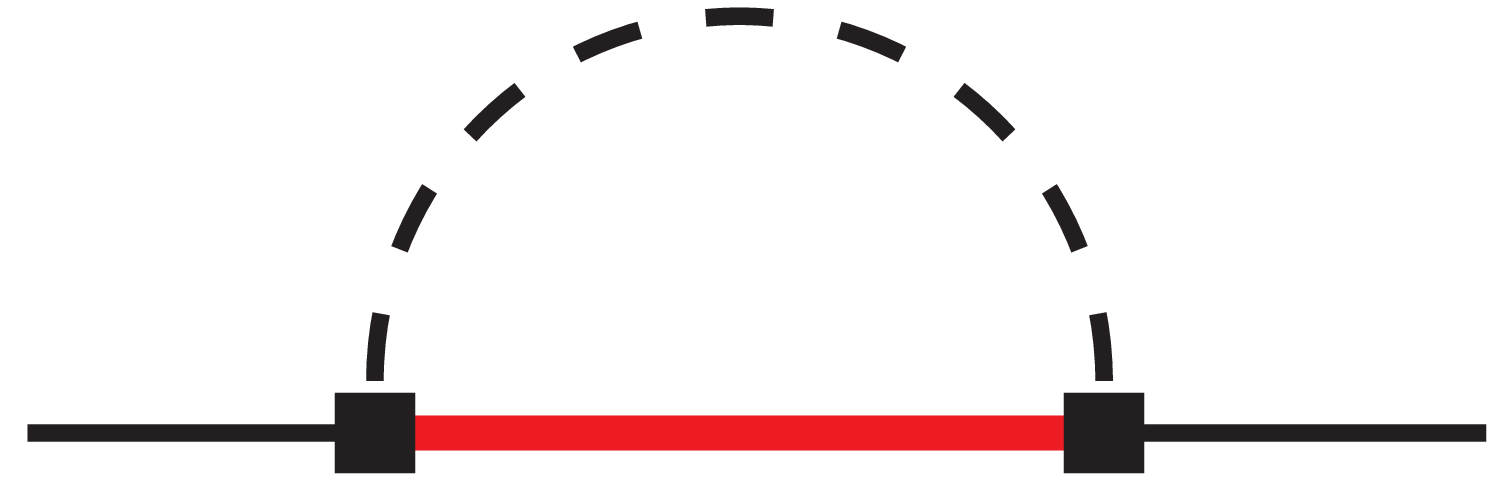} \\
(a) & (b) & (c)
\end{tabular}
\caption[Feynman diagrams depicting the leading order and next-to-leading order quark mass contribution to nucleon mass.]{\label{fig:LONmass}Leading order  and NLO quark mass dependence of nucleon mass.  The LO contribution, diagram (a), is given by the Lagrangian in Eq.~\eqref{eq:NucleonLO}, while the NLO pion-loop contributions, diagrams (b) and (c), arise from the axial coupling of the pions to the nucleons and deltas given in Eq.~\eqref{eq:NNpi}.  The solid black lines, thick-solid line (red online), and the dotted black lines denote nucleons, delta-resonances and pions respectively.}
\end{figure}

\bigskip
The leading quark mass dependence of the nucleon, $M_{N_i}^{(1)}$ comes from the diagram depicted in Figure~\ref{fig:LONmass}.
%%%%%%%%%%%%%%%%%%%%%%%%%%%%%%%%%%%%%%%%%%%%%%%%%%%
%
%	fig: LO nucleon mass
%
%%%%%%%%%%%%%%%%%%%%%%%%%%%%%%%%%%%%%%%%%%%%%%%%%%%
%%%%%%%%%%%%%%%%%%%%%%%%%%%%%%%%%%%%%%%%%%%%%%%%%%%
%
%%%%%%%%%%%%%%%%%%%%%%%%%%%%%%%%%%%%%%%%%%%%%%%%%%%
It is a simple tree-level diagram giving rise to the mass correction
\begin{equation}\label{eq:LONmass}
  M^{(1)}_{N_i} 
    = 2 \a_M m_i 
      +2 \sigma_M \, (m_u +m_d),
\end{equation}
where the nucleon dependent contribution is given by
\begin{equation}\label{eq:mB}
  m_i = \left\{
    \begin{array}{lc}
      m_u, & i=p \\
      m_d, & i=n
    \end{array}\right. .
\end{equation}

The leading pion loop graphs depicted in Fig.~\ref{fig:LONmass}(b) and \ref{fig:LONmass}(c) give rise to non-analytic dependence upon the quark mass.  It turns out when using dimensional regularization, the loop graph with an internal nucleon is completely finite, giving rise to the self energy correction
\begin{equation}
	\d \S_{N-loop} = -3 \pi\, g_A^2\, \frac{m_\pi^3}{\L_\chi^2}\, .
\end{equation}

\begin{figure}
\center
\includegraphics[width=0.8\textwidth]{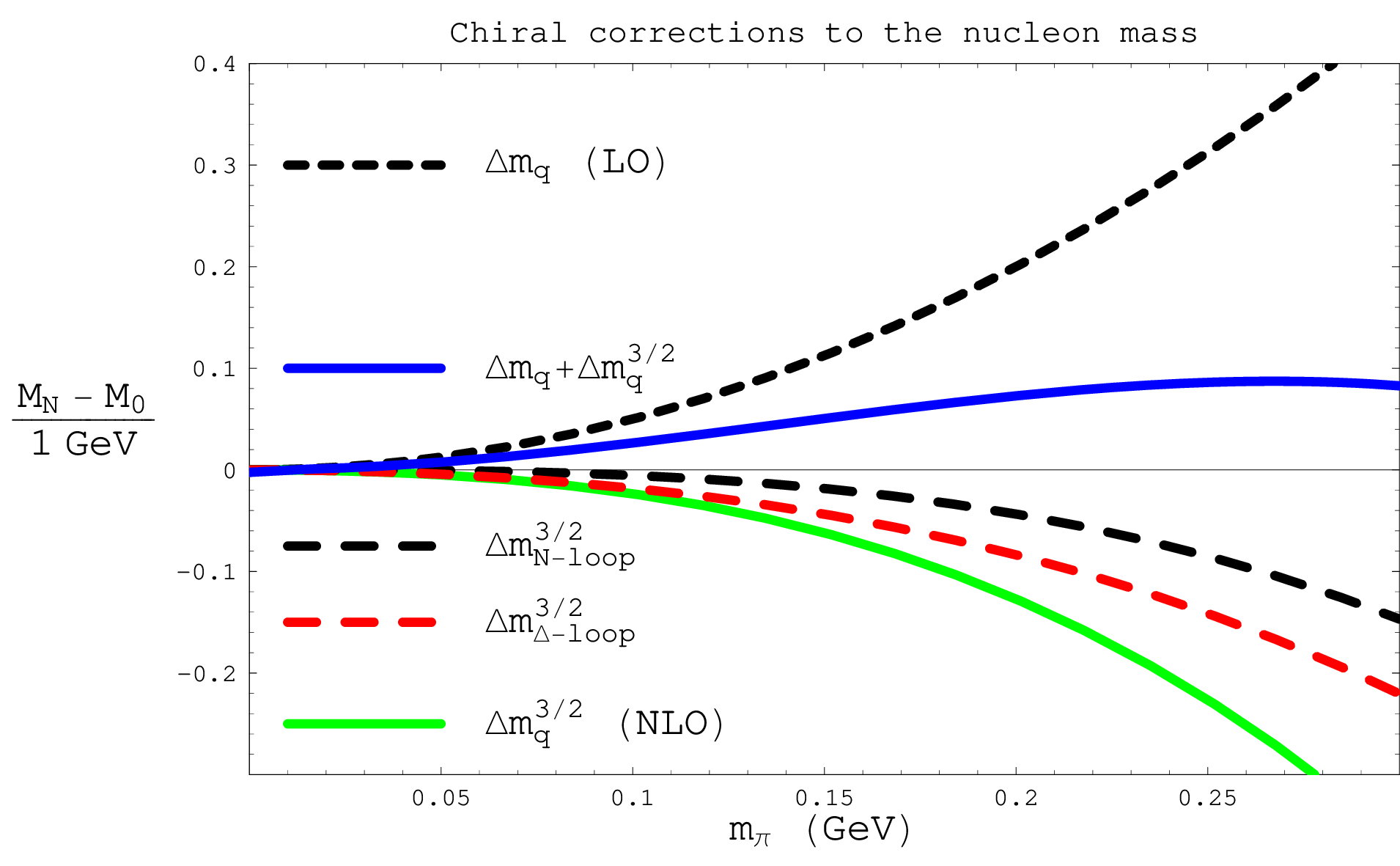}
\caption[Plot of the leading order and next-to leading order chiral corrections to the nucleon mass.]{\label{fig:NmassPlot} Here we plot the quark-mass dependent (chiral) corrections to the nucleon mass in the isospin limit.  The upper dashing curve (black dashing online) is the LO correction, $M_N^{(1)}$.  The next two longer dashing curves (longer black and red dashing online) are the NLO corrections coming from the nucleon-pion and delta-pion loops respectively.  The lowest curve (green on-line) is the total NLO mass correction, $M_N^{(3/2)}$.  The dark solid curve (blue online) is the total chiral correction, $M_N - M_0$, through NLO.}
\end{figure}

The loop-graphs with internal delta states are more interesting and give rise to the renormalization-scale dependence at this order.  The un-renormalized self energy correction arising from diagram~\ref{fig:LONmass}(c) is given by
\begin{multline}\label{eq:NmassNLOD}
	\d \S_{\D-loop}^{(3/2)} = \frac{4 g_{\D N}^2}{\L_\chi^2} \bigg\{
		\left[ \frac{2}{4-d} -\g_E +1 +\textrm{log} 4\pi \right] \left(\frac{2}{3}\D^3 -\D m_\pi^2 \right)
\\		+\frac{4}{9}\D^3 -\frac{1}{3}\D m_\pi^2
		-\frac{2}{3} \mc{F}(m_\pi,\D,\mu) \bigg\}
\end{multline}
where we have defined the function $\mc{F}$ by
\begin{multline}\label{eq:F}
	\mc{F} (m,\D,\mu) = ( \D^2-m^2 )^{3/2} 
		\log \left( 
			\frac{\D + \sqrt{\D^2 - m^2 + i \e}}{\D - \sqrt{\D^2 - m^2 + i \e}} \right)
\\
		- \frac{3}{2}\D\, m^2 \log \left( \frac{m^2}{\mu^2} \right)
		- \D^3 \log \left( \frac{4 \D^2}{m^2} \right)\, .
\end{multline}
This is slightly different than the function as defined in Ref.~\cite{Chen:2001yi}.  Here, we have chosen to subtract the chiral limit value of the function, $F(0,\D,\mu)$, as defined in Ref.~\cite{Chen:2001yi}, instead of absorbing this contribution into $M_0$ (as defined in Eq.~\eqref{eq:F}, $\mc{F}(0,\D,\mu)=0$).  This has the slight advantage that the decoupling of the deltas in the chiral limit~\cite{Gasser:1979hf} is more transparent.  After applying our renormalization prescription defined in Eq.~\eqref{eq:MSbar}, we observe that we can absorb all the analytic dependence upon $\D$ into a redefinition of the LECs of the LO chiral Lagrangian, 
Eq.~\eqref{eq:NucleonLO},
\begin{align}\label{eq:DeltaRenorm}
	M_0 &\rightarrow M_0^R(\D) = M_0 +\frac{16\, g_{\D N}^2}{9}\, \frac{\D^3}{\L_\chi^2}\, , \nonumber\\
	\a_M &\rightarrow \a_M^R = \a_M\, , \nonumber\\
	\s_M &\rightarrow \s_M^R(\D,\mu) = \s_M +\frac{2g_{\D N}^2}{3} \, \frac{B_0 \D}{\L_\chi^2}
		-2 g_{\D N}^2\, \frac{B_0 \D}{\L_\chi^2}\, \textrm{log} \left(\frac{\mu^2}{\mu_0^2} \right)\, ,
\end{align}
such that the nucleon mass to NLO is given by
\begin{multline}\label{eq:NmassNLO}
	M_{N_i} = M_0^R(\D) -2\, \a_M^R\, m_i 
		-2\, \s_M^R(\D,\mu)\, (m_u+m_d) 
\\		-3 \pi\, g_A^2\, \frac{m_\pi^3}{\L_\chi^2}
		-\frac{8 g_{\D N}^2}{3} \frac{\mc{F} (m_\pi,\D,\mu)}{\L_\chi^2} 
		+\mc{O}(m_q^2)\, .
\end{multline}
To help understand the importance of computing the next order chiral corrections, we will plot the contributions to $M_N$ in the above expression, setting $\mu = 4\pi f$.  Here, we are not concerned with precision, but rather are interested in the qualitative features of this expansion (of course a thorough understanding of the convergence of HB\CPT\ necessitates a precise determination of all the LEC's relevant to a given order).  Therefore, we estimate our values for $\a_M^R$ and $\s_M^R$ from Ref.~\cite{Bernard:1993nj} by matching onto $SU(2)$, arriving at the values $\frac{\a_M^R}{B_0} \simeq - 1 \textrm{ GeV}^{-1}$ and 
$\frac{\s_M^R}{B_0} \simeq -2 \textrm{ GeV}^{-1}$.  We also use the physical values of the nucleon-pion coupling, $g_A \simeq 1.25$, and the nucleon-delta-pion coupling, $g_{\D N} \simeq 1.5$, as the difference between using these values and the values at this order are higher order than we work.  The resulting contributions are show in Fig.~\ref{fig:NmassPlot}.
%%%%%%%%%%%%%%%%%%%%%%%%%%%%%%%%%%%%%%%%%%%%%%%%%%%
%
%	fig: chiral corrections to the nucleon mass
%
%%%%%%%%%%%%%%%%%%%%%%%%%%%%%%%%%%%%%%%%%%%%%%%%%%%
%%%%%%%%%%%%%%%%%%%%%%%%%%%%%%%%%%%%%%%%%%%%%%%%%%%
%
%%%%%%%%%%%%%%%%%%%%%%%%%%%%%%%%%%%%%%%%%%%%%%%%%%%
In this figure, we have plotted the NLO contributions arising from the nucleon-pion loops and the delta-pion loops separately as well as the sum.  This is simply to demonstrate the importance of including the delta as a dynamical degree of freedom in the effective field theory~\cite{Jenkins:1990jv,Jenkins:1991ne}, as can be seen by the comparable size of the nucleon loop as compared to the delta loop.
The important feature this figure highlights is that for pion masses of 
$m_\pi \sim 250$~MeV, the NLO contribution begins to compete with the LO contribution, as seen by a flattening of the total mass correction to NLO, the solid (blue) curve in Fig.~\ref{fig:NmassPlot}.  This clearly demonstrates the need for computing the NNLO mass correction; to have confidence in the fit for values of the pion mass $m_\pi \gtrsim 250$~MeV.  The precise value of the pion mass for which one needs the NNLO contribution to trust the expansion will be sensitive to the values of $\a_M^R$ and $\s_M^R$, which will in turn depend upon the size of the NNLO contributions.  Thus a detailed fitting of these parameters, comparing the NLO mass formula to the NNLO mass formula will not only provide information about the value of the LECs but also will help determine for which values of the pion mass we can trust the convergence of the chiral expansion.  The NNLO calculations are certainly necessary for the values of pion masses used in current lattice simulations, being typically $m_\pi > 300$~MeV.

%%%%%%%%%%%%%%%%%%%%%%%%%%%%%%%%%%%%%%%%%%%%%%%%%%%
%
%	HDO's for Nucleon Mass
%
%%%%%%%%%%%%%%%%%%%%%%%%%%%%%%%%%%%%%%%%%%%%%%%%%%%
\subsection*{\textbf{Higher dimensional operators}}

In order to compute the NNLO ($\mc{O}(m_q^2)$) correction to the nucleon and delta masses, we must first construct the relevant terms of the higher order chiral Lagrangian.
It is convenient to split these terms into three different categories.  The first set of higher-order terms are all operators whose coefficients are constrained
by reparameterization invariance (RPI)~\cite{Luke:1992cs}, and thus known. 
These terms must be included to insure the Lorentz invariance of HB\CPT\ 
at $\mc{O}(1/M)$, where $M \sim \Lambda_\chi$ is the average nucleon mass in the chiral limit. 
The momentum of the nucleon, as given in Eq.~\eqref{eq:RPImomentum}, is not a unique parameterization.  In particular, under the \textit{reparameterization}
\begin{equation}
	\vit \rightarrow \vit +\frac{\e}{M} \quad, \quad
	k \rightarrow k -\e\, ,
\end{equation}
the nucleon momentum, $p_\mu = M\vit_\mu +k_\mu$, is unchanged.  Reparameterization invariance is the requirement that the Lagrangian be invariant under such transformations, which ensures the Lorentz invariance of the theory to a given order in $1/M$.  Moreover, the implementation of reparameterization invariance~\cite{Luke:1992cs,Manohar:2000dt} is a useful tool to generate terms in the higher order Lagrangian, as the reparameterization automatically generates the higher terms needed for Lorentz invariance.  The result of applying these RPI techniques to Eq.~\eqref{eq:HBChPTLO} and using the LO equations of motion to eliminate terms~\cite{Georgi:1991ch,Arzt:1993gz} is the following fixed coefficient Lagrangian
\begin{align}\label{eq:RPIfixed}
	\mc{L} =& -\ol{N}  \frac{D_\perp^2}{2 M_0} N 
		+ \ol{T}^\mu \frac{D_\perp^2}{2 M_0} T_\mu 
		+ g_A \left( \ol{N} \frac{i \overleftarrow{D} \cdot S}{M_0} v \cdot \mc{A}\, N 
		- \ol{N}\, v \cdot \mc{A} \frac{S \cdot i \overrightarrow{D}}{M_0} N \right)
\nonumber\\ 
		&+ g_{\D \D} \left( \ol{T}^\mu \frac{i \overleftarrow{D} \cdot S}{M_0}
				\vit \cdot \mc{A}\, T_\mu 
			- \ol{T}^\mu \, \vit\cdot \mc{A} \frac{S\cdot i \overrightarrow{D}}{M_0} T_\mu \right),
\end{align}
where $D_\perp^2 = D^2 - (\vit \cdot D)^2$.  Note that by combining the kinetic term for the nucleon in the above Lagrangian and the kinetic term in Eq.~\eqref{eq:HBChPTLO}, the nucleon propagator becomes
\begin{equation}
	\mc{G}_N(p,\vit) = \frac{i}{p\cdot \vit +i\e}
		\rightarrow
		\frac{i}{p\cdot \vit -\frac{\vec{p}^2}{2M} +i\e}\, ,
\end{equation}
as promised.

The next set of operators are the standard terms in the $\mc{O}(q^2)$ and $\mc{O}(q^4)$ Lagrangian~\cite{Bernard:1995dp} with undetermined coefficients and which contribute to the masses at NNLO.  Here we use different notation for the LECs than in Ref.~\cite{Bernard:1995dp}, as in that work the delta's were not included as dynamical degrees of freedom, and thus the values of the LECs will be different.  These operators are~%
\footnote{One could also add operators of the form $\big( \ol{N} S\cdot \mc{A}\, S\cdot \mc{A} N \big)$, but it is straight forward to show that they are a linear combination of the operators in Eq.~(\ref{eq:NHDO}), and therefore not distinct.}
\begin{align}
	\mc{L}  =&
		\frac{1}{\L_\chi} \bigg\{
			b_1^M\, \ol{N}  \, \mc{M}_+^2  N
			+ b_5^M\, \ol{N} N \, \tr ( \mc{M}_+^2 )
			+ b_6^M\, \ol{N} \, \mc{M}_+ N \, \tr (\mc{M}_+) 
			+ b_8^M\, \ol{N} N \, [\tr (\mc{M}_+)]^2 \nonumber\\
		&\qquad\quad
		 	+ b^A\, \ol N N \, \tr ( \mc{A} \cdot \mc{A} )  
			+ b^{vA}\, \ol N N \, \tr ( v \cdot \mc{A} \, v \cdot \mc{A} ) 
		\bigg\}.
\label{eq:NHDO} 
\end{align}
The LECs $b_i^M$, $b^A$, and $b^{vA}$ are all dimensionless. 
The choice in numbering the coefficients was made to be consistent
with the partially quenched version of this Lagrangian~\cite{Walker-Loud:2004hf}, as will become clear in Section~\ref{sec:NDPQmasses}.  We have not written down operators of the form, 
$\ol{N} \mc{M}_-\, \mc{M}_-\, N$, even though they are formally of this order, as they do not contribute the nucleon masses until a higher order.  Also in this set of higher-dimensional operators are the equivalent terms for the delta-resonances,
\begin{align}\label{eq:DHDO}
	\mc{L} =& 
		\frac{1}{\L_\chi} \bigg\{
			t_2^A \, \ol{T}^{kji}_\mu (\mc{A}_\nu)_{i}{}^{i'} (\mc{A}^\nu)_{j}{}^{j'} T^\mu_{i'j'k} 
			+ t_3^A \, \left( \ol{T}_\mu T^\mu \right) \tr ( \mc{A}_\nu \mc{A}^\nu ) 
			+ t_2^{\tilde{A}} \, \ol{T}^{kji}_\mu (\mc{A}^\mu)_{i}{}^{i'} (\mc{A}_\nu)_{j}{}^{j'} T^\nu_{i'j'k} 
			\nonumber \\
		&\qquad
			+ t_3^{\tilde{A}} \, \ol{T}_\mu T^\nu \tr ( \mc{A}^\mu \mc{A}_\nu ) 
			+ t_2^{vA} \, \ol{T}^{kji}_\mu (v \cdot \mc{A})_{i}{}^{i'} (v \cdot \mc{A})_{j}{}^{j'} T^\mu_{i'j'k} 
			+ t_3^{vA} \, \ol{T}_\mu T^\mu  \tr ( v \cdot \mc{A} \, v \cdot \mc{A} ) \nonumber \\
		&\qquad
			+ t_1^M \, \ol{T}^{kji}_\mu (\mc{M}_+ \mc{M}_+)_{i}{}^{i'} T^\mu_{i'jk} 
			+ t_2^M  \, \ol{T}^{kji}_\mu (\mc{M}_+)_{i}{}^{i'} (\mc{M}_+)_{j}{}^{j'} T^\mu_{i'j'k} 
			+ t_3^M  \, \ol{T}_\mu T^\mu \tr (\mc{M}_+ \mc{M}_+) \nonumber \\
		&\qquad
			+ t_4^M \, \left( \ol{T}_\mu \mc{M}_+ T^\mu \right) \tr (\mc{M}_+)
			+ t_5^M  \, \ol{T}_\mu T^\mu  [ \tr(\mc{M}_+) ]^2
		\bigg\},
\end{align}
and all of the LECs $t_i^M$, $t_i^A$, $t_i^{vA}$, and $t_i^{\tilde{A}}$ are dimensionless. 

The last set of higher-dimensional operators with undetermined coefficients all involve 
the nucleon-delta mass-splitting parameter $\D$, which is a singlet under chiral transformations. 
Because of this, the inclusion of the spin-$\frac{3}{2}$ fields in \CPT\ requires the addition of operators involving powers of $\D / \Lambda_\chi$.  We shall not write out such 
operators explicitly. To account for the effects of these operators, all LECs in the calculation
must be treated as arbitrary polynomial functions of $\D / \Lambda_\chi$ and expanded out to the
required order. For example
\begin{align}\label{eq:DeltaExpnsn}
	\s_M \rightarrow \s_M \left(\frac{\D}{\L_\chi}\right)
		=& \s_M \left( 1+\s_1 \, \frac{\D}{\L_\chi} 
		+ \s_2 \, \frac{\D^2}{\L_\chi^2} +\dots \right)
		, \nonumber \\
	\gm \rightarrow \gm\left(\frac{\D}{\L_\chi}\right)
		=& \gm\left(1+\gamma_1 \, \frac{\D}{\L_\chi} 
		+ \gamma_2 \, \frac{\D^2}{\L_\chi^2} +\dots \right), \nonumber \\
	g_A \rightarrow g_A\left(\frac{\D}{\L_\chi}\right)
		=& g_A \left(1 + g_{A,1} \, \frac{\D}{\L_\chi} 
		+g_{A,2}\, \frac{\D^2}{\L_\chi^2} +\dots \right)\, .
\end{align}
Here we see the ambiguity in the treatment of $\D$.  In Eq.~\eqref{eq:DeltaRenorm} we showed how to renormalize the LECs of the LO Lagrangian to absorb the $\D$ dependence, in particular the LEC $\s_M$ was modified by a known term proportional to $\D$.  However, we see from Eq.~\eqref{eq:DeltaExpnsn} that we must additionally add a term to $\s_M$ linear in $\D$ but with an unknown coefficient.  And in fact every LEC in the HB\CPT\ Lagrangian has this unknown dependence upon $\D$.  However, determination of the LECs in Eq.~\eqref{eq:DeltaExpnsn} require the ability to tune the parameter $\D$, which we can not do in nature or in lattice QCD.

To resolve this ambiguity, we advocate the following method.  The only $\D$ dependence one should keep explicitly is that associated with non-analytic dependence upon $m_\pi$, for example in the function $\mc{F}(m_\pi, \D, \mu)$; this will keep all the important effects of the delta-resonances, the large chiral contributions and the cuts in graphs where the internal delta's can go on shell (for example in nucleon Compton scattering, Chapter~\ref{chap:Polarize}, nucleon pion scattering, and pion photo-production off the nucleon).  All the finite analytic dependence upon $\D$, whether known contributions as in Eq.~\eqref{eq:DeltaRenorm} or undetermined contributions as in Eq.~\eqref{eq:DeltaExpnsn}, should be absorbed in a renormalization of the LECs.  This was consistently done in Ref.~\cite{Chen:2001yi} to NLO, and in the following section we will demonstrate the consistency of this treatment in the nucleon masses through NNLO.%
\footnote{As this renormalization prescription is not guided by a symmetry, as is the case for the quark mass dependent contributions, it would be nice to check that this prescription holds for other nucleon observables as well, \textit{i.e.} the specific renormalization of the $\D$ dependence done for the nucleon masses gives an exact cancellation of the analytic $\D$ dependence in another nucleon observable like the magnetic moment.}
We now proceed to compute the NNLO nucleon and delta mass corrections.

%%%%%%%%%%%%%%%%%%%%%%%%%%%%%%%%%%%%%%%%%%%%%%%%%%%
%
%	Nucleon masses
%
%%%%%%%%%%%%%%%%%%%%%%%%%%%%%%%%%%%%%%%%%%%%%%%%%%%
\subsection{\textbf{Nucleon Masses through $\mc{O}(m_q^2)$}}

\begin{figure}
\center
\begin{tabular}{cccc}
\includegraphics[width=0.2\textwidth]{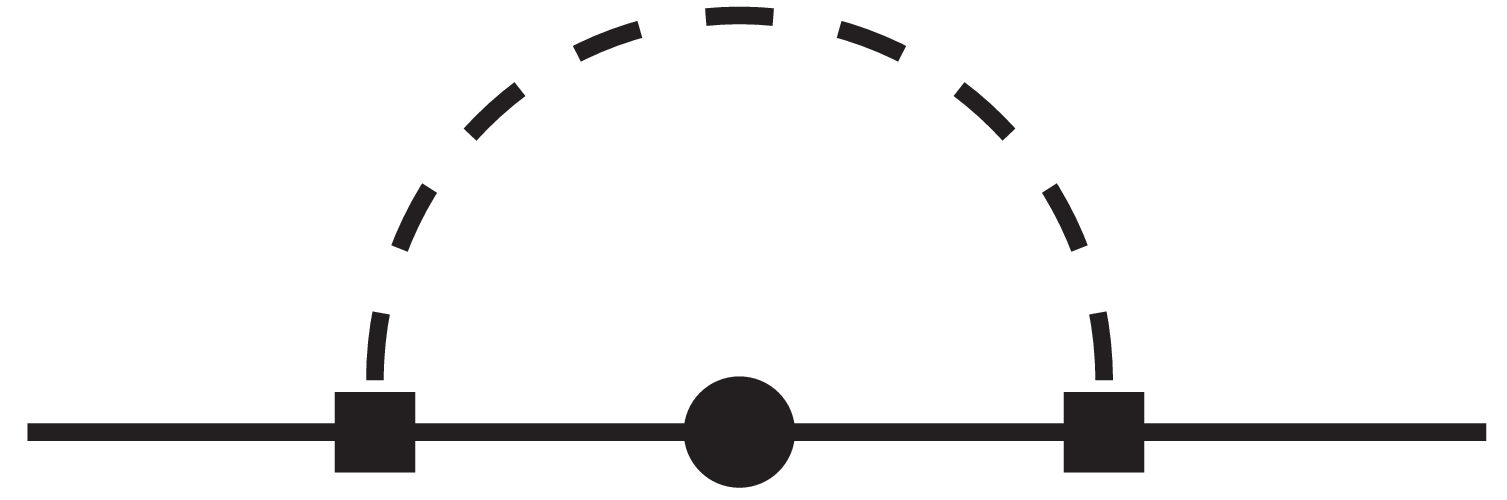}
&\includegraphics[width=0.2\textwidth]{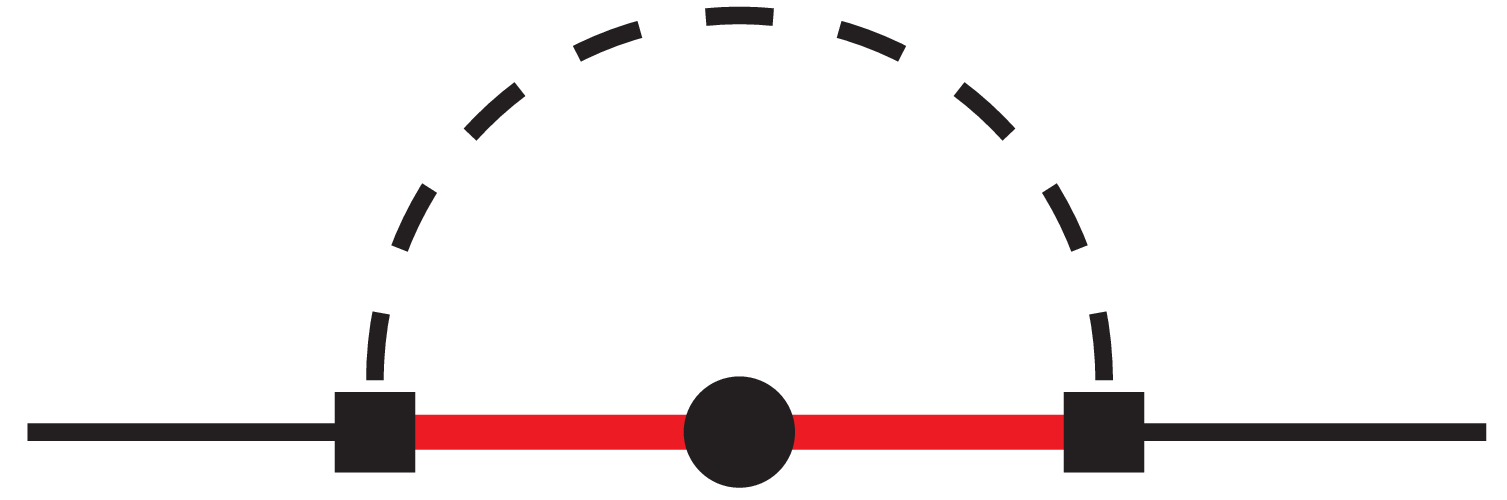}
&\includegraphics[width=0.2\textwidth]{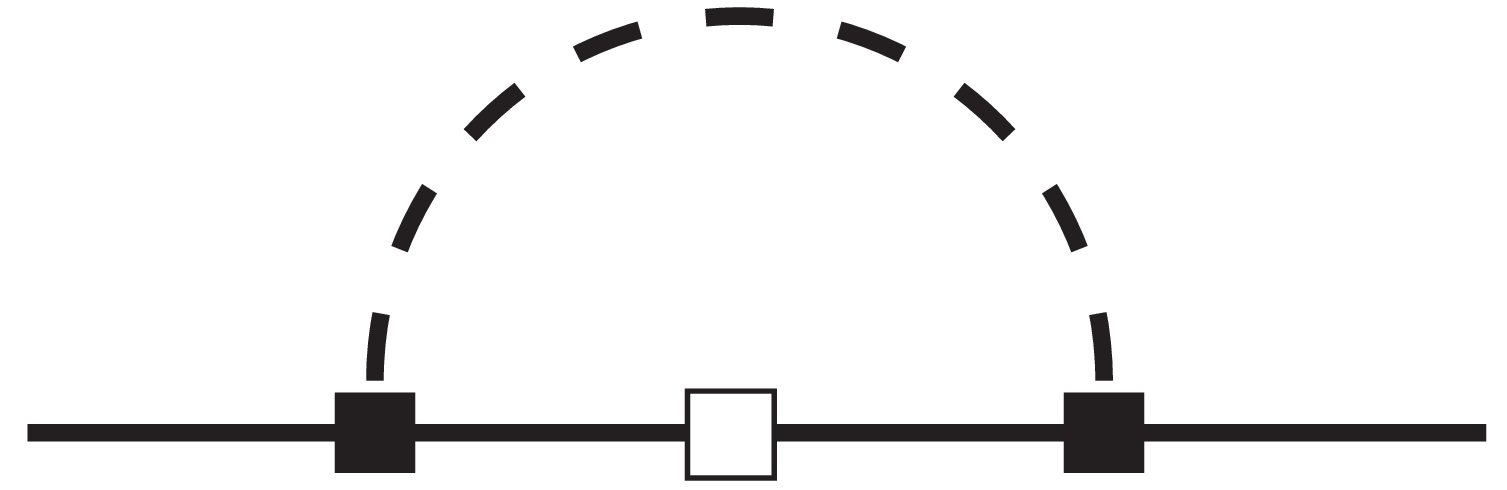}
&\includegraphics[width=0.2\textwidth]{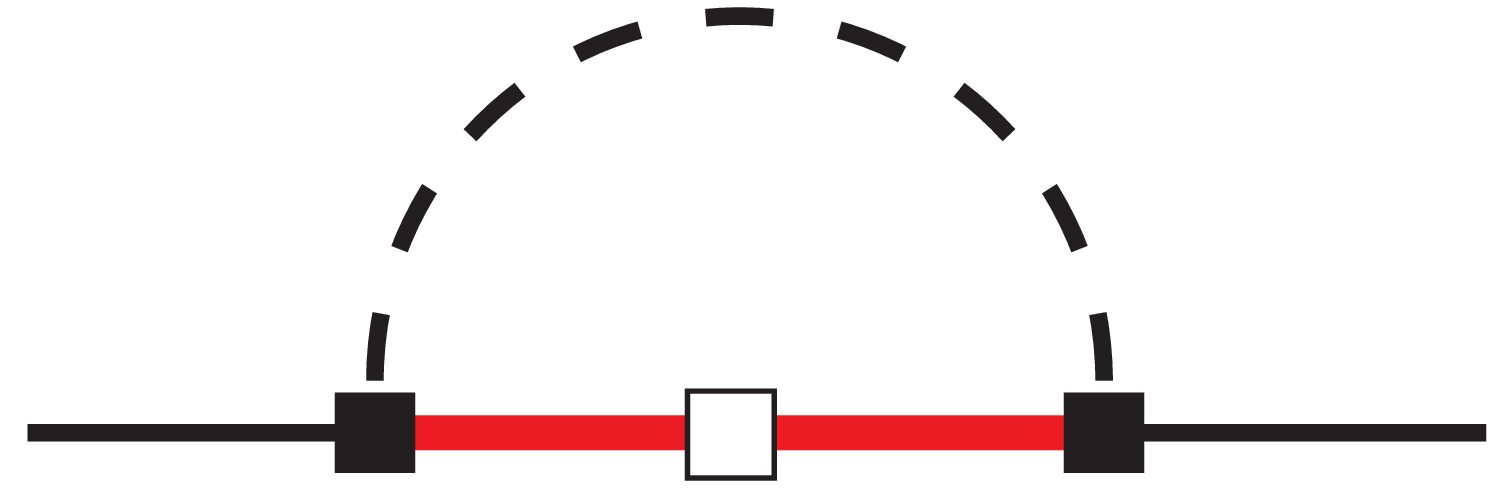}\\
(a) & (b) & (c) & (d)\\ \\
\includegraphics[width=0.2\textwidth]{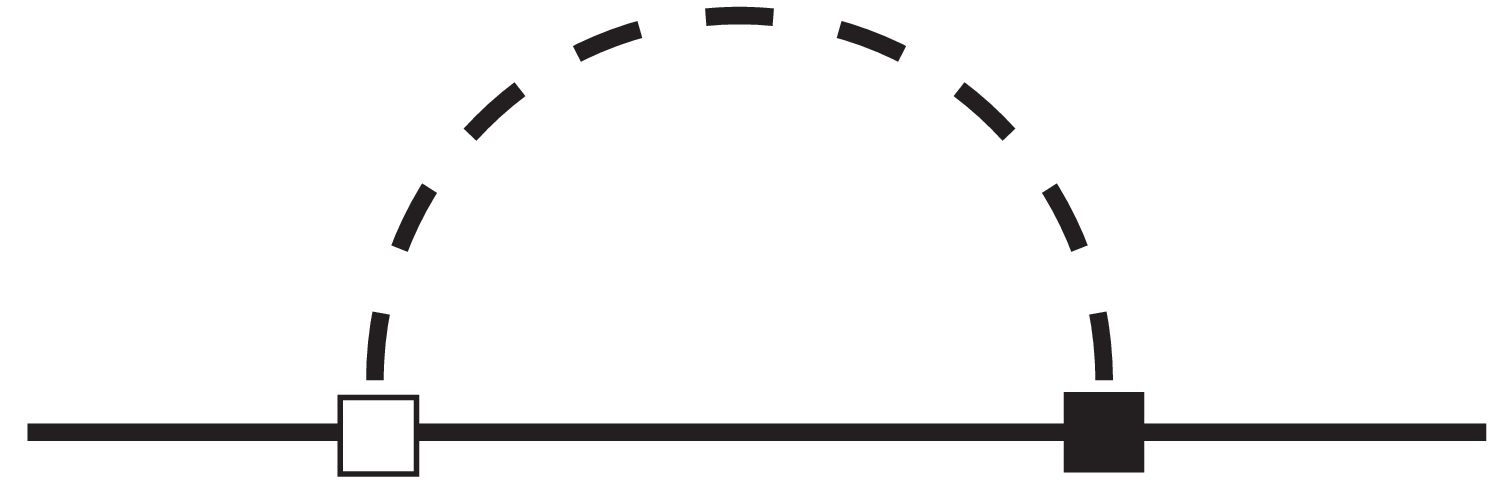}
&\includegraphics[width=0.2\textwidth]{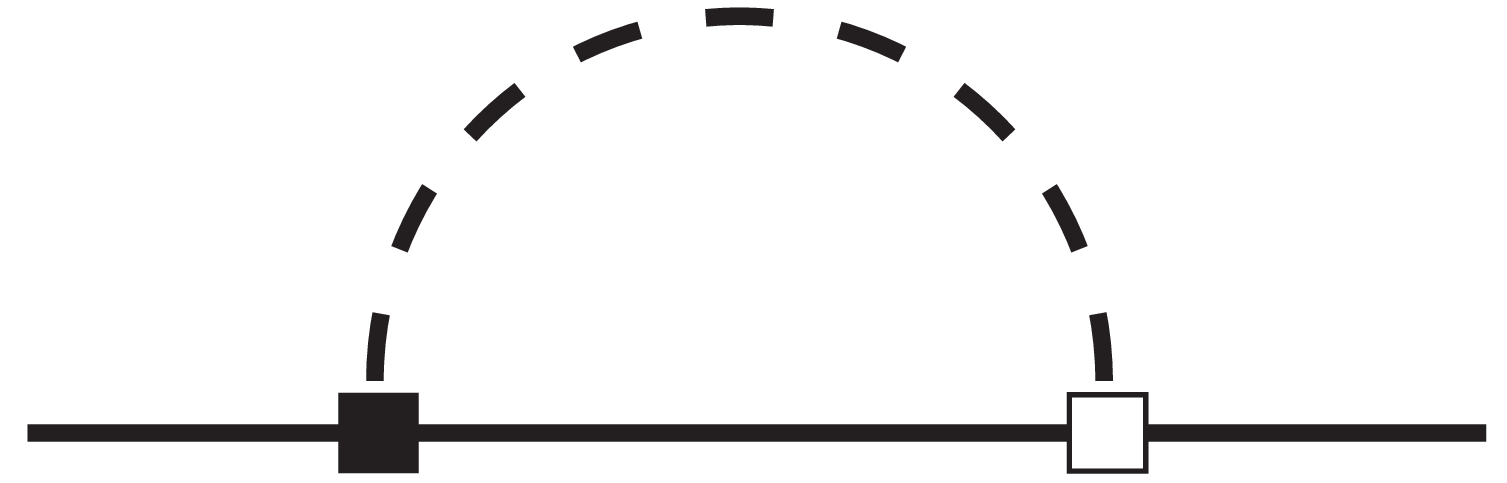} 
&\includegraphics[width=0.12\textwidth]{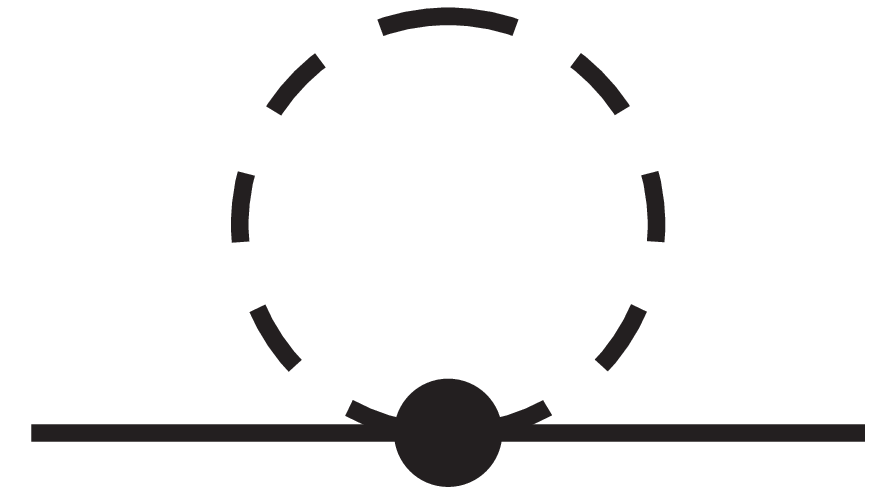}
&\includegraphics[width=0.12\textwidth]{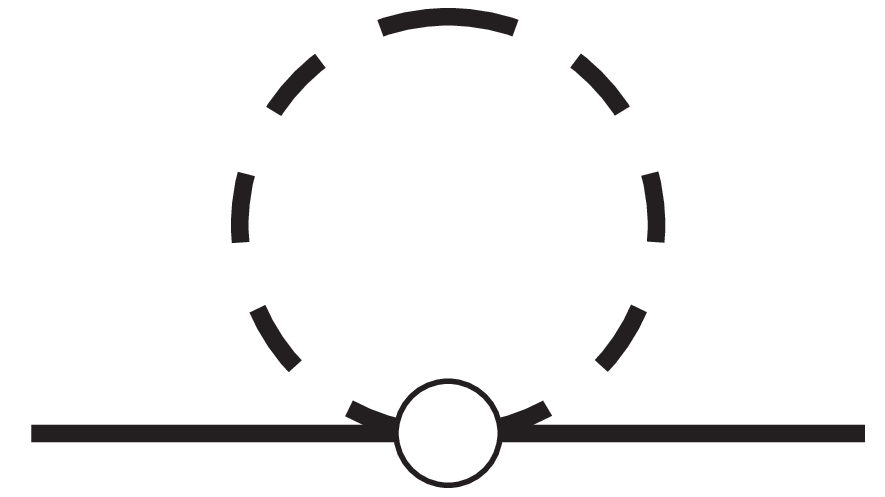}\\
(e) & (f) & (g) & (h)
\end{tabular}
\caption[Nucleon self energy diagrams contributing to the nucleon masses at $\mc{O}(m_q^2)$.]{\label{fig:NmassNNLO} Diagrams depicting the various contributions to the NNLO nucleon mass correction.  The black, thick grey (red online) and dashing lines correspond to nucleons, delta-resonances and pions respectively.  The filled circles and filled squares are insertions of operators from Eq.~\eqref{eq:HBChPTLO} and Eq.~\eqref{eq:NNpi} respectively.  The open squares are RPI fixed operators from Eq.~\eqref{eq:RPIfixed} and the open circles are higher dimensional operators from Eq.~\eqref{eq:NHDO}.}
\end{figure}

The chiral correction to the nucleon mass at $\mc{O}(m_q^2)$ is a bit more involved than the LO and NLO contributions, receiving corrections from the Lagrangians, Eqs.~\eqref{eq:HBChPTLO}, \eqref{eq:NNpi}, \eqref{eq:RPIfixed}, \eqref{eq:NHDO} and combinations of the first two.  The various contributions are depicted Figure~\ref{fig:NmassNNLO}.
%%%%%%%%%%%%%%%%%%%%%%%%%%%%%%%%%%%%%%%%%%%%%%%%%%%
%
%	figure: NNLO N mass corrections
%
%%%%%%%%%%%%%%%%%%%%%%%%%%%%%%%%%%%%%%%%%%%%%%%%%%%
%%%%%%%%%%%%%%%%%%%%%%%%%%%%%%%%%%%%%%%%%%%%%%%%%%%
%
%%%%%%%%%%%%%%%%%%%%%%%%%%%%%%%%%%%%%%%%%%%%%%%%%%%
Before giving the result, we first will demonstrate the consistency of the treatment of the nucleon-delta mass splitting parameter, $\D$.  To do this we will compare the mass correction of Figure~\ref{fig:NmassNNLO}(e-f), which involves the kinetic correction to the delta-resonances, to the contribution of Figure~\ref{fig:NmassPlot}(c), involving the bare LO delta propagator and resulting in Eq.~\eqref{eq:NmassNLOD}.  The mass correction from this diagram is
\begin{multline}\label{eq:NmassNNLOD}
	\d \S_{\D-loop}^{(2)} = \frac{5 g_{\D N}^2}{M_0 \L_\chi^2} \bigg\{
		\left[ \frac{2}{4-d} -\g_E +1 +\textrm{log} 4\pi \right] 
			\left(\frac{4}{3}\D^4 -2 \D^2 m_\pi^2 +\frac{1}{2} m_\pi^4 \right)
\\		+\frac{16}{45}\D^4 +\frac{2}{15} \D^2 m_\pi^2 
		-\frac{9}{20} m_\pi^4 -\frac{1}{2} m_\pi^4 \textrm{log} \left( \frac{m_\pi^2}{\mu^2} \right)
		-\frac{4}{3} \mc{F}(m_\pi,\D,\mu) \bigg\}\, .
\end{multline}
Note that if we shift the nucleon-delta-pion coupling by a known amount, 
\begin{equation}
	g_{\D N} \rightarrow g_{\D N} \left(1 + \frac{5}{4} \frac{\D}{M_0} +\dots \right)\, ,
\end{equation}
such that the difference between the self energy corrections at NNLO and NLO with the modified $g_{\D N}$ coupling (to higher orders in $\frac{\D}{M_0}$ dependence of $g_{\D N}$) becomes
\begin{multline}
	\d \S_{\D-loop}^{(2)} - \frac{5\D}{2M_0}\d \S_{\D-loop}^{(3/2)} 
	= \frac{5 g_{\D N}^2}{M_0 \L_\chi^2} \bigg\{
		\left[ \frac{2}{4-d} -\g_E +1 +\textrm{log} 4\pi \right] \frac{1}{2} m_\pi^4
\\
	-\frac{8}{15}\D^4 -\frac{8}{15} \D^2 m_\pi^2
	-\frac{9}{20} m_\pi^4 -\frac{1}{2} m_\pi^4 \textrm{log} \left( \frac{m_\pi^2}{\mu^2} \right)
	\bigg\}\, .
\end{multline}
Note that the important contributions from Figure~\ref{fig:NmassNNLO}(e-f) are exactly reproduced (the dimensional regularization divergences proportional to Eq.~\eqref{eq:MSbar} and the $\D$ dependence associated with non-analytic $m_\pi$ dependence as in $\mc{F}(m_\pi,\D,\mu)$).  There are additional analytic contributions, both finite and infinite, as well as a new non-analytic dependence which arise from this graph, but under the above stated treatment of the $\D$ dependence, it is redundant to keep all finite terms in Eq.~\eqref{eq:NmassNNLOD}.  We apply this analysis to all the nucleon mass corrections arising from the diagrams in Figure~\ref{fig:NmassNNLO} arriving at the following expression for the 
$\mc{O}(m_q^2)$ nucleon mass correction;
\begin{align}
	M_{N_i}^{(2)} =& ( Z_{N_i} - 1) M_{N_i}^{(1)} 
		+ \frac{1}{\L_\chi} \left\{
			b_1^M \, (m_i)^2 + b_5^M \, \tr (m_q^2) + b_6^M \, m_i \, \tr(m_q)
			+ b_8^M \, [\tr(m_q)]^2 \right\} \nonumber \\
		&- \Big[ C^{N_i}_\pi +6 \sigma_M \tr(m_q) \Big]
			\frac{m_\pi^2}{\L_\chi^2} \log \left( \frac{m_\pi^2}{\mu^2} \right) \nonumber \\
		& + \frac{m_\pi^4}{\L_\chi^3} \left[
			3\left( b^A +\frac{1}{4} b^{vA} \right)\, \log \left( \frac{m_\pi^2}{\mu^2} \right) 
			-\frac{3}{8}\, b^{vA} \right] \nonumber \\
		& +\frac{m_\pi^4}{M_0 \L_\chi^2} \left[
			\frac{27 g_A^2}{16} \left( \log \left( \frac{m_\pi^2}{\mu^2} \right) + \frac{5}{6} \right)
			+\frac{5 g_{\D N}^2}{2} \left( \log \left( \frac{m_\pi^2}{\mu^2} \right) + \frac{9}{10} \right)
			\right] \nonumber \\
		& + 9 g_A^2 \sigma_M \frac{m_\pi^2\, \tr (m_q)}{ \L_\chi^2} 
			\left[ \log \left( \frac{m_\pi^2}{\mu^2} \right) + \frac{2}{3} \right] 
		+ 8 g_{\D N}^2 \ol{\sigma}_M \frac{\tr(m_q)}{\L_\chi^2} 
			\Big[ \mc{J} (m_\pi, \D, \mu) + m_\pi^2 \Big] \nonumber \\
		& +3g_A^2\, F^{N_i}_\pi \frac{m_\pi^2}{ \L_\chi^2} 
			\left[ \log \left( \frac{m_\pi^2}{\mu^2} \right) + \frac{2}{3} \right] 
		- 2 g_{\D N}^2\, \gamma_M \frac{G^{N_i}_\pi}{\L_\chi^2} 
			\Big[ \mc{J} (m_\pi, \D, \mu) + m_\pi^2 \Big]\, . 
\label{e:MBNNLO} 
\end{align}
Here we have used $(m_i)^2$ which is the square of the tree-level
coefficients $m_i$ appearing in Eq.~\eqref{eq:mB}.
Above the wavefunction renormalization $Z_{N_i}$ is given by
\begin{equation}
	Z_{N_i} - 1 = - \frac{9 g_A^2}{2} \frac{m_\pi^2}{\L_\chi^2} 
			\left[  \log \left( \frac{m_\pi^2}{\mu^2} \right) +\frac{2}{3} \right]
		- \frac{4 g_{\D N}^2}{\L_\chi^2 } \left[ \mc{J}(m_\pi, \D, \mu) +
			m_\pi^2 \right].
\label{e:ZB} 
\end{equation}

The coefficients in the NNLO contribution, namely $C_\pi^{N_i}$, $F_\pi^{N_i}$, and $G_\pi^{N_i}$, are given in 
Table~\ref{t:NQCD-C} and depend on whether $i = p$ or $i = n$.
%%%%%%%%%%%%%%%%%%%%%%%%%%%%%%%%%%%%%%%%%%%%%%%%%%%
%
%	table: NNLO N mass coefficients
%
%%%%%%%%%%%%%%%%%%%%%%%%%%%%%%%%%%%%%%%%%%%%%%%%%%%
\begin{table}
\caption[Coefficients for nucleon mass correction at $\mc{O}(m_q^2)$.]{\label{t:NQCD-C}
The coefficients $C_\pi^{N_i}$, $F_\pi^{N_i}$, and $G_\pi^{N_i}$ in \CPT. Coefficients are
listed for the nucleons.}
\center
%\begin{ruledtabular}
\begin{tabular}{| l | c c c |}
\hline
 & $C_\pi^N$  
 & $F_\pi^N$ 
 & $G_\pi^N$ \\
\hline
$p$
 & $2 \a_M (2m_u +m_d)$ 
 & $\a_M (m_u + 2 m_d)$ 
 & $\frac{4}{9} ( 7 m_u + 2 m_d)$    \\
$n$
 & $2 \a_M (m_u + 2m_d)$ 
 & $\a_M (2 m_u + m_d)$ 
 & $\frac{4}{9} ( 2 m_u + 7 m_d)$    \\
\hline
\end{tabular}
%\end{ruledtabular}
\end{table}
%%%%%%%%%%%%%%%%%%%%%%%%%%%%%%%%%%%%%%%%%%%%%%%%%%%
%
%%%%%%%%%%%%%%%%%%%%%%%%%%%%%%%%%%%%%%%%%%%%%%%%%%%
The equations above, Eq.~\eqref{e:MBNNLO} and Eq.~\eqref{e:ZB}, also
employ abbreviations for non-analytic functions arising from 
loop contributions. The new functions is defined as
\begin{multline}
	\mc{J} (m,\D,\mu) = 2 \D^2 \log \left( \frac{4 \D^2}{m^2} \right)
		+m^2 \log \left( \frac{m^2}{\mu^2} \right)
\\		- 2 \D \sqrt{\D^2 - m^2} \log \left(
			\frac{\D + \sqrt{\D^2 - m^2 + i \varepsilon}}{\D - \sqrt{\D^2 - m^2 + i \varepsilon}} \right)\, ,
\end{multline}
where $\mc{J}(0,\D,\mu) = 0$, similarly to $\mc{F}(0,\D,\mu)$.

The expressions we have derived in this section, as well as those
throughout this work, are functions of the quark masses;
e.g. $m_\pi$ above is 
merely a replacement for the combination of quark masses given in
Eq.~\eqref{eq:piMasses}.  These expressions, thus require the
lattice practitioner to determine the quark masses.  
In $SU(2)$ \CPT\ away from the isospin limit, there are no equivalent
expressions in terms of the meson masses, as one cannot independently associate 
$m_u$ and $m_d$ to the pion masses, which are
degenerate to the order we are working.  One can only equate the
average value, $\frac{1}{2} (m_u +m_d)$, with the pion mass.
Therefore one cannot plot the 
baryon masses as a function of the physical meson masses, unless one
works in the isospin limit, $m_u = m_d$.  This is unlike the case in
the isospin limit of $SU(3)$, where there are only two independent
quark masses, but three independent meson masses, and one can always
convert from a quark mass expansion to a meson mass expansion via the
Gell-Mann--Okubo relation.%
\footnote{This problem can be avoided in $SU(4|2)$ \PQCPT, as there are more
independent meson masses than independent quark masses, so one can
algebraically convert from the quark mass expansion to the
meson mass expansion.  For example, to leading order,
$m_u = \frac{1}{2 B_0} \left(m_\pi^2 -m^2_{jd} +m^2_{ju} \right)$.
This would require one to know the mass of the mesons made of one
valence and one sea quark.  But if one is interested in the
non-isospin $SU(2)$ limit of the \PQCPT\ expressions, the problem is unavoidable.}
Lastly we remark that if one is using \CPT\ to determine the quark
masses from meson masses that are determined on the lattice, one must
use the one-loop expression in \CPT, else one looses contributions to
baryon masses that are of NNLO.  Thus the quark mass expansion is the one to
use in $SU(2)$ \CPT, at least away from the
isospin limit.

\begin{figure}
\center
\begin{tabular}{c}
\includegraphics[width=0.7\textwidth]{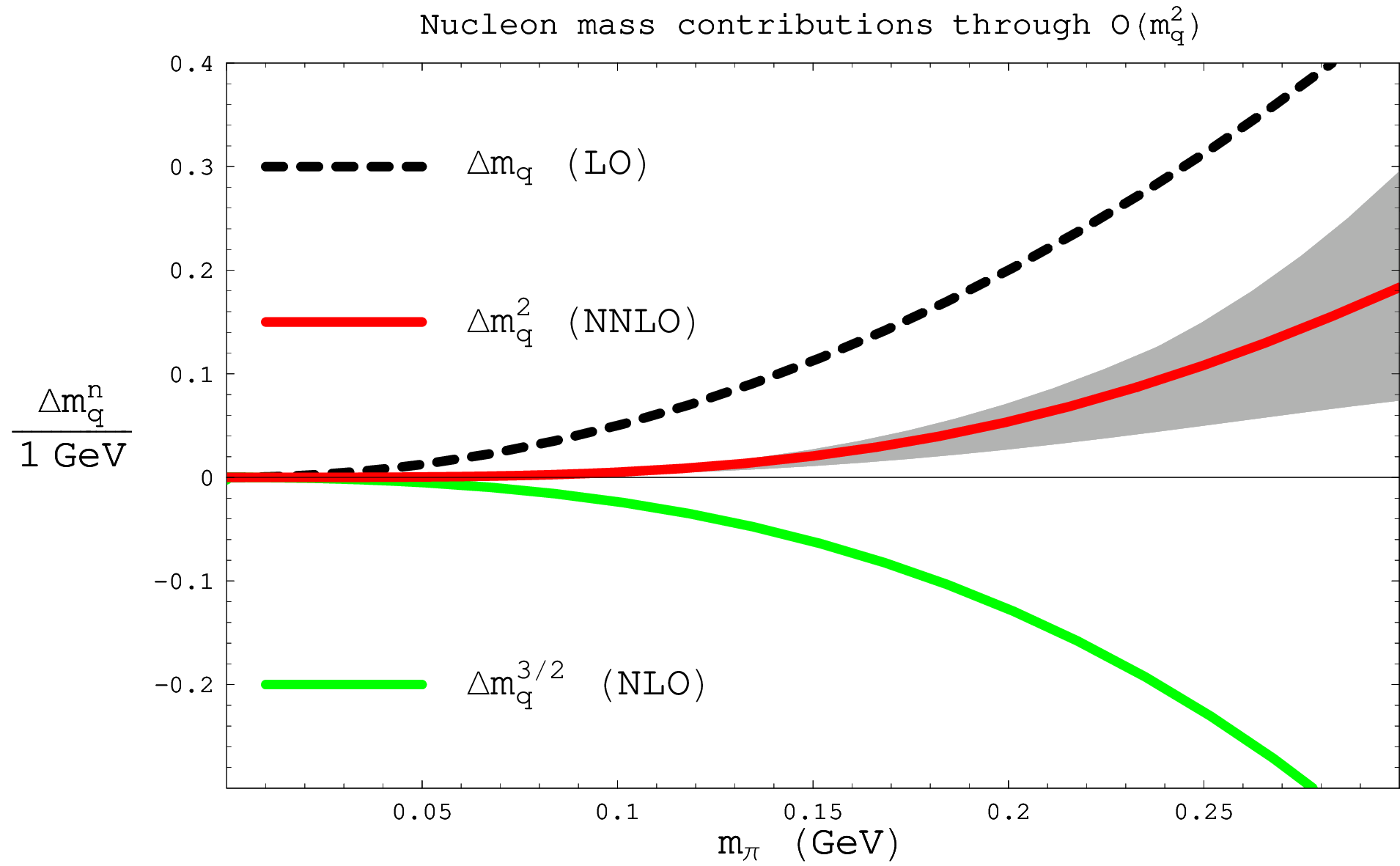} \\ (a)\\ \\
\includegraphics[width=0.7\textwidth]{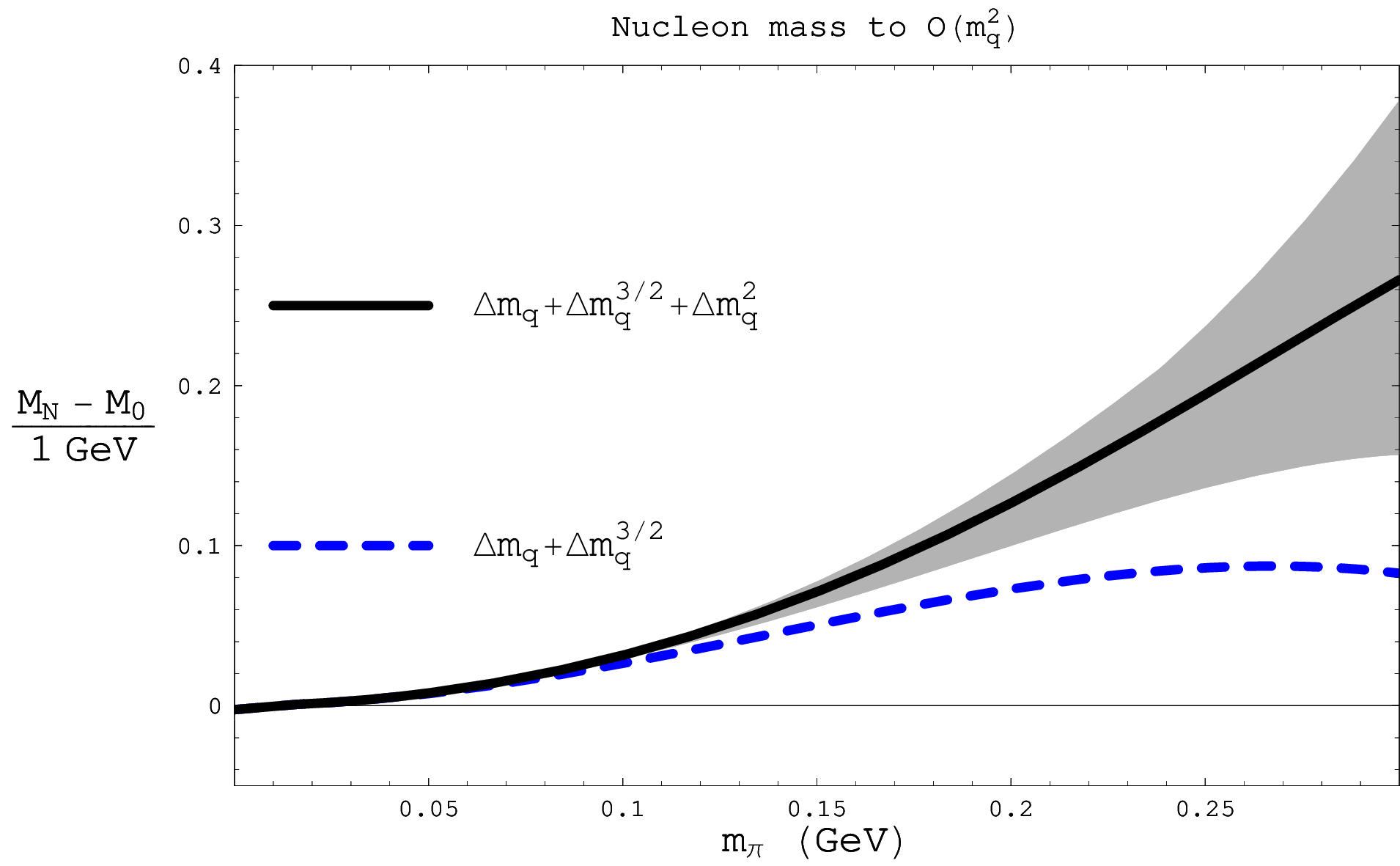} \\ (b)
\end{tabular}
\caption[Plot of LO, NLO and NNLO chiral corrections to nucleon mass.]{\label{fig:NmassError} In this figure, we plot (a) the first three chiral corrections to the nucleon mass in the isospin limit.  Although the NLO correction is not small compared to the NNLO contribution, we observe that for the range of LECs used, the NNLO corrections are definitely smaller than the LO and NLO corrections for the full range of pion masses plotted, and for most of the range, the NNLO contribution is significantly smaller.  The grey band is an estimate of the theoretical uncertainty at the NNLO order, as there are still undetermined LECs at this order.  In (b) we plot the total mass of the nucleon through $\mc{O}(m_q^2)$, and project the uncertainty in the NNLO contribution onto this plot as well.  For comparison, we also plot the total mass contribution through NLO, which is also plotted in Figure~\ref{fig:NmassPlot}.}
\end{figure}

We conclude this section by giving the nucleon mass in the isospin limit, expressed entirely in terms of the pion mass, as this is the most relevant and useful formula for lattice QCD simulations today.  The nucleon mass in this limit is
\begin{multline}\label{eq:NmassIsospin}
	M_N =
		M_0^R(\D) -\a_M^R\, \frac{m_\pi^2}{B_0} 
		-2\, \s_M^R(\D,\mu)\, \frac{m_\pi^2}{B_0}
		-3 \pi\, g_A^2\, \frac{m_\pi^3}{\L_\chi^2} 
		-\frac{8 g_{\D N}^2}{3} \frac{\mc{F} (m_\pi,\D,\mu)}{\L_\chi^2} 
\\		+\mc{C}^N (\D,\mu)\, m_\pi^4 
		+\mc{C}^N_l (\D,\mu)\, m_\pi^4\,  \textrm{log} \left( \frac{m_\pi^2}{\mu^2} \right) 
		+\mc{C}^N_J (\D,\mu)\, m_\pi^2\, \mc{J}(m_\pi,\D,\mu)
		+\mc{O}(m_q^{(5/2)})\, ,
\end{multline}
where
\begin{align}\label{eq:mq2LECs}
	\mc{C}^N =& \frac{4 g_{\D N}^2}{B_0 \L_\chi^2} \left(
			\a_M^R +2\s_M^R +\g_M^R -2\ol{\s}_M^R \right) 
		-\frac{9}{4 M_0 \L_\chi^2} \left( \frac{5}{8}g_A^2 +g_{\D N}^2 \right) \nonumber\\ 
		&\qquad\qquad
			+\frac{3 b^{vA}}{8 \L_\chi^3} -\frac{b_1^M}{4 B_0^2 \L_\chi} 
			-\frac{b_5^M}{2 B_0^2 \L_\chi} 
			-\frac{b_6^M}{2 B_0^2 \L_\chi} 
			-\frac{b_8^M}{B_0^2 \L_\chi} 
			+\frac{16\pi\, \ell_3}{B_0 \L_\chi^2} (\a_M^R + 2\s_M^R)\, ,\nonumber\\ \nonumber\\ 
	\mc{C}^N_l =& \frac{3 \a_M^R +6\s_M^R}{B_0 \L_\chi^2} 
		-\frac{3}{4 \L_\chi^3} \left( 4b^A +b^{vA} \right) 
		-\frac{1}{M_0 \L_\chi^2} \left( \frac{27 g_A^2}{16} +\frac{5 g_{\D N}^2}{2} \right)
		+\frac{1}{B_0 \L_\chi^2} (\a_M^R +2\s_M^R)\, , \nonumber\\
	\mc{C}^N_J =& \frac{4 g_{\D N}^2}{B_0 \L_\chi^2} \left(
		\a_M^R +2\s_M^R +\g_M^R -2\ol{\s}_M^R \right)\, .
\end{align}
Each LEC in the above coefficients is an implicit function of $\D$ as discussed in the previous sections.  The LECs of Eq.~\eqref{eq:mq2LECs} additionally have renormalization scale dependence, which we have suppressed, that exactly cancels the scale dependence of the logs in Eq.~\eqref{eq:NmassIsospin}.  At this order, in the isospin limit, there are these three linear combinations of LECs which contribute to the nucleon mass at this order, each of them containing LECs which are not well determined (if at all).  We use \textit{naive dimensional analysis} (NDA)~\cite{Manohar:1983md} to estimate the size of the new LECs, arriving at a central value set of $\g_M^R \sim \ol{\s}_M^R \sim b^A \sim b^{vA} \sim 1$, and $\frac{1}{4}b_M^1+\frac{1}{2}b_5^M +\frac{1}{2}b_6^M +b_8^M \sim 1$.  We then do an un-correlated variation of these LECs from $-5 \lesssim LEC \lesssim 5$.  The results are plotted in Figure~\ref{fig:NmassError}.
%%%%%%%%%%%%%%%%%%%%%%%%%%%%%%%%%%%%%%%%%%%%%%%%%%%
%
%	fig: chiral corrections to the nucleon mass
%
%%%%%%%%%%%%%%%%%%%%%%%%%%%%%%%%%%%%%%%%%%%%%%%%%%%
%%%%%%%%%%%%%%%%%%%%%%%%%%%%%%%%%%%%%%%%%%%%%%%%%%%
%
%%%%%%%%%%%%%%%%%%%%%%%%%%%%%%%%%%%%%%%%%%%%%%%%%%%
The important feature to note is that as can be seen by Figure~\ref{fig:NmassError}(a), the NNLO mass correction, for all values of the LECs used, is smaller than the LO and NLO mass contributions, lending confidence to the chiral expansion for this observable.  We anticipate that soon, all the LECs which contribute to the nucleon mass to this order will be determined from lattice QCD simulations, providing the first rigorous determination of HB\CPT\ parameters at this order, allowing a chiral extrapolation to the physical pion masses and a prediction of the nucleon masses from lattice QCD.

We now move on to compute the mass of the delta-resonances in the chiral expansion.

%%%%%%%%%%%%%%%%%%%%%%%%%%%%%%%%%%%%%%%%%%%%%%%%%%%
%
%	Delta masses
%
%%%%%%%%%%%%%%%%%%%%%%%%%%%%%%%%%%%%%%%%%%%%%%%%%%%
\subsection{\textbf{Delta Masses \label{sec:DeltaMasses}}}

Before embarking on the mass computation, it is first necessary to discuss what we mean by the mass of a state that can undergo strong decay.  The first simple answer is that for sufficiently large pion masses, the delta-resonances are actually stable particles.  This is because the nucleon delta mass splitting, as we have already mentioned, is independent of the quark masses, and thus for $m_\pi > \D = M_T -M_N$, the delta is kinematically forbidden from decaying.  This occurs for $m_\pi \sim 300$~MeV where the convergence of HB\CPT\ is questionable~\cite{Beane:2004ks} and thus in general it will be more difficult to extract observable quantities for the deltas than for nucleons.%
\footnote{There are other tricks one can use in lattice QCD to stabilize the deltas for $m_\pi < \D$, but we will not discuss them here.}
Nevertheless, the idea is to use the mass expressions derived in this section (or other observable quantities) to compare to lattice QCD simulations of delta observables for values of the pion (quark) masses such that the deltas are stable particles.  One then can determine the LECs which describe the properties of these ``resonance'' particles and then analytically continue to the physical quark masses of nature.  Using these techniques, for example, one can make a prediction with lattice QCD in conjunction with HB\CPT\ of the width of the deltas (this is well determined experimentally, so this is really a benchmark test).  The real test of this theory~\cite{Jenkins:1990jv,Jenkins:1991ne} however, will be to determine the LECs governing the delta properties by comparing HB\CPT\ calculations to lattice QCD correlation functions, and then make predictions of delta observables which play a noticeable role in experimentally measurable quantities.  The simplest quantity one can determine is the mass of the delta, which plays a large role in the majority of nucleon properties, see Figure~\ref{fig:NmassPlot} for example.  Thus in this section, we work out the mass of the deltas to $\mc{O}(m_q^2)$.

\bigskip

The computation of the delta-masses largely parallels that of the nucleon masses presented in the previous section.  Therefore we will mostly present the results, highlighting the differences from the nucleon mass computation.  Similar to the nucleons, the delta mass of the $T^{th}$ delta can be expressed in the chiral expansion as
\begin{equation}
     M_{T_i} = M_0 \left(\D \right) + \D +  M_{T_i}^{(1)} \left(\D, \mu \right)
                + M_{T_i}^{(3/2)}\left(\D, \mu \right)
                + M_{T_i}^{(2)}\left(\D, \mu \right) + \ldots
\label{eq:Tmassexp}
\end{equation}
Here, $M_0 \left(\D \right)$ is the renormalized nucleon mass 
in the chiral limit from Eq.~\eqref{eq:NucMassExp}, and $\D$ is the renormalized nucleon-delta mass splitting in the chiral limit. 
Both of these quantities are independent of $m_q$ and also of the $T_i$.
$M_{T_i}^{(n)}$ is the contribution to the 
$i^{th}$ delta baryon of the order $m_q^{(n)}$, and $\mu$ is the
renormalization scale.%
\footnote{As with the nucleon masses, the renormalization scale appears in each term 
contributing to the delta masses because we implicitly
treat the LECs as polynomial functions of the mass parameter $\Delta$. 
If we expand out the LECs, the $\mu$-dependence disappears at each order.}

\begin{figure}
\center
\begin{tabular}{ccc}
\includegraphics[width=0.15\textwidth]{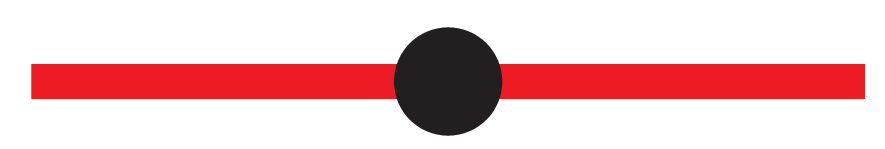} &
\includegraphics[width=0.3\textwidth]{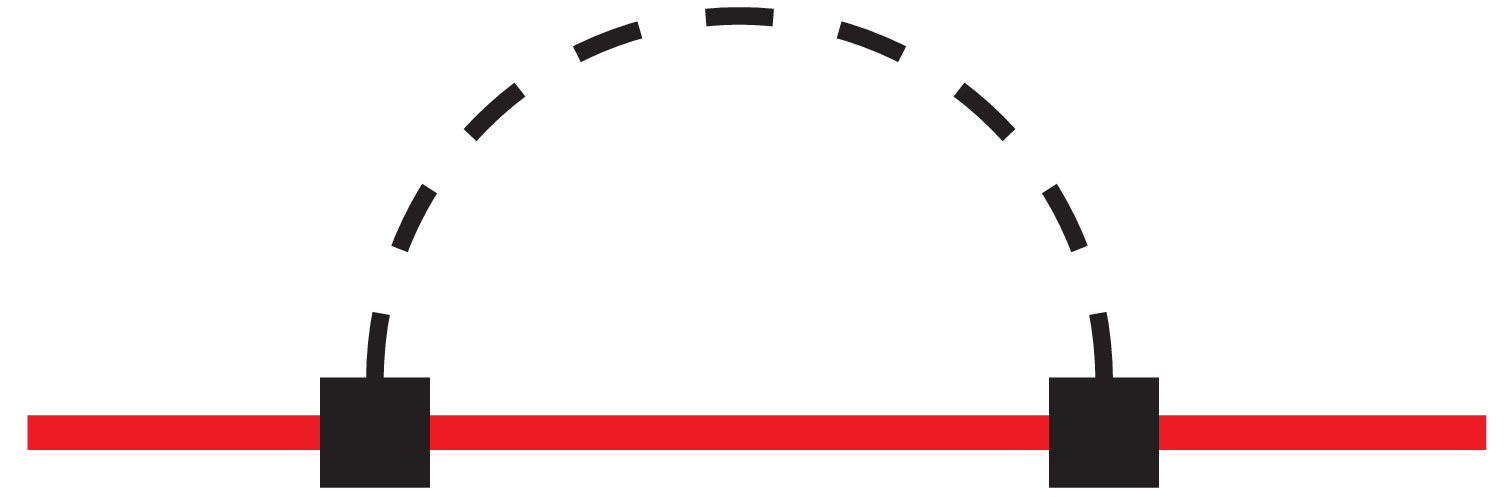} &
\includegraphics[width=0.3\textwidth]{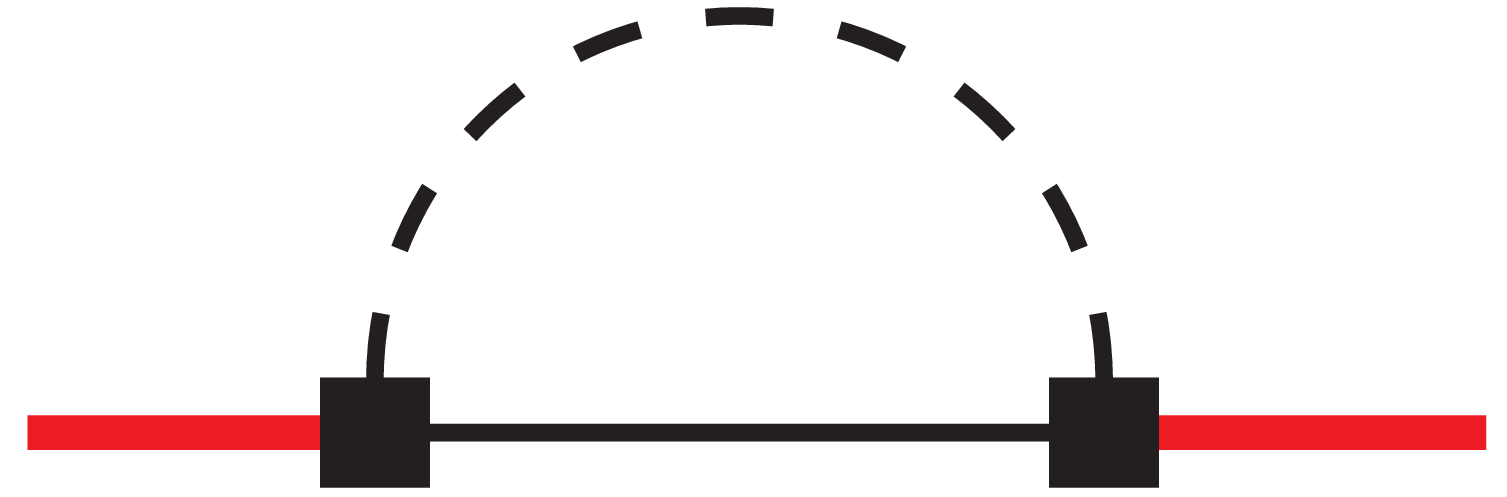} \\
(a) & (b) & (c)
\end{tabular}
\caption[Feynman diagrams depicting the leading order and next-to-leading order quark mass contribution to delta-resonance mass.]{\label{fig:LODmass}Leading order  and NLO quark mass dependence of delta-resonance mass.  The LO contribution, diagram (a), is given by the Lagrangian in Eq.~\eqref{eq:HBChPTLO}, while the NLO pion-loop contributions, diagrams (b) and (c), arise from the axial coupling of the pions to the nucleons and deltas given in Eq.~\eqref{eq:NNpi}.  The diagram (c), with an internal nucleon, is the diagram which gives rise to the imaginary contribution to the delta mass at this order, for $m_\pi < \D$.  The solid black lines, thick-solid line (red online), and the dotted black lines denote nucleons, delta-resonances and pions respectively.}
\end{figure}

The diagrams relevant to calculate the delta masses
to NNLO are depicted in Figure~\ref{fig:LODmass}. 
The leading-order contributions to the delta masses are
\begin{equation}
	M^{(1)}_{T_i} = \frac{2}{3} \, \gamma_M \, m^T_i  - 2 \ol \sigma_M \, \tr (m_q),
\label{eq:DmassLO}
\end{equation}
where the tree-level coefficients $m^T_i$ are given in Table \ref{t:mT} for the deltas $T$. 
%%%%%%%%%%%%%%%%%%%%%%%%%%%%%%%%%%%%%%%%%%%%%%%%%%%
%
%	table: LO delta mass coefficients
%
%%%%%%%%%%%%%%%%%%%%%%%%%%%%%%%%%%%%%%%%%%%%%%%%%%%
\begin{table}
\caption[Table of LO delta mass coefficients.]{\label{t:mT}The tree-level coefficients in \CPT. The coefficients $m^T_i$,
  $(m^2)^T_i$, and $(mm')^T_i$ are listed for the deltas $T$.}
\center
\begin{tabular}{| c | c c c |}
\hline
 & $m^T$ & $ \phantom{spac} (m^2)^T \phantom{spac}$ & $(mm')^T$ \\
\hline
$\D^{++}$ &  $3 m_u$ & $3 m_u^2$  & $3m_u^2$  \\
$\D^+$ &  $2 m_u + m_d$ & $2 m_u^2 + m_d^2$  & $m_u^2 + 2 m_u m_d$  \\
$\D^0$ &  $m_u + 2 m_d$ & $m_u^2 + 2 m_d^2$  & $2 m_u m_d + m_d^2$  \\
$\D^-$ &  $3 m_d$ & $3 m_d^2$  & $3 m_d^2$  \\
\hline
\end{tabular}
\end{table}
%%%%%%%%%%%%%%%%%%%%%%%%%%%%%%%%%%%%%%%%%%%%%%%%%%%
%
%	fig: LO delta mass
%
%%%%%%%%%%%%%%%%%%%%%%%%%%%%%%%%%%%%%%%%%%%%%%%%%%%
%%%%%%%%%%%%%%%%%%%%%%%%%%%%%%%%%%%%%%%%%%%%%%%%%%%
%
%%%%%%%%%%%%%%%%%%%%%%%%%%%%%%%%%%%%%%%%%%%%%%%%%%%

The NLO contributions to the delta masses are more interesting.  It is at this order that we first see the imaginary part of the delta mass (for $m_\pi < \D$), and see that it decays.  The mass corrections are depicted in Figure~\ref{fig:LODmass}(b)~and~\ref{fig:LODmass}(c), resulting in
\begin{equation}
	M^{(3/2)}_{T_i} = -\frac{25 g_{\D\D}^2}{432 \pi f^2} \, m_\pi^3 
		- \frac{2 g_{\D N}^2}{3(4 \pi f)^2} \, \mc{F} (m_\pi,-\D,\mu)\, ,
\label{eq:DmassNLO}
\end{equation}
where the function $\mc{F}(m_\pi, \D, \mu)$ is defined in Eq.~\eqref{eq:F}.  Note that in Eq.~\eqref{eq:DmassNLO}, it is $\mc{F}(m_\pi, -\D, \mu)$ which appears, arising from Figure~\ref{fig:LODmass}(c) with an internal nucleon loop.  For $m_\pi >\D$, this function is real valued,%
\footnote{For $m_\pi >0$, $\mc{F}(m_\pi, \D,\mu)$ is always real valued.}
but at $m_\pi = \D$, this function develops a cut, and thus has a real and imaginary component for $m_\pi < \D$, and thus the deltas become unstable and decays.%
\footnote{In lattice QCD simulations, the geometry of the box will also play a role in the decay of the delta (or other resonance states)~\cite{Luscher:1991cf}.  Because the lattice box only allows for quantized momentum modes, one can be in a region of pion masses where the delta would decay in infinite volume, but the geometry of the box (the box size) does not allow the necessary kinematics for the decay, and thus the deltas are stable.  Another way to see this is that in finite volume, the imaginary piece of the function $\mc{F}(m_\pi, -\D,\mu)$, comes from a series of poles determined by the kinematics allowed by the box, instead of a cut as in infinite volume.}
As with the nucleon masses, the LEC, $\ol{\s}_M$, becomes scale dependent to cancel the scale dependence introduced in Eq.~\eqref{eq:DmassNLO}.  The derivation of the scale dependence follows exactly from Eq.~\eqref{eq:DeltaRenorm}.  After applying the renormalization prescription of Eq.~\eqref{eq:MSbar}, the LO delta LECs become
\begin{align}
	M_0 \rightarrow&\ M_0^R(\D) 
		= M_0 +\frac{16 g_{\D N}^2}{9} \frac{\D^3}{\L_\chi^2}\, , \nonumber\\
	\D_0 \rightarrow&\ \D = \D_0 -\frac{26 g_{\D N}^2}{9} \frac{\D_0^3}{\L_\chi^2}\, , \nonumber\\
	\g_M \rightarrow&\ \g_M^R = \g_M\, , \nonumber\\
	\ol{\s}_M \rightarrow&\ \ol{\s}_M^R = \ol{\s}_M -\frac{2 g_{\D N}^2}{3} \frac{\D B_0}{\L_\chi^2}
		+\frac{g_{\D N}^2}{2} \frac{\D B_0}{\L_\chi^2}\, 
			\textrm{log} \left( \frac{\mu^2}{\mu_0^2} \right)\, .
\end{align}
Here we see that the parameter $M_0^R$ is the same as in Eq.~\eqref{eq:DeltaRenorm}.  We restate that this was a choice to make this equal to the nucleon mass in the chiral limit.  A different choice would simply result in a different value for $\D$ (but then also the mass splitting parameter in the $\mc{F}$ function would be $M_T - M_B$ instead of $\D$).  We also reiterate that the only important part of this renormalization prescription is the $\mu$ dependence, as there is arbitrary dependence upon $\D$ in all these LECs just as with the nucleon LECs.

The NNLO ($\mc{O}(m_q^2)$) corrections are given by diagrams similar to those in Figure~\ref{fig:NmassNNLO}, but with an interchange of the nucleon and delta resonance lines for both the internal and external states.  The mass correction is then
\begin{align}
	M_{T_i}^{(2)} =& (Z_{T_i} - 1) M_{T_i}^{(1)} \nonumber \\ 
	& + \frac{1}{\L_\chi} \left\{ \frac{1}{3} t_1^M \, (m^2)^T_i 
		+ \frac{1}{3} t_2^M \, (m m')^T_i 
		+ t_3^M \, \tr(m_q^2) 
		+ \frac{1}{3} t_4^M \,  m^T_i  \, \tr (m_q)
		+ t_5^M \, [\tr (m_q)]^2 \right\} \nonumber \\
	& - \Big[ 2 \,\g_M C_\pi^{T_i} -6\ol{\s}_M \, \tr(m_q) \Big] \frac{m_\pi^2}{\L_\chi^2}
			\textrm{log} \left( \frac{m_\pi^2}{\mu^2} \right) 
		-\left[ \frac{1}{2}\left(  t^{\tilde{A}}_2 +t^{vA}_2 \right) 
			+3 \left( t^{\tilde{A}}_3 +t^{vA}_3 \right)\right]\, \frac{m_\pi^4}{8\, \L_\chi^3} \nonumber\\
	& + \left[ \left( \frac{1}{2} t^A_2  + 3 t^A_3 +\frac{1}{8} t^{\tilde{A}}_2
		+\frac{1}{8} t^{vA}_2 +\frac{3}{4} t^{\tilde{A}}_3 +\frac{3}{4} t^{vA}_3 \right) \right]
			\frac{m_\pi^4}{\L_\chi^3}\, \textrm{log} \left( \frac{m_\pi^2}{\mu^2} \right) \nonumber \\
	& - \frac{m_\pi^4}{M_0 \L_\chi^2} \left[
		\frac{25 g_{\D\D}^2}{48} 
			\left( \textrm{log} \left( \frac{m_\pi^2}{\mu^2} \right) +\frac{19}{10} \right)
		+\frac{5 g_{\D N}^2}{8} \left( \textrm{log} \left( \frac{m_\pi^2}{\mu^2} \right) -\frac{1}{10} \right)
		\right] \nonumber \\
	& -\frac{25 g_{\D\D}^2\, \ol{\s}_M}{9} \frac{m_\pi^2 \tr(m_Q)}{\L_\chi^2} 
			\left[ \textrm{log} \left( \frac{m_\pi^2}{\mu^2} \right) +\frac{26}{15} \right]
		- 2 g_{\D N}^2 \sigma_M\frac{\tr(m_Q)}{\L_\chi^2} 
		\mc{J} (m_\pi,-\D,\mu)\nonumber \\
	& +\frac{10 g_{\D\D}^2 \g_M\, F_\pi^{T_i}}{9} \frac{m_\pi^2}{\L_\chi^2}
			 \left[ \textrm{log} \left( \frac{m_\pi^2}{\mu^2} \right) +\frac{26}{15} \right]
		- \frac{3 g_{\D N}^2\, G_\pi^{T_i} }{2}\, \frac{\mc{J} (m_\pi,-\D,\mu)}{\L_\chi^2}\, .
\label{eq:DmassNNLO}
\end{align}
Here we have used $(m^2)^T_i$, $(m m')^T_i$ to label the tree-level coefficients that
appear in Table~\ref{t:mT}. 
Above the wavefunction renormalization $Z_{T_i}$ is given by
\begin{equation}
Z_{T_i} - 1 = - \frac{25 g_{\D\D}^2}{18} \frac{m_\pi^2}{\L_\chi^2} 
           \left[ \textrm{log} \left( \frac{m_\pi^2}{\mu^2} \right) +\frac{26}{15} \right]
          - g_{\D N}^2 \frac{\mc{J} (m_\pi,-\D,\mu)}{\L_\chi^2}
\label{eq:Z}
.\end{equation}
The coefficients in the above expression, namely $C_\pi^{T_i}$, $F_\pi^{T_i}$, and $G_\pi^{T_i}$, are given in 
Table~\ref{t:QCD-C} and depend on the delta state $T$.
%%%%%%%%%%%%%%%%%%%%%%%%%%%%%%%%%%%%%%%%%%%%%%%%%%%
%
%	table: NNLO delta mass coefficients
%
%%%%%%%%%%%%%%%%%%%%%%%%%%%%%%%%%%%%%%%%%%%%%%%%%%%
\begin{table}
\caption[Coefficients for delta mass corrections at $\mc{O}(m_q^2)$.]{\label{t:QCD-C}The coefficients $C_\pi^{T_i}$, $F_\pi^{T_i}$, and $G_\pi^{T_i}$ in \CPT, for the various delta states, $T$.}
\center
\begin{tabular}{| l | c c c |}
\hline
 & $\quad \quad \quad C_\pi^T \quad \quad \quad $  
 & $\quad \quad \quad F_\pi^T \quad \quad \quad$ 
 & $\quad \quad \quad G_\pi^T  \quad  $ \\
\hline
$\D^{++}$       
 & $2 m_u + m_d$  
 & $\frac{13}{6} m_u + \frac{1}{3} m_d$ 
 & $\frac{4}{3} \a_M m_u$    \\
$\D^+$ 
 & $\frac{5}{3} m_u + \frac{4}{3} m_d$ 
 & $\frac{14}{9} m_u + \frac{17}{18} m_d$ 
 & $\frac{4}{9} \a_M ( 2 m_u + m_d )$\\
$\D^0$    
 & $\frac{4}{3} m_u + \frac{5}{3} m_d$  
 & $\frac{17}{18} m_u + \frac{14}{9} m_d$ 
 & $\frac{4}{9} \a_M (m_u + 2 m_d)$\\
$\D^-$ 
 & $m_u + 2 m_d$ 
 & $\frac{1}{3} m_u + \frac{13}{6} m_d$ 
 & $\frac{4}{3} \a_M m_d$  \\
\hline
\end{tabular}
\end{table}
%%%%%%%%%%%%%%%%%%%%%%%%%%%%%%%%%%%%%%%%%%%%%%%%%%%
%
%%%%%%%%%%%%%%%%%%%%%%%%%%%%%%%%%%%%%%%%%%%%%%%%%%%
Given these delta mass expressions, Eqs.~\eqref{eq:DmassLO}. \eqref{eq:DmassNLO} and \eqref{eq:DmassNNLO}, one can predict the width of the deltas with the knowledge of a small number of LECs, $f$, $B_0$, $\D$, $\a_M$, $\s_M$, $\g_M$, $\ol{\s}_M$ and $g_{\D N}$, most of which can be determined independently from the delta masses.  The expression for the delta width is given to this order by
\begingroup
\small
\begin{equation}
	\textrm{Im} (\ol M_T) = \left\{
		\begin{array}{lc}
		- \frac{g_{\D N}^2}{12 \pi f^2} \sqrt{\D^2 - m_\pi^2} 
			\bigg\{ \D^2 - m_\pi^2 %& \\
%        \qquad\qquad\qquad
		+ \D \frac{m_\pi^2 }{6 B_0}
			\left[ 3 \gamma_M  + 4  (\sigma_M - \ol \sigma_M )  + 2 \a_M \right]
		\bigg\}, & \text{for } m_\pi < \D \\
		\qquad\qquad\qquad 0\, , & \text{for } m_\pi > \D %\\
		\end{array}
\right.
\end{equation}
\endgroup
Additionally, one can compute the quark mass dependent nucleon delta mass splitting, which is the quantity which should be used in place of $\D$ in the functions, $\mc{F}$ and $\mc{J}$, as for a given quark mass, this is the parameter which governs the stability of the deltas.  Thus, the real component of the nucleon delta mass splitting is given through $\mc{O}(m_q^2)$ by the expression,%
\footnote{%
Here we have used the following replacements $A-G$ for particular combinations of the LECs: 
$A = \a_M + 2 (\sigma_M - \ol \sigma_M ) + \gamma_M$,
$B = \frac{3}{2} b_1^M + b_4^M + \frac{1}{2} t_1^M + t_3^M$,
$C = b_5^M + b_7^M + \frac{1}{3} t_2^M + \frac{1}{2} t_4^M + t_5^M$, 
$D = \frac{3}{2} b_1^A + 3 b_4^A + \frac{3}{8} b_1^{vA} + \frac{3}{4} b_4^{vA} + \frac{3}{2} t_1^A + \frac{1}{2} t_2^A + 3 t_3^A 
+ \frac{3}{8} (t_1^{\tilde{A}} + t_1^{vA}) + \frac{1}{8} (t_2^{\tilde{A}} + t_2^{vA}) + \frac{3}{4} (t_3^{\tilde{A}} + t_3^{vA})
$,
$E = \frac{27}{2} g_A^2 +  15 g_{\D N}^2 - \frac{25}{6} g_{\D \D}$, 
$F = - \frac{1}{8} \left[ \frac{3}{2} b_1^{vA} + 3 b_4^{vA} 
+ \frac{3}{2} (t_1^{\tilde{A}} + t_1^{vA}) + \frac{1}{2} (t_2^{\tilde{A}} + t_2^{vA}) + 3 (t_3^{\tilde{A}} + t_3^{vA})\right]
$, and
$G = \frac{45}{4} g_A^2 + \frac{37}{2} g_{\D N}^2 - \frac{95}{12} g_{\D \D}^2$.
}
\begin{eqnarray}
\textrm{Re} (\ol M_T - \ol M_B) &=& 
	\Delta 
	+ A \, \tr (m_q) 
	+ B \, \tr (m_q^2) 
	+ C \, [\tr(m_q)]^2  \notag \\  
	&& 
	+ \frac{m_\pi^3}{\L_\chi^2} \left[ 3 g_A^2 - \frac{25}{27} g_{\D \D}^2 \right] 
	+ \frac{10 g_{\D N}^2}{3 \L_\chi^2} \mc{F}(m_\pi, \D, \mu) \notag \\
	&& 
	+ \frac{A \, \tr(m_q)}{\L_\chi^2} \left\{ 3  m_\pi^2\, \textrm{log} \left( \frac{m_\pi^2}{\mu^2} \right) 
		- 5 g_{\D N}^2 \left[ \mc{J} (m_\pi, \D, \mu) + \frac{4}{5} m_\pi^2 \right]
	\right\} \notag \\
	&& + \frac{m_\pi^4}{\L_\chi^2}\, \textrm{log} \left( \frac{m_\pi^2}{\mu^2} \right) \left[ 
		\frac{D}{\L_\chi} + \frac{E}{8 M_B} \right]
	+ \frac{m_\pi^4}{\L_\chi^2} \left[ \frac{F}{\L_\chi} + \frac{G}{8 M_B} \right]. \notag \\
\end{eqnarray}

%%%%%%%%%%%%%%%%%%%%%%%%%%%%%%%%%%%%%%%%%%%%%%%%%%%
%
%	Strong Isospin contributions to nucleon and delta mass splittings
%
%%%%%%%%%%%%%%%%%%%%%%%%%%%%%%%%%%%%%%%%%%%%%%%%%%%
\subsection{\textbf{Strong isospin breaking of the nucleon and delta masses in lattice QCD}}

Having derived the nucleon and delta masses to NNLO in the chiral
expansion, we now focus on the mass splittings between these states.
To begin, we consider the nucleon mass splitting, which to our knowledge was first theoretically addressed in~\cite{Gasser:1974wd}. The degeneracy
between the proton and neutron is broken by leading-order effects
in the chiral theory, see Eq.~(\ref{eq:LONmass}). Beyond this order
pion loops contribute, but to the order we are working, all the pions
are degenerate, even away from the isospin limit.  Thus the
NLO contributions to the neutron and proton masses are the same,
see Eq.~(\ref{eq:NmassNLO}), and the mass splitting, $M_n
- M_p$, is given to NLO accuracy, entirely by the difference of the LO mass
contribution in Eq.~(\ref{eq:LONmass}), and is linear 
in $m_d - m_u$.   Any deviation from this linear mass splitting seen in
lattice simulations of the nucleon masses should be a signature of the NNLO
mass contributions and certain LECs that arise at this
order.  Additionally, the nucleon mass splitting can be enhanced from
that in nature by increasing the quark mass splitting on the lattice, $m_d - m_u$.
This enhancement, combined with the vanishing of the NLO contribution
to the mass splitting, provides us with a means of cleanly determining
the NNLO nucleon mass contributions and isolating certain LECs arising
at this order.  These effects are normally obscured by the NLO contributions.

We find the nucleon mass splitting is given to NNLO by%
\footnote{We note that during the completion of this thesis, the first determination of the strong isospin breaking in the nucleon masses from a fully dynamical lattice QCD simulation was completed~\cite{Beane:2006fk}, although unfortunately there were not enough different pion masses used to make use of this full mass splitting expression.}
\begin{align}\label{eq:NmassSplit}
	M_n - M_p =& - \frac{2\a_M}{B_0}\, \frac{1-\eta}{1+\eta}\, m_\pi^2 \nonumber\\
  		&+ \frac{1-\eta}{1+\eta}\, \frac{m_\pi^2}{B_0} \bigg\{ \frac{m_\pi^2}{\L_\chi^2} 
		\bigg[ 8\left( g_A^2 \a_M + g_{\D N}^2 \left( \a_M +\frac{5}{9}\g_M \right) \right)
			- (b^M_1 +b^M_6)\frac{\pi f}{B_0} \bigg] \nonumber\\
		&\qquad\qquad\qquad
		+ \frac{m_\pi^2}{\L_\chi^2}\ 2\a_M (6 g_A^2 +1)\, 
			\textrm{log} \left( \frac{m_\pi^2}{\mu^2} \right) \nonumber\\
		&\qquad\qquad\qquad
		+ \frac{\mc{J} (m_\pi,\D,\mu)}{\L_\chi^2}\ 8 g_{\D N}^2
			\left( \a_M + \frac{5}{9} \g_M \right) \bigg\}\, ,
\end{align}
where $\eta = \frac{m_u}{m_d}$.  An identical situation arises for the deltas, as their degeneracy
is broken at leading order in the chiral expansion, while the next contribution to their splittings occurs at NNLO. Their mass splittings are given by $\d M_T$, which stands for $M_{\D^-} -M_{\D^0}$\ ,\ $M_{\D^0} -M_{\D^+}$ and $M_{\D^+} -M_{\D^{++}}$.  We find,

\begin{align}\label{eq:DmassSplit}
	\d M_T =& \frac{2 \g_M}{3 B_0} \frac{1-\eta}{1+\eta}\, m_\pi^2 \nonumber\\
		&+ \frac{1-\eta}{1+\eta}\, \frac{m_\pi^2}{B_0}  \bigg\{ \frac{m_\pi^2}{\L_\chi^2} 
		\bigg[ \frac{\pi f}{3 B_0}(t^M_1 +t^M_4) -\frac{104}{243} g_{\D\D}^2 \g_M \bigg] \nonumber\\
		&\qquad\qquad\qquad
		- \frac{m_\pi^2}{\L_\chi^2} \frac{20\g_M}{81} \left( \frac{27}{10}+ g_{\D\D}^2 \right)\, 
			\textrm{log} \left( \frac{m_\pi^2}{\mu^2} \right) \nonumber\\
		&\qquad\qquad\qquad
		- \frac{\mc{J} (m_\pi,-\D,\mu)}{\L_\chi^2} \frac{2 g_{\D N}^2}{3} (\g_M +\a_M) \nonumber\\
		&\qquad\qquad\qquad
		+ \frac{1}{12 \pi f} t^M_2\ \d t^M  \bigg\},
\end{align}
where $\d t^M$ is given in Table~\ref{t:MassSplitViol}.  Considering these splittings, there are a few things to note.  We know from experiment that the neutron is more massive than the proton, and we expect that this is mostly due to $m_d > m_u$.  Similarly, we expect the deltas to follow a similar pattern with $\D^-$ being more massive than $\D^0$, and so on, although this is presently undetermined experimentally.  From these expectations and the expressions above, it is expected that $\a_M$ is
negative and $\g_M$ is positive.  Also, the delta masses, to a good approximation are expected to follow an equal spacing rule.  In HB\CPT, this rule is first violated at NNLO, and even then, only by the operator in Eq.~(\ref{eq:DHDO}) with coefficient $t^M_2$.  This coefficient can be isolated by taking successive differences.

\begin{table}
\center
\caption[The coefficient $\d t^M$, which encodes the violation of
the delta equal spacing rule at $\mc{O}(m_q^2)$.]{\label{t:MassSplitViol} The coefficient $\d t^M$, which encodes the violation of the delta equal spacing rule at $\mc{O}(m_q^2)$.}
\begin{tabular}{| c | c |}
\hline
$\d M_T$ & $\d t^M$\\
\hline
$M_{\D^-} -M_{\D^0}$	& $2m_d$\\
$M_{\D^0} -M_{\D^+}$	& $m_u +m_d$\\ 
$M_{\D^+} -M_{\D^{++}}$	& $2 m_u$\\
\hline
\end{tabular}
\end{table}

Lastly we should comment that in nature isospin violation in the
baryon masses has another source of the same size as the NNLO chiral
effects, namely electromagnetic contributions.  In lattice QCD calculations one can turn off the electric charges of the quarks.  This is the scenario for which our calculations are applicable.  There have been a few lattice computations of electromagnetic contributions to hadronic masses~%
\cite{Duncan:1996xy,Duncan:1996be, Duncan:2004ys}.  For a recent discussion of the electromagnetic effects in hadrons, see Ref.~\cite{Gasser:2003hk}, and for a comprehensive phenomenological review, 
\cite{Miller:1990iz}.

%%%%%%%%%%%%%%%%%%%%%%%%%%%%%%%%%%%%%%%%%%%%%%%%%%%
%
%	SU(3) Masses
%
%%%%%%%%%%%%%%%%%%%%%%%%%%%%%%%%%%%%%%%%%%%%%%%%%%%
\section{Octet and Decuplet Baryon Masses in $SU(3)$ \CPT \label{sec:SU3HbChPT}}

In this section and the following section, Sec.~\ref{sec:NDPQmasses}, we wish to extend the baryon mass computations to the $SU(3)$ theory as well as the partially quenched theories, $SU(4|2)$ and $SU(6|3)$.  There are many more baryons in each of these theories than the two nucleons and 4 deltas in $SU(2)$.  To highlight the differences between these theories we will only explicitly give the masses of the nucleons and deltas in the following sections and refer the reader to Refs.~\cite{Walker-Loud:2004hf,Tiburzi:2004rh,Tiburzi:2005na} for a complete listing of the formulae for all the baryon masses.  

There are no conceptual differences in the computation of the baryon masses in $SU(3)$ \textit{vs.} $SU(2)$.  The only differences are the addition of more operators in the chiral Lagrangian, due to the larger symmetry group of $SU(3)$, which leads to kaon and eta loops, in addition to the pion loops of the previous section.  We have already shown the Lagrangian necessary to compute the octet and decuplet baryon masses to NLO in Eqs.~\eqref{eq:HBChPTSU3} and \eqref{eq:BBpi}.  To compute the NNLO mass corrections, we must first construct the analogue of Eqs.~\eqref{eq:RPIfixed}, \eqref{eq:NHDO} and \eqref{eq:DHDO} in $SU(3)$.  The treatment of the octet-decuplet mass splitting parameter is identical to the discussion in Sec.~\ref{sec:Nmass}.  The reparameterization fixed Lagrangian is given by
\begin{align}
	\mc{L} =&\ -\tr \left( \ol{B} \frac{D_\perp^2}{2 M_0} B \right)
		+\tr \left( \ol{T}^\mu\, \frac{D_\perp^2}{2M_0}\, T_\mu \right)
		\nonumber\\
		&\ -(D+F)\, \tr \left( \ol{B}\, v\cdot \mc{A}\, \frac{S\cdot i \overrightarrow{D}}{M_0}\, B \right)
		+(D+F)\, \tr \left( \ol{B}\, \frac{i \overleftarrow{D}\cdot S}{M_0}\, v\cdot \mc{A}\, B \right)
		\nonumber\\
		&\ -(D-F)\, \tr \left( \ol{B}\, \frac{S\cdot i\overrightarrow{D}}{M_0}\, B\, v\cdot \mc{A} \right)
		+(D-F)\, \tr \left( \ol{B}\, \frac{i\overleftarrow{D}\cdot S}{M_0}\, B\, v\cdot \mc{A} \right)
		\nonumber\\
		&\ + \mc{H} \left( \ol{T}^\mu \frac{i \overleftarrow{D} \cdot S}{M_0}
				\vit \cdot \mc{A}\, T_\mu 
			- \ol{T}^\mu \, \vit\cdot \mc{A} \frac{S\cdot i \overrightarrow{D}}{M_0} T_\mu \right)\, .
\end{align}
The other terms in the Lagrangian with undetermined LECs are given by
\begin{align}\label{eq:BBAA}
	\mc{L} =&\ \frac{b_1^A}{\L_\chi}\, \tr \left( \ol{B}\, \mc{A}\cdot\mc{A}\, B \right)
		+\frac{b_2^A}{\L_\chi}\, \tr \left( \ol{B}\, \mc{A}_\mu\, B\, \mc{A}^\mu \right)
		+\frac{b_3^A}{\L_\chi}\, \tr \left( \ol{B}\, B\, \mc{A}\cdot\mc{A} \right)
		+\frac{b_4^A}{\L_\chi}\, \tr \left( \ol{B}\, B \right)\, \tr \left( \mc{A}\cdot\mc{A} \right)\nonumber\\
		&\ +\frac{b_1^{vA}}{\L_\chi} \tr \left( \ol{B}\, v\cdot \mc{A}\, v\cdot\mc{A}\, B \right)
		+\frac{b_2^{vA}}{\L_\chi}\, \tr \left( \ol{B}\, v\cdot\mc{A}\, B\, v\cdot\mc{A} \right)
		+\frac{b_3^{vA}}{\L_\chi}\, \tr \left( \ol{B}\, B\, v\cdot\mc{A}\, v\cdot\mc{A} \right) \nonumber\\
		&\ +\frac{b_4^{vA}}{\L_\chi}\, \tr \left( \ol{B}\, B \right)\, \tr \left( v\cdot\mc{A}\, v\cdot\mc{A} \right)
		+\frac{b_1^M}{\L_\chi}\, \tr \left(\ol{B}\, \mc{M}_+\, \mc{M}_+\, B \right)
		+\frac{b_2^M}{\L_\chi}\, \tr \left(\ol{B}\, \mc{M}_+\, B\, \mc{M}_+ \right) \nonumber\\
		&\ +\frac{b_3^M}{\L_\chi}\, \tr \left( \ol{B}\, B\, \mc{M}_+\, \mc{M}_+\right)
		+\frac{b_4^M}{\L_\chi}\, \tr \left( \ol{B}\, \mc{M}_+\, B\right)\, \tr \left(\mc{M}_+ \right)
		+\frac{b_5^M}{\L_\chi}\, \tr \left( \ol{B}\, B\, \mc{M}_+ \right)\, \tr \left(\mc{M}_+ \right) \nonumber\\
		&\ +\frac{b_6^M}{\L_\chi}\, \tr \left( \ol{B}\, B\right)\, \tr \left( \mc{M}_+\, \mc{M}_+ \right)
		+\frac{b_7^M}{\L_\chi}\, \tr \left( \ol{B}\, B\right)\, \tr\left(\mc{M}_+\right)\, \tr\left(\mc{M}_+\right)
		\nonumber\\
		&\ +\frac{b_8^M}{\L_\chi}\, \tr \left(\ol{B}\, \mc{M}_+\right)\, \tr\left( B\, \mc{M}_+\right)\, .
\end{align}
There are a few points worth making about the above Lagrangian.  Firstly, the operators with coefficients $b_{1-8}^M$ form an over-complete set.  One can make use of Cayley-Hamilton identities for $SU(3)$ matrices to reduce these operators by one~\cite{Borasoy:1996bx}.  However, there is no consensus in the literature about which 7 of the 8 operators to keep, and so we list them all for to form the complete basis.  This over-complete basis also has a more obvious extension to the partially quenched operators we will determine in Sec.~\ref{sec:NDPQmasses}.  Also, as in Sec.~\ref{sec:Nmass}, one can show that the operators of the form $\tr \left(\ol{B}\, S\cdot \mc{A}\, S\cdot \mc{A}\, B \right)$ are not linearly independent from the operators with coefficients $b_{1-4}^A$ and $b_{1-4}^{vA}$.

The operators of the form $\big( \Bbar v\cdot A v\cdot A B \big)$ have not been kept explicit in the literature so far for calculations of the octet baryon masses to $\mc{O}(m_q^2)$~\cite{Jenkins:1990jv,
  Jenkins:1992ts,Borasoy:1996bx}.  These operators have identical flavor structure to the corresponding
$\tr \left( \ol{B}\, \mc{A}\cdot \mc{A} B \right)$ operators.  However, these two sets of operators have different Lorentz structure, which gives rise to different finite $m_q^2$ contributions to the octet baryon self energy calculations.  One can choose a renormalization scheme such that the contributions to the octet baryon masses from these different operators can not be distinguished.  Therefore, with a suitable redefinition of the $b^A_{1-4}$ and $b^M_{1-7}$ coefficients, the operators with coefficients $b^{vA}_{1-4}$ can be neglected in the baryon mass calculations, as their contributions to the masses can be absorbed by the other operators to this order in the chiral expansion.

However, the operators with coefficients $b^A_{1-4}$ and $b^{vA}_{1-4}$ can be distinguished in $\pi N \rightarrow \pi N$ scattering at tree level, for example. It is therefore useful to keep both types of operators explicitly in the Lagrangian, allowing them to be distinguished in the octet baryon mass calculation.  This also provides a consistent means to determine these LECs, as they are the same coefficients which appear in the Lagrangian used for the calculation of other observables like $\pi N \rightarrow \pi N$ scattering.

In principle, additional $1/M_0$ operators with the same chiral symmetry properties as those contained in Eq.~(\ref{eq:BBAA}) can be generated.  However, these $1/M_0$ operators do not have their coefficients constrained by RI, therefore they can be absorbed by a re-definition of the $b^{A,vA,M}_i$ coefficients.  For example, the Rarita-Schwinger field used to describe the decuplet baryons contains
un-physical spin-$\frac{1}{2}$ degrees of freedom which can propagate when the decuplet baryons are off mass-shell.  However, these off-shell degrees of freedom are suppressed by $1/M_0$ and are implicitly included in the operators with coefficients $b^{A,vA}_{1-4}$.

There are a similar set of operators for the decuplet fields.  The greater flavor symmetry of the decuplet fields, however, reduces the number of operators as compared to the octet baryons but the Lorentz structure allows for more types of operators.  These operators are given by
\begin{align}\label{eq:TTAA}
	\mc{L} =&\ \frac{t_1^A}{\L_\chi}\, \ol{T}_\mu^{kji} \left( \mc{A} \cdot \mc{A} \right)_i^{\ l} T^\mu_{ljk}
		+\frac{t_2^A}{\L_\chi}\, \ol{T}_\mu^{kji} \mc{A}_{i}^{\ l}\cdot \mc{A}_{j}^{\ p} T^\mu_{lpk}
		+\frac{t_3^A}{\L_\chi}\, \left( \ol{T}^\mu T_\mu \right)\, \tr \left(\mc{A}\cdot \mc{A}\right)
		\nonumber\\
		&\ +\frac{t_1^{vA}}{\L_\chi}\, \ol{T}_\mu^{kji} \left( v\cdot\mc{A}\, v\cdot \mc{A} \right)_i^{\ l} 
			T^\mu_{ljk}
		+\frac{t_2^{vA}}{\L_\chi}\, \ol{T}_\mu^{kji} v\cdot\mc{A}_{i}^{\ l}\, v\cdot \mc{A}_{j}^{\ p} 
			T^\mu_{lpk}
		+\frac{t_3^{vA}}{\L_\chi}\, \left( \ol{T}^\mu\, T_\mu \right)\, 
			\tr \left(v\cdot \mc{A}\, v\cdot \mc{A}\right) \nonumber\\
		&\ +\frac{t_1^{\tilde{A}}}{\L_\chi}\, \ol{T}_\mu^{kji} \left( \mc{A}^\mu \cdot \mc{A}_\nu \right)_i^{\ l}
			 T^\nu_{ljk}
		+\frac{t_2^{\tilde{A}}}{\L_\chi}\, \ol{T}_\mu^{kji} \mc{A}_{i}^{\mu,l}\cdot \mc{A}_{\nu,j}^{\ p}\, 
			T^\nu_{lpk}
		+\frac{t_3^{\tilde{A}}}{\L_\chi}\, \left( \ol{T}^\mu\, T_\nu \right)\, 
			\tr \left(\mc{A}_\mu\, \mc{A}^\nu\right) \nonumber\\
		&\ +\frac{t_1^M}{\L_\chi}\, \ol{T}_\mu^{kji} \left( \mc{M}_+\, \mc{M}_+\right)_i^{\ l} T^\mu_{ljk}
		+\frac{t_2^M}{\L_\chi}\, \ol{T}_\mu^{kji} \left( \mc{M}_+\right)_i^{\ l} \left( \mc{M}_+\right)_j^{\ p}
			T^\mu_{lpk}
		+\frac{t_3^M}{\L_\chi}\, \left(\ol{T}_\mu\, T^\mu\right)\, \tr \left(\mc{M}_+\, \mc{M}_+ \right)
		\nonumber\\
		&\ +\frac{t_4^M}{\L_\chi}\, \left( \ol{T}_\mu \mc{M}_+ T^\mu \right)\, \tr \left( \mc{M}_+ \right)
		+\frac{t_5^M}{\L_\chi}\, \left( \ol{T}_\mu T^\mu \right)\, \tr \left( \mc{M}_+ \right)\, 
			\tr \left( \mc{M}_+ \right)\, .
\end{align}
All of the above LECs as well as those in Eq.~\eqref{eq:BBAA} are dimensionless.  We now have the complete set of operators needed to compute the masses of the octet and decuplet baryons through $\mc{O}(m_q^2)$.  As stated above, we shall only give the masses of the nucleons and deltas, and we shall restrict ourselves to the isospin limit, $m_u = m_d = \bar{m}$.  The quark mass expansion of the nucleon mass is given exactly as in Eq.~\eqref{eq:NucMassExp}.  For simplicity, we treat $m_K \sim m_\pi$ in the power counting, although phenomenologically this may not be the ideal counting.  The nucleon mass is then
\begin{align}\label{eq:NmassSU3}
	M_N =&\ M_0 - 2\bar{m} \left( \a_M+\b_M+2\s_M\right) -2m_s\, \s_M 
		-\frac{8\mc{C}^2}{3}\, \frac{\mc{F}(m_\pi,\D,\mu)}{\L_\chi^2}
		-\frac{2\mc{C}^2}{3}\, \frac{\mc{F}(m_K,\D,\mu)}{\L_\chi^2} \nonumber\\
		&\ 
		-3\pi(D+F)^2\, \frac{m_\pi^3}{\L_\chi^2}  -\frac{\pi(D-3F)^2}{3}\, \frac{m_\eta^3}{\L_\chi^2} 
		-\frac{2\pi(5D^2-6DF+9F^2)}{3}\, \frac{m_K^3}{\L_\chi^2} \nonumber\\
		&\ 
		-\left[ \frac{b_\pi^A+\frac{1}{4}b_\pi^{vA}}{\L_\chi^3} 
			+\frac{27(D+F)^2+40\mc{C}^2}{16M_0\L_\chi^2}\right]\,
			m_\pi^4\, \ln \left( \frac{m_\pi^2}{\mu^2} \right) \nonumber\\
		&\ -\left[ \frac{b_\eta^A+\frac{1}{4}b_\eta^{vA}}{\L_\chi^3} 
			+\frac{3(D-3F)^2}{16M_0\L_\chi^2}\right]\,
			m_\eta^4\, \ln \left( \frac{m_\eta^2}{\mu^2} \right) \nonumber\\
		&\ -\left[ \frac{b_K^A+\frac{1}{4}b_K^{vA}}{\L_\chi^3} 
			+\frac{3(5D^2-6DF+9F^2)+5\mc{C}^2}{8M_0\L_\chi^2}\right]\,
			m_K^4\, \ln \left( \frac{m_K^2}{\mu^2} \right) \nonumber\\
		&\ +6\bar{m}\, (\a_M+\b_M+2\s_M)\, 
			\frac{m_\pi^2}{\L_\chi^2}\, \ln \left( \frac{m_\pi^2}{\mu^2} \right) \nonumber\\
		&\ +\left[ \frac{2}{3}\bar{m}(\a_M+\b_M+2\s_M) +\frac{8}{3}m_s \right]
			\frac{m_\eta^2}{\L_\chi^2}\, \ln \left( \frac{m_\eta^2}{\mu^2} \right) \nonumber\\
		&\ +\bigg[ 2(\bar{m}+m_2)(\a_M+\b_M+4\s_M)
			+(5D^2-6DF+9F^2)M_N^{(1)} \nonumber\\
		&\qquad\qquad\qquad\qquad\qquad
			-\frac{9}{2}(D-F)^2M_\S^{(1)}
			-\frac{1}{2}(D+3F)^2M_\L^{(1)} \bigg]\,
		\frac{m_K^2}{\L_\chi^2}\, \ln \left( \frac{m_K^2}{\mu^2} \right) \nonumber\\
		&\ +4\mc{C}^2 \Big[ M_N^{(1)}+M_\D^{(1)} \Big] \frac{\mc{J}(m_\pi,\D,\mu)}{\L_\chi^2}
		+\mc{C}^2 \Big[ M_N^{(1)}+M_{\S^\star} \Big] \frac{\mc{J}(m_K,\D,\mu)}{\L_\chi^2} \nonumber\\
		&\ +\left[ \frac{b_\pi^{vA}}{8\L_\chi^3} 
			+\frac{1}{M_0\L_\chi^2}\left( \frac{45}{32}(D+F)^2 +\frac{9}{4}\mc{C}^2 \right) 
			\right]\, m_\pi^4
		+\frac{4\mc{C}^2 m_\pi^2}{\L_\chi^2}\, \Big[ M_N^{(1)} +M_\D^{(1)} \Big] \nonumber\\
		&\ +\left[ \frac{b_\eta^{vA}}{8\L_\chi^3} 
			+\frac{5(D-3F)^2}{32M_0\L_\chi^2} \right]\, m_\eta^4
		+\left[ \frac{b_K^{vA}}{8\L_\chi^3} 
			+\frac{5(5D^2-6DF+9F^2)+9\mc{C}^2}{16M_0\L_\chi^2} \right]\, m_K^4 \nonumber\\
		&\ +\bigg[ \frac{2}{3}(5D^2-6DF+9F^2)M_N^{(1)} -3(D-F)^2 M_\S^{(1)} \nonumber\\
		&\qquad\qquad\qquad\qquad
			-\frac{1}{3}(D+3F)^2 M_\L^{(1)} +\mc{C}^2 \left( M_N^{(1)}+M_\D^{(1)} \right)
			\bigg]\, \frac{m_K^2}{\L_\chi^2} \nonumber\\
		&\ +b^{\bar{m}\bar{m}} \frac{\bar{m}^2}{\L_\chi}
		+b^{\bar{m}m_s}\frac{\bar{m}m_s}{\L_\chi}
		+b^{m_sm_s}\frac{m_s^2}{\L_\chi}\, ,
\end{align}
where the leading order masses of the octet and decuplet baryons are
\begin{align}
	M_N^{(1)} &= 2\bar{m} \left( \a_M+\b_M+2\s_M\right) +2m_s\, \s_M \, ,\nonumber\\
	M_\S^{(1)} &= \bar{m} \left( \frac{5}{3}\a_M +\frac{2}{3}\b_M +4\s_M \right)
			+m_s \left( \frac{1}{3}\a_M+\frac{4}{3}\b_M +2\s_M \right) \, ,\nonumber\\
	M_\L^{(1)} &= \bar{m} (\a_M +2\b_M +4\s_M) +m_s(\a_M+2\s_M) \, ,\nonumber\\
	M_\D^{(1)} &= 2\bar{m}(\g_M-2\ol{\s}_M)-2m_s\ol{\s}_M \, ,\nonumber\\
	M_{\S^\star} &= \frac{2}{3}(2\bar{m}+m_s)(\g_M-3\ol{\s}_M)\, ,
\end{align}
and the linear combination of LECs is
\begin{align}
	b_\pi^{A,vA} &= \frac{3}{2}b_1^{A,vA}+\frac{3}{2}b_2^{A,vA} -b_3^{A,vA} +3b_4^{A,vA}\, 
		,\nonumber\\
	b_\eta^{A,vA} &= \frac{1}{6}b_1^{A,vA} +\frac{1}{6}b_2^{A,vA} +\frac{1}{6}b_3^{A,vA} +b_4^{A,vA}\, 
		,\nonumber\\
	b_K^{A,vA} &= b_1^{A,vA} +b_2^{A,vA}+4b_4^{A,vA}\, ,\nonumber\\
	b^{\bar{m}\bar{m}} &= b^M_1+2b^M_4 +2b^M_6+4b^M_7\, ,\nonumber\\
	b^{\bar{m}m_s} &= b^M_2 +b^M_4 +2b^M_5 +4b^M_7\, ,\nonumber\\
	b^{m_sm_s} &= b^M_3 +b^M_5 +b^M_6 +b^M_7\, .
\end{align}
In Eq.~\eqref{eq:NmassSU3}, the first two lines are the LO and NLO nucleon mass corrections, and the rest of the expression is the NNLO mass contribution.  Comparing to the nucleon mass in $SU(2)$, Eq.~\eqref{eq:NmassIsospin}, we can see that Eq.~\eqref{eq:NmassSU3} is significantly more complicated.  In the $SU(3)_V$ limit, in which the quark masses are all degenerate, this expression must reduce to a form more similar to Eq.~\eqref{eq:NmassIsospin}, but the group structure will still give rise to a splitting amongst the octet.  The full listing of the octet baryon masses to $\mc{O}(m_q^2)$ can be found in Ref.~\cite{Walker-Loud:2004hf}.

In a similar fashion, we can compute the mass of the delta baryons in the isospin limit of $SU(3)$.  The mass of the deltas are
\begin{align}
	M_\D =&\ M_0 + \D +2\g_M\, \bar{m} -2\ol{\s}_M (2\bar{m}+m_s)
		-\frac{2\mc{C}^2}{3} \frac{\mc{F}(m_\pi,-\D,\mu)}{\L_\chi^2}
		-\frac{2\mc{C}^2}{3} \frac{\mc{F}(m_K,-\D,\mu)}{\L_\chi^2} \nonumber\\
		&\ -\frac{5}{27}\, \mc{H}^2 \left( 5\frac{m_\pi^3}{\L_\chi^2}
		+2 \frac{m_K^3}{\L_\chi^3}+ \frac{m_\eta^3}{\L_\chi^3} \right)
		-\left( \frac{2}{3} \g_M\, \bar{m} - \frac{4}{3}\ol{\s}_M(\bar{m}+2m_s) \right)\, 
			\frac{m_\eta^2}{\L_\chi^2}\, \ln \left( \frac{m_\eta^2}{\mu^2} \right)
		\nonumber\\
		&\ -6(\g_M-2\ol{\s}_M)\, \bar{m} \frac{m_\pi^2}{\L_\chi^2}\, \ln \left( \frac{m_\pi^2}{\mu^2} \right)
		-2(\bar{m}+m_s)(\g_M -4\ol{\s}_M) \frac{m_K^2}{\L_\chi^2}\, 
			\ln \left( \frac{m_K^2}{\mu^2} \right) \nonumber\\
		&\ +\left[ \frac{t_\pi^A +\frac{t_\pi^{vA}}{4} +\frac{t_\pi^{\tilde{A}}}{4}}{\L_\chi^3}
			-\frac{25 \mc{H}^2}{48M_0 \L_\chi^2} -\frac{5\mc{C}^2}{8M_0\L_\chi^2}
			\right]\, m_\pi^4 \ln \left( \frac{m_\pi^2}{\mu^2} \right) \nonumber\\
		&\ +\left[ \frac{t_\eta^A +\frac{t_\eta^{vA}}{4} +\frac{t_\eta^{\tilde{A}}}{4}}{\L_\chi^3}
			-\frac{5 \mc{H}^2}{48M_0 \L_\chi^2}
			\right]\, m_\eta^4 \ln \left( \frac{m_\eta^2}{\mu^2} \right) \nonumber\\
		&\ +\left[ \frac{t_K^A +\frac{t_K^{vA}}{4} +\frac{t_K^{\tilde{A}}}{4}}{\L_\chi^3}
			-\frac{5 \mc{H}^2}{24M_0 \L_\chi^2} -\frac{5\mc{C}^2}{8M_0\L_\chi^2}
			\right]\, m_K^4 \ln \left( \frac{m_K^2}{\mu^2} \right) \nonumber\\
		&\ -\left[ \frac{ t_\pi^{vA} +t_\pi^{\tilde{A}}}{8\L_\chi^3}
			+\frac{95\mc{H}^2}{96M_0 \L_\chi^2} -\frac{3\mc{C}^2}{32M_0 \L_\chi^2}
			\right]\, m_\pi^4
		-\left[ \frac{ t_\eta^{vA} +t_\eta^{\tilde{A}}}{8\L_\chi^3}
			+\frac{19\mc{H}^2}{96M_0 \L_\chi^2} \right]\, m_\eta^4 \nonumber\\
		&\ -\left[ \frac{ t_K^{vA} +t_K^{\tilde{A}}}{8\L_\chi^3}
			+\frac{95\mc{H}^2}{48M_0 \L_\chi^2} -\frac{3\mc{C}^2}{32M_0 \L_\chi^2}
			\right]\, m_K^4 \nonumber\\
		&\ +\frac{25\mc{H}^2}{9\L_\chi^2}\Big( \g_M\, \bar{m}+\ol{\s}_M\, \tr(m_Q) \Big)\, 
			m_\pi^2 \left(\ln \left( \frac{m_\pi^2}{\mu^2} \right) +\frac{26}{15} \right) \nonumber\\
		&\ +\frac{10\mc{H}^2}{9\L_\chi^2}\Big( \frac{1}{3}\g_M(2\bar{m}+m_s) 
				+\ol{\s}_M\, \tr(m_Q) \Big)\, 
			m_K^2 \left(\ln \left( \frac{m_K^2}{\mu^2} \right) +\frac{26}{15} \right) \nonumber\\
		&\ +\frac{5\mc{H}^2}{9\L_\chi^2}\Big( \g_M\, \bar{m}+\ol{\s}_M\, \tr(m_Q) \Big)\, 
			m_\eta^2 \left(\ln \left( \frac{m_\eta^2}{\mu^2} \right) +\frac{26}{15} \right) \nonumber\\
		&\ -2\mc{C}^2 \Big( \bar{m}(\a_M+\b_M)+\s_M \tr(m_Q) \Big) 
			\frac{\mc{J}(m_\pi,-\D,\mu)}{\L_\chi^2} \nonumber\\
		&\ -\mc{C}^2 \Big( \frac{1}{3}\bar{m}(5\a_M+2\b_M)+\frac{1}{3}m_s(\a_M+4\b_M)
			+2\s_M \tr(m_Q) \Big) 
			\frac{\mc{J}(m_K,-\D,\mu)}{\L_\chi^2} \nonumber\\
		&\ -\Big( \g_M\, \bar{m} -2\ol{\s}_M\, \tr(m_Q) \Big)
			\bigg[ \frac{25 \mc{H}^2}{18\L_\chi^2}\, 
				m_\pi^2 \left(\ln \left( \frac{m_\pi^2}{\mu^2} \right) +\frac{26}{15} \right) \nonumber\\
		&\qquad\qquad
		+\frac{5 \mc{H}^2}{9\L_\chi^2}\, 
				m_K^2 \left(\ln \left( \frac{m_K^2}{\mu^2} \right) +\frac{26}{15} \right)
		+\frac{5 \mc{H}^2}{18\L_\chi^2}\, 
				m_\eta^2 \left(\ln \left( \frac{m_\eta^2}{\mu^2} \right) +\frac{26}{15} \right)
		\nonumber\\
		&\qquad\qquad
		+\mc{C}^2\left( \frac{\mc{J}(m_\pi,-\D,\mu)}{\L_\chi^2}
			+\frac{\mc{J}(m_K,-\D,\mu)}{\L_\chi^2} \right) \bigg] \nonumber\\
		&\ +t^{\bar{m}\bar{m}} \frac{\bar{m}^2}{\L_\chi}
		+t^{\bar{m}m_s} \frac{\bar{m}m_s}{\L_\chi}
		+t^{m_sm_s} \frac{m_s^2}{\L_\chi}\, ,
\end{align}
where the linear combination of LECs is
\begin{align}
	t_\pi^{A,vA,\tilde{A}} &= \frac{3}{2}t_1^{A,vA,\tilde{A}}+\frac{1}{2}t_2^{A,vA,\tilde{A}}
		+3t_3^{A,vA,\tilde{A}}\, ,\nonumber\\
	t_\eta^{A,vA,\tilde{A}} &= \frac{1}{6}t_1^{A,vA,\tilde{A}} +\frac{1}{6}t_2^{A,vA,\tilde{A}} 
		+t_3^{A,vA,\tilde{A}}\, ,\nonumber\\
	t_K^{A,vA,\tilde{A}} &= t_1^{A,vA,\tilde{A}} +4t_3^{A,vA,\tilde{A}}\, ,\nonumber\\
	t^{\bar{m}\bar{m}} &= t_1^M + t_2^M+2t_3^M+\frac{2}{3}t_4^M+4t_5^M\, , \nonumber\\
	t^{\bar{m}m_s} &= t_3^M+\frac{1}{3}t_4^M+4t_5^M\, , \nonumber\\
	t^{m_sm_s} &= t_3^M +t_5^M\, .
\end{align}

%%%%%%%%%%%%%%%%%%%%%%%%%%%%%%%%%%%%%%%%%%%%%%%%%%%
%
%	PQ Masses
%
%%%%%%%%%%%%%%%%%%%%%%%%%%%%%%%%%%%%%%%%%%%%%%%%%%%
\section{Baryon Masses in $SU(4|2)$ and $SU(6|3)$ PQ\CPT}\label{sec:NDPQmasses}

In Sec.~\ref{sec:PQChPT}, we showed the construction of the partially quenched heavy baryon Lagrangian to the order necessary to compute the baryon masses through NLO.  In this section we will construct the next order partially quenched heavy baryon Lagrangian necessary to compute the masses to NNLO.  Because of the symmetries of the partially quenched theories, at this order, the form of the PQ Lagrangian is independent of whether one is working with $SU(4|2)$ or $SU(6|3)$.  Therefore, when we construct the operators, we will do it simultaneously for both theories and show how to match these Lagrangian operators onto the $SU(2)$ and $SU(3)$ theories respectively.  

The PQ\CPT\ Lagrangians in general have more linearly independent operators than their respective \CPT\ Lagrangians~\cite{Chen:2001yi,Beane:2002vq,Sharpe:2003vy}.  Therefore, to determine the LECs of \CPT\ from partially quenched LQCD simulations~\cite{Sharpe:2000bc}, one must also match the operators order by order, which can be done by restricting the flavor indices of the partially quenched operators to the vector sub-space.  The only other computational difference from the nucleon mass computation outlined in Sec.~\ref{sec:Nmass} is the addition of the hairpin interactions of the flavor neutral mesons which give rise to the double pole pieces of the propagators of these mesons, Eq.~\eqref{eq:etaProp}.  There are various ways to handle these double pole terms and for these mass computations, we use the method in Refs.~\cite{Chen:2001yi,Beane:2002vq}.  To be specific, the two-point function of the flavor neutral propagators in momentum space, for three flavors of sea-quarks, is given by
\begin{equation}
	\mc{G}_{\eta_a\eta_b}(p^2) = \frac{i \d_{ab}\, \e_a}{p^2 -m_{\eta_a}^2 +i\e}
		-\frac{i}{3}\frac{(p^2 - m_{jj}^2)(p^2-m_{rr}^2)}{(p^2-m_{\eta_a}^2+i\e)
			(p^2-m_{\eta_b}^2+i\e)(p^2 -m_X^2 +i\e)}\, ,
\end{equation}
where the mass $m_X^2 = \frac{1}{3}(m_{jj}^2 +2m_{rr}^2)$ and the masses $m_{ff}^2$ are the masses of mesons composed of quark-antiquark flavor $f$.  We can rewrite this propagator as
\begin{equation}
	\mc{G}_{\eta_a\eta_b} = \e_a\, \d_{ab}\, P_a
		+ H_{ab} \left[ P_a,P_b,P_X \right]\, ,
\end{equation}
where the $H_{ab}$ operator is defined as
\begin{align}
	H_{ab}\left[A,B,C\right] = -\frac{1}{3} \bigg[&
		\frac{(m_{jj}^2-m_{\eta_a}^2)(m_{rr}^2-m_{\eta_a}^2)}
			{(m_{\eta_a}^2-m_{\eta_b}^2)(m_{\eta_a}^2-m_X^2)}\, A
		-\frac{(m_{jj}^2-m_{\eta_b}^2)(m_{rr}^2-m_{\eta_b}^2)}
			{(m_{\eta_a}^2-m_{\eta_b}^2)(m_{\eta_b}^2-m_X^2)}\, B \nonumber\\
		&\ +\frac{(m_{X}^2-m_{jj}^2)(m_{X}^2-m_{rr}^2)}
			{(m_{X}^2-m_{\eta_a}^2)(m_{X}^2-m_{\eta_b}^2)}\, C
		\bigg]\, .
\end{align}
This then allows one to simply evaluate the integrals arising in the mass computation (and other observables), as the double pole features are captured by the now mass dependent coefficients of the propagators in the $H_{ab}$ operator.  A similar procedure can be used in the $SU(4|2)$ theory~\cite{Beane:2002vq}.

%%%%%%%%%%%%%%%%%%%%%%%%%%%%%%%%%%%%%%%%%%%%%%%%%%%
%
%	PQ HBLagrangian
%
%%%%%%%%%%%%%%%%%%%%%%%%%%%%%%%%%%%%%%%%%%%%%%%%%%%
\subsection{\textbf{Partially Quenched Heavy Baryon Lagrangian}}
In Sec.~\ref{sec:LOPQHBChPT}, we showed the construction of the LO PQHB\CPT\ Lagrangian.  We now will construct the Lagrangian at the next order.  The various operator structures that enter will involve the $\mc{M}_+$ and $\mc{A}$ fields.  We shall use a labeling convention very similar to that in Secs.~\ref{sec:Nmass} and \ref{sec:SU3HbChPT}.  We shall then match the PQ Lagrangian onto the \CPT\ Lagrangians to make the labeling conventions clear.  The fixed coefficient PQ Lagrangian is
\begin{align}\label{eq:fixedPQ}
	\mc{L} =&\ - \left( \ol{\mc{B}} \frac{D_\perp^2}{2 M_0} \mc{B} \right)
		+ \a \left[ \left( \ol{\mc{B}}  \frac{i \loarrow{D} \cdot S}{M_0} \mc{B} \, v\cdot \mc{A} \right) 
		- \left( \ol{\mc{B}} \frac{S\cdot i \roarrow{D}} {M_0} \mc{B} \, v\cdot \mc{A} \right) \right] 
	\nonumber \\
	&\ +\b \left[ \left( \ol{\mc{B}}  \frac{i \loarrow D \cdot S}{M_0} v\cdot \mc{A}\, \mc{B} \right) 
	- \left( \ol{\mc{B}} \, v\cdot \mc{A} \frac{S\cdot i \roarrow D}{M_0} \mc{B} \right) \right] 
	+ \left( \ol{\mc{T}}^\mu \frac{D_\perp^2}{2 M_0} \mc{T}_\mu \right)
	\nonumber \\
	&\ +\mc{H} \left[\left(\ol{\mc{T}}^\mu \frac{i \loarrow D \cdot S}{M_0} v\cdot \mc{A}\, \mc{T}_\mu \right) 
	- \left( \ol{\mc{T}}^\mu \, v\cdot \mc{A} \frac{S\cdot i \roarrow D}{M_0} \mc{T}_\mu \right) \right]\, ,
\end{align}
where the flavor contractions are defined in Eq.~\eqref{eq:PQcontractions}.  The operators relevant to the mass of the spin-$\frac{1}{2}$ baryons with two insertions of the axial-vector pion fields are
\begin{align}\label{eq:NhdoPQ}
\mc{L} =&\ \frac{b^{A\pq}_1}{\L_\chi}\, \ol{\mc{B}}^{kji} \left( \mc{A}\cdot \mc{A} \right)_i^{\ n} \mc{B}_{njk}
	+ \frac{b^{A\pq}_2}{\L_\chi}\, (-)^{(\eta_i+\eta_j)(\eta_k+\eta_n)} \ol{\mc{B}}^{kji}
		\left( \mc{A}\cdot \mc{A} \right)_k^{\ n} \mc{B}_{ijn} \nonumber\\
	&\ + \frac{b^{A\pq}_3}{\L_\chi}\, (-)^{\eta_l(\eta_j+\eta_n)} \ol{\mc{B}}^{kji} (\mc{A}_\mu)_i^{\ l} 
		(\mc{A}^\mu )_j^{\ n} \mc{B}_{lnk} 
	+ \frac{b_4^{A\pq}}{\L_\chi} \, (-)^{\eta_j \eta_n + 1} \ol{\mc{B}}^{kji} (\mc{A}_\mu)_i^{\ n} 
		(\mc{A}^\mu )_j^{\ l} \mc{B}_{lnk} \nonumber \\
	&\ +\frac{b^{A\pq}_5}{\L_\chi}\, \ol{\mc{B}}^{kji} \mc{B}_{ijk} {\rm Tr}\left( \mc{A}\cdot \mc{A} \right)
	+ \frac{b^{vA \pq}_1}{\L_\chi}\, \ol{\mc{B}}^{kji} \left( v\cdot\, \mc{A}\, v\cdot \mc{A} \right)_i^{\ n} 
		\mc{B}_{njk} \nonumber \\
	&\ + \frac{b^{vA \pq}_2}{\L_\chi}\, (-)^{(\eta_i+\eta_j)(\eta_k+\eta_n)} \ol{\mc{B}}^{kji}
		\left( v\cdot \mc{A}\, v\cdot \mc{A} \right)_k^{\ n} \mc{B}_{ijn} \nonumber \\
	&\ +\frac{b^{vA \pq}_3}{\L_\chi}\, (-)^{\eta_l(\eta_j+\eta_n)} \ol{\mc{B}}^{kji} 
		(v\cdot \mc{A} )_i^{\  l} (v\cdot \mc{A} )_j^{\  n} \mc{B}_{lnk} \nonumber \\
	&\ +\frac{b_4^{vA \pq}}{\L_\chi} \, (-)^{\eta_j \eta_n + 1} \ol{\mc{B}}^{kji} 
		(v \cdot \mc{A})_i^{\  n} (v \cdot \mc{A} )_j^{\  l} \mc{B}_{lnk} 
	+ \frac{b^{vA \pq}_5}{\L_\chi}\, \ol{\mc{B}}^{kji} \mc{B}_{ijk} 
		\tr \left( v\cdot \mc{A}\, v\cdot \mc{A} \right) \, ,
\end{align}
and the operators with two insertions of the mass spurion are
\begin{align}
\mc{L} =&\ \frac{b^{M \pq}_1}{\L_\chi}\, \ol{\mc{B}}^{kji} \left( \mc{M}_+ \mc{M}_+ \right)_i^{\ n} \mc{B}_{njk}
	+ \frac{b^{M \pq}_2}{\L_\chi} \, (-)^{(\eta_i+\eta_j)(\eta_k+\eta_n)}
		\ol{\mc{B}}^{kji} \left( \mc{M}_+ \mc{M}_+ \right)_k^{\ n} \mc{B}_{ijn} \nonumber\\
	&\ + \frac{b^{M \pq}_3}{\L_\chi} \, (-)^{\eta_l(\eta_j+\eta_n)}
		\ol{\mc{B}}^{kji} (\mc{M}_+)_i^{\ l} (\mc{M}_+)_j^{\ n} \mc{B}_{lnk} 
	+ \frac{b_4^{M \pq}}{\L_\chi} \, (-)^{\eta_j \eta_n + 1 } 
		\ol{\mc{B}}^{kji} (\mc{M}_+)_i^{\ n} (\mc{M}_+)_j^{\ l} \mc{B}_{lnk} \nonumber \\
	&\ + \frac{b^{M \pq}_5}{\L_\chi}\, \ol{\mc{B}}^{kji} \mc{B}_{ijk}\ {\rm str}\left(\mc{M}_+
		\mc{M}_+ \right) 
	+ \frac{b^{M \pq}_6}{\L_\chi}\, \ol{\mc{B}}^{kji} (\mc{M}_+)_i^{\ n} \mc{B}_{njk}\ 
		{\rm str}\left( \mc{M}_+ \right) \nonumber \\
	&\ + \frac{b^{M \pq}_7}{\L_\chi}\, (-)^{(\eta_i+\eta_j)(\eta_k+\eta_n)}
		\ol{\mc{B}}^{kji} (\mc{M}_+)_k^{\ n} \mc{B}_{ijn}\ {\rm str}\left( \mc{M}_+ \right)
	+ \frac{b^{M \pq}_8}{\L_\chi} \, \ol{\mc{B}}^{kji} \mc{B}_{ijk}\ [{\rm str}\left( \mc{M}_+ \right)]^2\, .
\end{align}
We can then match these operators onto the \CPT\ operators which gives the following relations amongst the LECs,
\begin{align}
	b^{A,vA(2)} &= \frac{1}{2}b_1^{A,vA(4|2)} +\frac{1}{2}b_2^{A,vA(4|2)}
		-\frac{1}{3}b_3^{A,vA(4|2)} +\frac{5}{12}b_4^{A,vA(4|2)} +b_5^{A,vA(4|2)} \nonumber\\
	b_1^{M(2)} &= -\frac{1}{3} b_1^{M(4|2)} + \frac{2}{3} b_2^{M(4|2)} - \frac{1}{3} b_3^{M(4|2)} 
		+ \frac{1}{2} b_4^{M(4|2)}, \nonumber\\
	b_5^{M(2)} &=  \frac{2}{3} b_1^{M(4|2)} + \frac{1}{6} b_2^{M(4|2)} - \frac{1}{6} b_3^{M(4|2)} 
		+ \frac{1}{6} b_4^{M(4|2)} + b_5^{M(4|2)}, \nonumber\\
	b_6^{M(2)} &=  \frac{1}{2} b_3^{M(4|2)} - \frac{1}{3} b_4^{M(4|2)} - \frac{1}{3} b_6^{M(4|2)} 
		+ \frac{2}{3} b_7^{M(4|2)}, \nonumber\\
	b_8^{M(2)} &=  \frac{1}{6} b_3^{M(4|2)} - \frac{1}{6} b_4^{M(4|2)} + \frac{2}{3} b_6^{M(4|2)} 
		+ \frac{1}{6} b_7^{M(4|2)} + b_8^{M(4|2)} \, ,
\end{align}
and for the matching of $SU(6|3)$ to $SU(3)$ we find,
\begin{align}
	b_1^{A,vA(3)} &= -\frac{1}{3}b_1^{A,vA(6|3)} +\frac{2}{3}b_2^{A,vA(6|3)} \nonumber\\
	b_2^{A,vA(3)} &= -\frac{1}{6}b_3^{A,vA(6|3)} \nonumber\\
	b_3^{A,vA(3)} &= -\frac{2}{3}b_1^{A,vA(6|3)} -\frac{1}{6}b_2^{A,vA(6|3)} +\frac{2}{3}b_3^{A,vA(6|3)}
		\nonumber\\
	b_4^{A,vA(3)} &= \frac{2}{3}b_1^{A,vA(6|3)} +\frac{1}{6}b_2^{A,vA(6|3)}
		-\frac{1}{3}b_3^{A,vA(6|3)} +b_4^{A,vA(6|3)} \nonumber\\
	b_1^{M(3)} &= -\frac{1}{3}b_1^{M(6|3)}+\frac{2}{3}b_2^{M(6|3)} -\frac{1}{3}b_3^{M(6|3)}
		+\frac{1}{2}b_4^{M(6|3)} \nonumber\\
	b_2^{M(3)} &= -\frac{1}{2}b_3^{M(6|3)} +\frac{1}{3}b_4^{M(6|3)} \nonumber\\
	b_3^{M(3)} &= -\frac{2}{3}b_1^{M(6|3)} -\frac{1}{6}b_2^{M(6|3)} +\frac{1}{3}b_3^{M(6|3)}
		-\frac{1}{3}b_4^{M(6|3)} \nonumber\\
	b_4^{M(3)} &= \frac{1}{2}b_3^{M(6|3)} -\frac{1}{3}b_4^{M(6|3)} -\frac{1}{3}b_6^{M(6|3)}
		+\frac{2}{3}b_7^{M(6|3)} \nonumber\\
	b_5^{M(3)} &= -\frac{1}{3}b_3^{M(6|3)} +\frac{1}{3}b_4^{M(6|3)} -\frac{2}{3}b_6^{M(6|3)}
		-\frac{1}{6}b_7^{M(6|3)} \nonumber\\
	b_6^{M(3)} &= \frac{2}{3}b_1^{M(6|3)} +\frac{1}{6}b_2^{M(6|3)} -\frac{1}{6}b_3^{M(6|3)}
		+\frac{1}{6}b_4^{M(6|3)} +b_5^{M(6|3)} \nonumber\\
	b_7^{M(3)} &= \frac{1}{6}b_3^{M(6|3)} -\frac{1}{6}b_4^{M(6|3)} +\frac{2}{3}b_6^{M(6|3)}
		+\frac{1}{6}b_7^{M(6|3)} +b_8^{M(6|3)} \nonumber\\
	b_8^{M(3)} &= \frac{1}{3}b_3^{M(6|3)} -\frac{1}{2}b_4^{M(6|3)}\, .
\end{align}
One must similarly match the operators relevant for the spin-$\frac{3}{2}$ baryon masses.  The relevant PQ Lagrangian is
\begin{align}
\mc{L} =&\ \frac{t_1^{M\pq}}{\L_\chi}\, \ol{\mc{T}}^{kji}_\mu (\mc{M}_+ \mc{M}_+)_{i}^{\ i'} 
		\mc{T}^\mu_{i'jk}
	+ \frac{t_2^{M\pq}}{\L_\chi}(-)^{\eta_{i'} (\eta_j + \eta_{j'})} \ol{\mc{T}}^{kji}_\mu 
		(\mc{M}_+)_{i}^{\ i'} (\mc{M}_+)_{j}^{\ j'} \mc{T}^\mu_{i'j'k} \nonumber \\
	&\ + \frac{t_3^{M\pq}}{\L_\chi}\, \left( \ol{\mc{T}}_\mu \mc{T}^\mu \right) \str (\mc{M}_+ \mc{M}_+)
	+ \frac{t_4^{M\pq}}{\L_\chi}\, \left( \ol{\mc{T}}_\mu \mc{M}_+ \mc{T}^\mu \right) \str (\mc{M}_+)
	\nonumber\\
	&\ + \frac{t_5^{M\pq}}{\L_\chi}\, \left( \ol{\mc{T}}_\mu \mc{T}^\mu \right) [\str(\mc{M}_+)]^2
	+\frac{t_1^{A\pq}}{\L_\chi}\, \ol{\mc{T}}^{kji}_\mu (\mc{A}\cdot \mc{A})_{i}^{\ i'} \mc{T}^\mu_{i'jk} 
	\nonumber\\
	&\ + \frac{t_2^{A\pq}}{\L_\chi}\, (-)^{\eta_{i'} (\eta_j + \eta_{j'})}  \ol{\mc{T}}^{kji}_\mu 
		(\mc{A}_\nu)_{i}^{\ i'} (\mc{A}^\nu)_{j}^{j'} \mc{T}^\mu_{i'j'k} 
	+ \frac{t_3^{A\pq}}{\L_\chi}\, \left( \ol{\mc{T}}_\mu \mc{T}^\mu \right) \str ( \mc{A}\cdot \mc{A})
	\nonumber\\
	&\ + \frac{t_1^{\tilde{A}\pq}}{\L_\chi}\, \ol{\mc{T}}^{kji}_\mu (\mc{A}^\mu \mc{A}_\nu)_{i}^{i'} 
		\mc{T}^\nu_{i'jk} 
	+ \frac{t_2^{\tilde{A}}}{\L_\chi}\, \ol{\mc{T}}^{kji}_\mu (\mc{A}^\mu)_{i}^{\ i'} (\mc{A}_\nu)_{j}^{\ j'}
		\mc{T}^\nu_{i'j'k} \nonumber\\
	&\ +\frac{t_3^{\tilde{A}\pq}}{\L_\chi}\, \left( \ol{\mc{T}}_\mu \mc{T}^\nu \right) \str(\mc{A}\cdot \mc{A})
	+ \frac{t_1^{vA\pq}}{\L_\chi}\, \ol{\mc{T}}^{kji}_\mu (v \cdot \mc{A}\, v \cdot \mc{A})_{i}^{\ i'} 
		\mc{T}^\mu_{i'jk} \nonumber\\
	&\ + \frac{t_2^{vA\pq}}{\L_\chi}\, \ol{\mc{T}}^{kji}_\mu (v \cdot \mc{A})_{i}^{\ i'} 
		(v \cdot \mc{A})_{j}^{\ j'} \mc{T}^\mu_{i'j'k} 
	+ \frac{t_3^{vA\pq}}{\L_\chi}\, \left( \ol{\mc{T}}_\mu \mc{T}^\mu \right) 
		\str ( v \cdot \mc{A} \, v \cdot \mc{A} )\, .
\end{align}
Matching this Lagrangian onto the $SU(2)$ Lagrangian, one can show that all the coefficients involving the two mass spurions have the same numerical values as the QCD coefficients, $t_i^{M(4|2)} = t_i^{M(2)}$, \textit{i.e.} there is a one-to-one correspondence between these partially quenched operators and the $SU(2)$ operators.  This also holds for the matching of the $SU(6|3)$ operators onto the $SU(3)$ operators.  The partially quenched operators involving the axial-vector pion fields also match identically from $SU(6|3)$ to $SU(3)$, $t_i^{A,vA,\tilde{A}(6|3)} = t_i^{A,vA,\tilde{A}(3)}$.  However, because of the symmetries of the $SU(2)$ theory, there is one less operator, Eq.~\ref{eq:DHDO}, such that the matching is
\begin{align}
	t_2^{A,vA,\tilde{A}(2)} &= t_2^{A,vA,\tilde{A}(4|2)} \nonumber\\
	t_3^{A,vA,\tilde{A}(2)} &= \frac{1}{2}t_1^{A,vA,\tilde{A}(4|2)} +t_3^{A,vA,\tilde{A}(4|2)}\, .
\end{align}
We now proceed to compute the mass of the nucleon in partially quenched HB\CPT.

\newpage
%%%%%%%%%%%%%%%%%%%%%%%%%%%%%%%%%%%%%%%%%%%%%%%%%%%
%
%	N-Mass SU(4|2)
%
%%%%%%%%%%%%%%%%%%%%%%%%%%%%%%%%%%%%%%%%%%%%%%%%%%%
\subsection{\textbf{Nucleon Mass in $SU(4|2)$}}
We find that the mass of the nucleon, in the isospin limit of the sea and valence sectors, with the valence quark mass, $m_u$ and the sea quark mass, $m_j$, is given by
\begin{align*}
M_N^{(4|2)}=&\ M_0 -(2\a_M +\b_M)\, m_u -4\s_M\, m_j
	-\frac{2\pi\, m_\pi^3}{3\L_\chi^2}\left( 2g_A^2 +g_Ag_1-g_1^2 \right) \nonumber\\
	&\ -\frac{\pi\, m_{ju}^3}{3\L_\chi^2}\left(8g_A^2 +4g_Ag_1+5g_1^2\right) 
	-\frac{2\pi(g_A+g_1)^2}{\L_\chi^2}\, \mc{M}^3(m_\pi,m_\pi) \nonumber\\
	&\ -\frac{4g_{\D N}^2}{3}\frac{\mc{F}(m_\pi,\D,\mu)}{\L_\chi^2}
	-\frac{4g_{\D N}^2}{3}\frac{\mc{F}(m_{ju},\D,\mu)}{\L_\chi^2} \nonumber\\
	&\ +\Big( (2\a_M +\b_M)\, m_u +4\s_M\, m_j \Big) \Bigg[
		\frac{\left( 2g_A^2 +g_Ag_1-g_1^2 \right)\, m_\pi^2}{\L_\chi^2}
			\left( \ln \left( \frac{m_\pi^2}{\mu^2} \right) +\frac{2}{3} \right) \nonumber\\
	&\qquad\qquad
		+\frac{\left( 4g_A^2 +2g_Ag_1 +\frac{5}{2}g_1^2 \right)\, m_{ju}^2}{\L_\chi^2}
			\left( \ln \left( \frac{m_{ju}^2}{\mu^2} \right) +\frac{2}{3} \right) \nonumber\\
	&\qquad\qquad
		+\frac{(g_A+g_1)^2}{\L_\chi^2} \Big( 3L(m_\pi,m_\pi,\mu) + 2\mc{M}^2(m_\pi,m_\pi) \Big)
		\nonumber\\
	&\qquad\qquad
		+\frac{2g_{\D N}^2}{\L_\chi^2}	\Big(
			\mc{J}(m_\pi,\D,\mu) +\mc{J}(m_{ju}^2,\D,\mu) +m_\pi^2 +m_{ju}^2 \Big)
	\Bigg] \nonumber\\
	&\ +\frac{4(m_u+m_j)(\a_M+\b_M)}{\L_\chi^2}\, m_{ju}^2\, \ln \left( \frac{m_{ju}^2}{\mu^2} \right)
	+\frac{8m_u(\a_M+\b_M)}{\L_\chi^2}\, L(m_\pi,m_\pi,\mu) \nonumber\\
	&\ +\frac{16m_j\, \s_M}{\L_\chi^2}\, m_{jj}^2\, \ln \left( \frac{m_{jj}^2}{\mu^2} \right)
	+\frac{8m_j\, \s_M}{\L_\chi^2}\, L(m_{jj}^2,m_{jj}^2,\mu) \nonumber\\
	&\ -\frac{\left(b_\pi^A+\frac{1}{4}b_\pi^{vA}\right) m_\pi^4}{\L_\chi^3}\, 
		\ln \left( \frac{m_\pi^2}{\mu^2} \right)
	-\frac{\left(b_{ju}^{vA}+\frac{1}{4}b_{ju}^{vA}\right) m_{ju}^4}{\L_\chi^3}\, 
		\ln \left( \frac{m_{ju}^2}{\mu^2} \right) \nonumber\\
	&\ -\frac{\left(b_{jj}^{vA}+\frac{1}{4}b_{jj}^{vA}\right) m_{jj}^4}{\L_\chi^3}\, 
		\ln \left( \frac{m_{jj}^2}{\mu^2} \right)
	+\frac{b_\pi^{vA}}{8\L_\chi^3}\, m_\pi^4 +\frac{b_{ju}^{vA}}{8\L_\chi^3}\, m_{ju}^4
	+\frac{b_{jj}^{vA}}{8\L_\chi^3}\, m_{jj}^4 \nonumber\\
	&\ -\frac{\ol{b}_\pi^A+\frac{1}{4}\ol{b}_\pi^{vA}}{\L_\chi^3}\, \ol{L}(m_\pi,m_\pi,\mu)
	+\frac{\ol{b}_\pi^{vA}}{8\L_\chi^3}\, \mc{M}^4(m_\pi,m_\pi,\mu) \nonumber\\
	&\ -\frac{\ol{b}_{jj}^A+\frac{1}{4}\ol{b}_{jj}^{vA}}{\L_\chi^3}\, \ol{L}(m_{jj},m_{jj},\mu)
	+\frac{\ol{b}_{jj}^{vA}}{8\L_\chi^3}\, \mc{M}^4(m_{jj},m_{jj},\mu) \nonumber\\
	&\ -\frac{3(2g_A^2+g_Ag_1-g_1^2)}{8M_0}\frac{m_\pi^4}{\L_\chi^2}\left(
		\ln \left( \frac{m_\pi^2}{\mu^2} \right)+\frac{5}{6} \right) \nonumber\\
	&\ -\frac{3\left(2g_A^2 +g_Ag_A +\frac{5}{4}g_1^2 \right)}{4M_0}\frac{m_{ju}^4}{\L_\chi^2}
		\left( \ln \left( \frac{m_{ju}^2}{\mu^2} \right)+\frac{5}{6} \right) \dots
\end{align*}
\begin{align}\label{eq:NMass42Iso}
	&\ -\frac{9(g_A+g_1)^2}{8M_0\L_\chi^2}\left( \ol{L}(m_\pi,m_\pi,\mu) 
		+\frac{5}{6}\mc{M}^4(m_\pi,m_\pi) \right) \nonumber\\
	&\ -\frac{5g_{\D N}^2}{4M_0\L_\chi^2}\left[
		m_\pi^4\, \left( \ln \left( \frac{m_{\pi}^2}{\mu^2} \right)+\frac{9}{10} \right)
		+m_{ju}^4\,  \left( \ln \left( \frac{m_{ju}^2}{\mu^2} \right)+\frac{9}{10} \right) \right] \nonumber\\
	&\ -\frac{2\s_M \str(m_Q)(2g_A^2 +g_Ag_1 -g_1^2)}{\L_\chi^2}
		m_\pi^2\, \left( \ln \left( \frac{m_\pi^2}{\mu^2} \right)+\frac{2}{3} \right) \nonumber\\
	&\ -\frac{4\s_M \str(m_Q) \left(2g_A^2 +g_Ag_1 +\frac{5}{4}g_1^2\right)}{\L_\chi^2}
		m_{ju}^2\, \left( \ln \left( \frac{m_{ju}^2}{\mu^2} \right)+\frac{2}{3} \right) \nonumber\\
	&\ -\frac{6\s_M \str(m_Q) \left(g_A+g_1\right)^2}{\L_\chi^2}
		\left( L(m_\pi,m_\pi,\mu)+\frac{2}{3}\mc{M}^2(m_\pi,m_\pi) \right) \nonumber\\
	&\ -\frac{4g_{\D N}^2 \ol{\s}_M \str(m_Q)}{\L_\chi^2}\Big( 
		\mc{J}(m_\pi^2,\D,\mu) +\mc{J}(m_{ju}^2,\D,\mu) +m_\pi^2 +m_{ju}^2 \Big) \nonumber\\
	&\ -\frac{2 m_\pi^2\, m_u}{\L_\chi^2}(g_A+g_1)(2g_A-g_1)(\a_M+\b_M)
		\left( \ln \left( \frac{m_\pi^2}{\mu^2} \right)+\frac{2}{3} \right) \nonumber\\
	&\ -\frac{m_{ju}^2}{\L_\chi^2} \left( \ln \left( \frac{m_{ju}^2}{\mu^2} \right)+\frac{2}{3} \right)
		\Big[ m_u \Big( (\a_M+2\b_M)(4g_A^2+2g_Ag_1)+2g_1^2(2\a_M+\b_M) \Big) \nonumber\\
	&\qquad\qquad
		+m_j \Big(\a_M(4g_A^2+2g_Ag_1 +g_1^2(\a_M+3\b_M) \Big) \Big] \nonumber\\
	&\ -\frac{6(g_A+g_1)^2(\a_M+\b_M)\, m_u}{\L_\chi^2} \left[
		L(m_\pi,m_\pi,\mu) +\frac{2}{3}\mc{M}^2(m_\pi,m_\pi) \right] \nonumber\\
	&\ +\frac{8g_{\D N}^2\g_M\, m_u}{9\L_\chi^2}\Big[ 
		m_u \Big(3\mc{J}(m_\pi,\D,\mu) +2\mc{J}(m_{ju},\D,\mu) +3m_\pi^2 +2m_{ju}^2 \Big)
		\nonumber\\
	&\qquad\qquad\qquad
		+m_j \Big( \mc{J}(m_{ju},\D,\mu) +m_{ju}^2 \Big) \Big] \nonumber\\
	&\ -\frac{b^{uu}}{\L_\chi}m_u^2 -\frac{b^{uj}}{\L_\chi}m_um_j -\frac{b^{jj}}{\L_\chi}m_j^2\, ,
\end{align}
where in the above expression we have used various replacements,
\begin{align}
	\mc{M}^n(m_\phi,m_{\phi^\prime}) &= H[m_\phi^n,m_{\phi^\prime}^n,m_X^n]\, , \nonumber\\
	L[m_\phi,m_{\phi^\prime},\mu] &= 
		H\left[ m_\phi^2\ln \left(\frac{m_\phi^2}{\mu^2} \right),
			m_{\phi^\prime}^2\ln \left(\frac{m_{\phi^\prime}^2}{\mu^2} \right),
			m_X^2\ln \left( \frac{m_X^2}{\mu^2} \right) \right]\, , \nonumber\\
	\ol{L}[m_\phi,m_{\phi^\prime},\mu] &= 
		H\left[ m_\phi^4\ln \left(\frac{m_\phi^2}{\mu^2} \right),
			m_{\phi^\prime}^4\ln \left(\frac{m_{\phi^\prime}^2}{\mu^2} \right),
			m_X^4\ln \left( \frac{m_X^2}{\mu^2} \right) \right]\, ,
\end{align}
and we have used known linear combinations of the LECs, $b_\pi^{A,vA}$, $b^{uu}$, \textit{etc.}~\cite{Tiburzi:2005na}.  If we extend this computation to $SU(6|3)$, we will find a similar expression.  The difference is that now there are 3 more quark flavors in the theory, the strange $s$-quark, the ghostly $s$, the $\tilde{s}$-quark, and the sea strange quark, the $r$-quark.  Therefore this expression will simply be longer, as there will be more mesons which can propagate in the loops, but the general structure of the mass will be identical to Eq.~\eqref{eq:NMass42Iso}.  The full expression for the nucleon mass in $SU(6|3)$ as well as the other members of the octet-baryons can be found in Ref.~\cite{Walker-Loud:2004hf}.  Similarly, the mass of the deltas can be found for $SU(4|2)$ in Ref.~\cite{Tiburzi:2005na} and the full decuplet in Ref.~\cite{Tiburzi:2004rh}.

We would like to conclude this chapter by mentioning that while this thesis was being written, the NPLQCD collaboration has computed both the deviation from the Gell-Mann--Okubo baryon mass relations~\cite{Beane:2006pt} as well as the strong isospin breaking contributions to the nucleon-proton mass splitting~\cite{Beane:2006fk}.  Therefore, we anticipate that within one year, these expressions for the baryon masses to $\mc{O}(m_q^2)$ will be necessary for a more precise determination of these quantities as well as the masses themselves.  This is about 5 years earlier than the author anticipated when first computing these mass expressions.

% ========== Chapter 3: twisted Mass Baryons
\chapter{Including Baryons in Twisted Mass Chiral Perturbation Theory}\label{ch:tmChPT}

This chapter is based upon the work in Ref.~\cite{Walker-Loud:2005bt}.

%%%%%%%%%%%%%%%%%%%%%%%%%%%%%%%%%%%%%%%%%%%%%%%%%%%
%
%	Introduction
%
%%%%%%%%%%%%%%%%%%%%%%%%%%%%%%%%%%%%%%%%%%%%%%%%%%%
\section{\label{sec:intro} Introduction to Twisted Mass Lattice QCD}

Twisted mass lattice QCD (tmLQCD)~\cite{Frezzotti:2000nk} is an alternative fermion discretization technique for lattice QCD that has recently received considerable attention.%
\footnote{For recent reviews see Refs.~\cite{Frezzotti:2004pc,Shindler:2005vj}.}
It has the potential to match the attractive features of improved staggered fermions (efficient simulations~\cite{Kennedy:2004ae}, absence of ``exceptional configurations''~\cite{Frezzotti:1999vv}, $O(a)$ improvement at maximal twist~\cite{Frezzotti:2003ni}, operator mixing as in the continuum~\cite{Frezzotti:2000nk,Pena:2004gb,Frezzotti:2004wz}) while not sharing the disadvantage of needing to take roots of the determinant to remove unwanted degrees of freedom. Thus tmLQCD offers a promising and
interesting new way to probe the properties and interactions of hadrons non-perturbatively from first principles.

Due to the limitations in computational capabilities, the quark masses, $m_q$, used in current simulations are still unphysically large. Thus extrapolations in $m_q$ are necessary if physical predications are to be made from lattice calculations. This can be done in a systematic and model independent way through the use of chiral perturbation theory ($\chi$PT). Since $\chi$PT is derived in the continuum, it can be employed only after the continuum limit has been taken, where the
lattice spacing, $a$, is taken to zero. However, when close to the continuum, it can be extended to lattice QCD at non-zero $a$, where discretization errors arising from the finite lattice spacing are systematically included in a joint expansion in $a$ and $m_q$~\cite{Sharpe:1998xm}. For tmLQCD with mass-degenerate quarks, the resulting ``twisted mass chiral perturbation theory'' (tm$\chi$PT) has been formulated previously~\cite{Munster:2003ba,Scorzato:2004da,Sharpe:2004ps}, building on earlier work for the untwisted Wilson theory~\cite{Sharpe:1998xm,Rupak:2002sm,Bar:2003mh}.  

So far, tm$\chi$PT has only been applied to the mesonic (pionic) sector. There have been studies on pion masses and decay constants for $m_q\gg a\L_{\rm QCD}^2$~\cite{Munster:2003ba}, the phase structure of tmLQCD for $m_q\sim a^2\L_{QCD}^3$~\cite{Munster:2004am,Munster:2004wt,Scorzato:2004da,Sharpe:2004ps,Aoki:2004ta}, and quantities involving pions that do not involve final state interactions up to next-to-leading order (NLO) in the power counting scheme where $m_q\sim a\L_{\rm QCD}^2$~\cite{Sharpe:2004ny}. However, as pointed out in Ref.~\cite{Sharpe:2004ny}, many of the pionic quantities considered are difficult to calculate in numerical simulations because they involve quark-disconnected diagrams. This motivated us to extend tm$\chi$PT to the baryon sector, enabling us to analytically study baryonic quantities that do not involve quark-disconnected diagrams. Numerical studies of the baryons in tmLQCD are already underway, and the first results from quenched simulations studying the nucleon and delta spectra have been obtained recently in Ref.~\cite{Abdel-Rehim:2005gz}. 

In the baryon sector, the extension of $\chi$PT to the lattice at finite lattice spacing to $\mc{O}(a)$~\cite{Beane:2003xv}, and to $\mathcal{O}(a^2)$~\cite{Tiburzi:2005vy}, has been done for a theory with
untwisted Wilson fermions. We extend that work here to include the effects of ``twisting'', i.e. our starting underlying lattice theory is now tmLQCD. Specifically, we study the parity and flavor breaking
effects due to twisting in the masses and mass splittings of nucleons and deltas in an $SU(2)$ chiral effective theory. The mass splittings are of particular interest to us as they allow one to quantify the
size of the parity-flavor breaking effects in tmLQCD; furthermore, they present less difficulties to numerical simulations than their counterparts in the mesonic sector, which involve quark-disconnected
diagrams.   

We consider here tmLQCD with mass non-degenerate quarks~\cite{Frezzotti:2003xj}, which includes an additional parameter, the mass splitting, $\epsilon_q$. This allows us to consider the theory both in and away from the strong isospin limit. With simulations in the near future most likely able to access the region where $m_q \sim a \Lambda_{\rm QCD}^2$, the power counting scheme we will adopt is
\begin{equation}
1 \gg \varepsilon^2 \sim a \Lambda_{\rm QCD} 
\sim \frac{m_q}{\Lambda_{\rm QCD}} 
\sim \frac{\epsilon_q}{\Lambda_{\rm QCD}}
%\sim \frac{p^2}{\Lambda_\chi^2} \,,  
\end{equation} 
with $\varepsilon^2$ denoting the small dimensionless expansion parameters. In the following, we will work to $\mc{O}(\varepsilon^4)$ in this power counting. 

The remainder of this chapter is organized as follows. In Sec.~\ref{sec:EContL} and~\ref{sec:mesons}, we briefly review the definition of tmLQCD with mass non-degenerate quarks, and we show how the mass splitting can be included in the Symanzik Lagrangian and the $\mc{O}(a)$ meson chiral Lagrangian. Higher order corrections from the meson Lagrangian are not needed for the baryon observables to the order we work. In Sec.~\ref{sec:baryons}, we extend the heavy baryon $\chi$PT (HB$\chi$PT) to include the twisting effects to $\mc{O}(\varepsilon^4)$. In Sec.~\ref{sec:massiso} we present the nucleon and delta masses in tm$\chi$PT in the strong isospin limit, including lattice discretization errors and the flavor and parity breaking induced by the twisted mass term.  In Sec.~\ref{sec:mass} we extend the calculation to include isospin breaking effects and discuss the subtleties that arise. We conclude in Sec.~\ref{sec:conc}.

%%%%%%%%%%%%%%%%%%%%%%%%%%%%%%%%%%%%%%%%%%%%%%%%%%%
%
%	tmChPT
%
%%%%%%%%%%%%%%%%%%%%%%%%%%%%%%%%%%%%%%%%%%%%%%%%%%%
\section{\label{sec:tmChPT} Mass Non-Degenerate Twisted Mass Chiral Perturbation Theory}

In this section, we work out the extension of the baryon chiral Lagrangian to $\mathcal{O}(a^2)$ given in Ref.~\cite{Tiburzi:2005vy} in tmLQCD. We start by briefly outlining the construction of the Symanzik
Lagrangian in the mass non-degenerate case, which follows the same procedures as those in the mass-degenerate theory with minimal modifications. 

\subsection{\label{sec:EContL} \textbf{The effective continuum quark level Lagrangian}}

The fermionic part of the Euclidean lattice action of tmLQCD with two mass non-degenerate quarks is
\begin{align} \label{E:ActionTw}
S^L_F =& \; \sum_{x} \bar{\psi_l}(x) \Big[
 \frac{1}{2} \sum_{\mu} \gamma_\mu (\nabla^\star_\mu + \nabla_\mu)
-\frac{r}{2} \sum_{\mu} \nabla^\star_\mu \nabla_\mu 
+ m_0 + i \gamma_5 \tau_1 \mu_0 - \tau_3 \epsilon_0
\Big] \psi_l(x),
\end{align}
where we have written the action given in Ref.~\cite{Frezzotti:2003xj} for a general twist angle (not necessarily maximal), and in the so-called ``twisted basis''~\cite{Frezzotti:2003ni}. The quark (flavor) doublets $\psi_l$ and $\bar\psi_l$ are the dimensionless bare lattice fields (with ``$l$" standing for
lattice and not indicating left-handed), and $\nabla_\mu$ and $\nabla^\star_\mu$ are the usual covariant forward and backward dimensionless lattice derivatives, respectively. The matrices $\tau_i$ are the usual Pauli matrices acting in the flavor space, with $\tau_3$ the diagonal matrix. The bare normal mass, $m_0$, the bare twisted mass, $\mu_0$, and the bare mass splitting $\epsilon_0$, are all dimensionless parameters; an implicit identity matrix in flavor space multiplies the bare mass parameter $m_0$. The notation here is that both $m_0$ and $\epsilon_0$ are positive such that the upper component of the quark field is the lighter member of the flavor doublet with a positive bare mass. 

Note that in the mass-degenerate case, twisting can be done using any of the $\tau_i$, the choice of $\tau_3$ is merely for convenience. Given the identity
\begin{equation}
	\exp(-i\frac{\pi}{4}\tau_k)\,\tau_a\,\exp(i\frac{\pi}{4}\tau_k) 
	= \epsilon_{kab} \tau_b + \d_{ak}\, \t_a\,,
\end{equation}
one can always rotate from a basis where the twist is implemented by $\tau_a$, $a = 1,\,2$, to a basis where it is implemented by $\tau_3$ using the vector transformation 
\begin{equation}
	\bar{\psi} \rightarrow \bar{\psi} \exp(i\theta\tau_k) \,, \qquad
	\psi \rightarrow \exp(-i\theta\tau_k) \psi \,, \qquad\qquad
	k = 1,\,2,\,3 \,, \qquad \theta = \pm \frac{\pi}{4} \,,
\end{equation}
where the appropriate sign for $\theta$ is determined by the index $a$. However, with $\tau_3$ used here to split the quark doublet so that the mass term is real and flavor-diagonal, it can not be used again for twisting if the fermionic determinant is to remain real.%
\footnote{One way to see this is to note that the mass terms $m_0 + i\gamma_5\tau_3\mu_0 - \tau_3\epsilon_0$ can be written as $(x_0 - \tau_3 y_0)\exp(i\alpha\gamma_5)\exp(i\beta\gamma_5\tau_3)$,
where $x_0/y_0 = \tan\beta / \tan\alpha$. Thus, the twisted mass term can be transformed away leaving just the normal mass term and the mass splitting term. However, since this involves an $U(1)$ axial
transformation which is anomalous, an $i\alpha F\widetilde{F}$ term is introduced into the action which
we see now is complex (because of the factor of $i$). Thus, since the gauge action is real, this means that the fermionic action (before the transformation) must be complex, and so the fermionic determinant
obtained from it must also be complex. This also implies that a theory, where both the twist and the mass
splitting are implemented by $\tau_3$, is $\alpha$-dependent.}

Following the program of Symanzik~\cite{Symanzik:1983dc,Symanzik:1983gh}, and the same
enumeration procedure detailed in Ref.~\cite{Sharpe:2004ps}, one can obtain the
effective continuum Lagrangian at the quark level for mass
non-degenerate quarks that describes the long distance physics of the
underlying lattice theory. Its form is constrained by the symmetries
of the lattice theory. To $\mathcal{O}(\varepsilon^4)$ in our power
counting, in which we treat $a\Lambda_{\rm QCD}^2\sim m_q\sim\epsilon_q$,
we find that the Pauli term is again the only dimension five symmetry
breaking operator just as in the mass-degenerate case~\cite{Sharpe:2004ps}
(the details of this argument are provided in Appendix~\ref{sec:appD5SB}),  
\begin{align} \label{E:CLeff}
\mathcal{L}_{\rm eff} &= \mathcal{L}_g + \bar{\psi} 
(D \!\!\!\!/ + m + i \gamma_5 \tau_1 \mu - \epsilon_q \tau_3) \psi 
+ b_1 a \bar{\psi}\, i \sigma_{\mu\nu} F_{\mu\nu}\, \psi 
+ O(a^2) \,,
\end{align}
where $\mathcal{L}_g$ is the continuum gluon Lagrangian, $m$ is the physical
quark mass, defined in the usual way by 
\begin{equation} 
m = Z_m(m_0 - \widetilde m_c)/a \,,
\end{equation}
$\mu$ is the physical twisted mass 
\begin{equation}
\mu = Z_\mu \mu_0/a = Z_P^{-1} \mu_0/a \,,
\end{equation}
and $\epsilon_q$ is the physical mass splitting
\begin{equation}
\epsilon_q = Z_\epsilon \epsilon_0/a = Z_S^{-1} \epsilon_0/a \,. 
\end{equation}
The factors $Z_P$ and $Z_S$ are matching factors for the non-singlet
pseudoscalar and scalar densities respectively. Note that the
lattice symmetries forbid additive renormalization to both $\mu_0$ and
$\epsilon_0$~\cite{Frezzotti:2003xj}. The quantity $\tilde m_c$ is the
critical mass, aside from an $O(a)$ shift (see Ref.~\cite{Aoki:2004ta,Sharpe:2004ny,Sharpe:2005rq,Aoki:2006nv} and
discussion below). 

Anticipating the fact that the mesons contribute to the baryon masses
only through loops, and so will be of $\mc{O}(\varepsilon^3)$ or
higher, we only need to have a meson chiral Lagrangian to $\mc{O}(a)$
for the order we work; $\mc{L}_{\rm eff}$ as given in
Eq.~\eqref{E:CLeff} is sufficient for its construction. 
To build the effective chiral Lagrangian for baryons to
$\mc{O}(\varepsilon^4)$ on the other hand, terms of $\mc{O}(a^2)$ in
Eq.~\eqref{E:CLeff} are of the appropriate size to be
included. However, except for the operator which breaks $\mc{O}(4)$
rotation symmetry, $a^2\bar{\psi}\gamma_\mu D_\mu D_\mu D_\mu\psi$,  
the $\mc{O}(a^2)$ operators do not break the continuum symmetries in
a manner different than the terms explicitly shown in
Eq.~\eqref{E:CLeff}, and thus their explicit form is not needed. The
$\mc{O}(4)$ breaking term will lead to operators in the baryon chiral
Lagrangian at the order we work. However, it is invariant under
twisting and thus contributes as those in the untwisted
theory~\cite{Tiburzi:2005vy}.  

%%%%%%%%%%%%%%%%%%%%%%%%%%%%%%%%%%%%%%%%%%%%%%%%%%%
%
%	twisted Mesons
%
%%%%%%%%%%%%%%%%%%%%%%%%%%%%%%%%%%%%%%%%%%%%%%%%%%%
\subsection{\label{sec:mesons} \textbf{The $SU(2)$ Meson Sector}}

The low energy dynamics of the theory are described by a generalized
chiral Lagrangian found by matching from the continuum effective 
Lagrangian (\ref{E:CLeff}). As usual, the chiral Lagrangian is built
from the $SU(2)$ matrix-valued field $\Sigma$, which transforms under
the chiral group $SU(2)_L \times SU(2)_R$ as Eq.~\eqref{eq:SigTransform}
The vacuum expectation value, $\Sigma_0 = \langle \Sigma \rangle$,
breaks the chiral symmetry spontaneously down to an $SU(2)$
subgroup. The fluctuations around $\Sigma_0$ correspond to the
pseudo-Goldstone bosons (pions).

From a standard spurion analysis, the chiral Lagrangian at
$\mathcal{O}(\varepsilon^2)$ is (in Euclidean space%
\footnote{We will work in Euclidean space throughout this Chapter.}%
) 
\begin{equation} \label{eq:LOWChPT}
\mathcal{L}_\chi = 
 \frac{f^2}{8} 
 \mathrm{Tr}(\partial_\mu \Sigma \partial_\mu \Sigma^\dagger)
-\frac{f^2}{8} 
 \mathrm{Tr}(\chi^{\dagger} \Sigma + \Sigma^\dagger\chi) 
-\frac{f^2}{8} 
 \mathrm{Tr}(\hat{A}^{\dagger} \Sigma + \Sigma^\dagger\hat{A}) \,,
\end{equation}
where $f$ is the decay constant (normalized so that $f_\pi = 132$
MeV). The quantities $\chi$ and $\hat{A}$ are spurions for the quark
masses and discretization errors respectively. At the end of the
analysis they are set to the constant values 
\begin{equation}
\chi \longrightarrow 2 B_0 (m + i\mu \tau_1 - \epsilon_q \tau_3) 
\equiv \hat{m} + i\hat{\mu} \tau_1 - \hat{\epsilon}_q \tau_3
\,, \qquad\qquad 
\hat A \longrightarrow 2 W_0 a \equiv \hat{a} \,,
\end{equation}
where $B_0\sim\mc{O}(\L_{\rm QCD})$ and $W_0\sim\mc{O}(\L_{\rm QCD}^3)$ 
are unknown dimensionful constants, and we have defined the quantities
$\hat{m}$, $\hat{\mu}$ and $\hat{a}$.   

As explained in Ref.~\cite{Sharpe:2004ny}, since the Pauli term transforms
exactly as the quark mass term, they can be combined by using
the shifted spurion
\begin{equation}
\chi' \equiv \chi + \hat{A} \,,
\end{equation}
leaving the $\mathcal{O}(\varepsilon^2)$ chiral Lagrangian unchanged
from its continuum form. This corresponds at the quark level to a
redefinition of the untwisted component of the quark mass from $m$ to
\begin{equation} \label{E:m'def}
m' \equiv m + a W_0/B_0 \,.
\end{equation}
This shift corresponds to an $O(a)$ correction to the critical mass,
so that it becomes 
\begin{equation}
m_c = Z_m \tilde m_c / a - a W_0/B_0 \,.
\end{equation}

Since the $\mathcal{O}(\varepsilon^2)$ Lagrangian takes the continuum
form, and the mass splitting term does not contribute at this order,
the vacuum expectation value of $\Sigma$ at this order is that which
cancels out the twist in the shifted mass matrix, exactly as in the
mass-degenerate case:   
\begin{equation} \label{E:VacLO}
\langle0|\Sigma|0\rangle_{LO} \equiv \Sigma_{0} = 
% \frac{\chi'}{|\chi'|} = 
\frac{\hat{m} + \hat{a} + i \hat{\mu} \tau_1}{M'}
\equiv \exp(i \omega_0 \tau_1) \,,
\end{equation}
where
\begin{equation} 
\label{E:M'def}
M' = \sqrt{(\hat{m}+\hat{a})^2 +\hat{\mu}^2} \,.
\end{equation}
Note that $M'$ is the leading order result for the pion mass-squared,
i.e. $m_\pi^2 = M'$ at $\mathcal{O}(\varepsilon^2)$. If we define the
physical quark mass by 
\begin{equation}
m_q = \sqrt{m'^2 + \mu^2} \,,
\end{equation}
then it follows from (\ref{E:VacLO}) that
\begin{equation}
%c_0 \equiv 
\cos\omega_0 = m'/m_q \,,\qquad
%s_0 \equiv 
\sin\omega_0 = \mu /m_q \,.
\end{equation}
Details of the non-perturbative determination of the twist angle and the
critical mass can be found in~\cite{Aoki:2004ta,Sharpe:2004ny,Frezzotti:2005gi,Sharpe:2005rq,Aoki:2006nv}, and will not be repeated
here.

At $\mathcal{O}(\varepsilon^4)$, the mass non-degenerate chiral
Lagrangian for the pions retains the same form as that in the mass
degenerate case~\cite{Sharpe:2004ps,Sharpe:2004ny}, because the mass splitting does
not induce any additional symmetry breaking operators in
$\mathcal{L}_{\rm eff}$. The $\mathcal{O}(\varepsilon^4)$ pion
Lagrangian contains the usual Gasser-Leutwyler operators of 
$\mathcal{O}(\mathsf{m}^2,\,\mathsf{m}p^2,\,p^4)$ Eq.~\eqref{eq:ChPTNLO}, where $\mathsf{m}$ 
is a generic mass parameter that can be $m,\,\mu$, or $\epsilon_q$, as
well as terms of $\mathcal{O}(a\mathsf{m},\,a p^2,\,a^2)$ associated
with the discretization errors. Now as we stated earlier, since the
pions will enter only through loops in typical calculations of baryon
observables, keeping the pion masses to $\mathcal{O}(\varepsilon^4)$
will lead to corrections of $\mathcal{O}(\varepsilon^5)$, which is
beyond the order we work. As our concern is not in the meson sector,
the $\mathcal{O}(\varepsilon^2)$ pion Lagrangian~(\ref{eq:LOWChPT})
is thus sufficient for our purpose in this work.    

%%%%%%%%%%%%%%%%%%%%%%%%%%%%%%%%%%%%%%%%%%%%%%%%%%%
%
%	twisted Baryons
%
%%%%%%%%%%%%%%%%%%%%%%%%%%%%%%%%%%%%%%%%%%%%%%%%%%%
\subsection{\label{sec:baryons} \textbf{The $SU(2)$ Baryon Sector}}

With the effective continuum theory and the relevant part of the effective chiral theory describing the pions in hand, we now include the nucleon and delta fields into tm$\chi$PT by using HB$\chi$PT~\cite{Jenkins:1991ne,Jenkins:1990jv,Jenkins:1992ts}, which we will refer as the twisted mass HB$\chi$PT (tmHB$\chi$PT).  The construction of the tmHB\CPT\ Lagrangian parallels that in Sections~\ref{sec:NDLagrangian} and \ref{sec:Nmass}.  Recall that the nucleon is a doublet and the deltas a quartet under the $SU(2)_V$ transformations, with normalizations given in Eqs.~\eqref{eq:N} and \eqref{eq:Tnorm}.  The free Lagrangian for the nucleons and deltas to $\mathcal{O}(\varepsilon^2)$ consistent with the symmetries of the lattice theory is (in Euclidean space) 
\begin{align} \label{E:BLLOtw}
\mc{L}_\chi =&\,\ol{N} i v \cdot D N 
	-2\,\a_M\,\ol{N} \mc{M}_+^{tw} N
	-2\,\s_M\,\ol{N} N \,{\rm tr}(\mc{M}_+^{tw})
	-2\,\s_W\,\ol{N} N \,{\rm tr}(\mc{W}_+) \notag \\
	& + (\ol{T}_\mu i v \cdot D\,T_\mu) 
	+ \Delta\,(\overline{T}_\mu T_\mu) 
	+2\,\gamma_M\,(\overline{T}_\mu \mathcal{M}_+^{tw} T_\mu) \nonumber\\
	&-2\,\ol{\sigma}_M\,(\ol{T}_\mu T_\mu)\,{\rm tr}(\mathcal{M}_+^{tw})  
	-2\,\ol{\sigma}_W\,(\ol{T}_\mu T_\mu)\,{\rm tr}(\mathcal{W}_+) \, .
\end{align}
The ``twisted mass'' spurion field
is defined by    
\begin{equation}
	\mathcal{M}_{\pm}^{tw} = \frac{1}{2} \left[ 
		\xi^\dagger m_Q^{tw} \xi^\dagger \pm \xi (m_Q^{tw})^\dagger \xi
		\right] \,, \qquad\qquad
	m_Q^{tw} = \frac{\chi'}{2B_0} \,,
\end{equation}
with $m_Q^{tw}$ being the ``twisted'' mass spurion for the
baryons. The ``Wilson'' (discretization) spurion field is defined by
\begin{equation}\label{eq:wilsonSpurion}
	\mathcal{W}_{\pm} = \frac{1}{2} \left( 
		\xi^\dagger w_Q \xi^\dagger \pm \xi w_Q^\dagger \xi 
		\right) \,, \qquad\qquad
	w_Q = \frac{\Lambda_\chi^2}{2W_0}\hat{A} \,,
\end{equation}
with $w_Q$ being the Wilson spurion for the baryons. Note that we have made simplifications using the properties of $SU(2)$ matrices when writing down Eq.~\eqref{E:BLLOtw}. When setting the spurions to their constant values, $\mathcal{W}_+$ is proportional to the identity matrix in flavor space. Thus the operators $\ol{N}\mc{W}_+ N$ and $(\ol{T}_\mu\mc{W}_+ T_\mu)$, although allowed under the symmetries of tmLQCD, are not independent operators with respect to $\ol{N}N\,{\rm tr}(\mc{W}_+)$ and $(\ol{T}_\mu T_\mu)\,{\rm tr}(\mc{W}_+)$ respectively. This is also true of the nucleon and delta operators
involving $\mc{M}_+^{tw}$ in the isospin limit (but not away from it). The independent operators we choose to write down are those with the simplest flavor contractions, and this will be the case henceforth
whenever we make simplifications using the properties of $SU(2)$.   

In Eq.~(\ref{E:BLLOtw}), the four-vector, $v_\mu$, is the Euclidean heavy baryon four-velocity, and our conventional here is that in Euclidean space, $v \cdot v = -1$. The parameter, $\D$, is the mass splitting between the nucleons and deltas which is independent of the quark masses and we treat  $\Delta \sim m_\pi \sim \varepsilon^2$ following~\cite{Jenkins:1991ne,Jenkins:1990jv,Jenkins:1992ts}.  The dimensionless low energy constants (LECs), $\alpha_M$, $\sigma_M$, $\gamma_M$, and $\overline{\sigma}_M$ have the same numerical values as in the usual untwisted two-flavor HB$\chi$PT.  As was noted in Ref.~\cite{Sharpe:2004ny}, the shifting from $\chi$ to $\chi' = \chi + \hat{A}$, which corresponds to the shift of the physical mass $m$ to $m'$ at the quark level does not, in general, remove the discretization ($\hat{A}$) term, and this is seen explicitly here with the presence of the discretization terms. 

At this order, the Lagrangian describing the interactions of the nucleons and deltas with the pions is still given by Eq.~\eqref{eq:NNpi}.  Note that the $\mc{O}(\varepsilon^2)$ free Lagrangian (\ref{E:BLLOtw}) and interaction Lagrangian (\ref{eq:NNpi}) are the same as those given in Ref.~\cite{Tiburzi:2005vy} when the twist is removed, i.e. when $\mu = 0$. With non-vanishing twist, the mass operators carry a twisted component and the vacuum is ``twisted'' from the identity to point in the direction of the twist (the flavor $\tau_1$-direction here)~\cite{Sharpe:2004ny}.   

Following Ref.~\cite{Sharpe:2004ny}, we expand $\Sigma$ about its lattice vacuum expectation value, defining the physical lattice pion fields and the physical lattice
$\xi$ fields by  
\begin{align} \label{eq:LOpixi}
	\Sigma = \mc{T}\Sigma_{ph}\mc{T} \,, \qquad
	\xi = \mc{T}\xi_{ph} U(\xi_{ph}) \,, \qquad
	\mc{T} = \exp(i\omega_0 \tau_1 / 2) \,, \nonumber\\
	\Sigma_{ph} =
	\exp(i\sqrt{2}\;\boldsymbol{\pi}\cdot\boldsymbol{\tau}/f)\,, \qquad
	\mc{T} \,, U \in SU(2) \,,
\end{align}
If we make the following chiral transformation for which the fields transform as in Eqs.~\eqref{eq:SigTransform}, \eqref{eq:NucTransform}, \eqref{eq:xiTransform}, \eqref{eq:VandA} and \eqref{eq:Ttransform}, and use the particular $SU(2)$ matrices $L = R^\dagger = \mc{T}^\dagger$, we have in the transformed effective chiral Lagrangian
\begin{align} \label{eq:PhysFields}
	\Sigma &\rightarrow \Sigma_{ph} ,  
	&\xi \rightarrow \xi_{ph} \,, \nonumber\\
	\mathcal{A}_\mu &\rightarrow \frac{i}{2}\left(\xi_{ph} \partial_\mu \xi_{ph}^\dagger -
		\xi_{ph}^\dagger \partial_\mu \xi_{ph}\right) ,
	&\mathcal{V}_\mu \rightarrow \frac{1}{2}\left(\xi_{ph} \partial_\mu \xi_{ph}^\dagger +
		\xi_{ph}^\dagger \partial_\mu \xi_{ph}\right) \,,
\end{align}
and
\begin{align}
\mc{M}^{tw}_\pm &\rightarrow \mathcal{M}_\pm = 
	\frac{1}{2} \left(\xi_{ph}^\dagger m_Q \xi_{ph}^\dagger \pm \xi_{ph} m_Q^\dagger \xi_{ph} \right) \,, 
& m_Q^{tw} &\rightarrow m_Q = \mc{T}^\dagger \frac{\chi^\prime}{2B_0} \mc{T}^\dagger \,, \notag \\
\mc{W}_\pm &\rightarrow \mathcal{W}^{tw}_\pm = 
	\frac{1}{2} \left[\xi_{ph}^\dagger w_Q^{tw} \xi_{ph}^\dagger \pm 
		\xi_{ph} (w_Q^{tw})^\dagger \xi_{ph} \right] \,, 
& w_Q &\rightarrow w_Q^{tw} = 
	\mc{T}^\dagger \left (\frac{\L_\chi^2}{2W_0}\hat{A} \right) \mc{T}^\dagger \,.
\end{align}
Note that since $L = R^\dagger = \mc{T}^\dagger \in SU(2)$, and
$\xi = L^\dagger\xi_{ph}U \equiv U^\dagger\xi_{ph}R$,
\begin{equation}
\Sigma = L^\dagger\Sigma_{ph}R = 
 \xi^2 = (L^\dagger \xi_{ph} U)\cdot(U^\dagger \xi_{ph} R) 
       = L^\dagger\xi_{ph}^2 R
\,\Longrightarrow\,\xi_{ph}^2 = \Sigma_{ph} \,.
\end{equation}

We see that the $\xi$ field is now $\xi_{ph}$, the field associated with the physical pions, and the twist is transferred from the twisted mass ($\mathcal{M}^{tw}_\pm$) term to the ``twisted Wilson'' ($\mathcal{W}^{tw}_\pm$) term, making the mass term in the HB$\chi$PT now the same as that in the untwisted theory. The new mass spurion, $m_Q$, and the ``twisted Wilson'' spurion, $w_Q^{tw}$, now take constant values  
\begin{equation}
	m_Q \longrightarrow m_q - \epsilon_q \tau_3 
	\,, \qquad\qquad 
	w_Q^{tw} \longrightarrow 
	a\,\L_\chi^2 \exp(-i\omega_0 \tau_1) \,.
\end{equation}
We will call this the ``physical pion basis'' since this is the basis where the pions are physical as defined by the twisted lattice action~\cite{Sharpe:2004ny}, and we will work in this basis from now on, unless otherwise specified.%
\footnote{As detailed in Ref.~\cite{Sharpe:2004ny}, the twist angle that one determines
non-perturbatively in practice, call it $\omega$, will differ from $\omega_0$ by $O(a)$. This will give rise to a relative $O(a)$ contribution to the pion terms. But since the pions come into baryon calculations only through loops, the correction will be of higher order than we work. Thus to the accuracy we work, we may use either $\omega$ or $\omega_0$.}%
A technical point we note here is that, in the isospin limit where twisting can be implemented by any of the three Pauli matrices, say $\tau_k$, the physical pion basis can be found following the same recipe detailed above but with $\tau_1$ in $\mc{T}$ replaced by $\tau_k$ throughout. 

Rotating to the physical pion basis where the $\xi$ field is now the physical $\xi_{ph}$ field in all field quantities, the form of the interaction Lagrangian remains unchanged as given in (\ref{eq:NNpi}), while the $\mc{O}(\varepsilon^2)$ free heavy baryon chiral Lagrangian~(\ref{E:BLLOtw}) changes to   
\begin{align} \label{E:BLLOphy}
	\mc{L}_\chi = 
	&\,\ol{N} i v \cdot D N 
		-2\,\a_M\,\ol{N}\mc{M}_+ N
		-2\,\s_M\,\ol{N}N\,{\rm tr}(\mc{M}_+)
		-2\,\s_W\,\ol{N}N\,{\rm tr}(\mc{W}^{tw}_+) \notag \\
	& + (\ol{T}_\mu i v \cdot D\,T_\mu) 
		+\Delta\,(\ol{T}_\mu T_\mu) 
		+2\,\g_M\,(\ol{T}_\mu \mc{M}_+ T_\mu) \nonumber\\
	&  -2\,\ol{\s}_M\,(\ol{T}_\mu T_\mu)\,{\rm tr}(\mc{M}_+)
		-2\,\ol{\s}_W\,(\ol{T}_\mu T_\mu)\,{\rm tr}(\mc{W}^{tw}_+) \,.
\end{align}
Note that $\mc{W}^{tw}_+$ is also proportional to the identity matrix in flavor space when set to its constant value. Thus if we build the free chiral Lagrangian directly in the physical pion basis, the same simplifications due to $SU(2)$ we used in writing down Eq.~(\ref{E:BLLOtw}) apply. Note that at this point, one can not yet tell whether the nucleon ($N$) and the delta ($T_\mu$) fields are physical. This has to be determined by the theory itself. We will return to this point when calculating the nucleon and delta masses below. 

\bigskip

At $\mathcal{O}(\varepsilon^4)$, there are contributions from $\mathcal{O}(a \mathsf{m})$ and $\mathcal{O}(a^2)$ operators. The enumeration of the operators is similar to that set out in Ref.~\cite{Tiburzi:2005vy}, except now the Wilson spurion field carries a twisted component. The operators appearing in the $\mathcal{O}(\varepsilon^4)$ chiral Lagrangian will involve two insertions off the following: $\mathcal{M}_\pm$, $\mathcal{W}^{tw}_\pm$, and the axial current $\mathcal{A}_\mu$. Note that since parity combined with flavor is conserved in tmLQCD, any one insertion of $\mathcal{M}_-$ or $\mathcal{W}^{tw}_-$ must be accompanied by another insertion of $\mathcal{M}_-$ or $\mathcal{W}^{tw}_-$. Now operators with two insertions of $\mathcal{M}_+$ or $\mathcal{A}_\mu$, which contribute to baryon masses at tree and one-loop level respectively, have the same form as those in the untwisted theory (and so give the same contribution). These have been written down in~\cite{Tiburzi:2005na} and will not be repeated here. Operators with an insertion of either a combination of $v \cdot \mc{A}$ and $\mc{M}_-$ (which have the same form as in the untwisted theory), or a combination of $v \cdot \mathcal{A}$ and 
$\mathcal{W}^{tw}_-$, will also not contribute to the baryon masses at $\mathcal{O}(\varepsilon^4)$.    

At $\mathcal{O}(a \mathsf{m})$, there are two independent operators contributing to the masses in the nucleon sector
\begin{align} \label{E:Namq}
\mathcal{L}_\chi &= -\frac{1}{\Lambda_\chi} \bigg[
n_1^{WM_+}\,\overline{N}\mathcal{M}_+ N\,{\rm tr}(\mathcal{W}^{tw}_+) +
n_2^{WM_+}\,\overline{N}N
          \,{\rm tr} (\mathcal{M}_+) \,{\rm tr}(\mathcal{W}^{tw}_+)
\bigg] \,,
\end{align}
and two independent operators contributing to the masses in the delta sector 
\begin{align} \label{E:Tamq}
\mathcal{L}_\chi &= \frac{1}{\Lambda_\chi} \bigg[
t_1^{WM_+}\,(\overline{T}_\mu \mathcal{M}_+ T_\mu)
          \,{\rm tr}(\mathcal{W}^{tw}_+) +
t_2^{WM_+}\,(\overline{T}_\mu T_\mu)
          \,{\rm tr}(\mathcal{W}^{tw}_+)\,{\rm tr}(\mathcal{M}_+)
\bigg] \,, % \qquad i,\,j,\,k = 1,\,2 \,,
\end{align}
where we remind the reader $\L_\chi \equiv 4\pi f$.  Note that there are no operators involving the commutator, $[\mathcal{M}_+,\mathcal{W}^{tm}_+]$, because it is identically zero. There are also operators involving $\mc{M}_- \otimes\mc{W}^{tw}_-$ at $\mc{O}(a\mathsf{m})$, but these again do not contribute to the baryon masses at the order we work. 

At $\mathcal{O}(a^2)$, there are operators that do not break the chiral symmetry arising from the bilinear operators and four-quark operators (see e.g. Ref.~\cite{Bar:2003mh} for a complete listing) in the $mathcal{O}(a^2)$ part of $\mathcal{L}_{\rm eff}$. These give rise to the tmHB$\chi$PT operators
\begin{equation} \label{E:Na2chi}
	\mc{L}_\chi = a^2\L_\chi^3 \bigg[-b\,\ol{N}N+t\,(\ol{T}_\mu T_\mu) \bigg] \,.
\end{equation}
There are also chiral symmetry preserving but $O(4)$ rotation symmetry breaking operators which arise from the bilinear operator of the form $a^2\bar{\psi}\gamma_\mu D_\mu D_\mu D_\mu \psi$ in the
$\mc{O}(a^2)$ part of $\mathcal{L}_{\rm eff}$. These give rise to the tmHB$\chi$PT operators 
\begin{equation} \label{E:Na2O4}
	\mc{L}_\chi = a^2\L_\chi^3 \bigg[
		-b_v\,\ol{N} v_\mu v_\mu v_\mu v_\mu N
		+t_v\,(\ol{T}_\nu v_\mu v_\mu v_\mu v_\mu T_\nu)
		+t_{\bar{v}}\,(\ol{T}_\mu v_\mu v_\mu T_\mu) \bigg] \,.
\end{equation}
Note that these chiral symmetry preserving operators are clearly not affected by twisting (the $O(4)$ symmetry breaking operator at the quark level from which they arise involve only derivatives with no
flavor structure, and $\{\gamma_\mu,\gamma_5\} = 0$), and so they have the same form and contribute to the baryon masses in the same way as in the untwisted theory. The chiral symmetry breaking operators at $\mathcal{O}(a^2)$ are those with two insertions of the Wilson spurion fields. For the nucleons, there are two such independent operators     
\begin{equation} \label{E:Nasq}
\mathcal{L}_\chi = -\frac{1}{\L_\chi} \bigg[ n_1^{W_+}\,
		\overline{N}N\,{\rm tr}(\mathcal{W}^{tw}_+)\,{\rm tr}(\mathcal{W}^{tw}_+) 
	+ n_1^{W_-}\,\overline{N}N \,{\rm tr}(\mathcal{W}^{tw}_-\mathcal{W}^{tw}_-) 
\bigg] \,,
\end{equation}
and for the deltas, there are three such independent operators
\begin{multline} \label{E:Tasq}
	\mathcal{L}_\chi = \frac{1}{\L_\chi} \bigg[
		t_1^{W_+}\,(\overline{T}_\mu T_\mu) \,{\rm tr}(\mathcal{W}^{tw}_+)\,
			{\rm tr}(\mathcal{W}^{tw}_+) 
		+ t_1^{W_-}\,(\overline{T}_\mu T_\mu) \,{\rm tr}(\mathcal{W}^{tw}_- \mathcal{W}^{tw}_-) \\
		+t_2^{W_-}\,\overline{T}^{kji}_\mu (\mathcal{W}^{tw}_-)^{ii'}(\mathcal{W}^{tw}_-)^{jj'}
			T^{i^\prime j^\prime k}_\mu \bigg] \,.
\end{multline}

In the isospin limit where the mass splitting vanishes ($\epsilon_q \rightarrow 0$), more simplifications occur in the $\mc{O}(\varepsilon^4)$ chiral Lagrangian. The nucleon operators in Eq.~\eqref{E:Namq} with coefficients $n_1^{WM_+}$ and $n_2^{WM_+}$ are the same up to a numerical factor, and the same holds for delta operators in Eq.~\eqref{E:Tamq} with coefficients $t_1^{WM_+}$ and $t_2^{WM_+}$, and for operators in Eq.~\eqref{E:Tasq} with coefficient $t_1^{W_-}$ and $t_2^{W_-}$. 

Note that in the untwisted limit, the $\mathcal{O}(\varepsilon^4)$ chiral Lagrangian reduces to that given in Ref.~\cite{Tiburzi:2005vy}. In particular, with the twist set to zero, operators with two insertions of
$\mathcal{W}^{tw}_-$ will not contribute to the nucleon or the delta mass until $\mc{O}(a^2\mathsf{m}) \sim \mc{O}(\varepsilon^6)$, but for non-vanishing twist, they contribute at $\mathcal{O}(a^2)$.

%%%%%%%%%%%%%%%%%%%%%%%%%%%%%%%%%%%%%%%%%%%%%%%%%%%
%
%	N-D Masses in Isospin Limit
%
%%%%%%%%%%%%%%%%%%%%%%%%%%%%%%%%%%%%%%%%%%%%%%%%%%%
\section{\label{sec:massiso} Nucleon and Delta Masses in the Isospin
  Limit} 

In this work, we are concerned with corrections to the masses of the nucleons and the deltas due to the effect of the twisted mass parameter. We will therefore only give expressions for the mass corrections arising from the effects of lattice discretization and twisting in tmLQCD. A calculation of the nucleon and delta masses in the continuum in infinite volume to $\mc{O}(m_q^2)$ can be found in Ref.~\cite{Tiburzi:2005na}. The mass corrections due to finite lattice spacing to $\mc{O}(a^2)$ in the untwisted theory with Wilson quarks can be found in Ref.~\cite{Tiburzi:2005vy}, and the leading finite volume modifications to the nucleon mass can be found in Ref.~\cite{Beane:2004ks}.

In this section, we present the results of nucleon and delta masses calculated in tmHB$\chi$PT, in the isospin limit, where the quark doublet is mass-degenerate, and the twist is implemented by $\tau_3$. As we discussed in Sec.~\ref{sec:EContL}, in the isospin limit, the content of tmLQCD is the same regardless of which Pauli matrix is used to implement the twist -- the action for one choice is related to another by a flavor-vector rotation. This must also hold true of the effective chiral theory that arises from tmLQCD. Indeed, the heavy baryon Lagrangian constructed in Sec.~\ref{sec:baryons} with $\tau_1$-twisting can be rotated into that with $\tau_3$-twisting by making a vector transformation, which is given by $L = R = V = \exp(i\frac{\pi}{4}\tau_2)$. 

%%%%%%%%%%%%%%%%%%%%%%%%%%%%%%%%%%%%%%%%%%%%%%%%%%%
%
%	Nucleon mass in isospin limit
%
%%%%%%%%%%%%%%%%%%%%%%%%%%%%%%%%%%%%%%%%%%%%%%%%%%%
\subsection{\label{sec:NMassIso} \textbf{Nucleon Masses in the Isospin Limit}}

In the continuum, the mass of the nucleons in infinite volume HB$\chi$PT with two flavor-degenerate quarks are organized as an expansion in powers of the quark mass, which can be written as Eq.~\eqref{eq:NucMassExp}.  Here, we are interested in the corrections to this formula due to the effects of lattice discretization and twisting arising from tmLQCD. We denote these lattice corrections to the nucleon mass at $\mc{O}(\varepsilon^{2n}) \sim \mc{O}(m_q^n) \sim \mc{O}(a^n)\,$ (factors of $\L_{\rm QCD}$ needed to make the dimensions correct are implicit here) as $\d M_{N_i}^{(n)}$, and the nucleon mass in tmHB$\chi$PT is now written as 
\begin{equation}
	M_{N_i}^{tm} = M_0 - (M_{N_i}^{(1)} + \delta M_{N_i}^{(1)}) + \ldots
\end{equation}

\begin{figure}[t] 
\centering
\includegraphics[width=0.2\textwidth]{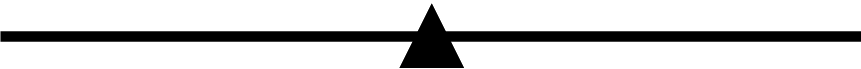}\\ (a)\\
\begin{tabular}{cccc}
\includegraphics[width=0.2\textwidth]{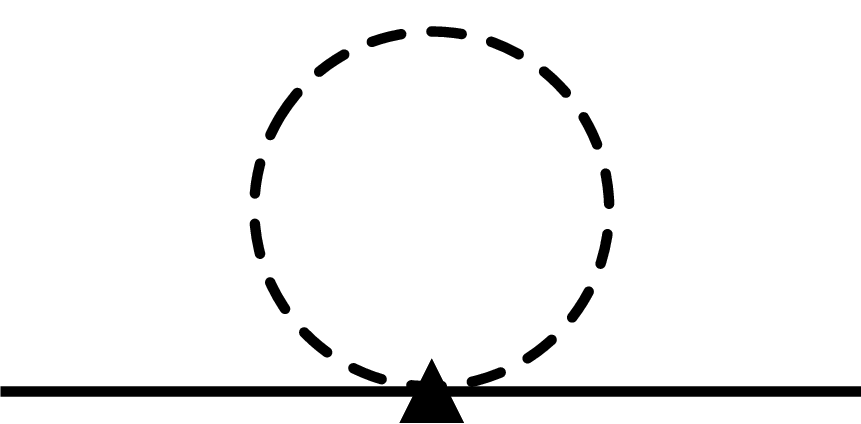}
&\includegraphics[width=0.2\textwidth]{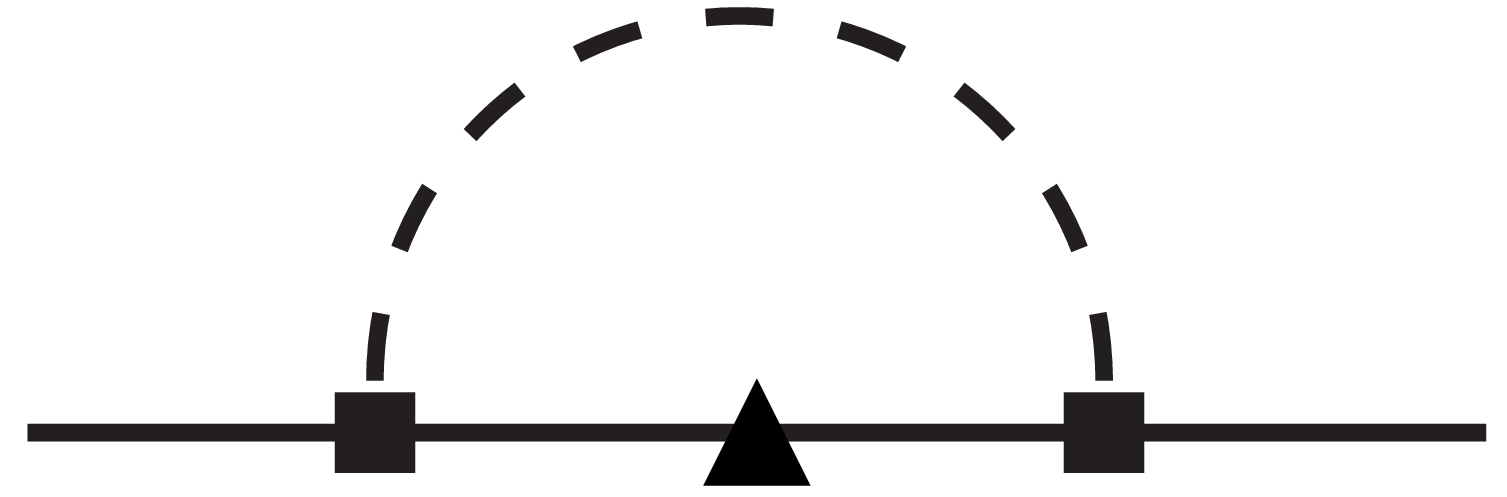}
&\includegraphics[width=0.2\textwidth]{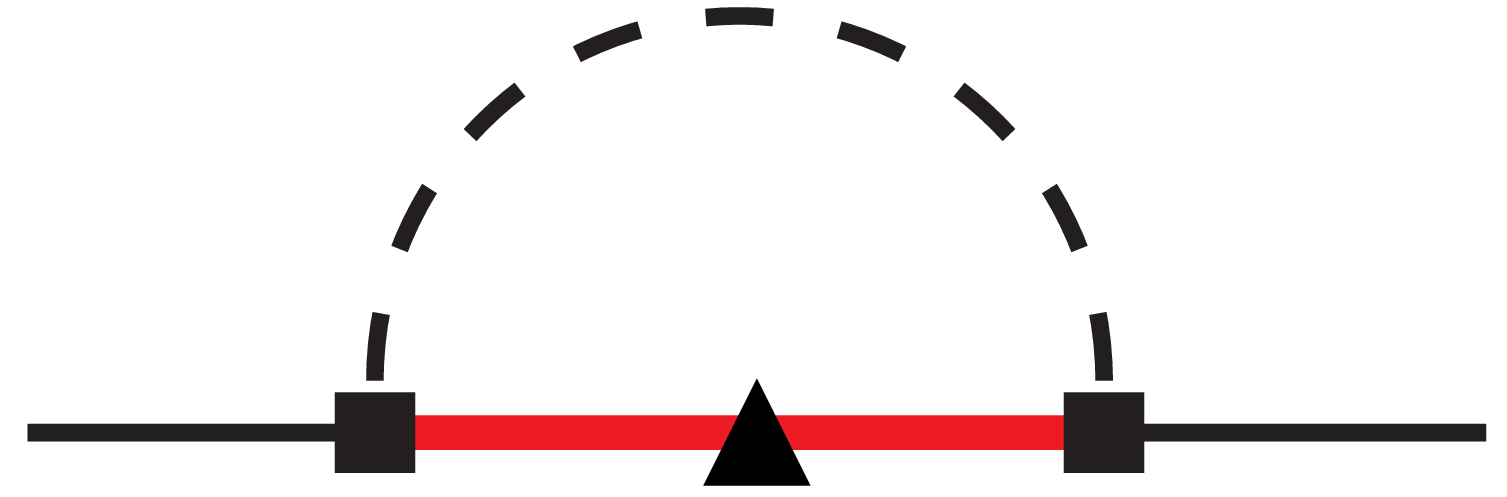}
&\includegraphics[width=0.2\textwidth]{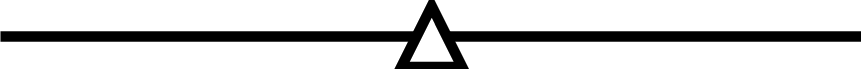}\\
(b) & (c) & (d) & (e)
\end{tabular}
\caption[Diagrams depicting the twisted mass corrections to the nucleon mass.]{\label{fig:NNLOtm} Diagrams depicting mass contributions to the nucleons through next-to-next-to-leading order (NNLO)
  in tmHB$\chi$PT in the physical pion basis. The solid, thick solid (red)
  and dashed lines denote nucleons, deltas and pions respectively. The
  solid triangle denotes an insertion of the twisted Wilson operator
  as given in~\eqref{E:BLLOphy}. The solid squares denote the coupling
  of the baryons to the axial current whose form is given
  in~\eqref{eq:NNpi}. The clear triangle denotes a tree level insertion
  of the operators given in Eqs.~\eqref{E:Namq} and
  \eqref{E:Na2chi}~--~\eqref{E:Nasq}.} 
\end{figure}

The leading correction in tm\CPT\ comes in at tree level, arising from the twisted Wilson nucleon operator in the free heavy baryon Lagrangian~(\ref{E:BLLOphy}) and is depicted in Fig.~\ref{fig:NNLOtm},
\begin{equation} \label{E:NBLO}
	\d M_{N_i}^{(1)}(\w) = 4\,\s_W\,a\,\L_\chi^2\cos(\w) \,,
\end{equation}
where to the accuracy we work, $\omega$ can either be $\omega_0$ or the twist angle non-perturbatively determined. Note that this correction is the same for both the proton and the neutron. At leading order, the nucleon mass is automatically $\mc{O}(a)$ improved, as $\d M_{N_i}^{(1)}$ vanishes at maximal twist, $\omega = \pi/2$. At zero twist, $\omega = 0$, it reduces to that in the untwisted
theory~\cite{Beane:2003xv,Tiburzi:2005vy}.  

The next contribution to the nucleon mass comes from the leading pion loop diagrams shown in Fig.~\ref{fig:LONmass}. However, at the order we work, the form of the $\mc{O}(\mathsf{m}^{3/2})$ nucleon mass
contribution is unchanged from the continuum, and given in Eq.~\eqref{eq:NmassNLO}.

%%%%%%%%%%%%%%%%%%%%%%%%%%%%%%%%%%%%%%%%%%%%%%%%%%%
%
%	figure: twisted mass corrections to nucleon
%
%%%%%%%%%%%%%%%%%%%%%%%%%%%%%%%%%%%%%%%%%%%%%%%%%%%
%%%%%%%%%%%%%%%%%%%%%%%%%%%%%%%%%%%%%%%%%%%%%%%%%%%
%
%%%%%%%%%%%%%%%%%%%%%%%%%%%%%%%%%%%%%%%%%%%%%%%%%%%
The corrections to $M_{N_i}^{(2)}$ come from both the tree level and the one-loop diagrams, as shown in Fig.~\ref{fig:NNLOtm}. The twisted Wilson operator in the free Lagrangian~(\ref{E:BLLOphy}) gives rise to a tadpole diagram, which produces a contribution of $\mc{O}(a \mathsf{m})$.  The leading Wilson spurions also contribute to $\mc{O}(a \mathsf{m})$ when inserted inside the pion-nucleon loops, and are
partly cancelled by wavefunction corrections. The tree level contributions come from the operators given in Eqs.~\eqref{E:Namq} and \eqref{E:Na2chi}~--~\eqref{E:Nasq}. Just as in the untwisted continuum
theory, these act both as the higher dimensional operators and as counter terms that renormalize divergences from the lower order loop contributions. For instance, coefficients $n_1^{WM_+}$ and
$n_2^{WM_+}$ are renormalized to absorb divergences from the tadpole and one-loop contributions mentioned above.  These coefficients are taken to be the renormalized coefficients (finite) in the mass
calculations, and to contain the counter terms needed in our renormalization scheme. The corrections to $M_{N_i}^{(2)}$ read
\begin{align} \label{eq:N2Mass}
	-\delta M_{N}^{(2)}(\omega) =&\ 12\,\s_W\,a\, m_\pi^2
			 \log\left(\frac{m_\pi^2}{\mu^2}\right)\cos(\w) \nonumber\\
		&+16\,g_{\D N}^2\,(\ol{\s}_W - \s_W)\, a\,
			\left[\mc{J}(m_\pi,\D,\mu) + m_\pi^2\right]\cos(\w) \notag \\ 
		& -2\left(n_1^{WM_+} + 2\,n_2^{WM_+}\right)
			a \L_\chi\, m_q\cos(\w) 
		-a^2\L_\chi^3\left(b + b_v\right) \notag \\ 
		& +a^2\L_\chi^3 \left( 2\,n_1^{W_-}\sin^2(\w)
		-4\,n_1^{W_+}\cos^2(\omega)\right) \,.
\end{align}
Note that the $\mc{O}(\varepsilon^4)$ corrections are again the same for both the proton and the neutron. At maximal twist, the $\mc{O}(\varepsilon^4)$ corrections are given by  
\begin{equation}
	-\delta M_{N_i}^{(2)}(\w=\pi/2) = 
		a^2 \L_\chi^3 \left(2n_1^{W_-} - b - b_v \right) \,,
\end{equation}
while at zero twist, these reduces to the corrections given in Ref.~\cite{Tiburzi:2005vy}. We see that the nucleon masses are also automatically $\mc{O}(a)$ improved at $\mc{O}(\varepsilon^4)$. 

To the order we work, the expressions for the nucleon mass corrections in tmHB$\chi$PT given in Eqs.~\eqref{E:NBLO} and~\eqref{eq:N2Mass}, together with the untwisted continuum HB$\chi$PT expressions for the nucleon masses, provide the functional form for the dependence of the nucleon masses on the twist and angle, $\w$, and the quark mass, $m_q$, which can be used to fit the lattice data.

%%%%%%%%%%%%%%%%%%%%%%%%%%%%%%%%%%%%%%%%%%%%%%%%%%%
%
%	Delta mass in isospin limit
%
%%%%%%%%%%%%%%%%%%%%%%%%%%%%%%%%%%%%%%%%%%%%%%%%%%%
\subsection{\label{sec:DMassIso} \textbf{Delta Masses in the Isospin Limit}}

\begin{figure}[tbp]
\centering
\includegraphics[width=0.2\textwidth]{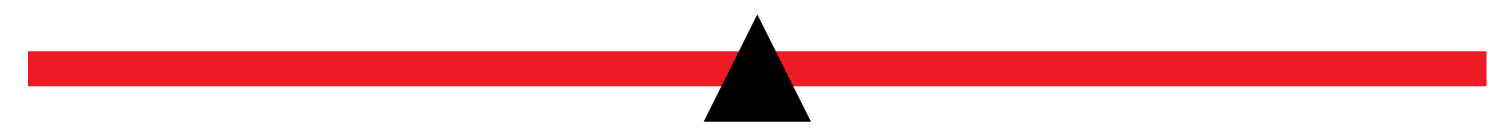} \\ (a) \\
\begin{tabular}{cccc}
\includegraphics[width=0.2\textwidth]{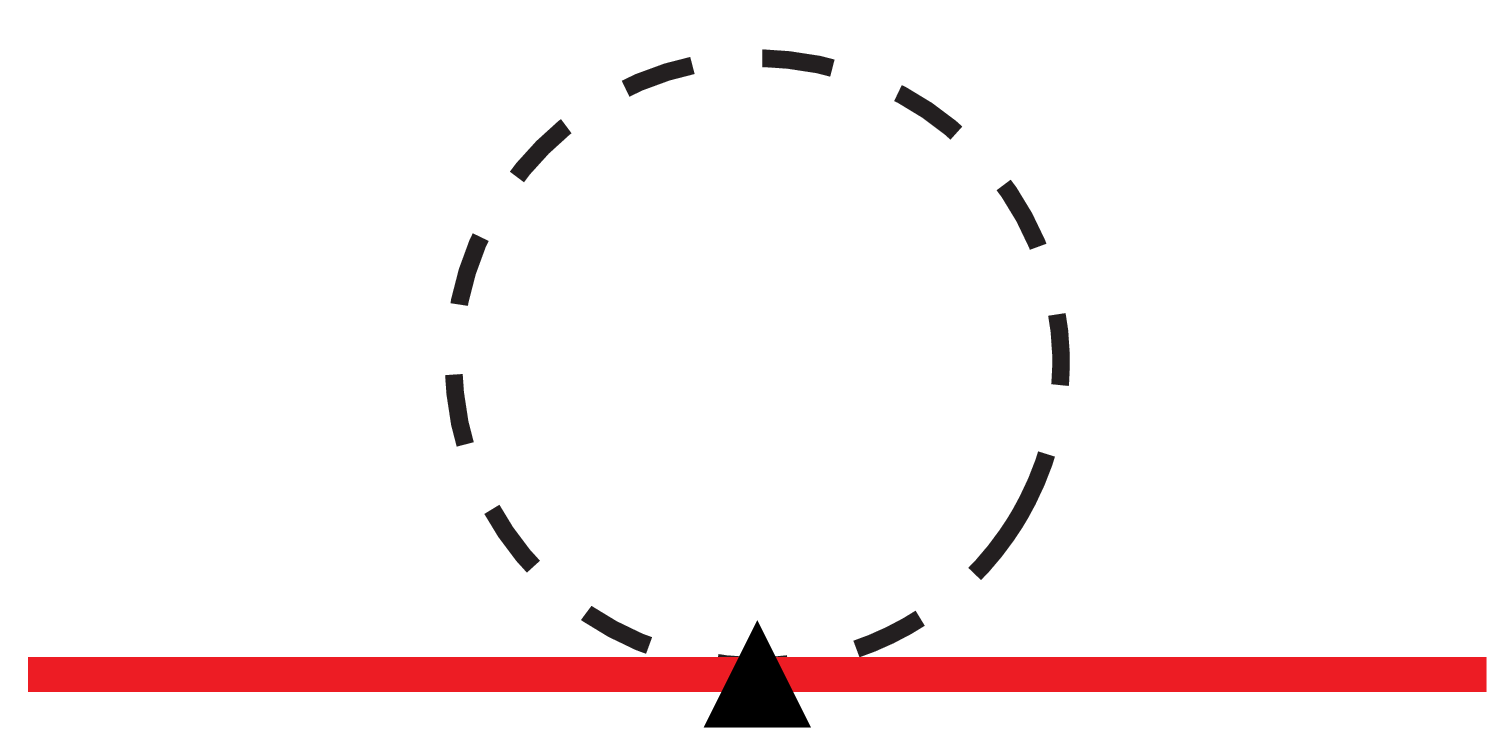}
&\includegraphics[width=0.2\textwidth]{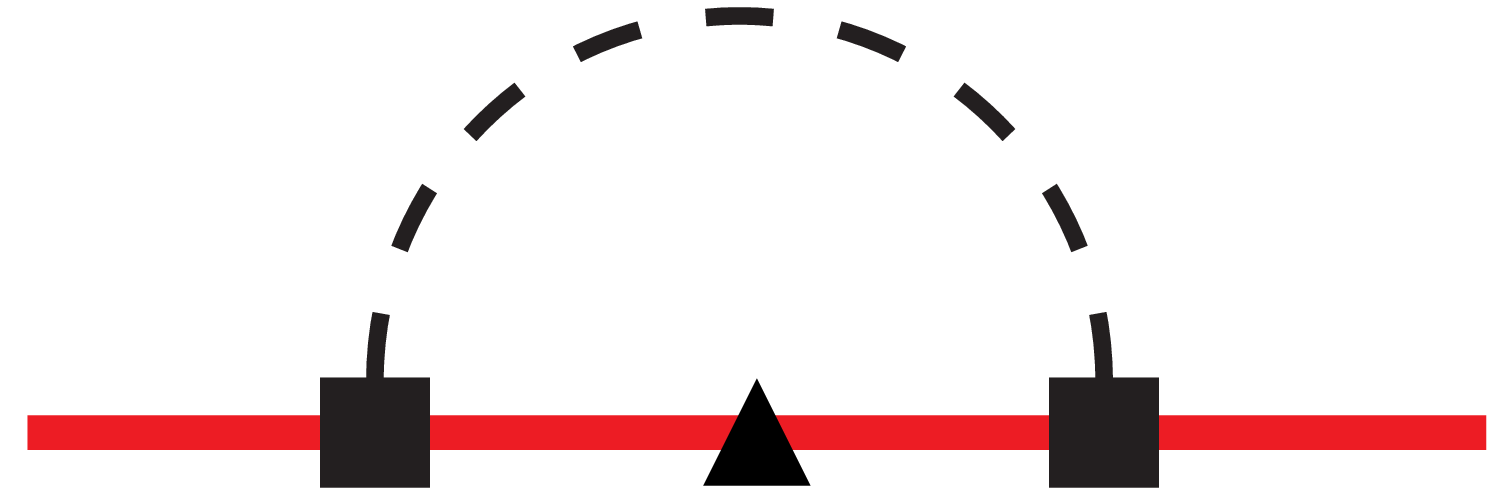}
&\includegraphics[width=0.2\textwidth]{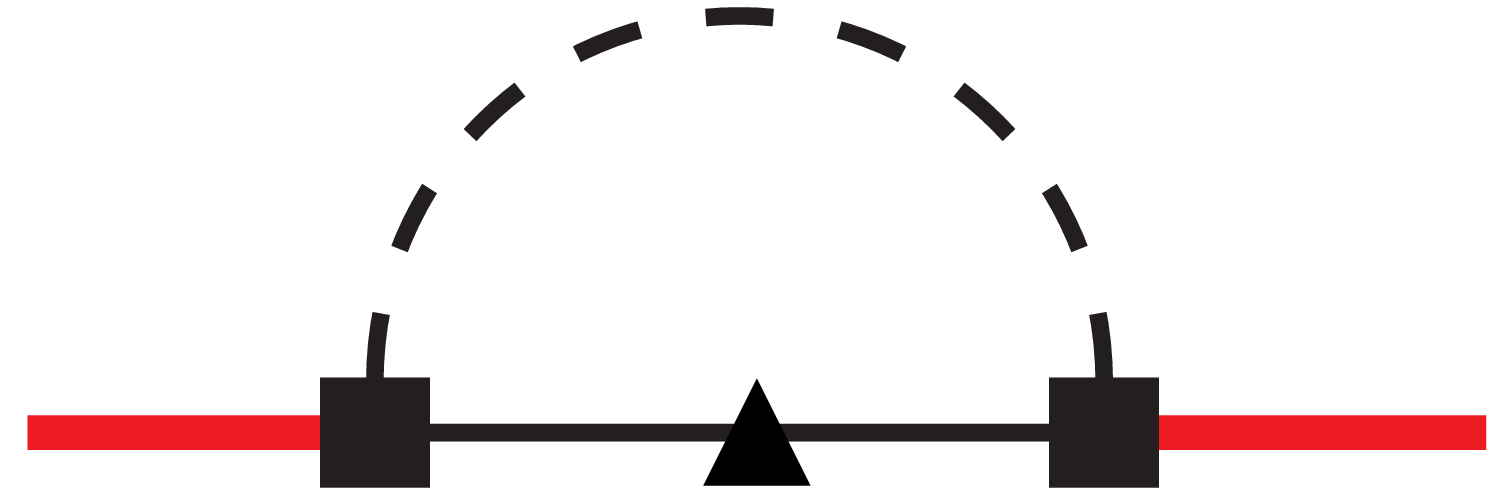}
&\includegraphics[width=0.20\textwidth]{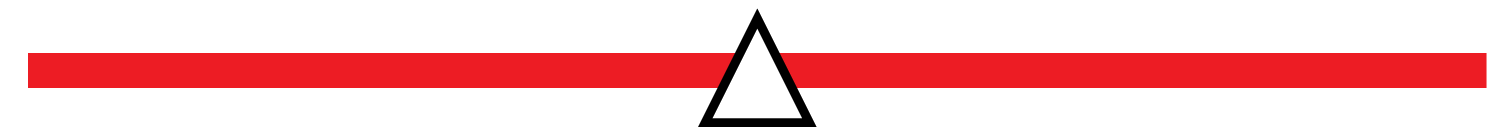}\\
(b) & (c) & (d) & (e)
\end{tabular}
\caption[Diagrams depicting the twisted mass corrections to the delta masses]{\label{fig:DNNLOtm} Diagrams depicting mass contributions to the deltas at NNLO
  in tmHB$\chi$PT in the physical pion basis. The solid, thick solid (red)
  and dashed lines denote nucleons, deltas and pions respectively. The
  solid triangle denotes an insertion of the discretization operator
  in Eq.~\eqref{E:BLLOphy}. The solid squares denote the coupling of
  the baryons to the axial current whose form is given in
  Eq.~\eqref{eq:NNpi}. The clear triangle denotes a tree level
  insertion of the operators given in Eqs.~\eqref{E:Tamq}~-~\eqref{E:Na2O4} and
  \eqref{E:Tasq}.} 
\end{figure}

We are again interested in the corrections to the delta masses arising from the twisted mass parameters, such that the delta mass expansion is written
\begin{equation}
M_{T_i}^{tm} = M_0  + \D + (M_{T_i}^{(1)} + \d M_{T_i}^{(1)}) + \ldots
\end{equation}
The leading mass correction arises at tree level from the twisted Wilson delta operator given in Eq.~\eqref{E:BLLOphy}, which we depict in Fig.~\ref{fig:DNNLOtm}
\begin{equation}\label{eq:DLO}
	\d M_{T}^{(1)}(\w) = -4\,\ol{\s}_W\,a\,\L_\chi^2\cos(\w) \,.
\end{equation}
Just as for the nucleons, this does not split the delta masses and vanishes at maximal twist. 

The $\mc{O}(\varepsilon^3)$ delta mass contributions are similarly given as for the nucleons.  They do not cause any splitting between the deltas, and receive no discretization corrections, and are shown in Fig.~\ref{fig:LODmass}.

%%%%%%%%%%%%%%%%%%%%%%%%%%%%%%%%%%%%%%%%%%%%%%%%%%%
%
%	figure: twisted delta mass corrections
%
%%%%%%%%%%%%%%%%%%%%%%%%%%%%%%%%%%%%%%%%%%%%%%%%%%%
%%%%%%%%%%%%%%%%%%%%%%%%%%%%%%%%%%%%%%%%%%%%%%%%%%%
%
%%%%%%%%%%%%%%%%%%%%%%%%%%%%%%%%%%%%%%%%%%%%%%%%%%%

At $\mc{O}(\varepsilon^4)$, contributions due to the effects of
twisting arise from similar diagrams as in the nucleon case, and are
shown in Fig.~\ref{fig:DNNLOtm}. A splitting in the delta masses first
arises at this order, which comes from the operator with coefficient
$t_2^{W_-}$ given in~\eqref{E:Tasq}. The mass corrections read
\begin{align} \label{eq:DNNLOtau1}
	\d M_{T_i}^{(2)}(\w) =&\ 
		12\,\ol{\s}_W\, a\,m_\pi^2
			\log\left(\frac{m_\pi^2}{\mu^2}\right)\cos(\w) %\nonumber\\
		+4\,g_{\D N}^2\,(\ol{\s}_W -\s_W)\,a\, \mc{J}(m_\pi,-\D,\mu)\cos(\w) \notag \\
	& +2\left(t_1^{WM_+} + 2\,t_2^{WM_+}\right) a\L_\chi\, m_q\cos(\w)
	+a^2\L_\chi^3\left(t+t_v\right) \notag \\
	& +a^2\L_\chi^3 \left(4\,t_1^{W_+}\cos^2(\w)-2\,t_1^{W_-}\sin^2(\w)
	+t_2^{W_-}\d_{T_i}\sin^2(\w)\right) \,,
\end{align}
where 
\begin{equation} \label{E:D2split}
	\d_{T_i} =  \begin{cases}
		-1 & \text{for $T_i = \D^{++}\,,\D^-$} \\
		\ \ \frac{1}{3} &\text{for $T_i = \D^+\,,\D^0$}
	\end{cases} \,.
\end{equation}
Note the appearance of the mass splitting, $\d_{T_i}$, in $\d M_{T_i}^{(2)}$. We see from above that starting at $\mc{O}(\varepsilon^4)$, the delta multiplet is split into two mass-degenerate pairs, with one pair containing $\D^{++}$ and $\D^-$, and the other, $\D^+$ and $\D^0$. At maximal twist, $\d M_{T_i}^{(2)}$ becomes 
\begin{equation}
	\d M_{T_i}^{(2)}(\w=\pi /2) = 
		a^2 \L_\chi^3 \left(t+t_v-2\,t_1^{W_-}+t_2^{W_-}\d_{T_i}\right) \,,
\end{equation}
while at zero twist, it reduces to that given in Ref.~\cite{Tiburzi:2005vy}. Just as in the nucleon case, the delta masses to $\mc{O}(\varepsilon^4)$ are also automatically $\mc{O}(a)$ improved. 

As is the case with the nucleons, to the order we work, the expressions for the delta mass corrections in tmHB$\chi$PT given in Eqs.~\eqref{eq:DLO} and~\eqref{eq:DNNLOtau1}, together with the untwisted continuum HB$\chi$PT expressions for the delta masses, provide the functional form for the dependence of the delta masses on the twist angle, $\w$, and the quark mass, $m_q$, which can be used to fit the lattice data.

%%%%%%%%%%%%%%%%%%%%%%%%%%%%%%%%%%%%%%%%%%%%%%%%%%%
%
%	Mass Splittings in the Isospin Limit
%
%%%%%%%%%%%%%%%%%%%%%%%%%%%%%%%%%%%%%%%%%%%%%%%%%%%
\subsection{\label{sec:split} \textbf{Mass Splittings}}

Having derived the expressions for the nucleon and delta masses in tmHB$\chi$PT to order $\mc{O}(\varepsilon^4)$ in the isospin limit, we now focus on the mass splittings between the nucleons and between the deltas. The mass contributions in the continuum, $M_{N_i}^{(n)}$ and $M_{T_i}^{(n)}$, clearly do not give rise to mass splittings for the nucleons and deltas, since they are calculated with degenerate quarks. Therefore, any mass splitting can only come from the mass corrections arising from tmLQCD. 

From the results of Sec.~\ref{sec:NMassIso} and Sec.~\ref{sec:DMassIso}, we find that to $\mc{O}(\varepsilon^4)$, the protons and neutrons remain degenerate, while the delta multiplet splits into two degenerate pairs, with $\D^{++}$ and $\D^-$ in one pair, and $\D^+$ and $\D^0$ in the other. The splitting between the degenerate pairs in the delta multiplet is given by  
\begin{equation}
	M_{\D^{+,0}} - M_{\D^{++,-}} = 
	\frac{4}{3}\,t_2^{W_-}a^2 \L_\chi^3\, \sin^2(\w) =
	\frac{4}{3}\,t_2^{W_-}a^2 \L_\chi^3 \,\frac{\mu^2}{m_q^2} \,.
\end{equation}
We reiterate here that the $\mc{O}(a)$ uncertainty inherent in the definition of the twist angle results in a correction to $M_{\D^{+,0}} - M_{\D^{++,-}}$ of $\mc{O}(a^3)\sim\mc{O}(\varepsilon^6)$, which is of higher order than we work. Hence to the accuracy we work, we may use $\w_0$ or any other
non-perturbatively determined twist angle for $\w$ above.

Just as the case of the pion mass splitting worked out in Ref.~\cite{Sharpe:2004ny}, this delta splitting must vanish quadratically in $a\mu = a m_q\sin(\w)$%
\footnote{Note: here $\mu$ is the twisted mass parameter, not to be confused with the renormalization scale.}%
~on general grounds, since the masses do not violate parity. One would therefore expect, naively, the splitting to be $\mc{O}(a^2m_q^2)\sim\mc{O}(\varepsilon^8)$. But as our results show, there is, in fact, a mass dependence in the denominator such that the effect is $\mc{O}(\varepsilon^4)$. Suppose we take $a^{-1} = 2$~GeV, then we would find a mass splitting 
\begin{equation}
	M_{\D^{+,0}} - M_{\D^{++,-}} \simeq 1.1\, t_2^{W_-}\ \text{GeV}.
\end{equation}
Using naive dimensional analysis, we expect $t_2^{W_-} \sim \mc{O}(1)$, giving a mass splitting of the delta pairs on the order of 1~GeV.  This is rather large, and in fact unexpected.  In fact, a recent study of the quenched tmLQCD spectrum found the splitting on the order of 50 to 100~MeV~\cite{Abdel-Rehim:2005gz}.  If however, we were to use $\L_{QCD}$ as the dimensionful parameter in Eqs.~\eqref{eq:wilsonSpurion} and \eqref{E:Tasq}, and guess some reasonable size of $\L_{QCD} \sim 500$~MeV, then we would have found the mass splitting to be $0.04 t_2^{W_-}$~GeV, and then naive dimensional analysis estimates for $t_2^{W_-}$ would place this LEC as $\mc{O}(1)$, in much better agreement with the twisted mass lattice data~\cite{Walker-Loud:2005bt}.  We can not make a direct comparison with the quenched data, as our computation is for full tmLQCD.  Nevertheless, the quenched spectrum is expected to be within 20\% of the full QCD result, and so it would be nice to determine this LEC, $t_2^{W_-}$.

Now, the degeneracies we found for the nucleons and the delta multiplet above hold not only at $\mc{O}(\varepsilon^4)$, but in fact they hold to all orders in tm$\chi$PT. This can be understood by considering the lattice Wilson-Dirac operator associated with the action of tmLQCD given in Eq.~(\ref{E:ActionTw}) in the isospin limit with $\tau_3$-twisting 
\begin{equation}
D_{WD} = \frac{1}{2} \sum_{\mu} \g_\mu (\nabla^\star_\mu + \nabla_\mu) 
        -\frac{r}{2} \sum_{\mu} \nabla^\star_\mu \nabla_\mu 
        + m_0 + i \gamma_5 \tau_3 \mu_0 \,,
\end{equation}
which has the self-adjointness property~\cite{Frezzotti:2003xj}
\begin{equation}
\t_1 \g_5\,D_{WD}\,\g_5 \t_1 = D_{WD}^\dagger \,.
\end{equation}
It follows then that the propagator for the upper and lower component of the quark doublet, $\psi_l(x)$, call them $S_u(x,y)$ and $S_d(x,y)$ respectively, satisfy the relations
\begin{equation}
	\g_5\,S_u(x,y)\,\g_5 = S_d^\dagger(y,x) \,,\qquad\qquad
	\g_5\,S_d(x,y)\,\g_5 = S_u^\dagger(y,x) \,.
\end{equation}
This means that any baryon two-point correlator which is invariant under the interchange of the quark states in the quark doublet combined with hermitian conjugation, leading to the degeneracies mentioned above.  An argument of this type has been given in Ref.~\cite{Abdel-Rehim:2005gz}.  

The same can also be shown in a chiral Lagrangian treatment, as must be the case. Now one of the symmetries of tmLQCD with two flavor-degenerate quarks and $\tau_3$-twisting is the pseudo-parity
transformation, $\mc{P}^1_F$, where ordinary parity is combined with a flavor exchange~\cite{Frezzotti:2003ni},
\begin{equation} \label{E:PF1}
	\mathcal{P}^1_F \colon
	\begin{cases}
		U_0(x) \rightarrow U_0(x_P) \,, \quad x_P = (-\mathbf{x},t) \\
		U_k(x) \rightarrow U^\dagger_k(x_P) \,, \quad k = 1,\,2,\,3 \\  
		\psi_l(x) \rightarrow i\tau_1 \gamma_0 \psi_l(x_P) \\
		\bar{\psi}_l(x) \rightarrow -i\bar{\psi}_l(x_P) \gamma_0 \tau_1
	\end{cases} \,,
\end{equation}
where $U_\mu$ are the lattice gauge fields. At the level of HB$\chi$PT, this is manifested as the invariance of the chiral Lagrangian under the transformations 
\begin{align} \label{E:chiP1f}
	p(x) \longleftrightarrow n(x_P) \quad,\quad 
	&\D^{++,-}(x) \longleftrightarrow \D^{-,++}(x_P) \quad,\quad 
	\D^{+,0}(x) \longleftrightarrow \D^{0,+}(x_P)\, , \nonumber\\
	&\otimes^{F}_{k=1}\mc{O}_k(x) \longrightarrow 
	\otimes^{F}_{k=1}\tau_1 \mc{O}_k(x_p)\tau_1 \,,
\end{align}    
where for an operator in the chiral Lagrangian, $\mc{O}_k$ is any operator matrix that contracts with the flavor indices of the the nucleon ($N$) or delta ($T_\mu$) fields in the operator.  If the $N$ or the $T_\mu$ fields contained in an operator have a total of 2F flavor indices,  $\otimes^{F}_{k=1}\mc{O}_k$ is the
tensor product of $F$ operator matrices which contract with the $F$ distinct pairs of these flavor indices. The degeneracies in the nucleons and the delta multiplets discussed above would then follow if all the operators in the chiral Lagrangian that contribute to the baryon masses have a structure that satisfies the condition 
\begin{equation} \label{E:symC}
	\otimes^{F}_{k=1}\mc{O}_k(x)
	= \otimes^{F}_{k=1}\tau_1\mc{O}_k(x_P)\tau_1 \,,
\end{equation}

Consider first the case for the nucleons. Since the nucleon fields are vectors in flavor space, we can take $F=1$ without loss of generality (the nucleon fields can only couple to one operator matrix). Since the chiral Lagrangian is built with just $\mc{M}_\pm$, $\mc{W}^{tw}_\pm$, $\mc{A}_\mu$, and $\mc{V}_\mu$, the operator matrix $\mc{O}_k$ can only be constructed from combinations of these fields. We need not consider combinations involving just $\mc{M}_+$ and $\mc{W}^{tw}_+$, since the flavor structure of both is trivial, i.e. proportional to the identity. We also need not consider mass contributions arising from pion loops, because they must have the same flavor structure as the tree level local counterterms used to cancel the divergences in these loops. Therefore, we do not have to consider operators involving $\mc{A}_\mu$ and $\mc{V}_\mu$, which give rise to mass contributions only through pion-nucleon interactions. This leaves us with only combinations involving $\mc{M}_-$ and $\mc{W}^{tw}_-$ as possible candidates to break the degeneracy in the nucleons. As was discussed in Sec.~\ref{sec:baryons}, because of the parity-flavor symmetry of tmLQCD, $\mc{O}_k$ can not contain just a single $\mc{M}_-$ or $\mc{W}^{tw}_-$, but must always have an even number from the set $\{\mc{M}_-,\mc{W}^{tw}_-\}$. Now any such combination would indeed have a pure tree level part, however, it is also proportional to the identity in flavor space. Thus there is no operator matrix, $\mc{O}_k$, that one can construct which violates the condition $\mc{O}_k(x) = \tau_1\mc{O}_k(x_P)\tau_1$.

The arguments for the case of the deltas runs similar to that for the nucleons. For the same reason given in the nucleon case, we need not consider operator structures that involve $\mc{M}_+$, $\mc{W}^{tw}_+$, $\mc{A}_\mu$, and $\mc{V}_\mu$. We need only consider operator structures involving an even number from the set $\{\mc{M}_-,\mc{W}^{tw}_-\}$. For the deltas, $F$ can be three since each delta field has three flavor indices. But since two of the $\mc{O}_k$ in $\otimes^{3}_{k=1}\mc{O}_k$ must come from the set $\{\mc{M}_-,\mc{W}^{tw}_-\}$ to satisfy the parity-flavor symmetry of tmLQCD, we can take $F$ to be at most two without loss of generality. Now each of $\mc{M}_-$ and $\mc{W}^{tw}_-$ has a tree level
part that is proportional to $\tau_3$, thus, under $\mc{P}^1_F$, $\mc{O}_1\otimes\mc{O}_2$ where $\mc{O}_k$ can be either $\mc{M}_-$ or $\mc{W}^{tw}_-$, satisfies the symmetry condition, Eq.~\eqref{E:symC}. Therefore, one can not construct operators for the deltas that break the degeneracy between the pairs in the delta multiplet.

%%%%%%%%%%%%%%%%%%%%%%%%%%%%%%%%%%%%%%%%%%%%%%%%%%%
%
%	N-D Masses with Isospin Breaking
%
%%%%%%%%%%%%%%%%%%%%%%%%%%%%%%%%%%%%%%%%%%%%%%%%%%%
\section{\label{sec:mass} Nucleon and Delta Masses Away from the Isospin Limit}   

In this section, we present results for mass corrections due to twisting away from the isospin limit, where the quarks are now mass non-degenerate. To the order we work, the corrections due to the mass splitting come in only at tree level. For clarity, we will only point out the change arising from the quark mass splitting; we will not repeat the discussion on the nucleon and delta masses that are the same both in and away from the isospin limit.

\subsection{\label{sec:Tdiag} \textbf{The Flavor-Diagonal Basis for the Mass Matrix at $\mathcal{O}(\varepsilon^4)$}}

The natural choice for splitting the quark doublet is to use the real and flavor-diagonal Pauli matrix, $\tau_3$, since the quark states one uses on the lattice correspond to the quarks in QCD in the continuum limit. But as was discussed in Sec.~\ref{sec:EContL} above, twisting can not be implemented with $\tau_3$ in this case (the fermionic determinant would be complex otherwise), and so $\tau_1$ is used instead.

Since the twist is implemented by a flavor nondiagonal Pauli matrix away from the isospin limit, flavor mixings are induced for non-zero twist: the quark states in tmLQCD are now linear combinations of the physical quarks of continuum QCD. At the level of the chiral effective theory, this manifests itself in that the hadronic states described by the tm$\chi$PT Lagrangian are linear combinations of the continuum QCD hadronic states we observe, viz. the pions, nucleons, deltas, etc. 

If the effects from twisting are perturbative as compared to the isospin breaking effects, the hadronic states described by tm$\chi$PT will be ``perturbatively close'' to their corresponding continuum QCD
states, i.e. the difference between them is small compared to the scales in the theory (see Appendix~\ref{sec:appDDM} for an explicit demonstration). In this case, we can still extract QCD observables
directly from tm$\chi$PT, as the corrections will be perturbative in the small expansion parameter. However, if the twisting effects are on the same order as the isospin breaking effects so that the flavor
mixings are large, these corrections will not be perturbative.%
\footnote{A qualitative guide to the size of the flavor mixings can be found in the ratios of two-point correlation functions. Define the ratio of QCD delta states by 
\begin{equation*}
	R_{ij} \equiv 
		\frac{\langle \D^i \ \D^j \rangle + \langle \D^j \ \D^i \rangle}
		{\langle \D^i \ \D^i \rangle + \langle \D^j \ \D^j \rangle} \,.
\end{equation*}
Flavor mixing should be small if the off diagonal elements of $R_{ij}$ are small. To determine the size of the flavor mixings quantitatively, one has to look at the splitting in the delta multiplet. We will discuss further in the text below.} 
Nevertheless, one can still extract information for the QCD observables: One can still measure the masses of these tmLQCD hadronic states in lattice simulations, and one can fit these to the analytic
expressions for these masses calculated in tm$\chi$PT to extract the values of the LECs. The LECs associated with the continuum $\chi$PT contributions have the same numerical values as in
tm$\chi$PT. Therefore, if one determines these from tmLQCD simulations, one knows the masses of the QCD hadronic states. 

At the order we work, flavor mixings are manifested in the appearance of flavor non-conserving pion-baryon vertices in the Feynman rules of tmHB$\chi$PT, and in that the baryon mass matrix is not
flavor-diagonal. Since we work in the physical pion basis where the twist is carried by the Wilson spurion (now flavor non-diagonal) instead of the mass spurion (now flavor-diagonal), flavor mixings can only arise from operators with one or more insertions of the Wilson spurion field. Because of this, the flavor non-conserving pion-baryon vertices and the non-diagonal terms in the mass matrix must be proportional to $a$, the lattice spacing, and so must vanish in the continuum limit where the effects of the twist are fake and can be removed by a suitable chiral change of variables~\cite{Frezzotti:2000nk,Frezzotti:1999vv}.%
\footnote{This shows again the convenience of the pion physical basis, where all the effects of symmetry breaking in the lattice theory are parametrized and contained in the Wilson spurion fields, which then vanish as the symmetries are restored in the continuum limit.} 

For the nucleons, flavor mixings induce only unphysical flavor non-conserving pion-nucleon vertices which vanish in the continuum limit; the nucleon matrix is still flavor diagonal at the order we work. In fact, this is true to all orders in tmHB$\chi$PT. The reason is the same as that given in Sec.~\ref{sec:split}. We need only consider the tree level part of the possible operator structures that one can construct from the spurion fields in tmHB$\chi$PT. Now the only spurion field that has a tree level part with non-diagonal
flavor structure is $\mc{W}^{tw}_-$, and as we discussed above, it must be paired either with another $\mc{W}^{tw}_-$ or with $\mc{M}_-$, which renders the flavor structure of the tree level part of the combination trivial. Thus we may take the basis of nucleons used in the tmHB$\chi$PT Lagrangian as the physical nucleon basis.  

For the deltas, not only are there flavor non-conserving pion-delta vertices, at the order we work, the delta mass matrix is already flavor nondiagonal at tree level. This happens for the deltas because the tensor nature of the $T_\mu$ field allows more freedom in the way the flavor structure of the delta operator can be constructed. Thus, in order to have only physical tree level mass terms for the deltas, we must change to a basis where the delta mass matrix is diagonal, which can now only be done order by order. 

When diagonalizing the delta mass matrix, we need, in principle, to diagonalize the mass matrix that contains all the mass contributions from both tree and loop level to the order that one works. But we find
the difference between diagonalizing the delta mass matrix including both tree and loop level contributions at the order we work, and diagonalizing that with only the tree level mass contributions, give rise to corrections only to the loop level mass contributions, which are higher order than we work. Thus, we will diagonalize the delta mass matrix containing just the tree level mass terms in our
calculation for the delta masses. 

To the order we work, if the tree level mass is given by 
\begin{equation} \label{E:Tbasis}
v_{\bar{\Delta}} M_\Delta v_\Delta \,, \qquad
v_{\bar{\Delta}} = 
\begin{pmatrix}
\bar{\Delta}^{++} & \bar{\Delta}^0 & \bar{\Delta}^+ & \bar{\Delta}^-
\end{pmatrix} \,, \qquad
v_\Delta = 
\begin{pmatrix}
\Delta^{++} & \Delta^0 & \Delta^+ & \Delta^-
\end{pmatrix}^T \,, 
\end{equation}
where $v_{\bar{\Delta}}$ and $v_\Delta$ are vectors of the delta
basis states used in the tmHB$\chi$PT Lagrangian, and $M_\Delta$ is
the tree level mass matrix, the physical delta basis is given by  
\begin{equation} \label{E:newT}
v'_\Delta = S^{-1} \cdot v_\Delta \,,\qquad 
v'_{\bar{\Delta}} = v_{\bar{\Delta}} \cdot S \,, 
\end{equation}
where $S$ is the matrix of eigenvectors of $M_\Delta$ such that
\begin{equation}
S \cdot M_\Delta \cdot S^{-1} = \mathcal{D} \,,
\end{equation}
with $\mc{D}$ the corresponding diagonal eigenvalue matrix. This
implies that
\begin{equation}
v_{\bar{\Delta}} M_\Delta v_\Delta =
(v'_{\bar{\Delta}} \cdot S^{-1}) \cdot
(S \cdot \mc{D} \cdot S^{-1}) \cdot (S \cdot v'_\Delta) = 
v'_{\bar{\Delta}} \mc{D} \, v'_\Delta \,.
\end{equation} 
The full details of the diagonalization are provided in Appendix~\ref{sec:appDDM}. In the following sections, we will work in this basis for calculating the delta masses.

The Feynman rules in the new basis are obtained from the same tmHB$\chi$PT Lagrangian given above in Sec.~\ref{sec:baryons} but with each of the delta flavor states now rewritten in terms of the new delta flavor states given by the defining relations Eq.~\eqref{E:newT}. Note that changing to the new delta basis induces new unphysical flavor non-conserving vertices in the delta interaction terms given in \eqref{eq:NNpi}, because in terms of the new basis states, flavors are mixed. However, these flavor mixing components are proportional to the off-diagonal elements of $S$, which are proportional to the lattice spacing as well as the twist angle (see Appendix~\ref{sec:appDDM}). Thus they vanish in the limit of vanishing twist or lattice spacing, and so the unphysical vertices arising from them also vanish in these limits. 

We note and reiterate here that in the isospin limit, this order by order mass matrix diagonalization is unnecessary as one can always rotate to a basis where the twist is flavor-diagonal from the outset, and issues of flavor nonconserving vertices and non-diagonal mass matrices due to flavor mixings do not arise.%
\footnote{In fact, as is shown in Appendix~\ref{sec:appDDM}, if one insists on remaining in the basis where the twist in flavor non-diagonal, one would find that the unphysical terms arising from flavor mixings do not vanish in the continuum limit.} 

%%%%%%%%%%%%%%%%%%%%%%%%%%%%%%%%%%%%%%%%%%%%%%%%%%%
%
%	Nucleons with Isospin Breaking
%
%%%%%%%%%%%%%%%%%%%%%%%%%%%%%%%%%%%%%%%%%%%%%%%%%%%
\subsection{\label{sec:tmNMass} \textbf{The Nucleon Masses}}

Away from the isospin limit, the first change caused by the mass splitting  occurs in the continuum mass contribution $M_{N_i}^{(1)}$, since the quark masses  
\begin{equation}
	m_u = m_q - \epsilon_q \,,\qquad\qquad m_d = m_q + \epsilon_q 
	\,,\qquad\qquad m_q\;,\epsilon_q > 0 \,,
\end{equation}  
are no longer equal.

At the order we work, the only other change due to the mass splitting appears at $\mc{O}(\varepsilon^4)$ in the contribution from the $\mc{O}(a\mathsf{m})$ nucleon operator with coefficient $n^{WM_+}$ given in Eq.~(\ref{E:Namq}). In the isospin limit, its contribution to $\d M_{N}^{(2)}(\w)$ is proportional to $m_q$, but away from the isospin limit, it becomes 
\begin{equation} \label{E:Nchange}
	2\,n_1^{WM_+}a\,\Lambda_\chi\,m_q\cos(\w) \longrightarrow
	2\,n_1^{WM_+}a\,\Lambda_\chi\,m_i\cos(\w) \,,
\end{equation} 
where $m_i$ is defined in Eq.~\eqref{eq:mB}.  The corrections to the nucleon masses from the effects of lattice discretization and twisting are otherwise the same as those given in Eqs.~\eqref{E:NBLO} and~\eqref{eq:N2Mass}.  Note that the nucleon masses are automatically $\mc{O}(a)$ improved, just as in the isospin limit.

%%%%%%%%%%%%%%%%%%%%%%%%%%%%%%%%%%%%%%%%%%%%%%%%%%%
%
%	Deltas with Isospin Breaking
%
%%%%%%%%%%%%%%%%%%%%%%%%%%%%%%%%%%%%%%%%%%%%%%%%%%%
\subsection{\label{sec:tmDmass} \textbf{The Delta Masses}} 

Away from the isospin limit ($\epsilon_q \neq 0$), we calculate the
delta mass and mass corrections in the basis where the delta mass
matrix is diagonal to the order we work. This diagonalization is
worked out in Appendix~\ref{sec:appDDM}, where we have obtained general
expressions for the new delta basis that are valid in the range from
$\epsilon_q = 0$ to $\epsilon_q \sim a\L_{\rm QCD}^2$. Here, we present
the case where $\epsilon_q > 0$ and $\epsilon_q \sim m_q \sim a\L_{\rm
  QCD}^2 \gg a^2\L_{\rm QCD}^3$, which is a regime that simulations
in the near future can probe. To the order we work, we may take the
new delta basis states in this regime to be   
\begin{align}
	T'_1 \leftrightarrow | \D_1 \rangle &= \mc{C}_1 \left[ 
		| \D^{++} \rangle + \frac{\sqrt{3} B}{4 A} | \D^0 \rangle \right]\, , \nonumber\\
	T'_3 \leftrightarrow | \D_3 \rangle &= \mc{C}_3 \left[ 
		\left( 1 + \frac{B}{4A} \right) | \D^{0} \rangle 
		- \frac{\sqrt{3} B}{4 A} | \D^{++} \rangle \right] \,, \nonumber\\
	T'_2 \leftrightarrow | \D_2 \rangle &= \mc{C}_2 \left[ 
		| \D^{+} \rangle + \frac{\sqrt{3} B}{4 A} | \D^- \rangle \right]\,, \nonumber\\
	T'_4 \leftrightarrow | \D_4 \rangle &= \mc{C}_4 \left[ 
		\left( 1 - \frac{B}{4A} \right) | \D^{-} \rangle - \frac{\sqrt{3} B}{4 A} | \D^+ \rangle \right] \,,
\end{align}
where $T'_i = \D_i$ denote the deltas in the new basis, $\mc{C}_i$
are normalization factors, and 
\begin{equation}\label{E:AandB}
	A = 2\,\e_q\, \left( \g_M + t_1^{WM_+} 
		a \L_\chi\, \cos(\w) \right) \,,\qquad 
	B = t_2^{W_-}a^2\L_\chi^3\sin^2(\w) \,.
\end{equation}
Note that $A\sim\mc{O}(\varepsilon^2)$ and $B\sim\mc{O}(\varepsilon^4)$ in our power counting, so $B/A \sim \mc{O}(\varepsilon^2)$ and the effects of the flavor mixings are perturbative.

The masses of these states are comprised of the continuum expressions given in Chapter~\ref{chap:BMasses}, Ref.~\cite{Tiburzi:2005na}, and corrections due to the effects of discretization and twisting. Note the continuum expressions for the delta masses here are necessarily changed from that in the isospin limit because $m_u \neq m_d$. The mass corrections due to the effects of lattice discretization and twisting, come in at tree level; the loop contributions remain unchanged from that in the isospin limit. The tree level mass contributions to the order we work have been worked out in  Eq.~\eqref{E:Dmtx}, in the process of diagonalization. We list
here the full delta mass corrections to $\mc{O}(\varepsilon^4)$, which we denote by $\d M_{T_i}$, to the mass of the delta state denoted by $T^\prime$: 
\begin{align} \label{E:TMassC}
\d M_{T_i}(\w) =&\  
	-4\,\ol{\s}_W\,a\,\L_\chi^2\cos(\w)	
	+12\,\ol{\s}_W\, a\,m_\pi^2 \log\left(\frac{m_\pi^2}{\mu^2}\right)\cos(\w) \nonumber\\
	&+4\,g_{\D N}^2\,(\ol{\s}_W -\s_W)\, a\,\mc{J}(m_\pi,-\D,\mu)\cos(\w) \nonumber\\
	&+2\,t_1^{WM_+}a\L_\chi\, \frac{m^\prime_{T_i}}{3}\cos(\w) 
	+4\,t_2^{WM_+}\, a\L_\chi\, m_q\cos(\w) 
	+a^2\L_\chi^3(t +t_v) \nonumber\\
	&+a^2\L_\chi^3\left(4\,t_1^{W_+}\cos^2(\w)-2\,t_1^{W_-}\sin^2(\w)
	+t_2^{W_-}\d'_{T_i}\, \sin^2(\w)\right)\, ,
\end{align}
for $T = 1 \dots 4$ and 
\begin{equation}
	m^\prime_{T_i} = 
		\begin{cases}
		3m_u      &\text{for $i=1$} \\
		2m_u+m_d  &\text{for $i=2$} \\
		 m_u+2m_d &\text{for $i=3$} \\
		3m_d      &\text{for $i=4$}
	\end{cases} \;,\qquad\qquad
	\d^\prime_{T_i} = 
		\begin{cases}
		\ \ 0        &\text{for $i=1\,,\,4$} \\
		-\frac{2}{3} &\text{for $i=2\,,\,3$} \\
	\end{cases} \;.
\end{equation}
Note that $\d M_{T_i}(\w)$ as given in Eq.~\eqref{E:TMassC}, is the same as the sum of $\d M_{T_i}^{(1)}$ and $\d M_{T_i}^{(2)}$ as given in Eqs.~\eqref{eq:DLO} and~\eqref{eq:DNNLOtau1} respectively, but with
the changes 
\begin{equation} \label{E:Tchange}
	2\,t_1^{WM_+}\, a\L_\chi\, m_q\cos(\w)
	\longrightarrow 
	2\,t_1^{WM_+} a\L_\chi\, \frac{m^\prime_{T_i}}{3}\cos(\w)
	\,, \qquad
	\d_{T_i} \longrightarrow \d^\prime_{T_i} \,.
\end{equation} 

The full expressions for the delta masses can be obtained when the continuum contributions are included. To the order we work, one can obtain the complete mass expression for delta denoted by $T^\prime$ to $\mc{O}(\varepsilon^4)$ in tmHB$\chi$PT by adding the mass corrections, $\d M_{T_i}(\w)$, to the continuum mass of the delta denoted by $T$, whose expression can be found in Section~\ref{sec:DeltaMasses}.  

We stress here that one can not take the isospin limit from any of the expressions give above in this subsection. They have been derived for $\epsilon_q \neq 0$ and with the assumption that the twisting effects are much smaller than the isospin breaking effects. One must use the general formulae given in Eqs.~\eqref{E:Smtx} and \eqref{E:Dmtx} when considering cases where these conditions are not true.

Observe that away from the isospin limit, the delta masses are also automatically $\mc{O}(a)$ improved at maximal twist ($\w = \pi/2$), as all terms proportional to $a$ in $\d M$ are proportional to $\cos(\w)$ as well. Hence, to the order we work, the contributions due to the isospin breaking are the same as that in the continuum at maximal twist.

%%%%%%%%%%%%%%%%%%%%%%%%%%%%%%%%%%%%%%%%%%%%%%%%%%%
%
%	Summary
%
%%%%%%%%%%%%%%%%%%%%%%%%%%%%%%%%%%%%%%%%%%%%%%%%%%%
\section{\label{sec:conc} Summary}

In this chapter we have studied the mass spectrum of the nucleons and the deltas in tmLQCD with mass non-degenerate quarks using effective field theory methods. We have extended heavy baryon chiral
perturbation theory for $SU(2)$ to include the effects of the twisted mass, and we have done so to $\mc{O}(\varepsilon^4)$ in our power counting, which includes operators of $\mc{O}(a\mathsf{m}^2,ap^2,a^2)$. Using the resulting tmHB$\chi$PT, we have calculated the nucleon and the delta masses to $\mc{O}(\varepsilon^4)$, and we found them to be automatically $\mc{O}(a)$ improved as expected from the properties of tmLQCD.

Because of the twisting, the vacuum is no longer aligned with the identity in flavor space, which has non-trivial effects on the physical excitations (pions) of the theory. Also, depending on whether the quarks are mass degenerate or not, the way twisting is implemented determines what the physical baryon states are in the theory. We have highlighted these subtleties when doing calculations in tmHB$\chi$PT. 

In order for the pions in the theory to be physical, we have to make a particular (non-anomalous) chiral change of variables to undo the twisting effects. This requires the knowledge of the twisting angle, but once that is determined, the physical pion basis can be determined {\it \`a priori}. However, whether or not the nucleons and deltas are physical must still be determined from the theory. In the isospin limit, both the nucleon and the delta mass matrices are diagonal, and so the nucleon and delta states contained in the $N$ and $T_\mu$ fields are physical. However, away from the isospin limit, only the
nucleon mass matrix remains diagonal. Thus, the $N$ field can still be regarded as physical, but the physical deltas are now linear combinations of the flavor states contained in the $T_\mu$ field. This
can be understood from the fact the at the quark level, the physical QCD states, the $u$ and $d$ quarks, are eigenstates of $\tau_3$ but not of $\tau_1$. So only in the isospin limit, where the twist can always be implemented by the flavor-diagonal Pauli matrix, $\tau_3$, are the states contained in the quark doublet physical quarks. Away from the isospin limit, the twist can not be implemented by $\tau_3$ anymore, and the eigenstates of the Hamiltonian of the theory are composed of linear combinations of the $u$ and $d$ quarks. 

The physical states in tm$\chi$PT are in general a mixture of those in the untwisted $\chi$PT. The size of the mixture is determined by the relative sizes of the discretization effects, which are $\mc{O}(a^2)$,
and the isospin splitting effects, which are $\mc{O}(\epsilon_q)$. In this chapter, we have given general expressions for the nucleon and delta masses with respect to this mixing of states that are valid in the
range from $\epsilon_q = 0$ to $\epsilon_q \sim a\L_{\rm QCD}$.

The quantities which provided the motivation for this work and turned out to be most interesting are the mass splittings between the nucleons and between the deltas. We found that in the isospin limit, the nucleon masses do not split to any order in tm$\chi$PT, while the delta multiplet splits into two degenerate pairs. This can be understood from the symmetries of tmLQCD at the quark level, and as we
have shown, also at the level of tm$\chi$PT. The mass splitting between the multiplets, $M_{\D^{+,0}} - M_{\D^{++,-}}$, first arises from a tree level contribution at $\mc{O}(a^2)$, and it gives an indication of
the size of the flavor breaking in tmLQCD. This splitting in the delta multiplet will be easier to calculate in lattice simulations than the corresponding quantity $m_{\pi_3}^2 - m_{\pi_{1,2}}^2$ in the meson
sector~\cite{Sharpe:2004ny}, since it involves no quark disconnected diagrams. 

Twisted mass HB$\chi$PT can also be extended to partially quenched theories (such extension of tm$\chi$PT for pions has recently been done~\cite{Munster:2004dj}).  This will be useful as unquenched twisted mass lattice QCD simulations are underway~\cite{Farchioni:2005bh}.

% ========== Chapter 4: Nucleon Polarizabilities
\chapter{Electromagnetic Polarizabilities of the Nucleon for lattice QCD}\label{chap:Polarize}

The work in this chapter is based upon Ref.~\cite{Detmold:2006vu}
%
%
%
%
%       Introduction
%
%

\section{Introduction}

Compton scattering at low energies is an invaluable tool with which to study the electromagnetic structure of hadrons. At very low photon energies, the Compton amplitude is dominated by point-like photon scattering from the total charge and magnetic moment of the target hadrons. As the frequency increases, contributions beyond point-like scattering enter and one begins to resolve the hadronic response to an applied electromagnetic field.  For unpolarized scattering on spin one-half objects, the first structure dependent contributions in this energy expansion of the amplitude are the electric polarisability, $\alpha$, and the magnetic polarisability, $\beta$. These quantities reflect the ability of the hadron's components to align or anti-align themselves in response to an applied electric or magnetic field. For the proton and neutron, the positivity of the accepted experimental values of these polarizabilities ($\alpha_p=12.0\pm0.6$, $\beta_p=1.9\mp0.6$, $\alpha_n=12.5\pm1.7$ and $ beta_n=2.7\mp1.8$ in units of $10^{-4}$ fm$^3$~\cite{Schumacher:2005an}) indicates that both nucleons are diamagnetic objects.  Recent experimental advances~\cite{Hyde-Wright:2004gh,Schumacher:2005an} have also allowed the extraction of certain combinations of target polarisation-dependent observables in Compton scattering. These involve the spin polarizabilities~\cite{Ragusa:1993rm}, conventionally labeled $\gamma_1$--$\gamma_4$, and they have consequently been investigated in numerous theoretical and further experimental studies. Although the classical interpretation of spin-dependent Compton scattering is less clear, the spin polarizabilities encode additional fundamental properties of the nucleon.  Compton scattering observables, however, are not limited to these six parameters. Higher order quasi-static properties of the nucleon appear from further terms in the energy expansion of the amplitude. These higher-order polarizabilities \cite{Holstein:1999uu}, as well as generalized polarizabilities \cite{Guichon:1995pu} (which arise in the singly (doubly) virtual Compton scattering process, $\gamma^* X \to \gamma^{(*)} X$) allow for an even finer resolution of the electromagnetic structure of hadrons at low energies.

While experimentally one is hoping to open further windows through
which to view hadronic electromagnetic structure, theoretically one
ultimately hopes to understand how hadronic polarizabilities arise
from the basic electromagnetic interaction of the photon with quarks
that are bound to form the hadrons. The electric and magnetic
polarizabilities should na{\"\i}vely scale with the volume of the
hadron. However, this expectation overestimates the observed
polarizabilities by four orders of magnitude, indicating that the
nucleon's constituents are strongly coupled. Lattice techniques
provide a method to investigate the non-perturbative structure of
hadrons directly from QCD.  In particular, the various hadron
polarizabilities can be computed.  Comparison of these results with
experimental determinations would provide stringent tests of the
lattice method's ability to reproduce the structure of physical
hadronic states; for the individual spin polarizabilities that have
not been measured, the lattice approach may be the only way to
determine them. On the lattice, direct calculations of the required
hadronic current-current correlators are difficult and so far have not
been attempted. However significant progress has been made
\cite{Fiebig:1988en,Christensen:2004ca,Lee:2005dq} in extracting the
electric and magnetic polarizabilities by performing quenched lattice
calculations in constant background electric and magnetic fields
respectively and studying the quadratic shift in the hadron mass that
is induced (essentially an application of the Feynman-Hellman
theorem). These studies have investigated the electric
polarizabilities of various neutral hadrons (in particular, the
uncharged vector mesons and uncharged octet and decuplet baryons), and
the magnetic polarizabilities of the baryon octet and decuplet, as
well as those of the non-singlet pseudo-scalar and vector mesons.  As
we shall discuss below, generalizations of these methods using
non-constant fields allow the extraction of the spin polarizabilities
from spin-dependent correlation functions and also allow the electric
polarizabilities to be determined for charged hadrons. More generally,
higher-order polarizabilities and generalized polarizabilities are
accessible using this technique.

As with all current lattice results, these calculations have a number
of limitations and so are not physical predictions that can be
directly compared to experiment. For the foreseeable future, lattice
QCD calculations will necessarily use quark masses that are larger
than those in nature because of limitations in the available
algorithms and computational power. Additionally, the volumes and
lattice-spacings used in these calculations will always be finite and
non-vanishing, respectively.  For sufficiently small masses and large
volumes, the effects of these approximations can be investigated
systematically using the effective field theory of the low energy
dynamics of QCD, chiral perturbation theory ($\chi$PT)
\cite{Weinberg:1966kf,Gasser:1983yg,Gasser:1984gg}.%
\footnote{The effects of the lattice discretization are short distance in nature, and while some of them can be analyzed in an extension of the effective field theory described here~\cite{Sharpe:1998xm,Rupak:2002sm,Bar:2003mh,Beane:2003xv,Tiburzi:2005vy,Tiburzi:2005is}. Here we will assume that a continuum extrapolation has been performed.}
In this chapter we shall perform an analysis of the nucleon electromagnetic and spin polarizabilities
at next-to-leading order (NLO) in the chiral expansion.  We do so to discuss the infrared effects of the quark masses and finite volume in two-flavour QCD and its quenched and partially-quenched analogues (QQCD and PQQCD).  The polarizabilities of the hadrons are particularly interesting in this regard since they are very sensitive to infrared physics and their quark mass and finite volume dependence is considerably stronger than that expected for hadron masses and magnetic moments.  This should be physically evident given that the polarizabilities scale with the volume.  In essence, chiral
perturbation theory provides a model independent analysis of the modification of the nucleon's pion cloud in a finite volume. When the charged pion cloud is influenced by to the periodic boundary conditions imposed on the lattice, the nucleon's response to external electromagnetic fields is altered compared to that at infinite volume, and in most cases the effects are dramatic.  A particularly striking
oddity that we find in this analysis is a modification of the Thomson cross section at finite volume.  This can be explained through the physics of chiral loop corrections to point-like hadron structure.

If future lattice QCD simulations are to provide physical predictions
for the electromagnetic and spin polarizabilities, careful attention
must be paid to both the chiral and infinite volume extrapolations. To
illustrate this point, we present our results at representative values
of the quark mass, finding significant effects.  We also use our
quenched chiral perturbation theory results to assess the volume
dependence of the quenched data at the lightest pion masses used in
Refs.~\cite{Christensen:2004ca,Lee:2005dq}.  While the quenched theory
contains unphysical low energy constants (LECs) and the convergence of
the chiral expansion is questionable at these pion masses, we can
still provide an estimate of the volume dependence of quenched data
for the nucleon polarizabilities using our results.  Such an estimate
is achievable because the corresponding polarizabilities in the
unquenched theory do not depend on phenomenologically undetermined
LECs at the order of the chiral expansion to which we work.  At the
lightest quark masses used in the existing quenched lattice
simulations, $m_\pi\sim0.5$~GeV, we find strong sensitivity to the
lattice volume (as large as 10\%).  The effects will only increase as
the pion mass is brought closer to that in nature.  Clearly careful
chiral and volume extrapolations of polarizabilities are mandated to
connect lattice calculations to real world QCD.

To begin our investigation of nucleon polarizabilities in lattice QCD, we discuss in Sec.~\ref{sec:gN_to_gN} the kinematics of Compton scattering and define the electromagnetic and spin polarizabilities that are the primary focus of this work.  In Sec.~\ref{sec:Compton}, we perform a general analysis of the external field method pertaining to all electromagnetic and spin polarizabilities.  We discuss how suitable background fields can be used in lattice QCD simulations to determine the spin polarizabilities and, more generally, generalized polarizabilities (though we limit our discussion of these in the present chapter). Following this we introduce the low-energy effective theories of QCD ($\chi$PT), quenched QCD (Q$\chi$PT) and partially-quenched QCD (PQ$\chi$PT). These effective theories provide the model independent input necessary for calculating the quark mass and lattice volume dependence of polarizabilities. We focus primarily on PQ$\chi$PT in Sec.~\ref{sec:hbxpt}, discussing the relation to $\chi$PT where relevant, and relegating the peculiarities of Q$\chi$PT to Appendix~\ref{A4}.  Our results for the dependence of the nucleon polarizabilities on quark masses and the lattice volume are presented in Sec.~\ref{sec:pols}.  We provide detailed plots relevant for full QCD simulations of polarizabilities showing the dependence on quark masses and lattice volumes. We also estimate the quenched QCD volume dependences of the polarizabilities at a pion mass typical of existing quenched lattice data.  A glossary of finite volume functions required to evaluate the polarizabilities in a periodic box appears in Appendix~\ref{FV_app}.  Lastly, Sec.~\ref{sec:discussion} consists of a concluding discussion of our results.

%
%
%
%
%
%
%       Compton Scattering
%
%
%
%
%
%
\section{Nucleon Compton Scattering and Electromagnetic Polarizabilities}
\label{sec:gN_to_gN}

The real Compton scattering amplitude describing the elastic
scattering of a photon on a spin-half target such as the proton or
neutron can be parameterized as
\begin{align}
T_{\gamma N} =&\  A_1(\omega,\theta)\, \vepsprime\cdot\veps
	+A_2(\omega,\theta)\, \vepsprime\cdot \hat{k} \, \veps\cdot \hat{k'}
	+i\,A_3(\omega,\theta)\, \vsigma\cdot (\vepsprime\times\veps) \nonumber\\
	&+i\,A_4(\omega,\theta)\, \vsigma\cdot (\hat{k'}\times\hat{k})\, \vepsprime\cdot\veps 
	+i\,A_5(\omega,\theta)\, \vsigma\cdot \left[(\vepsprime\times\hat{k})\,
		\veps\cdot\hat{k'} -(\veps\times\hat{k'})\, \vepsprime\cdot\hat{k}\right] \nonumber\\
	&+i\,A_6(\omega,\theta)\, \vsigma\cdot \left[(\vepsprime\times\hat{k'})\,
		\veps\cdot\hat{k'} -(\veps\times\hat{k})\, \vepsprime\cdot\hat{k}\right],
\label{eq:Ti}
\end{align}
where we have chosen to work in the Breit frame of the system%
\footnote{The Breit frame is not actually a reference frame, but a convenient choice of coordinates in momentum space for a fixed momentum transfer, such that the initial hadron momentum is given by $p_{i}=p-\frac{q}{2}$ and the final momentum by $p_{f}=p+\frac{q}{2}$.  Here, $p$ is the average of the initial and final momentum of the hadron and $p^2 = m^2 -\frac{q^2}{4}$, which highlights the fact that this is not a reference frame.  The total momentum transfer to the hadron is $q$.}
and the incoming and outgoing photons have momenta $k=(\omega,\vec{k}=\omega\,\hat{k})$ and $k^\prime=(\omega,\vec{k}^\prime=\omega\,\hat{k}^\prime)$, and polarisation vectors $\epsilon$ and $\epsilon^\prime$, respectively. The $A_i(\omega,\theta)$, $i=1\ldots 6$, are scalar functions of the photon energy and scattering angle, $\cos\theta=\hat{k}\cdot\hat{k}^\prime$.  It is convenient to work in Coulomb gauge throughout where $\epsilon_0=\epsilon_0^\prime=0$ (the physical amplitudes are gauge invariant).

The functions, $A_i$, determining the Compton amplitude can be separated into a number of pieces. The Born terms, or tree level graphs, describe the interaction of the photon with a point-like target with mass, $M_N$, charge, $e\, Z$ (where $e>0$), and magnetic moment, $\mu$.  These terms reproduce the Thomson-limit (the zero frequency limit) and quadratic frequency pieces \cite{RaiChoudhury68} of unpolarized scattering as well as the Low--Gell-Mann--Goldberger (LGMG) low energy theorems \cite{Gell-Mann:1954kc,Low:1954kd} for spin-dependent scattering.%
\footnote{The LGMG theorems relate for example, the two-photon-nucleon interactions which arise from two successive photon insertions to the square of the single photon insertion.  These are the one-nucleon reducible graphs, or the one-particle reducible graphs in quantum field theory.  In Figure~\ref{fig:EFT}, the third diagram is related to the square of the second diagram, while the fourth diagram is a new contribution, which gives rise to the nucleon polarizabilities.  To be completely explicit, at the second order in the nucleon-photon interactions, one contribution comes from the square of the nucleon magnetic dipole moment, which is related by the LGMG theorems to the square of the single magnetic dipole interaction, while the spin-dependent polarizabilities are a qualitatively new feature of the nucleon electromagnetic structure, and given by the interaction of the photons with the pion-cloud of the nucleon.}
The remaining parts of the amplitude describe the structural response of the target.  Expanding the amplitude for small photon energies relative to the target mass and keeping terms to ${\cal O}(\omega^3)$ one can write
\begin{eqnarray}
A_1(\omega,\theta) &=& -Z^2 \frac{e^2}{M_N}+\frac{e^2}{4M_N^3}
\left(\mu^2(1+\cos\theta)-Z^2\right)(1-\cos\theta)\,\omega^2
+ 4\pi(\alpha + \beta \, \cos\theta)\omega^2+
{\cal O}(\omega^4)\,,
\nonumber\\ 
A_2(\omega,\theta) &=& \frac{e^2}{4M_N^3} (\mu^2-Z^2) \omega^2
\cos\theta -4\pi\beta \omega^2 \,+ 
{\cal O}(\omega^4)\,,
\nonumber\\ 
A_3(\omega,\theta) &=&   \frac{e^2 \omega}{
2M_N^2}\left(Z(2\mu-Z)-\mu^2 \cos\theta\right) 
+ 4\pi\omega^3(\gamma_1 - (\gamma_2 + 2 \gamma_4) \, \cos\theta)+
{\cal O}(\omega^5)\,,
\nonumber\\ 
A_4(\omega,\theta) &=&  -\frac{e^2 \omega }{ 2
M_N^2 }\mu^2 + 4\pi\omega^3 \gamma_2 +
{\cal O}(\omega^5)\,,
\nonumber\\ 
A_5(\omega,\theta) &=&  \frac{e^2 \omega }{2 M_N^2 }\mu^2 +
4\pi\omega^3\gamma_4 + {\cal O}(\omega^5) \,,
\nonumber\\ 
A_6(\omega,\theta) &=&  -\frac{e^2 \omega}{2 M_N^2 } Z\mu +
4\pi\omega^3\gamma_3 + {\cal O}(\omega^5)\,,
\label{eq:amplitudes}
\end{eqnarray}
describing the target structure in terms of the electric, magnetic and
four spin polarizabilities, $\alpha$, $\beta$, and
$\gamma_{1\mbox{--}4}$, respectively. In the conventions above, the
spin polarizabilities receive contributions from the anomalous decay
$\pi^0\to\gamma\gamma$ (shown in Fig.~\ref{fig:anom} below). This
contribution varies rapidly with energy and is omitted from the
polarizabilities in some conventions. Higher order terms in the energy
expansion can be parameterized in terms of higher-order
polarizabilities \cite{Holstein:1999uu}. The more general process of
virtual (and doubly-virtual) Compton scattering at low energies can
similarly be described in terms of generalized
polarizabilities~\cite{Guichon:1995pu}. We will focus in the six
polarizabilities defined above.

The goal of this chapter is to determine the quark mass and volume
dependence of the polarizabilities defined above to allow accurate
extraction of their physical values from lattice calculations.  Before
we do this we shall discuss how these lattice calculations may be
implemented.

%
%
%
%
%
%
%       Lattice Compton Scattering
%
%
%
%
%
%
\section{Compton Scattering and Polarizabilities on the Lattice}
\label{sec:Compton}

Lattice QCD provides a way to study the polarizabilities of hadrons
from first principles. There are two ways to do this. The method most
reminiscent of the experimental situation is to study the (Euclidean
space) four point Green function defining the Compton scattering
tensor directly (the photon fields are amputated). By measuring the
large Euclidean time behaviour of this correlator, the hadron matrix
elements of the two vector currents can be extracted. In principle, by
calculating particular Lorentz components of the Compton tensor with
various different source and sink spin states, all six electromagnetic
and spin polarizabilities and their higher order and generalized
analogues can be extracted.  However, this is a complicated task,
requiring the evaluation a large number of quark propagator
contractions resulting from quark-line disconnected diagrams which are
statistically difficult to determine.  At present this approach is too
demanding for the available computational resources and has not been
attempted.

The second method is based on measuring the response of hadronic
states to fixed external fields. A number of exploratory quenched QCD
studies have been performed in this approach. The pioneering
calculations of
Refs.~\cite{Fucito:1982ff,Martinelli:1982cb,Bernard:1982yu,Aoki:1989rx,Aoki:1990ix,Shintani:2005du}
attempted to measure the nucleon axial couplings, magnetic moments and
electric dipole moments by measuring the linear shift in the hadron
energy as a function of an applied external weak or electromagnetic
field.  As discussed in the Introduction, various groups
\cite{Fiebig:1988en,Christensen:2004ca,Lee:2005dq}
have also used this approach to extract electric and magnetic
polarizabilities in quenched QCD by measuring a quadratic shift in the
hadron energy in external electric and magnetic fields. The method is
not limited to electroweak external fields and can be used to extract
many matrix elements such as those that determine the moments of
parton distributions and the total quark contribution to the spin of
the proton \cite{Detmold:2004kw}. Here we focus on the electromagnetic
case.

The Euclidean space ($x_4 \equiv \tau$) effective action describing the gauge and parity
invariant interactions of a non-relativistic spin-half hadron of mass
$M$ and charge $q$ with a classical U(1) gauge field,
$A^\mu(\vec{x},\tau)$, is
\begin{eqnarray}
  \label{eq:eff_act}
  S_{\rm eff}[A]=\int d^3 x\,d\tau\, {\cal L}_{\rm eff}(\vec{x},\tau;A)\,,
\end{eqnarray}
for the Lagrangian
\begin{multline}
  \label{eq:eff_L}
  {\cal L}_{\rm eff}(\vec{x},\tau;A) = \Psi^\dagger(\vec{x},\tau)
  \Bigg[\left(\frac{\partial}{\partial \tau}+i\,q\,A_4\right)
+\frac{(-i \vec\nabla-\,q\,\vec{A})^2}{2M}  - \mu\, \vec{\sigma}\cdot\vec{H}
  +{2\pi}\left(\alpha\, \vec{E}^2 -\beta\, \vec{H}^2\right) 
\\
	-2\pi i \left( -\gamma_{E_1E_1} \vec{\sigma}\cdot\vec{E}\times\dot{\vec{E}}
    + \gamma_{M_1M_1} \vec{\sigma}\cdot\vec{H}\times\dot{\vec{H}}
    + \gamma_{M_1E_2} \sigma^i E^{ij}H^j
    + \gamma_{E_1M_2} \sigma^i H^{ij}E^j
\right)\Bigg] \Psi(\vec{x},\tau)  +\ldots \,,
\end{multline}
where $\vec{E}=- \frac{\partial}{\partial \tau} \vec{A}(\vec{x},\tau) -\vec{\nabla} A_4(\vec{x},\tau)$ and $\vec{H}= \vec{\nabla}\times \vec{A}(\vec{x},\tau)$
are the corresponding electric and magnetic fields, $\dot{X} = \frac{\partial}{\partial \tau} X$ denotes the Euclidean time derivative, 
$X^{ij}=\frac{1}{2}(\partial^i X^j+\partial^j X^i)$, and the ellipsis
denotes terms involving higher dimensional operators.
By calculating one- and two-photon processes with this effective Lagrangian, it is
clear that the constants that appear in Eq.~(\ref{eq:eff_L}) are
indeed the relevant magnetic moment and electromagnetic and multipole
polarizabilities \cite{Babusci:1998ww} [these are simply related to
the polarizabilities defined in the previous section as:
$\gamma_{E_1E_1}=-(\gamma_1+\gamma_3)$, $\gamma_{M_1M_1}=\gamma_4$,
$\gamma_{E_1M_2}=\gamma_3$ and $\gamma_{M_1E_2}=\gamma_2+\gamma_4$].
The Schr\"odinger equation corresponding to Eq.~(\ref{eq:eff_L})
determines the energy of the particle in an external U(1) field in
terms of the charge, magnetic moment, and polarizabilities.  Higher
order terms in Eq.~(\ref{eq:eff_L}) (which contain in part the higher
order polarizabilities \cite{Holstein:1999uu}) can be neglected for
sufficiently weak external fields. For a magnetic field, the minimally
coupled terms generate towers of Landau levels and for a constant
electric field the same terms accelerate charged particles.

Lattice calculations of the energy of a hadron in an external U(1)
field are straight-forward. One measures the behaviour of the usual
two-point correlator on an ensemble of gauge configurations generated
in the presence of the external field. This changes the Boltzmann
weight used in selecting the field configurations from
$\det\left[\Dslash+m\right]\exp{\left[-S_g\right]}$ to
$\det\left[\Dslash +i\,\hat{Q}\,\slash\!\!\!\!A+m\right]\exp{\left[-S_g\right]}$,
where $\Dslash\,$ is the SU(3) gauge covariant derivative, $\hat{Q}$
is the quark electromagnetic charge operator, and $S_g$ is the usual
SU(3) gauge action.  Since calculations are required at a number of
different values of the field strength in order to correctly identify
shifts in energy from the external field, this is a relatively demanding
computational task (although it is at least conceptually simpler than
studying the four-point function). In general one must worry about the
positivity of the fermionic determinant calculated in the presence of
a background field, however for weak fields, positivity is preserved.
The exploratory studies of
Refs.~\cite{Fucito:1982ff,Martinelli:1982cb,Bernard:1982yu,Fiebig:1988en,Aoki:1989rx,Aoki:1990ix,Shintani:2005du,Christensen:2004ca,Lee:2005dq}
used quenched QCD in which the gluon configurations do not feel the
presence of the U(1) field as the quark determinant is absent. In this
case, the external field can be applied after the gauge configurations
had been generated and is simply implemented by multiplying the SU(3)
gauge links of each configuration by link variables corresponding to
the fixed external field: $\{U^{\mu}_{\alpha}(x)\}\longrightarrow
\{U^{\mu} _{\alpha}(x) \exp[i\, e\,a\, A^\mu]\}$, where $a$ is the
lattice spacing. These studies are interesting in that they provide a
proof of the method, however the values of the polarizabilities
extracted have no connection to those measured in experiment.

It is clear from Eq.~(\ref{eq:eff_L}) that all six polarizabilities
can be extracted using suitable space and time varying background
fields if the shift of the hadron energy at second order in the
strength of the field can be determined. One can also see this because
the Compton tensor appears explicitly as the second-order connected
term in the expansion of hadronic two-point correlation function in
weak background fields \cite{Detmold:2004kw}. Previous studies
\cite{Fiebig:1988en,Christensen:2004ca,Lee:2005dq}
have employed constant electric and magnetic fields to determine the
corresponding polarizabilities in quenched QCD. Here we perform a more
general analysis to show how the spin polarizabilities and the
electric polarizabilities of charged particles can be obtained.

In order to determine the polarizabilities, we consider lattice
calculations of the two-point correlation function
\begin{eqnarray}
  \label{eq:correlator}
  C_{s s^\prime}(\vec{p},\tau;A)=\int d^3x\, e^{i\vec{p}\cdot\vec{x}}
\langle 0| \chi_s(\vec{x},\tau)\chi^\dagger_{s^\prime}(0,0) |0\rangle_A\,,
\end{eqnarray}
where $\chi_s(\vec{x},\tau)$ is an interpolating field with the quantum
numbers of the hadron under consideration (we will focus on the
nucleons) with $z$ component of spin, $s$, and the correlator is
evaluated on the ensemble of gauge configurations generated with the
external field, $A^\mu$.

For uncharged hadrons at rest in constant electric and magnetic
fields, it is simple to show that this correlator falls off
exponentially at large times with an energy given by the appropriate
terms in Eq.~(\ref{eq:eff_L}) owing to the constancy of the effective
Hamiltonian.  However for space-time varying fields, charged particles
or states of non-zero $\vec{p}$, a more general analysis is needed.
This is most easily formulated using the effective field theory (EFT)
defined by Eq.~(\ref{eq:eff_L}).  For weak external fields (such that
higher order terms in Eq.~(\ref{eq:eff_L}) can be safely neglected),
the small $\vec{p}$ and large $\tau$ dependence of this QCD correlation
function is reproduced by the equivalent correlator calculated in the
effective theory corresponding to the Lagrangian,
Eq.~(\ref{eq:eff_L}). That is
\begin{eqnarray}
  \label{eq:eft_correlator}
  C_{s s^\prime}(\vec{p},\tau;A)&=&\int d^3x\, e^{i\vec{p}\cdot\vec{x}}
\frac{1}{{\cal Z}_{\rm eff}[A]}\int {\cal D}\Psi^\dagger {\cal D}\Psi
\,\Psi_s(\vec{x},\tau)\Psi^\dagger_{s^\prime}(0,0)
\exp\left(-S_{\rm
    eff}[A]\right)\,,
\end{eqnarray}
where ${\cal Z}_{\rm eff}[A]=\int {\cal D}\Psi^\dagger {\cal D}\Psi
\exp\left({-{S}_{\rm eff}[A]}\right)$.  Since the right-hand side of
Eq.~(\ref{eq:eft_correlator}) is completely determined in terms of the
charge, magnetic moment and polarizabilities that we seek to extract,
fitting lattice calculations of $C_{s s^\prime}(\vec{p},\tau;A)$ in a
given external field to the effective field theory expression will
enable us to determine the appropriate polarizabilities.  In the above
equation we have assumed that the ground state hadron dominates the
correlator at the relevant times. For weak fields this will be the
case. However one can consider additional terms in the effective
Lagrangian that describe the low excitations of the hadron spectrum
that have the same quantum numbers as the hadron under study. This
will lead to additive terms in Eq.~(\ref{eq:eft_correlator}) that
depend on the mass, magnetic moment and polarizabilities of the
excited hadron instead of those of the ground state. With precise
lattice data, the properties of these excited states can also be
determined.

In many simple cases such as constant or plane-wave external fields,
the EFT version of $C_{s s^\prime}(\vec{p},\tau;A)$ can be determined
analytically in the infinite volume, continuum limit
\cite{Schwinger:1951nm}. However in finite lattice spacing and at
finite volume, calculating $C_{s s^\prime}(\vec{p},\tau;A)$ in the EFT
becomes more complicated. In order to determine the EFT correlator, we
must invert the matrix ${\cal K}$ defined by
\begin{eqnarray}
  \label{eq:S_latt}
  S_{\rm
  latt}[A]=\sum_{\vec{x},\tau_x}\sum_{\vec{y},\tau_y}\sum_{s,s^\prime}\Psi^\dagger_s(\vec{x},\tau_x)
{\cal K}_{ss^\prime}[\vec{x},\tau_x,\vec{y},\tau_y;A] \Psi_s(\vec{y},\tau_y)\,,
\end{eqnarray}
where $S_{\rm latt}[A]$ is a discretization of the EFT action in which
derivatives are replaced by finite differences (the time derivative is
given by a forward difference as we can ignore anti-particles). ${\cal
  K}$ has dimension $4N_{l}^2$ where $N_l$ is the number of lattice
sites.  For the most general space-time varying external field, this
must be inverted numerically; given a set of lattice results for the
correlator, Eq.~(\ref{eq:eft_correlator}) is repeatedly evaluated for
varying values of the polarizabilities until a good description of the
lattice data is obtained.

\begin{figure}[!t]
  \centering
  \includegraphics[width=0.8\columnwidth]{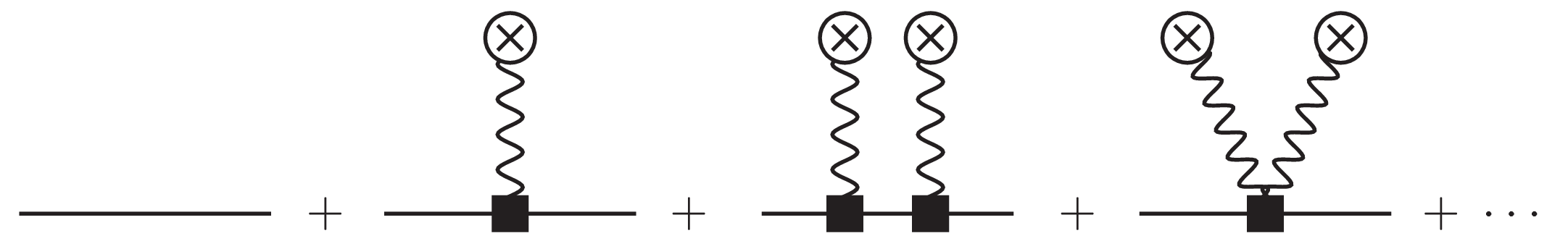}
  \caption[Perturbative expansion of a hadron propagator in a background field]{\label{fig:EFT}Perturbative expansion of the hadron propagator in an external field. }
\end{figure}

For weak fields such that $|A^\mu(\vec{x},\tau)|^2\ll\L_{\rm QCD}^2$ for
all $\vec{x}$ and $\tau$, a perturbative expansion of ${\cal K}^{-1}$ in
powers of the field can be used. This corresponds to the series of
diagrams in Fig.~\ref{fig:EFT}. 
To extract all six polarizabilities using such an analysis, we need to
consider a number of different fields; lattice calculations of the
correlators in Eq.~(\ref{eq:correlator}) using% the fields
\begin{eqnarray}
  \label{eq:fields}
  A^\mu_{(1)}(x) =\begin{pmatrix} i a_1 \tau \\0\\0\\0\end{pmatrix},\quad
  A^\mu_{(2)}(x) =\begin{pmatrix}-\frac{a_2}{2} x_2\\ \frac{a_2}{2} x_1\\ 0
    \\0\end{pmatrix},\quad
  A^\mu_{(3)}(x) =\begin{pmatrix}0 \\ -  \frac{i}{a} a_3 \tau \,x_3\\ -b_3 x_1\\0
    \end{pmatrix}, 
\\ \nonumber
  A^\mu_{(4)}(x) =\begin{pmatrix} - \frac{i}{a}a_4 \tau \,x_2 \\ 0\\
    -\frac{1}{2}\,b_4 x_1 \\0\end{pmatrix},\quad 
  A^\mu_{(5)}(x) =\begin{pmatrix} \frac{1}{a}a_5\, x_2\, x_1 \\
    \frac{1}{2a}a_5\, x_2^2 \\ i b_5 \tau
    \\0\end{pmatrix},\quad 
  A^\mu_{(6)}(x) =\begin{pmatrix} - \frac{1}{2a}a_6 \tau^2 \\ -  i\frac{1}{2}
    b_6 \tau \\ 0 \\0\end{pmatrix}\,,
\end{eqnarray}
for a number of different choices for the strength parameters, $a_i$
and $b_i$ (with $|a_i|,\,|b_i|\ll\L_{\rm QCD}^2$), are sufficient to
determine the full set of polarizabilities.%
%	footnote
\footnote{These fields correspond to real $\vec{E}$ and $\vec{H}$ fields in Minkowski-space for real-valued $a_i$ and $b_i$.  Since periodic spatial
  boundary conditions are envisaged for the link variables, there are
  quantization conditions that must be satisfied by the $a_i$ 
  \cite{Martinelli:1982cb,Bernard:1982yu}. For example, $q_i
  a_2=\frac{2\pi\, n}{a\,L}$ for each of the quark charges $q_i$. 
  The more complicated fields in Eq.~(\ref{eq:fields}) require two
  parameters to satisfy these conditions.  
}  
By measuring correlators for different spin
configurations (including those that flip spin), we can reduce the
number of fields required to extract the polarizabilities.

As an example, the behaviour of the correlator in the field
$A^\mu_{(1)}(x)$ (which corresponds to a constant electric field in
the $x_1$ direction) is given by
\begin{eqnarray}
  C_{s s^\prime}(\vec{p},\tau;A_{(1)}) &=&
  \delta_{s,s^\prime}\exp\left\{
- \frac{a_1\,\tau}{6M}\left[ a_1
  \left(q^2 \tau^2 + 12 M \pi  \alpha \right) - 3 i q\, \tau\,p_1\right]\right\}
e^{-M\,\tau} e^{-\frac{\tau}{2M}|\vec{p}|^2} +{\cal O}(a_1^3)
\notag
\\
&\stackrel{|\vec{p}|\to0}{\longrightarrow}&
\delta_{s,s^\prime}\exp\left[-(M+2\pi\alpha
  a_1^2)\tau-\frac{q^2a_1^2}{6M}\tau^3\right] +{\cal O}(a_1^3) \,.
  \label{eq:A1field}
\end{eqnarray}
In this case, the perturbative series has been resummed exactly in the
continuum, infinite volume limit and the higher order corrections come
from terms omitted in Eq.~(\ref{eq:eff_L}). For electrically neutral
particles, the exponential fall-off of this correlator determines the
polarisability $\alpha$ once the mass $M$ has been measured in the
zero-field case.  When a charged particle is placed in such a field it
undergoes continuous acceleration in the $x_1$ direction (this is
described by the $\tau^3$ term in the exponent).  However at times small
compared to $\frac{\sqrt{6M}}{q\, a_1}$, the correlator essentially
falls off exponentially. Matching the behaviour of
Eq.~(\ref{eq:A1field}) to lattice data for a charged hadron will again
enable us to determine the electric polarisability, $\alpha$.

As a second analytic example, we consider one of the multipole
polarizabilities. In the presence of the field $A_{(6)}^\mu(x)$, which
corresponds to a more complicated electric field
$\vec{E}_{(6)}(x)=(\frac{a_6}{a} \tau, i\frac{b_6}{2},0)$, we find that
\begin{eqnarray}
  \label{eq:A6exp}
  \frac{C_{\uparrow\uparrow}(\vec{p},\tau;A_{(6)})}
  {C_{\downarrow\downarrow}(\vec{p},\tau;A_{(6)})}
  &=& \exp\left[\frac{2\pi}{a}\,a_6\,b_6\,\gamma_{E_1E_1}\,  \tau \right]+\ldots \,,
\end{eqnarray}
independent of $\vec{p}$ and the ellipsis denotes terms cubic in the
field that have been neglected in Eq.~(\ref{eq:eff_L}).  Whilst the
individual correlators, $C_{\uparrow\uparrow}$ and
$C_{\downarrow\downarrow}$, have relatively complicated time-momentum
behaviour involving $q$ and $\alpha$ as well as $\gamma_{E_1E_1}$,
this becomes very simple in the ratio and $\gamma_{E_1E_1}$ can be
determined cleanly.

Analogous results can be derived for the other fields in
Eq.~(\ref{eq:fields}), however to take into account the finite lattice
spacing and periodic finite volume nature of the underlying lattice
simulations to which the EFT description is matched, the correlator is
most easily calculated by inverting the matrix ${\cal K}$ numerically.
This also allows for more general choices of fields. If we seek to
extract higher order polarizabilities, the Lagrangian in
Eq.~(\ref{eq:eff_L}) must be extended to include higher dimension
operators \cite{Holstein:1999uu}.  At this order, relativistic
corrections and three-photon couplings also need to be included.
Correlation functions similar to those in Eq.~(\ref{eq:correlator})
involving two different external momenta will allow us to also extract
the generalized polarizabilities \cite{Guichon:1995pu}.

%
%
%
%
%
%
%       Heavy Baryon ChPT
%
%
%
%
%
%
\section{Heavy Baryon Lagrangian with Electromagnetic Interactions}
\label{sec:hbxpt}

To calculate the quark mass and volume dependence of the nucleon
polarizabilities, we use heavy baryon chiral perturbation theory
(HB$\chi$PT) as was first constructed in
Refs.~\cite{Jenkins:1990jv,Jenkins:1991ne,Bernard:1993nj,Jenkins:1991es}.
In current lattice calculations, valence and sea quarks are often
treated differently, with sea quarks either absent (quenched QCD) or
having different masses than the valence quarks (partially-quenched
QCD).\footnote{At finite lattice spacing, different actions can even
  be used for the different quark sectors (e.g., staggered sea quarks
  and domain wall valence quarks).  As was shown in
  Refs.~\cite{Beane:2003xv,Arndt:2003vd,Tiburzi:2005is}, the lattice
  spacing corrections to baryon electromagnetic properties are
  expected to be small, as they can not enter at tree level, and for
  current simulations with $a \L_{QCD}^2 \sim m_q$, they generally
  enter at leading loop order through valence-sea meson masses.  In
  our work we assume a continuum extrapolation has been performed.}
The extensions of \hbxpt\ to quenched \hbxpt\ 
\cite{Labrenz:1996jy,Savage:2001dy} and partially quenched \hbxpt\ 
\cite{Chen:2001yi,Beane:2002vq} to accommodate these modifications are
also well established and have been used to calculate many baryon
properties.  In this section, we will primarily focus on the two
flavour partially-quenched theory and briefly introduce the relevant
details following the conventions set out in Ref.~\cite{Beane:2002vq}.
Since QCD is a special limit of the partially-quenched theory, our
discussion also encompasses two flavour \xpt.  Additional
complications in quenched \xpt\ are relegated to Appendix~\ref{A4}.

%
%
%
%
%
%
%       Mesons
%
%
%
%
%
%
\subsection{\textbf{Pseudo-Goldstone mesons}}

We consider a partially-quenched theory of valence ($u$, $d$), sea
($j$, $l$) and ghost ($\tilde u,\,\tilde d$) quarks with masses
contained in the matrix
\begin{equation}
\label{eq:Mq_def}
m_Q = {\rm  diag}(m_u,m_d,m_j,m_l,m_{\tilde u},m_{\tilde d})\,,
\end{equation}
where $m_{\tilde u,\tilde d}=m_{u,d}$ such that the path-integral
determinants arising from the valence and ghost quark sectors exactly
cancel. The corresponding low-energy meson dynamics are described by
the \pqxpt\ Lagrangian, Eq.~\eqref{eq:PQChPT}, with the addition of minimally coupled electromagnetism (the U(1) gauge field is again denoted by $A^\mu$ and its field
strength tensor $F^{\mu\nu}=\partial^\mu A^\nu-\partial^\nu A^\mu$) to
the theory through the chiral, and U(1) gauge covariant derivative
\begin{equation}
\label{eq:cov_deriv_def}
  {\cal D}^{\mu} = \partial^{\mu}
  + \left [ {\cal V}^{\mu},\quad \right ] \,,
\end{equation}
with the vector current
\begin{eqnarray}
  \label{eq:vector_current}
  {\cal V}^{\mu} &=& \frac{1}{2}\left[\ \xi\left(\partial^\mu-i e {\cal
        Q} A^\mu\right)\xi^\dagger \ + \  
\xi^\dagger\left(\partial^\mu-i e {\cal Q} A^\mu\right)\xi \ \right],
\end{eqnarray}
depending on the quark charge matrix, ${\cal Q}$. In coupling
electromagnetism to this theory, we must specify how the quark charges
are extended to the partially-quenched theory. We choose:
\begin{eqnarray}
  \label{eq:7}
  {\cal Q} &=& {\rm diag}(q_u,q_d,q_j,q_l,q_u,q_d)\,,
\end{eqnarray}
though other arrangements are possible.  However, one must set
$q_j+q_l\ne0$ in order to retain sensitivity to the full set of LECs
that appear in two flavour \xpt~\cite{Tiburzi:2004mv,Detmold:2005pt}.
In addition to the Lagrangian, Eq.~(\ref{eq:PQChPT}), the
anomalous couplings of the Wess-Zumino-Witten Lagrangian
\cite{Wess:1971yu,Witten:1983tw} will also contribute to the spin
polarizabilities. These terms are described below.

%
%
%
%
%
%
%       Baryons
%
%
%
%
%
%
\subsection{\textbf{Baryons}}

The baryon Lagrangian is given in Eq.~\eqref{eq:leadlagPQ}, again with the addition of minimally coupled electromagnetism,
\begin{eqnarray}
{\cal A}^\mu \ =\  \frac{i}{2}\left[\ \xi\left(\partial^\mu-i e {\cal
        Q} A^\mu\right)\xi^\dagger \ - \  
\xi^\dagger\left(\partial^\mu-i e {\cal
        Q} A^\mu\right)\xi \ \right]
\, .
\label{eq:Amu}
\end{eqnarray}

As with the mesons, at leading order the photon is minimally coupled
to the baryons with fixed coefficients.  At the next order in the
expansion there are a number of new electromagnetic gauge invariant
operators which contribute to the Compton amplitude and the
polarizabilities.  Here, we display the relevant terms at this order,
\begin{eqnarray}
        {\cal L}_{B}^{(1)}& = &
                \frac{i\,e}{ 2 M_N} F_{\mu\nu}
                \Big[
                        \mu_\alpha \left( \ol{\cal B} \left[S^{\mu},S^{\nu}\right]  {\cal B}  
                                {\cal Q}_{\xi+} \right) 
                        + \mu_\beta \left( \ol{\cal B}
                          \left[S^{\mu},S^{\nu}\right]
                                {\cal Q}_{\xi+} {\cal B} \right)
                         \nonumber \\
                &&+  \mu_\gamma  {\rm str}\left[ {\cal Q}_{\xi+} \right]
                 \left( \ol{\cal B} \left[S^{\mu},S^{\nu}\right] {\cal B} \right)
                 \Big] \nonumber\\
                 &&+ \sqrt{\frac{3}{2}} \mu_T \frac{i e}{ 2 M_N} F_{\mu\nu}
                \left[
                        \left( \ol{\cal B} S^\mu {\cal Q}_{\xi+} {\cal T}^\nu \right) 
                +\left( \ol{\cal T} {}^\mu S^\nu  {\cal Q}_{\xi+} {\cal B} \right) 
                \right],
\label{L1}
\end{eqnarray}
where $\mu_{\alpha,\beta,\gamma}$ are magnetic moment coefficients
\cite{Savage:2001dy,Beane:2002vq}, $\mu_T$ is the coefficient of the
M1 transition {\bf 70}--{\bf 44} operator
\cite{Arndt:2003we,Arndt:2003vd} and
\begin{eqnarray}
  \label{eq:1}
  {\cal Q}_{\xi^\pm} &=& \frac{1}{2}
\left( \xi^\dagger {\cal Q}\xi \pm \xi {\cal Q}\xi^\dagger
\right).
\end{eqnarray}
The partially quenched magnetic moment coefficients are related to the
isoscalar and isovector magnetic coefficients, $\mu_0$ and $\mu_1$, in
standard two flavour \CPT\ as
\begin{equation}
        \mu_0 = \frac{1}{6} \Big( \mu_\a +\mu_\b +2\mu_\g \Big), \quad\quad
        \mu_1= \frac{1}{6} \Big( 2\mu_\a -\mu_\b \Big),
\end{equation}
where the \CPT\ Lagrangian describing the magnetic moments of the
nucleons [the proton and neutron magnetic moments are
$\mu_{p,n}=\frac{1}{2}(\mu_0\pm \mu_1)$] is given by
\begin{eqnarray}
        {\cal L} &=& \frac{i e}{2 M_N} F_{\mu\nu} \left( \mu_0
          \ol{N} \left[S^{\mu},S^{\nu}\right] N 
                                +\mu_1 \ol{N} \left[S^{\mu},S^{\nu}\right]
                                \t^3_{\xi+} N \right),
\end{eqnarray}
for $\t^a_{\xi\pm} = \frac{1}{2} \left( \xi^\dagger \t^a \xi \pm \xi
  \t^a \xi^\dagger \right)$.  

There are other operators formally at this order which do not
contribute to the polarizabilities at the order to which we work.
There are kinetic operators and higher dimensional couplings of the
baryons to the axial current whose coefficients are exactly fixed by
the reparameterisation invariance of the baryon
four-momentum~\cite{Luke:1992cs,Tiburzi:2005na}. These operators give
the $Z$ dependent pieces of the Compton amplitudes in
Eq.~(\ref{eq:amplitudes}). There are also additional operators with
unconstrained coefficients such as $ \left( \bar{\cal B}\, {\cal A}
  \cdot {\cal A}\, {\cal B} \right)$ that contribute to the Compton
amplitude at higher order.  In two-flavor \CPT\ there are two such
operators, and in the SU(4$|$2) case there are ten
\cite{Tiburzi:2005na}.

The leading operators which contribute to the electromagnetic
polarizabilities at tree level occur at ${\cal O}(Q^4)$ and are given
by the general form,
\begin{equation*}
        \frac{e^2F_{\mu\rho}F_{\nu}^{\; \rho}}{\L_\chi^3}
                \left( \ol{\cal B}\, \Gamma^{\mu\nu} {\cal
                    Q}_{\xi \pm}^2 {\cal B} \right), 
\end{equation*}
(where the $\Gamma^{\mu\nu}$ are spin structures) while the leading
tree-level contributions to the spin polarizabilities occur at ${\cal
  O}(Q^5)$. The complete set of such operators in the case of
two-flavour \CPT\ is given in Ref.~\cite{Fettes:2000gb}. Again there
are significantly more such operators in \qxpt\ and \pqxpt. We do not
explicitly show these operators, as they do not contribute at the
order we are working and will not modify volume dependence until
${\cal O}(Q^6)$.

%
%
%
%
%
%
%       Nucleon polarizabilities
%
%
%
%
%
%
\section{Nucleon Compton Scattering in Finite Volume}
\label{sec:pols}

\begin{figure}[!t]
\centering
\begin{tabular}{cc}
\includegraphics[width=0.35\columnwidth]{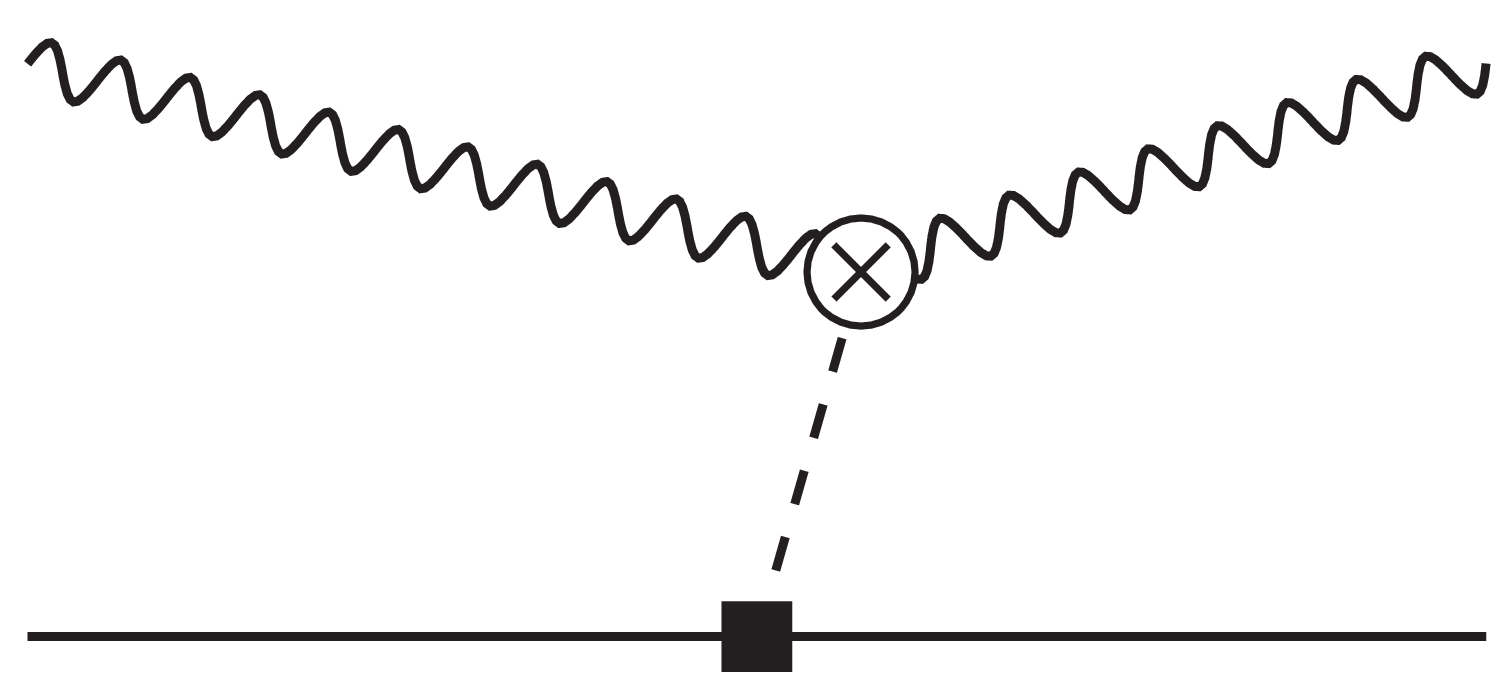}
&\includegraphics[width=0.35\columnwidth]{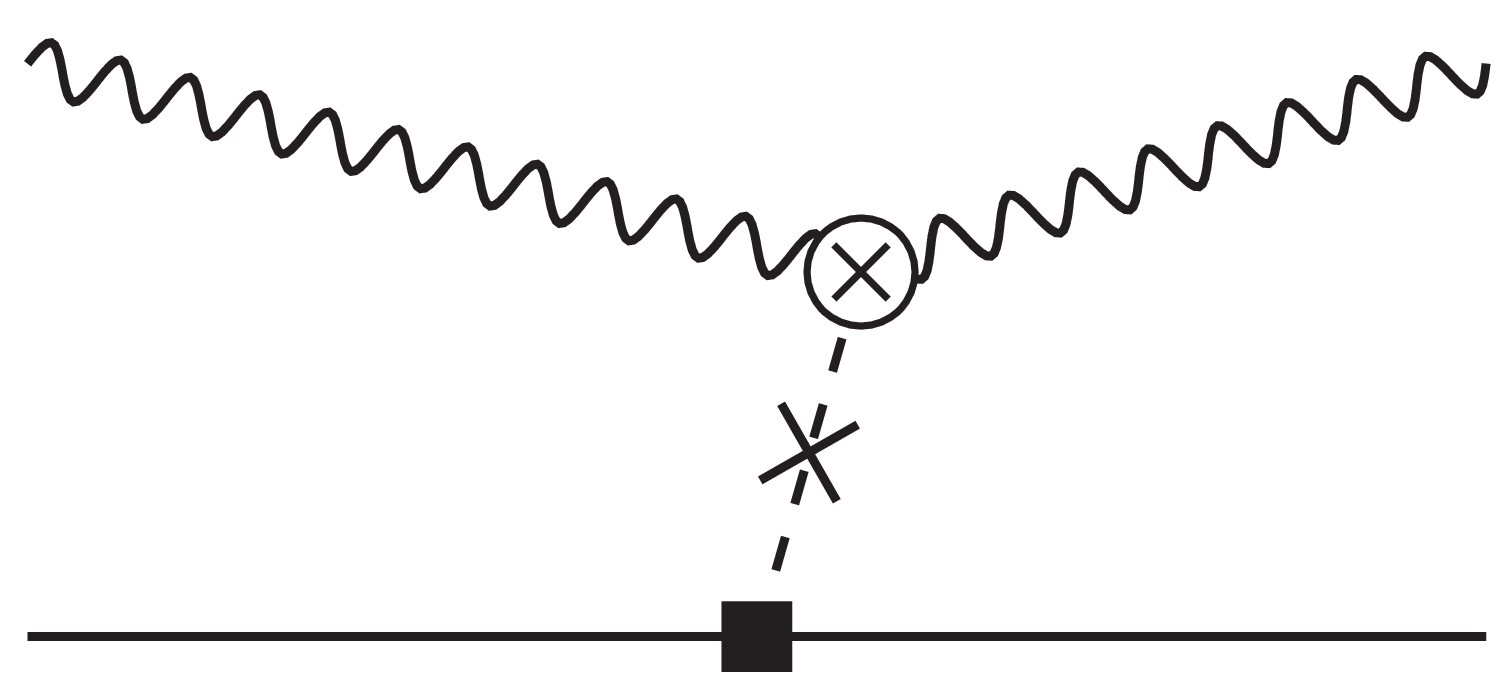}
\end{tabular}
\caption[Anomalous $\pi^0 \rightarrow \g\g$ contribution to nucleon polarizabilities.]{\label{fig:anom} Anomalous contributions to the polarizabilities. The crossed
  circle corresponds to the insertion of an operator from the
  Wess-Zumino-Witten Lagrangian, Eq.~(\ref{eq:LWZWPQ}), and the
  crossed meson line corresponds to a hairpin interaction
  \protect\cite{Sharpe:1992ft}.}
\end{figure}

\begin{figure}[!t]
\centering
\begin{tabular}{cc}
        \includegraphics[width=0.35\columnwidth]{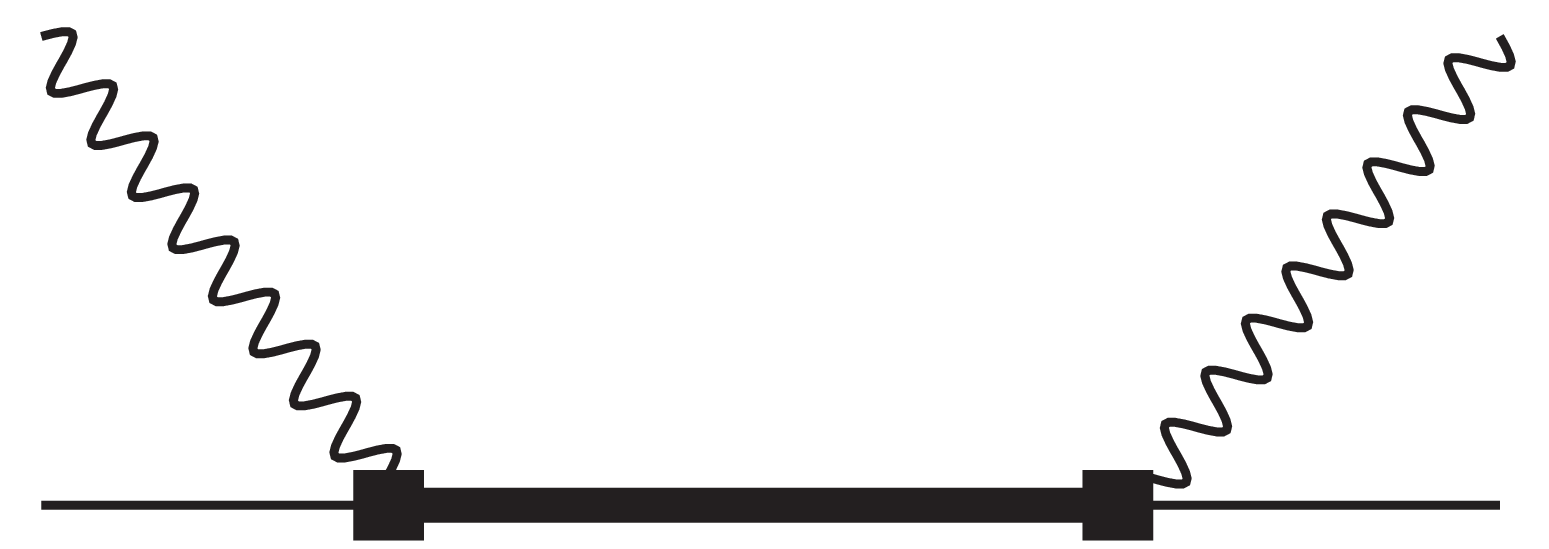} 
        &\includegraphics[width=0.35\columnwidth]{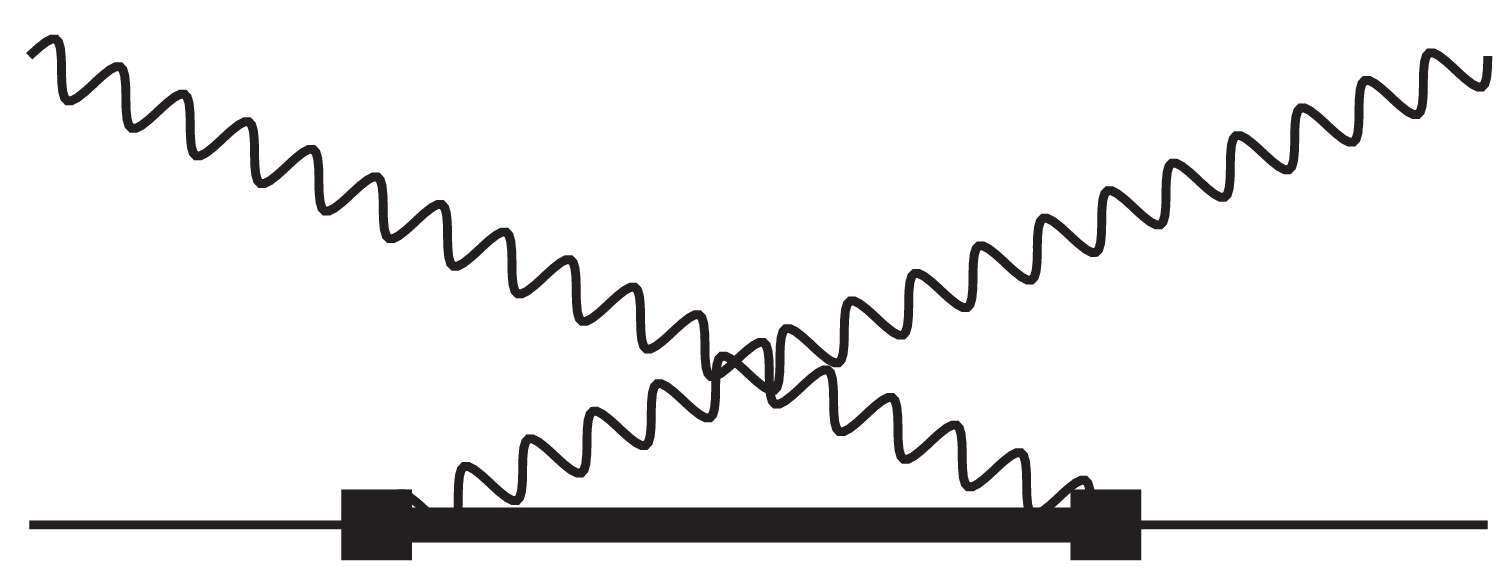} \\
\end{tabular}
\caption[Delta-resonance contributions to Born scattering]{\label{fig:Born} Born diagrams involving internal {\bf 44}-plet states that give contribution to the polarizabilities.}
\end{figure}

\begin{figure}[!t]
  \centering
\begin{tabular}{ccc}
        \includegraphics[width=0.3\columnwidth]{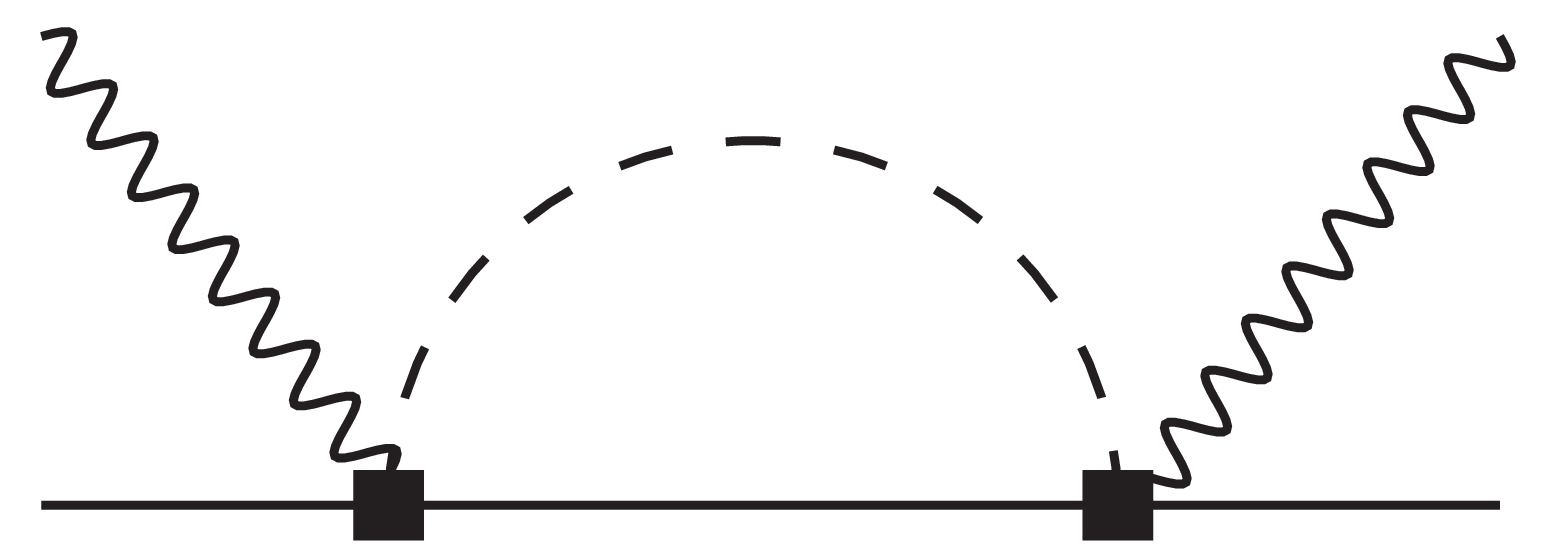}
        & \includegraphics[width=0.3\columnwidth]{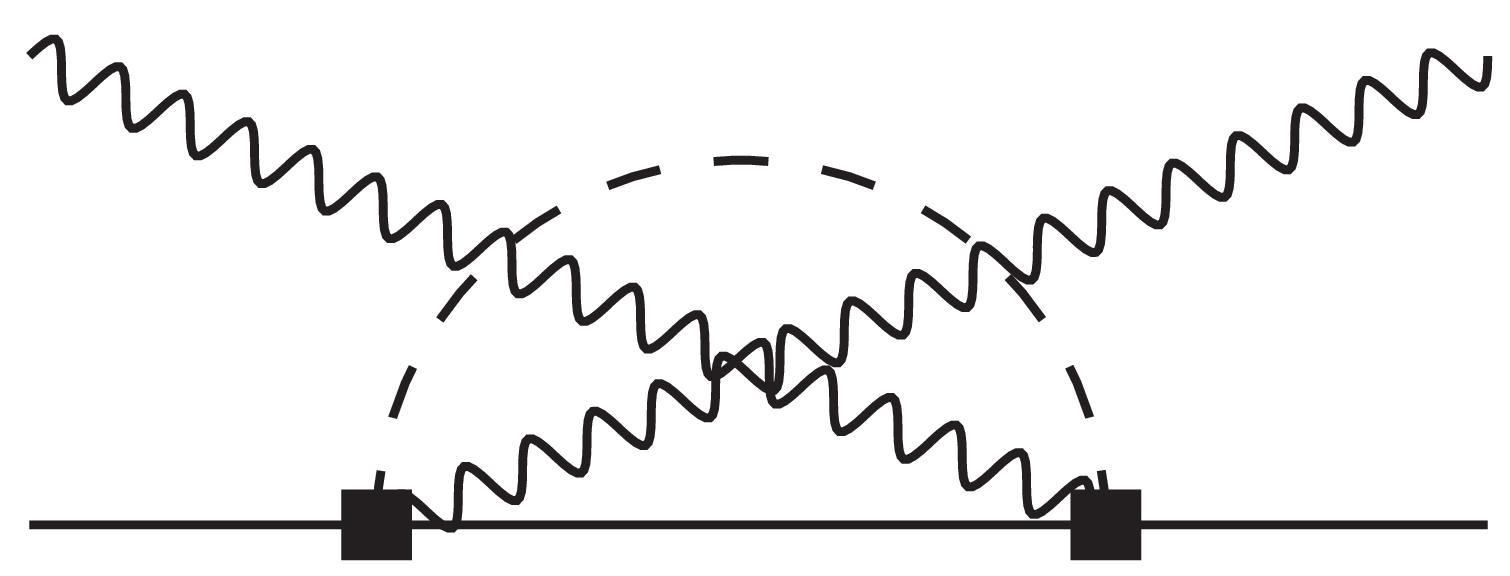} 
        & \includegraphics[width=0.3\columnwidth]{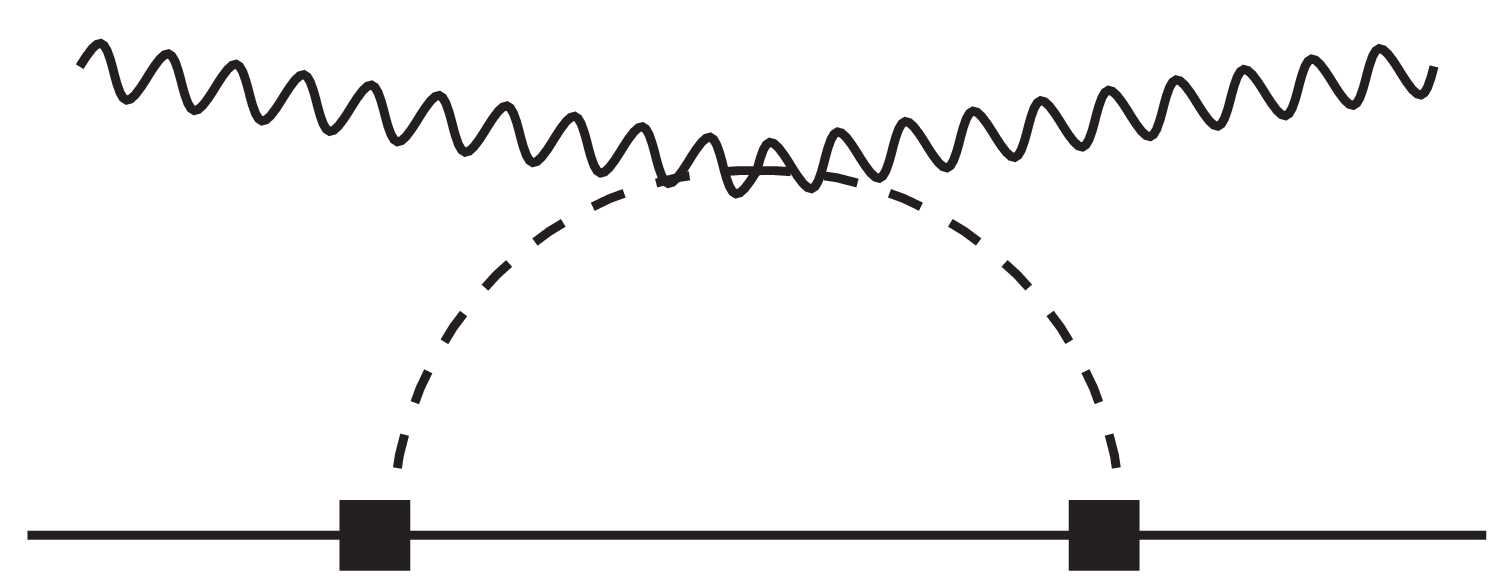} \\
        (a) & (b) &(c) \\ \\
        \includegraphics[width=0.3\columnwidth]{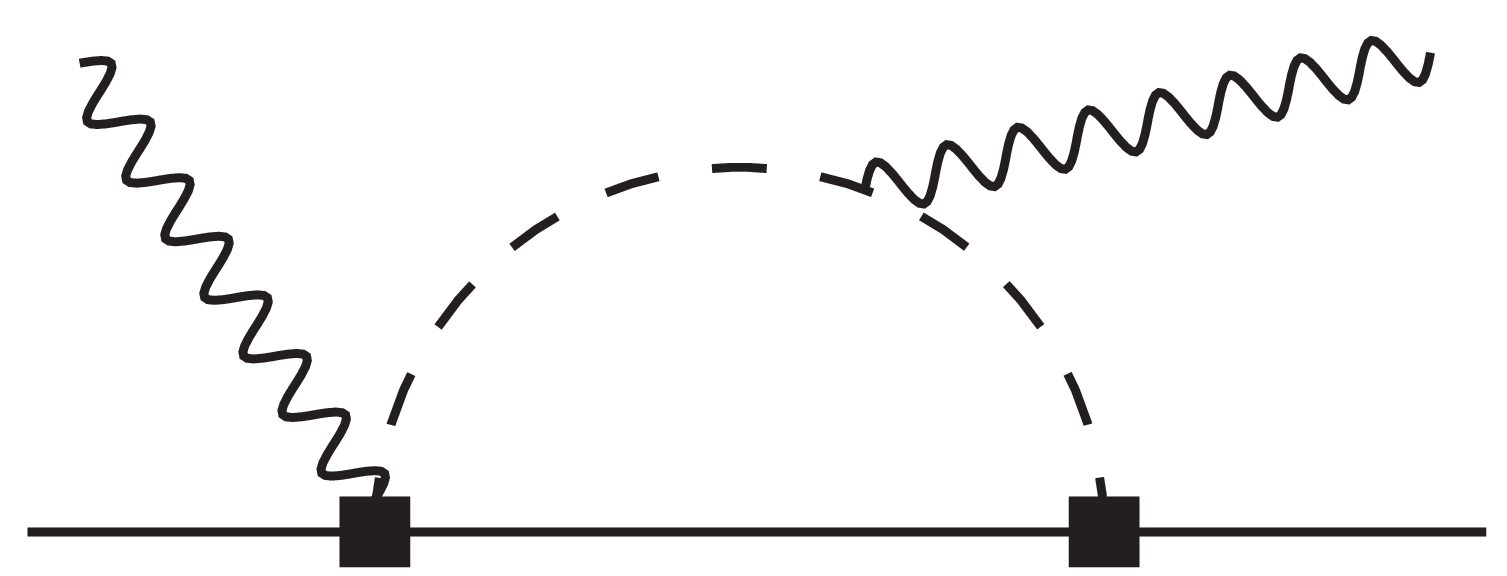} 
        &\includegraphics[width=0.3\columnwidth]{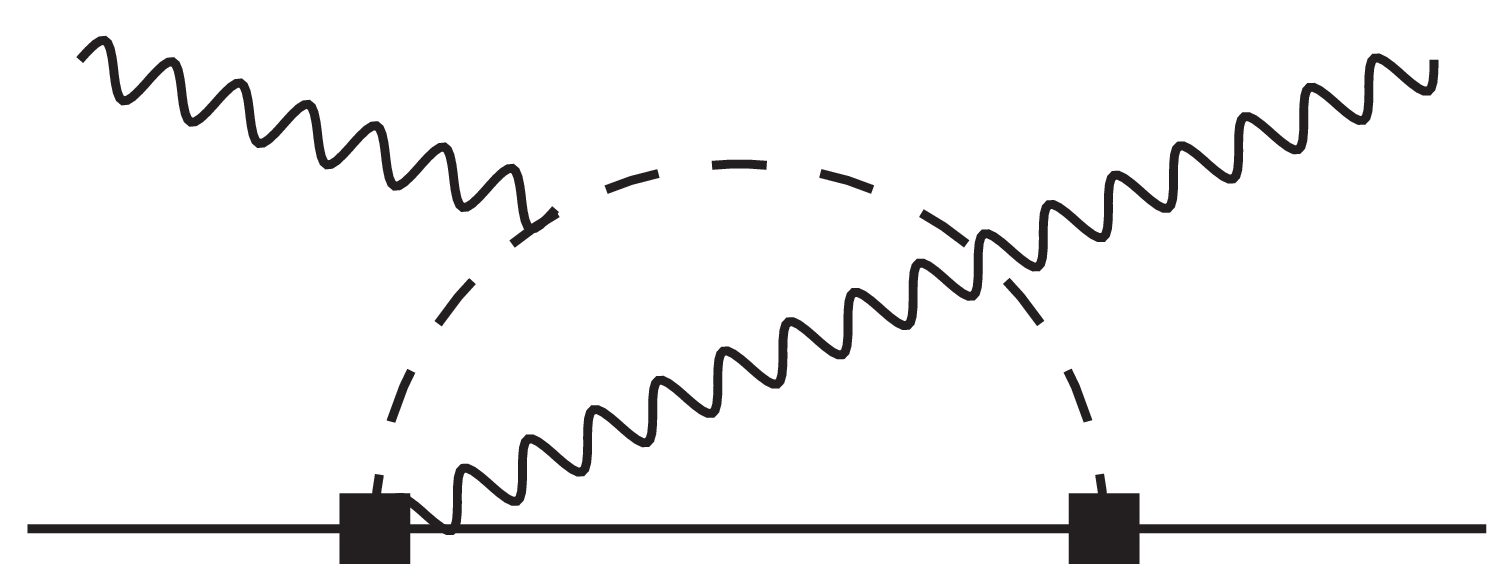} 
        & \includegraphics[width=0.3\columnwidth]{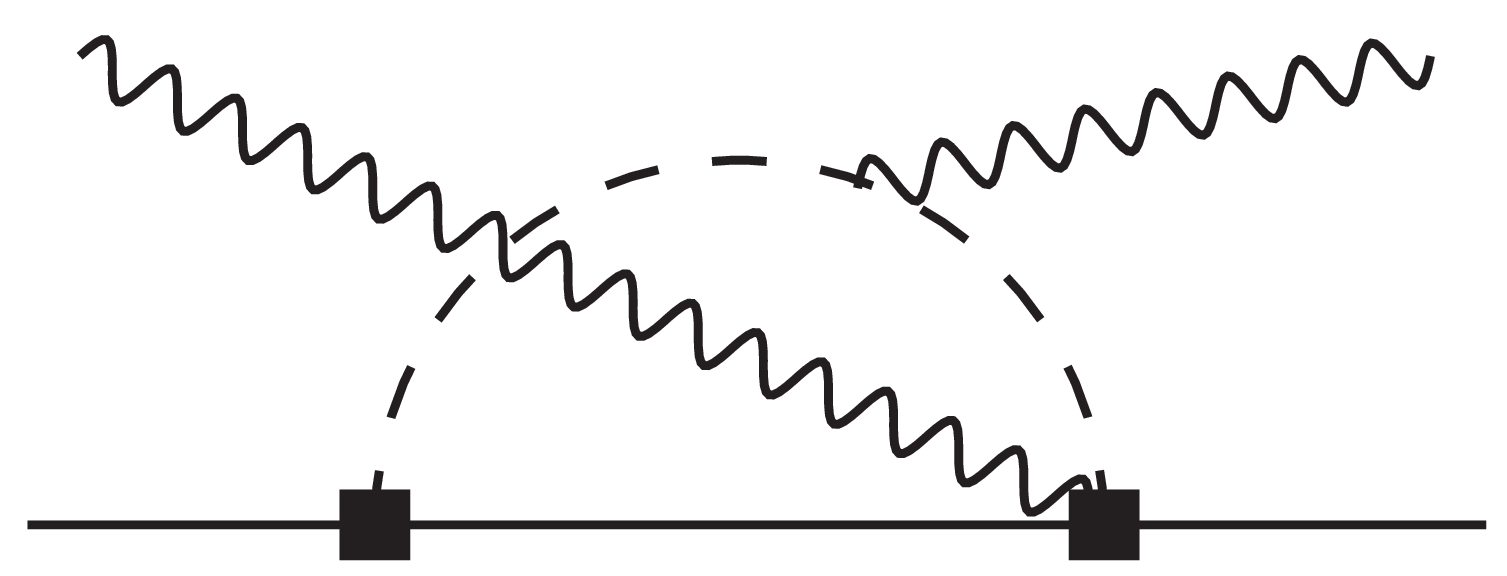} \\
        (d) & (e) &(f) \\ \\
        \includegraphics[width=0.3\columnwidth]{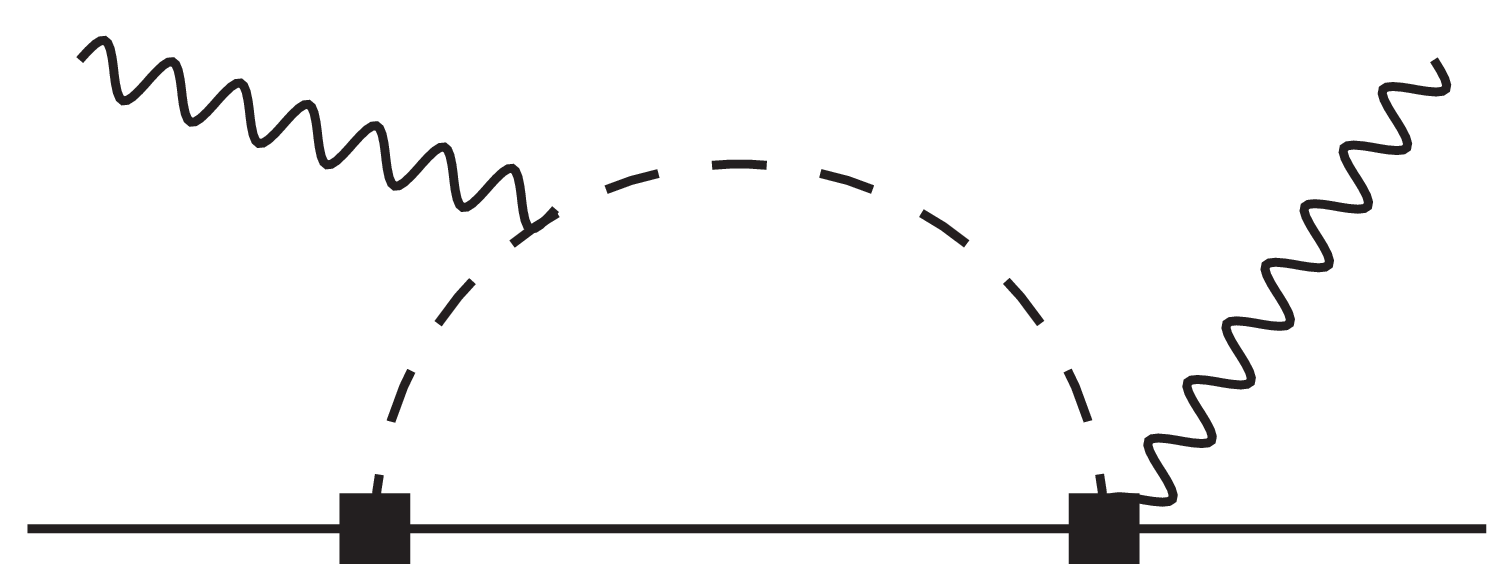} 
        & \includegraphics[width=0.3\columnwidth]{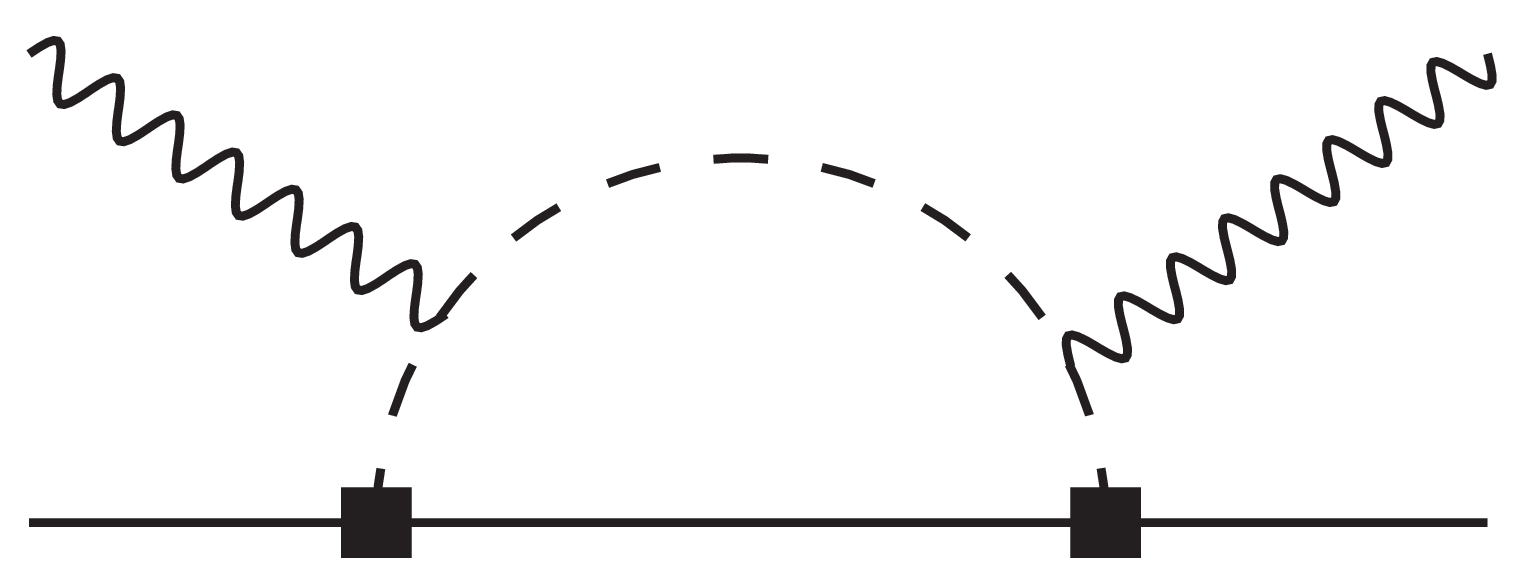} 
        &\includegraphics[width=0.3\columnwidth]{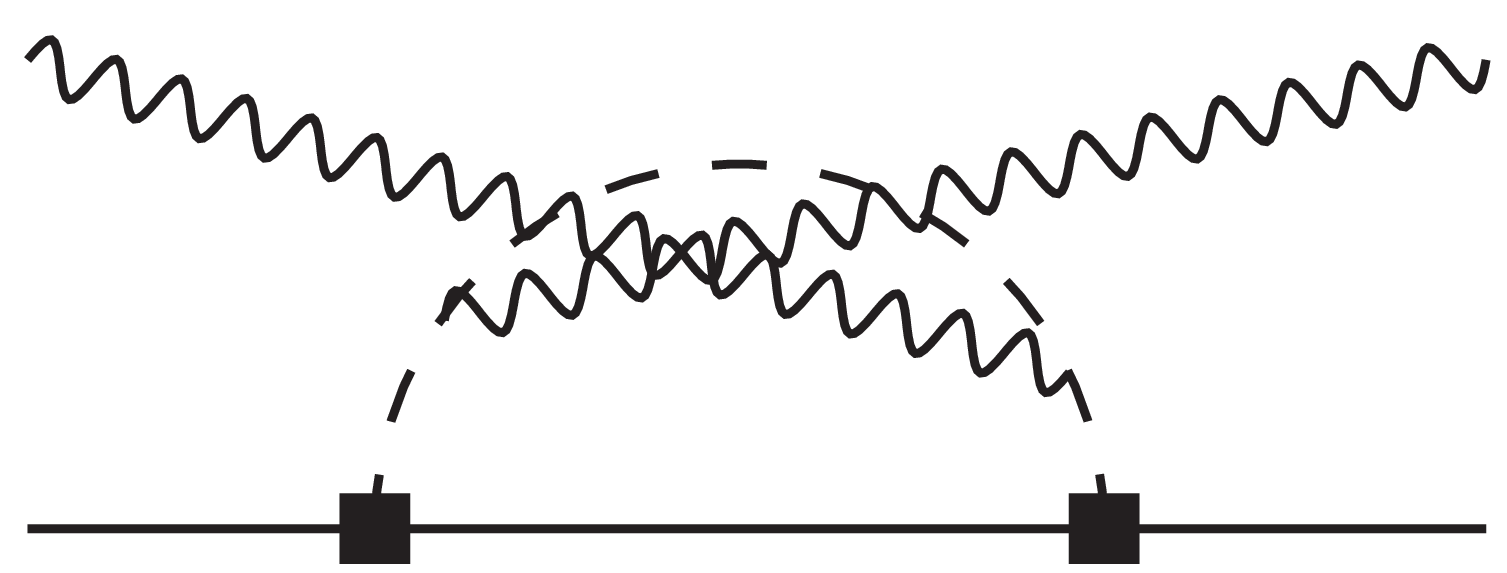} \\
        (g)& (h) & (i)
\end{tabular}
\caption[Various diagrams contributing to nucleon polarizabilities at $\mc{O}(Q^3)$]{\label{fig:Q3} Diagrams contributing to nucleon polarizabilities at order
  $Q^3$. The solid and dashed lines denote nucleons, and pions
  respectively. Additionally, our results include a similar set of
  diagrams in which the internal {\bf 70}-plet propagator is replaced
  by a {\bf 44}-plet resonance.}
\end{figure}

Using the Lagrangian of the preceding section, we can calculate the
amplitudes defined in Eq.~(\ref{eq:Ti}) for Compton scattering from a
nucleon (extensions to full octet and decuplet of baryons are
straight-forward although the convergence of \hbxpt\ with three-flavours
is not clear).  We work with a power counting such that
\begin{equation}
        Q \sim e \sim \frac{|\vec{p}|}{\L_\chi} 
                \sim \frac{m_\pi}{\Lambda_\chi} 
                \sim \frac{\w}{\L_\chi}\,
\end{equation}
(it is also convenient to count $\Delta/\Lambda_\chi$ as the same as $Q$
as it is numerically similar at the masses relevant for current
lattice calculations).\footnote{Loop and pole~\cite{Butler:1992pn}
  contributions with {\bf 44}-plet intermediate states must be
  included since $\Delta$ is a small-scale. %$^{\bf TM}$
  Any $\Delta$ dependent terms analytic in $m_\pi$ arising from the
  loop diagrams, and additional operators proportional to powers of
  $\Delta/\L_\chi$ can be resummed into the appropriate LECs of $\Delta$
  independent operators (the LECs then depend on
  $\Delta$)~\cite{Tiburzi:2004kd,Tiburzi:2005na}.  Keeping these
  contributions explicit is redundant as $\Delta$ can not be varied in
  a controlled manner.}  Below, we will also restrict ourselves to the
low frequency limit $\w \ll m_\pi$ in order to extract the
polarizabilities from the Compton scattering amplitudes defined in
Eqs.~(\ref{eq:Ti}) and (\ref{eq:amplitudes}).  For larger energies,
the concept of polarizabilities breaks down and the target essentially
becomes a dispersive medium. Working to order $Q^3$ in the chiral
expansion, Compton scattering requires the calculation of the diagrams
shown in Figs.~\ref{fig:anom}, \ref{fig:Born} and \ref{fig:Q3} (and a
corresponding set involving internal {\bf 44}-plet baryons). By
definition, tree level contributions from nucleon pole diagrams do not
contribute to the polarizabilities; their contribution to the
amplitudes are given explicitly in Eq.~(\ref{eq:amplitudes}).  For
each polarisability $X=\alpha,\,\beta,\,\gamma_{1\mbox{--}4}$, it is
convenient to separate the different contributions as
\begin{eqnarray}
  \label{eq:5}
  X= X^{\rm anomaly} + X^\Delta +X^{\rm loop} \,,
\end{eqnarray}
corresponding to the contributions from Figs \ref{fig:anom},
\ref{fig:Born} and \ref{fig:Q3}, respectively. We discuss these
contributions in the following subsections. At order $Q^3$, all
contributions are expressible in terms of a small set of LECs that
contribute in many other processes and are thus reasonably well
determined (at least in the \xpt\ case).  The total ${\cal O}(Q^3)$
loop contribution is finite, but loop-contributions at higher orders
are divergent; as discussed in the preceding section, the
counter-terms specific to Compton scattering that absorb these
divergences and the associated scale dependence enter at ${\cal
  O}(Q^4)$ for the electric and magnetic polarizabilities and ${\cal
  O}(Q^5)$ for the spin polarizabilities.
%
%	figure: Anomalous WZW diagrams
%

\subsection{\textbf{Volume independent contributions to polarizabilities}\label{sec:VolumeIndependent}}

%
%
%
%       Anomalous pi --> gamma gamma 
%
%
%
%
%
%
\subsubsection{\textbf{Anomalous contribution to $\g N\rightarrow \g N$:
  $\pi^0 \rightarrow \g\g$}\label{sec:Anomaly} }

The anomalous decay of flavour neutral mesons to two photons
\cite{Bell:1969ts,Adler:1969gk} has important consequences in Compton
scattering in non-forward directions. These contributions arise from
the meson pole diagram shown in Fig.~\ref{fig:anom}.  Anomalous decays
are well understood in \CPT, entering through the Wess-Zumino-Witten
(WZW) Lagrangian \cite{Wess:1971yu,Witten:1983tw}.  However, these
effects have not been investigated in the quenched and partially
quenched theories and some interesting subtleties arise.

In SU(2) \CPT, the one-pion, two-photon piece of the WZW Lagrangian,
is given by
\begin{eqnarray}
        {\cal L}_{\pi^0 \g\g} &=& - \frac{3 e^2}{16 \pi^2 f} 
                \tr \left[\phi \hat{\cal Q}^2\right]
                \epsilon^{\mu\nu\rho\sigma}F_{\mu\nu}F_{\rho\sigma}, 
\label{eq:LWZW}
\end{eqnarray}
where
\begin{equation}
        \phi = \begin{pmatrix}
                        \frac{\pi^0}{\sqrt{2}} & \pi^+ \\
                        \pi^- & -\frac{\pi^0}{\sqrt{2}}
                \end{pmatrix}\quad, \quad
        \hat {\cal Q} = \begin{pmatrix}
                        \frac{2}{3} & 0 \\
                        0 & -\frac{1}{3}
                \end{pmatrix}.
\end{equation}
This Lagrangian is completely determined as its coefficient can be
fixed by directly matching to the perturbative QCD calculation of the
relevant triangle diagram (the one loop calculation is exact
\cite{Adler:1969er}, in accordance with Witten's geometric
quantization condition \cite{Witten:1983tw}).  At higher orders,
additional anomalous operators appear \cite{Bijnens:2001bb} but they
do not contribute to Compton scattering until ${\cal O}(Q^5)$.

It is well known that quenched and partially-quenched chiral
perturbation theories generally have more complicated operator
structure than in the case of QCD ({\it e.g.}, one can not use
Cayley-Hamilton identities \cite{Sharpe:2003vy}). Thus, in order to
generalize Eq.~(\ref{eq:LWZW}) to the partially-quenched cases, we
might imagine the extended $\pi^0\to\g\g$ Lagrangian to be of the form
\begin{multline}
\label{eq:PQWZWwrong}
        {\cal L}_{\pi^0 \g\g}^{PQ} \propto 
                \epsilon^{\mu\nu\rho\sigma}F_{\mu\nu}F_{\rho\sigma}
                \Big[ 
                        c_1\, \str \left[ \Phi {\cal Q}^2 \right]  
                        +c_2\, \str \left[ \Phi {\cal Q} \right] \str
                        \left[ {\cal Q} \right]  \\
                        +c_3\, \str \left[ \Phi \right] \str
                          \left[ {\cal Q} \right] ^2  
                        +c_4\, \str \left[ \Phi \right] \str \left[
                          {\cal Q}^2 \right] \Big], 
\end{multline}
(in the quenched case only the first operator is non-vanishing, but a
similar discussion applies).  With the condition that in the QCD limit
where the sea-quark and ghost-quark masses and charges are set equal
to those of the valence-quarks, matrix elements of
Eq.~\eqref{eq:PQWZWwrong} reproduce the matrix elements of
Eq.~\eqref{eq:LWZW}. As discussed in Section~\ref{sec:hbxpt}, the
singlet field, $\Phi_0$ acquires a large mass from the strong U(1)$_A$
anomaly \cite{Sharpe:2000bn} and can be integrated out of the
partially-quenched theory; consequently, the operators proportional to
$a_3$ and $a_4$ can be ignored.  Additionally, from the multiple
super-trace structure, one can deduce that the operators
\begin{equation*}
        \str \left[ \Phi {\cal Q} \right] \str \left[ {\cal Q}
        \right], 
\quad
\str \left[ \Phi \right] \str \left[ {\cal Q} \right]^2, 
\quad 
\str \left[ \Phi \right] \str \left[
                          {\cal Q}^2 \right]\,,
\end{equation*}
have at least two closed loops at the quark level.  Following the
arguments in Refs.~\cite{Bell:1969ts,Adler:1969gk,Adler:1969er}, one
can show that these operators do not correspond to anomalous quark
level processes.  Moreover, the leading dependence of the underlying
quark-level diagrams is proportional to the quark mass, and thus the
coefficient of these operators must scale as, $a_{2,3,4} \sim
m_q/\Lambda_\chi^2$.  Although these operators contribute to $\eta_a
\to \g\g$, they are not anomalous, and only contribute at higher
orders in the chiral expansion.  We can thus conclude that the only
operator in the Lagrangian, Eq.~\eqref{eq:PQWZWwrong}, which
contributes to the anomalous decay of the neutral mesons at leading
order is $\str\left[\Phi{\cal Q}^2\right]$. The coefficient is easily
determined by matching to either perturbative partially-quenched QCD
or to the \CPT\ expression in the QCD limit.\footnote{We can thereby
  bypass the need to extend Witten's global quantization condition
  \cite{Witten:1983tw} to non-compact graded Lie groups.}  The
appropriate Lagrangian is therefore
\begin{eqnarray}
        {\cal L}_{\pi^0 \g\g}^{PQ} &=& - \frac{3 e^2}{16 \pi^2 f} 
                \str \left[\Phi {\cal Q}^2\right]
                \epsilon^{\mu\nu\rho\sigma}F_{\mu\nu}F_{\rho\sigma}\,. 
\label{eq:LWZWPQ}
\end{eqnarray}

From the above Lagrangian, it is apparent that all of the flavour
diagonal fields in Eq.~(\ref{eq:mesons}), have anomalous couplings to
two photons. Calculating the diagrams in Fig~\ref{fig:anom} leads to
the following anomalous contribution to Compton scattering on the
proton\footnote{The anomalous contribution to neutron-Compton
  scattering is given by Eq.~\eqref{eq:TAnom} with the interchange of
  $u \leftrightarrow d$.} in partially-quenched \CPT
\begin{multline}\label{eq:TAnom}
        T^{PQ,{\rm anomaly}}_{\mu\nu} = -i\ \epsilon_{\mu\nu\alpha\beta}
       \, k^{\prime\alpha}\,k^\beta\, r \cdot S\, \frac{24 e^2}{(4\pi f)^2}
                  \Bigg\{
                        2 g_A \bigg[
                        \left(q_u^2-\frac{1}{2}q_j^2-\frac{1}{2}q_l^2
                        \right) \frac{1}{r^2-m_{uu}^2} \\
                        +\frac{\left( q_j^2-q_l^2 \right)
                        }{4}\frac{\D_{lj}^2}{(r^2-m_{uu}^2)(r^2-m_X^2)}\bigg] \\ 
%       &\qquad\qquad\qquad
                 +g_1 \bigg[ \left( q_u^2 -\frac{1}{2}q_j^2
                   -\frac{1}{2}q_l^2 \right) \frac{1}{r^2-m_{uu}^2} 
                        -\left( \frac{1}{2}q_j^2
                          +\frac{1}{2}q_l^2-q_d^2 \right)
                        \frac{1}{r^2-m_{dd}^2} \\ 
                        -\frac{ \left( q_j^2-q_l^2 \right)}{4} 
                                \frac{\D_{lj}^2}{(r^2-m_X^2)} \left(
                                  \frac{1}{(r^2-m_{uu}^2)} 
                                                +\frac{1}{(r^2-m_{dd}^2)} \right) 
                \bigg] \Bigg\}.
\end{multline}
In the above expression, $r=q^\prime-q$, is the momentum transfer to
the nucleon and $\D_{lj}^2 = m_{ll}^2-m_{jj}^2$ is a measure of the
isospin breaking in the sea sector.  In the sea isospin limit ($m_l
\to m_j$), the double pole structure of the amplitude vanishes, and in
the QCD limit all dependence on $g_1$ vanishes.

Expanding Eq.~(\ref{eq:TAnom}) in frequency and comparing with
Eq.~(\ref{eq:Ti}) leads to the following anomalous contributions to
the polarizabilities:
\begin{eqnarray}
        \alpha^{\rm anomaly}&=& 0\,,\\
        \beta^{\rm anomaly}&=& 0\,,\\
        \gamma_1^{\rm anomaly}&=& -\frac{3e^2G_{\rm anom}}{8\pi^3
                f^2m_\pi^2}\,, \\
        \gamma_2^{\rm anomaly}&=& 0\,,\\
        \gamma_3^{\rm anomaly}&=& \frac{3e^2G_{\rm anom}}{16\pi^3
                f^2m_\pi^2} \,,\\
        \gamma_4^{\rm anomaly}&=& -\frac{3e^2G_{\rm anom}}{16\pi^3
                f^2m_\pi^2} \,,
\end{eqnarray}
where the coefficients, $G_{\rm anom}$, are given in
Table~\ref{tab:couplings} (at the end of this chapter) for the different theories under
consideration.  These contributions vanish in the iso-scalar
combination of proton and neutron targets in the QCD limit.

\subsubsection{\textbf{$\Delta$ resonance contributions}}
\label{sec:Deltares}
The contributions to the amplitude from the Born-terms involving the
{\bf 44}-plet resonance (which contains the $\Delta$-isobar),
Fig.~\ref{fig:Born}, are identical in \CPT, \PQCPT\ and \QCPT\ as they
are purely valence quark processes. They are given by
\begin{eqnarray}
\alpha^\Delta &=& 0\,, \\
\beta^\Delta &=&   \mu_T^2\frac{e^2(q_u-q_d)^2}{36 \pi ( 2 M_N)^2 \Delta} \,,\\
\gamma_1^\Delta &=& 0 \,,\\
\gamma_2^\Delta &=&  -  \mu_T^2\frac{e^2(q_u-q_d)^2}{72 \pi ( 2 M_N)^2 \Delta^2} \,,\\
\gamma_3^\Delta &=& 0 \,,\\
\gamma_4^\Delta &=&  \mu_T^2\frac{e^2(q_u-q_d)^2}{72 \pi ( 2 M_N)^2 \Delta^2}\,,
\end{eqnarray}
where $\mu_T$ is the magnetic dipole transition coupling of
Eq.~(\ref{L1}).

%
%
%
%
%
% Loops
%
%
%

\subsection{\textbf{Infinite volume}\label{sec:inf_vol}}

The loop contributions to infinite volume chiral expansion of the
polarizabilities in \xpt\ are well known at order $Q^3$
\cite{Bernard:1991rq,Bernard:1991ru,Butler:1992ci,Hemmert:1996rw,Hemmert:1997tj,Pascalutsa:2002pi,Pascalutsa:2003zk}
and at ${\cal O}(Q^4)$
\cite{Bernard:1993ry,Bernard:1993bg,Ji:1999sv,VijayaKumar:2000pv,Gellas:2000mx,McGovern:2001dd,Beane:2002wn,Beane:2004ra}
(at this order, the $\Delta$-resonances have not been included as
dynamical degrees of freedom, restricting the range of applicability
to $m_\pi\ll\Delta$). Since the photon only couples to charged mesons,
the results in the quenched and partially-quenched theories are
similar to those in \xpt.  In particular, no quenched or
partially-quenched sicknesses (double pole contributions from neutral
meson propagators) enter expression for the loop diagrams.  In
general, the quenched power counting presents differences for
electromagnetic observables
\cite{Savage:2001dy,Arndt:2003vd,Arndt:2003ww,Arndt:2003we}, however
no new contributions appear at the order we work.

Using the effective couplings $G_B$, $G_B^\prime$, $G_T$ and
$G_T^\prime$ given in Table~\ref{tab:couplings}, we find that the loop
contributions to the polarizabilities are
\begin{sidewaystable}[h!]
\caption[Various electromagnetic couplings for contributions to nucleon polarizabilities]{\label{tab:couplings} Effective couplings for the various contributions to the polarizabilities.}
\centering
\begin{tabular}{| c | c | c | c |}
\hline
& QCD & QQCD & PQQCD \\ 
\hline
        $G_{\rm anom}$ 
        & $g_A\, (2Z-1)(q_u^2-q_d^2)$ & $2 g_A \left( Z q_u^2+ (1-Z) q_d^2 \right)$
        & $g_A \Big[ 2\Big( Z q_u^2+ (1-Z) q_d^2 \Big) -q_j^2-q_l^2\Big]$ 
\\
        & & $+g_1 \left(q_d^2+q_u^2\right)$ 
        & $+g_1\left(q_d^2-q_j^2-q_l^2+q_u^2\right)$    
\\ \hline 
      $G_B$ & $4g_A^2\left(q_d-q_u\right)^2$ 
      & $\frac{1}{3}\left(4g_A^2-4g_A g_1-5g_1^2\right)\left(q_d-q_u\right)^2$ 
      & $-\frac{1}{3} \left(5 g_1^2+4 g_A g_1-4
        g_A^2\right) \left(q_d-q_u\right)^2$
\\ \hline
      $G_B^\prime$ & 0 & 0 
      & $\begin{array}{l}
        \frac{1}{3} \Big[\left(6 q_d^2-6
          \left(q_j+q_l\right) q_d+5 q_j^2+5 q_l^2+4 q_u^2-4 q_j q_u-4
          q_l q_u\right) g_1^2 \\
        \hspace*{5mm}+4 g_A \left(q_j^2-2 q_u 
          q_j+q_l^2+2 q_u^2-2 q_l q_u\right) g_1 \\
        \hspace*{5mm}+8 g_A^2 \left(q_j^2-2 q_u
          q_j+q_l^2+2 q_u^2-2 q_l q_u\right)\Big]\end{array}$ 
\\ \hline
      $G_T$ 
      & $\frac{4}{3}g_{\Delta N}^2\left(q_d-q_u\right)^2$
      & $\frac{5}{6}g_{\text{$\Delta $N}}^2\left(q_d-q_u\right)^2$
      & $\frac{5}{6}
      g_{\text{$\Delta $N}}^2 \left(q_d-q_u\right)^2$ 
\\ \hline
      $G_T^\prime$ &  0 & 0 & $\frac{1}{6} g_{\text{$\Delta $N}}^2
      \left(4 q_d^2-4 \left(q_j+q_l\right) q_d+3 q_j^2+3 q_l^2+2
        q_u^2-2 q_j q_u-2 q_l q_u\right)$ \\
\hline
\end{tabular}
\end{sidewaystable}

\begingroup
\small
\begin{eqnarray}\label{eq:alpha_inf}
	\alpha^{\rm loop}&=&
		\frac{e^2}{4\pi f^2}\left[\frac{5 G_B}{192 \pi}\frac{1}{m_\pi} 
		+\frac{5 G_B^\prime}{192 \pi}\frac{1}{m_{uj}} 
		+\frac{G_T}{72 \pi^2} F_\alpha(m_\pi,\D)
		+\frac{G_T^\prime}{72 \pi^2}F_\alpha(m_{uj},\D) \right],
\end{eqnarray}
\begin{eqnarray}\label{eq:beta_inf}
	\beta^{\rm loop}&=&
		\frac{e^2}{4\pi f^2}\left[\frac{ G_B}{384 \pi}\frac{1}{m_\pi} 
		+\frac{G_B^\prime}{384 \pi}\frac{1}{m_{uj}} 
		+\frac{G_T}{72 \pi^2}F_\beta(m_\pi,\D)
		+\frac{G_T^\prime}{72 \pi^2}F_\beta(m_{uj},\D)\right],
\end{eqnarray}
\begin{eqnarray}\label{eq:gamma1_inf}
	\gamma_1^{\rm loop}&=&
		\frac{e^2}{4\pi f^2}\left[\frac{ G_B}{48 \pi^2}\frac{1}{m_\pi^2} 
		+\frac{G_B^\prime}{48 \pi^2}\frac{1}{m_{uj}^2} 
		+\frac{G_T}{72 \pi^2} F_{\gamma_1}(m_\pi,\D)
		+\frac{G_T^\prime}{72 \pi^2}F_{\gamma_1}(m_{uj},\D) \right],
\end{eqnarray}
\begin{eqnarray}
	\label{eq:gamma2_inf}
		\gamma_2^{\rm loop}&=&
		\frac{e^2}{4\pi f^2}\left[\frac{ G_B}{96 \pi^2}\frac{1}{m_\pi^2} 
		+\frac{G_B^\prime}{96 \pi^2}\frac{1}{m_{uj}^2} 
		+\frac{G_T}{72 \pi^2} F_{\gamma_2}(m_\pi,\D)
		+\frac{G_T^\prime}{72 \pi^2}F_{\gamma_2}(m_{uj},\D) \right],
\end{eqnarray}
\begin{eqnarray}\label{eq:gamma3_inf}
	\gamma_3^{\rm loop}&=&
		\frac{e^2}{4\pi f^2}\left[\frac{ G_B}{192 \pi^2}\frac{1}{m_\pi^2} 
		+\frac{G_B^\prime}{192 \pi^2}\frac{1}{m_{uj}^2} 
		+\frac{G_T}{144 \pi^2} F_{\gamma_3}(m_\pi,\D)
		+\frac{G_T^\prime}{144 \pi^2}F_{\gamma_3}(m_{uj},\D) \right],
\end{eqnarray}
\begin{eqnarray}\label{eq:gamma4_inf}
	\gamma_4^{\rm loop}&=&
		-\frac{e^2}{4\pi f^2}\left[\frac{ G_B}{192 \pi^2}\frac{1}{m_\pi^2} 
		+\frac{G_B^\prime}{192 \pi^2}\frac{1}{m_{uj}^2} 
		+\frac{G_T}{144 \pi^2} F_{\gamma_4}(m_\pi,\D)
		+\frac{G_T^\prime}{144 \pi^2}F_{\gamma_4}(m_{uj},\D) \right],
\end{eqnarray}
\endgroup
where
\begin{eqnarray}\label{eq:8}
	F_\alpha(m,\D) &=& \frac{9\D}{\D^2-m^2}
		-\frac{\D^2-10m^2}{2 (\D^2-m^2)^{3/2}}\ln 
		\left[\frac{\D -\sqrt{\D^2 - m^2 + i \epsilon}}{\D + \sqrt{\D^2 - m^2 + i \e}} \right] \,, \\
	F_\beta(m,\D) &=& - \frac{1}{2(\D^2-m^2)^{1/2}}\ln 
		\left[\frac{\D -\sqrt{\D^2 - m^2 + i \e}}{\D + \sqrt{\D^2 - m^2 + i \e}} \right]\,, \\
	F_{\gamma_1}(m,\D) &=& - \frac{\D^2 + 2 m^2}{(\D^2 - m^2)^2} -
		\frac{3 \D m^2}{2 (\D^2 - m^2)^{5/2}} 
		\ln \left[\frac{\D - \sqrt{\D^2 - m^2 + i \e}}{\D + \sqrt{\D^2 - m^2 + i \e}}\right] \,,\\
	F_{\gamma_{2,3,4}}(m,\D) &=& 
	\frac{1}{\D^2 - m^2} + \frac{\D}{2 (\D^2 - m^2)^{3/2}} \ln
		\left[\frac{\D -\sqrt{\D^2 - m^2 + i \e}}{\D + 
			\sqrt{\D^2 - m^2 + i \e}}\right] \,.
\end{eqnarray}
Here we have used dimensional regularization, however the results are
finite and hence independent of the regulator without the addition of
counterterms. These loop contributions vanish at zero photon
frequency, preserving the Thomson limit. They are identical for both
proton and neutron targets, though isospin breaking effects from
loops enter at ${\cal O}(Q^4)$ in the expansion.  In the \xpt\ case,
these results reproduce those of
Refs.~\cite{Hemmert:1996rw,Hemmert:1997tj}.

%
%
%
%
%
%
%       Nucleon polarizabilities: Finite Volume
%
%
%
%
%
%
\subsection{\textbf{Finite volume}\label{sec:p_regime}}

In momentum space, the finite volume of a lattice simulation restricts the available momentum modes and consequently the results differ from their infinite volume values. These long-distance effects can be
accounted for in the low-energy effective theory. Here we shall consider a hyper-cubic box of dimensions $L^3\times T$ with $T\gg L$.  Imposing periodic boundary conditions on mesonic fields leads to quantized momenta $k=(k_0,{\vec k})$, ${\vec k}=\frac{2\pi}{L} {\vec j}=\frac{2\pi}{L} (j_1,j_2,j_3)$ with $j_i\in \mathbb{Z}$, but $k_0$ treated as continuous.  On such a finite volume, spatial momentum
integrals are replaced by sums over the available momentum modes. This leads to modifications of the infinite volume results presented in the previous section; the various functions arising from loop integrals
are replaced by their finite volume (FV) counterparts. In a system where $m_\pi L\gg 1$, the power counting of the infinite volume low-energy effective theory remains valid and finite volume effects are predominantly from Goldstone mesons propagating to large distances where they are sensitive to boundary conditions and can even ``wrap around the world''.  Since the lowest momentum mode of the Goldstone propagator is $\sim \exp(-m_\pi L)$ in position space, finite volume effects will behave as a polynomial in $1/L$ times this exponential if no multi-particle thresholds are reached in the loop (as is the case in these calculations provided the photon energy is small enough, $\omega\lesssim m_\pi$).

Repeating the calculation of the loop diagrams using finite volume sums rather than integrals leads to the following expressions for the loop contributions to the polarizabilities:
%
%\begingroup
%\small
\begin{multline}\label{eq:alpha_fv}
	\alpha^{\rm loop}(L) = \frac{e^2}{1152 \pi f^2} \int_0^\infty
		d\lambda\Big[ 3G_B {\cal F}_\alpha({\cal M}_{uu})
		+ 3G_B^\prime {\cal F}_\alpha({\cal M}_{uj}) \\
		+ 8G_T {\cal F}_\alpha({\cal M}_{uu}^\Delta)
		+ 8G_T^\prime {\cal F}_\alpha({\cal M}_{uj}^\Delta) \Big]\,,
\end{multline}
\begin{multline}\label{eq:beta_fv}
	\beta^{\rm loop}(L) = \frac{e^2}{1152 \pi f^2 }\int_0^\infty
		d\lambda\Big[
		3G_B {\cal F}_\beta({\cal M}_{uu})
		+ 3G_B^\prime {\cal F}_\beta({\cal M}_{uj}) \\
		+ 8G_T {\cal F}_\beta({\cal M}_{uu}^\Delta)
		+ 8G_T^\prime {\cal F}_\beta({\cal M}_{uj}^\Delta) \Big]\,,
\end{multline}
\begin{multline}\label{eq:gamma1_fv}
	\gamma_1^{\rm loop}(L) =  \frac{7e^2}{576 \pi  f^2}\int_0^\infty
		d\lambda\Big[
		3G_B {\cal F}_{\gamma_1}({\cal M}_{uu})
		+ 3G_B^\prime {\cal F}_{\gamma_1}({\cal M}_{uj}) \\
		-4G_T {\cal F}_{\gamma_1}({\cal M}_{uu}^\Delta)
		-4 G_T^\prime {\cal F}_{\gamma_1}({\cal M}_{uj}^\Delta) \Big]\,,
\end{multline}
\begin{multline}\label{eq:gamma2_fv}
	\gamma_2^{\rm loop}(L) =  \frac{7e^2}{64 \pi f^2}\int_0^\infty
		d\lambda\Big[
		3G_B {\cal F}_{\gamma_2}({\cal M}_{uu})
		+ 3G_B^\prime {\cal F}_{\gamma_2}({\cal M}_{uj}) \\
		-4 G_T {\cal F}_{\gamma_2}({\cal M}_{uu}^\Delta)
		-4 G_T^\prime {\cal F}_{\gamma_2}({\cal M}_{uj}^\Delta) \Big]\,,
\end{multline}
\begin{multline}\label{eq:gamma3_fv}
	\gamma_3^{\rm loop}(L) = \frac{7e^2}{1152 \pi f^2}\int_0^\infty
		d\lambda\Big[
		3G_B {\cal F}_{\gamma_3}({\cal M}_{uu})
		+3 G_B^\prime {\cal F}_{\gamma_3}({\cal M}_{uj}) \\
		-4 G_T {\cal F}_{\gamma_3}({\cal M}_{uu}^\Delta)
		-4 G_T^\prime {\cal F}_{\gamma_3}({\cal M}_{uj}^\Delta) \Big]\,,
\end{multline}
\begin{equation}
  \label{eq:gamma4_fv}
  \gamma_4^{\rm loop}(L) = -\gamma_3^{\rm loop}(L)\,,
\end{equation}
%\endgroup
%
where ${\cal M}_{ab}=\sqrt{m_{ab}^2+\lambda^2}$ and ${\cal
  M}_{ab}^\Delta=\sqrt{m_{ab}^2+2\lambda \Delta +\lambda^2}$ and
\begin{multline}\label{eq:F_alpha}
	\mc{F}_\alpha(m) = 
		180 \lambda ^2 {\cal I}_{\frac{7}{2}}(m)
		+190 \mc{J}_{\frac{7}{2}}(m)
		-280 \lambda ^2 {\cal J}_{\frac{9}{2}}(m) \\
		-455 \mc{K}_{\frac{9}{2}}(m)
		+315 \lambda ^2 {\cal K}_{\frac{11}{2}}(m)
		+252 \mc{L}_{\frac{11}{2}}(m) \,,
\end{multline}
\begin{eqnarray}\label{eq:F_beta}
  {\cal F}_\beta(m) &=& 60 {\cal J}_{\frac{7}{2}}(m)-224
  {\cal K}_{\frac{9}{2}}(m)+189 {\cal L}_{\frac{11}{2}}(m) \,,
\\
  \label{eq:F_gamma1}
  {\cal F}_{\gamma_1}(m) &=& 30 \lambda ^3
  {\cal I}_{\frac{9}{2}}(m)+10 \lambda {\cal J}_{\frac{9}{2}}(m)-45
\lambda ^3
  {\cal J}_{\frac{11}{2}}(m)-18 \lambda{\cal K}_{\frac{11}{2}}(m)  \,,
\\
  \label{eq:F_gamma2}
  {\cal F}_{\gamma_2}(m) &=&\lambda{\cal K}_{\frac{11}{2}}(m) \,,
\\
  \label{eq:F_gamma34}
  {\cal F}_{\gamma_3}(m) &=& 10\lambda
  {\cal J}_{\frac{9}{2}}(m)-9 \lambda{\cal K}_{\frac{11}{2}}(m)  \,,
\end{eqnarray}
and the finite volume sums ${\cal I}_\beta(m)$, \ldots, ${\cal
  L}_\beta(m)$ are defined in Appendix \ref{FV_app}. These 
expressions reduce to the results of
Eqs.~(\ref{eq:alpha_inf})--(\ref{eq:gamma4_inf}) above in the infinite
volume limit. 

\begin{figure}[!t]
\centering
        \includegraphics[width=\columnwidth]{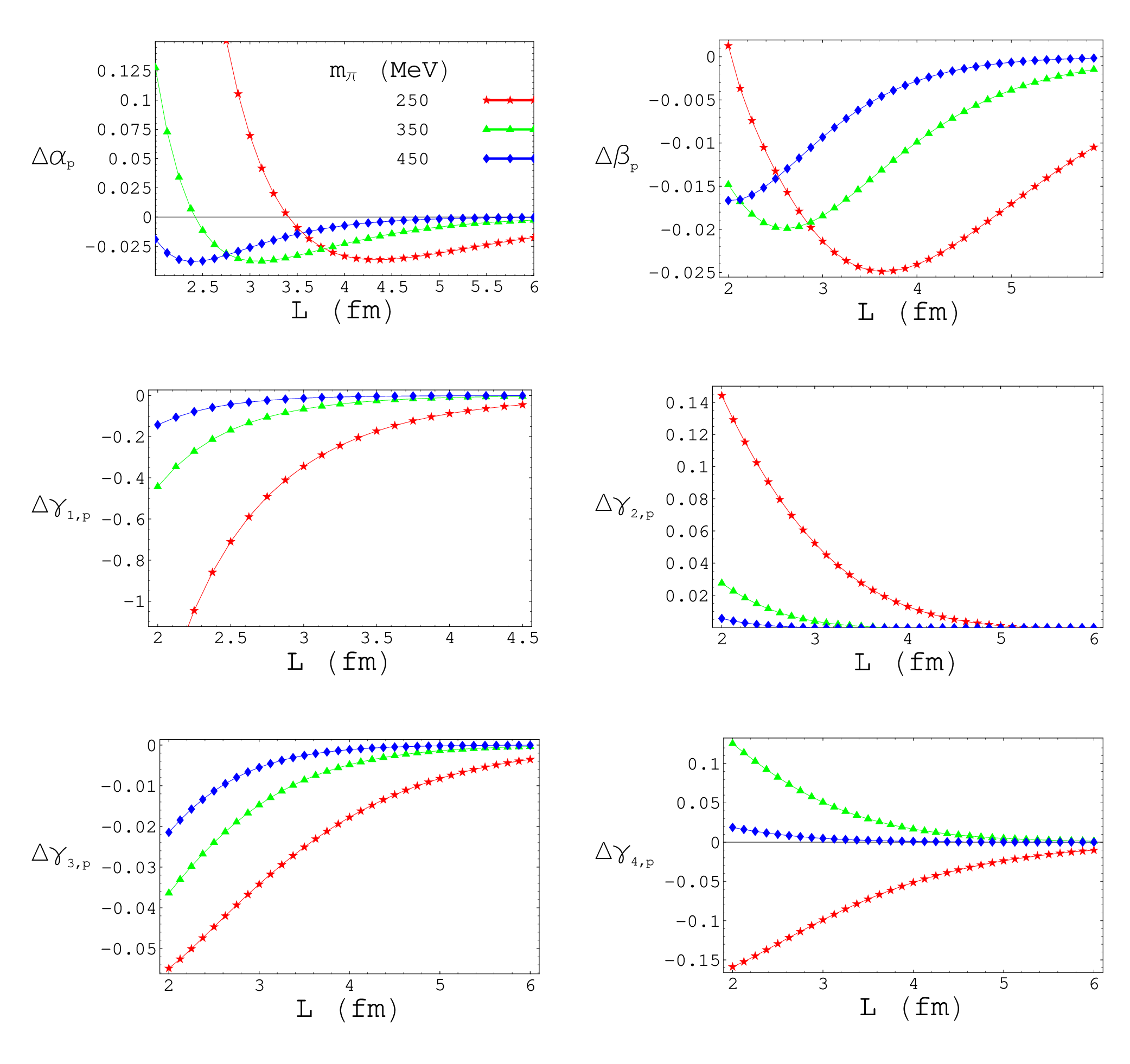}
\caption[Volume dependence of the proton polarizabilities]{\label{fig:FV1}Volume dependence of the proton polarizabilities. Here we show the ratio of the difference of the finite and infinite volume results to the infinite volume results for three values of the pion mass using the parameters described in the text. The curves terminate at $m_\pi\ L=3$.}
\end{figure}

\begin{figure}[!th]
\centering
        \includegraphics[width=\columnwidth]{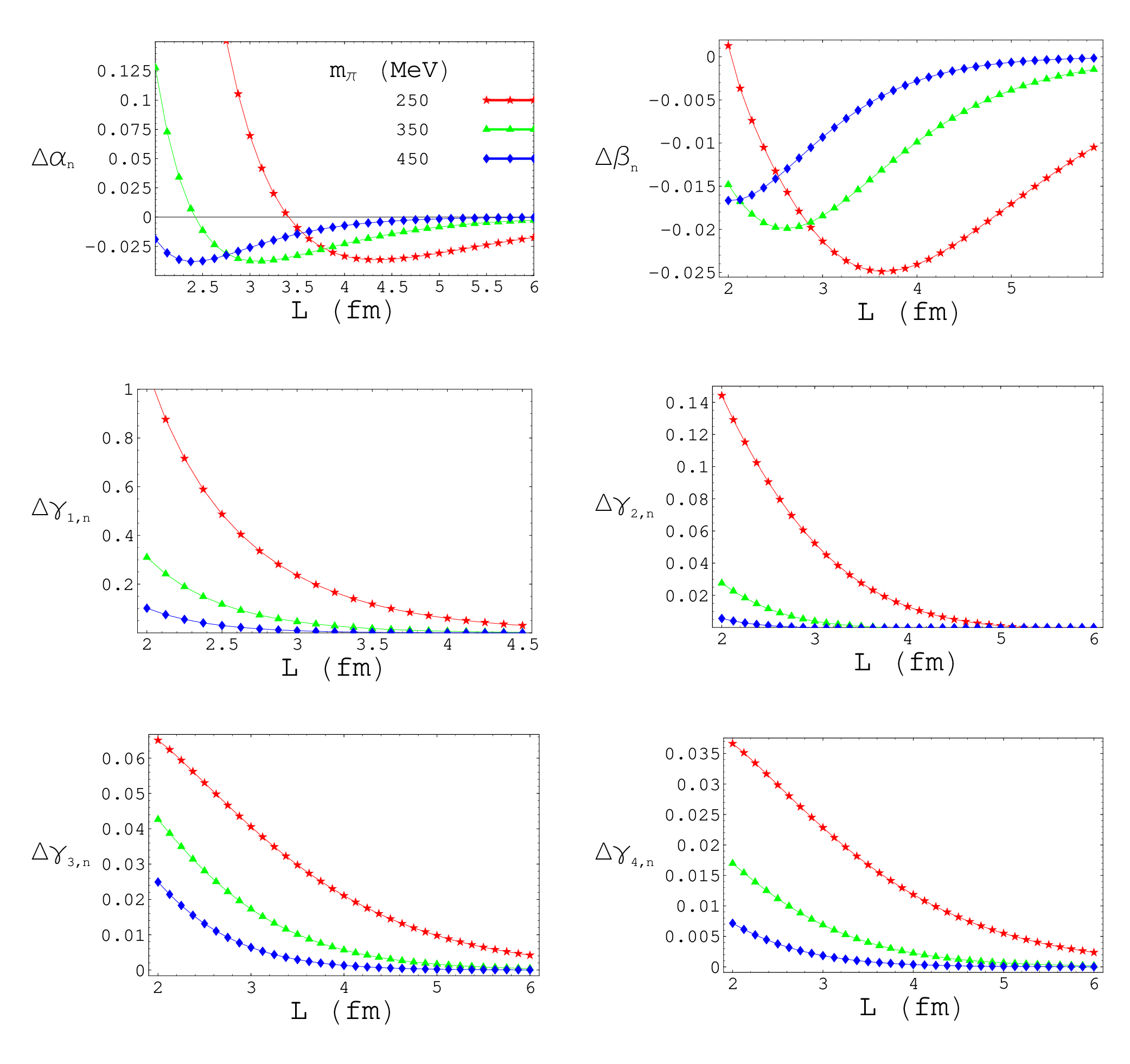}
\caption[Volume dependence of the neutron polarizabilities]{\label{fig:FV2}Volume dependence of the neutron polarizabilities. The various curves are as in Fig.~\protect\ref{fig:FV1}. }
\end{figure}

To illustrate these effects, Figs. \ref{fig:FV1} and \ref{fig:FV2}
show the volume dependence of the various polarizabilities in the
proton and the neutron, respectively. Here we have specialized to QCD,
setting $q_u=2/3$, $q_d=-1/3$, $g_A=1.25$, $|g_{N\Delta}|=1.5$,
$\mu_T=10.9$, $f=0.132$~GeV, $M_N=0.938$~GeV and
$\Delta=0.3$~GeV.%
\footnote{The value of $\mu_T$ is chosen to
  correspond to that found in analysis of Ref.~\cite{Hemmert:1997tj}
  ($\mu_T=2\sqrt{2}b_1$ of that reference).  In principle this LEC can
  be determined from an analysis of lattice polarizabilities or
  $N$--$\Delta$ transition matrix elements.}
In each plot we show results for the ratio
\begin{equation}
  \Delta X(L)=\frac{X(L)-X(\infty)}{X(\infty)}\,,
\end{equation}
for the six polarizabilities at three different pion masses,
$m_\pi=0.25,\,0.35,\,0.50$~GeV. The overall magnitude of these shifts
varies considerably; generally volume effects are at the level of
5--10\% for $m_\pi=0.25$~GeV and smaller for larger masses. Larger
effects are seen in a number of the spin polarizabilities but there
are as yet no lattice calculations of these quantities. The magnetic
polarisability has a particularly small volume dependence which can be
understood from the large decuplet resonance contribution that is
independent of the volume.

\begin{figure}[!t]
\centering
        \includegraphics[width=\columnwidth]{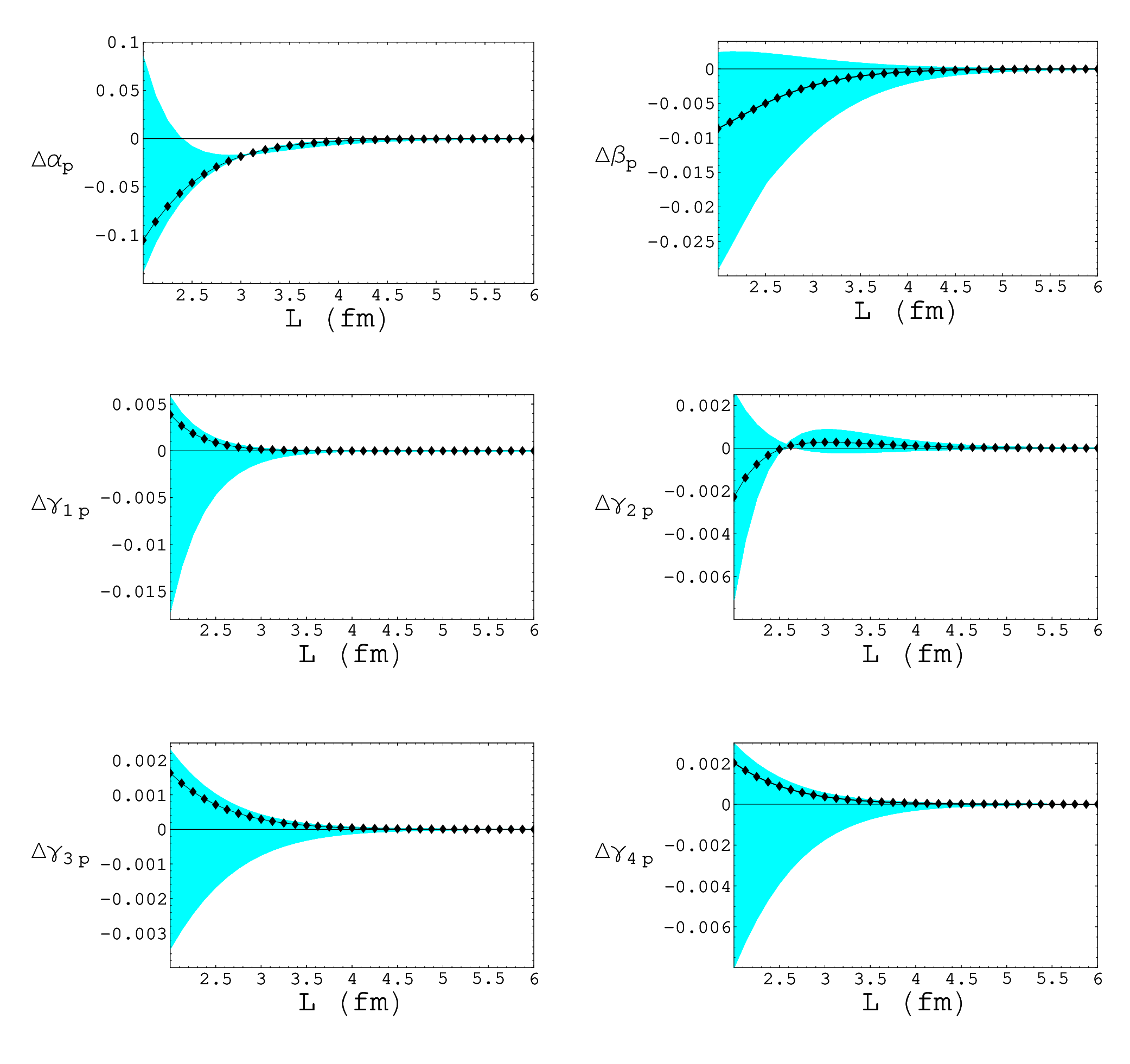}
\caption[Leading volume dependence of the proton polarizabilities in quenched QCD.]{\label{fig:FVQ_p} Volume dependence of the proton polarizabilities in quenched
  QCD at the lightest quark mass used in the lattice calculations of
  Refs.~\protect{\cite{Christensen:2004ca,Lee:2005dq}}. The central
  curves and shaded region correspond to the parameters quoted in the
  text.}

\end{figure}

\begin{figure}[!t]
\centering
        \includegraphics[width=\columnwidth]{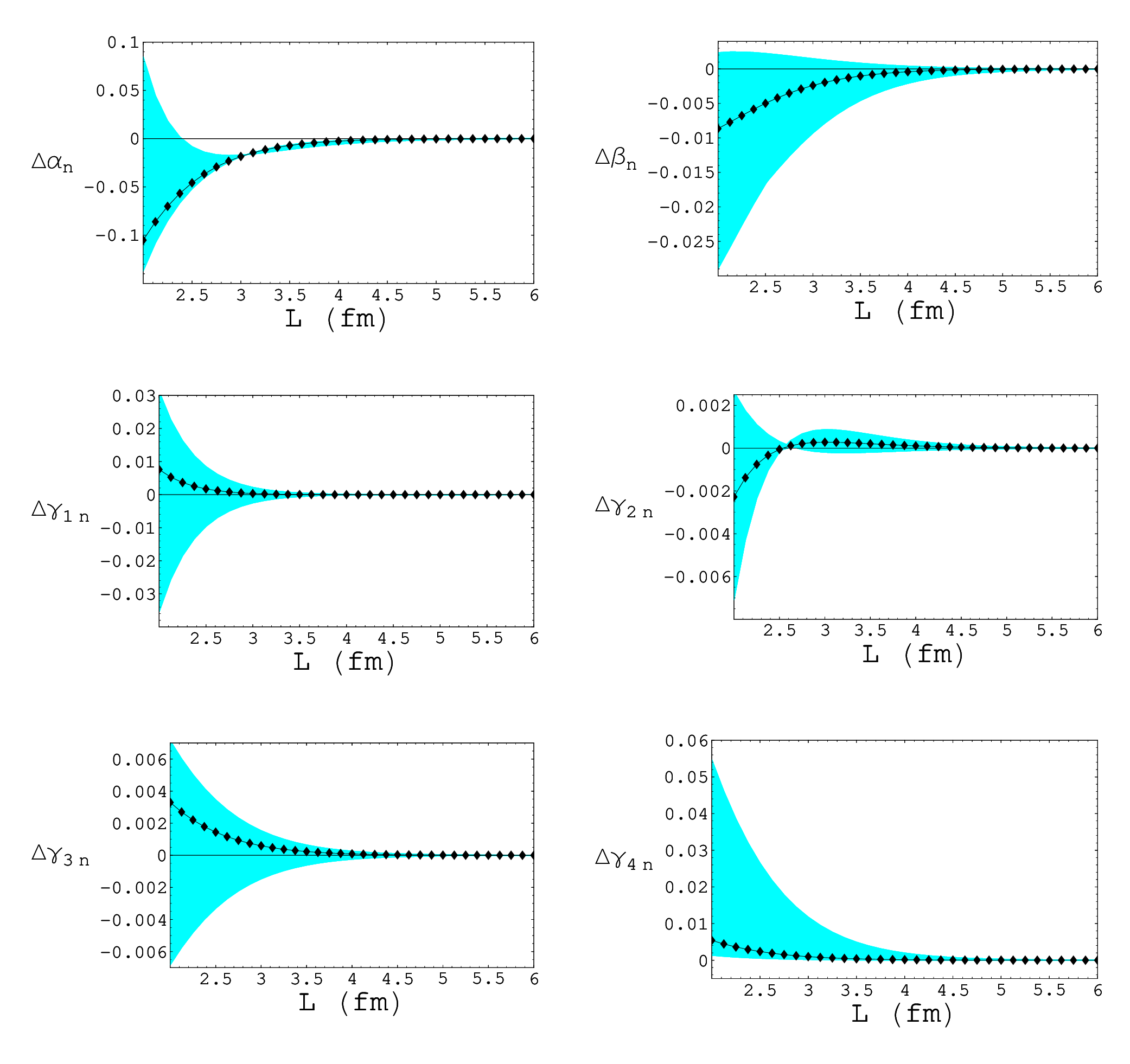}
\caption[Leading volume dependence of neutron polarizabilities in quenched QCD]{\label{fig:FVQ_n}As in Fig.~\protect\ref{fig:FVQ_p}, but for the neutron.}

\end{figure}

The above expressions also allow us to calculate the finite volume
effects in the quenched data on the various polarizabilities
calculated in Refs.~\cite{Christensen:2004ca,Lee:2005dq}. The quenched
expressions involve a number of undetermined LECs (quenched $g_A$,
$g_1$, $g_{N\Delta}$ and $\mu_T$ are unrelated to their \pqxpt/\xpt\ 
values), so we can only estimate the volume effects. To do so, we
choose $q_u=2/3$, $q_d=-1/3$, $g_A=1.25$, $g_1=1$,
$|g_{N\Delta}|=1.5$, $\mu_T=5.85$, $f=0.132$~GeV, $M_N=0.938$~GeV and
$\Delta=0.3$~GeV and take the pion masses corresponding to the
lightest used in these lattice calculations, $m_\pi\sim 0.5$~GeV (we
ignore the issue of the convergence of \xpt\ at such masses). The
results for the volume dependence of the various polarizabilities of
the proton and neutron are shown in Figs.~\ref{fig:FVQ_p} and
\ref{fig:FVQ_n}. In each plot, the shaded region corresponds to
reasonable variation of the unknown couplings, $1<g_A<1.75$, $-1<g_1<1$,
$0.8<|g_{N\Delta}|<2$ and $2.8<|\mu_T|<8.5$.  From these figures, we
see that the calculations on a (2.4 fm)$^3$ lattice with $m_\pi=$0.5~GeV
may differ from their infinite volume values by 5--10\% in the case of
the electric polarisability and a few percent for the magnetic and
spin polarizabilities.

\begin{figure}[!t]
\centering
        \includegraphics[width=\columnwidth]{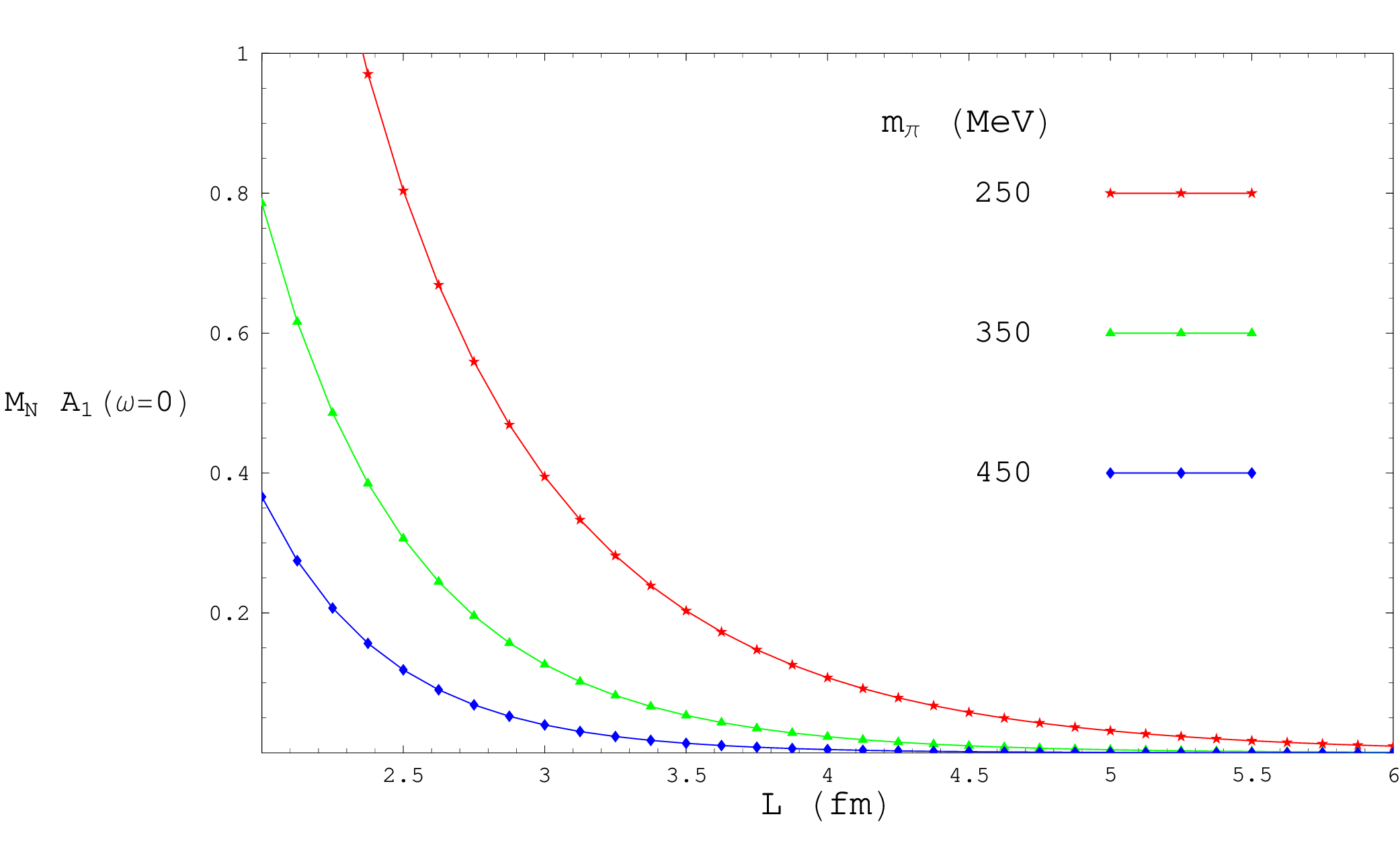}
\caption[Thomson limit of photon neutron scattering]{\label{fig:FVThomson} Volume dependence of the Thomson limit of photon neutron scattering.}
\end{figure}

An interesting effect that arises at finite volume is that the Thomson-limit and other Born terms in the frequency expansion of the scattering amplitude (terms in Eq.~(\ref{eq:amplitudes}) that are not polarizabilities) receive finite volume contributions from the loop diagrams in Fig.~\ref{fig:Q3} that vanish exponentially as the volume is increased. As an example, the amplitude for Thomson-limit (zero
frequency) scattering on the neutron (which is identically zero at infinite volume) is shown in Fig~\ref{fig:FVThomson}. This result is somewhat counter-intuitive, but arises from the effects of the periodic boundary conditions on the long range charge distribution of the hadron. It does not imply the non-conservation of charge.  One way to understand this feature is that in a finite box with periodic boundary conditions, only quantized momentum are allowed, including the momentum of the photon field.  The Thomson limit is defined as the limit of a zero-frequency photon scattering off a charged particle, and thus can only resolve the total charge of an object.  However, to take the limit of zero-frequency, one must first take the box size to infinity, such that one can smoothly vary the photon energy to zero.  But at infinite volume, as seen in Fig.~\ref{fig:FVThomson}, the deviation of the zero-frequency limit scattering from the na{\"i}eve infinite volume Thomson limit goes to zero.  Thus, this is another example where the order of limits is important, and in this case one must take the infinite volume limit before the zero-frequency limit.

The results presented here all assume that the higher order terms in the $Q$ expansion provide small contributions to the volume dependence of the polarizabilities. This may or may not be the case as diagrams that are formally of higher order in the infinite volume \xpt\ power-counting can have volume effects that are enhanced over those at lower infinite volume order (see Ref.~\cite{Detmold:2005pt} for a
detailed discussion). Such issues may be particularly relevant for the polarizabilities where the convergence of the chiral expansion is tenuous. In this regard, studying the FV behaviour of the lattice
results may in fact be a useful diagnostic tool with which to determine if or why the convergence is poor.

%
%
%
%
%
%
%       Discussion
%
%
%
%
%
%
\section{Conclusion \label{sec:discussion}}

We have investigated Compton scattering from spin-half targets from
the point of view of lattice QCD. We first discussed how external
field methods can be used to probe all six polarizabilities of real
Compton scattering for both charged and uncharged targets. Such
calculations will tell us a lot about the low energy QCD structure of
hadrons and will be of much use in phenomenological studies requiring
the full set of polarizabilities as only certain linear combinations
are available from current
experiments~\cite{Schumacher:2005an,Hyde-Wright:2004gh}. The
techniques discussed here also allow us to extract other electric
properties of charged particles using external fields including the
electric dipole moment of the proton and the quadrupole moment of the
deuteron.

Our second major focus was on the effects of the finite volume used in
lattice calculations on the polarizabilities. Since polarizabilities
are infrared-sensitive observables (they scale as inverse powers of
the pion mass near the chiral limit), the are expected to have strong
volume dependence. This is indeed borne out in the explicit
calculations presented here. In QCD, we generically find that the
polarizabilities experience volume shifts of 5--10\% from the infinite
volume values for lattice volumes $\sim$(2.4~fm)$^3$ and pions of mass
0.25~GeV. The electric and first spin polarizabilities are
particularly sensitive. In the case of quenched QCD (relevant to the
only existing lattice data), we find significant effects even at pion
masses $\sim0.5$~GeV. Future lattice studies of the polarizabilities
should take these effects into account in order to present physically
relevant results.

As extensions of this work, one can also consider the generalized
polarizabilities, higher-order polarizabilities and parity violating
polarizabilities (see Ref.~\cite{Bedaque:1999dh}) all of which can be
extracted from appropriate lattice calculations similar to those
detailed in Section~\ref{sec:Compton}. Such information would lead to
a further-improved understanding of the low-energy structure of the
hadrons and prove very useful in directing the next generation of
precision Compton scattering experiments. The lattice provides a novel
opportunity to study the neutron polarizabilities directly instead of
from nuclear targets and extending the lattice methods of Section
\ref{sec:Compton} to the deuteron (along similar lines to those
discussed in Ref.~\cite{Detmold:2004qn}) will also prove useful for
comparison to experiment.

% ========== Chapter 5: 2-hadron systems - FV
\chapter{Two-Hadron Interactions:\\ $\pi\pi$ Scattering in a Finite Volume}\label{chap:pipiFV}

The work in this chapter is based upon Refs.~\cite{Bedaque:2006yi}

%
%
%
%
%
%
%	Introduction
%
%
%
%
%
%
\section{Introduction}
Except in the case of infinitely heavy baryons, where an adiabatic potential can be defined, the interaction between two hadrons is studied with lattice QCD by numerically calculating energy levels of the system in a finite box. This is because in the infinite volume limit and away from kinematical thresholds, the two-hadron Euclidean correlator gives no information about the Minkowski space amplitude~\cite{Maiani:1990ca}. The  alternative is to consider the system in a universe with finite spatial extent (a \textit{finite box}), as is the case with numerical calculations anyway. The energy levels of a system composed of two hadrons are not simply the sum of the energies carried by each hadron, but there is an additional (usually small) shift that arises due to the interaction between them. The smaller the box, the larger the shift in energy levels.  This volume dependence is inversely proportional to the volume and furthermore, there is a relation between the energy level shifts and the scattering phase shifts~\cite{Hamber:1983vu,Luscher:1990ux,Luscher:1986pf}. This relation, valid for energies below the first inelastic threshold is a consequence of unitarity and is thus model 
independent.%
\footnote{By model independent relation we mean  a relation valid whether one is considering QCD or some other theory, as long as this theory obeys unitarity, locality, \textit{etc}.}
In addition to this power law shift in the energy levels, there are exponentially suppressed corrections which are {\it not} model independent and are the analogue of the exponentially suppressed corrections to the mass, decay constants, \textit{etc}., in the single-hadron sector~%
\cite{Luscher:1985dn,Gasser:1986vb,Colangelo:2003hf,Beane:2004tw,Colangelo:2005gd}. These exponential volume effects arise because the off-shell propagation of intermediate states is altered by the presence of the finite box, which allows them, for instance, to ``wrap around" the lattice.  As such, these effects are dominated by the lightest particle, the pion in QCD, and are proportional to $e^{-m_\pi L}$ with $m_\pi$ the pion mass and $L$ the linear dimension of the box. For simulations done with small enough quark masses such that the pions are within the chiral regime, these soft pion effects can be computed using the chiral perturbation theory (\CPT)~\cite{Gasser:1983yg,Gasser:1984gg}.
The $\pi\pi$ scattering phase shifts have been computed using lattice QCD following the universal finite volume method mentioned above~\cite{Sharpe:1992pp,Gupta:1993rn,Kuramashi:1993ka,Fukugita:1994na,Fukugita:1994ve,Liu:2001zp,Aoki:2002in,Aoki:2002ny,Yamazaki:2004qb,Du:2004ib,Aoki:2005uf,Beane:2005rj}. As the chiral limit %(or realistic quark masses) 
is approached~\cite{Beane:2005rj} and more precise calculations appear, these exponentially suppressed corrections will need to be understood. 
% Our goal in this paper is to compute the dominant exponential volume 
% dependence for the two-pion system, which occurs through the leading 
% loop-order of 2-flavor \CPT.

Our goal in this paper is first to show the modification of 
the universal scattering formula for a hadron-hadron system in a box
due to the exponentially suppressed finite volume corrections.
Second, we compute the dominant exponential volume dependence explicitly
for the two-pion system in $I=2$ channel near threshold by use of the 
leading loop-order two-flavor \CPT.

%
%	Volume Dependence
%
%
\section{Finite Volume $\pi\pi$ Scattering}

\subsection{\textbf{Power law and exponential volume dependence}}

As discussed above there  are two types of volume dependence of the energy levels of two hadrons in a box: power law (proportional to $1/L^3$) and exponential (proportional to $e^{-m_\pi L}$). The first is exploited by the finite volume method to extract information about scattering parameters~\cite{Hamber:1983vu,Luscher:1990ux,Luscher:1986pf}. The second, usually numerically smaller, appears as a correction to the relation between energy levels in a box and scattering parameters.  In order to compute the exponentially suppressed terms we need to separate them from the larger power law contribution.  

\begin{figure}[t]\label{fig:feynman}
\center
\begin{tabular}{ccccc}
	\includegraphics[width=0.25\textwidth]{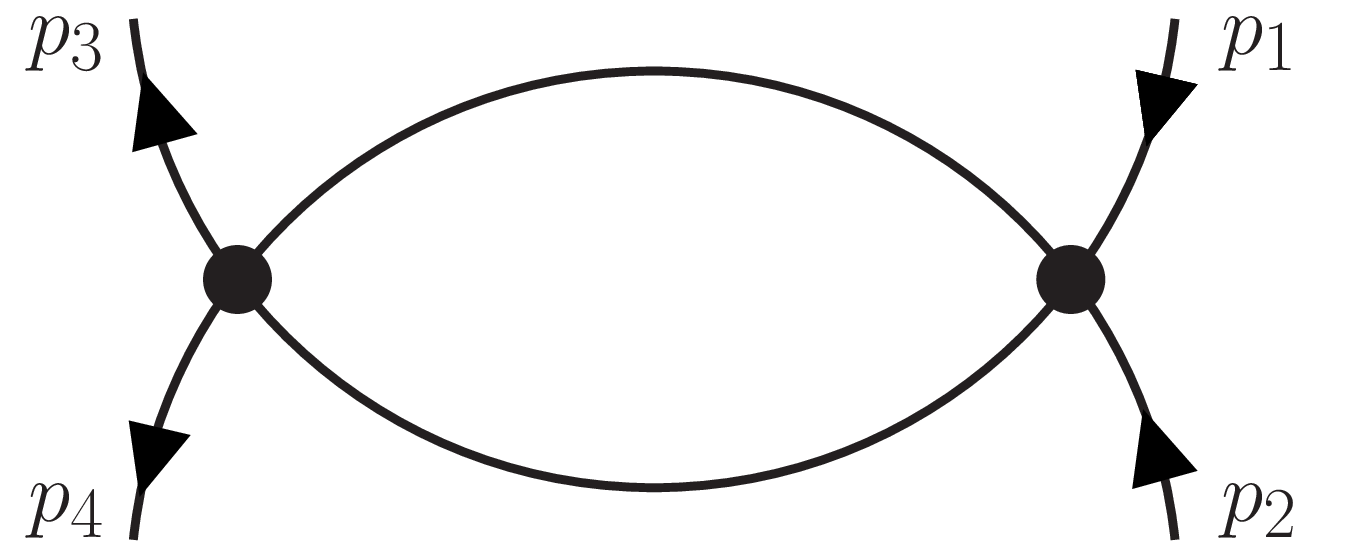} & $\;\;\;\;$ & \includegraphics[width=0.25\textwidth]{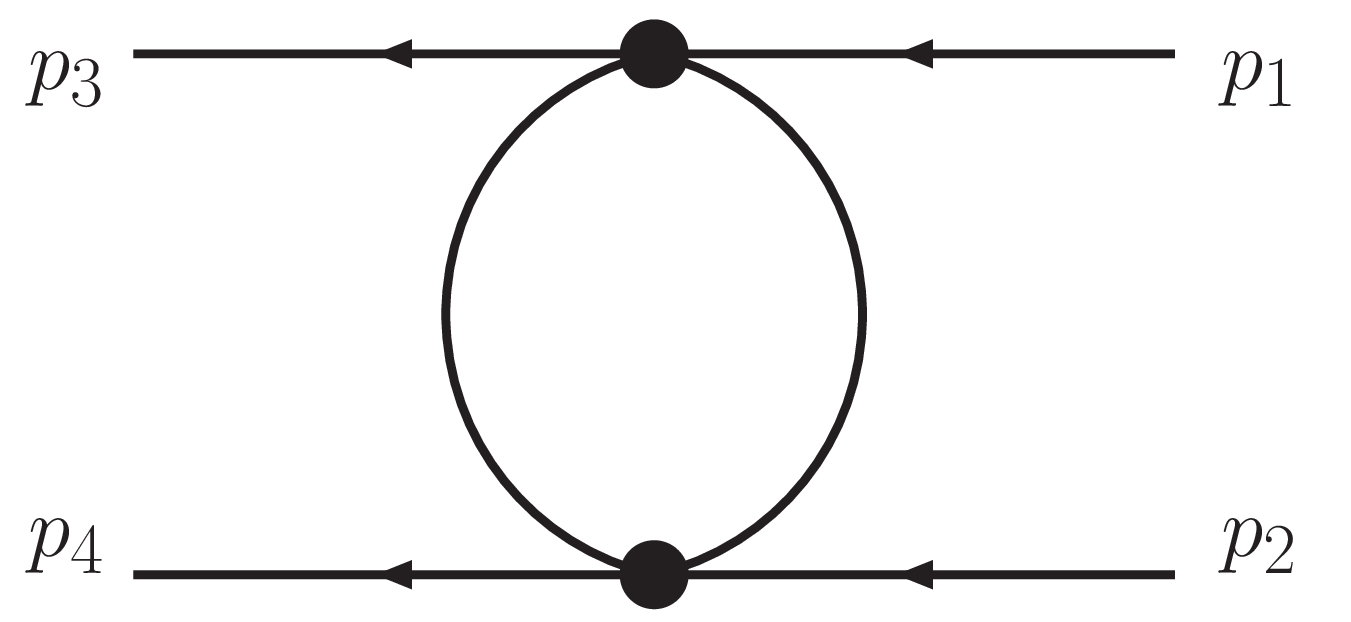} & $\;\;\;\;$ & \includegraphics[width=0.25\textwidth]{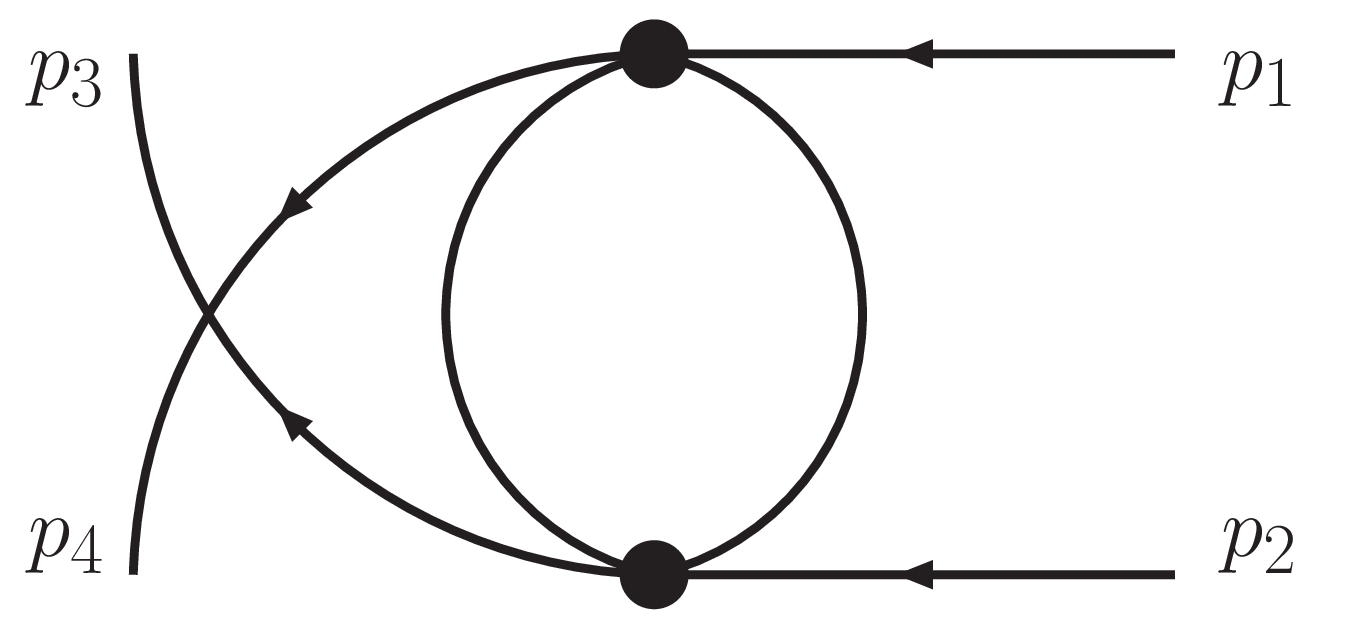} \\
	(a) & $\;\;$ & (b) & $\;\;$ & (c) \\
\end{tabular}
\vspace{2mm}
\begin{tabular}{cccc}
	\includegraphics[width=0.19\textwidth]{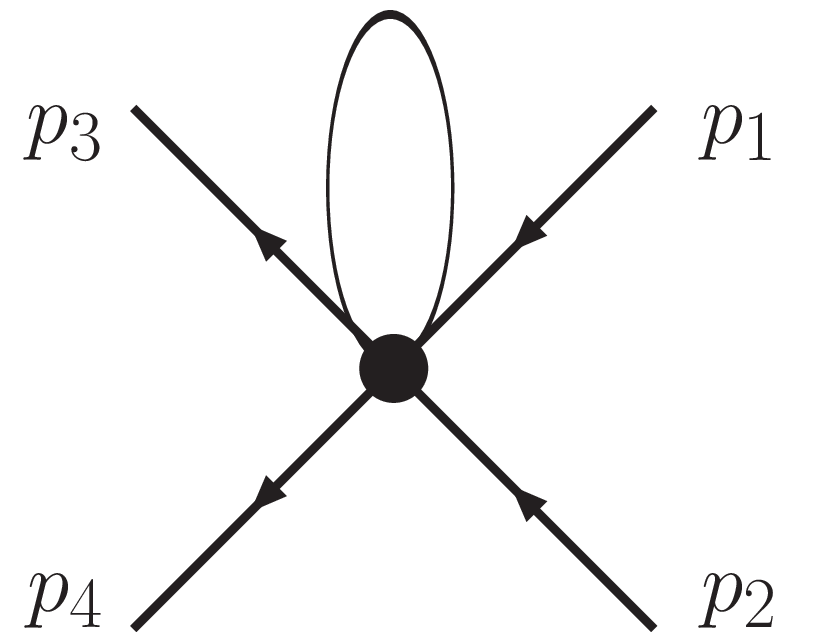} & $\;\;\;\;\;\;$ & \includegraphics[width=0.19\textwidth]{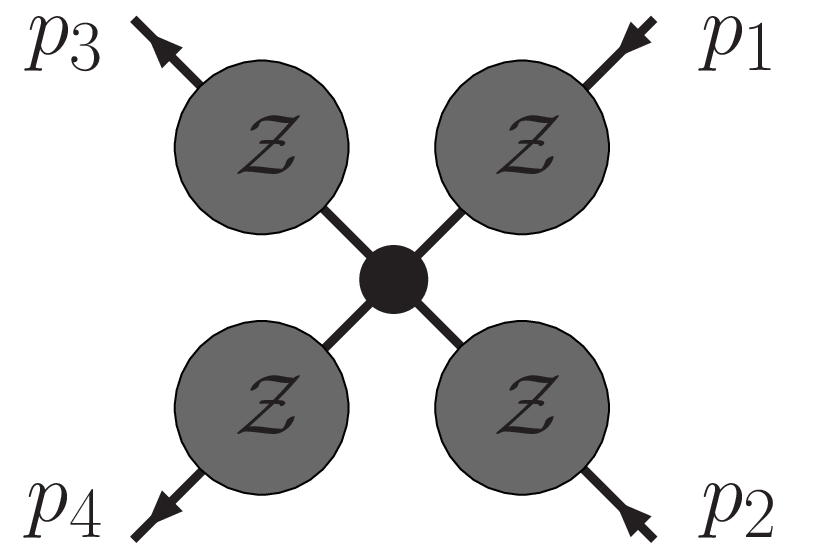}  \\
	(d) & $\;\;\;\;\;\;\;\;\;\;\;\;\;\;$ & (e)  \\
	\end{tabular}
	\caption[The one loop diagrams contributing to $\pi\pi$ scattering]{\label{fig:one-loop}The one-loop diagrams which contributing to the $\pi\pi$ scattering amplitude.  Only the $s$-channel diagram, (a), contributes to the power-law volume dependence.  Diagrams (b) and (c) are the $t$-, and $u$-channel diagrams, respectively, while diagram (e) represents wavefunction renormalization.  All these diagrams contribute to the exponential volume dependence.}

\end{figure}

Figure~\ref{fig:one-loop} shows all $\pi\pi$ scattering diagrams which contribute at one-loop order. As we will discuss in more detail in the next section, the power law corrections arise only from 
\nobreak{$s$-channel} diagrams as shown in Fig.~\ref{fig:one-loop}~(a), where the intermediate particles can be on-shell, and thus propagate far and ``feel'' the finiteness of the box. In all other diagrams the intermediate particles are very off-shell, cannot propagate farther than a distance of order $1/m_\pi$ and therefore have only small, exponentially suppressed sensitivity to the size of the box. 

%%%%%%%%%%%%%%%%%%%%%%%%%%%%%%%%%%
%\bigskip
%%%%%%%%%%%%%%%%%%%%%%%%%%%%%%%%%%%%%%%%%%%%%
Let us now discuss the general form of finite volume corrections.
Consider first the pion propagator at finite volume which is a function of the spatial momentum $\vec{k}=2\pi \vec{n}/L$ and the energy $E$. It will have poles for values of $E$ corresponding to the values of the energy of a pion in the box. In particular, for $\vec{k}=0$ the pole will be at $m_\pi(L)$ (the ``finite volume mass"), differing from the (infinite volume) mass of the pion $m_\pi$ by an exponentially small quantity proportional to $(m_\pi^2 / (4\pi f)^2)\, e^{-m_\pi L} / L\sqrt{mL}$~%
\cite{Gasser:1986vb,Colangelo:2003hf,Colangelo:2005gd}. 
The extra suppression factor $m_\pi^2/(4\pi f_\pi)^2$ is due to the fact that only loop diagrams contribute to the finite volume corrections.

The volume corrections for systems with more than one hadron are more subtle. The reason is that there are two kinds of volume corrections to the energy levels: a power law one described by the L\"{u}scher formula and the exponentially suppressed ones. To understand how to separate them let us first look at the infinite-volume, S-wave scattering amplitude, $T(s)$, with on-shell external pions. It is given at one loop by%	footnote
\footnote{We are considering the s-wave projected amplitude and disregarding the mixing with higher partial waves induced by the breaking of rotational symmetry.}:
\begin{eqnarray}\label{eq:Tgeneral0}
	T(s) &\simeq& T^{(0)}(s) +T^{(1)} _{t,u}(s) + T^{(1)}_{s,R} (s) + i T^{(1)}_{s,I}(s)\, ,
\end{eqnarray}
and unitarity allows us to re-sum this amplitude as a bubble series,
\begin{eqnarray}\label{eq:Tgeneral}
	T(s) &\simeq& \frac{(T^{(0)}(s))^2}{T^{(0)}(s) -T^{(1)}_{t,u}(s) - T^{(1)}_{s,R} (s) - i T^{(1)}_{s,I}(s)},
\end{eqnarray}
where $T^{(n)}$ is $n$-th loop contribution, the $s$-channel contribution is separated into its real part, $T^{(1)}_{s,R}(s)$, and imaginary part, $T^{(1)}_{s,I}(s)$, and all other contributions at one-loop including $t$- and $u$-channels are denoted by $T^{(1)}_{t,u}(s)$. The imaginary part, $T^{(1)}_{s,I}(s)$, comes from picking in the loop integration, the particle poles in both pion propagators, such that both the loop pions are on-shell. The loop integral is then proportional to the phase space volume and is given by 
\begin{equation}
T^{(1)}_{s,I}(s) = \frac{(T^{(0)}(s))^2}{32\pi\sqrt{s}} \sqrt{s-4m_\pi^2}. 
\end{equation}
The fact that the imaginary part is determined by the tree level amplitude is a consequence of the optical theorem.

It is useful to define the $K$-matrix~\cite{taylor}, which at one-loop is given by:
\beq\label{eq:K}
K(s) \simeq T^{(0)}(s)  +T^{(1)} _{t,u}(s) + T^{(1)}_{s,R} (s)
\simeq  \frac{(T^{(0)}(s))^2}{T^{(0)}(s) -T^{(1)} _{t,u}(s) - T^{(1)}_{s,R} (s)}.
\eeq 
Since the scattering amplitude can be written in terms of the phase shift, $\delta(s)$, as
\beq
T(s) = \frac{32\pi\sqrt{s}}{\sqrt{s-4m_\pi^2}}\frac{1}{\cot\delta(s)-i}
= \frac{32\pi\sqrt{s}}{\sqrt{s-4m_\pi^2}}\frac{e^{i\delta(s)}-1}{2i},
\eeq 
the relation between the $K$-matrix and the phase shift is then given by
\beq\label{eq:kcot}
\frac{1}{K(s)} = \frac{1}{32\pi}\sqrt{\frac{s-4m_\pi^2}{s}} \cot\delta(s).
\eeq  

Now we look at the finite volume amplitude $\mathcal{T}(s)$\footnote{By finite volume scattering amplitude we mean the amputated four-point correlator since, of course, there is no scattering at finite volume.}. It is computed in the same way as the infinite volume amplitude, except the loop integrals are substituted by sums over the  momenta allowed in a finite box.
The important point to keep in mind is that sums where the summand is regular are, at large enough $L$, well approximated by the analogous integral, up to exponentially small terms. If the summand, however, contains a singularity, power law dependence on the volume arises.
As mentioned before, only the kinematics of the $s$-channel diagram allows for both of the intermediate pions to be on-shell simultaneously. This implies that the summand in the sum over the loop momentum contains a singularity and leads to power law volume corrections. For the remaining diagrams no singularities are present and only exponentially suppressed corrections can arise.
As it will be shown explicitly below, the finite volume amplitude then has the following form: the tree term remains the same, the $t$-, $u$- and the real part of the $s$-channels pick only exponential corrections but the imaginary part turns into the term with power law $L$-dependence (and is real at finite $L$):
\bea\label{eq:finiteT}
\mathcal{T}(s) &\simeq& T^{(0)}(s) + T^{(1)} _{t,u}(s) + T^{(1)}_{s,R} (s) +\Delta T_{\rm exp}^{(1)}(s) + \frac{(T^{(0)}(s))^2}{16\pi^2 L\sqrt{s}}\mathcal{S}\left( \frac{k^2L^2}{4\pi^2} \right) \nn\\
&\simeq& \frac{(T^{(0)}(s))^2}{T^{(0)}(s)-  T^{(1)} _{t,u}(s) - T^{(1)}_{s,R} (s) -\Delta T_{\rm exp}^{(1)}(s) - \frac{(T^{(0)}(s))^2}{16\pi^2 L\sqrt{s}}\mathcal{S}\left( \frac{k^2L^2}{4\pi^2} \right) }.
\eea 
where $s=4(k^2+m_\pi^2)$, $\Delta T_{\rm exp}^{(1)}(s)$ is the finite volume correction to $T^{(1)} _{t,u}(s) + T^{(1)}_{s,R} (s)$, and $\mathcal{S}$ is a universal (independent of the interaction) function of 
$s$~\cite{Beane:2003yx,Beane:2003da}:
\beq
\mathcal{S} \left( \frac{k^2L^2}{4\pi^2} \right) 
	= 4\pi^2 L \left[ \frac{1}{ L^3}\sum_{\vec{q}=\frac{2\pi \vec{n}}{L}} -\int \frac{d^3q}{(2\pi)^3} \right] \frac{1}{\vec{q}^2-k^2}
	= \lim_{\Lambda_n\rightarrow\infty}\sum_{|\vec{n}|<\Lambda_n}\frac{1}{\vec{n}^2-\frac{k^2L^2}{4\pi^2}} -4\pi\Lambda_n.
\eeq
Combining Eqs.~\eqref{eq:Tgeneral}, \eqref{eq:K}, \eqref{eq:kcot} and \eqref{eq:finiteT}, we have
\bea
\mathcal{T}(s) &\simeq& \frac{1}{\frac{1}{K(s)} - \frac{\Delta T_{\rm exp}^{(1)}(s)}{(T^{(0)}(s))^2} - \frac{1}{16\pi^2 L\sqrt{s}}\mathcal{S}\left(\frac{(s-4m_\pi^2)L^2}{16\pi^2}\right)}\nn\\
&=& \frac{16\pi \sqrt{s}}{k\cot\delta(s)- 16\pi\sqrt{s}\frac{\Delta T_{\rm exp}^{(1)}(s)}{(T^{(0)}(s))^2} -\frac{1}{\pi L}\mathcal{S}\left(\frac{(s-4m_\pi^2)L^2}{16\pi^2}\right) }.
\eea 
The energy of the states in the box are determined by the location of the poles of the finite volume amplitude  determined by the solution of
\beq\label{eq:FVluscher}
k\cot\delta(s)- 16\pi\sqrt{s}\frac{\Delta T_{\rm exp}^{(1)}(s)}{(T^{(0)}(s))^2} = \frac{1}{\pi L}\mathcal{S}\left(\frac{(s-4m_\pi^2)L^2}{16\pi^2}\right).
\eeq 
We recognize in Eq.~(\ref{eq:FVluscher}) the familiar form of the L\"{u}scher relation modified by the
finite volume correction: the quantity $- 16\pi\sqrt{s}\frac{\Delta T_{\rm exp}^{(1)}(s)}{(T^{(0)}(s))^2} $ is the sought-after (exponentially small) correction to $k\cot\delta(s)$. 
% Added by Andre
In this work, we will focus on the 2-pion correlator near threshold in the center of mass frame, for which the infinite volume energy is given by $\sqrt{s}=2m_\pi$.
The solution of Eq.~(\ref{eq:FVluscher}), $s^*$, will be away from threshold by an amount given by $\sqrt{s^*}-2m_\pi$~$\approx \sqrt{s^*}-2m_\pi(L)$~$\sim 1 / f_\pi^2L^3$. Therefore, for $s\approx s^*$ the correction term, 
$\D(k \cot\delta(s)) = - 16\pi\sqrt{s}\frac{\Delta T_{\rm exp}^{(1)}(s)}{(T^{(0)}(s))^2} $,  can be approximated by $- 32\pi m_\pi\frac{\Delta T_{\rm exp}^{(1)}(4m_\pi^2)}{(T^{(0)}(4m_\pi^2))^2} $, the difference being suppressed by $\sim 1/L^3$.

It is customary to expand Eq.~(\ref{eq:FVluscher}) in  powers of $k^2 \sim 1/L^3$. Up to  the first three orders of this expansion (near threshold), $k\cot\delta$ can be approximated by the inverse scattering length, $1/a$, resulting in 
\beq\label{eq:luscher_approx}
	\sqrt{s^*}-2 m_\pi = \frac{4\pi a}{m_\pi L^3}
				\left( 1 + c_1 \frac{a}{L} + c_2\left( \frac{a}{L}\right)^2 +\cdots\right),
\eeq 
where $c_{1,2}$ are known numerical factors. A generalization of this formula  including the exponentially suppressed corrections, however, is not useful. The error in using Eq.~(\ref{eq:luscher_approx}) instead of Eq.~(\ref{eq:FVluscher}), that is, the error in the extrapolation from $s^*$ to $4m_\pi^2$, is of order $1/L^3$, which is parametrically larger than the exponential corrections we are interested in.  Numerically,  it may be the case that, for a set of simulation parameters,  the exponential term is larger than the $1/L^3$ terms. But the analogue of  Eq.~(\ref{eq:luscher_approx}) one would obtain by formally counting, for instance $e^{-m_\pi L} \sim 1/L^2$, would involve the effective range in addition to the scattering length. In any case, it is unclear {\it a priori} that a set of simulation parameters exist where this kind of expansion is useful and we will not pursue this line of thought in this paper.
% Added by Andre
We now will compute the exponential corrections to $k\cot \delta(s)$.
%

%%%%%%%%%%%%%%%%%%%%%%%%%%%%%%%%%%%%%%%%%%%%%%%%

%
%	Finite Volume Amplitude
%
%
\subsection{\textbf{The $\pi\pi$ scattering amplitude}\label{sec:fvT}}

The  $\pi\pi$ finite volume correlator in the $I=2$ isospin channel, for arbitrary momentum (in the chiral regime) is given by
\begin{align}\label{eq:Ttensor}
	\mc{T}_2 
	&= -\frac{2}{f^2} \bigg\{
			\frac{(3s -2m^2-\sum_{i=1}^4 p_i^2)}{3} 
		+\Big[ \frac{10s}{9f ^2} -\frac{6m ^2}{ f ^2}
			\Big]\, i\mc{I} 
		+\frac{4}{9 f ^2} p_s^\mu p_s^\nu\, i\mc{J}_{\mu\nu} ( p_s) \nonumber\\
	&\qquad\qquad
		+ \Big[ \frac{4}{9 f ^2} p_t^\mu p_t^\nu 
			+\frac{2}{9 f ^2} \Big( p_t +3 (p_1 +p_3) \Big)^\mu 
				\Big( p_t +3 (p_2 +p_4) \Big)^\nu \Big]\, i\mc{J}_{\mu\nu}( p_t) \nonumber\\
	&\qquad\qquad
		+ \Big[ \frac{4}{9 f ^2} p_u^\mu p_u^\nu 
			+\frac{2}{9 f ^2} \Big( p_u +3 (p_1 +p_4) \Big)^\mu 
				\Big( p_u +3 (p_2 +p_3) \Big)^\nu \Big]\, i\mc{J}_{\mu\nu}( p_u) \nonumber\\
	&\qquad\qquad
		+\frac{4(s-3m ^2)^2}{9 f ^2}\, i\mc{J}( p_s) %\nonumber\\
%	&\qquad\qquad
		+\Big[ \frac{m ^4}{f ^2} -\frac{4 t m ^2}{3 f ^2} +\frac{2 t^2}{3 f ^2}
			\Big]\, i\mc{J}( p_t) \nonumber\\
	&\qquad\qquad
		+\Big[ \frac{m ^4}{f ^2} -\frac{4 u m ^2}{3 f ^2} +\frac{2 u^2}{3 f ^2}
			\Big]\, i\mc{J}( p_u) %\nonumber\\
%	&\qquad\qquad
		-\frac{8(s-3m ^2)}{9f ^2}\, p_s^\mu\, i\mc{J}_\mu( p_s) 
		+\frac{4 (m ^2-t)}{3 f ^2} p_t^\mu \, i\mc{J}_\mu( p_t) \nonumber\\
	&\qquad\qquad
		+\frac{4 (m ^2-u)}{3 f ^2} p_u^\mu \, i\mc{J}_\mu( p_u) %\nonumber\\
%	&\qquad\qquad
		-\frac{4 \ell_1}{f ^2} \Big[ (t-2m^2)^2 +(u- 2m^2)^2 \Big] \nonumber\\
	&\qquad\qquad
		-\frac{2 \ell_2}{f ^2} \Big[ 2(s-2m^2)^2 +(t-2m^2)^2 +(u-2m^2)^2 \Big] 
		-\frac{32}{3}\ell_3 \frac{m ^4}{f ^2}
		\bigg\}.
\end{align}
In the above expression, $m$ is the (volume independent) tree level pion mass, $f\approx 132$~MeV is the (volume independent) tree level decay constant and the $\ell_i$'s are the Gasser-Leutwyler coefficients of counter terms appearing in the chiral Lagrangian at next-to-leading order (NLO)~\cite{Gasser:1983yg}.  The loop integrals/sums are given by $\mc{I}$, $\mc{J}(P)$, $\mc{J}_\mu(P)$, and $\mc{J}_{\mu\nu}(P)$ and will be defined below shortly.  The external momenta $p_i,\;i=1\cdots4$ are described in Fig.~\ref{fig:one-loop}, and the Mandelstam variables are employed: $s=p_s^2$ with $p_s=p_1+p_2$, $t=p_t^2$ with $p_t = p_1-p_3$, and $u=p_u^2$ with $p_u=p_1-p_4$.  In the above equation, the first term is the leading-order (LO) tree level contribution, and the remaining terms come from the one-loop diagrams shown in Fig.~\ref{fig:one-loop} and tree level diagrams with vertices of the $\mc{O}(p^4)$ Lagrangian~%
\cite{Gasser:1983yg,Gasser:1984gg}. In the NLO contributions, we have approximated $p_i^2 = m_\pi^2$, as the corrections to this at finite volume are beyond the order we are working.
%
\begin{comment} %commented out by Andre
The volume dependence in the amplitude we are interested in come only from loop integrals/sums.
\end{comment}
%
In the one-loop terms, we have expressed all contributions in terms of the bare pion mass as the difference is higher order in the chiral expansion.
One can, of course, choose to express the scattering amplitude in terms of the ``lattice quantities'' such as $m_\pi(L)$ and $f_\pi(L)$, which are measured directly from Euclidean correlation functions by lattice simulations.  Converting the bare quantities into lattice quantities involves additional tadpole loops, which will affect the form of $\D \mc{T}_{\rm exp}^{(1)}$, the finite volume corrections to the two-pion amplitude.  The exponential volume dependence of course doesn't depend upon whether one expresses the amplitude in terms of either the bare or physical parameters, and so it is useful to use the form which is simplest, that in terms of bare parameters.  In what follows, we will be interested in the 2-pion correlator near threshold, for which the external pion momentum are given by 
$p_i \simeq (\frac{1}{2}\sqrt{s}, \vec{0})$.

%
%
%	Loop integrals - sums
%
%
\subsection{\textbf{Loop integrals/sums at one-loop}}

The loop integrals/sums appearing in Eq.~(\ref{eq:Ttensor}) are defined by
\begin{align}
 \mathcal{I}  &= \int \frac{dq_0}{2\pi} \frac{1}{L^3}\sum_{\vec{q} = \frac{2\pi \vec{n}}{L}} 
                         \frac{1}{q^2-m ^2}, \label{eq:intI}   \\ 
\mathcal{J}(P) &= \int \frac{dq_0}{2\pi} \frac{1}{L^3}\sum_{\vec{q} = \frac{2\pi \vec{n}}{L}} 
                         \frac{1}{q^2-m ^2}   \frac{1}{(P+q)^2-m ^2}, \label{eq:intJ} \\
\mathcal{J}_\mu(P) &= \int \frac{dq_0}{2\pi} \frac{1}{L^3}\sum_{\vec{q} = \frac{2\pi \vec{n}}{L}} 
                         \frac{q_\mu}{q^2-m ^2}   \frac{1}{(P+q)^2-m ^2}, \label{eq:intJmu}
\end{align}
and
\begin{equation}
\mathcal{J}_{\mu\nu}(P) = \int \frac{dq_0}{2\pi} \frac{1}{L^3}
\sum_{\vec{q} = \frac{2\pi \vec{n}}{L}} \frac{q_\mu q_\nu}{q^2-m ^2} \frac{1}{(P+q)^2-m ^2}. 
\label{eq:intJmunu}
\end{equation}
Note that an integral is taken along the $0^{\rm th}$ component whereas sums over discrete momenta are taken with cubic symmetry.  Finite volume effects in the loop integrals/sums in Eqs.~\eqref{eq:intI}-\eqref{eq:intJmunu} can be computed by first evaluating the $q_0$ contour integral and then using the Poisson resummation formula,
\beq\label{eq:poisson}
	\frac{1}{L^3}\sum_{\vec{q} =
		\frac{2\pi \vec{n}}{L}} f(\vec{q}) =
		\int\frac{d^3q}{(2\pi)^3}f(\vec{q})
			+\sum_{\vec{n}\neq 0, \vec{n}\in \mathbb{N}^3} \int\frac{d^3q}{(2\pi)^3} f(\vec{q}) e^{iL\vec{q}\cdot\vec{n}}.
\eeq 
The difference between the finite volume and infinite volume loop integrals/sums is given by the second term in the right-hand side of Eq.~(\ref{eq:poisson}), and is always ultraviolet finite.  If the function $f(\vec{q})$ is regular, this difference is exponentially suppressed in the large $L$ limit. Power law dependence on $L$ can however appear if $f(\vec{q})$ has a singularity, {\it i.e.}, the case when $P=p_s$.  

Let us now evaluate the \textit{difference} between the finite and infinite volume integrals/sums given in Eqs.~(\ref{eq:intI})-\eqref{eq:intJmunu}.  We shall define the difference between the finite and infinite volume integrals/sums as
\begin{equation}
	\D f \equiv f(FV) - f(\infty) =
		 \int \frac{dq_0}{2\pi} \left[ \frac{1}{L^3}\sum_{\vec{q}= \frac{2\pi \vec{n}}{L}}
				- \int \frac{d^3 q}{(2\pi)^3} \right] 
			f(\vec{q}),
\end{equation}
where it is implicit that we regulate the sum and integral in the same manner such that the UV divergences cancel.  The tadpole integral in Eq.~\eqref{eq:intI}, which contributes to $m_\pi$ and $f_\pi$ and the loop diagrams of Fig.~\ref{fig:one-loop}, has the following volume correction;
\begin{align} \label{eq:deltaI}
	i\Delta \mc{I} &= \int \frac{dq_0}{2\pi} \left[ \frac{1}{L^3}\sum_{\vec{q}= \frac{2\pi \vec{n}}{L}}
					- \int \frac{d^3 q}{(2\pi)^3} \right] 
				\frac{i}{q^2-m ^2} \nonumber\\
			&= \left[\frac{1}{L^3}\sum_{\vec{q}}-\int\frac{d^3 q}{(2\pi)^3}\right]\frac{1}{2\omega_q},
		\nonumber\\
			&= \frac{m}{4\pi^2 L} \sum_{\vec{n}\neq 0} 
				\frac{1}{|\vec{n}|} K_1(|\vec{n}| m  L).
\end{align}
where $\omega_q=\sqrt{\vec{q}^{\:2}+m^2}$.  The mass and decay constant measured in lattice simulations are thus given to NLO by~\cite{Luscher:1985dn,Colangelo:2003hf}%
\footnote{This relation is known up to two loops~\cite{Colangelo:2005gd,Bijnens:2005ne}.}
\begin{align}
	m_\pi^2(L) &= m_\pi^2 \left[ 1+\frac{i \Delta\mc{I}}{f_\pi^2} \right] 
			= m^2 \left[ 1+\frac{i \mc{I}(L=\infty) + i \D \mc{I}}{f^2} 
				+ \frac{4 \ell_3 m^2}{f^2} \right], \\
	f_\pi(L) &= f_\pi \left[ 1- \frac{2 i \D \mc{I}}{f_\pi^2} \right]
			= f \left[ 1-\frac{2i\mc{I}(L=\infty) + 2i \D \mc{I}}{f^2}
				+\frac{2 \ell_4 m^2}{f^2}\right].
\end{align}
Using the asymptotic form of the Bessel function, one can see that for large $m L$, the volume shift of the pion mass is exponential~\cite{Gasser:1986vb,Colangelo:2003hf},
\begin{equation}
	\frac{\D m_\pi^2}{m_\pi^2} = \frac{i\D \mc{I}}{f_\pi^2} 
		= \frac{1}{2^{5/2} \pi^{3/2}} \frac{m_\pi}{L f_\pi^2} \sum_{n=|\vec{n}| \neq 0} 
			\frac{e^{-n\, m_\pi L}}{n^{3/2}} \frac{c(n)}{\sqrt{m_\pi L}}
			\left[ 1 + \frac{3}{8} \frac{1}{n\, m_\pi L} + \ldots \right],
\end{equation}
where the ellipses denote more terms in the asymptotic expansion of the Bessel function and $c(n)$ is the multiplicity factor counting the number of times $n=|\vec{n}|$ appears in the 3-dimensional sum.  Note, this sum is not over integers, but rather over the square-roots of integers.  In Table~\ref{t:c(n)}, we list the first few values of the multiplicity factors.

\begin{table}[t]
\center
\caption[Multiplicity factors for converting 3-D sums to scalar sums]{\label{t:c(n)} Here we list the first few multiplicity factors which arise when converting the three-dimensional sum to a scalar sum.}
\begin{tabular}{| c | c c c c c c c c c c |}
\hline
$n$
 & $1$ 
 & $\sqrt{2}$ 
 & $\sqrt{3}$ 
 & $\sqrt{4}$ 
 & $\sqrt{5}$ 
 & $\sqrt{6}$
 & $\sqrt{7}$
 & $\sqrt{8}$
 & $\sqrt{9}$
 & $\sqrt{10}$\\
\hline
$c(n)$
 & $6$ 
 & $12$ 
 & $8$ 
 & $6$ 
 & $24$
 & $24$
 & $0$
 & $12$
 & $30$
 & $24$ \\
\hline
\end{tabular}
\end{table}

\bigskip

Power law $L$-dependence can only occur through the integrals/sums in Eqs.~\eqref{eq:intJ}-\eqref{eq:intJmunu} when $P^2>0$. For the  center-of-mass scattering kinematics we are considering here this can only occur for $P=p_s$, since $p_s^2 = s>0$. As argued above we will only need the amplitude at threshold, \textit{i.e.}, $p_s=(2m_\pi,\vec{0})$ and $p_t=p_u=0$, except for the terms with power law $L$-dependence. Consequently, we will need only the values of $\Delta \mathcal{J}(P=0)$, $\Delta \mathcal{J}_0(P=0)$, and $\Delta \mathcal{J}_{00}(P=0)$ for $t$- and $u$-channels as well as  $\Delta \mathcal{J}(P=p_s)$, $\Delta \mathcal{J}_{0}(P=p_s)$, and $\Delta \mathcal{J}_{00}(P=p_s)$ for $s$-channels. The $\mc{J}$ integrals/sums at $P=0$ can be shown to be related to $\mathcal{I}$, giving the volume difference:
\begin{align}
i\Delta\mc{J}(0) &= -\frac{1}{4} \left[ 
\frac{1}{L^3}\sum_{\vec{q}} -\int\frac{d^3q}{(2\pi)^3} \right] 
\frac{1}{\omega_q^3} = \frac{d}{dm ^2} \left(i\Delta \mathcal{I}\right),\\
i\Delta \mathcal{J}_0(0) &= 0,
\end{align}
and
\begin{align}
i\Delta \mc{J}_{00} (0) &= i\int \frac{dq_0}{2\pi} 
\left[ \frac{1}{L^3}\sum_{\vec{q}} -\int\frac{d^3q}{(2\pi)^3} \right] 
\left[\frac{1}{q^2-m ^2}+\frac{m^2}{(q^2-m ^2)^2}+\frac{\vec{q}^{\:2}}{(q^2-m ^2)^2}\right]\nn\\
&=i\Delta\mc{I}+m^2 i\Delta\mc{J}(0)+3\left(-\frac{1}{6} i\Delta\mc{I}-\frac{1}{3}m^2 \frac{d}{dm^2} i\Delta\mc{I}\right) \nn\\
&=\frac{1}{2} i\Delta \mc{I}.
\end{align}

The power law volume dependence appears in the remaining integrals/sums. In those we keep $s$ away from the threshold value and take $\sqrt{s}=2\sqrt{k^2+m^2}$.  After performing the $q_0$ integral,  we separate the singular piece of the summand from the rest as
\beq
i\mc{J}(p_s) = -\frac{1}{4L^3} \sum_{\vec{q}} \frac{1}{\omega_q} \frac{1}{\vec{q}^{\:2}-k^2}
= -\frac{1}{4\omega_k L^3} \sum_{\vec{q}}  \frac{1}{\vec{q}^{\:2}-k^2}
  +\frac{1}{4L^3} \sum_{\vec{q}} \frac{1}{\omega_q\omega_k} \frac{\omega_q-\omega_k}{\vec{q}^{\:2}-k^2}.
\eeq
The first term contains a singularity when the internal momentum coincides with the external momentum, while the second term is regular. The difference $\Delta \mathcal{J}(p_s)$ is then
\beq\label{eq:I}
	i\D \mc{J}(p_s) = -\frac{1}{8\pi^2L\sqrt{s}} \mc{S} \left( \frac{k^2L^2}{4\pi^2} \right)
				+\underbrace{\sum_{\vec{n}\neq 0} \int\frac{d^3q}{(2\pi)^3} 
					e^{iL\vec{q} \cdot \vec{n}}\, 
					\frac{\omega_q-\omega_k}{\vec{q}^{\:2}-k^2}
					\frac{1}{4\omega_k\omega_q}}_{i\Delta\mathcal{J}_{\rm exp}(p_s)} 
\eeq 
The first piece above is the promised universal function containing the power law volume dependence.   The summand in the second term contains only exponential finite volume corrections. This term, contributing to $\Delta \mc{T}^{(1)}_{\rm exp}$, can be computed at the $s=4m^2, \vec{k}=0$ threshold point, 
\begin{align}\label{eq:Jexp}
	i\D \mc{J}_{\rm exp}(p_s) 
		&= \frac{1}{16\pi^2} \frac{1}{L \sqrt{m^2 +k^2}}
			 \sum_{\vec{n}\neq 0} \frac{1}{|\vec{n}|} \int_{-\infty}^{\infty} d y
			 \frac{y\, {\rm Im} e^{i2\pi y |\vec{n}|}}
			 	{\sqrt{y^2+\frac{m^2L^2}{4\pi^2}} \left(\sqrt{y^2+\frac{m^2L^2}{4\pi^2}}
					+\sqrt{\frac{k^2L^2}{4\pi^2}+\frac{m^2L^2}{4\pi^2}}\right)} \nn \\
		&\simeq -\frac{1}{16\pi} \sum_{\vec{n}\neq 0} 
			\left[ K_0(\left| \vec{n} \right| m  L) \bar{L}_{-1}(\left| \vec{n} \right| m  L) 
  				+K_1(\left| \vec{n} \right| m  L) \bar{L}_{0}(\left| \vec{n} \right| m  L) 
  				-\frac{1}{\left| \vec{n} \right| m  L} \right],
\end{align}
where  $\bar{L}_\nu$ is the Struve function.  To get the second line of Eq.\eqref{eq:Jexp}, we have neglected terms which are suppressed by $\mc{O}(k^2 / m^2)$ relative to the first.  For the two-pion system, this is approximately given by $\frac{k^2}{m^2} \simeq \frac{4 \pi |a|}{m^2 L^3} \ll 1$.  
The asymptotic expansion of $i\D\mc{J}_{\rm exp}(p_s)$ is given by 
\begin{equation}
	i\D\mc{J}_{\rm exp}(p_s) \simeq \frac{\sqrt{2\pi}}{(4\pi)^2} \frac{1}{(mL)^{3/2}}
				\sum_{n=|\vec{n}| \neq 0} 
					c(n)\, \frac{e^{-n\, m_\pi L}}{n^{3/2}} 
					\left[ 1 - \frac{5}{8} \frac{1}{n\, m_\pi L} + \ldots \right].
\end{equation}
Again, one can see that these volume corrections to the integral are exponentially suppressed.

The finite volume dependence of the other $s$-channel loop integral functions, 
$i\Delta\mathcal{J}_0(p_s)$ and $i\Delta\mathcal{J}_{00}(p_s)$ become simpler to evaluate by first observing that the summands can be separated into the following pieces:
\begin{align}
\frac{q_0}{q_0^2-\omega_q^2}\frac{1}{(q_0+p_{s0})^2-\omega_q^2} &=
\frac{1}{2 p_{s0}}\left[ \frac{1}{q_0^2-\omega_q^2}-\frac{1}{(p_{s0}+q_0)^2-\omega_q^2}
-\frac{(p_{s0})^2}{q_0^2-\omega_q^2}\frac{1}{(q_0+p_{s0})^2-\omega_q^2} \right], \nn \\
\frac{(q_0)^2}{q_0^2-\omega_q^2}\frac{1}{(q_0+p_{s0})^2-\omega_q^2} &=
\left( 1+\frac{\omega_q^2}{q_0^2-\omega_q^2} \right)\frac{1}{(q_0+p_{s0})^2-\omega_q^2}.\nn
\end{align}
One then obtains,
\begin{equation}\label{eq:I0}
i\D \mathcal{J}_0(p_s)
= -\frac{\sqrt{s}}{2} i\D \mathcal{J}(p_s)
\end{equation}
and
\begin{align}\label{eq:J00}
	i\D \mc{J}_{00}(p_s) &= i\D\mc{I} - \frac{1}{4L^3} \sum_{\vec{q}} \frac{1}{\omega_q} 
			\left( 1+\frac{\omega_k^2}{\vec{q}^{\:2}-k^2} \right) \nonumber\\
		&= \frac{1}{2} i\D \mathcal{I} + \frac{s}{4}\, i\D \mathcal{J}(p_s).
\end{align}

Having these tensor integrals/sums written in terms of the scalar integrals/sums $i\D\mc{J}(p_s)$ and $i\D\mc{I}$ and using the scattering amplitude in Eqs.~(\ref{eq:Tgeneral})
and~\eqref{eq:Ttensor}, one can now verify that the coefficient of $\mathcal{S}(\frac{k^2L^2}{4\pi^2})$ in the amplitude is what was promised in Eq.~(\ref{eq:Tgeneral}).  

%
%
%	Finite Volume Corrections to scattering length
%
%
%
\subsection{\textbf{Exponential volume correction to the $I=2$ $\pi\pi$ correlator}}

\begin{figure}[t]
\center
\includegraphics[width=0.7\textwidth]{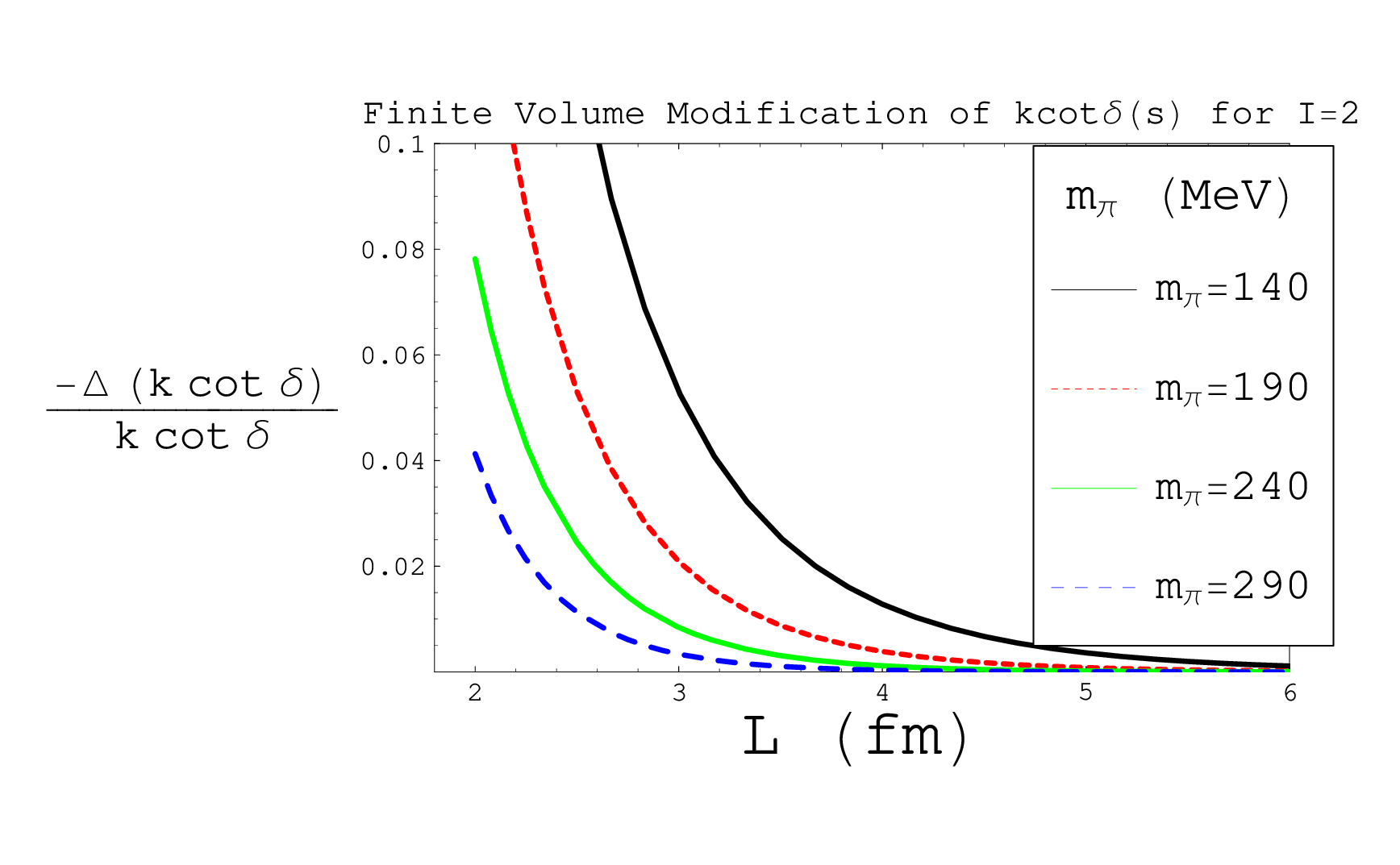} 
\caption[Finite volume modification to near threshold I=2 $\pi\pi$ scattering.]{\label{f:dma} Ratio of the magnitude of the exponential correction term $ \Delta (k\cot\delta) = -32\pi m_\pi\frac{\Delta T_{\rm exp}^{(1)}(4m_\pi^2)}{(T^{(0)}(4m_\pi^2))^2}$ to $k\cot\delta$ for different values of the pion mass.}
\label{fig:avsLP}
\end{figure}

Collecting the results for the sums/integrals in Eqs.~(\ref{eq:deltaI}-\ref{eq:J00}) and using the amplitude in Eq.~(\ref{eq:Ttensor}) we can now compute the correction term in Eq.~(\ref{eq:FVluscher}) for $I=2$ two-pion system near threshold. We find
\bea\label{eq:result}
\Delta (k\cot\delta(s)) &=& - 32\pi m_\pi\frac{\Delta T_{\rm exp}^{(1)}(4m_\pi^2)}{(T^{(0)}(4m_\pi^2))^2}
	    = \frac{8\pi}{m_\pi} \left[
		\frac{11}{3}i\Delta\mathcal{I}
		+ m_\pi^2 \frac{\partial}{\partial m_\pi^2} i\D\mc{I} 
		+ 2   i\D\mc{J}_{\rm exp}(4m_\pi^2) \right]\nn\\
	  &=& -\frac{m_\pi}{\sqrt{2\pi}} 
		\sum_{n=|\vec{n}| \neq 0} 
			c(n) \frac{e^{-n m_\pi L}}{\sqrt{n m_\pi L}} \left[
				1 -\frac{227}{24} \frac{1}{n m_\pi L} +\dots \right]. 
\eea
Equation~(\ref{eq:result}) is our main result (the first line being the exact one-loop answer and the second line the asymptotic expansion in $m_\pi L$).  In this expression, one can use either the bare parameters, the physical parameters or the finite volume parameters as the difference is higher order than we work in either the chiral expansion, or in the exponential dependence.  It is most convenient to use the values of $m_\pi(L)$ and $f_\pi(L)$ directly measured in a given lattice simulation.
\begin{comment}% commented out by Andre
The leading term of the expansion in Eq.~(\ref{eq:result}) is a poor approximation to the whole expression for realistic values of $mL$. 
\end{comment}
%
In Fig.~\ref{fig:avsLP} we plot the ratio of $\Delta (k\cot\delta(s))$ to the one-loop value of $k\cot\delta(s)$  using Eq.~(\ref{eq:result}) as a function of $L$ for some reasonable values of $m_\pi$. 
We find the finite volume corrections to be relatively small, a few times smaller than the statistical and systematic errors quoted in recent simulations. An error of about $10\%$ was quoted in reference~\cite{Beane:2005rj} for the determination of the scattering length for a pion mass of $m_\pi \simeq 290\;{\rm MeV}$ and a box size of $L\simeq 2.5\;{\rm fm}$.%
\footnote{In Ref.~\cite{Beane:2005rj}, Beane \emph{et. al.} determined the $I=2$ $\pi\pi$ scattering length for various pion masses using a mixed action simulation with Domain-Wall valence quarks and staggered sea quarks~\cite{Bar:2005tu}.  Because of the mixed-action, the mesons composed of sea quarks and the mesons composed of valence quarks receive different mass shifts from the finite lattice spacing.  This means that even when the sea and valence quark masses are tuned equal, there are still partial quenching effects in the simulation.  In Ref.~\cite{Chen:2005ab}, the partial quenching and lattice spacing corrections to the $I=2$ $\pi\pi$ scattering length were worked out for this mixed action theory.  It was shown that the these two lattice artifacts were largely suppressed, and almost non-existent for the mass tunings used in Ref.~\cite{Beane:2005rj}.  However, as shown in Ref.~\cite{Chen:2005ab}, there are still partial quenching effects and in particular, in the $t$- and $u$-channel diagrams the hairpin contributions can be significantly more sensitive to the boundary effects.  For $I=2$ these effects are only exponential, and for the pion masses and box sizes used in Ref.~\cite{Beane:2005rj}, we have found they are the same order of magnitude as the corrections of this paper, and thus not-significant to the work of Beane \emph{et. al.}}
 The finite volume correction from Eq.~(\ref{eq:result}) for these parameters is approximately $1\%$. These corrections however grow fast with the approach to the chiral limit, and they become non-negligible as smaller pion masses are used and statistical errors are reduced in simulations.

%
%
%	Discussion
%
%
%
\section{Discussion}
We have described the leading exponential volume dependence explicitly for the scattering parameter of a $I=2$ two-pion system near threshold in a box, by extending the one-loop \CPT\ calculation of pion scattering~\cite{Gasser:1983yg} to include the volume dependence.  The exponentially suppressed volume corrections can distort the universal relation between the infinite volume scattering parameters and the power-law volume dependence of the two-particle system, especially as the chiral limit is approached.  An important point we want to stress is that the useful way to add the exponential volume dependence to the relation between infinite volume scattering parameters and the energy of the two-particle system in a box, is via Eq.~\eqref{eq:FVluscher}, which allows an understanding of the leading exponential volume dependence to $k\cot\delta(s)$.  This is contrast to the notion of studying the exponential volume dependence of the scattering length, the effective range etc., separately.

It is important to stress the limits of validity of the present calculation. On one hand, the pion masses should be small enough so that the chiral expansion is converging. From the experience acquired in the three flavor case, where kaon loops are a borderline case for the convergence of the expansion, one expects chiral perturbation theory to be useful for $m_\pi < 500 $ MeV (of course the exponential volume dependence for a $500$ MeV pion, or kaon will be negligible). Also, the box size has to be large enough so the usual power counting used here (the so-called ``$p$-counting") is valid. When $L$ is much smaller than the inverse pion mass, another power counting is required such as the $\epsilon$-~\cite{Gasser:1987ah} or the $\epsilon^\prime$-regime~\cite{Detmold:2004ap}.  Additionally, we have neglected corrections which occur from higher loops, all of which are suppressed by additional factors of $(m_\pi / 4\pi f_\pi)^2$ and some of which are suppressed by additional exponential factors of $e^{-m_\pi L}$.  The diagrams with this extra exponential suppression result from two-loop diagrams where intermediate states in both loops are purely off-shell and hence ``going around the box".

We have focussed on the exponential corrections to phase shifts close to threshold.  One can easily extend this work to include the exponential volume dependence of the phase shifts at higher energies.  
Alternatively, one can access non zero momenta by using twisted or partially twisted boundary conditions to probe the low-momentum dependence of the scattering amplitude~\cite{Bedaque:2004kc,deDivitiis:2004kq,Sachrajda:2004mi,Bedaque:2004ax,Tiburzi:2005hg,Mehen:2005fw,Flynn:2005in,Guadagnoli:2005be}.  This method boosts the entire two-particle system, however, requiring the extraction of scattering parameters in a boosted frame~\cite{Rummukainen:1995vs,Kim:2005gf,Christ:2005gi}. Our methods generalize trivially to this case.  These methods can also be extended to other interesting two-hadron systems~\cite{Beane:2003da,Beane:2003yx}, where these exponential volume effects may be more significant.

% ========== Chapter 6: 2-hadron systems - MA
\chapter{Two-Hadron Interactions:\\ Ginsparg-Wilson Pions Scattering in a Sea of Staggered Quarks}\label{chap:pipiMA}

The work in this Chapter is based upon Ref.~\cite{Chen:2005ab}.
%
%
%
%
%
%
%	Introduction
%
%
%
%
%
%
\section{Introduction}

Lattice simulations with staggered fermions~\cite{Susskind:1976jm} can at present reach significantly lighter quark masses than other fermion discretizations%
\footnote{We note that recently there have been significant improvements in the last few years for developing numerically fast Wilson fermions~\cite{Luscher:2003vf,Luscher:2003qa,Luscher:2005rx,Luscher:2005mv}, in addition to the twisted mass discretization method discussed briefly in Chapter~\ref{ch:tmChPT}.}
and have proven extremely successful in accurately reproducing experimentally measurable quantities~\cite{Davies:2003ik,Aubin:2004fs}.  Staggered fermions, however, have the disadvantage that each quark flavor comes in four tastes.  Because these species are degenerate in the continuum, one can formally remove them by taking the fourth root of the quark determinant.  In practice, however, the fourth root must be taken before the continuum limit;  thus it is an open theoretical question whether or not this fourth-rooted theory becomes QCD in the continuum limit.%
\footnote{See Ref.~\cite{Durr:2004ta,Durr:2005ax,Shamir:2005sv,Creutz:2006ys,Bernard:2006vv,Durr:2006ze,Hasenfratz:2006nw,Bernard:2006ee} for a recent discussion of staggered fermions and the fourth-root trick.}
Even if one assumes the validity of the fourth-root trick, which we do in the rest of this chapter, staggered fermions have other drawbacks.  On the lattice, the four tastes of each quark flavor are no longer degenerate, and this taste symmetry breaking is numerically significant in current simulations~\cite{Aubin:2004fs}.  Thus one must use staggered chiral perturbation theory (S$\chi$PT), which accounts for taste-breaking discretization effects, to extrapolate correctly staggered lattice calculations to the
continuum ~\cite{Lee:1999zx,Aubin:2003mg,Aubin:2003uc,Sharpe:2004is}.
Fits of S$\chi$PT expressions for meson masses and decay constants
have been remarkably successful.  Nevertheless, the large number
of operators in the next-to-leading order (NLO) staggered chiral
Lagrangian~\cite{Sharpe:2004is} and the complicated form of the kaon
B-parameter in S$\chi$PT~\cite{VandeWater:2005uq} both show that S$\chi$PT
expressions for many physical quantities will contain a daunting number
of undetermined fit parameters.  Another practical hindrance to the
use of staggered fermions as valence quarks is the construction of
lattice interpolating fields.  Although the construction of a staggered
interpolating field is straightforward for mesons since they are spin 0
objects~\cite{Golterman:1984cy,Golterman:1985dz}, this is not in general
the case for vector mesons, baryons or multi-hadron states since the
lattice rotation operators mix the spin, angular momentum and taste of
a given interpolating field~\cite{Golterman:1984dn,Golterman:1986jf,Bailey:2005ss}.

The use of Ginsparg-Wilson (GW) fermions~\cite{Ginsparg:1981bj} evades both
the practical and theoretical issues associated with staggered fermions.
Because GW fermions are tasteless, one can simply construct
interpolating operators with the right quantum numbers for the desired
meson or baryon.  Moreover, massless GW fermions possess an exact
chiral symmetry on the lattice~\cite{Luscher:1998pq} which protects
expressions in $\chi$PT from becoming unwieldy.\footnote{In practice,
the degree of chiral symmetry is limited by how well the domain-wall
fermion~\cite{Kaplan:1992bt,Shamir:1993zy,Furman:1994ky}
is realized or the overlap
operator~\cite{Narayanan:1993sk,Narayanan:1993ss,Narayanan:1994gw} is
approximated.}  Unfortunately, simulations with dynamical GW quarks are approximately 10
to 100 times slower than those with staggered quarks~\cite{Kennedy:2004ae}
and thus are not presently practical for realizing light quark masses.

A practical compromise is therefore the use of GW valence quarks and
staggered sea quarks.  This so-called ``mixed action" theory
is particularly appealing because the MILC improved staggered
field configurations are publicly available.  Thus one only needs
to calculate correlation functions on top of these background
configurations, making the numerical cost the same as that
of quenched GW simulations. Several lattice calculations using
domain-wall or overlap valence quarks with the MILC configurations
are underway~\cite{Renner:2004ck,Bowler:2004hs,Bonnet:2004fr},
including a determination of the isospin 2 ($I=2$)
$\pi\pi$ scattering length~\cite{Beane:2005rj}.  Although
this is not the first $I=2$ $\pi\pi$ scattering lattice
simulation~\cite{Sharpe:1992pp,Gupta:1993rn,Aoki:2002in,Yamazaki:2004qb,Aoki:2005uf},
it is the only one with pions light enough to be in the chiral
regime~\cite{Bernard:2002yk,Beane:2004ks}.  Its precision is limited,
however, without the appropriate mixed action $\chi$PT expression for use
in continuum and chiral extrapolation of the lattice data.  With this
motivation we calculate the $I=2$ $\pi\pi$ scattering length in chiral perturbation theory for
a mixed action theory with GW valence quarks and staggered sea quarks.

Mixed action chiral perturbation theory (MA$\chi$PT) was first introduced
in Refs.~\cite{Bar:2002nr,Bar:2003mh,Tiburzi:2005vy} and was extended to include GW
valence quarks on staggered sea quarks for both mesons and baryons
in Refs.~\cite{Bar:2005tu} and~\cite{Tiburzi:2005is}, respectively.
$\pi\pi$ scattering is well understood in continuum, infinite-volume
\CPT~\cite{Weinberg:1966kf,Gasser:1983yg,Gasser:1984gg,Knecht:1995tr,Bijnens:1995yn,Bijnens:1997vq,Bijnens:2004eu},
and is the simplest two-hadron process that one can study
numerically with LQCD.  We extend the NLO \CPT\ calculations
of Refs.~\cite{Gasser:1983yg,Gasser:1984gg} to MA\CPT.
A mixed action simulation necessarily involves partially quenched QCD
(PQQCD)~\cite{Bernard:1993sv,Sharpe:1997by,Golterman:1997st,Sharpe:2000bc,Sharpe:2001fh,Sharpe:2003vy},
in which the valence and sea quarks are treated differently. Consequently,
we provide the PQ\CPT\ $\pi\pi$ scattering amplitude by taking an
appropriate limit of our MA\CPT\ expressions. In all of our computations,
we work in the isospin limit both in the sea and valence sectors.

\bigskip

This chapter is organized as follows.  We first comment on the determination of infinite volume scattering parameters from lattice simulations in Section~\ref{sec:malqcd}, focusing on the applicability of L\"{u}scher's method~\cite{Luscher:1986pf,Luscher:1990ux} to mixed action lattice simulations.  We then review mixed action LQCD
and MA\CPT\ in Section~\ref{sec:mixed}.  In Section~\ref{sec:mixedScatt}
we calculate the $I=2\ \pi\pi$ scattering amplitude in MA\CPT, first
by reviewing $\pi\pi$ scattering in continuum $SU(2)$ \CPT\ and then
by extending to partially quenched mixed action theories with $N_f=2$
and $N_f=2+1$ sea quarks.  We discuss the role of the double poles in
this process~\cite{Bernard:1993ga} and parameterize the partial quenching
effects in a particularly useful way for taking various interesting and
important limits.  Next, in section~\ref{sec:mixedLength}, we present results for the pion scattering length
in both 2 and $2+1$ flavor MA$\chi$PT.  
These expressions show that it is advantageous to fit to partially
quenched lattice data using the lattice pion mass and pion decay
constant measured on the lattice rather than the LO parameters in the chiral Lagrangian.
We also give expressions for the corresponding continuum PQ$\chi$PT scattering amplitudes, which do not already appear in the literature.    
Finally, in Section~\ref{sec:summary} we briefly discuss how to use our MA$\chi$PT formulae to determine the physical scattering length in QCD from mixed action lattice data and conclude.

%
%
%
%
%
%
%	Mixed Actions lattice simulations
%
%
%
%
%
%
\section{Determination of Scattering Parameters from Mixed Action Lattice Simulations}\label{sec:malqcd}

Lattice QCD calculations are performed in Euclidean spacetime,
thereby precluding the extraction of S-matrix elements from
infinite volume~\cite{Maiani:1990ca}.  L\"{u}scher, however,
developed a method to extract the scattering phase shifts of two
particle scattering states in quantum field theory by studying the
volume dependence of two-point correlation functions in Euclidean
spacetime~\cite{Luscher:1986pf,Luscher:1990ux}.  In particular, for two particles of equal mass $m$ in an $s$-wave state with zero total 3-momentum in a finite volume, the difference between the energy of the two particles and twice their rest mass is related to the $s$-wave scattering length:\footnote{Here we use
the ``particle physics" definition of the scattering length which is
opposite in sign to the ``nuclear physics" definition.}
\begin{equation}\label{eq:LusForm}
	\Delta E_0 = -\frac{4\pi a_0}{m \textrm{L}^3} \left[ 1 + c_1 \frac{a_0}{\textrm{L}} + c_2 \left(\frac{a_0}{\textrm{L}}\right)^2 + \mc{O}\left(\frac{1}{\textrm{L}^3}\right) \right]\,.
\end{equation}
In the above expression, $a_0$ is the scattering length (not to be confused with the lattice spacing, $a$), L is the length of one side of the spatially symmetric lattice, and $c_1$ and $c_2$ are known geometric coefficients.%
%	footnote
\footnote{This expression generalizes to scattering parameters of higher partial waves and non-stationary particles~\cite{Luscher:1986pf,Luscher:1990ux,Rummukainen:1995vs,Kim:2005gf}.}
Thus, even though one cannot directly calculate scattering amplitudes with lattice simulations, 
Eq.~(\ref{eq:LusForm}), which we will refer to as L\"{u}scher's formula, allows one to determine the infinite volume scattering length.  One can then use the expression for the scattering length computed in infinite volume \CPT\ to extrapolate the lattice data to the physical quark masses.  

Because L\"{u}scher's method requires the extraction of energy levels, it relies upon the existence of a Hamiltonian for the theory being studied.  This has not been demonstrated (and is likely false) for partially quenched and mixed action QCD, both of which are nonunitary.  Nevertheless, 
one can calculate the ratio of the two-pion correlator to the square of the single-pion correlator in lattice simulations of these theories and extract the coefficient of the term which is linear in time, which becomes the energy shift in the QCD (and continuum) limit.  We claim that in certain scattering channels, despite the inherent sicknesses of partially quenched and mixed action QCD, this quantity is still related to the infinite volume scattering length via Eq.~(\ref{eq:LusForm}), \emph{i.e.} the volume dependence is identical to Eq.~\eqref{eq:LusForm} up to exponentially suppressed corrections.%
\footnote{Here, and in the following discussion, we restrict ourselves to a perturbative analysis.}
This is what we mean by ``L\"{u}scher's method" for nonunitary theories.  We will expand upon this point in the following paragraphs.

It is well known that L\"{u}scher's formula does not hold for many scattering channels in quenched theories because unitarity-violating diagrams give rise to enhanced finite volume effects~\cite{Bernard:1995ez}.  For certain scattering channels, however,  quenched $\chi$PT calculations in finite volume show that, at 1-loop order, the volume dependence is identical in form to L\"{u}scher's formula~\cite{Bernard:1995ez,Colangelo:1997ch,Lin:2002aj}.  Chiral perturbation theory calculations additionally show that the same sicknesses that generate enhanced finite volume effects in quenched QCD also do so in partially quenched and mixed action theories~%
\cite{Sharpe:2000bc,Sharpe:2001fh,Beane:2002np,Bar:2002nr,Bar:2003mh,Lin:2003tn,Bar:2005tu,Golterman:2005xa}.  It then follows that if a given scattering channel has the same volume dependence as Eq.~\eqref{eq:LusForm} in quenched QCD, the corresponding partially quenched (and mixed action) two-particle process will also obey Eq.~\eqref{eq:LusForm}.  Correspondingly, scattering channels which have enhanced volume dependence in quenched QCD also have enhanced volume dependence in partially quenched and mixed action theories.  We now proceed to discuss in some detail why 
L\"{u}scher's formula does or does not hold for various 2$\rightarrow$2 scattering channels.  

Finite volume effects in lattice simulations come from the ability of particles to propagate over long distances and feel the finite extent of the box through boundary conditions.  Generically, they are proportional either to inverse powers of L or to exp(-$m$L), but L\"{u}scher's formula neglects exponentially suppressed corrections.  Calculations of scattering processes in effective field theories at finite volume show that the power-law corrections only arise from $s$-channel diagrams~%
\cite{Bernard:1995ez,Lin:2002nq,Lin:2002aj,Lin:2003tn,Beane:2003da}.  This is because all of the intermediate particles can go on-shell simultaneously, and thus are most sensitive to boundary effects.  Consequently, when there are no unitarity-violating effects in the $s$-channel diagrams for a particular scattering process, the volume dependence will be identical to Eq.~\eqref{eq:LusForm}, up to exponential corrections.  Unitarity-violating \emph{hairpin} propagators in $s$-channel diagrams, however,  give rise to enhanced volume corrections because they contain double poles which are more sensitive to boundary effects~\cite{Bernard:1995ez}.%
\footnote{We note that, while the enhanced volume corrections in quenched QCD invalidate the extraction of scattering parameters from certain scattering channels, \emph{e.g.} $I=0$~\cite{Bernard:1995ez,Lin:2002aj}, this is not the case in principle for partially quenched QCD, since QCD is a subset of the theory.  Because the enhanced volume contributions must vanish in the QCD limit, they provide a ``handle" on the enhanced volume terms.  In practice, however, these enhanced volume terms may dominate the correlation function, making the extraction of the desired (non-enhanced) volume dependence impractical.}
Thus all violations of L\"{u}scher's formula come from on-shell hairpins in the $s$-channel.  

Let us now consider $I=2$ $\pi\pi$ scattering in the mixed action theory.  
All intermediate states must have isospin 2 and $s\geq 4m^2$.  If one cuts an arbitrary graph connecting the incoming and outgoing pions, there is only enough energy for two of the internal pions to be on-shell, and, by conservation of isospin, they must be valence $\pi^+$'s.\footnote{We restrict the incoming pions to be below the inelastic threshold;  this is necessary for the validity of L\"{u}scher's formula even in QCD.}  Thus no hairpin diagrams ever go on-shell in the $s$-channel, and the structure of the integrals which contribute to the power-law volume dependence in the partially quenched and mixed action theories is identical to that in continuum $\chi$PT.   This insures that L\"uscher's formula is correctly reproduced to all orders in 1/L with the correct ratios between coefficients of the various terms.  Moreover, this holds to all orders in $\chi$PT, PQ$\chi$PT, MA$\chi$PT, and even quenched $\chi$PT.  The sicknesses of the partially quenched and mixed action theories only alter the exponential volume dependence of the $I=2$ scattering amplitude.%
\footnote{In fact, hairpin propagators will give larger exponential dependence than standard propagators because they are more chirally sensitive.}
This is in contrast to the $I=0$ $\pi\pi$ amplitude, which suffers from enhanced volume corrections away from the QCD limit.  In general, the argument which protects L\"{u}scher's formula from enhanced power-like volume corrections holds for all ``maximally-stretched" states at threshold in the meson sector, i.e. those with the maximal values of all conserved quantum numbers;  other examples include $K^+K^+$ and $K^+\pi^+$ scattering.  We expect that a similar argument will hold for certain scattering channels in the baryon sector.  

Therefore the $s$-wave $I=2$ $\pi\pi$ scattering length can be extracted from mixed action lattice simulations using L\"{u}scher's formula and then extrapolated to the physical quark masses and to the continuum using the infinite volume MA$\chi$PT expression for the scattering length.%
\footnote{For a related discussion, see Ref.~\cite{Bedaque:2004ax}}

%
%
%
%
%
%
%	Mixed Actions
%
%
%
%
%
%
\section{Mixed Action Lagrangian and Partial Quenching}\label{sec:mixed}

Mixed action theories use different discretization techniques in
the valence and sea sectors and are therefore a natural extension
of partially quenched theories.  We consider a theory with
$N_f$ staggered sea quarks and $N_v$ valence quarks (with $N_v$
corresponding ghost quarks) which satisfy the Ginsparg-Wilson
relation~\cite{Ginsparg:1981bj,Luscher:1998pq}.  In particular we
are interested in theories with two light dynamical quarks ($N_f
= 2$) and with three dynamical quarks where the two light quarks are degenerate (commonly
referred to as $N_f = 2+1$).  To construct the continuum effective
Lagrangian which includes lattice artifacts one follows the two
step procedure outlined in Ref.~\cite{Sharpe:1998xm}.  First one
constructs the Symanzik continuum effective Lagrangian at the quark
level~\cite{Symanzik:1983gh,Symanzik:1983dc} up to a given order in the
lattice spacing, $a$:
\begin{equation}
	\mc{L}_{Sym} = \mc{L} + a \mc{L}^{(5)} + a^2 \mc{L}^{(6)} + \ldots,
\end{equation}
where $\mc{L}^{(4+n)}$ contains higher dimensional operators of dimension
$4+n$. Next one uses the method of spurion analysis to map the Symanzik
action onto a chiral Lagrangian, in terms of pseudo-Goldstone mesons,
which now incorporates the lattice spacing effects. This has been done
in detail for a mixed GW-staggered theory in Ref.~\cite{Bar:2005tu};
here we only describe the results.

The leading quark level Lagrangian is given by
\begin{equation}\label{eq:LOqLag}
	\mc{L} = \sum_{a,b=1}^{4N_f +2N_v} 
		\bar{Q}^{a} \left[ i \Dslash -m_Q \right]_a^{\ b} Q_b,
\end{equation}
where the quark fields are collected in the vectors
\begin{align}
	Q^{N_f = 2} &= (\underbrace{u,d}_\textrm{valence},
		\underbrace{j_1, j_2, j_3, j_4, l_1, l_2, l_3, l_4}_\textrm{sea}, 
		\underbrace{\tilde{u},\tilde{d}}_\textrm{ghost})^T, \\
	Q^{N_f = 2+1} &= (\underbrace{u,d,s}_\textrm{valence}, 
		\underbrace{j_1,j_2,j_3,j_4, l_1,l_2,l_3,l_4, r_1,r_2,r_3,r_4}_\textrm{sea}, 
		\underbrace{\tilde{u},\tilde{d},\tilde{s}}_{\textrm{ghost}})^T
\end{align}
for the two theories. There are 4 tastes for each flavor of sea
quark, $j,l,r$.%
\footnote{Note that we use different labels for the
valence and sea quarks than Ref.~\cite{Bar:2005tu}.  Instead we use
the labeling convention consistent with
Ref.~\cite{Tiburzi:2005is}.}%
We work in the isospin limit in both the
valence and sea sectors so the quark mass matrix in the 2+1 sea flavor
theory is given by
\begin{equation}\label{eq:Masses}
	m_Q = \text{diag}(\underbrace{m_u, m_u, m_s}_\textrm{valence}, 
			\underbrace{m_j, m_j, m_j, m_j, m_j, m_j, m_j, m_j, m_r, m_r, m_r, m_r}_\textrm{sea}, 
			\underbrace{m_u, m_u, m_s}_\textrm{ghost}).
\end{equation}
The quark mass matrix
in the two flavor theory is analogous but without strange valence,
sea and ghost quark masses. The leading order mixed action Lagrangian,
Eq.~\eqref{eq:LOqLag}, has an approximate graded chiral symmetry,
$SU(4N_f+N_v|N_v)_L~\otimes~SU(4N_f+N_v|N_v)_R$, which is exact in the
massless limit.~\footnote{This is a ``fake" symmetry of PQQCD. However,
it gives the correct Ward identities and thus can be used to understand
the symmetries and symmetry breaking of PQQCD~\cite{Sharpe:2001fh}.} In
analogy to QCD, we assume that the vacuum spontaneously breaks this
symmetry down to its vector subgroup, $SU(4N_f +N_v | N_v)_V$, giving
rise to $(4N_f +2N_v)^2 -1$ pseudo-Goldstone mesons. These mesons are
contained in the field
\begin{equation}
	\S = {\rm exp} \left( \frac{2 i \Phi}{f} \right) ,  \;\;\; 
	\Phi = \begin{pmatrix}
			M & \chi^\dagger\\
			\chi & \tilde{M}\\
		\end{pmatrix}.
\label{eq:sigmaMA}
\end{equation}
The matrices $M$ and $\tilde{M}$ contain bosonic mesons while $\chi$
and $\chi^\dagger$ contain fermionic mesons. Specifically,
\begin{align}\label{eq:mesonsMA}
	M &=\begin{pmatrix}
			\eta_u & \pi^+ & \ldots & \phi_{uj} & \phi_{ul} & \ldots \\
			\pi^- & \eta_d & \ldots & \phi_{dj} & \phi_{dl} & \ldots \\
			\vdots & \vdots & \ddots & \ldots & \ldots & \ldots \\
			\phi_{ju} & \phi_{jd} & \vdots & \eta_j & \phi_{jl} & \ldots \\
			\phi_{lu} & \phi_{ld} & \vdots & \phi_{lj} & \eta_l & \ldots \\
			\vdots & \vdots & \vdots & \vdots & \vdots & \ddots \\
		\end{pmatrix}\quad ,\quad 
	\tilde M = \begin{pmatrix}
			{\tilde \eta}_u & {\tilde \pi}^+ & \ldots \\
			{\tilde \pi}^- & {\tilde \eta}_d & \ldots \\
			\vdots & \vdots & \ddots \\
		\end{pmatrix}
	\nonumber\\
	\chi &= \begin{pmatrix}
		\phi_{\tilde{u} u} & \phi_{\tilde{u} d} & \ldots & \phi_{\tilde{u} j} & \phi_{\tilde{u} l} & \ldots \\
		\phi_{\tilde{d} u} & \phi_{\tilde{d} d} & \ldots & \phi_{\tilde{d} j} & \phi_{\tilde{d} l} & \ldots \\
		\vdots & \vdots & \vdots & \vdots & \vdots & \ddots \\
		\end{pmatrix}.
\end{align}
In Eq.~\eqref{eq:mesonsMA} we only explicitly show the mesons needed in
the two flavor theory. The ellipses indicate mesons containing strange
quarks in the 2+1 theory. The upper $N_v \times N_v$ block of $M$
contains the usual mesons composed of a valence quark and anti-quark.
The fields composed of one valence quark and one sea anti-quark, such as
$\phi_{uj}$, are $1 \times 4$ matrices of fields where we have suppressed
the taste index on the sea quarks. Likewise, the sea-sea mesons such as
$\phi_{jl}$ are $4 \times 4$ matrix-fields. Under chiral transformations,
$\Sigma$ transforms as
\begin{equation}
	\S \longrightarrow L\ \S\ R^\dagger \quad ,\quad
		L,R \in SU(4N_f +N_v |N_v)_{L,R}.
\end{equation}

In order to construct the chiral Lagrangian it is useful to first define
a power-counting scheme. Continuum \CPT\ is an expansion in
powers of the pseudo-Goldstone meson momentum and mass squared
\cite{Gasser:1983yg,Gasser:1984gg}:
\begin{equation}\label{eq:smallScale}
	\varepsilon^2 \sim p_\pi^2 / \Lambda_\chi^2 \sim m_\pi^2 / \Lambda_\chi^2\,,
\end{equation}
where $m_\pi^2 \propto m_Q$ and $\Lambda_\chi$ is the cutoff of $\chi$PT.
In a mixed theory (or any theory which incorporates lattice spacing
artifacts) one must also include the lattice spacing in the power
counting. Both the chiral symmetry of the Ginsparg-Wilson valence
quarks and the remnant $U(1)_A$ symmetry of the staggered sea quarks
forbid operators of dimension five; therefore the leading lattice spacing
correction for this mixed action theory arises at $\mc{O}(a^2)$. Moreover,
current staggered lattice simulations indicate that taste-breaking
effects (which are of $\mc{O}(a^2)$) are numerically of the same size
as the lightest staggered meson mass~\cite{Aubin:2004fs}. We therefore
adopt the following power-counting scheme:
\begin{equation}
 	\varepsilon^2 \sim p_\pi^2 / \Lambda_\chi^2 \sim m_Q / \Lambda_\textrm{QCD} 
		\sim a^2 \Lambda_\textrm{QCD}^2\,.
\label{eq:epsilons}
\end{equation}
The leading order (LO), $\mc{O}(\varepsilon^2)$, Lagrangian is then given
in Minkowski space by~\cite{Bar:2005tu}
\begin{equation}
	\mc{L} = \frac{f^2}{8} \text{str} \left( \partial_\mu \S\, \partial^\mu \S^\dagger \right)
		+ \frac{f^2 B}{4} \text{str} \left( \S m_Q^\dagger + m_Q \S^\dagger \right)
		- a^2 \left( \mc{U}_S +\mc{U}^\prime_S + \mc{U}_V \right),
\end{equation}
where we use the normalization $f \sim 132$~MeV and have already
integrated out the taste singlet $\Phi_0$ field, which is proportional
to str$(\Phi)$~\cite{Sharpe:2001fh}. $\mc{U}_S$ and $\mc{U}^\prime_S$ are
the well-known taste breaking potential arising from the staggered sea
quarks~\cite{Lee:1999zx,Aubin:2003mg}. The staggered
potential only enters into our calculation through an additive shift to the
sea-sea meson masses; we therefore do not write out its explicit form.
The enhanced chiral properties of the mixed action theory are illustrated
by the fact that only one new potential term arises at this order:
\begin{equation}\label{eq:MixPotential}
	\mc{U}_V = -C_{Mix}\ \text{str} \left( T_3 \S T_3 \S^\dagger \right),
\end{equation}
where
\begin{equation}
	T_3 = \mc{P}_S -\mc{P}_V = \text{diag}(-I_V, I_t \otimes I_S, -I_V).
\end{equation}
The projectors, $\mc{P}_S$ and $\mc{P}_V$, project onto the sea and
valence-ghost sectors of the theory, $I_V$ and $I_S$ are the valence and
sea flavor identities, and $I_t$ is the taste identity matrix. From this
Lagrangian, one can compute the LO masses of the various pseudo-Goldstone
mesons in Eq.~\eqref{eq:mesonsMA}. For mesons composed of only valence
(ghost) quarks of flavors $a$ and $b$,
\begin{equation}
	m_{ab}^2 = B (m_a +m_b).
\label{eq:m_tree}\end{equation}
This is identical to the continuum LO meson mass because the chiral
properties of Ginsparg-Wilson quarks protect mesons composed of only
valence (ghost) quarks from receiving mass corrections proportional
to the lattice spacing. However, mesons composed of only sea quarks
of flavors $s_1$ and $s_2$ and taste $t$, or mixed mesons with one
valence ($v$) and one sea ($s$) quark both receive lattice spacing mass
shifts.  Their LO masses are given by
\begin{align}\label{eq:ssvsmasses}
	\tilde{m}_{s_1 s_2,t}^2 &= B(m_{s_1} +m_{s_2}) +a^2 \D(\xi_t), \\
	\tilde{m}_{vs}^2 &= B(m_v +m_s) +a^2 \D_{Mix}.
\end{align}
From now on we use tildes to indicate masses that include
lattice spacing shifts.  The only sea-sea mesons that enter $\pi\pi$ scattering to the order at
which we are working are the taste-singlet mesons (this is because the
valence-valence pions that are being scattered are tasteless), which are
the heaviest; we therefore drop the taste label, $t$.  The splittings
between meson masses of different tastes have been determined numerically
on the MILC configurations~\cite{Aubin:2004fs}, so $\D(\xi_I)$ should
be considered an input rather than a fit parameter. The mixed mesons
all receive the same $a^2$ shift given by
\begin{equation}
	\D_{Mix} = \frac{16 C_{Mix}}{f^2}\,,
\end{equation}
which has yet to be determined numerically.

\begin{comment}%	Comented out by Ruth
When chiral perturbation theory is used to describe nature, the LO
formulae for meson masses, Eq.~\eqref{eq:m_tree}, are not equal to the
physical values of meson masses measured in experiments.  For example, the
physical pion mass $m_\pi^2$ is not equal to $m_{uu}^2$.  This is because
the pion mass is renormalized by loop graphs in chiral perturbation
theory. Similarly, in applications of chiral perturbation theory to the
lattice, the LO mass $m_{uu}^2$ does not fully describe the mass of a pion
propagating on the lattice, as determined, for example, by studying the
exponential decay of a correlator.%
%
\footnote{Notice that once the lattice
spacing $a$ has been determined, the lattice-physical pion mass can be
unambiguously determined in this way. We assume the lattice spacing $a$
has been determined, for example by studying the heavy quark potential
or quarkonium spectrum.}
%
Therefore, it is useful to define
$m_\pi^2$ to be the renormalized pion mass on the lattice. By analogy
with nature, we will describe this pion mass as the lattice-physical pion
mass. We denote the lattice-physical mass of a meson $M$ by the symbol
$m_M^2$. In this chapter, we work consistently to second order in chiral
perturbation theory; therefore, we will only need the NLO formulae for
the meson masses to relate the lattice-physical mass to the LO mass.
Similarly, the lattice-physical pion decay constant $f_\pi$ is related
to the constant $f$ which appears in the chiral Lagrangian by computing
a one-loop correction.
\end{comment}

After integrating out the $\Phi_0$ field, the two point correlation
functions for the flavor-neutral states deviate from the simple single
pole form. The momentum space propagator between two flavor neutral
states is found to be at leading order~\cite{Sharpe:2001fh}
\begin{equation}\label{eq:etaPropMA}
	\mc{G}_{\eta_a \eta_b}(p^2) =
		\frac{i \e_a \d_{ab}}{p^2 -m_{\eta_a}^2 +i\e}
		- \frac{i}{N_f} \frac{\prod_{k=1}^{N_f}(p^2 -\tilde{m}_{k}^2 +i\e)}
			{(p^2 -m_{\eta_a}^2 +i\e)(p^2 -m_{\eta_b}^2 +i\e) \prod_{k^\prime=1}^{N_f-1}
				(p^2 -\tilde{m}_{k^\prime}^2 +i\e)},
\end{equation}
where
\begin{equation}
	\e_a = \left\{ \begin{array}{ll}
			+1& \text{for a = valence or sea quarks}\\
			-1 & \text{for a = ghost quarks}\,.
			\end{array} \right.
\end{equation}
In Eq.~\eqref{eq:etaPropMA}, $k$ runs over the flavor neutral states
($\phi_{jj}, \phi_{ll}, \phi_{rr}$) and $k^\prime$ runs over the
mass eigenstates of the sea sector. For $\pi\pi$ scattering, it will
be useful to work with linear combinations of these $\eta_a$ fields.
In particular we form the linear combinations
\begin{equation}
	\pi^0 = \frac{1}{\sqrt{2}} \left( \eta_u - \eta_d \right)\quad , \quad
	\bar{\eta} = \frac{1}{\sqrt{2}} \left( \eta_u +\eta_d \right),
\end{equation}
for which the propagators are
\begin{eqnarray}
	\mc{G}_{\pi^0}(p^2) &=& \frac{i}{p^2 -m_\pi^2 +i\e},\label{eq:PionPropMA} \\
	\mc{G}_{\bar{\eta}}(p^2) &=& \frac{i}{p^2 -m_\pi^2 +i\e}
		- \frac{2i}{N_f} \frac{\prod_{k=1}^{N_f}(p^2 -\tilde{m}_{k}^2 +i\e)}
			{(p^2 -m_\pi^2 +i\e)^2 \prod_{k^\prime=1}^{N_f-1}
				(p^2 -\tilde{m}_{k^\prime}^2 +i\e)}\label{eq:EtaBarPropMA}.
\end{eqnarray}
Specifically,
\begin{align}
	\mc{G}_{\bar{\eta}}(p^2) &= 
		\frac{i}{p^2 -m_\pi^2} -i \frac{p^2 -\tilde{m}_{jj}^2}{(p^2 -m_\pi^2)^2}, 
			 &\text{for $N_f = 2$}, \\
		&= 
		\frac{i}{p^2 - m_\pi^2} -\frac{2i}{3} \frac{(p^2 - \tilde m_{jj}^2)(p^2 - \tilde m_{rr}^2)}
				{(p^2 - m_\pi^2)^2\, (p^2 - \tilde m_{\eta}^2)},
			&\text{for $N_f = 2+1$,} \label{eq:EtaBarProp63}
\end{align}
where $\tilde{m}_{\eta}^2 = \frac{1}{3}(\tilde{m}_{jj}^2 +2\tilde{m}_{rr}^2)$.

%
%
%
%
%
%
%	Domain Wall Pions Scattering on a Staggered Sea
%
%
%
%
%
%
\section{Calculation of the $I=2$ Pion Scattering Amplitude}\label{sec:mixedScatt}

Our goal in this work is to calculate the $I=2$ $\pi\pi$ scattering
length in chiral perturbation theory for a partially quenched, mixed
action theory with GW valence quarks and staggered sea quarks, in order
to allow correct continuum and chiral extrapolation of mixed action
lattice data. We begin, however, by reviewing the pion scattering
amplitude in continuum $SU(2)$ chiral perturbation theory. We next
calculate the scattering amplitude in $N_f = 2$ PQ$\chi$PT and
MA$\chi$PT, and finally in $N_f = 2+1$ PQ$\chi$PT and MA$\chi$PT.
When renormalizing divergent 1-loop integrals, we use dimensional regularization 
and a modified minimal subtraction scheme ($\ol{MS}$) where we consistently subtract all terms proportional 
to~\cite{Gasser:1983yg}:
\begin{equation*}
\frac{2}{4-d} -\gamma_E + \log 4\pi +1,
\end{equation*} 
where $d$ is the number of space-time dimensions.  The scattering amplitude can
be related to the scattering length and other scattering parameters,
as we discuss in Section~\ref{sec:mixedLength}.

%
%
%
%
%
%
%	SU(2)
%
%
%
%
%
%
\subsection{\textbf{Continuum $SU(2)$}}

The tree-level $I=2$ pion scattering amplitude at threshold is
well known to be~\cite{Weinberg:1966kf}
\begin{equation}
	\mc{T} = - \frac{4 m_\pi^2}{f_\pi^2} .
\end{equation}
It is corrected at $\mc{O}({\varepsilon^4})$ by loop diagrams and
also by tree level terms from the NLO (or Gasser-Leutwyler)
chiral Lagrangian~\cite{Gasser:1983yg}.%
\footnote{The
continuum $\pi\pi$ scattering amplitude is known to
two-loops~\cite{Knecht:1995tr,Bijnens:1995yn,Bijnens:1997vq,Bijnens:2004eu}.}
The diagrams that
contribute at one loop order are shown in Figure~\ref{fig:one-loop};
they lead to the following NLO expression for the scattering amplitude:
\begin{equation}
        \mc{T}_{\vec{p_i}=0} = -\frac{4 m_{uu}^2}{f^2} \Bigg\{ 1 
                +\frac{m_{uu}^2}{(4\pi f)^2} \bigg[
                        8 \ln \left( \frac{m_{uu}^2}{\mu^2} \right) -1 + l^\prime_{\pi\pi}(\mu) \bigg] \Bigg\} ,
\label{eq:su2amplbare}
\end{equation}
where $m_{uu}$ is the tree-level expression given in
Eq.~(\ref{eq:m_tree}) and $f$ is the LO pion decay constant which appears
in Eq.~(\ref{eq:sigmaMA}).  The coefficient $l^\prime_{\pi\pi}$ is a linear
combination of low energy constants appearing in the Gasser-Leutwyler
Lagrangian whose scale dependence exactly cancels the scale dependence
of the logarithmic term. One can re-express the amplitude, however,
in terms of the physical pion mass and decay constant using the NLO
formulae for $m_\pi$ and $f_\pi$ to find:
\begin{align}
        \mc{T}_{\vec{p_i}=0} &= -\frac{4m_\pi^2}{f_\pi^2} \Bigg\{ 1 
                +\frac{m_\pi^2}{(4\pi f_\pi)^2} \bigg[ 
                        3 \ln \left( \frac{m_\pi^2}{\mu^2} \right) - 1 +l_{\pi\pi}(\mu) \bigg] \Bigg\},
\label{eq:su2ampl}
\end{align}
where $l_{\pi \pi}$ is a different linear combination of
low energy constants. The expression for $l_{\pi\pi}$ can
be found in Ref.~\cite{Bijnens:1997vq}. We do not, however, include
it here because we do not envision either using the known values of the
Gasser-Leutwyler parameters in the the fit of the scattering length or
using the fit to determine them. The simple expression~\eqref{eq:su2ampl}
has already been used in extrapolation of lattice data from mixed action
simulations~\cite{Beane:2005rj}, but it neglects lattice spacing effects
from the staggered sea quarks which are known from other simulations to be
of the same order as the leading order terms in the chiral expansion of
some observables~\cite{Aubin:2004fs}. We therefore proceed to calculate
the scattering amplitude in a partially quenched, mixed action theory
relevant to simulations.

%
%
%
%
%
%
%	SU(4|2)
%
%
%
%
%
%
\subsection{\textbf{Mixed GW-Staggered Theory with two Sea Quarks}}

The scattering amplitude in the partially quenched theory differs
from the unquenched theory in three important respects. First, more
mesons propagate in the loop diagrams. Second, some of the
mesons have more complicated propagators due to hairpin diagrams at
the quark level~\cite{Bernard:1993ga,Sharpe:2001fh}. Third, there are
additional terms in the NLO Lagrangian which arise from partial quenching~\cite{Sharpe:2003vy}, and lattice spacing effects~\cite{Bar:2005tu,Sharpe:2004is}.

\begin{figure}[t]
\center
\includegraphics[width=0.4\textwidth]{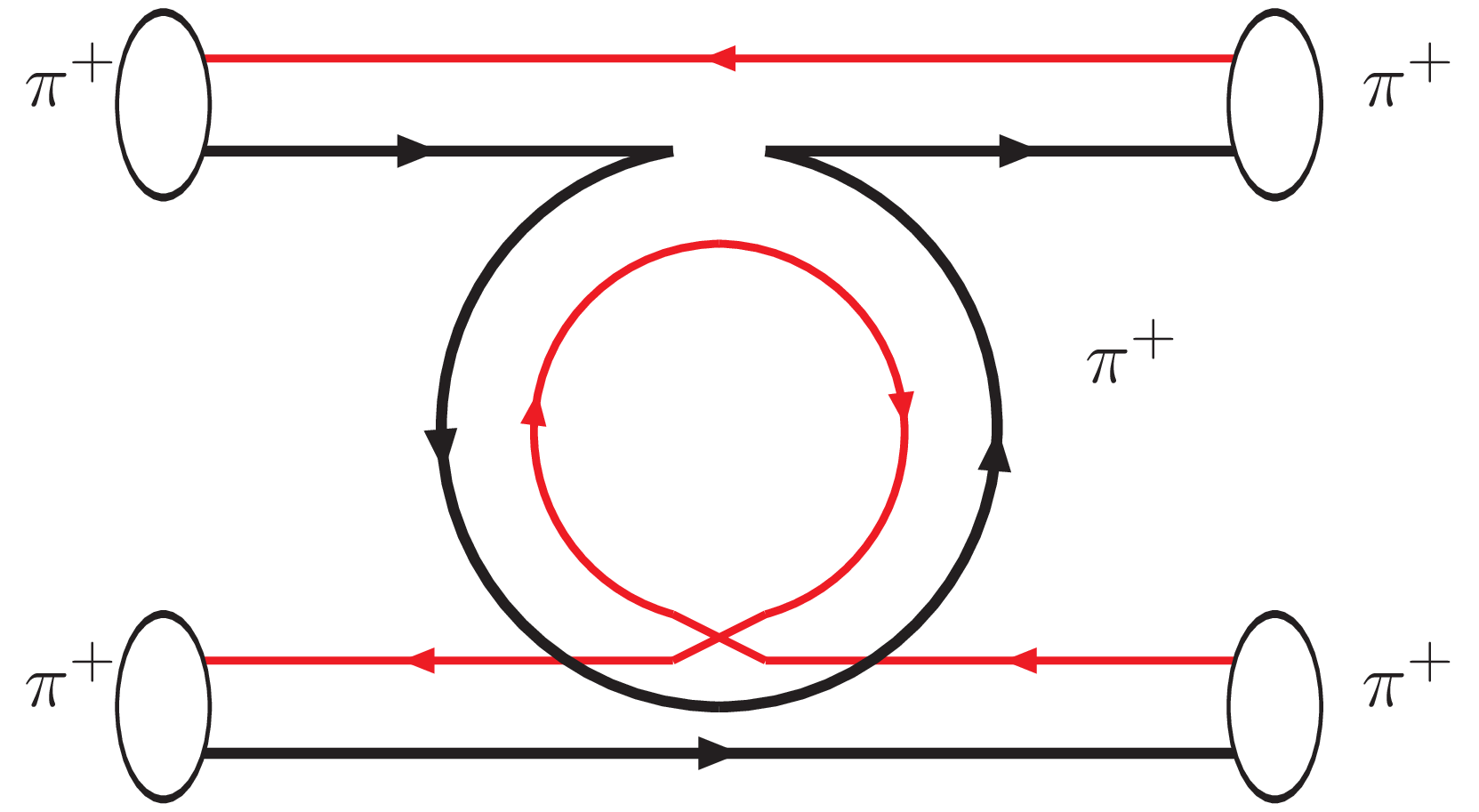}
\caption[$t$-channel $\pi\pi$ scattering graph with only valence quarks]{\label{fig:crossedline}Example quark flow for a one-loop $t$-channel graph. This diagram illustrates the presence of meson loops composed of purely valence-valence mesons which are not canceled by valence-ghost loops. Different colors (shades of grey) represent different quark flavors.} 
\end{figure}

At the level of quark flow, there are diagrams such as
Figure~\ref{fig:crossedline}, which route the valence quarks through
the diagram in a way which has no ghostly counterpart. Consequently,
the ghosts do not exactly cancel the valence quarks in loops. Of
course, this is simply a reflection of the fact that the initial and
final states --- valence pions --- are themselves not symmetric under
the interchange of ghost and valence quarks, and therefore the graded
symmetry between the valence and ghost pions has already been violated.
This is well known in quenched and partially quenched heavy baryon
\CPT~\cite{Labrenz:1996jy,Chen:2001yi,Beane:2002vq}.  This fact also
partly explains the success of quenched $\pi\pi$ scattering in the
$I=2$ channel~\cite{Sharpe:1992pp,Gupta:1993rn};  quenching does not
eliminate \emph{all} loop graphs like it does in many other processes,
and in particular, the $s$-channel diagram is not modified by (partial)
quenching effects.  As a consequence, it is necessary to compute all
the graphs contributing to this process in order to determine the scattering
amplitude.

%%	Figure 2

\begin{figure}[t]
\center
\includegraphics[width=0.28\textwidth]{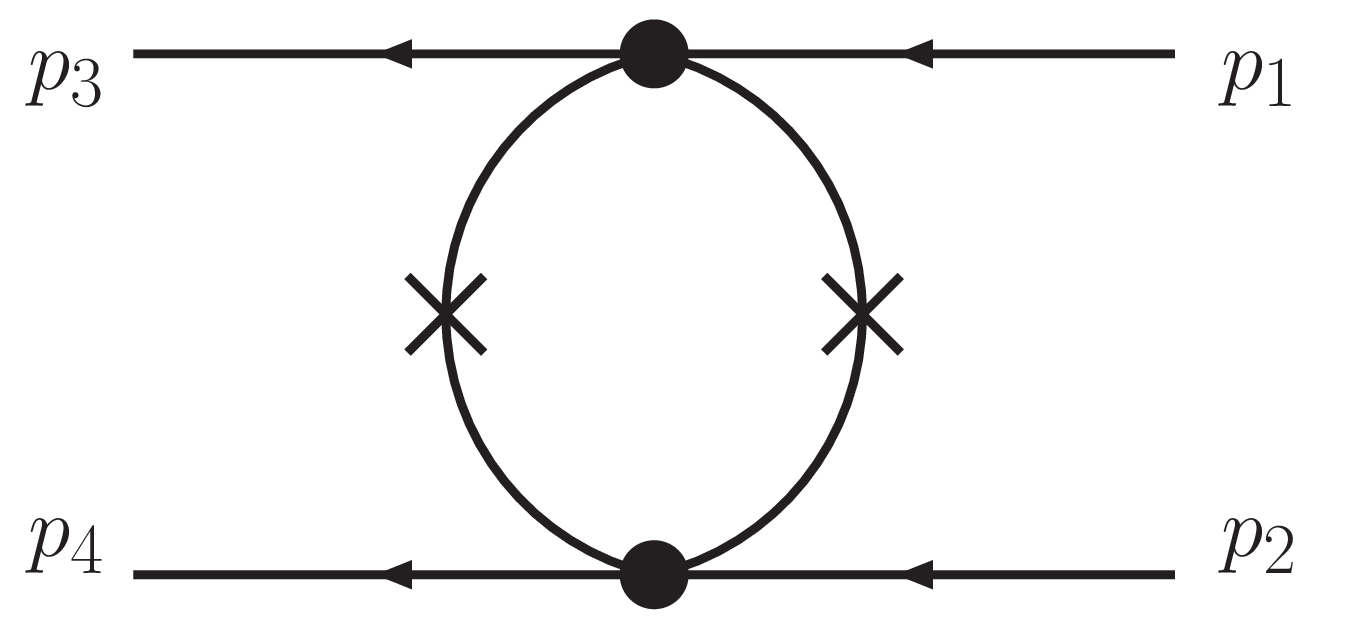}
\hspace{0.05\textwidth}
\includegraphics[width=0.28\textwidth]{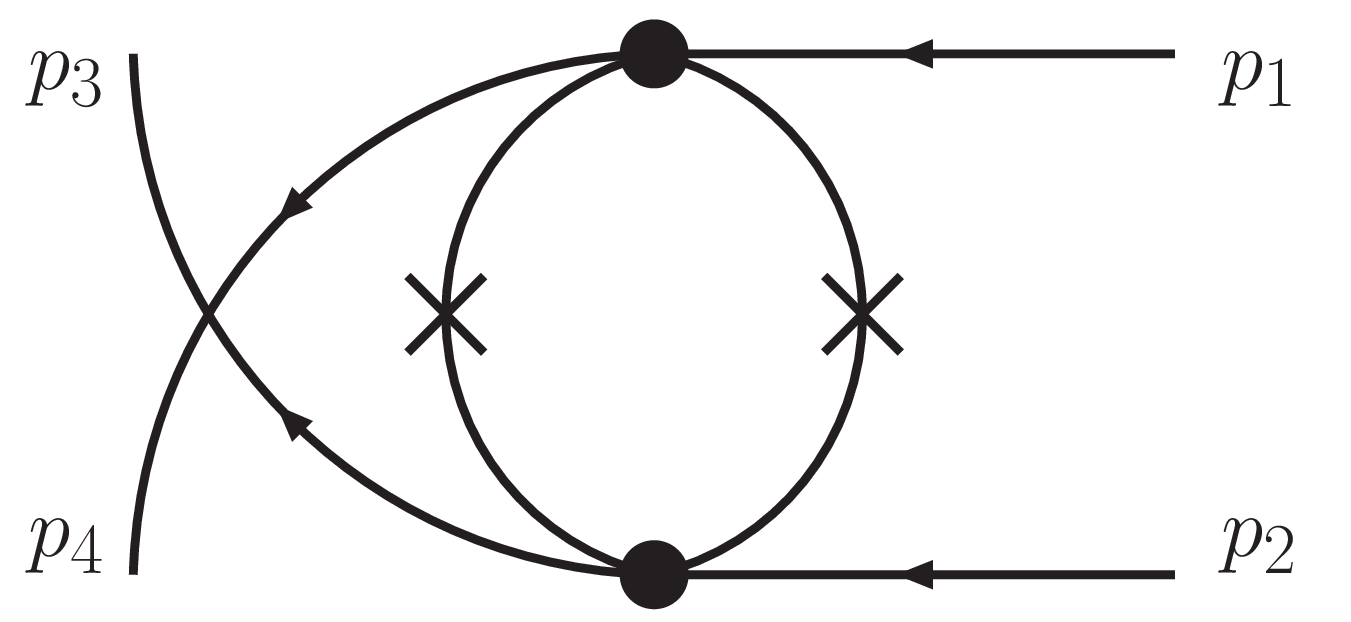}
\caption[$t$- and $u$-channel diagrams with hairpin interactions]{\label{fig:Hairpins} Example hairpin diagrams contributing to pion scattering. The propagator
with a cross through it indicates the quark-disconnected piece of the $\bar{\eta}$ propagator, Eq.~\eqref{eq:su42hairprop}.}
\end{figure}

Quark level disconnected (hairpin) diagrams lead to higher order poles in the propagator of any particle which has the quantum numbers of the vacuum~\cite{Bernard:1993ga,Sharpe:2001fh}. In the isospin limit of the $N_f = 2$ partially quenched theory, conservation of isospin prevents the $\pi^0$ from suffering any hairpin effects at leading order.  At higher orders, the $\pi^0$ mass (as well as the other mesons) will receive hairpin contributions but the $\pi^0$ propagator will never develop double poles, the characteristic unitarity violating feature of partially quenched and mixed action theories.  Hence only the $\bar \eta$ acquires a disconnected propagator. Moreover, in the $m_0 \rightarrow \infty$ limit, the $\bar \eta$ propagator (given for a general PQ theory in Eq.~\eqref{eq:EtaBarPropMA}) is given by the simple expression
\begin{align}
	G_{\bar\eta}(p^2) &= 
		\frac{i}{p^2 -m_\pi^2} -i \frac{p^2 - \tilde m^2_{jj}}{(p^2 - m^2_\pi)^2} \nonumber\\
		&= \frac{i \tilde\D_{PQ}^2}{(p^2 - m_\pi^2)^2},
\label{eq:su42hairprop}
\end{align}
where the parameter
\begin{equation}
        \tilde\D_{PQ}^2 = \tilde{m}_{jj}^2 -m_\pi^2
\end{equation}
quantifies the partial quenching. (Recall that $\tilde{m}_{jj}$ is the
physical mass of a taste \emph{singlet} sea-sea meson.)  Notice that when
$\tilde\D_{PQ} \rightarrow 0$ the propagator~\eqref{eq:su42hairprop} also goes
to zero; this is what we expect since, in the $SU(2)$ theory, the
only neutral propagating state is the $\pi^0$.
The propagator in Eq.~\eqref{eq:su42hairprop} can appear
in loops, thereby producing new diagrams such as those in
Fig.~\ref{fig:Hairpins}.%
%	footnote
\footnote{We note that there are also similar
contributions to the four particle vertex with a loop and to the mass
correction. We do not show them, however, because they cancel against
one another in the amplitude expressed in lattice-physical parameters, which we will show in the following pages.}
After adding all such hairpin diagrams, one finds that the contribution of the $\bar\eta$ to the amplitude is
%	footnote
\footnote{We note that this contribution does not vanish in the limit that $m_\pi^2
\rightarrow 0$ with $\tilde{m}_{jj}^2 \neq 0$.  Similar effects have
been observed in quenched computations of pion scattering
amplitudes~\cite{Colangelo:1997ch,Bernard:1995ez}.  This non-vanishing contribution is the
$I=2$ remnant of the divergences that are known to occur in the $I=0$ amplitude at threshold.  These divergences give rise to enhanced volume corrections to the $I=0$ amplitude with respect to the one-loop $I=2$ amplitude and prevent the use of L\"{u}scher's formula.  Moreover, it is
known~\cite{Sharpe:1997by,Sharpe:2000bc} that PQ\CPT\ is singular in the limit $m_u
\rightarrow 0$ with nonzero sea quark masses, so the behavior of
the amplitude in this limit is meaningless.}
%	end footnote
\begin{equation}\label{eq:su42amplhairpin}
	\mc{T}_{\bar \eta} = 
		\frac{4}{(4 \pi f_\pi)^2} \frac{\tilde\D_{PQ}^4}{6 f_\pi^2} .
\end{equation}

%%	Figure 3

In addition to 1-loop contributions, the NLO scattering amplitude receives tree-level analytic contributions from operators of $\mc{O}(\epsilon^4)$ in the chiral Lagrangian.  At this order, the mixed action Lagrangian contains the same $\mc{O}(p^4)$, $\mc{O}(p^2 m_q)$, and $\mc{O}(m_q^2)$ operators as in the continuum partially quenched chiral Lagrangian, plus additional $\mc{O}(a^4)$, $\mc{O}(a^2 m_q)$, and $\mc{O}(a^2 p^2)$ operators arising from discretization effects.  We can now enumerate the generic forms of analytic contributions from these NLO operators.  Because of the chiral symmetry of the GW valence sector, all tree-level contributions to the scattering length must vanish in the limit of vanishing valence quark mass.%
%	footnote
\footnote{As we discussed in the previous footnote, this condition need not hold for loop contributions to the scattering amplitude.}
Thus there are only three possible forms, each of which must be multiplied by an undetermined coefficient:  $m^4_{uu}$, $m^2_{uu}m^2_{jj}$, and $m^2_{uu}a^2$.  It may, at first, seem surprising that operators of $\mc{O}(a^2m_q)$, which come from taste-symmetry breaking and contain projectors onto the sea sector, can contribute at tree-level to a purely valence quantity.  Nevertheless, this turns out to be the case.  These $\mc{O}(a^2m_q)$ mixed action operators can be determined by first starting with the NLO staggered chiral Lagrangian~\cite{Sharpe:2004is}, and then inserting a sea projector, $\mc{P}_S$, next to every taste matrix.  One example of such an operator is $\left[\text{\str} \left(\Sigma m_Q^\dagger \right) \text{str}\left( \mc{P}_S \xi_5 \Sigma \xi_5 \Sigma^\dagger\right) + \textrm{p.c.}\right]$, where, $\xi_5$ is the $\g_5$ matrix acting in taste-space and p.c. indicates parity-conjugate.  This double-trace operator will contribute to the lattice pion mass, decay constant, and 4-point function at tree-level because one can place all of the valence pions inside the first supertrace, and the second supertrace containing the projector $\mc{P}_S$ will just reduce to the identity.  

Putting everything together, the total mixed action scattering amplitude to NLO is
\begin{multline}\label{eq:2seaBare}
        \mc{T}_{\vec{p_i}=0} = -\frac{4m_{uu}^2}{f^2} \Bigg\{ 1 
                +\frac{m_{uu}^2}{(4\pi f)^2} \Bigg[
                        4 \ln \left( \frac{m_{uu}^2}{\mu^2} \right) 
                +4 \frac{\tilde{m}_{ju}^2}{m_{uu}^2} \ln \left( \frac{\tilde{m}_{ju}^2}{\mu^2} \right) 
                -1 +l^\prime_{\pi\pi}(\mu) \Bigg]
                \\
		- \frac{m_{uu}^2}{(4\pi f)^2} \Bigg[
             	   	\frac{\tilde\D_{PQ}^4}{6 m_{uu}^4}
			+\frac{\tilde\D_{PQ}^2}{m_{uu}^2} \left[ \ln \left( \frac{m_{uu}^2}{\mu^2} \right) +1 \right]
		\Bigg] 
		\\
                + \frac{\tilde\D_{PQ}^2}{(4\pi f)^2}l^\prime_{PQ}(\mu) 
                + \frac{a^2}{(4\pi f)^2} l^\prime_{a^2}(\mu)
                \Bigg\}.
\end{multline}
The first line of Eq.~\eqref{eq:2seaBare} contains those terms which remain in the continuum and full QCD limit, Eq.~\eqref{eq:su2amplbare}, while the second and third lines account for the effects of partial quenching and of the nonzero lattice spacing.  Note that, for consistency with the 1-loop terms, we chose to re-express the analytic contribution proportional to the sea quark mass as $m^2_{uu}\tilde\Delta_{PQ}^2$.  In Eq.~\eqref{eq:2seaBare} we have multiplied every contribution
from diagrams which contain a sea quark loop by $1/4$, thus making our expression applicable to lattice simulations in which the fourth root of the staggered sea quark determinant is taken.

It is useful, however, to  re-express the scattering amplitude in terms of the quantities that one measures in a lattice simulation:   $m_\pi$ and $f_\pi$.  Throughout this chapter, we will refer to these renormalized measured quantities as the lattice-physical pion mass and decay constant.%
\footnote{Notice that once the lattice spacing $a$ has been determined, the lattice-physical pion mass can be unambiguously determined by measuring the exponential decay of a pion-pion correlator. We assume that the lattice spacing $a$ has been determined, for example, by studying the heavy quark potential or quarkonium spectrum.}
Because we are working consistently to second order in chiral perturbation theory, we can equate the lattice-physical pion mass to the 1-loop chiral perturbation theory expression for the pion mass, and likewise for the lattice-physical decay constant.  Thus, in terms of lattice-physical parameters, the mixed action $I=2$ $\pi\pi$ scattering amplitude is
\begin{equation}\label{eq:su42ampl}
        \mc{T}^{MA{\chi}PT}_{\vec{p_i}=0} = -\frac{4 m_\pi^2}{f_\pi^2} \Bigg\{ 1 
		+ \frac{m_\pi^2}{(4\pi f_\pi)^2} \bigg[ 
			3\ln \left( \frac{m_\pi^2}{\mu^2} \right) 
			-1 +l_{\pi\pi}(\mu) \bigg]
		-\frac{m_\pi^2}{(4\pi f_\pi)^2} \frac{\tilde\D_{PQ}^4}{6\, m_\pi^4} 
	\Bigg\},
\end{equation}
where the first few terms are identical in form to the full QCD amplitude, Eq.~\eqref{eq:su2ampl}.  This expression for the scattering amplitude is vastly simpler than the one in terms of the bare parameters.  First, the hairpin contributions from all diagrams except those in Fig.~\ref{fig:Hairpins} have exactly cancelled, removing the enhanced chiral logs and leaving the last term in Eq.~\eqref{eq:su42ampl} as the only explicit modification arising from the partial quenching and discretization effects.  Second, 
all contributions from mixed valence-sea mesons in loops have cancelled, thereby removing the new mixed action parameter, $C_{Mix}$, completely.%
\footnote{Another consequence of the exact cancellation of the loops with mixed valence-sea quarks is that one does not have to implement the ``fourth-root trick'' through this order.}
Third, all tree-level contributions proportional to the sea quark mass have also cancelled from this expression.  And finally, most striking is the fact that an explicit computation of the $\mc{O}(a^2 m_q)$ contributions to the amplitude arising from the NLO mixed action Lagrangian show that these effects exactly cancel when the amplitude is expressed in lattice-physical parameters.  This result will be discussed in detail in Ref.~\cite{Chen:noasqd}.  Thus to reiterate, the only  partial quenching and lattice spacing dependence in the amplitude comes from the hairpin diagrams of Fig.~\ref{fig:Hairpins}, which produce contributions proportional to $\tilde{\D}_{PQ}^4 = (m_{jj}^2 + a^2 \D(\xi_I) - m_\pi^2)^2$, where $m^2_{jj}+a^2\D(\xi_1)$ is the mass-squared of the taste-singlet sea-sea meson.  Moreover, we presume that anyone performing a mixed action lattice simulation will separately measure the taste-singlet sea-sea meson mass and use it as an input to fits of other quantities such as the $\pi\pi$ scattering length.  Thus we do not consider it to be an undetermined parameter.

It is now clear that one should fit $\pi\pi$ scattering lattice data in terms of the
lattice-physical pion mass and decay constant
rather than in terms of the LO pion mass and LO decay constant.  By doing this, one eliminates three undetermined fit parameters:  $C_{Mix}$, $l'_{PQ}$, and $l'_{a^2}$, as well as the enhanced chiral logs.

%
%
%
%
%
%
%	SU(6|3)
%
%
%
%
%
%
\subsection{\textbf{Mixed GW-Staggered Theory with $2+1$ Sea Quarks}}

The $2+1$ flavor theory has three additional quarks -- the strange
valence and ghost and strange sea quarks -- which can lead to new
contributions to the scattering amplitude.  Because we only consider
the scattering of valence pions, however, strange valence quarks cannot
appear in this process.  Thus all new contributions to the scattering
amplitude necessarily come only from the sea strange quark, $r$. Because
the $r$ quark is heavier than the other sea quarks there is $SU(3)$
symmetry breaking in the sea. This symmetry breaking only affects the
pion scattering amplitude, expressed in lattice-physical quantities,
through the graphs with internal $\bar{\eta}$ propagators because the
masses of the mixed valence-sea mesons cancel in the final amplitude as
they did in the earlier two flavor case. In addition, the only signature
of partial quenching in the amplitude comes from these same diagrams. It
is therefore worthwhile to investigate the physics of the neutral meson
propagators further.

There are more hairpin graphs in the $2+1$ flavor theory since the
$\eta_s$ may propagate as well as the $\eta_u$ and the $\eta_d$. Because
these mesons mix with one another, the flavor basis is not the most
convenient basis for the computation. Rather, a useful basis of states
is $\pi^0$, $\bar\eta = (\eta_u + \eta_d)/\sqrt 2$ and $\eta_s$. Since
we work in the isospin limit, the $\pi^0$ cannot mix with $\bar \eta$
or $\eta_s$; in addition, there is no vertex between the $\eta_s$
and $\pi^+ \pi^-$ at this order, so we never encounter a propagating
$\eta_s$. Thus all the PQ effects are absorbed into the $\bar \eta$
propagator, which is given by
\begin{equation}
	G_{\bar\eta}(p^2) = \frac{i}{p^2 - m_\pi^2} 
		-\frac{2i}{3} \frac{(p^2 - \tilde m_{jj}^2)(p^2 - \tilde m_{rr}^2)}
			{(p^2 - m_\pi^2)^2\, (p^2 - \tilde m_{\eta}^2)} .
\label{eq:su63hairpinprop}
\end{equation}
In $SU(3)$ chiral perturbation theory, the neutral mesons are the $\pi^0$
and the $\eta_8$. Therefore, in the PQ theory, we know that there will be
a contribution from the $\bar\eta$ graphs that does not result from
partial quenching or $SU(3)$ symmetry breaking.  Therefore the extra PQ
graphs arising from the internal $\bar{\eta}$ fields must not vanish in
the $\tilde{\D}_{PQ} \rightarrow 0$ limit, in contrast to the two flavor
case of Eq.~\eqref{eq:su42amplhairpin}. 

To make this clear, we can re-express the propagator of
Eq.~\eqref{eq:su63hairpinprop} in terms of $\tilde{\D}_{PQ}$ as
\begin{equation}
	G_{\bar\eta}(p^2) = 
		i \left[ \frac{\tilde{\D}_{PQ}^2}{(p^2 - m_\pi^2)^2} 
		+\frac{1}{3} \frac{1}{p^2 - \tilde m_{\eta}^2} 
			\left(1 - \frac{\tilde{\D}_{PQ}^2}{p^2 -m_\pi^2} \right)^2 \right] .
\label{eq:su63hairprop}
\end{equation}
This propagator has a single pole which is independent of
$\tilde{\D}_{PQ}$, as well as higher order poles that are at least
quadratic in $\tilde{\D}_{PQ}$. It is interesting to consider the large
$m_r$ limit of this propagator. In this limit, $\tilde m_\eta^2 \approx
\frac{4}{3} B m_r$ is also large. For momenta that are small compared
to $\tilde m_{\eta}$, the second term of this equation goes to zero in
the large $m_r$ limit, and the $2+1$ flavor propagator reduces to the
2 flavor propagator, Eq.~\eqref{eq:su42hairprop}, as expected.

While the above expression clarifies the $\tilde{\D}_{PQ}$
dependence of the propagator and the large $m_r$ limit, it obscures the
$SU(3)_{sea}$ limit. An equivalent form of the propagator is
\begin{equation}
	G_{\bar\eta}(p^2) = i \left[ \frac{\tilde{\D}_{PQ}^2}{(p^2 - m_\pi^2)^2} 
		+\frac{1}{3} \left(1 + \frac{\tilde{\D}_3^2}
			{p^2-\tilde m_{\eta}^2}\right) \frac{1}{p^2 - m_\pi^2} 
			\left(1 - \frac{\tilde{\D}_{PQ}^2}{p^2 -m_\pi^2} \right) \right] ,
\label{eq:su63hairpropdelta3}
\end{equation}
where the quantity $\tilde{\D}_3 = \sqrt{\tilde{m}_{\eta}^2 -
\tilde{m}_{jj}^2}$ parametrizes the $SU(3)_{sea}$ breaking. When
$\tilde{\D}_3= 0$ this propagator is similar in form to the corresponding
2 flavor propagator, Eq.~\eqref{eq:su42hairprop}, but it has an additional
single pole due to the extra neutral meson in the $SU(3)$ theory.

Having considered the new physics of the hairpin propagator, we can now
calculate the scattering amplitude. For our purposes here, it is most
convenient to express the total $I=2$ $\pi\pi$ scattering amplitude in
terms of $\tilde{\D}_{PQ}$. Just as in the 2-flavor computation, the NLO analytic contributions due to partial quenching and finite lattice spacing effects exactly cancel when the amplitude is expressed in lattice-physical parameters.  All sea quark mass and lattice spacing dependence comes from the hairpin diagrams, which produce terms proportional to powers of $\tilde\Delta_{PQ}$ with known coefficients.
The amplitude is
\begin{multline}
        \mc{T}^{MA{\chi}PT}_{\vec{p_i}=0} = -\frac{4 m_\pi^2}{f_\pi^2} \Bigg\{ 1 
                +\frac{m_\pi^2}{(4\pi f_\pi)^2} \Bigg[
                        3 \ln \left( \frac{m_\pi^2}{\mu^2} \right) -1 
                        +\frac{1}{9}\left[ \ln \left( \frac{\tilde{m}_{{\eta}}^2}{\mu^2} \right) +1 \right]
                        + \bar{l}_{\pi\pi}(\mu) \Bigg]
                         \\
		+\frac{1}{(4\pi f_\pi)^2} \Bigg[
%			m_{jr}^2\, l_{\pi K} 
			-\frac{\tilde{\D}_{PQ}^4}{6 m_\pi^2} 
%			+\tilde{\D}_{PQ}^2\, \bar{l}_{PQ} 
%	\\
		+m_\pi^2 \sum_{n=1}^4 \left( \frac{\tilde{\D}_{PQ}^2}{m_\pi^2} \right)^n\, 
                        	\mc{F}_n \left( \tilde{m}_{\eta}^2/m_\pi^2 \right)
	\Bigg] \Bigg\},
\label{eq:su63ampl}
\end{multline}
where $\tilde{\D}_{PQ}^2 = m_{jj}^2 + a^2 \D(\xi_I) - m_\pi^2$ and
\begin{subequations}
\begin{align}
        \mc{F}_1(x) &= -\frac{2}{9(x-1)^2} \left[ 5(x-1) -(3x +2)\ln (x) \right], \\ %\nonumber\\
        \mc{F}_2(x) &= \frac{2}{3(x-1)^3} \left[ (x-1)(x+3) -(3x +1)\ln(x) \right], \\ %\nonumber\\
        \mc{F}_3(x) &= \frac{1}{9(x-1)^4} \left[ (x-1) (x^2 -7x -12) +2(7x+2) \ln(x) \right], \\ %\nonumber\\
        \mc{F}_4(x) &= -\frac{1}{54 (x-1)^5} \left[ (x-1) (x^2 -8x -17) +6(3x+1)\ln(x) \right] .
\end{align}\label{eq:coolFs}\end{subequations}
The functions $\mc{F}_i$ have the property that $\mc{F}_i(x) \rightarrow 0$
in the limit that $x \rightarrow \infty$. Therefore, when
the strange sea quark mass is very large, i.e. $\tilde m_{\eta}^2
/ m_\pi^2 \gg 1$, the $2+1$ flavor amplitude reduces to the 2 flavor amplitude, Eq.~\eqref{eq:su42ampl}, with the exception of terms that can be absorbed into the analytic terms.  The low energy constants have a scale dependence which exactly cancels the scale dependence in the logarithms.  The coefficient $\bar{l}_{\pi\pi}$ is the same linear combination of Gasser-Leutwyler coefficients that appear in the $SU(3)$ scattering amplitude expressed in terms of the physical pion mass and decay constant~\cite{Knecht:1995tr,Bijnens:2004eu}. 

Because the functions $\mc{F}_i$ depend logarithmically on
$x$, the $2+1$ flavor scattering amplitude features enhanced chiral logarithms~\cite{Sharpe:1997by}
that are absent from the 2 flavor amplitude. 
This is a useful observation, as we will now explain. Because there
is a strange quark in nature and its mass is less than the
QCD scale, $\Lambda_{\textrm QCD}$, lattice simulations must
use $2+1$ quark flavors. It is often practical to fix the strange quark mass
at a constant value near its physical value in these simulations. This
circumstance is helpful because, just as $SU(2)$ chiral perturbation
theory is useful to describe nature at scales smaller than the strange
quark mass, the 2 flavor amplitude given in Eq.~\eqref{eq:su42ampl}
can be used to extrapolate $2+1$ flavor lattice data at energy scales smaller than the
strange sea quark mass used in the simulation (provided, of course, there are no strange valence quarks)~\cite{Chen:2002bz}. This
is valid because, at energy scales smaller than the strange quark mass (or actually twice the strange quark mass, since the purely pionic systems have no valence strange quarks),
one can integrate out the strange quark. This is not an approximation,
because all of the effects of the strange quark are absorbed into a
renormalization of the parameters of the chiral Lagrangian.
Moreover, since the 2 flavor amplitude does not exhibit
enhanced chiral logarithms, signatures of partial quenching can be reduced
by extrapolating lattice data with the 2 flavor, rather than the $2+1$
flavor, expression. We note that in this case the effects of the strange quark are absorbed in the coefficients of the analytic terms appearing in Eq.~\eqref{eq:su42ampl}, and thus they are not constant,
but rather depend logarithmically upon the strange sea quark mass.

%
%
%
%
%
%
%	Scattering Length
%
%
%
%
%
%
\section{$I=2$ Pion Scattering Length Results}\label{sec:mixedLength}

In this section we  present our results for the $s$-wave $I=2\ \pi\pi$ scattering
length in the two theories most relevant to current mixed action lattice
simulations: those with GW valence quarks and either $N_f=2$ or $N_f =
2+1$ staggered sea quarks.  We only present results for the scattering length expressed in lattice-physical parameters.  The $s$-wave scattering length is trivially related
to the full scattering amplitude at threshold by an overall prefactor:

\begin{equation}
	a_{l=0}^{(I=2)} = \frac{1}{32 \pi m_\pi}  \mc{T}^{I=2}  \bigg|_{\vec{p_i}=0}.
\end{equation}

%
%
%
%
%
%
%	SU(4|2)
%
%
%
%
%
%
\subsection{\textbf{Scattering Length with 2 Sea Quarks}}

The $I=2\ \pi\pi$ $s$-wave scattering length in a MA\CPT\ theory with 2 sea quarks is given by
\begin{equation}\label{eq:2seaScattLength}
	{a_{0}^{(2)}}^{MA\chi PT} = -\frac{m_\pi}{8 \pi f_\pi^2} \Bigg\{ 1 
                + \frac{m_\pi^2}{(4\pi f_\pi)^2} \bigg[ 
                        3\ln \left( \frac{m_\pi^2}{\mu^2} \right) 
                        -1 +l_{\pi\pi}(\mu) \bigg]
		-\frac{m_\pi^2}{(4\pi f_\pi)^2} \frac{\tilde\D_{PQ}^4}{6\, m_\pi^4} 
	\Bigg\},
\end{equation}
where $\tilde{\D}_{PQ}^2 = m_{jj}^2 + a^2 \D(\xi_I) - m_\pi^2$.
The first two terms are the result one obtains in $SU(2)$ \CPT~\cite{Bijnens:1997vq} and the last term is the only new effect arising from the partial quenching and mixed action.  All other possible partial quenching terms, enhanced chiral logs and additional linear combinations of the $\mc{O}(p^4)$ Gasser-Leutwyler coefficients, exactly cancel when the scattering length is expressed in terms of lattice-physical parameters.  And, most strikingly, the pion mass, decay constant and the 4-point function all receive 
$\mc{O}(a^2 m_q)$ corrections from the lattice, but they exactly cancel in the scattering length expressed in terms of the lattice -physical parameters~\cite{Chen:noasqd}.  It is remarkable that the only artifact of the nonzero lattice spacing, $m^2_{jj} + a^2 \D_I$, can be separately determined simply by measuring the exponential fall-off of the taste-singlet sea-sea meson 2-point function.  Thus there are no undetermined fit parameters in the mixed action scattering length expression from either partial quenching or lattice discretization effects;  there is only the unknown continuum coefficient, $l_{\pi\pi}$.  

One can trivially deduce the continuum PQ scattering
length from Eq.~\eqref{eq:2seaScattLength}: simply let $a \rightarrow 0$, reducing $\tilde
m_{jj} \rightarrow m_{jj} = 2 B m_j$ in $\tilde{\D}_{PQ}$, resulting in
\begin{equation}
      {a_{0}^{(2)}}^{PQ{\chi}PT} = -\frac{m_\pi}{8 \pi f_\pi^2} \Bigg\{ 1 
                + \frac{m_\pi^2}{(4\pi f_\pi)^2} \bigg[ 
                        3\ln \left( \frac{m_\pi^2}{\mu^2} \right) 
                        -1 +l_{\pi\pi}(\mu) \bigg]
                        -\frac{\D_{PQ}^4}{6(4\pi m_\pi f_\pi)^2} %\bigg[
%                        	l_{PQ} -
%			\frac{\D_{PQ}^4}{6 m_\pi^2} \bigg]
		 \Bigg\} .
\label{eq:PQsu42Length}
\end{equation}

%
%
%
%
%
%
%	SU(6|3)
%
%
%
%
%
%
\subsection{\textbf{Scattering Length with 2+1 Sea Quarks}}

The $I=2\ \pi\pi$ $s$-wave scattering length in a MA\CPT\ theory with 2+1 sea quarks is given by
\begin{multline}\label{eq:2P1seaScattLength}
	{a_{0}^{(2)}}^{MA\chi PT} =  -\frac{m_\pi}{8 \pi f_\pi^2} \Bigg\{ 1 
                +\frac{m_\pi^2}{(4\pi f_\pi)^2} \Bigg[
                        3 \ln \left( \frac{m_\pi^2}{\mu^2} \right) -1 
                        +\frac{1}{9}\left[ \ln \left( \frac{\tilde{m}_{{\eta}}^2}{\mu^2} \right) +1 \right]
                        + \bar{l}_{\pi\pi}(\mu) \Bigg]
                         \\
		+\frac{1}{(4\pi f_\pi)^2} \Bigg[
%			m_{jr}^2\, l_{\pi K} 
			-\frac{\tilde{\D}_{PQ}^4}{6 m_\pi^2} 
%			+\tilde{\D}_{PQ}^2\, \bar{l}_{PQ} \\
%
		+m_\pi^2 \sum_{n=1}^4 \left( \frac{\tilde{\D}_{PQ}^2}{m_\pi^2} \right)^n\, 
                        	\mc{F}_n \left( \tilde{m}_{\eta}^2/m_\pi^2 \right)
	\Bigg] \Bigg\},
\end{multline}
where the functions $\mc{F}_i$ are defined in Eq.~\eqref{eq:coolFs}.  As in the 2-flavor
MA\CPT\  expression, Eq.~\eqref{eq:2seaScattLength}, the only undetermined parameter is the linear combination of Gasser-Leutwyler coefficients, $\bar{l}_{\pi\pi}$, which also appears in the continuum \CPT\ expression.   

We note as an aside that this suppression of lattice spacing counterterms is in contrast to the larger number of terms that one would need in order to correctly fit data from simulations with Wilson valence quarks on Wilson sea
quarks.  Because the Wilson action breaks chiral symmetry at $\mc{O}(a)$, even for massless quarks, there will be terms proportional to all powers of the lattice spacing in the expression for the scattering length in Wilson $\chi$PT~\cite{Chen:noasqd,Rupak:2002sm,Aoki:2005mb}.  Moreover, such lattice spacing corrections begin at $\mc{O}(a)$, rather than $\mc{O}(a^2)$.  If one uses $\mc{O}(a)$ improved Wilson quarks, then the leading discretization effects are of $\mc{O}(a^2)$, as for staggered quarks; however, this does not remove the additional chiral symmetry-breaking operators.  
Another practical issue is whether or not one can perform simulations with
Wilson sea quarks that are light enough to be in the chiral regime.

%
%
%
%
%
%
%	Summary
%
%
%
%
%
%
\section{Discussion}\label{sec:summary}

Considerable progress has recently been made in fully dynamical simulations of pion scattering in the $I=2$ channel~\cite{Yamazaki:2004qb,Beane:2005rj}. We have considered $I=2$ scattering of pions composed of Ginsparg-Wilson quarks on a staggered sea. We have calculated the scattering length in both this mixed action theory and in continuum PQ$\chi$PT for theories with either 2 or $2+1$ dynamical
quarks. These expressions are necessary for the correct continuum and chiral extrapolation of PQ and mixed action lattice data to the physical pion mass.

\begin{figure}
\center
\includegraphics[width=0.75\textwidth]{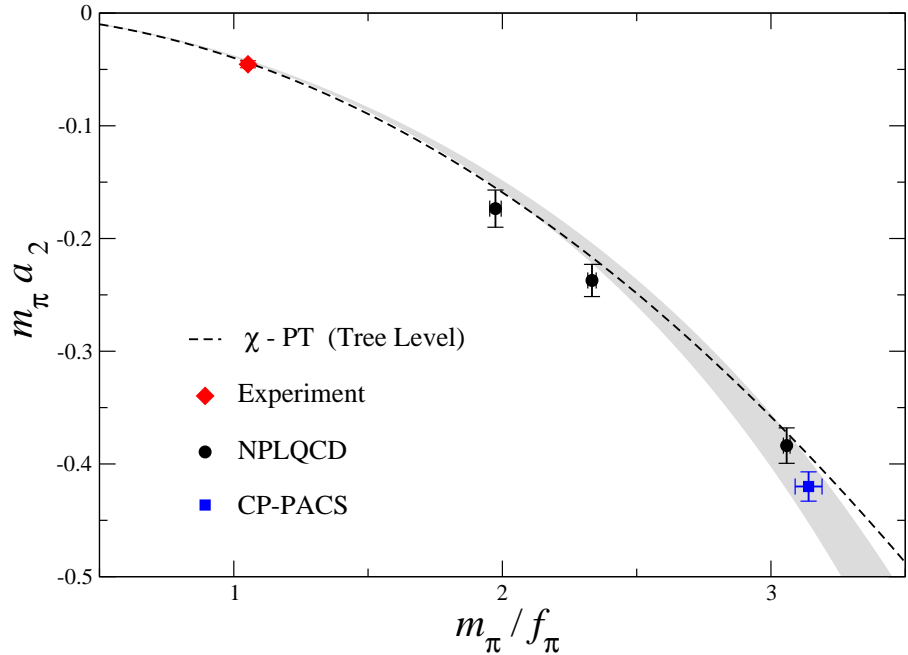}
\caption[Plot of $m_\pi a_2^{(0)}$ vs. $m_\pi / f_\pi$ from NPLQCD collaboration.]{\label{fig:mpia2} This is an updated plot of the $I=2\ \pi\pi$ scattering length determined by the NPLQCD collaboration~\cite{Beane:2005rj}.  The scattering length is plotted as a function of $m_\pi / f_\pi$ as measured in the lattice simulation, or in terms of the lattice parameters.  The dashed line is the tree level prediction.  The experimental point was not included in the fit.}
\end{figure}

Our formulae, Eqs.~\eqref{eq:2seaScattLength},~\eqref{eq:2P1seaScattLength}, not only provide the form for the mixed action scattering length, but also contain two predictions relevant to the recent work of Ref.~\cite{Beane:2005rj}.  Beane \emph{et.~al.}~calculated the $I=2$ $s-$wave $\pi\pi$ scattering length using domain wall valence quarks and staggered sea quarks, but used the continuum $\chi$PT expression to extrapolate to the physical quark masses.  In Figure~2 of Ref.~\cite{Beane:2005rj} (see Fig.~\ref{fig:mpia2}), which plots  $m_\pi a_2^{(0)}$ versus $m_\pi / f_\pi$, the fit of the $\chi$PT expression to the lattice data does a remarkably good job.  However, Eq.~\eqref{eq:2seaScattLength} predicts a \emph{known}, positive shift to $m_\pi a_2^{(0)}$ of size $\tilde\Delta_{PQ}^4/(768 f_\pi^4 \pi^3)$.  Accounting for this positive shift is equivalent to lowering the entire curve.  In Ref.~\cite{Beane:2005rj}, the valence and sea quark masses are tuned to be equal, so $\tilde\D_{PQ}^2 = a^2 \D_I \simeq (446 \textrm{ MeV})^2$~\cite{Aubin:2004fs}.  Despite the large value of $\tilde\D_{PQ}$, the predicted shift is insignificant, being an order of magnitude less than the statistical error. In Table~\ref{t:ashifts}, we collect the predicted shifts to $m_\pi a_2^{(0)}$ at the three pion masses used in Ref.~\cite{Beane:2005rj}.    We also list the magnitude of the ratio of these predicted shifts to the leading contribution to the scattering length, which turn out to be small, lending confidence to the power counting we have used, Eq.~\eqref{eq:epsilons}.  The other more important prediction is that there are no unknown corrections to the \CPT\ formula for the scattering length arising from lattice spacing corrections or partial quenching through the order $\mc{O}(m_q^2)$, $\mc{O}(a^2 m_q)$ and $\mc{O}(a^4)$.  Therefore, to within statistical and systematic errors, the continuum \CPT\ expression used by Beane \emph{et.~al.}~to fit their numerical $\pi\pi$ scattering data~\cite{Beane:2005rj} receives no corrections through the 1-loop level.

The central result of this chapter is that the appropriate way to extrapolate lattice $\pi\pi$ scattering data
is in terms of the lattice-physical pion mass and decay constant rather than in terms of the LO parameters which appear in the chiral Lagrangian.  When expressed in terms of the LO parameters, the scattering length depends upon 4 undetermined parameters, $l^\prime_{\pi\pi}$, $l^\prime_{PQ}$, $l^\prime_{a^2}$ and $C_{Mix}$.  In contrast, the scattering length expressed in terms of the lattice-physical parameters depends upon only one unknown parameter, $l_{\pi\pi}$, the same linear combination of Gasser-Leutwyler coefficients which contributes to the scattering length in continuum \CPT.

\newpage

%%%%%%%%%%%%%%%%%%%%%%%%%%%%%%%%%%%%%%%%%%%%%%%%%%%
%
%	table: pipi values of NPLQCD
%
%%%%%%%%%%%%%%%%%%%%%%%%%%%%%%%%%%%%%%%%%%%%%%%%%%%
\begin{table}
\center
\caption[Predicted lattice spacing shifts to $I=2\ \pi\pi$ scattering length]{\label{t:ashifts}Predicted shifts to the scattering length computed in Ref.~\cite{Beane:2005rj} arising from finite lattice spacing effects in the mixed action theory.  The first two rows show the approximate values of $m_\pi$ and $f_\pi$  while the third shows $m_\pi a_2^{(0)}$ plus the statistical error calculated in \cite{Beane:2005rj}.  In the fourth row, we give the predicted shifts in the scattering length (times $m_\pi$) and, in the fifth row, we give the ratio of the predicted shift to the leading order contribution to the scattering length.}
%\begin{ruledtabular}
\begin{tabular}{| c | c c c |}
\hline
 $m_\pi$ (MeV) & $294$ & $348$ & $484$   \\
 $f_\pi$ (MeV) & $145$ & $149$ & $158$ \\
$ m_\pi a_2^{(0)}$ 
& $-0.212 \pm 0.024$ 
& $-0.222 \pm 0.014$ 
& $-0.38 \pm 0.03$ \\
\hline 
$\frac{\tilde\D_{PQ}^4}{768 \pi^3 f_\pi^4}$
 & $0.00374$ 
 & $0.00336$ 
 & $0.00266$  \\    
$\frac{\tilde\D_{PQ}^4}{6 (4\pi f_\pi m_\pi)^2}$
 & $0.0229$ 
 & $0.0155$ 
 & $0.00711$ \\
\hline
\end{tabular}
%\end{ruledtabular}
\end{table}
%%%%%%%%%%%%%%%%%%%%%%%%%%%%%%%%%%%%%%%%%%%%%%%%%%%
%
%%%%%%%%%%%%%%%%%%%%%%%%%%%%%%%%%%%%%%%%%%%%%%%%%%%

%%%%%%%%%%%%%%%%%%%%%%%%%%%%%%%%%%%%%%%%%%%%%%%%%%%
%
%	figure: pipi from Martin
%
%%%%%%%%%%%%%%%%%%%%%%%%%%%%%%%%%%%%%%%%%%%%%%%%%%%
%%%%%%%%%%%%%%%%%%%%%%%%%%%%%%%%%%%%%%%%%%%%%%%%%%%
%
%%%%%%%%%%%%%%%%%%%%%%%%%%%%%%%%%%%%%%%%%%%%%%%%%%%

 % ========== Chapter 7: Mixed Actions with GW-valence
\chapter{Mixed Action Lattice QCD with Ginsparg-Wilson Valence Quarks}\label{chap:MAGW}

This chapter is based upon work that is in progress~\cite{Chen:noasqd,Tiburzi:noasqd}, and which provides an understanding of the results found in the previous chapter.  This chapter makes extensive use of Refs.~\cite{Sharpe:2000bc,Sharpe:2001fh,Chen:2001yi,Beane:2002vq,Sharpe:2003vy,Bar:2003mh,Tiburzi:2005vy,Tiburzi:2005is,Bar:2005tu,Sharpe:2004is,Bar:2002nr,Golterman:2005xa}.  We discuss in some generality the features of a mixed action theory with Ginsparg-Wilson valence quarks, and arbitrary sea quarks.  We will use staggered sea quarks at times for specific examples, as this combination of valence and sea quarks is the most relevant for current lattice simulations.

%%%%%%%%%%%%%%%%%%%%%%%%%%%%%%%%%%%%%%%%%%%%%%%%%%%
%
%	Mixed Actions: quark level
%
%%%%%%%%%%%%%%%%%%%%%%%%%%%%%%%%%%%%%%%%%%%%%%%%%%%
\section{Mixed Action Quark Level Lagrangian}\label{sec:MAQuarks}

An understanding of mixed action lattice QCD must involve the effective theory for the mixed action simulation, but to construct this effective theory we must first understand the mixed action quark-level Lagrangian.  As we have discussed several times in this work, the procedure is to first construct the effective continuum theory of the lattice action, the Symanzik Lagrangian~\cite{Symanzik:1983gh,Symanzik:1983dc}, and then to construct the low-energy effective field theory which is consistent with the lattice symmetries~\cite{Sharpe:1998xm}.  The Symanzik Lagrangian for a mixed action theory has the form~\cite{Bar:2003mh}
\begin{equation}\label{eq:MASymanzik}
\mc{L} = \mc{L}_{PQQCD} 
		+a \mc{L}^{(5)} +a^2 \mc{L}^{(6)} +\dots
\end{equation}
We see that the leading Lagrangian is simply the partially quenched Lagrangian which contains QCD as a limit.  We comment again that a mixed action theory is automatically partially quenched, even if the bare quark mass parameters in the valence and sea sector are set equal.  This is simply a consequence of the different lattice spacing modifications from the sea and valence Dirac operators, which force one to break the partially quenched symmetry and treat these quarks differently.  However, we will show that it is useful to think of a mixed action theory with Ginsparg-Wilson valence quarks not as a mixed action theory, but rather as a full QCD or partially quenched theory of Ginsparg-Wilson quarks with a perturbative breaking of the symmetry between the valence and sea sectors, which can be seen by Eq.~\eqref{eq:MASymanzik}.  To determine the mixed action effects, we must understand the operators in $\mc{L}^{(5)}$ and beyond.  We will in fact begin with $\mc{L}^{(6)}$, as typically, Wilson sea quarks are generally implemented with $\mc{O}(a)$ improvement and staggered quark discretization effects begin at $\mc{O}(a^2)$.  To construct $\mc{L}^{(6)}$, it is convenient to group the different operators into a few categories.  Firstly, there are those operators which consist of only sea-quarks, there are those of only valence quarks and then there are operators which mix the valence and sea quarks.  We can further decompose these three sets into operators which break chiral symmetry, operators which do not break chiral symmetry and those operators which respect the lattice symmetry but break the continuum Lorentz symmetry.  The mixed action symmetry forbids quark bi-linear operators which mix valence and sea quarks and therefore the only type of operators which can mix the quarks are products of bi-linears.  The Lorentz breaking operators are of the form%
\footnote{In Euclidean spacetime, as with a lattice simulation, it is an $O(4)$ rotational symmetry which is broken.}
\begin{equation}\label{eq:O4break}
	\ol{Q}_S\, \gamma_\mu D_\mu D^\mu D^\mu\, Q_S \quad,\quad
	\ol{Q}_V\, \gamma_\mu D_\mu D^\mu D^\mu\, Q_V\, ,
\end{equation}
in which there is a sum over the Lorentz indices.  In fact these operators can be distinguished in the dispersion relation for the various hadrons.  For the mesons, these effects will be of $\mc{O}(p^4 a^2)$ and thus beyond the order we are interested, but for the baryons, these operators can enter at $\mc{O}(a^2)$~\cite{Tiburzi:2005is}, and for a zero momentum projection they act as a simple mass shift.

There are additional quark bi-linears which do not break chiral symmetry, and are of the form
\begin{equation}
	\ol{Q}_S\, \Dslash\, D^2\, Q_S \quad, \quad \ol{Q}_V\, \Dslash\, D^2\, Q_V\, .
\end{equation}
These operators provide some difficulty from an effective field theory point of view.  Under chiral transformations, these operators transform identically to the (PQ)QCD quark kinetic operator, and thus there is no way to distinguish these operators by their symmetry transformations.  Thus, these operators lead to an $\mc{O}(a^2)$ shift to all the LECs of the continuum chiral Lagrangian.  For example, these operators will contribute a perturbative correction to the chiral condensate which drives chiral symmetry breaking~\cite{Bar:2003mh}.  Thus to disentangle these effects from the physics of interest (the continuum LECs), one must vary the lattice spacing in combination with the quark masses.  However, unlike with the continuum LECs, we do not necessarily need or want to know the precise values of these LECs associated with finite lattice spacing, as they are lattice artifacts and not physical.  Rather we want to be able parameterize their total contributions to a given correlation function or observable, such that there effects can be removed and we can extract the physics of interest.

From a mixed action perspective, the most interesting operators are the 4-quark operators which allow an interaction of the valence and sea quarks through products of quark bi-linears.  These 4-quark operators will have the general form
\begin{align}
	\left( \ol{Q}_S\, \Gamma_{A,B}\, Q_S \right) \left( \ol{Q}_S\, \Gamma_{A,B}\, Q_S \right) \quad&,\quad
	\left( \ol{Q}_V\, \Gamma_{A,B}\, Q_V \right) \left( \ol{Q}_V\, \Gamma_{A,B}\, Q_V \right)\, ,\nonumber\\
	\left( \ol{Q}_S\, \Gamma_{A,B}\, Q_S \right) & \left( \ol{Q}_V\, \Gamma_{A,B}\, Q_V \right)\, ,
\end{align}
where the matrix $\Gamma_{A,B}$ acts on the internal quantum numbers of the quark fields; the spin and for example in the case of staggered quarks, the taste.  Perhaps a more illuminating form to write these operators is with the valence and sea projectors,
\begin{align}
	\ol{Q}_S\, \Gamma_{A,B}\, Q_S &= \ol{Q}\, \mc{P}_S\, \Gamma_{A,B}\, \mc{P}_S\, Q \nonumber\\
	\ol{Q}_V\, \Gamma_{A,B}\, Q_V &= \ol{Q}\, \mc{P}_V\, \Gamma_{A,B}\, \mc{P}_V\, Q\, .
\end{align}
For the unmixed operators, one has all the possibilities discussed in Ref.~\cite{Bar:2003mh}.
The Ginsparg-Wilson symmetry of the valence quarks prevents the addition of any operator which breaks chiral symmetry except for operators involving the quark mass matrix.  Thus, for the dimension-6 operators with Ginsparg-Wilson valence quarks, there are only two mixed action operators,
\begin{align}
	\mc{L}_{Mix}^{(6)} = C \left( \ol{Q}\, \g_\mu \mc{P}_V\, Q \right)
			\left( \ol{Q}\, \g^\mu \mc{P}_S\, Q \right)
		+C^\prime \left( \ol{Q}\, \g_\mu\g_5 \mc{P}_V\, Q \right)
			\left( \ol{Q}\, \g^\mu\g_5 \mc{P}_S\, Q \right)\, .
\end{align}
These operators give rise to the mixed term in Eq.~\eqref{eq:MixPotential} for the example of staggered sea quarks, and in the following we will be interested in the next higher order effective terms arising from these mixed quark level operators.

Why go through all this effort to construct the Lagrangian which includes all these lattice spacing corrections?  Why not simply treat all the continuum LECs as polynomial functions of the lattice spacing?  There are several answers to this question.  Firstly, not all of the lattice spacing dependence can be captured in this way.  For example, with a mixed action theory, there will be hairpin contributions such as those in Eq.~\eqref{eq:su42amplhairpin} which can not be parameterized with the continuum LECs.  For lattice actions which give rise to a splitting amongst hadron masses which are degenerate in the continuum limit, such as twisted mass LQCD or staggered LQCD, these mass splittings can only be understood with the appropriate effective theories~\cite{Walker-Loud:2005bt,Sharpe:2004ny,Aubin:2003mg}, and again can not be parameterized with a simple polynomial dependence of the continuum LECs.  More importantly, with lattice simulations to date and for the foreseeable future, these lattice spacing contributions are not negligible and therefore there effects must be understood.  Additionally, as a great example, the form of the $I=2\ \pi\pi$ scattering length worked out in the previous chapter shows that there was a remarkable simplification of the scattering length expressed in terms of the lattice pion mass and lattice decay constant, in which all the explicit lattice spacing dependence cancelled from the expression up to a known shift from the hairpin interactions, Eq.~\eqref{eq:2seaScattLength}.  We contrast this with the amplitude expressed in terms of the LO \CPT\ parameters, Eq.~\eqref{eq:2seaBare} in which all of the lattice spacing dependence and partially quenched effects which one would expect, contribute to the amplitude.  We then see that the naive guess of the form of the amplitude would add 3 extra undetermined fit parameters to the $I=2$ scattering length, but that in reality, through the one-loop order there are no extra undetermined parameters.  We now move on to discuss the mixed action effective theory which will allow us to understand this simple continuum like behavior of the $\pi\pi$ scattering length and for which other processes this behavior holds, and more importantly how it breaks down.

%%%%%%%%%%%%%%%%%%%%%%%%%%%%%%%%%%%%%%%%%%%%%%%%%%%
%
%	Mixed Actions: Effective Field Theory
%
%%%%%%%%%%%%%%%%%%%%%%%%%%%%%%%%%%%%%%%%%%%%%%%%%%%
\section{Mixed Action Effective Field Theory}\label{sec:MACPT}

In this section we will show that for a mixed action theory with Ginsparg-Wilson valence quarks, most of the lattice spacing dependence for purely valence quantities can be parameterized as a multiplicative renormalization of the continuum \CPT\ LECs, through the leading loop order.  This is in contrast to a theory with Wilson valence quarks, in which every LEC receives additive as well multiplicative renormalizations.  The benefits of this are obvious, and in general the additive corrections tend to be significant for Wilson and staggered quarks.  Of course, for a dynamical simulation with GW valence and sea quarks, all of the lattice spacing dependence, aside from the $O(4)$ rotational breaking operators as in Eq.~\eqref{eq:O4break}, can be parameterized as multiplicative renormalizations of the continuum LECs.  It is in this sense that it is useful to think of a mixed action theory with GW valence quarks as the full GW theory with perturbative breakings of symmetry between the valence and sea quarks.

As has been shown in Refs.~\cite{Bar:2003mh,Bar:2005tu}, the LO Lagrangian is given by the partially quenched Lagrangian, and thus is invariant under the full partially quenched symmetry.  The first correction arising from the mixed action is given in Eq.~\eqref{eq:MixPotential} and leads to an $\mc{O}(a^2)$ shift to all the mesons composed of one valence and one sea quark.  This does not depend upon the type of sea quark used, as this operator does not break the taste symmetry of the staggered quarks.  The value of $C_{Mix}$ will of course depend upon the type of sea quarks however.  Through the leading loop order, the effect of this operator is to simply transform all valence-sea meson masses to their mixed action form
\begin{equation}
	B_0(m_V + m_S) \rightarrow B_0(m_V + m_S) + a^2\D_{Mix}\, 
\end{equation}
not just for the meson masses, but also for all interaction-vertices in $\pi\pi$ scattering and for the contributions to the meson decay constants.

At the next order, things get more interesting, but as we will show, the continuum like behavior still holds.  It is instructive to first consider what happens to the Gasser-Leutwyler operators.  For partially quenched \CPT\ the Gasser-Leutwyler operators retain their exact form with the replacement of $\tr \rightarrow \str$, and the LECs retain their exact values as in QCD.  There is one additional operator which must be added to the mesonic chiral Lagrangian, but it has been shown that it does not contribute to physical observables until one order higher~\cite{Sharpe:2003vy}, and so we shall ignore this operator in the following discussion.  When considering a mixed action theory with Ginsparg-Wilson valence quarks, we find that at $\mc{O}(m_q^2)$ the standard Gasser-Leutwyler Lagrangian splits into more operators.  For example, single super-trace operators split into two operators, one for the valence sector and one for the sea sector.  The double super-trace operators split into three or four operators, depending upon whether the two super-traces are over the same or different sets of fields.  To be precise, we give here two examples,
\begin{align}\label{eq:MAGasserLeutwyler}
2 B_0\, L_5\, &\str \left( \partial_\mu \S \partial^\mu \S^\dagger 
		\left( m_Q\S^\dagger +\S m_Q^\dagger \right) \right)
	\underrightarrow{\quad PQ \rightarrow MA\quad} \nonumber\\ \nonumber\\
	&\quad
	2 B_0^V\, C_5^{VV} L_5\, \str \left( \mc{P}_V \partial_\mu \S \partial^\mu \S^\dagger\, 
		\mc{P}_V\, \left( m_Q\S^\dagger +\S m_Q^\dagger \right) \right) \nonumber\\
	&\quad
	+2 B_0^S\, C_5^{SS} L_5\, \str \left( \mc{P}_S \partial_\mu \S \partial^\mu \S^\dagger \, 
		\mc{P}_S\, \left( m_Q\S^\dagger +\S m_Q^\dagger \right) \right)\, , \nonumber\\ \nonumber\\
4 B_0^2\, L_6\, &\left[ \str \left( m_Q \S^\dagger +\S m_Q^\dagger \right) \right]^2
	\underrightarrow{\quad PQ \rightarrow MA\quad} \nonumber\\ \nonumber\\
	&\quad 4 L_6\, C_6^{VV}\, \left[ \str \left( \mc{P}_V B_0^V 
		\left( m_Q \S^\dagger +\S m_Q^\dagger \right) \right) \right]^2 \nonumber\\
	&\quad +8 L_6\, C_6^{VS}\, 
		\str \left( \mc{P}_V B_0^V \left( m_Q \S^\dagger +\S m_Q^\dagger \right) \right)
		\str \left( \mc{P}_S B_0^S \left( m_Q \S^\dagger +\S m_Q^\dagger \right) \right) \nonumber\\
	&\quad +4 L_6\, C_6^{SS}\, \left[ \str \left( \mc{P}_S B_0^S 
		\left( m_Q \S^\dagger +\S m_Q^\dagger \right) \right) \right]^2\, ,
\end{align}
where in the above equation we have distinguished between the valence and sea quark condensates, $B_0^V = B_0^S + \mc{O}(a^2)$.  We have written the Lagrangian in this way because in the continuum limit, we know these operators have to recombine to form the Gasser-Leutwyler Lagrangian, as these operators are not distinct under the (PQ)QCD symmetries.  In fact, we can write each of the new coefficients as
\begin{align}\left.
	\begin{array}{c}
		C_{i}^{VV} \\
		C_i^{VS} \\
		C_i^{SV} \\
		C_i^{SS}
	\end{array}  \right\}= 1 + \mc{O}(a^2 \L_\chi^2)\, ,
\end{align}
where $i$ is for all relevant Gasser-Leutwyler LEC.  We see also that assuming the lattice spacing corrections to these LECs are perturbatively small, these mixed action effects which break the PQ symmetry are of $\mc{O}(p^4a^2)$ and thus do not enter until a higher order than we considered in Chapter~\ref{chap:pipiMA}.  Understanding how these different operators contribute to $m_\pi$, $f_\pi$ and $\mc{T}_{\pi\pi}$ provides valuable insight into how all the lattice spacing corrections enter mixed action theories.  We now track how the sea quark contributions from the operators in Eq.~\eqref{eq:MAGasserLeutwyler}, as well as the equivalent operators related to the Gasser-Leutwyler operator with coefficient $L_4$, contribute to these observables and then show how all the lattice spacing dependence enters.

Computing the contributions to the above mentioned observables at $\mc{O}(p^4)$ is a straightforward exercise as this only involves tree level graphs.  To be specific, we consider a 2-flavor theory and find that
\begin{align}
	\d m_\pi^2 &= -C_4^{VS} L_4 m_\pi^2\, \frac{32 B_0^S}{f^2}\, \str(m_Q)
		+C_6^{VS} L_6 m_\pi^2\, \frac{64 B_0^S}{f^2}\, \str(m_Q)\, , \nonumber\\
	\d f_\pi &= f\, C_4^{VS} L_4 \frac{16 B_0^S}{f^2}\, \str(m_Q)\, ,\nonumber\\
	\d \mc{Z}_\pi &= -C_4^{VS} L_4 \frac{32 B_0^S}{f^2}\str(m_Q)\, , \nonumber\\
	\d \mc{T}^{I=2}_{bare} &= -\frac{4 m_\pi^2}{f^2}\, C^{VS}_4 L_4 \frac{128 B_0^S}{f^2}\str(m_Q)
		+\frac{4m_\pi^2}{f^2}\, C^{VS}_6 L_6 \frac{64 B_0^S}{3 f^2}\, \str(m_Q)\, ,
\end{align}
and when we add all these effects to the $I=2\ \pi\pi$ scattering length expressed in terms of the lattice parameters, we find
\begin{equation}
	\d\mc{T}^{I=2}_{Mix} = 0\, ,
\end{equation}
as claimed in Chapter~\ref{chap:pipiMA}.  Operationally, we understand this result as coming from a factorization of sea-quark effects, which was made explicit by the mixed action breaking of the Gasser-Leutwyler operators.  Of course this same analysis explains why there is also no contribution at this order arising from partial quenching, aside from the hairpin contributions in the $t$- and $u$-channel graphs of Fig.~\ref{fig:Hairpins}.  The real power of this analysis comes when one then constructs the operators at this order which encode the lattice spacing effects.  Regardless of the type of sea-quarks used, when building the mixed action effective Lagrangian with Ginsparg-Wilson valence quarks, one can determine that all of the operators which can give rise to lattice spacing effects must have those effects come from the sea quarks, and thus be of the form of the above mentioned operators with coefficients $C_i^{VS}$.  In other words, the lattice spacing effects factorize in the same fashion as the sea quark mass effects shown above.

One then realizes that all the lattice spacing operators one can construct which will contribute to quantities composed of Ginsparg-Wilson valence quarks, must all behave as multiplicative renormalizations to the continuum \CPT\ operators, aside from the lattice spacing corrections which enter in the hairpin interactions.  Not only are the lattice spacing corrections then multiplicative renormalizations to the continuum expressions, in contrast to all other lattice actions to date (aside from a full Ginsparg-Wilson theory), this also implies that for sufficiently small lattice spacings, these corrections are also small.  There is now growing numerical evidence in support of these ideas given by the recent mixed action LQCD simulations which employ Ginsparg-Wilson valence quarks with staggered sea quarks~\cite{Beane:2005rj,Bonnet:2004fr,Edwards:2005ym,Beane:2006mx,Beane:2006pt,Beane:2006fk}.  We would then like to understand how this simple picture breaks down, which we address in the next section.

%%%%%%%%%%%%%%%%%%%%%%%%%%%%%%%%%%%%%%%%%%%%%%%%%%%
%
%	Mixed Actions: Taste breaking 
%
%%%%%%%%%%%%%%%%%%%%%%%%%%%%%%%%%%%%%%%%%%%%%%%%%%%
\section{Taste Breaking in Mixed Action Lattice QCD}\label{sec:MAbreakdown}

\begin{figure}
\center
\includegraphics[width=0.5\textwidth]{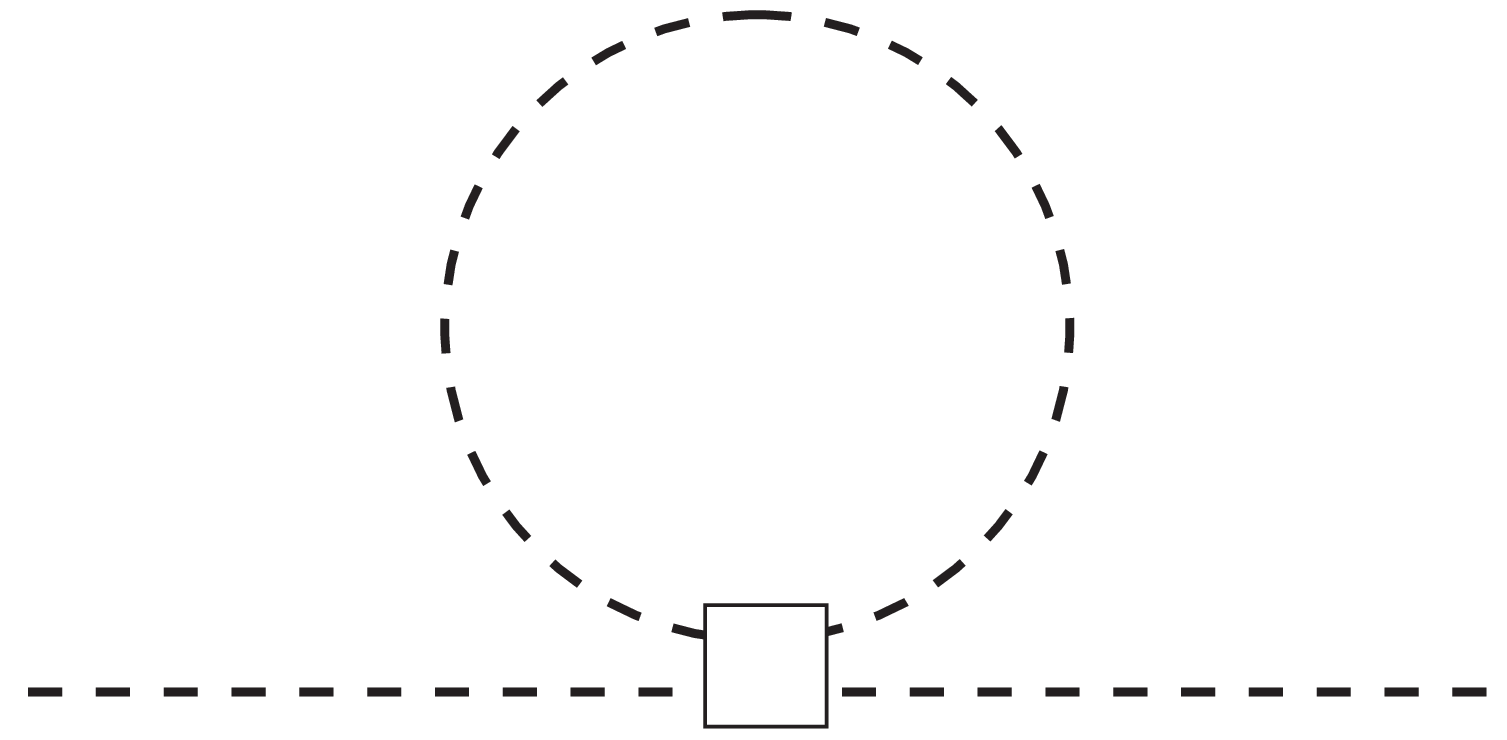}
\caption[Diagram contributing to the pion mass correction at NNLO]{\label{fig:mpi_MA_NNLO} This graph represents corrections to the pion mass from the Lagrangian, Eq.~\eqref{eq:MAGasserLeutwyler}.  For staggered sea quarks, this contribution is taste breaking.}
\end{figure}

In this section, we describe how this continuum like picture of the mixed action theories with Ginsparg-Wilson quarks breaks down.  We will use the specific example of staggered sea quarks, both because currently, mixed action simulations are predominantly performed with staggered sea quarks, and because the oddities of the staggered theory help to highlight the points we are making.  In fact the analysis of the previous section directly leads to an understanding of how the taste breaking of the staggered potential gives rise to contributions which can not be absorbed by multiplicative renormalizations of the continuum Gasser-Leutwyler LECs.  To be very specific, we shall focus on the operator
\begin{align}\label{eq:MAtastebreak}
	\mc{L} = 8 C_6^{VS} L_6\, \str \left( \mc{P}_V \left( \S m_Q + m_Q \S^\dagger \right) \right)
		\str \left( \mc{P}_S \left( \S m_Q + m_Q \S^\dagger \right) \right)\, ,
\end{align}
and work to $\mc{O}(p^6)$.  We shall compute the pion mass and consider the contribution from the Figure~\ref{fig:mpi_MA_NNLO}, in which the pion loop comes from the second $\str$ in Eq.~\eqref{eq:MAtastebreak}.
%%%%%%%%%%%%%%%%%%%%%%%%%%%%%%%%%%%%%%%%%%%%%%%%%%%
%
%	Figure: Taste breaking for pion mass
%
%%%%%%%%%%%%%%%%%%%%%%%%%%%%%%%%%%%%%%%%%%%%%%%%%%%
%%%%%%%%%%%%%%%%%%%%%%%%%%%%%%%%%%%%%%%%%%%%%%%%%%%
%
%%%%%%%%%%%%%%%%%%%%%%%%%%%%%%%%%%%%%%%%%%%%%%%%%%%
This operator will give a contribution to the pion mass,
\begin{equation}
	\d m_\pi^2 = - \frac{32 C_6^{VS}L_6\, m_\pi^2}{f^2}\, \sum_{t} \frac{n_t}{16}
		\frac{4B_0(m_q +m_{q^\prime})}{(4\pi f)^2}\, 
		\ln \left( \frac{m_t^2}{\mu^2} \right)\, ,
\end{equation}
where the sum is over the various staggered taste mesons with $n_t$ counting the weighting of the $t^{th}$ taste meson for this loop, whose masses are given by Eq.~\eqref{eq:ssvsmasses}.  Here we then see a contribution to the meson masses which can not be treated as a multiplicative renormalization of the continuum LECs.  We note as a final comment that this feature will also enter in the same manner in the baryon spectrum~\cite{Tiburzi:noasqd}.

We then conclude this chapter by re-emphasizing that for quantities which are protected by chiral symmetry, eg. the pion mass, the scattering length, etc., the good chiral properties of Ginsparg-Wilson quarks, even when only used in the valence sector, protect these quantities from receiving chiral symmetry breaking corrections from the lattice spacing through the one loop order, irrespective of the sea-quark type.

 % ========== Chapter 8: Concluding Remarks
\chapter{Concluding Remarks}\label{chap:Conclusion}

Presently, lattice QCD simulations are performed with lattice spacings which are not negligibly small, lattice volumes which are not significantly larger than the typical hadron systems of interest and quark masses which are larger than those of nature.  In order to make a rigorous connection between these lattice QCD simulations and nature, we need to understand how these lattice artifacts modify the computed correlation functions and thus the observable quantities of interest.  The tool to do this is effective field theory.  In this work, we have presented developments and applications of effective field theory techniques to various observable quantities which are necessary for the lattice QCD simulations of today and tomorrow.

In particular, in Chapter~\ref{chap:BMasses}, we have extended the partially quenched heavy baryon Lagrangian necessary to compute the masses of the lightest spin-$\frac{1}{2}$ and spin-$\frac{3}{2}$ baryons to $\mc{O}(m_q)^2$.  This is necessary as the convergence of the effective field theory expansion for baryons is not as good as with the mesons, and so must be pushed to higher orders to test the convergence of the theory and hopefully provide more accurate theoretical knowledge of the baryon masses.  In Chapter~\ref{ch:tmChPT}, we showed how to include baryons into twisted mass chiral perturbation theory and then use this Lagrangian to determine the effects the twisted mass discretization technique has on the spectrum of the nucleons and deltas, which we were able to qualitatively compare to the available quenched twisted mass lattice QCD data of the baryon spectrum.  We also explored the very interesting effects of simultaneously including both the isospin breaking of the quark mass parameter and the twisted mass parameter which leads to a mixing of quark flavors and obscures the definition of isospin.

In Chapter~\ref{chap:Polarize}, we showed how one can use background electromagnetic fields to determine the electromagnetic and spin polarisabilities of hadrons.  This is particularly interesting because currently, we can only measure two of the four spin polarisabilities of the nucleon and therefore we are in a position to make a prediction of these polarisabilities with lattice QCD.  From a theoretical standpoint, the polarisabilities are very interesting because of there sensitivity to the chiral physics.  Unlike many baryon quantities, the leading contribution to the nucleon polarisabilities comes from the pion cloud and not a local point interaction.  Therefore these quantities are particularly sensitive to finite volume modifications, and even for the present pion masses used in determinations of these quantities, $m_\pi \sim 500$~MeV, the leading finite volume modifications can be on the order of 10\% corrections.

In Chapters~\ref{chap:pipiFV} and \ref{chap:pipiMA}, we explored both the finite volume as well as the lattice spacing corrections to the two pion system.  We first determined the exponential volume dependence of the $I=2\ \pi\pi$ system and showed that these effects will become particularly important in the next generation of lattice simulations as the pion masses are brought significantly down from 300~MeV.  We then determined the partial quenching and lattice spacing effects on the $I=2\ \pi\pi$ scattering length which arise from a mixed action simulation with Ginsparg-Wilson valence quarks and staggered sea quarks.  We showed that if one expresses the scattering length in terms of the lattice parameters measured from correlation functions, then the form of the answer is identical to continuum chiral perturbation theory up to a computably small shift arising from hairpin interactions of the flavor neutral mesons; all other possible partial quenching and lattice spacing corrections to the scattering length cancel in the final answer in terms of these lattice parameters.

In Chapter~\ref{chap:MAGW}, we showed that lattice QCD simulations which utilize mixed actions with Ginsparg-Wilson valence quarks exhibit continuum like behavior for quantities protected by chiral symmetry through the leading loop order, such that one can treat all the lattice spacing dependence as a multiplicative renormalization of the continuum \CPT\ operators, which is contrast to all other types of fermion discretization methods presently used (aside from dynamical Ginsparg-Wilson sea quarks with Ginsparg-Wilson valence quarks).  We then showed how this behavior breaks down at the next order, using the example of staggered sea quarks to show how the taste breaking of the staggered mesons disrupts this continuum like behavior of mixed action lattice QCD.

\printendnotes

%
% ==========   Bibliography
%
%\nocite{*}   % include everything in the uwthesis.bib file
\bibliographystyle{jhep}
\bibliography{NNEFT,EFT,general,hadron_structure,lattice_general,lattice_fermions,lattice_FV,lattice_PQ,lattice_physics}

\providecommand{\href}[2]{#2}\begingroup\raggedright\begin{thebibliography}{10%
0}

\bibitem{Yang:1954ek}
C.-N. Yang and R.~L. Mills, {\it Conservation of isotopic spin and isotopic
  gauge invariance},  {\em Phys. Rev.} {\bf 96} (1954) 191--195.

\bibitem{Fritzsch:1972jv}
H.~Fritzsch and M.~Gell-Mann, {\it Current algebra: Quarks and what else?},
  \href{http://xxx.lanl.gov/abs/hep-ph/0208010}{{\tt hep-ph/0208010}}.

\bibitem{Fritzsch:1973pi}
H.~Fritzsch, M.~Gell-Mann, and H.~Leutwyler, {\it Advantages of the color octet
  gluon picture},  {\em Phys. Lett.} {\bf B47} (1973) 365--368.

\bibitem{Politzer:1973fx}
H.~D. Politzer, {\it Reliable perturbative results for strong interactions?},
  {\em Phys. Rev. Lett.} {\bf 30} (1973) 1346--1349.

\bibitem{Gross:1973id}
D.~J. Gross and F.~Wilczek, {\it Ultraviolet behavior of non-{A}belian gauge
  theories},  {\em Phys. Rev. Lett.} {\bf 30} (1973) 1343--1346.

\bibitem{Bjorken:1969ja}
J.~D. Bjorken and E.~A. Paschos, {\it Inelastic electron proton and gamma
  proton scattering, and the structure of the nucleon},  {\em Phys. Rev.} {\bf
  185} (1969) 1975--1982.

\bibitem{Panofsky:1968pb}
W.~K.~H. Panofsky, {\it Electromagnetic interactions: Low $q^2$
  electrodynamics: Elastic and inelastic electron (and muon) scattering}, .
  Presented at 14th Int. Conf. on High Energy Physics, Vienna, Aug 1968.

\bibitem{Feynman:1969ej}
R.~P. Feynman, {\it Very high-energy collisions of hadrons},  {\em Phys. Rev.
  Lett.} {\bf 23} (1969) 1415--1417.

\bibitem{Wilson:1974sk}
K.~G. Wilson, {\it Confinement of quarks},  {\em Phys. Rev.} {\bf D10} (1974)
  2445--2459.

\bibitem{Fermi:1934sk}
E.~Fermi, {\it Trends to a theory of beta radiation. (in {I}talian)},  {\em
  Nuovo Cim.} {\bf 11} (1934) 1--19.

\bibitem{Fermi:1934hr}
E.~Fermi, {\it An attempt of a theory of beta radiation. 1},  {\em Z. Phys.}
  {\bf 88} (1934) 161--177.

\bibitem{Lee:1956qn}
T.~D. Lee and C.-N. Yang, {\it Question of parity conservation in weak
  interactions},  {\em Phys. Rev.} {\bf 104} (1956) 254--258.

\bibitem{Wu:1957my}
C.~S. Wu, E.~Ambler, R.~W. Hayward, D.~D. Hoppes, and R.~P. Hudson, {\it
  Experimental test of parity conservation in beta decay},  {\em Phys. Rev.}
  {\bf 105} (1957) 1413--1414.

\bibitem{Feynman:1958ty}
R.~P. Feynman and M.~Gell-Mann, {\it Theory of the {Fermi} interaction},  {\em
  Phys. Rev.} {\bf 109} (1958) 193--198.

\bibitem{Sudarshan:1958vf}
E.~C.~G. Sudarshan and R.~e. Marshak, {\it Chirality invariance and the
  universal {Fermi} interaction},  {\em Phys. Rev.} {\bf 109} (1958)
  1860--1862.

\bibitem{Sakurai:1958}
J.~J. Sakurai, {\it Mass reversal and weak interactions},  {\em Nuovo Cim.}
  {\bf 7} (1958) 649--.

\bibitem{Glashow:1961tr}
S.~L. Glashow, {\it Partial symmetries of weak interactions},  {\em Nucl.
  Phys.} {\bf 22} (1961) 579--588.

\bibitem{Weinberg:1967tq}
S.~Weinberg, {\it A model of leptons},  {\em Phys. Rev. Lett.} {\bf 19} (1967)
  1264--1266.

\bibitem{Salam:1964ry}
A.~Salam and J.~C. Ward, {\it Electromagnetic and weak interactions},  {\em
  Phys. Lett.} {\bf 13} (1964) 168--171.

\bibitem{Higgs:1964ia}
P.~W. Higgs, {\it Broken symmetries, massless particles and gauge fields},
  {\em Phys. Lett.} {\bf 12} (1964) 132--133.

\bibitem{Higgs:1964pj}
P.~W. Higgs, {\it Broken symmetries and the masses of gauge bosons},  {\em
  Phys. Rev. Lett.} {\bf 13} (1964) 508--509.

\bibitem{Higgs:1966ev}
P.~W. Higgs, {\it Spontaneous symmetry breakdown without massless bosons},
  {\em Phys. Rev.} {\bf 145} (1966) 1156--1163.

\bibitem{Arnison:1983rp}
{\bf UA1} Collaboration, G.~Arnison {\em et.~al.}, {\it Experimental
  observation of isolated large transverse energy electrons with associated
  missing energy at $\sqrt{s} = 540$~{GeV}},  {\em Phys. Lett.} {\bf B122}
  (1983) 103--116.

\bibitem{Arnison:1983mk}
{\bf UA1} Collaboration, G.~Arnison {\em et.~al.}, {\it Experimental
  observation of lepton pairs of invariant mass around 95~{GeV} / $c^2$ at the
  {CERN SPS} collider},  {\em Phys. Lett.} {\bf B126} (1983) 398--410.

\bibitem{'tHooft:1971rn}
G.~'t~Hooft, {\it Renormalizable {Lagrangians} for massive {Yang-Mills}
  fields},  {\em Nucl. Phys.} {\bf B35} (1971) 167--188.

\bibitem{'tHooft:1972fi}
G.~'t~Hooft and M.~J.~G. Veltman, {\it Regularization and renormalization of
  gauge fields},  {\em Nucl. Phys.} {\bf B44} (1972) 189--213.

\bibitem{Jenkins:1990jv}
E.~Jenkins and A.~V. Manohar, {\it Baryon chiral perturbation theory using a
  heavy fermion lagrangian},  {\em Phys. Lett.} {\bf B255} (1991) 558--562.

\bibitem{Jenkins:1991ne}
E.~Jenkins and A.~V. Manohar, {\it Baryon chiral perturbation theory}, . Talk
  presented at the Workshop on Effective Field Theories of the Standard Model,
  Dobogoko, Hungary, Aug 1991.

\bibitem{Kaplan:1986yq}
D.~B. Kaplan and A.~E. Nelson, {\it Strange goings on in dense nucleonic
  matter},  {\em Phys. Lett.} {\bf B175} (1986) 57--63.

\bibitem{Nelson:1987dg}
A.~E. Nelson and D.~B. Kaplan, {\it Strange condensate realignment in
  relativistic heavy ion collisions},  {\em Phys. Lett.} {\bf B192} (1987) 193.

\bibitem{Frezzotti:2000nk}
{\bf Alpha} Collaboration, R.~Frezzotti, P.~A. Grassi, S.~Sint, and P.~Weisz,
  {\it Lattice {QCD} with a chirally twisted mass term},  {\em JHEP} {\bf 08}
  (2001) 058, [\href{http://xxx.lanl.gov/abs/hep-lat/0101001}{{\tt
  hep-lat/0101001}}].

\bibitem{Ji:1996ek}
X.-D. Ji, {\it Gauge invariant decomposition of nucleon spin},  {\em Phys. Rev.
  Lett.} {\bf 78} (1997) 610--613,
  [\href{http://xxx.lanl.gov/abs/hep-ph/9603249}{{\tt hep-ph/9603249}}].

\bibitem{Ji:1996nm}
X.-D. Ji, {\it Deeply-virtual compton scattering},  {\em Phys. Rev.} {\bf D55}
  (1997) 7114--7125, [\href{http://xxx.lanl.gov/abs/hep-ph/9609381}{{\tt
  hep-ph/9609381}}].

\bibitem{Ragusa:1993rm}
S.~Ragusa, {\it Third order spin polarizabilities of the nucleon},  {\em Phys.
  Rev.} {\bf D47} (1993) 3757--3767.

\bibitem{Guichon:1995pu}
P.~A.~M. Guichon, G.~Q. Liu, and A.~W. Thomas, {\it Virtual {Compton}
  scattering and generalized polarizabilities of the proton},  {\em Nucl.
  Phys.} {\bf A591} (1995) 606--638,
  [\href{http://xxx.lanl.gov/abs/nucl-th/9605031}{{\tt nucl-th/9605031}}].

\bibitem{Holstein:1999uu}
B.~R. Holstein, D.~Drechsel, B.~Pasquini, and M.~Vanderhaeghen, {\it Higher
  order polarizabilities of the proton},  {\em Phys. Rev.} {\bf C61} (2000)
  034316, [\href{http://xxx.lanl.gov/abs/hep-ph/9910427}{{\tt
  hep-ph/9910427}}].

\bibitem{Hamber:1983vu}
H.~W. Hamber, E.~Marinari, G.~Parisi, and C.~Rebbi, {\it Considerations on
  numerical analysis of {QCD}},  {\em Nucl. Phys.} {\bf B225} (1983) 475.

\bibitem{Luscher:1986pf}
M.~Luscher, {\it Volume dependence of the energy spectrum in massive quantum
  field theories. 2. {S}cattering states},  {\em Commun. Math. Phys.} {\bf 105}
  (1986) 153--188.

\bibitem{Luscher:1990ux}
M.~Luscher, {\it Two particle states on a torus and their relation to the
  scattering matrix},  {\em Nucl. Phys.} {\bf B354} (1991) 531--578.

\bibitem{Maiani:1990ca}
L.~Maiani and M.~Testa, {\it Final state interactions from {Euclidean}
  correlation functions},  {\em Phys. Lett.} {\bf B245} (1990) 585--590.

\bibitem{Rummukainen:1995vs}
K.~Rummukainen and S.~A. Gottlieb, {\it Resonance scattering phase shifts on a
  nonrest frame lattice},  {\em Nucl. Phys.} {\bf B450} (1995) 397--436,
  [\href{http://xxx.lanl.gov/abs/hep-lat/9503028}{{\tt hep-lat/9503028}}].

\bibitem{Beane:2003yx}
S.~R. Beane, P.~F. Bedaque, A.~Parreno, and M.~J. Savage, {\it Exploring
  hyperons and hypernuclei with lattice {QCD}},  {\em Nucl. Phys.} {\bf A747}
  (2005) 55--74, [\href{http://xxx.lanl.gov/abs/nucl-th/0311027}{{\tt
  nucl-th/0311027}}].

\bibitem{Beane:2003da}
S.~R. Beane, P.~F. Bedaque, A.~Parreno, and M.~J. Savage, {\it Two nucleons on
  a lattice},  {\em Phys. Lett.} {\bf B585} (2004) 106--114,
  [\href{http://xxx.lanl.gov/abs/hep-lat/0312004}{{\tt hep-lat/0312004}}].

\bibitem{Kim:2005gf}
C.~H. Kim, C.~T. Sachrajda, and S.~R. Sharpe, {\it Finite-volume effects for
  two-hadron states in moving frames},  {\em Nucl. Phys.} {\bf B727} (2005)
  218--243, [\href{http://xxx.lanl.gov/abs/hep-lat/0507006}{{\tt
  hep-lat/0507006}}].

\bibitem{Ginsparg:1981bj}
P.~H. Ginsparg and K.~G. Wilson, {\it A remnant of chiral symmetry on the
  lattice},  {\em Phys. Rev.} {\bf D25} (1982) 2649.

\bibitem{Susskind:1976jm}
L.~Susskind, {\it Lattice fermions},  {\em Phys. Rev.} {\bf D16} (1977)
  3031--3039.

\bibitem{Beane:2005rj}
{\bf NPLQCD} Collaboration, S.~R. Beane, P.~F. Bedaque, K.~Orginos, and M.~J.
  Savage, {\it {$I = 2\ \pi\pi$} scattering from fully-dynamical mixed-action
  lattice {QCD}},  {\em Phys. Rev.} {\bf D73} (2006) 054503,
  [\href{http://xxx.lanl.gov/abs/hep-lat/0506013}{{\tt hep-lat/0506013}}].

\bibitem{Sharpe:1993wt}
S.~R. Sharpe, {\it Introduction to lattice field theory}, . Prepared for
  Uehling Summer School on Phenomenology and Lattice QCD, Seattle, WA, 21 Jun -
  2 Jul 1993.

\bibitem{Rothe:1997kp}
H.~J. Rothe, {\em Lattice gauge theories: An introduction}.
\newblock World Sci., 2005.

\bibitem{Feynman:1948ur}
R.~P. Feynman, {\it Space-time approach to nonrelativistic quantum mechanics},
  {\em Rev. Mod. Phys.} {\bf 20} (1948) 367--387.

\bibitem{Bernard:2002yk}
C.~Bernard {\em et.~al.}, {\it Panel discussion on chiral extrapolation of
  physical observables},  {\em Nucl. Phys. Proc. Suppl.} {\bf 119} (2003)
  170--184, [\href{http://xxx.lanl.gov/abs/hep-lat/0209086}{{\tt
  hep-lat/0209086}}].

\bibitem{Beane:2004ks}
S.~R. Beane, {\it In search of the chiral regime},  {\em Nucl. Phys.} {\bf
  B695} (2004) 192--198, [\href{http://xxx.lanl.gov/abs/hep-lat/0403030}{{\tt
  hep-lat/0403030}}].

\bibitem{Sharpe:2000bc}
S.~R. Sharpe and N.~Shoresh, {\it Physical results from unphysical
  simulations},  {\em Phys. Rev.} {\bf D62} (2000) 094503,
  [\href{http://xxx.lanl.gov/abs/hep-lat/0006017}{{\tt hep-lat/0006017}}].

\bibitem{Bernard:2001av}
C.~W. Bernard {\em et.~al.}, {\it The {QCD} spectrum with three quark flavors},
   {\em Phys. Rev.} {\bf D64} (2001) 054506,
  [\href{http://xxx.lanl.gov/abs/hep-lat/0104002}{{\tt hep-lat/0104002}}].
  http://qcd.nersc.gov/.

\bibitem{Negele:2004iu}
J.~W. Negele {\em et.~al.}, {\it Insight into nucleon structure from lattice
  calculations of moments of parton and generalized parton distributions},
  {\em Nucl. Phys. Proc. Suppl.} {\bf 128} (2004) 170--178,
  [\href{http://xxx.lanl.gov/abs/hep-lat/0404005}{{\tt hep-lat/0404005}}].

\bibitem{Bonnet:2004fr}
{\bf Lattice Hadron Physics} Collaboration, F.~D.~R. Bonnet, R.~G. Edwards,
  G.~T. Fleming, R.~Lewis, and D.~G. Richards, {\it Lattice computations of the
  pion form factor},  {\em Phys. Rev.} {\bf D72} (2005) 054506,
  [\href{http://xxx.lanl.gov/abs/hep-lat/0411028}{{\tt hep-lat/0411028}}].

\bibitem{Edwards:2005ym}
{\bf LHPC} Collaboration, R.~G. Edwards {\em et.~al.}, {\it The nucleon axial
  charge in full lattice {QCD}},  {\em Phys. Rev. Lett.} {\bf 96} (2006)
  052001, [\href{http://xxx.lanl.gov/abs/hep-lat/0510062}{{\tt
  hep-lat/0510062}}].

\bibitem{Beane:2006mx}
S.~R. Beane, P.~F. Bedaque, K.~Orginos, and M.~J. Savage, {\it Nucleon nucleon
  scattering from fully-dynamical lattice {QCD}},
  \href{http://xxx.lanl.gov/abs/hep-lat/0602010}{{\tt hep-lat/0602010}}.

\bibitem{Beane:2006pt}
S.~R. Beane, K.~Orginos, and M.~J. Savage, {\it The {Gell-Mann - Okubo} mass
  relation among baryons from fully-dynamical mixed-action lattice {QCD}},
  \href{http://xxx.lanl.gov/abs/hep-lat/0604013}{{\tt hep-lat/0604013}}.

\bibitem{Beane:2006fk}
S.~R. Beane, K.~Orginos, and M.~J. Savage, {\it Strong-isospin violation in the
  neutron-proton mass difference from fully-dynamical lattice {QCD} and
  {PQQCD}},  \href{http://xxx.lanl.gov/abs/hep-lat/0605014}{{\tt
  hep-lat/0605014}}.

\bibitem{Manohar:1995xr}
A.~V. Manohar, {\it Effective field theories},
  \href{http://xxx.lanl.gov/abs/hep-ph/9508245}{{\tt hep-ph/9508245}}.

\bibitem{Kaplan:1995uv}
D.~B. Kaplan, {\it Effective field theories},
  \href{http://xxx.lanl.gov/abs/nucl-th/9506035}{{\tt nucl-th/9506035}}.

\bibitem{Manohar:1996cq}
A.~V. Manohar, {\it Effective field theories},
  \href{http://xxx.lanl.gov/abs/hep-ph/9606222}{{\tt hep-ph/9606222}}.

\bibitem{Phillips:2002da}
D.~R. Phillips, {\it Building light nuclei from neutrons, protons, and pions},
  {\em Czech. J. Phys.} {\bf 52} (2002) B49,
  [\href{http://xxx.lanl.gov/abs/nucl-th/0203040}{{\tt nucl-th/0203040}}].

\bibitem{Scherer:2002tk}
S.~Scherer, {\it Introduction to chiral perturbation theory},  {\em Adv. Nucl.
  Phys.} {\bf 27} (2003) 277,
  [\href{http://xxx.lanl.gov/abs/hep-ph/0210398}{{\tt hep-ph/0210398}}].

\bibitem{Kaplan:2005es}
D.~B. Kaplan, {\it Five lectures on effective field theory},
  \href{http://xxx.lanl.gov/abs/nucl-th/0510023}{{\tt nucl-th/0510023}}.

\bibitem{Weinberg:1967kj}
S.~Weinberg, {\it Precise relations between the spectra of vector and axial
  vector mesons},  {\em Phys. Rev. Lett.} {\bf 18} (1967) 507--509.

\bibitem{Weinberg:1968de}
S.~Weinberg, {\it Nonlinear realizations of chiral symmetry},  {\em Phys. Rev.}
  {\bf 166} (1968) 1568--1577.

\bibitem{Weinberg:1978kz}
S.~Weinberg, {\it Phenomenological {L}agrangians},  {\em Physica} {\bf A96}
  (1979) 327.

\bibitem{Gasser:1983yg}
J.~Gasser and H.~Leutwyler, {\it Chiral perturbation theory to one loop},  {\em
  Ann. Phys.} {\bf 158} (1984) 142.

\bibitem{Gasser:1984gg}
J.~Gasser and H.~Leutwyler, {\it Chiral perturbation theory: Expansions in the
  mass of the strange quark},  {\em Nucl. Phys.} {\bf B250} (1985) 465.

\bibitem{Harris:1999jx}
P.~G. Harris {\em et.~al.}, {\it New experimental limit on the electric dipole
  moment of the neutron},  {\em Phys. Rev. Lett.} {\bf 82} (1999) 904--907.

\bibitem{Crewther:1979pi}
R.~J. Crewther, P.~Di~Vecchia, G.~Veneziano, and E.~Witten, {\it Chiral
  estimate of the electric dipole moment of the neutron in quantum
  chromodynamics},  {\em Phys. Lett.} {\bf B88} (1979) 123.

\bibitem{Pich:1991fq}
A.~Pich and E.~de~Rafael, {\it Strong {CP} violation in an effective chiral
  {L}agrangian approach},  {\em Nucl. Phys.} {\bf B367} (1991) 313--333.

\bibitem{Borasoy:2000pq}
B.~Borasoy, {\it The electric dipole moment of the neutron in chiral
  perturbation theory},  {\em Phys. Rev.} {\bf D61} (2000) 114017,
  [\href{http://xxx.lanl.gov/abs/hep-ph/0004011}{{\tt hep-ph/0004011}}].

\bibitem{Hockings:2005cn}
W.~H. Hockings and U.~van Kolck, {\it The electric dipole form factor of the
  nucleon},  {\em Phys. Lett.} {\bf B605} (2005) 273--278,
  [\href{http://xxx.lanl.gov/abs/nucl-th/0508012}{{\tt nucl-th/0508012}}].

\bibitem{O'Connell:2005un}
D.~O'Connell and M.~J. Savage, {\it Extrapolation formulas for neutron {EDM}
  calculations in lattice {QCD}},  {\em Phys. Lett.} {\bf B633} (2006)
  319--324, [\href{http://xxx.lanl.gov/abs/hep-lat/0508009}{{\tt
  hep-lat/0508009}}].

\bibitem{Aoki:1989rx}
S.~Aoki and A.~Gocksch, {\it The neutron electric dipole moment in lattice
  {QCD}},  {\em Phys. Rev. Lett.} {\bf 63} (1989) 1125.

\bibitem{Aoki:1990ix}
S.~Aoki, A.~Gocksch, A.~V. Manohar, and S.~R. Sharpe, {\it Calculating the
  neutron electric dipole moment on the lattice},  {\em Phys. Rev. Lett.} {\bf
  65} (1990) 1092--1095.

\bibitem{Guadagnoli:2002nm}
D.~Guadagnoli, V.~Lubicz, G.~Martinelli, and S.~Simula, {\it Neutron electric
  dipole moment on the lattice: {A} theoretical reappraisal},  {\em JHEP} {\bf
  04} (2003) 019, [\href{http://xxx.lanl.gov/abs/hep-lat/0210044}{{\tt
  hep-lat/0210044}}].

\bibitem{Shintani:2005xg}
E.~Shintani {\em et.~al.}, {\it Neutron electric dipole moment from lattice
  {QCD}},  {\em Phys. Rev.} {\bf D72} (2005) 014504,
  [\href{http://xxx.lanl.gov/abs/hep-lat/0505022}{{\tt hep-lat/0505022}}].

\bibitem{Berruto:2005hg}
F.~Berruto, T.~Blum, K.~Orginos, and A.~Soni, {\it Calculation of the neutron
  electric dipole moment with two dynamical flavors of domain wall fermions},
  {\em Phys. Rev.} {\bf D73} (2006) 054509,
  [\href{http://xxx.lanl.gov/abs/hep-lat/0512004}{{\tt hep-lat/0512004}}].

\bibitem{Vafa:1983tf}
C.~Vafa and E.~Witten, {\it Restrictions on symmetry breaking in vector - like
  gauge theories},  {\em Nucl. Phys.} {\bf B234} (1984) 173.

\bibitem{Nambu:1960xd}
Y.~Nambu, {\it Axial vector current conservation in weak interactions},  {\em
  Phys. Rev. Lett.} {\bf 4} (1960) 380--382.

\bibitem{Goldstone:1961eq}
J.~Goldstone, {\it Field theories with `superconductor' solutions},  {\em Nuovo
  Cim.} {\bf 19} (1961) 154--164.

\bibitem{Goldstone:1962es}
J.~Goldstone, A.~Salam, and S.~Weinberg, {\it Broken symmetries},  {\em Phys.
  Rev.} {\bf 127} (1962) 965--970.

\bibitem{Coleman:1969sm}
S.~R. Coleman, J.~Wess, and B.~Zumino, {\it Structure of phenomenological
  {Lagrangians.} 1},  {\em Phys. Rev.} {\bf 177} (1969) 2239--2247.

\bibitem{Callan:1969sn}
J.~Callan, Curtis~G., S.~R. Coleman, J.~Wess, and B.~Zumino, {\it Structure of
  phenomenological {Lagrangians.} 2},  {\em Phys. Rev.} {\bf 177} (1969)
  2247--2250.

\bibitem{Gell-Mann:1968rz}
M.~Gell-Mann, R.~J. Oakes, and B.~Renner, {\it Behavior of current divergences
  under {$SU(3) \otimes SU(3)$}},  {\em Phys. Rev.} {\bf 175} (1968)
  2195--2199.

\bibitem{Ananthanarayan:2000ht}
B.~Ananthanarayan, G.~Colangelo, J.~Gasser, and H.~Leutwyler, {\it Roy equation
  analysis of $\pi\pi$ scattering},  {\em Phys. Rept.} {\bf 353} (2001)
  207--279, [\href{http://xxx.lanl.gov/abs/hep-ph/0005297}{{\tt
  hep-ph/0005297}}].

\bibitem{Colangelo:2000jc}
G.~Colangelo, J.~Gasser, and H.~Leutwyler, {\it The $\pi\pi$ s-wave scattering
  lengths},  {\em Phys. Lett.} {\bf B488} (2000) 261--268,
  [\href{http://xxx.lanl.gov/abs/hep-ph/0007112}{{\tt hep-ph/0007112}}].

\bibitem{Colangelo:2001sp}
G.~Colangelo, J.~Gasser, and H.~Leutwyler, {\it The quark condensate from
  {K(e4)} decays},  {\em Phys. Rev. Lett.} {\bf 86} (2001) 5008--5010,
  [\href{http://xxx.lanl.gov/abs/hep-ph/0103063}{{\tt hep-ph/0103063}}].

\bibitem{Colangelo:2001df}
G.~Colangelo, J.~Gasser, and H.~Leutwyler, {\it $\pi\pi$ scattering},  {\em
  Nucl. Phys.} {\bf B603} (2001) 125--179,
  [\href{http://xxx.lanl.gov/abs/hep-ph/0103088}{{\tt hep-ph/0103088}}].

\bibitem{Pislak:2001bf}
{\bf BNL-E865} Collaboration, S.~Pislak {\em et.~al.}, {\it A new measurement
  of {K(e4)} decay and the s-wave $\pi\pi$ scattering length $a^{(0)}_{(0)}$},
  {\em Phys. Rev. Lett.} {\bf 87} (2001) 221801,
  [\href{http://xxx.lanl.gov/abs/hep-ex/0106071}{{\tt hep-ex/0106071}}].

\bibitem{Pislak:2003sv}
S.~Pislak {\em et.~al.}, {\it High statistics measurement of {K(e4)} decay
  properties},  {\em Phys. Rev.} {\bf D67} (2003) 072004,
  [\href{http://xxx.lanl.gov/abs/hep-ex/0301040}{{\tt hep-ex/0301040}}].

\bibitem{Stern:1993rg}
J.~Stern, H.~Sazdjian, and N.~H. Fuchs, {\it What $\pi\pi$ scattering tells us
  about chiral perturbation theory},  {\em Phys. Rev.} {\bf D47} (1993)
  3814--3838, [\href{http://xxx.lanl.gov/abs/hep-ph/9301244}{{\tt
  hep-ph/9301244}}].

\bibitem{Knecht:1995tr}
M.~Knecht, B.~Moussallam, J.~Stern, and N.~H. Fuchs, {\it The low-energy
  $\pi\pi$ amplitude to one and two loops},  {\em Nucl. Phys.} {\bf B457}
  (1995) 513--576, [\href{http://xxx.lanl.gov/abs/hep-ph/9507319}{{\tt
  hep-ph/9507319}}].

\bibitem{Knecht:1995ai}
M.~Knecht, B.~Moussallam, J.~Stern, and N.~H. Fuchs, {\it Determination of
  two-loop $\pi\pi$ scattering amplitude parameters},  {\em Nucl. Phys.} {\bf
  B471} (1996) 445--470, [\href{http://xxx.lanl.gov/abs/hep-ph/9512404}{{\tt
  hep-ph/9512404}}].

\bibitem{Manohar:1983md}
A.~Manohar and H.~Georgi, {\it Chiral quarks and the nonrelativistic quark
  model},  {\em Nucl. Phys.} {\bf B234} (1984) 189.

\bibitem{Weinberg:1966kf}
S.~Weinberg, {\it Pion scattering lengths},  {\em Phys. Rev. Lett.} {\bf 17}
  (1966) 616--621.

\bibitem{Georgi:1990um}
H.~Georgi, {\it An effective field theory for heavy quarks at low-energies},
  {\em Phys. Lett.} {\bf B240} (1990) 447--450.

\bibitem{Manohar:2000dt}
A.~V. Manohar and M.~B. Wise, {\em Heavy quark physics}.
\newblock Cambridge University Press, 2000.

\bibitem{Becher:1999he}
T.~Becher and H.~Leutwyler, {\it Baryon chiral perturbation theory in
  manifestly {Lorentz} invariant form},  {\em Eur. Phys. J.} {\bf C9} (1999)
  643--671, [\href{http://xxx.lanl.gov/abs/hep-ph/9901384}{{\tt
  hep-ph/9901384}}].

\bibitem{Fuchs:2003qc}
T.~Fuchs, J.~Gegelia, G.~Japaridze, and S.~Scherer, {\it Renormalization of
  relativistic baryon chiral perturbation theory and power counting},  {\em
  Phys. Rev.} {\bf D68} (2003) 056005,
  [\href{http://xxx.lanl.gov/abs/hep-ph/0302117}{{\tt hep-ph/0302117}}].

\bibitem{Jenkins:1992ts}
E.~Jenkins, {\it Baryon masses in chiral perturbation theory},  {\em Nucl.
  Phys.} {\bf B368} (1992) 190--203.

\bibitem{Butler:1992ci}
M.~N. Butler and M.~J. Savage, {\it Electromagnetic polarizability of the
  nucleon in chiral perturbation theory},  {\em Phys. Lett.} {\bf B294} (1992)
  369--374, [\href{http://xxx.lanl.gov/abs/hep-ph/9209204}{{\tt
  hep-ph/9209204}}].

\bibitem{Butler:1992pn}
M.~N. Butler, M.~J. Savage, and R.~P. Springer, {\it Strong and electromagnetic
  decays of the baryon decuplet},  {\em Nucl. Phys.} {\bf B399} (1993) 69--88,
  [\href{http://xxx.lanl.gov/abs/hep-ph/9211247}{{\tt hep-ph/9211247}}].

\bibitem{Bernard:1993nj}
V.~Bernard, N.~Kaiser, and U.~G. Meissner, {\it Critical analysis of baryon
  masses and sigma terms in heavy baryon chiral perturbation theory},  {\em Z.
  Phys.} {\bf C60} (1993) 111--120,
  [\href{http://xxx.lanl.gov/abs/hep-ph/9303311}{{\tt hep-ph/9303311}}].

\bibitem{Butler:1994ej}
M.~N. Butler, M.~J. Savage, and R.~P. Springer, {\it Electromagnetic moments of
  the baryon decuplet},  {\em Phys. Rev.} {\bf D49} (1994) 3459--3465,
  [\href{http://xxx.lanl.gov/abs/hep-ph/9308317}{{\tt hep-ph/9308317}}].

\bibitem{Lebed:1994yu}
R.~F. Lebed, {\it Baryon decuplet mass relations in chiral perturbation
  theory},  {\em Nucl. Phys.} {\bf B430} (1994) 295--318,
  [\href{http://xxx.lanl.gov/abs/hep-ph/9311234}{{\tt hep-ph/9311234}}].

\bibitem{Lebed:1994gt}
R.~F. Lebed and M.~A. Luty, {\it Baryon masses at second order in chiral
  perturbation theory},  {\em Phys. Lett.} {\bf B329} (1994) 479--485,
  [\href{http://xxx.lanl.gov/abs/hep-ph/9401232}{{\tt hep-ph/9401232}}].

\bibitem{Banerjee:1995bk}
M.~K. Banerjee and J.~Milana, {\it Baryon mass splittings in chiral
  perturbation theory},  {\em Phys. Rev.} {\bf D52} (1995) 6451--6460,
  [\href{http://xxx.lanl.gov/abs/hep-ph/9410398}{{\tt hep-ph/9410398}}].

\bibitem{Hemmert:1996rw}
T.~R. Hemmert, B.~R. Holstein, and J.~Kambor, {\it Delta(1232) and the
  polarizabilities of the nucleon},  {\em Phys. Rev.} {\bf D55} (1997)
  5598--5612, [\href{http://xxx.lanl.gov/abs/hep-ph/9612374}{{\tt
  hep-ph/9612374}}].

\bibitem{Hemmert:1997tj}
T.~R. Hemmert, B.~R. Holstein, J.~Kambor, and G.~Knochlein, {\it {Compton}
  scattering and the spin structure of the nucleon at low energies},  {\em
  Phys. Rev.} {\bf D57} (1998) 5746--5754,
  [\href{http://xxx.lanl.gov/abs/nucl-th/9709063}{{\tt nucl-th/9709063}}].

\bibitem{Beane:2002wn}
S.~R. Beane, M.~Malheiro, J.~A. McGovern, D.~R. Phillips, and U.~van Kolck,
  {\it Nucleon polarizabilities from low-energy {Compton} scattering},  {\em
  Phys. Lett.} {\bf B567} (2003) 200--206,
  [\href{http://xxx.lanl.gov/abs/nucl-th/0209002}{{\tt nucl-th/0209002}}].

\bibitem{Pascalutsa:2002pi}
V.~Pascalutsa and D.~R. Phillips, {\it Effective theory of the delta(1232) in
  {Compton} scattering off the nucleon},  {\em Phys. Rev.} {\bf C67} (2003)
  055202, [\href{http://xxx.lanl.gov/abs/nucl-th/0212024}{{\tt
  nucl-th/0212024}}].

\bibitem{Pascalutsa:2003zk}
V.~Pascalutsa and D.~R. Phillips, {\it Model-independent effects of delta
  excitation in nucleon spin polarizabilities},  {\em Phys. Rev.} {\bf C68}
  (2003) 055205, [\href{http://xxx.lanl.gov/abs/nucl-th/0305043}{{\tt
  nucl-th/0305043}}].

\bibitem{Beane:2004ra}
S.~R. Beane, M.~Malheiro, J.~A. McGovern, D.~R. Phillips, and U.~van Kolck,
  {\it {Compton} scattering on the proton, neutron, and deuteron in chiral
  perturbation theory to $\mathcal{O}(q^4)$},  {\em Nucl. Phys.} {\bf A747}
  (2005) 311--361, [\href{http://xxx.lanl.gov/abs/nucl-th/0403088}{{\tt
  nucl-th/0403088}}].

\bibitem{Wies:2006rv}
N.~Wies, J.~Gegelia, and S.~Scherer, {\it Consistency of the pi delta
  interaction in chiral perturbation theory},  {\em Phys. Rev.} {\bf D73}
  (2006) 094012, [\href{http://xxx.lanl.gov/abs/hep-ph/0602073}{{\tt
  hep-ph/0602073}}].

\bibitem{Witten:1979kh}
E.~Witten, {\it Baryons in the {1/N} expansion},  {\em Nucl. Phys.} {\bf B160}
  (1979) 57.

\bibitem{Rarita:1941mf}
W.~Rarita and J.~S. Schwinger, {\it On a theory of particles with half integral
  spin},  {\em Phys. Rev.} {\bf 60} (1941) 61.

\bibitem{Gell-Mann:1962xb}
M.~Gell-Mann, {\it Symmetries of baryons and mesons},  {\em Phys. Rev.} {\bf
  125} (1962) 1067--1084.

\bibitem{Gell-Mann:1964nj}
M.~Gell-Mann, {\it A schematic model of baryons and mesons},  {\em Phys. Lett.}
  {\bf 8} (1964) 214--215.

\bibitem{Okubo:1961jc}
S.~Okubo, {\it Note on unitary symmetry in strong interactions},  {\em Prog.
  Theor. Phys.} {\bf 27} (1962) 949--966.

\bibitem{Bernard:1993sv}
C.~W. Bernard and M.~F.~L. Golterman, {\it Partially quenched gauge theories
  and an application to staggered fermions},  {\em Phys. Rev.} {\bf D49} (1994)
  486--494, [\href{http://xxx.lanl.gov/abs/hep-lat/9306005}{{\tt
  hep-lat/9306005}}].

\bibitem{Sharpe:1997by}
S.~R. Sharpe, {\it Enhanced chiral logarithms in partially quenched {QCD}},
  {\em Phys. Rev.} {\bf D56} (1997) 7052--7058,
  [\href{http://xxx.lanl.gov/abs/hep-lat/9707018}{{\tt hep-lat/9707018}}].

\bibitem{Golterman:1997st}
M.~F.~L. Golterman and K.-C. Leung, {\it Applications of partially quenched
  chiral perturbation theory},  {\em Phys. Rev.} {\bf D57} (1998) 5703--5710,
  [\href{http://xxx.lanl.gov/abs/hep-lat/9711033}{{\tt hep-lat/9711033}}].

\bibitem{Sharpe:2001fh}
S.~R. Sharpe and N.~Shoresh, {\it Partially quenched chiral perturbation theory
  without {$\Phi_0$}},  {\em Phys. Rev.} {\bf D64} (2001) 114510,
  [\href{http://xxx.lanl.gov/abs/http://arXiv.org/abs/hep-lat/0108003}{{\tt
  http://arXiv.org/abs/hep-lat/0108003}}].

\bibitem{Savage:2001jw}
M.~J. Savage, {\it Heavy-meson observables at one-loop in partially quenched
  chiral perturbation theory},  {\em Phys. Rev.} {\bf D65} (2002) 034014,
  [\href{http://xxx.lanl.gov/abs/hep-ph/0109190}{{\tt hep-ph/0109190}}].

\bibitem{Chen:2001yi}
J.-W. Chen and M.~J. Savage, {\it Baryons in partially quenched chiral
  perturbation theory},  {\em Phys. Rev.} {\bf D65} (2002) 094001,
  [\href{http://xxx.lanl.gov/abs/http://arXiv.org/abs/hep-lat/0111050}{{\tt
  http://arXiv.org/abs/hep-lat/0111050}}].

\bibitem{Beane:2002vq}
S.~R. Beane and M.~J. Savage, {\it Nucleons in two-flavor partially-quenched
  chiral perturbation theory},  {\em Nucl. Phys.} {\bf A709} (2002) 319--344,
  [\href{http://xxx.lanl.gov/abs/hep-lat/0203003}{{\tt hep-lat/0203003}}].

\bibitem{Sharpe:2003vy}
S.~R. Sharpe and R.~S. Van~de Water, {\it Unphysical operators in partially
  quenched {QCD}},  {\em Phys. Rev.} {\bf D69} (2004) 054027,
  [\href{http://xxx.lanl.gov/abs/hep-lat/0310012}{{\tt hep-lat/0310012}}].

\bibitem{Beane:2002nu}
S.~R. Beane and M.~J. Savage, {\it Nucleon nucleon interactions on the
  lattice},  {\em Phys. Lett.} {\bf B535} (2002) 177--180,
  [\href{http://xxx.lanl.gov/abs/hep-lat/0202013}{{\tt hep-lat/0202013}}].

\bibitem{Beane:2002np}
S.~R. Beane and M.~J. Savage, {\it Partially-quenched nucleon nucleon
  scattering},  {\em Phys. Rev.} {\bf D67} (2003) 054502,
  [\href{http://xxx.lanl.gov/abs/hep-lat/0210046}{{\tt hep-lat/0210046}}].

\bibitem{Damgaard:2000gh}
P.~H. Damgaard and K.~Splittorff, {\it Partially quenched chiral perturbation
  theory and the replica method},  {\em Phys. Rev.} {\bf D62} (2000) 054509,
  [\href{http://xxx.lanl.gov/abs/hep-lat/0003017}{{\tt hep-lat/0003017}}].

\bibitem{BahaBalantekin:1981qy}
A.~B. Balantekin and I.~Bars, {\it Dimension and character formulas for {L}ie
  supergroups},  {\em J. Math. Phys.} {\bf 22} (1981) 1149.

\bibitem{Chen:2005ab}
J.-W. Chen, D.~O'Connell, R.~S. Van~de Water, and A.~Walker-Loud, {\it
  {Ginsparg-Wilson} pions scattering in a sea of staggered quarks},  {\em Phys.
  Rev.} {\bf D73} (2006) 074510,
  [\href{http://xxx.lanl.gov/abs/hep-lat/0510024}{{\tt hep-lat/0510024}}].

\bibitem{Labrenz:1996jy}
J.~N. Labrenz and S.~R. Sharpe, {\it Quenched chiral perturbation theory for
  baryons},  {\em Phys. Rev.} {\bf D54} (1996) 4595--4608,
  [\href{http://xxx.lanl.gov/abs/http://arXiv.org/abs/hep-lat/9605034}{{\tt
  http://arXiv.org/abs/hep-lat/9605034}}].

\bibitem{Savage:2001dy}
M.~J. Savage, {\it The magnetic moments of the octet baryons in quenched chiral
  perturbation theory},  {\em Nucl. Phys.} {\bf A700} (2002) 359--376,
  [\href{http://xxx.lanl.gov/abs/nucl-th/0107038}{{\tt nucl-th/0107038}}].

\bibitem{Eidelman:2004wy}
{\bf Particle Data Group} Collaboration, S.~Eidelman {\em et.~al.}, {\it Review
  of particle physics},  {\em Phys. Lett.} {\bf B592} (2004) 1.

\bibitem{Ne'eman:1961cd}
Y.~Ne'eman, {\it Derivation of strong interactions from a gauge invariance},
  {\em Nucl. Phys.} {\bf 26} (1961) 222--229.

\bibitem{Walker-Loud:2004hf}
A.~Walker-Loud, {\it Octet baryon masses in partially quenched chiral
  perturbation theory},  {\em Nucl. Phys.} {\bf A747} (2005) 476--507,
  [\href{http://xxx.lanl.gov/abs/hep-lat/0405007}{{\tt hep-lat/0405007}}].

\bibitem{Tiburzi:2004rh}
B.~C. Tiburzi and A.~Walker-Loud, {\it Decuplet baryon masses in partially
  quenched chiral perturbation theory},  {\em Nucl. Phys.} {\bf A748} (2005)
  513--536, [\href{http://xxx.lanl.gov/abs/hep-lat/0407030}{{\tt
  hep-lat/0407030}}].

\bibitem{Tiburzi:2005na}
B.~C. Tiburzi and A.~Walker-Loud, {\it Strong isospin breaking in the nucleon
  and delta masses on the lattice},  {\em Nucl. Phys.} {\bf A764} (2006)
  274--302, [\href{http://xxx.lanl.gov/abs/hep-lat/0501018}{{\tt
  hep-lat/0501018}}].

\bibitem{Gasser:1979hf}
J.~Gasser and A.~Zepeda, {\it Approaching the chiral limit in {QCD}},  {\em
  Nucl. Phys.} {\bf B174} (1980) 445.

\bibitem{Luke:1992cs}
M.~E. Luke and A.~V. Manohar, {\it Reparametrization invariance constraints on
  heavy particle effective field theories},  {\em Phys. Lett.} {\bf B286}
  (1992) 348--354, [\href{http://xxx.lanl.gov/abs/hep-ph/9205228}{{\tt
  hep-ph/9205228}}].

\bibitem{Georgi:1991ch}
H.~Georgi, {\it On-shell effective field theory},  {\em Nucl. Phys.} {\bf B361}
  (1991) 339--350.

\bibitem{Arzt:1993gz}
C.~Arzt, {\it Reduced effective lagrangians},  {\em Phys. Lett.} {\bf B342}
  (1995) 189--195, [\href{http://xxx.lanl.gov/abs/hep-ph/9304230}{{\tt
  hep-ph/9304230}}].

\bibitem{Bernard:1995dp}
V.~Bernard, N.~Kaiser, and U.-G. Meissner, {\it Chiral dynamics in nucleons and
  nuclei},  {\em Int. J. Mod. Phys.} {\bf E4} (1995) 193--346,
  [\href{http://xxx.lanl.gov/abs/hep-ph/9501384}{{\tt hep-ph/9501384}}].

\bibitem{Luscher:1991cf}
M.~Luscher, {\it Signatures of unstable particles in finite volume},  {\em
  Nucl. Phys.} {\bf B364} (1991) 237--254.

\bibitem{Gasser:1974wd}
J.~Gasser and H.~Leutwyler, {\it Implications of scaling for the proton -
  neutron mass - difference},  {\em Nucl. Phys.} {\bf B94} (1975) 269.

\bibitem{Duncan:1996xy}
A.~Duncan, E.~Eichten, and H.~Thacker, {\it Electromagnetic splittings and
  light quark masses in lattice {QCD}},  {\em Phys. Rev. Lett.} {\bf 76} (1996)
  3894--3897, [\href{http://xxx.lanl.gov/abs/hep-lat/9602005}{{\tt
  hep-lat/9602005}}].

\bibitem{Duncan:1996be}
A.~Duncan, E.~Eichten, and H.~Thacker, {\it Electromagnetic structure of light
  baryons in lattice {QCD}},  {\em Phys. Lett.} {\bf B409} (1997) 387--392,
  [\href{http://xxx.lanl.gov/abs/hep-lat/9607032}{{\tt hep-lat/9607032}}].

\bibitem{Duncan:2004ys}
A.~Duncan, E.~Eichten, and R.~Sedgewick, {\it Computing electromagnetic effects
  in fully unquenched {QCD}},
  \href{http://xxx.lanl.gov/abs/hep-lat/0405014}{{\tt hep-lat/0405014}}.

\bibitem{Gasser:2003hk}
J.~Gasser, A.~Rusetsky, and I.~Scimemi, {\it Electromagnetic corrections in
  hadronic processes},  {\em Eur. Phys. J.} {\bf C32} (2003) 97--114,
  [\href{http://xxx.lanl.gov/abs/hep-ph/0305260}{{\tt hep-ph/0305260}}].

\bibitem{Miller:1990iz}
G.~A. Miller, B.~M.~K. Nefkens, and I.~Slaus, {\it Charge symmetry, quarks and
  mesons},  {\em Phys. Rept.} {\bf 194} (1990) 1--116.

\bibitem{Borasoy:1996bx}
B.~Borasoy and U.-G. Meissner, {\it Chiral expansion of baryon masses and
  sigma-terms},  {\em Annals Phys.} {\bf 254} (1997) 192--232,
  [\href{http://xxx.lanl.gov/abs/hep-ph/9607432}{{\tt hep-ph/9607432}}].

\bibitem{Walker-Loud:2005bt}
A.~Walker-Loud and J.~M.~S. Wu, {\it Nucleon and delta masses in twisted mass
  chiral perturbation theory},  {\em Phys. Rev.} {\bf D72} (2005) 014506,
  [\href{http://xxx.lanl.gov/abs/hep-lat/0504001}{{\tt hep-lat/0504001}}].

\bibitem{Frezzotti:2004pc}
R.~Frezzotti, {\it Twisted mass lattice {QCD}},  {\em Nucl. Phys. Proc. Suppl.}
  {\bf 140} (2005) 134--140,
  [\href{http://xxx.lanl.gov/abs/hep-lat/0409138}{{\tt hep-lat/0409138}}].

\bibitem{Shindler:2005vj}
A.~Shindler, {\it Twisted mass lattice {QCD}: {Recent} developments and
  results},  {\em PoS} {\bf LAT2005} (2005) 014,
  [\href{http://xxx.lanl.gov/abs/hep-lat/0511002}{{\tt hep-lat/0511002}}].

\bibitem{Kennedy:2004ae}
A.~D. Kennedy, {\it Algorithms for lattice {QCD} with dynamical fermions},
  {\em Nucl. Phys. Proc. Suppl.} {\bf 140} (2005) 190--203,
  [\href{http://xxx.lanl.gov/abs/hep-lat/0409167}{{\tt hep-lat/0409167}}].

\bibitem{Frezzotti:1999vv}
R.~Frezzotti, P.~A. Grassi, S.~Sint, and P.~Weisz, {\it A local formulation of
  lattice {QCD} without unphysical fermion zero modes},  {\em Nucl. Phys. Proc.
  Suppl.} {\bf 83} (2000) 941--946,
  [\href{http://xxx.lanl.gov/abs/hep-lat/9909003}{{\tt hep-lat/9909003}}].

\bibitem{Frezzotti:2003ni}
R.~Frezzotti and G.~C. Rossi, {\it Chirally improving wilson fermions. {I}:
  $\mc{O}(a)$ improvement},  {\em JHEP} {\bf 08} (2004) 007,
  [\href{http://xxx.lanl.gov/abs/hep-lat/0306014}{{\tt hep-lat/0306014}}].

\bibitem{Pena:2004gb}
C.~Pena, S.~Sint, and A.~Vladikas, {\it Twisted mass {QCD} and lattice
  approaches to the {$\Delta I = 1/2$} rule},  {\em JHEP} {\bf 09} (2004) 069,
  [\href{http://xxx.lanl.gov/abs/hep-lat/0405028}{{\tt hep-lat/0405028}}].

\bibitem{Frezzotti:2004wz}
R.~Frezzotti and G.~C. Rossi, {\it Chirally improving wilson fermions. {II:
  Four-quark} operators},  {\em JHEP} {\bf 10} (2004) 070,
  [\href{http://xxx.lanl.gov/abs/hep-lat/0407002}{{\tt hep-lat/0407002}}].

\bibitem{Sharpe:1998xm}
S.~R. Sharpe and R.~J. Singleton, {\it Spontaneous flavor and parity breaking
  with {Wilson} fermions},  {\em Phys. Rev.} {\bf D58} (1998) 074501,
  [\href{http://xxx.lanl.gov/abs/hep-lat/9804028}{{\tt hep-lat/9804028}}].

\bibitem{Munster:2003ba}
G.~Munster and C.~Schmidt, {\it Chiral perturbation theory for lattice {QCD}
  with a twisted mass term},  {\em Europhys. Lett.} {\bf 66} (2004) 652--656,
  [\href{http://xxx.lanl.gov/abs/hep-lat/0311032}{{\tt hep-lat/0311032}}].

\bibitem{Scorzato:2004da}
L.~Scorzato, {\it Pion mass splitting and phase structure in twisted mass
  {QCD}},  {\em Eur. Phys. J.} {\bf C37} (2004) 445--455,
  [\href{http://xxx.lanl.gov/abs/hep-lat/0407023}{{\tt hep-lat/0407023}}].

\bibitem{Sharpe:2004ps}
S.~R. Sharpe and J.~M.~S. Wu, {\it The phase diagram of twisted mass lattice
  {QCD}},  {\em Phys. Rev.} {\bf D70} (2004) 094029,
  [\href{http://xxx.lanl.gov/abs/hep-lat/0407025}{{\tt hep-lat/0407025}}].

\bibitem{Rupak:2002sm}
G.~Rupak and N.~Shoresh, {\it Chiral perturbation theory for the {Wilson}
  lattice action},  {\em Phys. Rev.} {\bf D66} (2002) 054503,
  [\href{http://xxx.lanl.gov/abs/hep-lat/0201019}{{\tt hep-lat/0201019}}].

\bibitem{Bar:2003mh}
O.~Bar, G.~Rupak, and N.~Shoresh, {\it Chiral perturbation theory at
  $\mathcal{O}(a^2)$ for lattice {QCD}},  {\em Phys. Rev.} {\bf D70} (2004)
  034508, [\href{http://xxx.lanl.gov/abs/hep-lat/0306021}{{\tt
  hep-lat/0306021}}].

\bibitem{Munster:2004am}
G.~Munster, {\it On the phase structure of twisted mass lattice {QCD}},  {\em
  JHEP} {\bf 09} (2004) 035,
  [\href{http://xxx.lanl.gov/abs/hep-lat/0407006}{{\tt hep-lat/0407006}}].

\bibitem{Munster:2004wt}
G.~Munster, C.~Schmidt, and E.~E. Scholz, {\it Chiral perturbation theory for
  twisted mass {QCD}},  {\em Nucl. Phys. Proc. Suppl.} {\bf 140} (2005)
  320--322, [\href{http://xxx.lanl.gov/abs/hep-lat/0409066}{{\tt
  hep-lat/0409066}}].

\bibitem{Aoki:2004ta}
S.~Aoki and O.~Bar, {\it Twisted-mass {QCD}, $\mc{O}(a)$ improvement and wilson
  chiral perturbation theory},  {\em Phys. Rev.} {\bf D70} (2004) 116011,
  [\href{http://xxx.lanl.gov/abs/hep-lat/0409006}{{\tt hep-lat/0409006}}].

\bibitem{Sharpe:2004ny}
S.~R. Sharpe and J.~M.~S. Wu, {\it Twisted mass chiral perturbation theory at
  next-to-leading order},  {\em Phys. Rev.} {\bf D71} (2005) 074501,
  [\href{http://xxx.lanl.gov/abs/hep-lat/0411021}{{\tt hep-lat/0411021}}].

\bibitem{Abdel-Rehim:2005gz}
A.~M. Abdel-Rehim, R.~Lewis, and R.~M. Woloshyn, {\it Spectrum of quenched
  twisted mass lattice {QCD} at maximal twist},  {\em Phys. Rev.} {\bf D71}
  (2005) 094505, [\href{http://xxx.lanl.gov/abs/hep-lat/0503007}{{\tt
  hep-lat/0503007}}].

\bibitem{Beane:2003xv}
S.~R. Beane and M.~J. Savage, {\it Nucleons properties at finite lattice
  spacing in chiral perturbation theory},  {\em Phys. Rev.} {\bf D68} (2003)
  114502, [\href{http://xxx.lanl.gov/abs/hep-lat/0306036}{{\tt
  hep-lat/0306036}}].

\bibitem{Tiburzi:2005vy}
B.~C. Tiburzi, {\it Baryon masses at $\mathcal{O}(a^2)$ in chiral perturbation
  theory},  {\em Nucl. Phys.} {\bf A761} (2005) 232--258,
  [\href{http://xxx.lanl.gov/abs/hep-lat/0501020}{{\tt hep-lat/0501020}}].

\bibitem{Frezzotti:2003xj}
R.~Frezzotti and G.~C. Rossi, {\it Twisted-mass lattice {QCD} with mass
  non-degenerate quarks},  {\em Nucl. Phys. Proc. Suppl.} {\bf 128} (2004)
  193--202, [\href{http://xxx.lanl.gov/abs/hep-lat/0311008}{{\tt
  hep-lat/0311008}}].

\bibitem{Symanzik:1983dc}
K.~Symanzik, {\it Continuum limit and improved action in lattice theories. 1.
  principles and $\phi^4$ theory},  {\em Nucl. Phys.} {\bf B226} (1983) 187.

\bibitem{Symanzik:1983gh}
K.~Symanzik, {\it Continuum limit and improved action in lattice theories. 2.
  $\mathcal{O}(n)$ nonlinear sigma model in perturbation theory},  {\em Nucl.
  Phys.} {\bf B226} (1983) 205.

\bibitem{Sharpe:2005rq}
S.~R. Sharpe, {\it Observations on discretization errors in twisted-mass
  lattice {QCD}},  {\em Phys. Rev.} {\bf D72} (2005) 074510,
  [\href{http://xxx.lanl.gov/abs/hep-lat/0509009}{{\tt hep-lat/0509009}}].

\bibitem{Aoki:2006nv}
S.~Aoki and O.~Bar, {\it Automatic $\mathcal{O}(a)$ improvement for
  twisted-mass {QCD} in the presence of spontaneous symmetry breaking},
  \href{http://xxx.lanl.gov/abs/hep-lat/0604018}{{\tt hep-lat/0604018}}.

\bibitem{Frezzotti:2005gi}
R.~Frezzotti, G.~Martinelli, M.~Papinutto, and G.~C. Rossi, {\it Reducing
  cutoff effects in maximally twisted lattice {QCD} close to the chiral limit},
   \href{http://xxx.lanl.gov/abs/hep-lat/0503034}{{\tt hep-lat/0503034}}.

\bibitem{Munster:2004dj}
G.~Munster, C.~Schmidt, and E.~E. Scholz, {\it Chiral perturbation theory for
  partially quenched twisted mass lattice {QCD}},  {\em Europhys. Lett.} {\bf
  86} (2004) 639--644, [\href{http://xxx.lanl.gov/abs/hep-lat/0402003}{{\tt
  hep-lat/0402003}}].

\bibitem{Farchioni:2005bh}
F.~Farchioni {\em et.~al.}, {\it Numerical simulations with two flavours of
  twisted-mass {Wilson} quarks and {DBW2} gauge action},
  \href{http://xxx.lanl.gov/abs/hep-lat/0512017}{{\tt hep-lat/0512017}}.

\bibitem{Detmold:2006vu}
W.~Detmold, B.~C. Tiburzi, and A.~Walker-Loud, {\it Electromagnetic and spin
  polarisabilities in lattice {QCD}},
  \href{http://xxx.lanl.gov/abs/hep-lat/0603026}{{\tt hep-lat/0603026}}.

\bibitem{Schumacher:2005an}
M.~Schumacher, {\it Polarizability of the nucleon and {Compton} scattering},
  {\em Prog. Part. Nucl. Phys.} {\bf 55} (2005) 567--646,
  [\href{http://xxx.lanl.gov/abs/hep-ph/0501167}{{\tt hep-ph/0501167}}].

\bibitem{Hyde-Wright:2004gh}
C.~E. Hyde-Wright and K.~de~Jager, {\it Electromagnetic form factors of the
  nucleon and compton scattering},  {\em Ann. Rev. Nucl. Part. Sci.} {\bf 54}
  (2004) 217--267, [\href{http://xxx.lanl.gov/abs/nucl-ex/0507001}{{\tt
  nucl-ex/0507001}}].

\bibitem{Fiebig:1988en}
H.~R. Fiebig, W.~Wilcox, and R.~M. Woloshyn, {\it A study of hadron electric
  polarizability in quenched lattice {QCD}},  {\em Nucl. Phys.} {\bf B324}
  (1989) 47.

\bibitem{Christensen:2004ca}
J.~Christensen, W.~Wilcox, F.~X. Lee, and L.-m. Zhou, {\it Electric
  polarizability of neutral hadrons from lattice {QCD}},  {\em Phys. Rev.} {\bf
  D72} (2005) 034503, [\href{http://xxx.lanl.gov/abs/hep-lat/0408024}{{\tt
  hep-lat/0408024}}].

\bibitem{Lee:2005dq}
F.~X. Lee, L.~Zhou, W.~Wilcox, and J.~Christensen, {\it Magnetic polarizability
  of hadrons from lattice {QCD} in the background field method},
  \href{http://xxx.lanl.gov/abs/hep-lat/0509065}{{\tt hep-lat/0509065}}.

\bibitem{Tiburzi:2005is}
B.~C. Tiburzi, {\it Baryons with ginsparg-wilson quarks in a staggered sea},
  {\em Phys. Rev.} {\bf D72} (2005) 094501,
  [\href{http://xxx.lanl.gov/abs/hep-lat/0508019}{{\tt hep-lat/0508019}}].

\bibitem{RaiChoudhury68}
S.~Rai~Choudhury and D.~Z. Freedman, {\it Higher-order low-energy theorems for
  nucleon compton scattering},  {\em Phys. Rev.} {\bf 168} (1968).

\bibitem{Gell-Mann:1954kc}
M.~Gell-Mann and M.~L. Goldberger, {\it Scattering of low-energy photons by
  particles of spin 1/2},  {\em Phys. Rev.} {\bf 96} (1954) 1433--1438.

\bibitem{Low:1954kd}
F.~E. Low, {\it Scattering of light of very low frequency by systems of spin
  1/2},  {\em Phys. Rev.} {\bf 96} (1954) 1428--1432.

\bibitem{Fucito:1982ff}
F.~Fucito, G.~Parisi, and S.~Petrarca, {\it First evaluation of g(a) / g(v) in
  lattice {QCD} in the quenched approximation},  {\em Phys. Lett.} {\bf B115}
  (1982) 148--150.

\bibitem{Martinelli:1982cb}
G.~Martinelli, G.~Parisi, R.~Petronzio, and F.~Rapuano, {\it The proton and
  neutron magnetic moments in lattice {QCD}},  {\em Phys. Lett.} {\bf B116}
  (1982) 434.

\bibitem{Bernard:1982yu}
C.~W. Bernard, T.~Draper, K.~Olynyk, and M.~Rushton, {\it Lattice {QCD}
  calculation of some baryon magnetic moments},  {\em Phys. Rev. Lett.} {\bf
  49} (1982) 1076.

\bibitem{Shintani:2005du}
E.~Shintani {\em et.~al.}, {\it Neutron electric dipole moment on the lattice},
   {\em PoS} {\bf LAT2005} (2005) 128,
  [\href{http://xxx.lanl.gov/abs/hep-lat/0509123}{{\tt hep-lat/0509123}}].

\bibitem{Detmold:2004kw}
W.~Detmold, {\it Flavour singlet physics in lattice {QCD} with background
  fields},  {\em Phys. Rev.} {\bf D71} (2005) 054506,
  [\href{http://xxx.lanl.gov/abs/hep-lat/0410011}{{\tt hep-lat/0410011}}].

\bibitem{Babusci:1998ww}
D.~Babusci, G.~Giordano, A.~I. L'vov, G.~Matone, and A.~M. Nathan, {\it
  Low-energy {Compton} scattering of polarized photons on polarized nucleons},
  {\em Phys. Rev.} {\bf C58} (1998) 1013--1041,
  [\href{http://xxx.lanl.gov/abs/hep-ph/9803347}{{\tt hep-ph/9803347}}].

\bibitem{Schwinger:1951nm}
J.~S. Schwinger, {\it On gauge invariance and vacuum polarization},  {\em Phys.
  Rev.} {\bf 82} (1951) 664--679.

\bibitem{Jenkins:1991es}
E.~Jenkins and A.~V. Manohar, {\it Chiral corrections to the baryon axial
  currents},  {\em Phys. Lett.} {\bf B259} (1991) 353--358.

\bibitem{Arndt:2003vd}
D.~Arndt and B.~C. Tiburzi, {\it Baryon decuplet to octet electromagnetic
  transitions in quenched and partially quenched chiral perturbation theory},
  {\em Phys. Rev.} {\bf D69} (2004) 014501,
  [\href{http://xxx.lanl.gov/abs/hep-lat/0309013}{{\tt hep-lat/0309013}}].

\bibitem{Tiburzi:2004mv}
B.~C. Tiburzi, {\it Baryon electromagnetic properties in partially quenched
  heavy hadron chiral perturbation theory},  {\em Phys. Rev.} {\bf D71} (2005)
  054504, [\href{http://xxx.lanl.gov/abs/hep-lat/0412025}{{\tt
  hep-lat/0412025}}].

\bibitem{Detmold:2005pt}
W.~Detmold and C.~J.~D. Lin, {\it Twist-two matrix elements at finite and
  infinite volume},  {\em Phys. Rev.} {\bf D71} (2005) 054510,
  [\href{http://xxx.lanl.gov/abs/hep-lat/0501007}{{\tt hep-lat/0501007}}].

\bibitem{Wess:1971yu}
J.~Wess and B.~Zumino, {\it Consequences of anomalous {Ward} identities},  {\em
  Phys. Lett.} {\bf B37} (1971) 95.

\bibitem{Witten:1983tw}
E.~Witten, {\it Global aspects of current algebra},  {\em Nucl. Phys.} {\bf
  B223} (1983) 422--432.

\bibitem{Arndt:2003we}
D.~Arndt and B.~C. Tiburzi, {\it Electromagnetic properties of the baryon
  decuplet in quenched and partially quenched chiral perturbation theory},
  {\em Phys. Rev.} {\bf D68} (2003) 114503,
  [\href{http://xxx.lanl.gov/abs/hep-lat/0308001}{{\tt hep-lat/0308001}}].

\bibitem{Fettes:2000gb}
N.~Fettes, U.-G. Meissner, M.~Mojzis, and S.~Steininger, {\it The chiral
  effective pion nucleon {Lagrangian} of $\mathcal{O}(p^4)$},  {\em Ann. Phys.}
  {\bf 283} (2000) 273--302,
  [\href{http://xxx.lanl.gov/abs/hep-ph/0001308}{{\tt hep-ph/0001308}}].

\bibitem{Tiburzi:2004kd}
B.~C. Tiburzi, {\it Baryon masses in partially quenched heavy hadron chiral
  perturbation theory},  {\em Phys. Rev.} {\bf D71} (2005) 034501,
  [\href{http://xxx.lanl.gov/abs/hep-lat/0410033}{{\tt hep-lat/0410033}}].

\bibitem{Sharpe:1992ft}
S.~R. Sharpe, {\it Quenched chiral logarithms},  {\em Phys. Rev.} {\bf D46}
  (1992) 3146--3168, [\href{http://xxx.lanl.gov/abs/hep-lat/9205020}{{\tt
  hep-lat/9205020}}].

\bibitem{Bell:1969ts}
J.~S. Bell and R.~Jackiw, {\it A {PCAC} puzzle: $\pi^0 \to \gamma\gamma$ in the
  sigma model},  {\em Nuovo Cim.} {\bf A60} (1969) 47--61.

\bibitem{Adler:1969gk}
S.~L. Adler, {\it Axial vector vertex in spinor electrodynamics},  {\em Phys.
  Rev.} {\bf 177} (1969) 2426--2438.

\bibitem{Adler:1969er}
S.~L. Adler and W.~A. Bardeen, {\it Absence of higher order corrections in the
  anomalous axial vector divergence equation},  {\em Phys. Rev.} {\bf 182}
  (1969) 1517--1536.

\bibitem{Bijnens:2001bb}
J.~Bijnens, L.~Girlanda, and P.~Talavera, {\it The anomalous chiral
  {Lagrangian} of $\mathcal{O}(p^6)$},  {\em Eur. Phys. J.} {\bf C23} (2002)
  539--544, [\href{http://xxx.lanl.gov/abs/hep-ph/0110400}{{\tt
  hep-ph/0110400}}].

\bibitem{Sharpe:2000bn}
S.~R. Sharpe and N.~Shoresh, {\it Physical results from partially quenched
  simulations},  {\em Int. J. Mod. Phys.} {\bf A16S1C} (2001) 1219--1224,
  [\href{http://xxx.lanl.gov/abs/http://arXiv.org/abs/hep-lat/0011089}{{\tt
  http://arXiv.org/abs/hep-lat/0011089}}].

\bibitem{Bernard:1991rq}
V.~Bernard, N.~Kaiser, and U.~G. Meissner, {\it Chiral expansion of the
  nucleon's electromagnetic polarizabilities},  {\em Phys. Rev. Lett.} {\bf 67}
  (1991) 1515--1518.

\bibitem{Bernard:1991ru}
V.~Bernard, N.~Kaiser, and U.~G. Meissner, {\it Nucleons with chiral loops:
  Electromagnetic polarizabilities},  {\em Nucl. Phys.} {\bf B373} (1992)
  346--370.

\bibitem{Bernard:1993ry}
V.~Bernard, N.~Kaiser, U.~G. Meissner, and A.~Schmidt, {\it Aspects of nucleon
  {Compton} scattering},  {\em Z. Phys.} {\bf A348} (1994) 317,
  [\href{http://xxx.lanl.gov/abs/hep-ph/9311354}{{\tt hep-ph/9311354}}].

\bibitem{Bernard:1993bg}
V.~Bernard, N.~Kaiser, A.~Schmidt, and U.~G. Meissner, {\it Consistent
  calculation of the nucleon electromagnetic polarizabilities in chiral
  perturbation theory beyond next- to-leading order},  {\em Phys. Lett.} {\bf
  B319} (1993) 269--275, [\href{http://xxx.lanl.gov/abs/hep-ph/9309211}{{\tt
  hep-ph/9309211}}].

\bibitem{Ji:1999sv}
X.-D. Ji, C.-W. Kao, and J.~Osborne, {\it The nucleon spin polarizability at
  order $\mathcal{O}(p^4)$ in chiral perturbation theory},  {\em Phys. Rev.}
  {\bf D61} (2000) 074003, [\href{http://xxx.lanl.gov/abs/hep-ph/9908526}{{\tt
  hep-ph/9908526}}].

\bibitem{VijayaKumar:2000pv}
K.~B. Vijaya~Kumar, J.~A. McGovern, and M.~C. Birse, {\it Spin polarisabilities
  of the nucleon at {NLO} in the chiral expansion},  {\em Phys. Lett.} {\bf
  B479} (2000) 167--172, [\href{http://xxx.lanl.gov/abs/hep-ph/0002133}{{\tt
  hep-ph/0002133}}].

\bibitem{Gellas:2000mx}
G.~C. Gellas, T.~R. Hemmert, and U.-G. Meissner, {\it Complete one-loop
  analysis of the nucleon's spin polarizabilities},  {\em Phys. Rev. Lett.}
  {\bf 85} (2000) 14--17, [\href{http://xxx.lanl.gov/abs/nucl-th/0002027}{{\tt
  nucl-th/0002027}}].

\bibitem{McGovern:2001dd}
J.~A. McGovern, {\it {Compton} scattering from the proton at fourth order in
  the chiral expansion},  {\em Phys. Rev.} {\bf C63} (2001) 064608,
  [\href{http://xxx.lanl.gov/abs/nucl-th/0101057}{{\tt nucl-th/0101057}}].

\bibitem{Arndt:2003ww}
D.~Arndt and B.~C. Tiburzi, {\it Charge radii of the meson and baryon octets in
  quenched and partially quenched chiral perturbation theory},  {\em Phys.
  Rev.} {\bf D68} (2003) 094501,
  [\href{http://xxx.lanl.gov/abs/hep-lat/0307003}{{\tt hep-lat/0307003}}].

\bibitem{Bedaque:1999dh}
P.~F. Bedaque and M.~J. Savage, {\it Parity violation in $\gamma$
  polarized-proton {Compton} scattering},  {\em Phys. Rev.} {\bf C62} (2000)
  018501, [\href{http://xxx.lanl.gov/abs/nucl-th/9909055}{{\tt
  nucl-th/9909055}}].

\bibitem{Detmold:2004qn}
W.~Detmold and M.~J. Savage, {\it Electroweak matrix elements in the
  two-nucleon sector from lattice {QCD}},  {\em Nucl. Phys.} {\bf A743} (2004)
  170--193, [\href{http://xxx.lanl.gov/abs/hep-lat/0403005}{{\tt
  hep-lat/0403005}}].

\bibitem{Bedaque:2006yi}
P.~F. Bedaque, I.~Sato, and A.~Walker-Loud, {\it Finite volume corrections to
  $\pi\pi$ scattering},  {\em Phys. Rev.} {\bf D73} (2006) 074501,
  [\href{http://xxx.lanl.gov/abs/hep-lat/0601033}{{\tt hep-lat/0601033}}].

\bibitem{Luscher:1985dn}
M.~Luscher, {\it Volume dependence of the energy spectrum in massive quantum
  field theories. 1. {S}table particle states},  {\em Commun. Math. Phys.} {\bf
  104} (1986) 177.

\bibitem{Gasser:1986vb}
J.~Gasser and H.~Leutwyler, {\it Light quarks at low temperatures},  {\em Phys.
  Lett.} {\bf B184} (1987) 83.

\bibitem{Colangelo:2003hf}
G.~Colangelo and S.~Durr, {\it The pion mass in finite volume},  {\em Eur.
  Phys. J.} {\bf C33} (2004) 543--553,
  [\href{http://xxx.lanl.gov/abs/hep-lat/0311023}{{\tt hep-lat/0311023}}].

\bibitem{Beane:2004tw}
S.~R. Beane, {\it Nucleon masses and magnetic moments in a finite volume},
  {\em Phys. Rev.} {\bf D70} (2004) 034507,
  [\href{http://xxx.lanl.gov/abs/hep-lat/0403015}{{\tt hep-lat/0403015}}].

\bibitem{Colangelo:2005gd}
G.~Colangelo, S.~Durr, and C.~Haefeli, {\it Finite volume effects for meson
  masses and decay constants},  {\em Nucl. Phys.} {\bf B721} (2005) 136--174,
  [\href{http://xxx.lanl.gov/abs/hep-lat/0503014}{{\tt hep-lat/0503014}}].

\bibitem{Sharpe:1992pp}
S.~R. Sharpe, R.~Gupta, and G.~W. Kilcup, {\it Lattice calculation of {$I = 2$}
  pion scattering length},  {\em Nucl. Phys.} {\bf B383} (1992) 309--356.

\bibitem{Gupta:1993rn}
R.~Gupta, A.~Patel, and S.~R. Sharpe, {\it I = 2 pion scattering amplitude with
  {Wilson} fermions},  {\em Phys. Rev.} {\bf D48} (1993) 388--396,
  [\href{http://xxx.lanl.gov/abs/hep-lat/9301016}{{\tt hep-lat/9301016}}].

\bibitem{Kuramashi:1993ka}
Y.~Kuramashi, M.~Fukugita, H.~Mino, M.~Okawa, and A.~Ukawa, {\it Lattice {QCD}
  calculation of full pion scattering lengths},  {\em Phys. Rev. Lett.} {\bf
  71} (1993) 2387--2390.

\bibitem{Fukugita:1994na}
M.~Fukugita, Y.~Kuramashi, H.~Mino, M.~Okawa, and A.~Ukawa, {\it An exploratory
  study of nucleon-nucleon scattering lengths in lattice {QCD}},  {\em Phys.
  Rev. Lett.} {\bf 73} (1994) 2176--2179,
  [\href{http://xxx.lanl.gov/abs/hep-lat/9407012}{{\tt hep-lat/9407012}}].

\bibitem{Fukugita:1994ve}
M.~Fukugita, Y.~Kuramashi, M.~Okawa, H.~Mino, and A.~Ukawa, {\it Hadron
  scattering lengths in lattice {QCD}},  {\em Phys. Rev.} {\bf D52} (1995)
  3003--3023, [\href{http://xxx.lanl.gov/abs/hep-lat/9501024}{{\tt
  hep-lat/9501024}}].

\bibitem{Liu:2001zp}
C.-A. Liu, J.-H. Zhang, Y.~Chen, and J.~P. Ma, {\it I = 2 pion scattering
  length from a coarse anisotropic lattice calculation},
  \href{http://xxx.lanl.gov/abs/hep-lat/0109010}{{\tt hep-lat/0109010}}.

\bibitem{Aoki:2002in}
{\bf JLQCD} Collaboration, S.~Aoki {\em et.~al.}, {\it I = 2 pion scattering
  length with the wilson fermion},  {\em Phys. Rev.} {\bf D66} (2002) 077501,
  [\href{http://xxx.lanl.gov/abs/hep-lat/0206011}{{\tt hep-lat/0206011}}].

\bibitem{Aoki:2002ny}
{\bf CP-PACS} Collaboration, S.~Aoki {\em et.~al.}, {\it I = 2 pion scattering
  phase shift with {Wilson} fermions},  {\em Phys. Rev.} {\bf D67} (2003)
  014502, [\href{http://xxx.lanl.gov/abs/hep-lat/0209124}{{\tt
  hep-lat/0209124}}].

\bibitem{Yamazaki:2004qb}
{\bf CP-PACS} Collaboration, T.~Yamazaki {\em et.~al.}, {\it I = 2 pi pi
  scattering phase shift with two flavors of $\mathcal{O}(a)$ improved
  dynamical quarks},  {\em Phys. Rev.} {\bf D70} (2004) 074513,
  [\href{http://xxx.lanl.gov/abs/hep-lat/0402025}{{\tt hep-lat/0402025}}].

\bibitem{Du:2004ib}
X.~Du, G.-w. Meng, C.~Miao, and C.~Liu, {\it I = 2 pion scattering length with
  improved actions on anisotropic lattices},  {\em Int. J. Mod. Phys.} {\bf
  A19} (2004) 5609--5614, [\href{http://xxx.lanl.gov/abs/hep-lat/0404017}{{\tt
  hep-lat/0404017}}].

\bibitem{Aoki:2005uf}
{\bf CP-PACS} Collaboration, S.~Aoki {\em et.~al.}, {\it I = 2 pion scattering
  length from two-pion wave functions},  {\em Phys. Rev.} {\bf D71} (2005)
  094504, [\href{http://xxx.lanl.gov/abs/hep-lat/0503025}{{\tt
  hep-lat/0503025}}].

\bibitem{taylor}
J.~R. Taylor, {\em Scattering Theory}.
\newblock John Wiley \& Sons, 1972.

\bibitem{Bijnens:2005ne}
J.~Bijnens, N.~Danielsson, K.~Ghorbani, and T.~Lahde, {\it Two loop partially
  quenched and finite volume chiral perturbation theory results},  {\em PoS}
  {\bf LAT2005} (2005) 058,
  [\href{http://xxx.lanl.gov/abs/hep-lat/0509042}{{\tt hep-lat/0509042}}].

\bibitem{Bar:2005tu}
O.~Bar, C.~Bernard, G.~Rupak, and N.~Shoresh, {\it Chiral perturbation theory
  for staggered sea quarks and {Ginsparg-Wilson} valence quarks},  {\em Phys.
  Rev.} {\bf D72} (2005) 054502,
  [\href{http://xxx.lanl.gov/abs/hep-lat/0503009}{{\tt hep-lat/0503009}}].

\bibitem{Gasser:1987ah}
J.~Gasser and H.~Leutwyler, {\it Thermodynamics of chiral symmetry},  {\em
  Phys. Lett.} {\bf B188} (1987) 477.

\bibitem{Detmold:2004ap}
W.~Detmold and M.~J. Savage, {\it Nucleon properties at finite volume: The
  $\epsilon^\prime$-regime},  {\em Phys. Lett.} {\bf B599} (2004) 32--42,
  [\href{http://xxx.lanl.gov/abs/hep-lat/0407008}{{\tt hep-lat/0407008}}].

\bibitem{Bedaque:2004kc}
P.~F. Bedaque, {\it {Aharonov-Bohm} effect and nucleon nucleon phase shifts on
  the lattice},  {\em Phys. Lett.} {\bf B593} (2004) 82--88,
  [\href{http://xxx.lanl.gov/abs/nucl-th/0402051}{{\tt nucl-th/0402051}}].

\bibitem{deDivitiis:2004kq}
G.~M. de~Divitiis, R.~Petronzio, and N.~Tantalo, {\it On the discretization of
  physical momenta in lattice {QCD}},  {\em Phys. Lett.} {\bf B595} (2004)
  408--413, [\href{http://xxx.lanl.gov/abs/hep-lat/0405002}{{\tt
  hep-lat/0405002}}].

\bibitem{Sachrajda:2004mi}
C.~T. Sachrajda and G.~Villadoro, {\it Twisted boundary conditions in lattice
  simulations},  {\em Phys. Lett.} {\bf B609} (2005) 73--85,
  [\href{http://xxx.lanl.gov/abs/hep-lat/0411033}{{\tt hep-lat/0411033}}].

\bibitem{Bedaque:2004ax}
P.~F. Bedaque and J.-W. Chen, {\it Twisted valence quarks and hadron
  interactions on the lattice},  {\em Phys. Lett.} {\bf B616} (2005) 208--214,
  [\href{http://xxx.lanl.gov/abs/hep-lat/0412023}{{\tt hep-lat/0412023}}].

\bibitem{Tiburzi:2005hg}
B.~C. Tiburzi, {\it Twisted quarks and the nucleon axial current},  {\em Phys.
  Lett.} {\bf B617} (2005) 40--48,
  [\href{http://xxx.lanl.gov/abs/hep-lat/0504002}{{\tt hep-lat/0504002}}].

\bibitem{Mehen:2005fw}
T.~Mehen and B.~C. Tiburzi, {\it Quarks with twisted boundary conditions in the
  $\epsilon$-regime},  {\em Phys. Rev.} {\bf D72} (2005) 014501,
  [\href{http://xxx.lanl.gov/abs/hep-lat/0505014}{{\tt hep-lat/0505014}}].

\bibitem{Flynn:2005in}
{\bf UKQCD} Collaboration, J.~M. Flynn, A.~Juttner, and C.~T. Sachrajda, {\it A
  numerical study of partially twisted boundary conditions},  {\em Phys. Lett.}
  {\bf B632} (2006) 313--318,
  [\href{http://xxx.lanl.gov/abs/hep-lat/0506016}{{\tt hep-lat/0506016}}].

\bibitem{Guadagnoli:2005be}
D.~Guadagnoli, F.~Mescia, and S.~Simula, {\it Lattice study of semileptonic
  form factors with twisted boundary conditions},
  \href{http://xxx.lanl.gov/abs/hep-lat/0512020}{{\tt hep-lat/0512020}}.

\bibitem{Christ:2005gi}
N.~H. Christ, C.~Kim, and T.~Yamazaki, {\it Finite volume corrections to the
  two-particle decay of states with non-zero momentum},  {\em Phys. Rev.} {\bf
  D72} (2005) 114506, [\href{http://xxx.lanl.gov/abs/hep-lat/0507009}{{\tt
  hep-lat/0507009}}].

\bibitem{Luscher:2003vf}
M.~Luscher, {\it Lattice {QCD} and the {Schwarz} alternating procedure},  {\em
  JHEP} {\bf 05} (2003) 052,
  [\href{http://xxx.lanl.gov/abs/hep-lat/0304007}{{\tt hep-lat/0304007}}].

\bibitem{Luscher:2003qa}
M.~Luscher, {\it Solution of the {Dirac} equation in lattice {QCD} using a
  domain decomposition method},  {\em Comput. Phys. Commun.} {\bf 156} (2004)
  209--220, [\href{http://xxx.lanl.gov/abs/hep-lat/0310048}{{\tt
  hep-lat/0310048}}].

\bibitem{Luscher:2005rx}
M.~Luscher, {\it Schwarz-preconditioned {HMC} algorithm for two-flavour lattice
  {QCD}},  {\em Comput. Phys. Commun.} {\bf 165} (2005) 199--220,
  [\href{http://xxx.lanl.gov/abs/hep-lat/0409106}{{\tt hep-lat/0409106}}].

\bibitem{Luscher:2005mv}
M.~Luscher, {\it Lattice {QCD} with light {Wilson} quarks},  {\em PoS} {\bf
  LAT2005} (2006) 002, [\href{http://xxx.lanl.gov/abs/hep-lat/0509152}{{\tt
  hep-lat/0509152}}].

\bibitem{Davies:2003ik}
{\bf HPQCD} Collaboration, C.~T.~H. Davies {\em et.~al.}, {\it High-precision
  lattice {QCD} confronts experiment},  {\em Phys. Rev. Lett.} {\bf 92} (2004)
  022001, [\href{http://xxx.lanl.gov/abs/hep-lat/0304004}{{\tt
  hep-lat/0304004}}].

\bibitem{Aubin:2004fs}
{\bf MILC} Collaboration, C.~Aubin {\em et.~al.}, {\it Light pseudoscalar decay
  constants, quark masses, and low energy constants from three-flavor lattice
  {QCD}},  {\em Phys. Rev.} {\bf D70} (2004) 114501,
  [\href{http://xxx.lanl.gov/abs/hep-lat/0407028}{{\tt hep-lat/0407028}}].

\bibitem{Durr:2004ta}
S.~Durr and C.~Hoelbling, {\it Scaling tests with dynamical overlap and rooted
  staggered fermions},  {\em Phys. Rev.} {\bf D71} (2005) 054501,
  [\href{http://xxx.lanl.gov/abs/hep-lat/0411022}{{\tt hep-lat/0411022}}].

\bibitem{Durr:2005ax}
S.~Durr, {\it Theoretical issues with staggered fermion simulations},  {\em
  PoS} {\bf LAT2005} (2005) 021,
  [\href{http://xxx.lanl.gov/abs/hep-lat/0509026}{{\tt hep-lat/0509026}}].

\bibitem{Shamir:2005sv}
Y.~Shamir, {\it Renormalization-group blocking the fourth root of the staggered
  determinant},  {\em PoS} {\bf LAT2005} (2006) 240,
  [\href{http://xxx.lanl.gov/abs/hep-lat/0509163}{{\tt hep-lat/0509163}}].

\bibitem{Creutz:2006ys}
M.~Creutz, {\it Flavor extrapolations and staggered fermions},
  \href{http://xxx.lanl.gov/abs/hep-lat/0603020}{{\tt hep-lat/0603020}}.

\bibitem{Bernard:2006vv}
C.~Bernard, M.~Golterman, Y.~Shamir, and S.~R. Sharpe, {\it Comment on 'flavor
  extrapolations and staggered fermions'},
  \href{http://xxx.lanl.gov/abs/hep-lat/0603027}{{\tt hep-lat/0603027}}.

\bibitem{Durr:2006ze}
S.~Durr and C.~Hoelbling, {\it Lattice fermions with complex mass},
  \href{http://xxx.lanl.gov/abs/hep-lat/0604005}{{\tt hep-lat/0604005}}.

\bibitem{Hasenfratz:2006nw}
A.~Hasenfratz and R.~Hoffmann, {\it Validity of the rooted staggered
  determinant in the continuum limit},
  \href{http://xxx.lanl.gov/abs/hep-lat/0604010}{{\tt hep-lat/0604010}}.

\bibitem{Bernard:2006ee}
C.~Bernard, M.~Golterman, and Y.~Shamir, {\it Observations on staggered
  fermions at non-zero lattice spacing},
  \href{http://xxx.lanl.gov/abs/hep-lat/0604017}{{\tt hep-lat/0604017}}.

\bibitem{Lee:1999zx}
W.-J. Lee and S.~R. Sharpe, {\it Partial flavor symmetry restoration for chiral
  staggered fermions},  {\em Phys. Rev.} {\bf D60} (1999) 114503,
  [\href{http://xxx.lanl.gov/abs/hep-lat/9905023}{{\tt hep-lat/9905023}}].

\bibitem{Aubin:2003mg}
C.~Aubin and C.~Bernard, {\it Pion and kaon masses in staggered chiral
  perturbation theory},  {\em Phys. Rev.} {\bf D68} (2003) 034014,
  [\href{http://xxx.lanl.gov/abs/hep-lat/0304014}{{\tt hep-lat/0304014}}].

\bibitem{Aubin:2003uc}
C.~Aubin and C.~Bernard, {\it Pseudoscalar decay constants in staggered chiral
  perturbation theory},  {\em Phys. Rev.} {\bf D68} (2003) 074011,
  [\href{http://xxx.lanl.gov/abs/hep-lat/0306026}{{\tt hep-lat/0306026}}].

\bibitem{Sharpe:2004is}
S.~R. Sharpe and R.~S. Van~de Water, {\it Staggered chiral perturbation theory
  at next-to-leading order},  {\em Phys. Rev.} {\bf D71} (2005) 114505,
  [\href{http://xxx.lanl.gov/abs/hep-lat/0409018}{{\tt hep-lat/0409018}}].

\bibitem{VandeWater:2005uq}
R.~S. Van~de Water and S.~R. Sharpe, {\it B(k) in staggered chiral perturbation
  theory},  {\em Phys. Rev.} {\bf D73} (2006) 014003,
  [\href{http://xxx.lanl.gov/abs/hep-lat/0507012}{{\tt hep-lat/0507012}}].

\bibitem{Golterman:1984cy}
M.~F.~L. Golterman and J.~Smit, {\it Selfenergy and flavor interpretation of
  staggered fermions},  {\em Nucl. Phys.} {\bf B245} (1984) 61.

\bibitem{Golterman:1985dz}
M.~F.~L. Golterman, {\it Staggered mesons},  {\em Nucl. Phys.} {\bf B273}
  (1986) 663.

\bibitem{Golterman:1984dn}
M.~F.~L. Golterman and J.~Smit, {\it Lattice baryons with staggered fermions},
  {\em Nucl. Phys.} {\bf B255} (1985) 328.

\bibitem{Golterman:1986jf}
M.~F.~L. Golterman, {\it Irreducible representations of the staggered fermion
  symmetry group},  {\em Nucl. Phys.} {\bf B278} (1986) 417.

\bibitem{Bailey:2005ss}
J.~A. Bailey and C.~Bernard, {\it Staggered lattice artifacts in 3-flavor heavy
  baryon chiral perturbation theory},  {\em PoS} {\bf LAT2005} (2005) 047,
  [\href{http://xxx.lanl.gov/abs/hep-lat/0510006}{{\tt hep-lat/0510006}}].

\bibitem{Luscher:1998pq}
M.~Luscher, {\it Exact chiral symmetry on the lattice and the {Ginsparg-Wilson}
  relation},  {\em Phys. Lett.} {\bf B428} (1998) 342--345,
  [\href{http://xxx.lanl.gov/abs/hep-lat/9802011}{{\tt hep-lat/9802011}}].

\bibitem{Kaplan:1992bt}
D.~B. Kaplan, {\it A method for simulating chiral fermions on the lattice},
  {\em Phys. Lett.} {\bf B288} (1992) 342--347,
  [\href{http://xxx.lanl.gov/abs/hep-lat/9206013}{{\tt hep-lat/9206013}}].

\bibitem{Shamir:1993zy}
Y.~Shamir, {\it Chiral fermions from lattice boundaries},  {\em Nucl. Phys.}
  {\bf B406} (1993) 90--106,
  [\href{http://xxx.lanl.gov/abs/hep-lat/9303005}{{\tt hep-lat/9303005}}].

\bibitem{Furman:1994ky}
V.~Furman and Y.~Shamir, {\it Axial symmetries in lattice {QCD} with {Kaplan}
  fermions},  {\em Nucl. Phys.} {\bf B439} (1995) 54--78,
  [\href{http://xxx.lanl.gov/abs/hep-lat/9405004}{{\tt hep-lat/9405004}}].

\bibitem{Narayanan:1993sk}
R.~Narayanan and H.~Neuberger, {\it Chiral determinant as an overlap of two
  vacua},  {\em Nucl. Phys.} {\bf B412} (1994) 574--606,
  [\href{http://xxx.lanl.gov/abs/hep-lat/9307006}{{\tt hep-lat/9307006}}].

\bibitem{Narayanan:1993ss}
R.~Narayanan and H.~Neuberger, {\it Chiral fermions on the lattice},  {\em
  Phys. Rev. Lett.} {\bf 71} (1993) 3251--3254,
  [\href{http://xxx.lanl.gov/abs/hep-lat/9308011}{{\tt hep-lat/9308011}}].

\bibitem{Narayanan:1994gw}
R.~Narayanan and H.~Neuberger, {\it A construction of lattice chiral gauge
  theories},  {\em Nucl. Phys.} {\bf B443} (1995) 305--385,
  [\href{http://xxx.lanl.gov/abs/hep-th/9411108}{{\tt hep-th/9411108}}].

\bibitem{Renner:2004ck}
{\bf LHP} Collaboration, D.~B. Renner {\em et.~al.}, {\it Hadronic physics with
  domain-wall valence and improved staggered sea quarks},  {\em Nucl. Phys.
  Proc. Suppl.} {\bf 140} (2005) 255--260,
  [\href{http://xxx.lanl.gov/abs/hep-lat/0409130}{{\tt hep-lat/0409130}}].

\bibitem{Bowler:2004hs}
{\bf UKQCD} Collaboration, K.~C. Bowler, B.~Joo, R.~D. Kenway, C.~M. Maynard,
  and R.~J. Tweedie, {\it Lattice {QCD} with mixed actions},  {\em JHEP} {\bf
  08} (2005) 003, [\href{http://xxx.lanl.gov/abs/hep-lat/0411005}{{\tt
  hep-lat/0411005}}].

\bibitem{Bar:2002nr}
O.~Bar, G.~Rupak, and N.~Shoresh, {\it Simulations with different lattice
  {Dirac} operators for valence and sea quarks},  {\em Phys. Rev.} {\bf D67}
  (2003) 114505, [\href{http://xxx.lanl.gov/abs/hep-lat/0210050}{{\tt
  hep-lat/0210050}}].

\bibitem{Bijnens:1995yn}
J.~Bijnens, G.~Colangelo, G.~Ecker, J.~Gasser, and M.~E. Sainio, {\it Elastic
  $\pi\pi$ scattering to two loops},  {\em Phys. Lett.} {\bf B374} (1996)
  210--216, [\href{http://xxx.lanl.gov/abs/hep-ph/9511397}{{\tt
  hep-ph/9511397}}].

\bibitem{Bijnens:1997vq}
J.~Bijnens, G.~Colangelo, G.~Ecker, J.~Gasser, and M.~E. Sainio, {\it $\pi\pi$
  scattering at low energy},  {\em Nucl. Phys.} {\bf B508} (1997) 263--310,
  [\href{http://xxx.lanl.gov/abs/hep-ph/9707291}{{\tt hep-ph/9707291}}].

\bibitem{Bijnens:2004eu}
J.~Bijnens, P.~Dhonte, and P.~Talavera, {\it $\pi\pi$ scattering in three
  flavour $\chi$pt},  {\em JHEP} {\bf 01} (2004) 050,
  [\href{http://xxx.lanl.gov/abs/hep-ph/0401039}{{\tt hep-ph/0401039}}].

\bibitem{Bernard:1993ga}
C.~W. Bernard and M.~Golterman, {\it Partially quenched {QCD} and staggered
  fermions},  {\em Nucl. Phys. Proc. Suppl.} {\bf 34} (1994) 331--333,
  [\href{http://xxx.lanl.gov/abs/hep-lat/9311070}{{\tt hep-lat/9311070}}].

\bibitem{Bernard:1995ez}
C.~W. Bernard and M.~F.~L. Golterman, {\it Finite volume two pion energies and
  scattering in the quenched approximation},  {\em Phys. Rev.} {\bf D53} (1996)
  476--484, [\href{http://xxx.lanl.gov/abs/hep-lat/9507004}{{\tt
  hep-lat/9507004}}].

\bibitem{Colangelo:1997ch}
G.~Colangelo and E.~Pallante, {\it Quenched chiral perturbation theory to one
  loop},  {\em Nucl. Phys.} {\bf B520} (1998) 433--468,
  [\href{http://xxx.lanl.gov/abs/hep-lat/9708005}{{\tt hep-lat/9708005}}].

\bibitem{Lin:2002aj}
C.~J.~D. Lin, G.~Martinelli, E.~Pallante, C.~T. Sachrajda, and G.~Villadoro,
  {\it Finite-volume two-pion amplitudes in the {$I = 0$} channel},  {\em Phys.
  Lett.} {\bf B553} (2003) 229--241,
  [\href{http://xxx.lanl.gov/abs/hep-lat/0211043}{{\tt hep-lat/0211043}}].

\bibitem{Lin:2003tn}
C.~J.~D. Lin, G.~Martinelli, E.~Pallante, C.~T. Sachrajda, and G.~Villadoro,
  {\it Finite-volume partially-quenched two-pion amplitudes in the {$I = 0$}
  channel},  {\em Phys. Lett.} {\bf B581} (2004) 207--217,
  [\href{http://xxx.lanl.gov/abs/hep-lat/0308014}{{\tt hep-lat/0308014}}].

\bibitem{Golterman:2005xa}
M.~Golterman, T.~Izubuchi, and Y.~Shamir, {\it The role of the double pole in
  lattice {QCD} with mixed actions},  {\em Phys. Rev.} {\bf D71} (2005) 114508,
  [\href{http://xxx.lanl.gov/abs/hep-lat/0504013}{{\tt hep-lat/0504013}}].

\bibitem{Lin:2002nq}
C.~J.~D. Lin, G.~Martinelli, E.~Pallante, C.~T. Sachrajda, and G.~Villadoro,
  {\it {$K \rightarrow \pi^+\pi^0$} decays on finite volumes and at next-to-
  leading order in the chiral expansion},  {\em Nucl. Phys.} {\bf B650} (2003)
  301--355, [\href{http://xxx.lanl.gov/abs/hep-lat/0208007}{{\tt
  hep-lat/0208007}}].

\bibitem{Chen:noasqd}
J.-W. Chen, D.~O'Connell, and A.~Walker-Loud \textit{To be published.}

\bibitem{Chen:2002bz}
J.-W. Chen and M.~J. Savage, {\it The chiral extrapolation of strange matrix
  elements in the nucleon},  {\em Phys. Rev.} {\bf D66} (2002) 074509,
  [\href{http://xxx.lanl.gov/abs/hep-lat/0207022}{{\tt hep-lat/0207022}}].

\bibitem{Aoki:2005mb}
S.~Aoki, O.~Bar, S.~Takeda, and T.~Ishikawa, {\it Pseudo scalar meson masses in
  {Wilson} chiral perturbation theory for 2+1 flavors},  {\em Phys. Rev.} {\bf
  D73} (2006) 014511, [\href{http://xxx.lanl.gov/abs/hep-lat/0509049}{{\tt
  hep-lat/0509049}}].

\bibitem{Tiburzi:noasqd}
B.~C. Tiburzi and A.~Walker-Loud \textit{To be published.}

\bibitem{Chen:2002mj}
J.-W. Chen, {\it Connecting the quenched and unquenched worlds via the large
  {N(c)} world},  {\em Phys. Lett.} {\bf B543} (2002) 183--188,
  [\href{http://xxx.lanl.gov/abs/hep-lat/0205014}{{\tt hep-lat/0205014}}].

\bibitem{Shoresh:2001ha}
N.~Shoresh, {\em Applications of chiral perturbation theory}.
\newblock PhD thesis, University of Washington, Seattle, 2001.
\newblock UMI-30-36529.

\bibitem{Arndt:2004as}
D.~Arndt, {\em Chiral perturbation theory on the lattice and its applications}.
\newblock PhD thesis, University of Washington, Seattle, 2004.
\newblock \href{http://xxx.lanl.gov/abs/hep-lat/0406011}{{\tt
  hep-lat/0406011}}.

\bibitem{VandeWater:2005ns}
R.~S. Van~de Water, {\em Applications of chiral perturbation theory to lattice
  QCD}.
\newblock PhD thesis, University of Washington, Seattle, 2005.
\newblock UMI-31-83434.

\end{thebibliography}\endgroup
%
% ==========   Appendices
%
\appendix
\raggedbottom\sloppy
 
% ========== Appendix A
\chapter{\label{app:tm} Twisted Mass Appendices}

\section{\label{sec:appD5SB} Absence of additional dimension five
  symmetry breaking operators induced by the mass splitting}

In this appendix, we show that the mass splitting does not induce any
symmetry breaking terms in the effective continuum Lagrangian at the
quark level at quadratic order. The (mass) dimension six operators in
the Symanzik Lagrangian we can drop for the same reason given in
Ref.~\cite{Sharpe:2004ps}, since they are either of too high order (cubic or
higher in our expansion) or they do not break the symmetries further
than those of lower dimensions. For dimension five operators, we will
show that the only allowable terms by the symmetries of the lattice
theory are those that either vanish by the equations of motion, or can
be removed by suitable $\mathcal{O}(a)$ redefinitions of the parameters 
in $\mathcal{L}_0$, the effective Lagrangian in the continuum limit
(the lowest order effective Lagrangian). 

In the mass-degenerate case~\cite{Sharpe:2004ps}, the only dimension five
operator that appears is the Pauli term. Since in the limit of
vanishing mass splitting (the isospin limit) the mass non-degenerate
theory must be the same as the mass-degenerate theory, any additional
operators induced by the mass splitting must be proportional to the mass
splitting. These can only be of the form
\begin{align} \label{E:dim5op}
&\epsilon_q^2 \bar{\psi} \mathcal{O}_0 \psi: \quad
\mathcal{O}_0 = \Gamma_0\backslash\{\mathbb{1}\}\,, \quad
\Gamma_0 = \{\mathbb{1} \,,\, \tau_k \,,\, \gamma_5 \,,\,\gamma_5\tau_k\}
\,, \quad k = 1,\,2,\,3,  
&{\rm dim}\,[\mathcal{O}_0] = 0 \,, \notag \\ 
&\epsilon_q \bar{\psi} \mathcal{O}_1 \psi: \quad
\mathcal{O}_1 =
\{D\!\!\!\!/\,\Gamma_0\,,\,m\,\Gamma_0\,,\,\mu\,\Gamma_0\,\} 
\backslash\{D\!\!\!\!/\;\tau_3\,,\,m\tau_3\}\,, 
&{\rm dim}\,[\mathcal{O}_1] = 1 \,,
\end{align}
where the notation ``$P \backslash Q$'' means ``the set P excluding
the set Q''. The quantities $\mathcal{O}_0$ and $\mathcal{O}_1$ are
all the possible independent structures with the correct dimension,
which do not lead to dimension five operators vanishing by the
equations of motion, or are not removable by redefinitions of parameters
in $\mathcal{L}_0$. However, none of these operators are allowed under
the symmetries of the lattice theory. Specifically, they are forbidden
by charge conjugation ($\mathcal{C}$) and the pseudo-parity
transformations that combine the ordinary parity transformation
($\mathcal{P}$) with a parameter sign change
\begin{equation}
\widetilde{\mathcal{P}} \equiv 
\mathcal{P} \times (\mu \rightarrow -\mu) \,,
\end{equation}
or a flavor exchange or both
\begin{equation} \label{E:pseudoP}
\mathcal{P}^2_{F,\,\epsilon_q} \equiv
\mathcal{P}^2_F \times (\epsilon_q \rightarrow -\epsilon_q) \,,\quad 
\mathcal{P}^3_F \,,
\end{equation}
where
\begin{equation} \label{E:PF23}
\mathcal{P}^{2,3}_F \colon
\begin{cases}
U_0(x) \rightarrow U_0(x_P) \,, \quad x_P = (-\mathbf{x},t) \\
U_k(x) \rightarrow U^\dagger_k(x_P) \,, \quad k = 1,\,2,\,3 \\  
\psi(x) \rightarrow i\tau_{2,3} \gamma_0 \psi(x_P) \\
\bar{\psi}(x) \rightarrow -i\bar{\psi}(x_P) \gamma_0 \tau_{2,3}
\end{cases} \,,
\end{equation}
and $U_\mu$ are the lattice link fields.  Note that we have displayed the symmetries of the lattice 
theory~\cite{Frezzotti:2003ni,Frezzotti:2003xj} in the form which applies to the effective
continuum theory.  

In Table~1, we show explicitly which symmetry forbids each of the
possible structures of $\mathcal{O}_0$ and $\mathcal{O}_1$ listed
in~(\ref{E:dim5op}).\footnote{Most of what we show can be readily
  inferred from~\cite{Frezzotti:2003xj}. What is new here is the need for
  $\widetilde{\mathcal{P}}$, and the use of
  $\mathcal{P}^{2}_{F,\,\epsilon_q}$.} We group the operators in
columns according to the symmetry under which they are forbidden. 
\begin{table}[tbp]
\center
\begin{tabular}{|c|c|c|c|c|}
\hline \hline
Structure & 
\hspace{1.5cm}$\mathcal{C}$\hspace{1.5cm} & 
\hspace{1.5cm}$\mathcal{P}^3_F$\hspace{1.5cm} & 
\hspace{0.8cm}$\widetilde{\mathcal{P}}$\hspace{0.8cm} &
\hspace{0.5cm}$\mathcal{P}^2_{F,\,\epsilon_q}$\hspace{0.5cm}
\\ \hline
$\mc{O}_0$ & $\tau_2$, $\gamma_5\tau_2$ & 
$\tau_1$, $\gamma_5$, $\gamma_5\tau_3$ & $\gamma_5\tau_1$ & $\tau_3$
\\ \hline
$\mc{O}_1$ & $D\!\!\!\!/\;\gamma_5\times\{\mathbb{1},\,\tau_1,\,\tau_3\}$ &
$D\!\!\!\!/\;\tau_1$, $D\!\!\!\!/\;\tau_2$ &
$D\!\!\!\!/\;\gamma_5\tau_2$ & \\ 
& $m\tau_2$, $m\gamma_5\tau_2$ & 
$m\times\{\tau_1,\,\gamma_5,\,\gamma_5\tau_3\}$ & $m\gamma_5\tau_1$ &\\
& $i\mu\tau_2$, $i\mu\gamma_5\tau_2$ &
$i\mu\times\{\tau_1,\,\gamma_5,\,\gamma_5\tau_3\}$ & $i\mu\tau_3$ &
$i\mu\gamma_5\tau_1$ 
\\ \hline \hline 
\end{tabular}
\caption[Dimension five operators contributing the effective continuum theory of twisted mass LQCD]{\label{T:OpStr} The structures of the dimension five operators that are
  non-vanishing by the equations of motion and non-removable by 
  parameter redefinitions. They are classified by the symmetries that
  forbid them.}  
\end{table}

The conclusion of the above discussion is that the mass splitting does
not induce any additional operators that do not vanish by the
equations of motion, or can not be removed by redefinitions of the
parameters in the theory. Thus beyond $\mathcal{L}_0$, the effective
continuum Lagrangian contains only the Pauli term to the order we
work, exactly as in the mass-degenerate case.

%
%
%
%
%
%
%	Diagonalization of Deltas
%
%
%
%
%
%
\section{\label{sec:appDDM} Diagonalization of the delta mass matrix}

Here we diagonalize the tree level mass matrix for the delta states.
We reiterate that the difference between first diagonalizing the tree
level mass contributions, then calculating loop effects, versus
calculating the loop contributions then diagonalizing, is of higher
order than we work. To proceed, first we list all the independent
operators to $\mc{O}(\varepsilon^4)$ that have tree level mass
contributions,  
\begin{align} \label{E:Mtree}
\mc{O}(\varepsilon^2): \quad &
(\ol{T}_\mu \mc{M}_+ T_\mu) \,, \quad 
(\ol{T}_\mu T_\mu)\,{\rm tr}(\mc{M}_+) \,, \quad 
(\ol{T}_\mu T_\mu)\,{\rm tr}(\mc{W}^{tw}_+) \notag \\
\mc{O}(a\mathsf{m}): \quad &
(\ol{T}_\mu \mc{M}_+ T_\mu)\,{\rm tr}(\mc{W}^{tw}_+) \,, \quad  
(\ol{T}_\mu T_\mu)\, {\rm tr}(\mc{W}^{tw}_+)\,{\rm tr}(\mc{M}_+)  \notag \\
\mc{O}(a^2): \quad &
(\ol{T}_\mu T_\mu)\,
{\rm tr}(\mc{W}^{tw}_+)\,{\rm tr}(\mc{W}^{tw}_+) \,, \quad
(\ol{T}_\mu T_\mu)\,
{\rm tr}(\mc{W}^{tw}_- \mc{W}^{tw}_-) \,, \quad
\ol{T}^{kji}_\mu(\mc{W}^{tw}_-)^{ii'}(\mc{W}^{tw}_-)^{jj'} T^{i'j'k}_\mu
\end{align} 
The tree level delta mass matrix at the order we work, $M_\Delta$, is
then given by 
\begin{align}
v_{\bar{\Delta}} M_\Delta v_\Delta &=
v_{\bar{\Delta}} 
\begin{pmatrix}
-A+C                & -\frac{B}{\sqrt{3}} & 0 & 0 \\
-\frac{B}{\sqrt{3}} & \frac{1}{3}(A-2B)+C & 0 & 0 \\
0 & 0 & -\frac{1}{3}(A + 2B)+C  & -\frac{B}{\sqrt{3}} \\
0 & 0 & -\frac{B}{\sqrt{3}}     & A+C
\end{pmatrix}
v_\Delta \,, \notag \\
&= v_{\bar{\Delta}}\bigg\{C\,\mathbb{1}_{4\times4} + K_\Delta\bigg\}
   v_\Delta \,,
\end{align}
where the vectors $v_{\bar{\Delta}}$ and $v_\Delta$ are vectors of the
(QCD) delta basis states, 
\begin{equation}\label{E:QCDeigenV}
v_{\bar{\Delta}} = 
\begin{pmatrix}
\bar{\D}^{++} & \bar{\D}^0 & \bar{\D}^+ & \bar{\D}^-
\end{pmatrix} \,, \qquad\qquad
v_\D = 
\begin{pmatrix}
\D^{++} & \D^0 & \D^+ & \D^-
\end{pmatrix}^T \,, 
\end{equation}
and
\begin{equation}
K_\Delta = 
\begin{pmatrix}
-A & -\frac{B}{\sqrt{3}} & 0 & 0 \\
-\frac{B}{\sqrt{3}} & \frac{1}{3}(A-2B) & 0 & 0 \\
0 & 0 & -\frac{1}{3}(A + 2B) & -\frac{B}{\sqrt{3}} \\
0 & 0 &  -\frac{B}{\sqrt{3}} & A
\end{pmatrix} \,.
\end{equation}
The entries in $M_\D$ are given by
\begin{align}
A &= 2\,\e_q\, \left( \g_M + t_1^{WM_+} 
      \frac{a \L_{QCD}^2}{\L_\chi}\cos(\w) \right) \,,
\qquad\qquad
B = t_2^{W_-}a^2\L_{QCD}^3\sin^2(\w) \,, 
\notag\\
C &= 2\,m_q\left(\g_M - 2\,\ol{\s}_M\right)
    -4\,\ol{\s}_W a\,\L_{QCD}^2\cos(\w)
    +2\,m_q\left(t_1^{WM_+} + 2\,t_2^{WM_+}\right)
     \frac{a \L_{\rm QCD}^2}{\L_\chi}\cos(\w) 
\notag\\
&\quad
+a^2\L_{QCD}^3\,(t +t_v) 
+a^2\L_{QCD}^3\left( 4\,t_1^{W_+}\cos^2(\w)
                    -2\,t_1^{W_-}\sin^2(\w)\right) \,. 
\end{align}
Note that to the accuracy we work, $\w$ can be either $\w_0$ or the
non-perturbatively determined twist angle. 

Except for the operators
\begin{equation*}
(\ol{T}_\mu \mc{M}_+ T_\mu) \,,\qquad 
(\ol{T}_\mu\mc{M}_+ T_\mu)\,{\rm tr}(\mc{W}^{tw}_+) \,,\qquad
\ol{T}^{kji}_\mu(\mc{W}^{tw}_-)^{ii'}(\mc{W}^{tw}_-)^{jj'}T^{i'j'k}_\mu
\,, 
\end{equation*}
which contribute to $K_\Delta$, all other operators listed in
(\ref{E:Mtree}) above have trivial flavor structure, and so contribute
to the identity part of $M_\Delta$. Hence, to diagonalize $M_\Delta$,
we need only diagonalize $K_\Delta$. The orthogonal matrix that
accomplishes this is
\begingroup
\small
\begin{equation} \label{E:Smtx}
S = 
\begin{pmatrix}
\frac{(2A-B+2X_-)^{1/2}}{2X_-^{1/2}} &
-\frac{(-2A+B+2X_-)^{1/2}}{2X_-^{1/2}} & 0 & 0 \\
\frac{\sqrt{3}B}{2X_-^{1/2}(2A-B+2X_-)^{1/2}} &
\frac{\sqrt{3}B}{2X_-^{1/2}(-2A+B+2X_-)^{1/2}} & 0 & 0 \\
0 & 0 & \frac{(2A+B+2X_+)^{1/2}}{2X_+^{1/2}} & 
       -\frac{(-2A-B+2X_+)^{1/2}}{2X_+^{1/2}} \\
0 & 0 & \frac{\sqrt{3}B}{2X_+^{1/2}(2A+B+2X_+)^{-1/2}} &
        \frac{\sqrt{3}B}{2X_+^{1/2}(-2A-B+2X_+)^{-1/2}} 
\end{pmatrix}
\end{equation}
\endgroup
where $X_\pm = \sqrt{A^2 \pm AB + B^2}$, and each column of $S$ is a 
normalized eigenvector of $M_\Delta$ (and hence $K_\Delta$ also). The
diagonal matrix one obtains after diagonalizing $M_\Delta$ is then
\begin{equation} \label{E:Dmtx}
\mc{D} = S^{-1} \cdot M_\D \cdot S = 
\frac{1}{3}\,{\rm diag}
\begin{pmatrix}
-A - B - 2X_- + 3C \\ 
-A - B + 2X_- + 3C \\ 
\!\quad A - B - 2X_+ + 3C \\ 
\!\quad A - B + 2X_+ + 3C 
\end{pmatrix} \,,
\end{equation}
where each entry in $\mc{D}$ is an eigenvalue of $M_\Delta$.

Now if $\epsilon_q \neq 0$, $A \neq 0$. Hence, since in our power
counting $A \sim \mathcal{O}(\varepsilon^2)$ and 
$B \sim \mathcal{O}(\varepsilon^4)$, we may expand $X_\pm$ in the
ratio of $B/A \sim \mathcal{O}(\varepsilon^2) \ll 1$ as
\begin{equation}
X_\pm = A\sqrt{1 \pm \frac{B}{A} + \frac{B^2}{A^2}} 
      = A\left(1 \pm \frac{1}{2}\frac{B}{A} + \frac{3}{8}\frac{B^2}{A^2}
        +\mc{O}(\varepsilon^6)\right) \,,
\end{equation}
from which it follows that
\begin{equation} \label{E:Spert}
S = 
\begin{pmatrix}
1 & -\frac{\sqrt{3}B}{4A} & 0 & 0 \\
\frac{\sqrt{3}B}{4A} & 1+\frac{B}{4A} & 0 & 0 \\
0 & 0 & 1 & -\frac{\sqrt{3}B}{4A} \\
0 & 0 & \frac{\sqrt{3}B}{4A} & 1-\frac{B}{4A}
\end{pmatrix} \,, \qquad\qquad
\mathcal{D} = {\rm diag}
\begin{pmatrix}
-A + C \\
\!\quad\frac{A}{3} - \frac{2B}{3} + C \\
-\frac{A}{3} - \frac{2B}{3} + C \\
\!\quad A + C \\
\end{pmatrix} \,, 
\end{equation}
up to corrections of $\mathcal{O}(\varepsilon^4)$ for $S$ and
$\mathcal{O}(\varepsilon^6)$ for $\mc{D}$.  

If $\e_q = 0$, i.e. in the isospin limit, $A = 0$ and 
$X_\pm = B$. In this case, one can not find $S$ and $\mc{D}$ in the
isospin limit by taking the limit $A \rightarrow 0$ in~\eqref{E:Spert},
since expansion in the ratio of $B/A$ is clearly not valid. Instead,
one has to go back to Eq.~\eqref{E:Smtx} and Eq.~\eqref{E:Dmtx}, which
in the isospin limit reduce to   
\begin{equation} \label{E:zeroA}
S = 
\begin{pmatrix}
\frac{1}{2} & -\frac{\sqrt{3}}{2} & 0 & 0 \\
\frac{\sqrt{3}}{2} & \frac{1}{2} & 0 & 0 \\
0 & 0 & \frac{\sqrt{3}}{2} & -\frac{1}{2} \\
0 & 0 & \frac{1}{2} & \frac{\sqrt{3}}{2} 
\end{pmatrix} \,, \qquad\qquad
\mathcal{D} = {\rm diag}
\begin{pmatrix}
-B + C \\ 
\;\;\,\frac{B}{3} + C \\ 
-B + C \\ 
\;\;\,\frac{B}{3} + C
\end{pmatrix} \,,
\end{equation}
and the eigenvalues contained in $\mc{D}$ given in Eq.~\eqref{E:zeroA}
are the masses of the deltas at tree level in the isospin limit given in
Eq.~\eqref{eq:DLO} and Eq.~\eqref{eq:DNNLOtau1}. Note that as
discussed in the text, in the isospin limit, we need not perform any
mass matrix diagonalization at all, since we can simply rotate from
the outset to the basis where the twist is implemented by the diagonal
$\tau_3$.  

The new delta basis states are defined by 
\begin{equation}
v'_\Delta = S^{-1} \cdot v_\Delta \,,\qquad 
v'_{\bar{\Delta}} = v_{\bar{\Delta}} \cdot S \,,
\end{equation} 
in which the delta mass matrix is diagonal to the order we work.
%($\transpose v_{\bar{\Delta}} M_\Delta v_\Delta = 
% (\transpose v'_{\bar{\Delta}}\cdot S)\,M_\Delta\,(S^{-1} v'_\Delta) = 
%  \transpose v'_{\bar{\Delta}}\,D_\Delta\,v'_\Delta$, where $D_\Delta$
%  is the diagonal eigenvalue matrix of $M_\Delta$).
By writing the old (unprimed) delta basis states in terms of the new
(primed) basis states using the defining relations given above, i.e.
\begin{equation}
v_\Delta = S \cdot v'_\Delta \,,\qquad 
v_{\bar{\Delta}} = v'_{\bar{\Delta}} \cdot S^{-1} \,,
\end{equation} 
the Lagrangian in the new delta basis can be obtained. Note that in
the case where $\epsilon_q \neq 0$, the new basis states contained in $v'_\Delta$ are ``perturbatively close'' to those contained in the $v_\Delta$, i.e. the
difference is $\mc{O}(\varepsilon^2)$ as can be easily seen from
Eq.~\eqref{E:Spert}. This is of course not true if we are in a region
where $B \sim A$, or $\epsilon_q \sim a^2\L_{\rm QCD}^3$.
 
\chapter{\label{app:EM} Compton Scattering Appendices}

%
%
%
%
%
%
%       Quenched
%
%
%
%
%
%
\section{Quenched chiral Lagrangian}
\label{A4}

In this Appendix, we display the relevant pieces of the quenched
chiral Lagrangian in the meson and baryon sectors and note particular
pathologies of the quenched theory.  In a quenched two flavor theory,
we have valence ($u$, $d$) and ghost ($\tilde u,\,\tilde d$) quarks
with masses contained in the matrix
\begin{equation}
\label{eq:Mq_def2}
\ol{m}_Q = {\rm  diag}(m_u,m_d,m_{\tilde u},m_{\tilde d})\,,
\end{equation}
where $m_{\tilde u,\tilde d}=m_{u,d}$ to maintain the exact
cancellation from the path-integral determinants arising from the
valence and ghost quark sectors.  The corresponding low-energy meson
dynamics are described by the \qxpt\ Lagrangian. At leading order, the
form of this Lagrangian is the same as in Eq.~\eqref{eq:PQChPT}
where the pseudo-Goldstone mesons are embedded non-linearly in
$\Sigma$ with the matrix $\Phi$ now given by
\begin{equation}
  \label{eq:Phi_def2}
\Phi = \begin{pmatrix} M & \chi^\dagger \cr \chi &\tilde{M} \end{pmatrix}
\,,
\end{equation}
where
\begin{eqnarray}
  \label{eq:Mchi_def2}
  M=\begin{pmatrix} \eta_u    & \pi^+        \\
                    \pi^-     & \eta_d      
                    \end{pmatrix}\,, 
& 
\hspace*{15mm}
\tilde{M}=\begin{pmatrix} \tilde\eta_u    & \tilde\pi^+        \\
                          \tilde\pi^-     & \tilde\eta_d       
                          \end{pmatrix}\,, 
 \hspace*{15mm}
&
\chi=\begin{pmatrix} \phi_{\tilde{u}u} & \phi_{\tilde{u}d}       \\
                     \phi_{\tilde{d}u}  & \phi_{\tilde{d}d}
                     \end{pmatrix}\,.
\end{eqnarray}
The matrix $M$ contains the usual valence--valence mesons, while
mesons in $\tilde{M}$ are composed of ghost quarks and anti-quarks,
and finally those in $\chi$ of ghost--valence quark--anti-quark pairs.
Unlike the partially quenched theory, there is no strong U(1)$_A$
anomaly, and the flavor-singlet field, $\Phi_0=\str[\Phi]/\sqrt{2}$
(along with its couplings $m_0$ and $\alpha_\Phi$), must be retained
in the theory. For the electromagnetic and spin polarisabilities in
QQCD, no loop contributions from the singlet are needed to the order
we work as flavor-neutral mesons are not present in loop diagrams at
this order. Despite flavour neutral mesons being absent in loop
graphs, the anomalous tree-level term couples the quenched singlet to
the nucleon. Cancellations, however, lead to final results that are
independent of $m_0$ and $\alpha_\Phi$.

For the quenched electric charge matrix of the valence and ghost
quarks, we choose
\begin{eqnarray}
  \label{eq:88}
  \hat{\cal Q} &=& {\rm diag}(q_u,q_d,q_u,q_d)\,.
\end{eqnarray}
Notice the peculiarity that $\str \, \hat{\cal Q} = 0$ is unavoidable
in the quenched theory.  In the quenched theory, there are anomalous
decays of flavour neutral mesons into two photons. In terms of
SU(2$|$2) QQCD quark fields, contributions to the anomaly from the
valence and ghost sectors come weighted with squares of the quark
charges, and we are thus not restricted to only the flavor singlet
current (as is the case for the strong U(1)$_A$ anomaly).  The
relevant term of the anomalous quenched chiral Lagrangian is the same
as has been detailed above in Sec.~\ref{sec:Anomaly}.

In SU(2$|$2) \hbxpt, the nucleons (those composed of three valence
quarks) enter as part of a {\bf 20}-dimensional representation
described by a three index flavour-tensor, ${\cal B}$.  The quenched
$\Delta$-isobar is contained in the totally symmetric three index
flavour-tensor ${\cal T}^\mu$ transforming in the {\bf 12}-dimensional
representation of SU(2$|$2).  The leading-order Lagrangian describing
these baryons and their interactions with Goldstone mesons is
\begin{align}
\label{eq:free_lagrangian2}
{\cal L}_{B\;Q}^{(0)} = &\  
i\left(\ol{\cal B} v\cdot {\cal D} {\cal B}\right)
+ 2\alpha \left(\ol{\cal B} S^\mu {\cal B} {\cal A}_\mu\right)
+ 2\beta  \left(\ol{\cal B} S^\mu {\cal A}_\mu {\cal B} \right)
+ 2\gamma \left(\ol{\cal B} S^\mu {\cal B} \right) \, \str \mathcal{A}_\mu \nonumber\\
& - i \left(\ol{\cal T}^\mu v\cdot {\cal D} {\cal T}_\mu\right) 
+ \Delta\ \left(\ol{\cal T}^\mu {\cal T}_\mu\right)
+  2{\cal H} \left(\ol{\cal T}^\nu S^\mu {\cal A}_\mu {\cal T}_\nu \right) \nonumber\\
&+  2 \gamma' \left(\ol{\cal T}^\nu S^\mu {\cal T}_\nu \right) \, \str \mathcal{A}_\mu
+ \sqrt{\frac{3}{2}}{\cal C}  \left[
	\left( \ol{\cal T}^\nu {\cal A}_\nu {\cal B}\right) + 
	\left(\ol{\cal B} {\cal A}_\nu {\cal T}^\nu\right) \right].
\end{align}
In contrast to partially-quenched and unquenched chiral perturbation
theory, there are two additional axial couplings $\gamma$ and
$\gamma'$ due to the presence of the flavour-singlet field. One should
keep in mind that although we use the same notation for simplicity,
all of the coefficients in the quenched Lagrangian have distinct
numerical values from those of the partially-quenched Lagrangian. In
the large $N_c$ limit, the coefficients of the two theories are
related~\cite{Chen:2002mj}.

Again the photon is minimally coupled in the above Lagrangian with
fixed coefficients. At the next order in the expansion, the relevant
terms that appear are
\begin{eqnarray}
{\cal L}_{B\;Q}^{(1)}& = &
\frac{i\,e}{ 2 M_N} F_{\mu\nu}
\left[ 
\mu_\alpha \left( \ol{\cal B} \left[S^{\mu},S^{\nu}\right]  {\cal B}  
{\cal Q}_{\xi+} \right) +
\mu_\beta \left( \ol{\cal B} \left[S^{\mu},S^{\nu}\right]
{\cal Q}_{\xi+} 
{\cal B} \right)
 \right]
 \nonumber \\
&&
+ \sqrt{\frac{3}{2}} \mu_T \frac{i e}{ 2 M_N} F_{\mu\nu}
\left[
\left( \ol{\cal B} S^\mu {\cal Q}_{\xi+} {\cal T}^\nu  
 \right) +
\left( \ol{\cal T} {}^\mu S^\nu  {\cal Q}_{\xi+} {\cal B} \right) 
\right]
.\label{L12}
\end{eqnarray}
The \pqxpt\ term with coefficient $\mu_\gamma$ is absent in the
quenched theory.  This only affects the Born terms of the Compton
amplitude, which are essentially unknown because they depend on the
quenched magnetic moment.  Finally, the leading two-photon operators
that give completely local contributions to the Compton scattering
tensor appear in quenched chiral perturbation theory in essentially
the same form as \pqxpt.
However, there are fewer operators per spin structure compared to the
partially quenched case because of the super-tracelessness of the
electric charge matrix.  Our computation is unchanged since these
terms do not contribute at the order we work.

\section{Finite volume functions}
\label{FV_app}

The sums required in the evaluation of the polarisabilities at finite
volume are ($\vec{k}=\frac{2\pi\vec{n}}{L}$, with $\vec{n}$ a triplet
of integers)
\begin{eqnarray}
  \label{eq:7a}
  {\cal I}_{\beta}(M)&=&\frac{1}{L^3}\sum
  _{\vec{k}}\frac{1}{\left[|\vec{k}|^2+M^2\right]^\beta}\,,
\\
  {\cal J}_\beta(M)&=&{\cal I}_{\beta-1}(M)-M^2 {\cal I}_\beta(M)\,,
  \\
  {\cal K}_\beta(M) &=& {\cal I}_{\beta-2}(M)
  - 2M^2 {\cal  I}_{\beta-1}(M)+M^4{\cal I}_\beta(M)\,,
  \\
  {\cal L}_\beta(M) &=&{\cal I}_{\beta-3}(M)
  -3M^2 {\cal I}_{\beta-2}(M)+3M^4{\cal I}_{\beta-1}(M)
  -M^6{\cal I}_\beta(M) \,.
\end{eqnarray}
At infinite volume these can be simplified using,
\begin{eqnarray}
  \label{eq:10}
  {\cal I}_\beta(M,L\to\infty) =
  \frac{1}{(4\pi)^{\frac{3}{2}}} 
  \frac{\Gamma(\beta -\frac{3}{2})}{\Gamma(\beta)}
  \frac{1}{(M^2)^{\beta-\frac{3}{2}}}  \,,
\end{eqnarray}
for $\beta>3/2$.

In numerically evaluating these sums, it is useful to note that
\begin{multline}
	\sum_{\vec{n}}\frac{1}{(|\vec{n}|^2+x^2)^\beta} =
	\sum_{\vec{n}}\frac{E_{1-\beta}(|\vec{n}|^2+x^2)}{\Gamma(\beta)} \\
		+\frac{\pi^{\frac{3}{2}}}{\Gamma(\beta)} 
		\int_0^1dt\, t^{\beta-5/2}e^{-t\, x^2}\left[\sum_{\vec{n}\ne0}e^{-\frac{\pi^2 |\vec{n}|^2}{t}}+1\right]
\end{multline}
where $E_n(x)$ is the exponential integral function. This form is
valid for $\beta>\frac{3}{2}$, $x\in\mathbb{R}$ and the remaining sums
converge exponentially fast in $|\vec{n}|$.

% ========== VITA
\vita{Andr\'{e} Walker-Loud was born at his parents' home one rainy day in Seattle, Washington, a few too many years ago.  A while later, this would-be flower-child with a black belt in karate (a good thing given the area) effortlessly navigated his way through the overcrowded Seattle public schools where he graduated as 1 of 16 valedictorians (much to his amusement and his teachers amazement) from Garfield High School in 1995.  Andr\'{e} then moved the extraordinary distance of 5 miles to attend the University of Washington, fully funded by the Washington State taxpayers - your tax dollars hard at work.  The thrill of dorm life became all too distressing and he ultimately decided to move back home and just commute - and came to earn a \textit{Bachelor of Science} degree in physics in 2001.  Immediately afterwards he ventured about 150 feet to become a graduate student in physics at the University of Washington where he earned his \textit{Master of Science} degree in physics in 2002 and, thanks to his advisor and mentor, was able to recognize his one driving motivation: mainly the absolute and paralyzing fear of failure.  One girlfriend later, and unable to resist the University District's tantalizing restaurants, Andr\'{e} traveled those same 5 miles he previously found too ominous, to live in the heart of the University.  From there he was eventually able to work his way 3 floors up to complete his research, and therefore this thesis, which earned him a \textit{Ph.D.} in physics with the Nuclear Theory group in 2006 (and NO, you can't call him Dr. Dr\'{e} - the joke is too old).  In writing this thesis, Andr\'{e} learned that the use of a pion two-point correlation function is a popular example from the University of Washington to explain lattice QCD concepts~\cite{Shoresh:2001ha,Arndt:2004as,VandeWater:2005ns}, and that not one of his family/friends/future or past acquaintances will ever look at his work right-side up without first turning it in all directions and squinting a lot.  In parting, the aforementioned doctor would like to add that if you can actually read more in this 150+ page thesis than this paragraph and the Acknowledgement section: He hopes you were able to find the last 5 years of his life (in research) useful and/or interesting.}

\end{document}